%% file: preprintSecond.tex
\documentclass[12pt]{article}

\usepackage[dvips]{color}
\usepackage{amsmath} 
\usepackage{graphicx} 
\usepackage[dvips]{color}
\usepackage{graphicx}
\usepackage{tabularx}
\usepackage{subfig}
\usepackage[outercaption]{sidecap}

\newcommand{\bigTwelve}{{\ttfamily{'big12'}}}

\begin{document}
\title{Cell-type-specific transcriptomes and the Allen Atlas (II):
 discussion of the linear model of brain-wide densities of cell types}
\author{}
\author{Pascal Grange$^{1,\ast}$, Jason W. Bohland$^{2}$, Benjamin Okaty$^{3}$,\\
 Ken  Sugino$^{4}$, Hemant Bokil$^{5}$, Sacha Nelson$^{6}$,\\
 Lydia Ng$^{7}$, Michael Hawrylycz$^{7}$, Partha P. Mitra$^{5}$\\
{\normalsize{$^{1}$ Xi'An Jiaotong-Liverpool University,  Department of Mathematical Sciences,}}
{\normalsize{Suzhou 215021, China}}\\
{\normalsize{$^{2}$ Boston University, College of Health \& Rehabilitation Sciences,}}\\
 {\normalsize{ Boston, MA 02215, United States}}\\
{\normalsize{$^{3}$ Department of Genetics, Harvard Medical School, Boston, MA 02215, United States}}\\
{\normalsize{$^{4}$ Janelia Farm Research Campus, Howard Hughes Medical Institute,}}\\
 {\normalsize{ Ashburn, {\hbox{VA 20147}}, United States}}\\
{\normalsize{$^{5}$  Cold Spring Harbor Laboratory, Cold Spring Harbor, NY 11724, United States}}\\
{\normalsize{$^{6}$ Department of Biology and Center for Behavioral Genomics,}}\\
  {\normalsize{Brandeis University, Waltham, Massachusetts, United States}}\\
{\normalsize{$^{7}$ Allen Institute for Brain Science, Seattle, WA 98103, United States}}\\
\normalsize{$^\ast$E-mail: {\ttfamily{pascal.grange@polytechnique.org}}}}
\date{}

\maketitle
\begin{abstract}
The voxelized Allen Atlas of the adult mouse brain (at a resolution of 200 microns) 
 has been used  in [arXiv:1303.0013]
 to estimate the
region-specificity of 64 cell types whose transcriptional profile in
the mouse brain has been measured in microarray experiments. In particular, 
 the model yields estimates for the brain-wide density of each of these 
 cell types. We 
conduct numerical experiments to estimate the errors 
 in the estimated density profiles. First of all, we check that a simulated thalamic profile
 based on 200 well-chosen genes can transfer signal from cerebellar 
Purkinje cells to the thalamus. This inspires us to sub-sample the 
 atlas of genes by repeatedly drawing random sets of 200 genes and refitting the model.
 This results in a random distribution of density profiles, that can be compared 
 to the predictions of the model. 
 This results in a ranking of cell types
 by the overlap between the original and sub-sampled density profiles. Cell types with high rank include 
medium spiny neurons, several samples of cortical pyramidal neurons,
 hippocampal pyramidal neurons, granule cells and cholinergic neurons from the brain stem.
 In some cases with lower rank, the average sub-sample can have
 better contrast properties than the original model (this is the case for amygdalar neurons
 and dopaminergic neurons from the ventral midbrain). Finally, we 
 add some noise to the cell-type-specific transcriptomes by mixing them
  using a scalar parameter weighing a random matrix. After refitting the model, we observe than
 a mixing parameter of 5\% leads to modifications of density profiles that span
 the same interval as the ones resulting from sub-sampling.
\end{abstract}

\clearpage

\tableofcontents
\input{introductionNotations.tex}

\input{appendSimulationMissing.tex}
\clearpage

\input{appendErrorEstimates.tex}

\clearpage
\input{appendSubsampleFigures}
\clearpage
\input{appendSubsampleFiguresExtra.tex}
\clearpage
\section{Figures for sub-sampled profiles (II): random splitting of the genes}
\input{appendSplittingFigures.tex}

\clearpage
\section*{Acknowledgments}
We thank Nicolai Meinshausen and Gregor Havkin for correspondence.

\end{document}

%% file: introductionNotations.tex
\section{Introduction and notations}
The present note is a quantitative discussion of the 
 model presented in \cite{preprintFirstAnalysis}.
 Let us briefly review the data and computations. The Allen Brain Atlas
(ABA), the first Web-based, genome-wide atlas of gene expression in
the adult mouse brain (eight-week old C57BL/6J male mouse brain), was
obtained using an unified automated experimental pipeline
\cite{AllenGenome,images,BrainAtlasInsights,
  neufoAllen,digitalAtlasing,AllenFiveYears,corrStructureAllen}. The resulting data
set consists of {\emph{in situ}} hybridization (ISH) digitized image
series for thousands of genes.  These image series are co-registered
to the Allen Reference Atlas (ARA) \cite{AllenAtlas}. 
 In the digitized version of the ARA we use, the mouse brain is partitioned into $V = 49,742$ cubic voxels of
side 200 microns. For a voxel labeled $v$, the {\it{expression energy}} of
the gene labeled $g$ is defined \cite{AllenAtlasMol} as a weighted sum of the greyscale-value
intensities of pixels intersecting the voxel.
 We developed the Brain Gene Expression Analysis (BGEA)
 MATLAB toolbox \cite{toolboxManual,BGEA}, which
 allows to manipulate the gene-expression energies of the 
 brain-wide ABA on the desktop as matrices \cite{qbCoExpression,markerGenes}.\\
 
  A complementary (cell-based) approach to the study of gene-expression 
 energy in the brain uses microarray
 experiments to study co-expression patterns in a small set of brain
 cells of the same type. We studied cell-type-specific microarray gathered from
  different studies
  \cite{OkatyCells,RossnerCells,CahoyCells,DoyleCells,ChungCells,ArlottaCells,HeimanCells,foreBrainTaxonomy}, 
   analyzed comparatively in \cite{OkatyComparison}, for $T=64$ cell
  types.\\

We determined the set of genes that are represented in
\textit{both} data sets (there are $G=2,131$ such genes).  The
 ISH data of the Allen Atlas are arranged in a {\emph{voxel-by-gene}} matrix $E$,
 and the cell-type-specific microarray data are arranged in a  {\emph{type-by-gene}}
 matrix $C$. The columns of both matrices correspond to the same set of
genes, ordered in the same way:
\begin{equation}
E(v,g) = {\mathrm{expression\;of\;gene\;labeled\;}}g\;{\mathrm{in\;voxel\;labeled\;}}v,
\label{voxelByGene}
\end{equation}
\begin{equation}
C(t,g) = {\mathrm{expression\;of\;gene\;labeled\;}}g\;{\mathrm{in\;cell\;type\;labeled\;}}t.
\label{typeByGene}
\end{equation} 

We proposed a linear model 
 to attempt a decomposition of the signal at each voxel (taking all genes into account), 
over cell-type-specific samples\footnote{We will sometimes use a dot to denote an index
 that takes all the possible values in a range that is well-defined from the context.
 For example, the symbol $E(v,.)$ denotes the vector $\left(E(v,g)\right)_{1\leq g \leq G}$, an
 element of $\mathbf{R}^G$, 
and the symbol $\rho_.(v)$ denotes the vector $\left(\rho_t(v)\right)_{1\leq t \leq T}$,
 an element of $\mathbf{R}^T$.}:
\begin{equation}
\forall v \in [1..V],\;\;(\rho_t( v ))_{1\leq t \leq T}= {\mathrm{argmin}}_{\nu\in {\mathbf{R}}_+^T}\sum_{g=1}^G \left( E(v,g) - \sum_{t=1}^T\nu(t) C_t(g)\right)^2.
\label{voxelByVoxel}
\end{equation}
We solved these quadratic optimization problems with positivity constraints (one per voxel),
using the CVX toolbox for convex optimization \cite{cvxLink,cvxLecture}.\\

The results were presented graphically in the tables of \cite{preprintFirstAnalysis}
 for all cell types. It is clear from visual examination that some cell types are predicted to have density profiles with striking 
 anatomical features, while others are more amorphous. Moreover, there is 
 intrinsic heterogeneity between the two data sets (ISH and microarrays),
 missing cell types, and missing genes due to the fact that the coronal ABA
 does not cover the whole genome. We therefore have to estimate 
 how stable our estimates are against these sources of errors.\\

The present note is organized as follows. In section 2 we simulate 
 a missing cell type in a thalamus in order to study the errors induced
 by one missing cell type. It turns out that a deviation from 
 the average transcriptome profile for only 200 genes (roughy 10 \% of our data set) is enough 
 to induce the expected transfer of density from a class of Purkinje cells (labeled $t=52$)
 to the simulated cell type. In section 2 we repeatedly 
 draw random samples of 200 genes from the data set, refit the model, 
 thereby simulating the distribution
 of densities of cell types in sub-sampled models. We study the distribution of overlaps
 with the densities predicted in the original model, which gives rise to a ranking 
 of cell types. The most visually striking anatomical (and least sparse) patterns tend 
 to rank higher than the very sparse and amorphous patterns. The sub-sampling procedure also 
 gives rise to a simulation of localization scores of the density profiles in the brain regions 
 defined by the ARA. Cumulative distribution functions of the overlaps, distributions of 
 localization scores and heat maps of average density profiles
 from sub-samples are presented graphically in section 6. However, this approach does 
 not take into account the variation of stability properties from voxel to voxel for a given cell type.
   We perform random splitting of the data set into two 
 equal parts (section 3.2), and compute the probability at each voxel of 
 detecting each cell type from one sample, conditional on detecting it from 
 the complementary sub-sample. This gives rise to spatial profiles estimating 
 the stability properties of the predictions for each cell type. The results are presented 
 graphically in the form of heat maps in section 7 (where density profiles threshold 
 at 99\% and 75\% of conditional probability are plotted for each cell type.
 In section 4 we add noise to the cell-type-specific transcriptome profiles 
 by mixing them in a Gaussian way, weighted by a scalar noise parameter.
 Adjusting the value of the parameter allows us to approxinmately recover
 the distribution of overlaps obtained from sub-sampling, which gives rise to a self-consistent estimate
 of the level of noise giving rise to the same amount of instability
 as missing genes.\\

{\bf{Convention and notation.}} In this note we will work with 
 the type-by-gene matrix of cell-type-specific  transcriptomes obtained by
subtracting the smallest entry of $C$ from $C$, as this matrix has been 
 shown in section 4.3 \cite{preprintFirstAnalysis} to give gise to lower residual terms than the
 original microarray data (moreover, none of the original microarray data are zero, whereas
 all genes and voxels in the Allen Atlas contain some zeroes. For the sake of brevity
of notation, this matrix will be denoted by $C$ again.

%% file: appendSimulationMissing.tex
\section{Simulations with missing cell types}
\subsection{Simulation scenario}
 One of the surprising features of the results 
 comes from the Purkinje cells labeled $t=52$ (see section 5 for tables containing
 the correspondence between cell types and labels), that were 
 dissected from the cerebellum, but whose estimated profile 
 $\rho_{52}$ is mostly supported in the thalamus (the correlation
 between $C_{52}$ and thalamic voxels is also remarkably high).
 There are two other samples
 of Purkinje cells from the cerebellum in our data set. 
 We verified that refitting the model with only one Purkinje cell sample left (the one labeled $t=52$),
 leads to a transfer of the cerebellar signal of the other Purkinje cell samples to this cell type,
 while the large density spot in the thalamus is conserved. Although the
  high value of the correlation is independent from our linear model, we
 conjecture that the high estimated density of Purkinje cells ($\rho_{52}$) in the thalamus
 is due to a missing transcriptome from the thalamus: given the choice  
 between $C_{52}$ and a genuine thalamic transcriptome profile $C_{\mathrm{thalamus}}$, we expect that most 
 thalamic voxels would choose $C_{\mathrm{thalamus}}$ rather than $C_{52}$.\\

 We do not have such data available, but the model suggests
a simulation strategy: using ISH data for $G=2,131$ or roughly 10 percent of the genome
 allowed us to recover some striking anatomical density patterns.
 We can use this fraction of the genes at the scale of our data set (i.e. 200 genes)
 to simulate missing transcriptomes. Let us assume that a thalamic transcriptome\footnote{A trivial simulation would involve choosing $C_{T+1}= E(v_{Th},:)$, where
 $v_{Th}$ is a voxel from thalamus such that $\rho_{52}(v_{Th})$ is large. 
This guarantees a perfect fit at this voxel: 
 $\rho_{t}(v_{Th}) \sim \delta_{t,T+1}$; we checked that this is the case numerically, and that the other voxels 
 with positive densities at cell type labeled  $T+1$ are all in the thalamus, and that the densities of other
  cell types are well-conserved, apart from $t=52$ loosing its density in the thalamus. But 
  given the heterogeneity between microarray and ISH data, this check is more akin 
 to a clustering property of the rows of $E$ than to a test of the model, and we have to build $C_{T+1}$ out of 
 microarray data to run a more realistic simulation.} 
 would only need to deviate from the average cell type $\bar{C}$ in a 
 suitably chosen set of 200 genes, {\emph{and in a competitive way with $C_{52}$}},
  to inherit the thalamic signal from $C_{52}$. If the deviation 
 is well chosen, this cell type should be able to 
 inherit thalamic signal from $t=52$ upon refitting the model, providing a test of our conjecture\footnote{See the next section
 on random sub-sampling for further tests of this idea using random samples of 200 genes from the data set.}.\\

 To simulate the missing transcriptome $C_{\mathrm{thalamus}}$, we appended a simulated transcriptome $C_{T+1}$ to the 
 fitting panel, defined from mixtures of transcriptomes by the following pseudocode:\\
{\ttfamily{ 1. Initialize $C_{T+1}$ at the average transcriptome, $C_{T+1} = \bar{C} = \sum_{t=1}^C(t,.)/T$.\\
 2. Find the voxel $v_{th}$ in the thalamus with highest value of $\rho_{52}$:\\
  $v_{th} := {\mathrm{argmax}}_{v\in\mathrm{thalamus}}\rho_{52}( v ).$\\
 3. Rank the $G$ entries in $E(v_{th},.)$ by deacreasing value:\\
   $E(v_{th}, g_1)\geq E(v_{th}, g_2)\geq \dots \geq E(v_{th}, g_{T-1})\geq E(v_{th}, g_T).$\\
 4. For each of the 200 highest-ranking genes, replace the entry in $C_{T+1}$ by the entry
  in $C_{52}$, modified by some dilation parameter $\phi$ taken between 0 and 1:\\
 \begin{equation}
  \forall i \in [ 1..200 ], C^{(\phi)}_{T+1}(g_i) = (1+ \phi) C_{52}(g_i).
 \label{simulProfile}
\end{equation}
}}
The last instruction in the pseudocode aims to ensure that the simulated profile $C^{(\phi)}_{T+1}$  deviates from the average transcriptome in the $10\%$ of genes that matter most in some thalamic voxels, and that it does so in 
  a more intense way than $C_{52}$.
 Fitting the model to the above-defined panel of $T+1 = 65$ transcriptomes yields 
 a new estimated voxel-by-type density matrix, denoted by $\rho^{(\phi)}$.
 If $\phi$ grows from $0$ to some critical value,
 we expect the thalamic density to transfer smoothly 
  from $t=52$ to $t=T+1$, and the thalamic voxels should
 be mostly affected.\\

\subsection{Results}
\begin{figure}   
\includegraphics[width=1\textwidth,keepaspectratio]{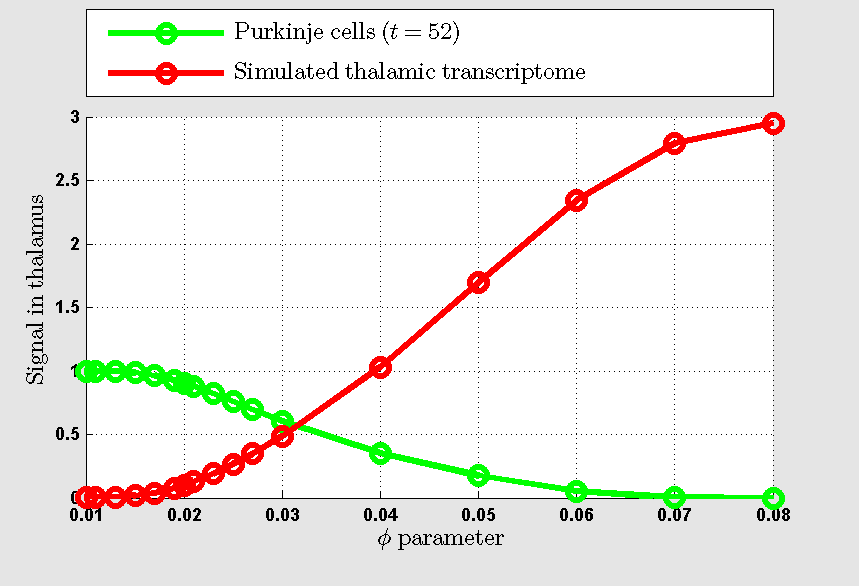}
\caption{{\bf{Thalamic signals for Purkinje cells and simulated thalamic transcriptome, as a function of the
 simulation parameter $\phi$ (Eq. \ref{simulProfile}).}} The thalamic
 signal is transferred from Purkinje cells to the simulated thalamic
 transcriptome $C_{T+1}$. See Fig. \ref{thalamicVisu} for
 visualization of the entire profiles at some values of $\phi$.  The
 signals are normalized by the value
 $\sum_{v\in{\mathrm{thalamus}}}\rho_{52}(v)$ of the thalamic signal
 in Purkinje cells $t=52$ at $\phi = 0$, hence the start of the green
 curve at 1.}
\label{intensityCorrelCumul}
\end{figure}

\begin{figure}   
\includegraphics[width=1\textwidth,keepaspectratio]{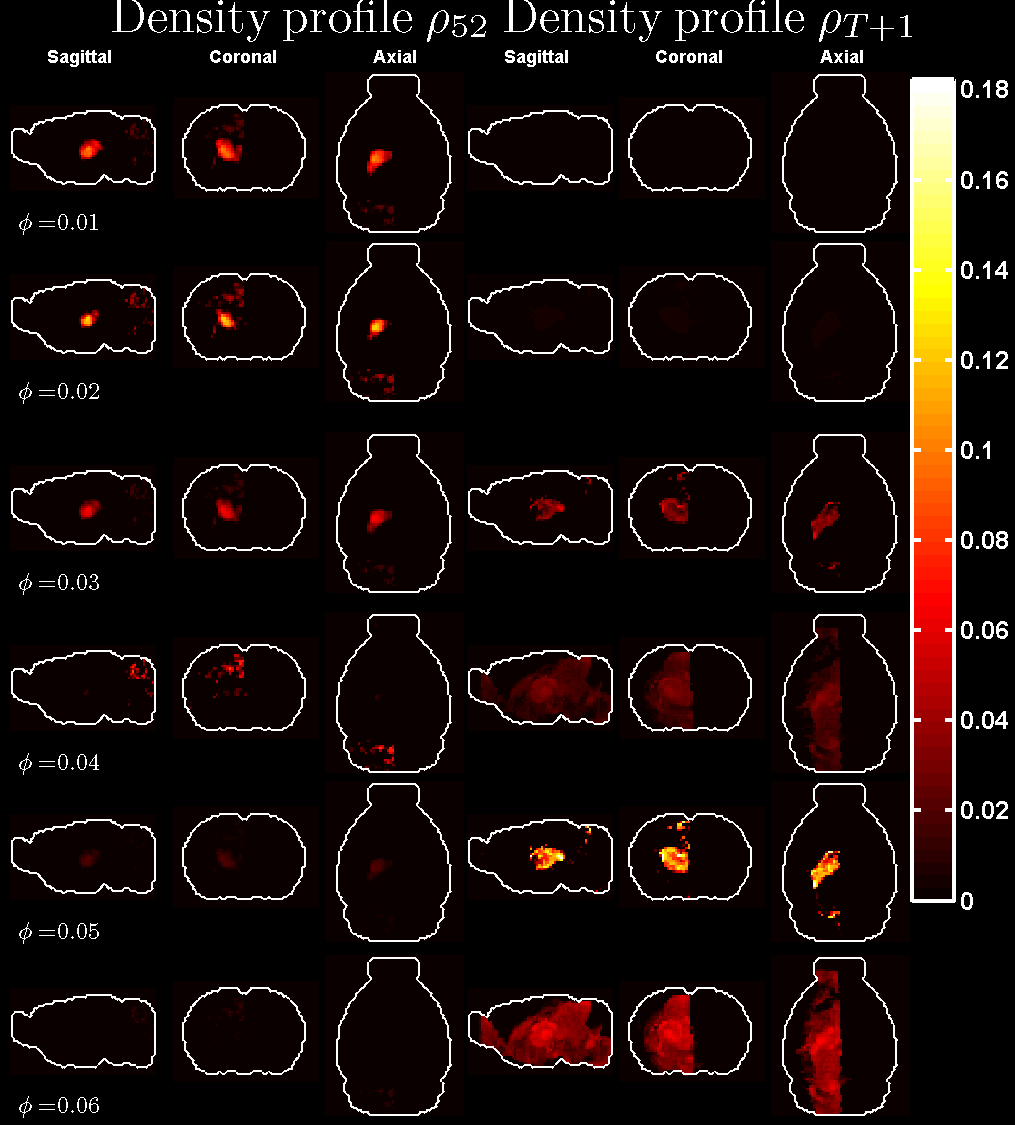}
\caption{{\bf{Maximal-intensity projections of the profiles $\rho_{52}^\phi$ (first three columns) and $\rho_{T+1}^\phi$ (last three columns).}} The signal in thalamic voxels is transferred from $\rho_{52}^\phi$ to $\rho_{T+1}^\phi$ as $\phi$ grows.}
\label{thalamicVisu}
\end{figure}

 To test this conjecture, we ran simulations for growing values 
 of $\phi$, starting from 0.01. The thalamic signal is indeed 
 gradually transferred from $t=52$ to $t=65$ when $\phi$ grows (see Fig. \ref{intensityCorrelCumul}),
 and for sufficiently low values of $\phi$, the support of 
 $\rho^\phi_{T+1}$ is confined to the thalamus. However, as 
 $\phi$ continues to grow beyond 0.05, $\rho_{T+1}$ develops positive
 density in other areas of the brain (see Fig. \ref{thalamicVisu}), and its 
 values in the thalamus become larger than the values of $\rho_{52}$ at $\phi=0$.\\


This simulation of a missing cell type extends the qualitative logic
 that was unveiled by refitting the model to a panel containing a single
 composite pyramidal cell type \cite{preprintFirstAnalysis}, but it does so in a more refined way in the sense that
the new fitting panel contains more transcriptomes than the original one. 
 In the particular case of missing thalamic cell types, we showed that 
 adding a cell type that is closer than the available ones
 to some voxels leads to a continuous transfer of signal 
  to the new cell type (and 200 genes are enough to observe 
 this regime of transfer). Hence a class of errors inherent to our model (the abusive
 prediction of a positive density of a given cell type in a region of the brain 
 far from where it was extracted), can probably be checked against 
 when new cell-type-specific transcriptomes become available and are included in the 
 fitting panel.

%% file: appendErrorEstimates.tex
\section{Error estimates for the predicted density profiles}

 Subtracting a uniform
term from the matrix of microarray data  led to results \cite{preprintFirstAnalysis} that present similar anatomical aspects for many 
cell types, and fit the ABA in a closer way. This raises the issue of the
 accuracy of the model. 
  Some of the $T=64$ cell types have striking neuroanatomical profiles in our results,
 some others have non-zero density at very few voxels, but how vulnerable are these anatomical properties
 to errors introduced by missing data and noise?
This is a difficult statistical problem, but we can perform a few numerical experiments to address it,
 and work out statistical bounds from theoretical results. In this section we will 
 repeteadly refit the model to sub-sampled sets of genes
 in order to simulate the impact of missing genes on density estimates.\\

\subsection{Theoretical bounds on the error as a function of noise}
 Bounds have been provided recently by Meinshausen in \cite{Meinshausen2013}
 on the discrepancy  between the signed-contrained estimates $\rho$ obtained by 
  minimizing quadratic discrepancies (as we did), and the quantity $\rho^\ast$ such that
\begin{equation}
E(v,.) = \sum_{t=1}^T C(t, .) \rho_t^\ast( v ) + \sigma \Xi(v,.),
\label{betaStarDef}
\end{equation}
where the last term is drawn from a Gaussian distribution:
\begin{equation} 
 \Xi(v,.) \sim \mathcal{N}( 0, 1 ),
\end{equation}
The voxel index $v$ is fixed in Eq. \ref{betaStarDef}. The vector $E(v,.)$ 
corresponds to the vector $\mathbf{Y}$ in the notations of \cite{Meinshausen2013}, 
 and the matrix $C$ corresponds to $X^T$, while our $\rho$ corresponds to $\hat{\beta}$,
 and our $\rho^\ast$ corresponds to $\beta^\ast$.
 Before stating the theorem and checking the hypotheses, let us note that the upper bound
 on $||\rho(v) - \rho^\ast(v)||_1$ proven in \cite{Meinshausen2013} scales as the
inverse square root of the number of genes $G$, which makes the bounds narrower 
 when more genes are taken into account. As we are taking genetic data into account collectively, 
 we are in an encouraging regime.\\

 The covariance matrix $\hat{\Sigma}$ is built from the 
matrix $X$ obtained by $L^2$-normalizing  the columns of $C^T$:
\begin{equation}
X(g,t) = \frac{C(t,g)}{\sqrt{\sum_{g=1}^G C(t,g)^2}},
\end{equation}
\begin{equation}
\hat{\Sigma} = \frac{1}{G} X^TX.
\end{equation}
 An upper bound on the $L^1$-norm of the difference between the vectors $\left(\rho_t(v)\right)_{1\leq t \leq T},$
and $\left(\rho^\ast_t(v)\right)_{1\leq t \leq T}$  is given by Theorem 3 in \cite{Meinshausen2013} as a linear function of the noise
parameter $\sigma$, with a coefficient given in terms of the properties of the matrix $\hat{\Sigma}$,
 which depends on the number of non-zero entries in $\rho^\ast_.(v)$, 
denoted by 
\begin{equation}
S := \left\{  k \in [1..T], \beta^\ast_k \neq 0 \right\},
\label{SDef}
\end{equation}
which depends on the voxel label $v$ in our case.\\

 The first hypothesis of the theorem is the following {\emph{compatibility condition}}:
\begin{equation} 
\exists \phi_\infty > 0,\;\; {\mathrm{min}}\left\{ |S| \frac{\beta^T \hat{\Sigma}\beta}{||\beta||_1^2}, \beta \in \mathcal{R}( 0,S ) \right\}\geq \phi_\infty,
\end{equation}
where $||\;||_1$ denotes the $L^1$-norm, and
 $\mathcal{R}( 0,S )$ is the set of vectors in $\mathbf{R}^T$ that have zero entries outside the support  $S$:
\begin{equation}
\forall L > 0,\;\;\;  \mathcal{R}( L,S ):=\left\{ \beta, \sum_{k\in \bar{S}}|\beta_k|\leq L\sum_{k\in S}|\beta_k|\right\}.
\end{equation}
 We computed the matrix $\hat{\Sigma}$, and found that
  its entries are all strictly positive (even though there are some zeroes in $C$ due to the 
 subtraction of the minimum value of microarray data across all genes and cell types, all the entries sit
 comfortably above zero):
\begin{equation}
\forall s,t \in [1..T], \;\;\; 4.51\times10^{-4}\leq \hat{\Sigma}(s,t) \geq 4.70\times10^{-4},
\end{equation}
 hence the existence of a lower bound $\phi_\infty=|S|\mathrm{min}(\hat{\Sigma})$ proportional to the minimum entry of 
$\hat{\Sigma}$.\\

 The second hypothesis constrains {\emph{minimal positive eigenvalue}} of $\hat{\Sigma}$, defined 
as\footnote{The symbol $||\;||_1$ denotes the $L^1$-norm in $\mathbf{R}^T$, so $||x||_1 = \sum_{k=1}^T |x_t|$.}:
\begin{equation}
\phi^2_{pos,S}(\hat{\Sigma}) = {\mathrm{min}}\left\{ \frac{\beta^T \hat{\Sigma}\beta}{||\beta||_1^2},\;\beta \in {\mathbf{R}}^T,\; {\mathrm{min}}_{k\in\bar{S}}\beta_k\geq 0\right\},
\end{equation}
where $\bar{S}$ denotes the complement in $[1..T]$ of the support $S$ of $\beta^\ast$ defined in Eq. \ref{SDef}.
If there exists some $\kappa > 0$ such that $\phi^2_{pos,S}(\hat{\Sigma}) \geq\kappa$, then (by Theorem 3 in \cite{Meinshausen2013})
 for some $\eta$ chosen in the interval $]0,1/5[$, the following upper bound on the $ L^1$-norm of the difference between 
 $\rho(v)$ and $\rho^\ast(v)$ holds with probability at least 
 $1-\eta$:
\begin{equation}
|| \rho(v) - \rho^\ast(v) ||_1 \leq \frac{8Ks\sigma}{\kappa\sqrt{G\phi_\infty}}.
\label{theorem3bound}
\end{equation}

\begin{figure}
\includegraphics[width=1\textwidth,keepaspectratio]{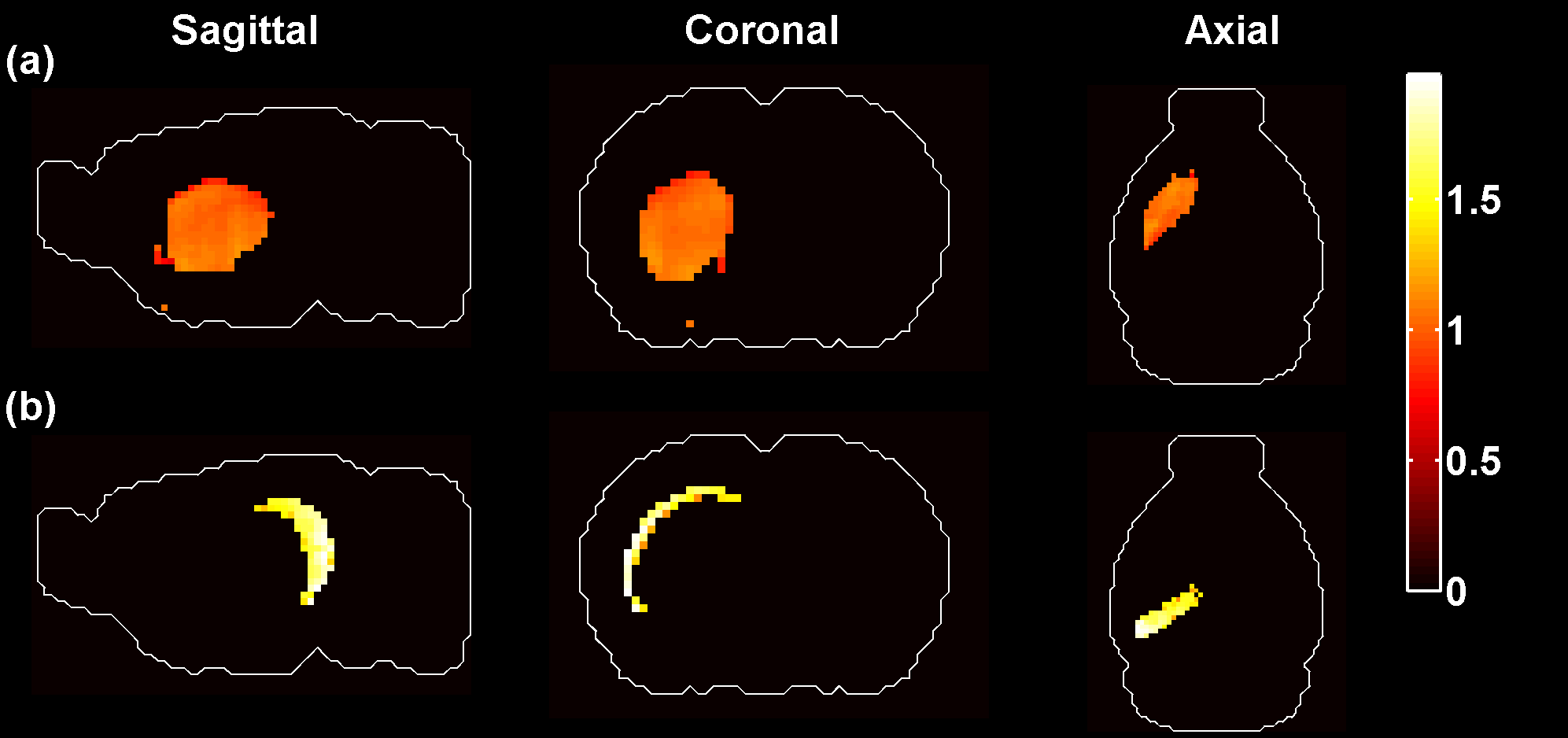}
\caption{{\bf{Maximal-intensity projections of density profiles of $\rho^{single}_{16}$ and $\rho^{single}_{49}$, 
as defined in Eq. \ref{rhoSingleDef}.}} {\bf{(a), Medium spiny neurons $t=16$}}  and {\bf{(b) hippocampal pyramidal neurons  $t=49$}},
 restricted to voxels at which each of them is the only cell type detected by the model.}
\label{uniqueVoxProj}
\end{figure}

 Since the hypotheses are formulated in terms of the support $S$ of $\rho^\ast(v)$,
 not $\rho(v)$,
we cannot check them directly in terms of the quantities we computed. However,
 the inspection of our results across voxels and cell types yields a couple of simple
 cases (special values of $v$, with low values of $|S|$). Indeed there 
 are a few  hundreds of voxels $v$ at which 
 only one cell type is predicted by the model (i.e. $\rho_t(v)>0$ for only one 
 cell-type index $t$). Moreover, the two cell types that are detected in this 
 situation are the medium spiny neurons
 (index $t=16$) and the hippocampal pyramidal neurons (index $t = 49$),
 and the corresponding voxels consist of subsets of the striatum and the hippocampus respectively,
 which are visually striking even if they contain only part of the 
 signal of the two cell types. We plotted the following densities on Fig. \ref{uniqueVoxProj}
 for $t=16$ and $t=49$:
\begin{equation}
\rho^{single}_t(v) = \rho_t(v) \mathbf{1}\left( | s \in [ 1..T ], \rho_s(v) > 0 | = 1 \right).
\label{rhoSingleDef}
\end{equation}
\clearpage

 To get an idea of the 
 values of the upper bounds yielded by \ref{theorem3bound}, let us imagine that the 
 underlying quantities $\rho^\ast(v)$ are also supported by just one cell type, with index $t=16$ (for
 some values of  $v$  corresponding to striatal voxels), and with index $t=49$ (for some  values of  $v$  corresponding to 
 hippocampal voxels). Given that the anatomical origin of these 
 two cell-type-specific transcriptomes (striatum and hippocampus) are guessed correctly by the model, this
 assumption seems reasonable.\\

\begin{figure}
\includegraphics[width=1\textwidth,keepaspectratio]{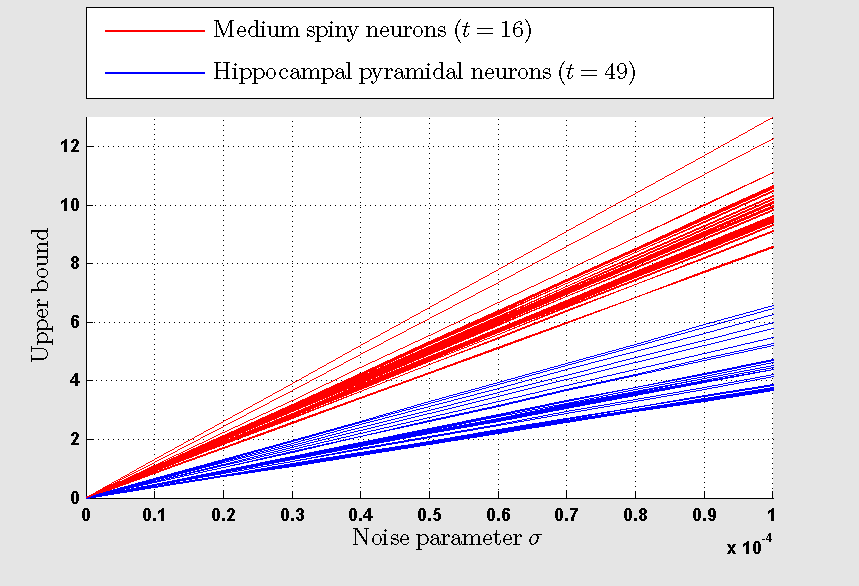}
\caption{{\bf{The upper bound on the discrepancy between $\rho_t$ and $\rho^\ast_t$ as a function of noise, for the two values
 of $t$ for which the model predicts the largest number of voxels to have support with $s=1$.}} We normalized the bounds
 by $||\rho_.(v)||_1$, so the interesting regime of noise is the one for which the values in the vertical 
 axis are below 1.} 
\label{figLinearNoise}
\end{figure}

 For a voxel $v$ at which $\rho^\ast(v)$  is supported by only one cell type, labeled $s$ 
 (i.e. $S = \left\{ s\right\}$),
  the minimal positive eigenvalue condition can be checked by a solving  a simple 
 quadratic programming problem. Indeed, if the support consists only of 
cell-type labeled $s$ at a given voxel, the quantity to minimize 
 to check the hypothesis on the minimal positive eigenvalue 
 is the following:
 \begin{equation}
\mathcal{Q}(\beta) = \frac{\beta^T \hat{\Sigma}\beta}{||\beta||_1^2},\;\;\; 
\label{QDef}
 \end{equation}
 on the space of vectors $\beta$ with positive entries at all indices except $s$.
 We can take care of the denominator in Eq. \ref{QDef} by rewriting $\mathcal{Q}(\beta)$
 as a quadratic form in the $L^1$-normalized vector 
\begin{equation}
\beta^{norm} = \left(\beta_1, \dots, \beta_{s-1}, \pm(1 - \sum_{t\neq s}\beta_t), \beta_{s+1},\dots, \beta_T\right),
\end{equation}
where the expression at the $s$-th entry follows from the fact that $\beta_t = |\beta_t|$ for any $t\neq s$. Hence,
we can bound the expression in Eq. \label{QDef} from below as follows
\begin{equation}
\mathcal{Q}(\beta) = {\beta^{norm}}^T\hat{\Sigma}\beta^{norm} \geq \mathcal{Q'}(\beta_1, \dots, \beta_{s-1}, \beta_{s+1},\dots, \beta_T),
\label{QWorkout}
\end{equation}
where the expression $\mathcal{Q'}(\beta_1, \dots, \beta_{s-1}, \beta_{s+1},\dots, \beta_T)$ is a quadratic
form in the vector 
\begin{equation}
\beta^{\bar{s}}= (\beta_1, \dots, \beta_{s-1}, \beta_{s+1},\dots, \beta_T),
\end{equation}
with positive entries. As the entries of the (symmetric) covariance matrix $\hat{\Sigma}$
 are all positive, the expression of  $\mathcal{Q'}$ in the lower bound of Eq. \ref{QWorkout} is as follows:
\begin{equation}
\mathcal{Q'}( \beta^{\bar{s}} ) = {\beta^{\bar{s}}}^T \hat{\Sigma}_{\bar{s}\bar{s}}\beta^{\bar{s}} + \hat{\Sigma}_{ss}\left(1-\sum_{t\neq s}\beta_t\right)^2
-2\left(1-\sum_{t\neq s}\beta_t\right) \sum_{t\neq s} \hat{\Sigma}_{st} \beta_t,
\end{equation}
where $\hat{\Sigma}_{\bar{s}\bar{s}}$ is the $(T-1)$-by-$(T-1)$ matrix obtained by destroying the $s$-th column
 and the $s$-th row in $\hat{\Sigma}$.
We minimized $\mathcal{Q'}$ over positive vectors  $\beta^{\bar{s}}$ in  $\mathbf{R}_+^{T-1}$, and found 
 it to be strictly positive for all values of $s$. In particular, with $s = 16$ and $s = 49$, choosing 
 $\eta= 0.15$, the upper bound of \cite{Meinshausen2013} are linear functions of the noise parameter $\sigma$
plotted on Fig. \ref{figLinearNoise} (divided by the value of $||\rho(v)||_1$ for each voxel $v$, hence the voxel-dependent 
 lines on the figure).\\

The regime of noise corresponding to narrow bounds even in this simple 
case of single-supported voxel is quite low (at $\sigma<10^{-5}$ ), whereas the average entry 
 in $E(v,.)$ at the voxels plotted in Fig. \ref{uniqueVoxProj} is larger than $4$.
Hence we found the theoretical upper bounds to be very large,
 assuming the simplest possible support ($s=1$). This confirms the difficulty of
theoretical approaches to the problem, even though our data set verifies the strong assumptions
 of Theorem 3 of \cite{Meinshausen2013} in the case of $|S|=1$. 
 This triggers us to conduct  simulations to assess more quantitatively with which probability  the support
 of densites $\rho^{\ast}(v)$  in voxel space (index $v$) and 
 in the space of cell types (index $t$) is correctly identified. 

\subsection{Repeated sub-sampling of gene space} 
\subsubsection{Simulation scenario}
As our data set contains only $G=2,131$
genes (those with data both the coronal ABA and in all transcriptomes), 
our results are based on roughly 10\% of the mouse genome. 
One obvious source of error is therefore the incompleteness of the coverage of gene space. 
 Given a cell type with a striking predicted density pattern, it is easy to imagine ways
to append thousands of columns to the matrices $E$ and $C$ (collectively labeled $g_{new}$), 
 such that the new entries $E(v,g_{new})$ and $C(t,g_{new})$ are anticorrelated, for voxels $v$ where cell
 type $t$ has a positive predicted density, $\rho_t(v)>0$. Refitting 
 the model to this larger data set  could destroy the anatomical pattern of the estimated density (bringing $\rho_t(v)$ to 0) if 
 the uncorrelated entries in the appended columns are large enough.\\

  However, detailed neurobiological knowledge presided over the choice of the genes included in the coronal ABA
 \cite{AllenGenome}:
 genes found to have striking expression patterns in preliminary data, and genes of particular interest in the
 neuroscience literature, were prioritized. There must therefore
 be more genes with region-specific expression pattern amoung our $G$
 genes than expected by chance. The mathematical scenario with large
 anti-correlated extra columns in matrices $E$ and $C$
  should therefore be made improbable by the design of the coronal ABA, 
but this scenario suggests a simulation technique to assess how badly our 
anatomical conclusions can be affected by missing genes.
 Indeed we can sub-sample the available gene space, taking 
 random sets of genes into account, each containing only 200 genes (a little less than 10\% of the data set,
which is approximately the scaling of the data set compared to the complete genome).
 Repeating this operation yields a random family of density estimates for each cell type. The results of 
 the sub-samplings
 are all of the same voxel-by-type 
 format, and their distribution in the space of matrices ${\mathcal{M}}_{\mathbf{R}_+}(V,T)$
can be studied to estimate confidence intervals on the densities, one per transcriptome profile (i.e. one per column).\\

The following pseudocode helps introduce notations:\\
{\ttfamily{for $s$ in [ 1..S ]}}\\
 $\;\;\;\;$  {\ttfamily{1. draw a random set of  200  integers from [1..G], without replacement;}}\\ 
 $\;\;\;\;$ {\ttfamily{2. construct the matrices $C^{(s)}$ and $E^{(s)}$ by concatenating the columns of $C$ and $E$}}\\
  $\;\;\;\;$            {\ttfamily{   corresponding to these integers;}}\\ 
 $\;\;\;\;$  {\ttfamily{3. compute  $(\rho^{(s)}_t(v))_{1\leq t \leq T}$ at each voxel:}}\\
$\;\;\;\;$     $(\rho^{(s)}_t(v))_{1\leq t \leq T} = {\mathrm{argmin}}_{\phi\in\mathbf{R}_+^T}\left(|||E^{(s)}(v,.)-\sum_t \phi_t C^{(s)}(t, .)||^2\right).$\\
{\ttfamily{end}}\\

\subsubsection{Variability of density profiles between the original and sub-sampled models}

The situation studied theoretically in the article \cite{Meinshausen2013} is identical
 to ours (we have one instance of this situation at each voxel). We 
 observed that it is difficult to predict the values of the densities,
 and even their support in type space (the number of indices $t$ for 
which $\rho_t(v)>0$ at a fixed value of the voxel index $v$). When we consider
  all values of the voxel index $v$ for a fixed value of the cell-type index $t$,
 which is what we do when we plot estimated densities
of a cell type, it is therefore difficult to know which voxels have positive 
 densities for type $t$ (the support of cell-type labeled $t$ in gene space).
 However, the estimated profile $\rho_t$ can be  visualized compared to of the average
 sub-sampled profile
\begin{equation}
\bar{\rho}_t(v) = \frac{1}{S}\sum_{s=1}^S \rho^{(s)}_t(v).
\end{equation}

\begin{figure}   
\includegraphics[width=1\textwidth,keepaspectratio]{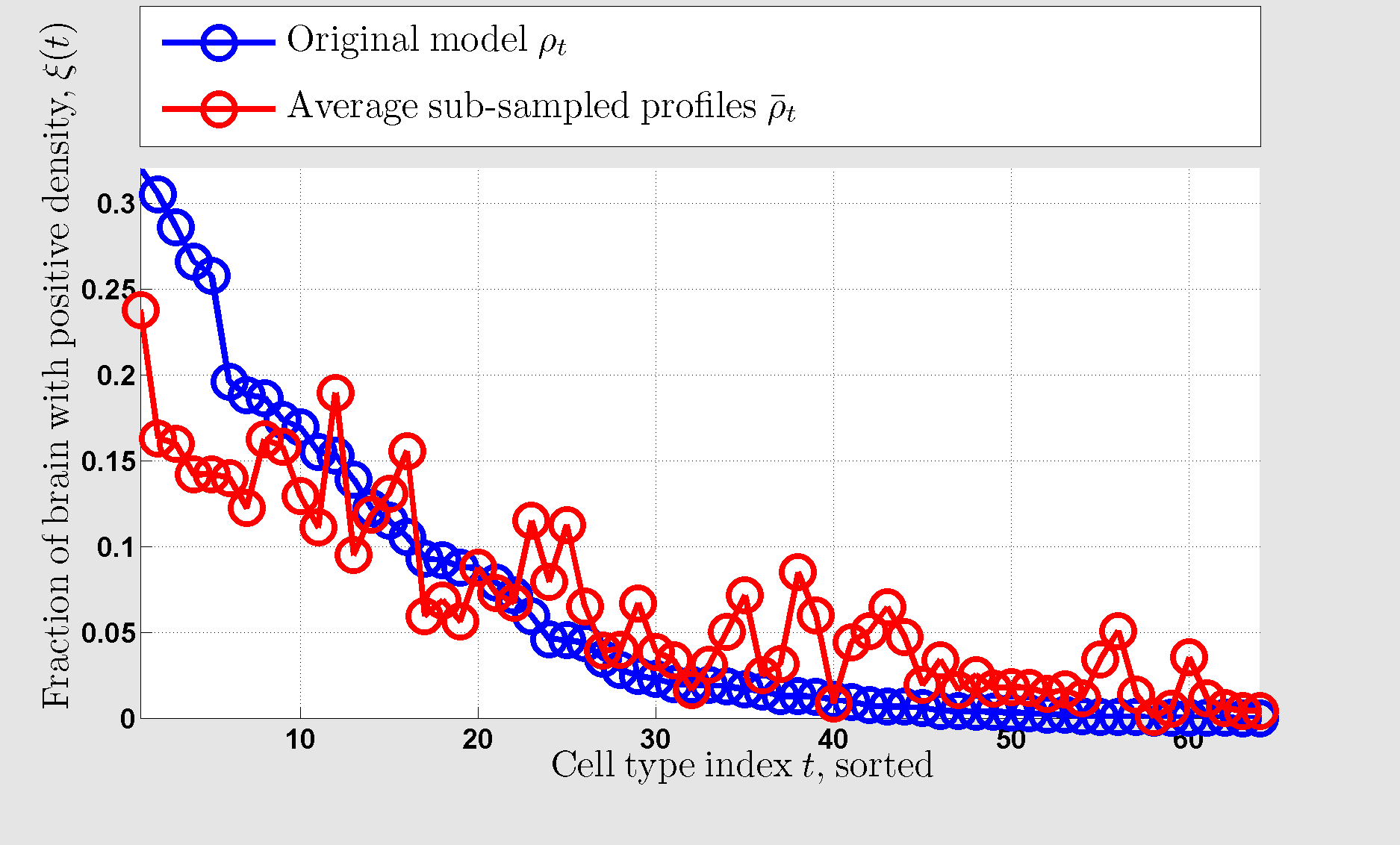}
\caption{Sorted values of the fraction of the volume of the brain annotation in the ARA (Eq. \ref{fracVoxels}) 
 supporting the total density of each of the $T=64$ cell types in the study. The largest value corresponds to
 $t=40$, cortical pyramidal neurons, whose support occupies $32\%$ of the left hemisphere, and corresponds 
 roughly to the entire left cerebral cortex.}
\label{voxelCounts}
\end{figure}

   
 To compare density profiles in the original and sub-sampled models, 
  we first observe that the density profiles in the original model,
  denoted by $(\rho_t(v))_{1\leq t \leq T, 1\leq v \leq V}$ are rather
 sparse in voxel space, in the sense that for all values of $t$, the number of voxels 
  in the support ${\mathrm{Supp}}(t)$ of cell type labeled $t$, defined\footnote{We used 
 the set of voxels belonging to the set of voxels annotated in the digitized Allen Reference Atlas 
 to belong to the left hemisphere. This set consists of 25,155 voxels. This convention corresponds to the one
 used to determine the {\emph{top region by density}} in the coarsest annotation of the mouse brain in the ARA,
 which covers the left hemisphere. Moreover, we chose to study the left hemisphere in order
 to be able to conduct more simulations, as the model involves the solution of one quadratic optimization
 problem per voxel. As the results of the model can be observed to have 
 a large degree of left-right symmetry, at least for the cell types with large predicted densities,
 this choice of annotation should give a reasonable simulation 
 of brain-wide prediction errors, assuming these errors are also left-right symmetric. The 
 analysis can easily be extended to the entire brain by doubling the alloted computation time.} as follows:
 \begin{equation}
{\mathrm{Supp}}(t) = \left\{  v\in {\mathrm{Brain\; Annotation}}, \rho_t(v)>0 \right\},
\label{suppDef}
 \end{equation}
 corresponds to $6.52\%$ of the brain on average, and never exceed $32\%$ . The sorted 
 values of the fraction
 \begin{equation}
 \xi(t) = \frac{|{\mathrm{Supp}}(t)|}{\sum_{v\in{\mathrm{Brain\;Annotation}}} 1 }
\label{fracVoxels}
 \end{equation} 
    are plotted in Fig. \ref{voxelCounts}. The values of the fractions defined in Eq. \ref{fracVoxels} are also typical 
 of the densities obtained after sub-sampling.\\

We can use the 
 support ${\mathrm{Supp}}(t)$ of cell type labeled $t$ as a control set,
 and compute how much of the signal of the second profile
  (which for us  will be the profile $\rho^{(s)}_t$ of the 
 same cell type in the $s$-th sample) is supported in ${\mathrm{Supp}}(t)$ .
 We computed the following quantity $\mathcal{I}(s,t)$ for each
 sub-sample $s$ and each cell type $t$, which is an overlap
 ranging from 0 (when the two profiles have disjoint supports), to $1$ 
 (when the signal of the sample $s$ is entirely supported by the  
 support of $\rho_t$):
\begin{equation}
 {\mathcal{I}}(s,t):= \frac{1}{\sum_v \rho^{(s)}_t(v)}\sum_v {\mathbf{1}}(\rho_t(v)>0) \rho^{(s)}_t(v).
\label{overlapDef}
\end{equation}


 For each cell type labeled $t$, the repeated sub-sampling approach simulates the distribution of the overlaps 
 $\mathcal{I}(.,t)$.
  The closer to 1 this distribution is concentrated, the more stable on average the estimate
 of the density of cell type labeled $t$ is under sub-sampling.
 This simulation of the distributions of the overlaps $\mathcal{I}(.,t)$  for all values
 of $t$ can be used to rank the transcriptome profiles by decreasing 
level of average overlap between $\rho_t$ and sub-samples density profiles:
 \begin{equation}
 \overline{\mathcal{I}}(t):= \frac{1}{S}\sum_{s=1}^S{\mathcal{I}}(s,t).
\label{meanoverlap}
 \end{equation}
  Let us denote the corresponding ranking by $(r^{signal}_t)_{1\leq t \leq T}$,
  which is defined by the following inequalities:
\begin{equation}
\overline{\mathcal{I}}(r^{signal}_1)\geq\overline{\mathcal{I}}(r^{signal}_2)\geq\dots
 \geq\overline{\mathcal{I}}(r^{signal}_T).
\label{rankingSubsampling}
\end{equation}

\begin{figure}
\includegraphics[width=1\textwidth,keepaspectratio]{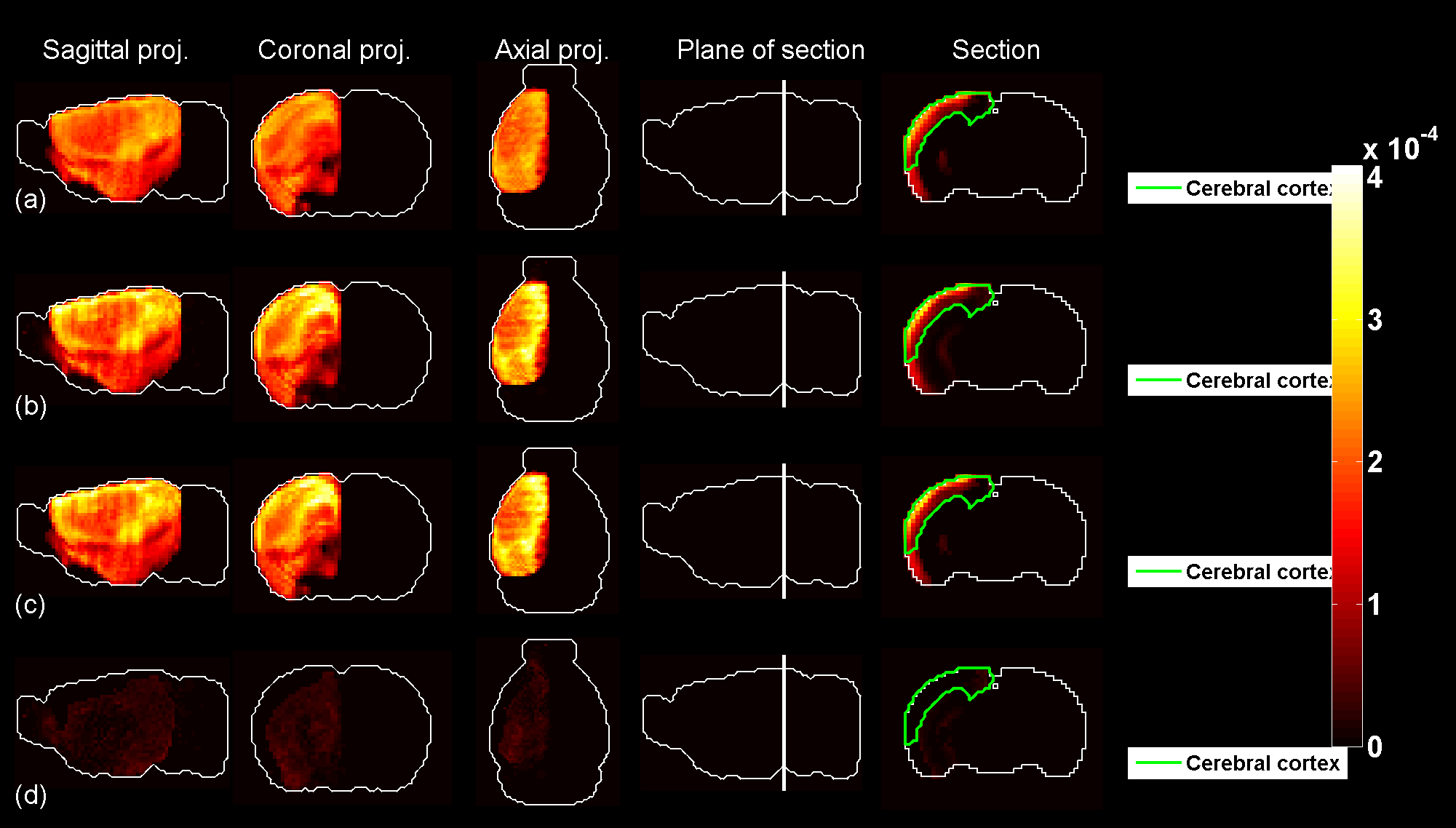}
\caption{{\bf{Density profiles for cell type labeled $t=40$, cortical pyramidal neurons,
 which has the highest overlap ($r_1^{signal}=40$) between the predicted profilel $\rho_{t}$ and the average sub-sampled 
 profile $\bar{\rho}_t$.}} (a) The predicted density profile $\rho_{40}$. (b) The average sub-sampled profile $\bar{\rho}_{40}$.
 (c) The part of $\bar{\rho}_{40}$ supported
 in the same voxels as $\rho_{40}$, Eq. \ref{rhoBarSupported}. (d) The part of $\bar{\rho}_{40}$ that does not
 overlap with the support of $\rho_{40}$, Eq. \ref{rhoBarUnsupported}. The ranking  of cell-type-specific transcriptomes induced 
by the sub-sampling procedure and Eq. \ref{rankingSubsampling} is supposed to rank highly the cell types for which 
 the profiles (a), (b) and (c) look very similar, while (d) looks close to zero.}
\label{figBestOverlap40}
\end{figure}
\begin{figure}
\includegraphics[width=1\textwidth,keepaspectratio]{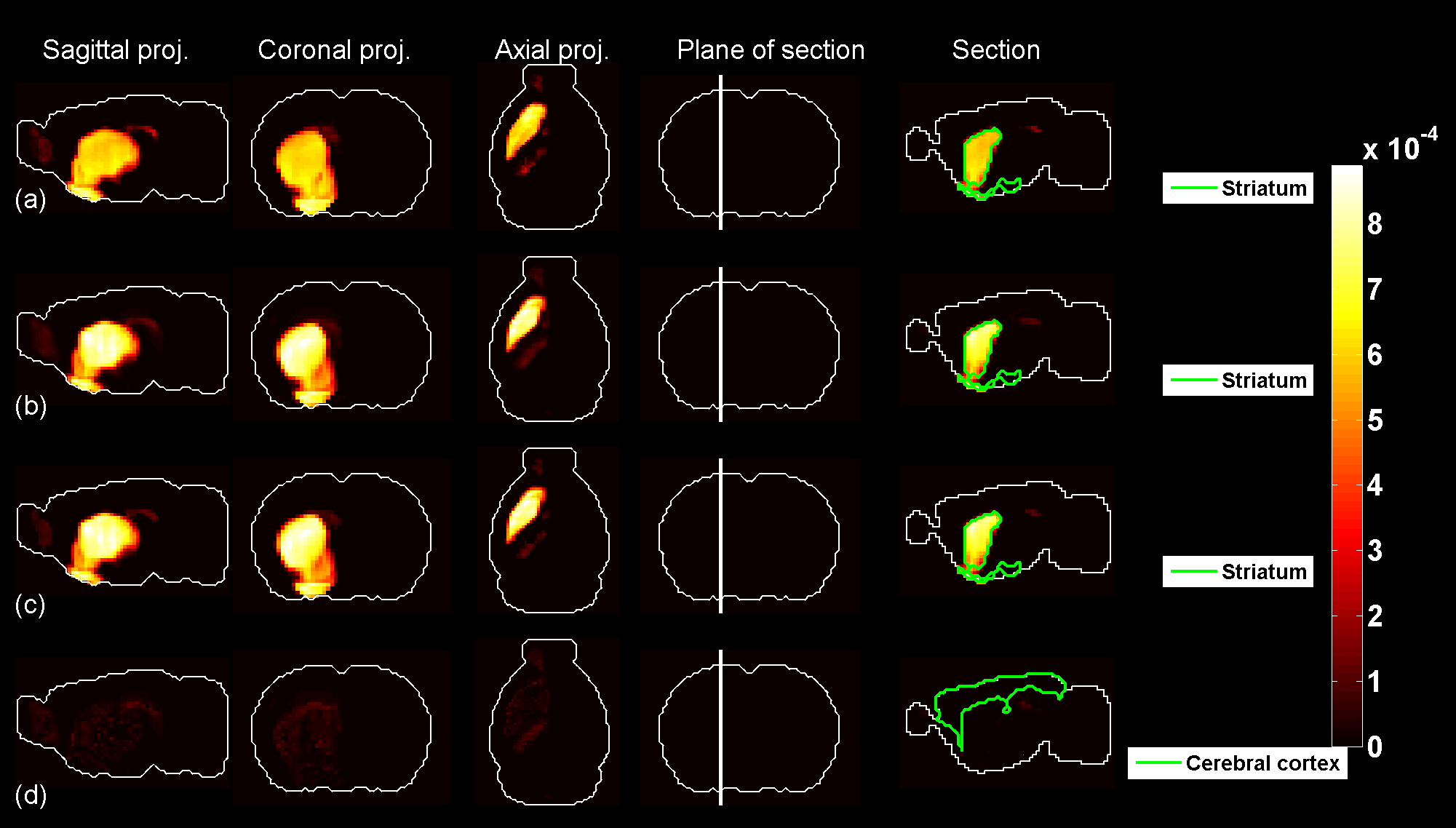}
\caption{{\bf{Density profiles for cell type labeled $t=16$ medium spiny neurons,
 which has the second highest overlap ($r_2^{signal}=16$) between the predicted profilel $\rho_{t}$ and the average sub-sampled 
 profile $\bar{\rho}_t$.}} (a) The predicted density profile $\rho_{16}$. (b) The average sub-sampled profile $\bar{\rho}_{16}$.
 (c) The part of $\bar{\rho}_{16}$ supported
 in the same voxels as $\rho_{16}$, Eq. \ref{rhoBarSupported}. (d) The part of $\bar{\rho}_{16}$ that does not
 overlap with the support of $\rho_{40}$, Eq. \ref{rhoBarUnsupported}.}
\label{figBestOverlap}
\end{figure}

The index $t= 40$ is ranked first: $r^{signal}_1 =40 $, followed by $t=16$, medium 
 spiny neurons. The results of the simulations are presented graphically in the 64 figures of section 6 (the cell-type indices $t$ are ordered from 1 to $T$ as in the 
 other tables for the sake of consistency, but the ranking of each cell type
  induced by Eq. \ref{rankingSubsampling} can be found in Tables \ref{tableCompare1} and \ref{tableCompare2}),
for which the following four densities are plotted, following the format
 of Figs. \ref{figBestOverlap40} and \ref{figBestOverlap}:\\
(a) As a reminder, the original density $\rho_t$.\\
(b) The average sub-sampled profile $\bar{\rho}_t$.\\
(c) The part of the average sub-sampled profile that is supported by the support of $\rho_t$, obtained by applying a 
 Boolean mask to it:
  \begin{equation}
 \bar{\rho}^{(supported)}_t(v) =\bar{\rho}_t(v) {\mathbf{1}}( \rho_t( v ) > 0 ).
 \label{rhoBarSupported}
  \end{equation} 
(d) The part of the average sub-sampled profile that is supported outside the support of $\rho_t$:
  \begin{equation}
   \bar{\rho}^{(unsupported)}_t(v) =\bar{\rho}_t(v) {\mathbf{1}}( \rho_t( v ) = 0 ).
 \label{rhoBarUnsupported}
  \end{equation}
 If the overlap $\overline{\mathcal{I}}(t)$ is close to 1, the density profiles (a), (b) and (c) will look
 very similar, and the density profile (c) will look empty. Examples taken from the top two cell types 
 of the ranking $r^{signal}$ are shown on Figs. \ref{figBestOverlap40} and \ref{figBestOverlap}.\\

\clearpage

\subsubsection{Confidence thresholds for the density of cell types}

Having simulated the distribution of the sub-sampled densities of all the $T$ cell-type
specific transcriptomes in our data set, we can estimate confidence thresholds
 in two ways, for a cell type labeled $t$.\\

 (1) Impose a threshold $\alpha$ in the interval $[0,1]$ 
on the overlap with the density $\rho_t$ estimated in the linear model,
 and work out the probability $p_{t,\alpha}$ of reaching that threshold from the 
sub-samples:\\
\begin{equation}
p_{t,\alpha} := P( \mathcal{I}(.,t) \geq \alpha) = \frac{1}{S}\left| s\in [1..S],   \mathcal{I}(s,t) > \alpha\right|.
\label{ptAlpha}
\end{equation}
 For a cell type labeled $t$, the distribution of the overlaps ${\mathcal{I}}(.,t)$ 
 can be visualized using the cumulative distribution function ${\mathrm{\sc{CDF}}}_t$ (in the space $[0,1]$ of the
 values of the overlap between $\rho_t$ sub-sampled profiles $\rho_t^{(s)}$):
 \begin{equation}
 {\mathrm{\sc{CDF}}}_t( \alpha ) = \frac{1}{S}\left|\left\{ s \in [1..S],  {\mathcal{I}}(s,t) \leq \alpha \right\} \right|.
 \end{equation}
 The value ${\mathrm{\sc{CDF}}}_t(u)$ is related 
 to the probability defined in Eq. \ref{ptAlpha} as follows:
\begin{equation}
p_{t,\alpha} = 1 - {\mathrm{\sc{CDF}}}_t( \alpha ).\\
\end{equation}

 (2) Impose a threshold $\beta$ in the interval $[0,1]$ on the fraction of sub-samples, 
and work out which overlap $\mathcal{I}_{thresh}(t,\beta)$ with the estimated density $\rho_t$  is reached by that fraction of the 
sub-samples. The threshold value of the intercept $\mathcal{I}_{thresh}(t,\beta)$ is readily expressed
in terms of the inverse of the cumulative distribution function:
\begin{equation}
\mathcal{I}_{thresh}(t,\beta) = {\mathrm{\sc{CDF}}}_{t}^{-1}(\beta).
\end{equation} 

 The more stable the prediction $\rho_t$ is again sub-sampling,
 the more concentrated the values of  ${\mathcal{I}}(.,t)$ are at high values (close to 1),
 the slower the take-off of the cumulative function  ${\mathrm{\sc{CDF}}}_t$ is, the lower the
 value of ${\mathrm{\sc{CDF}}}_t( \alpha )$ is,  and the 
 larger the probablity $p_{t,\alpha}$ is (for a fixed value of $\alpha$ in [0,1]).\\
 
\begin{figure}  
\includegraphics[width=1\textwidth,keepaspectratio]{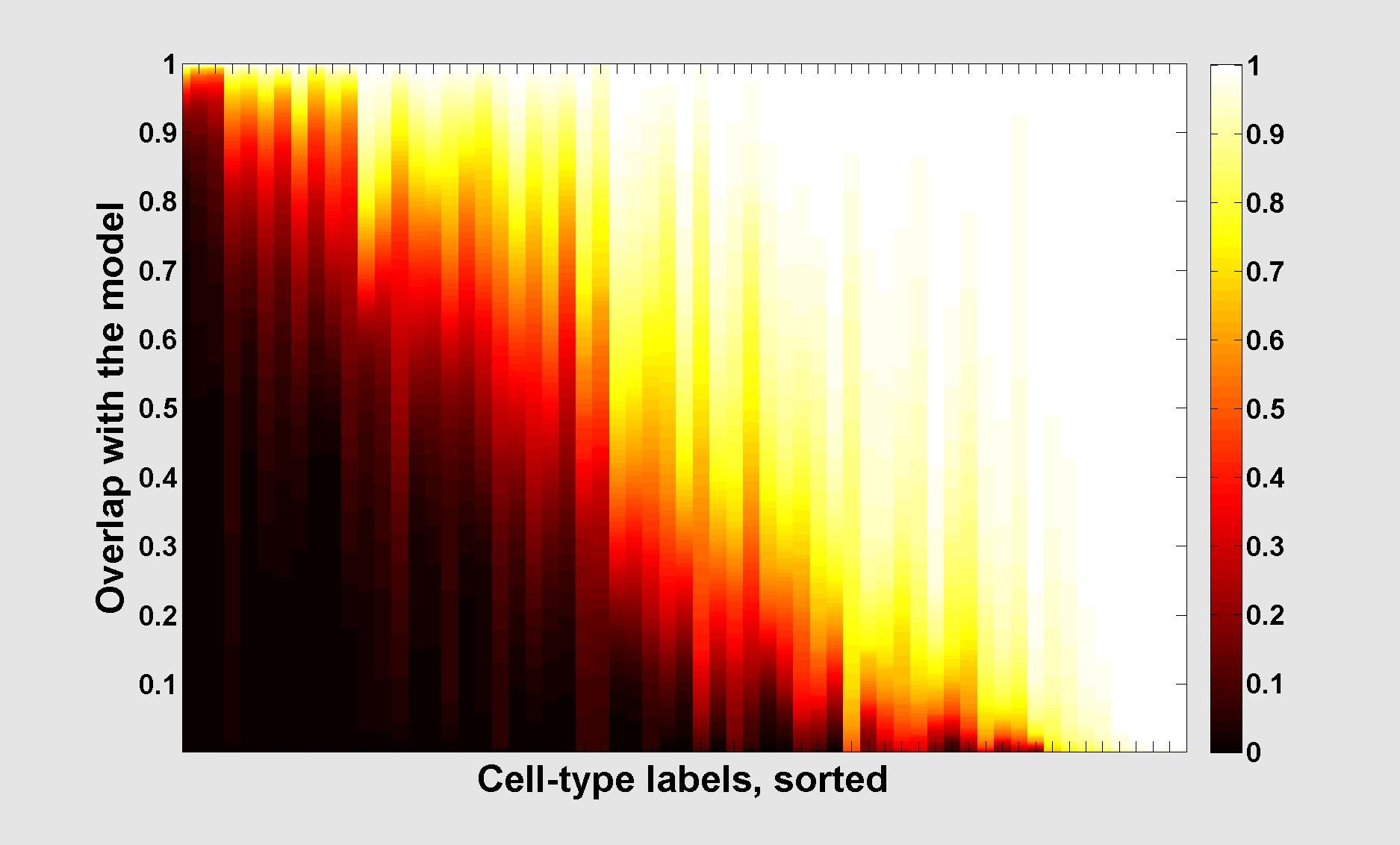}
\caption{{\bf{Heat map of the CDFs of the overlaps between sub-sampled densities
 and the result of the model, ordered as in the matrix $\mathcal{C}$ (Eq. \ref{cdfHeatMapDef})}}. One cell type per column, the columns are 
sorted by decreasing order of the average value of the overlap defined 
 in \ref{rankingSubsampling}. Each of the columns of this heat map is plotted as a function
 at the end of this note (in the figures of section 6), next to the visual rendering of the average sub-sampled profile.}
\label{figureCDF}
\end{figure}

For a fixed cell type labeled $t$, the values of $p(t,.)$ and  $\mathcal{I}_{thresh}(t,.)$
can therefore be readily read off from a plot of the cumulative distribution 
 function $\mathrm{\sc{CDF}}_t$ (this plot is in the $\alpha\beta$ plane in our notations).
 For the sake of visualization of results for all cell types, we constructed 
 the matrix ${\mathcal{C}}$, whose columns correspond to cell types,
 and whose rows correspond to values of the threshold $\alpha$:
\begin{equation}
 {\mathcal{C}}( \alpha, t ) := {\mathrm{\sc{CDF}}}_{r_t^{signal}}( \alpha ),
\label{cdfHeatMapDef}
\end{equation}
 where the index $r_t^{signal}$ in the r.h.s. means that the 
cell types are ordered by decreasing order of overlap between the
 average sub-sampled profile and the predicted profile.
 If the entries of the matrix $\mathcal{C}$ are plotted 
as a heat map (see Fig. \ref{figureCDF}), 
 the hot colors will be more concentrated in the left-most 
 part of the image. For a fixed column, the more concentrated 
 the hot colors are in the heat map, the more stable the corresponding cell type is. 

\subsubsection{Discussion of  errors estimates obtained through sub-sampling}
 The large bright area in the heat map of Fig. \ref{figureCDF} makes it
 clear that quite a few cell types in the present data set (at the bottom of the
 ranking by average overlap) have a very unstable signal across sub-samples. In particular, 
 nearly all the sub-samples for types such that $r_t^{signal}>45$ have profiles with empty 
 overlap with the original profile $\rho_t$ predicted by optimizing over 
 the full gene space.
 The ranked list of cell types is given in tables, but we can study 
 how the rankings $r^{signal}$ and error bars correlate with features 
 of the predicted profiles, and with the metadata.\\ 
  
\begin{figure}
\includegraphics[width=1\textwidth,keepaspectratio]{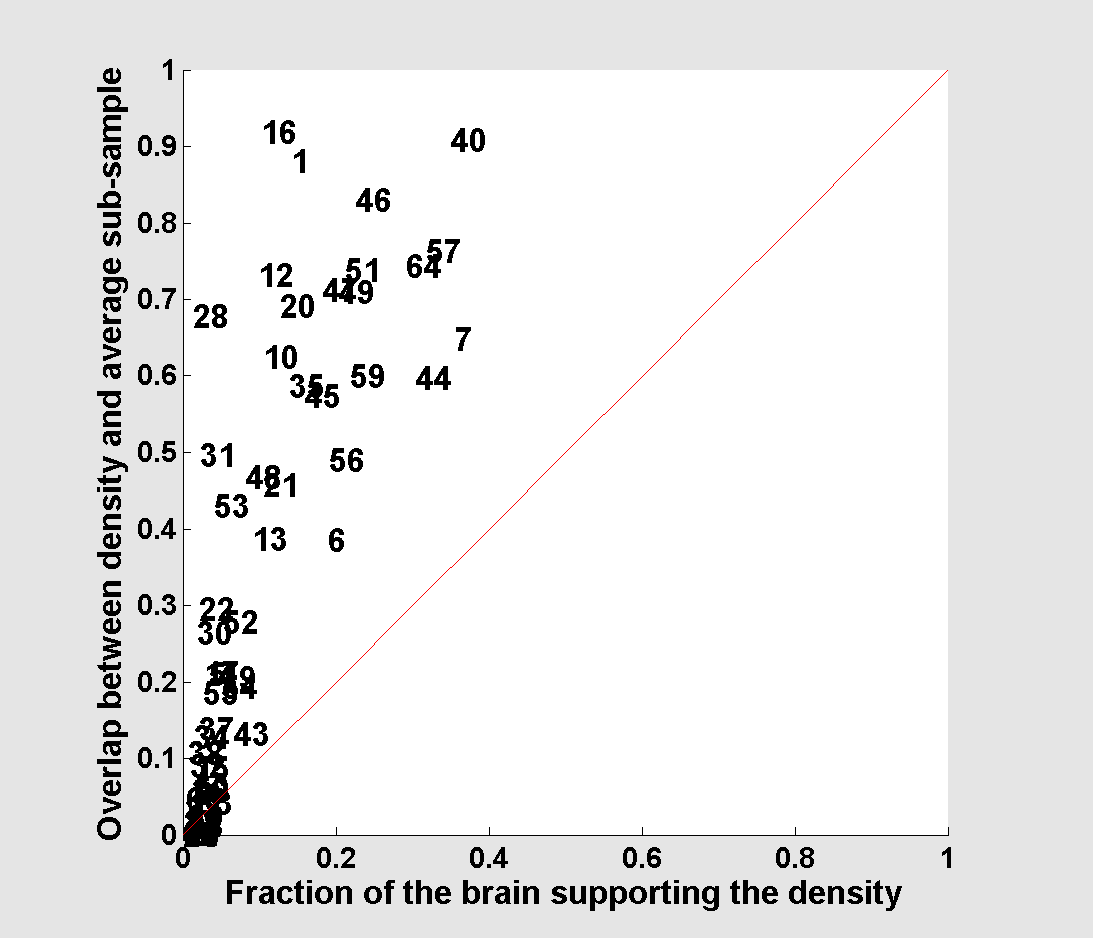}
\caption{Scatter-plot of overlaps $\bar{\mathcal{I}}(t)$ against fraction of voxels supporting the predicted density, $\xi(t)$.}
\label{overlapsVersusFracs}
\end{figure}

\begin{figure}
\includegraphics[width=1\textwidth,keepaspectratio]{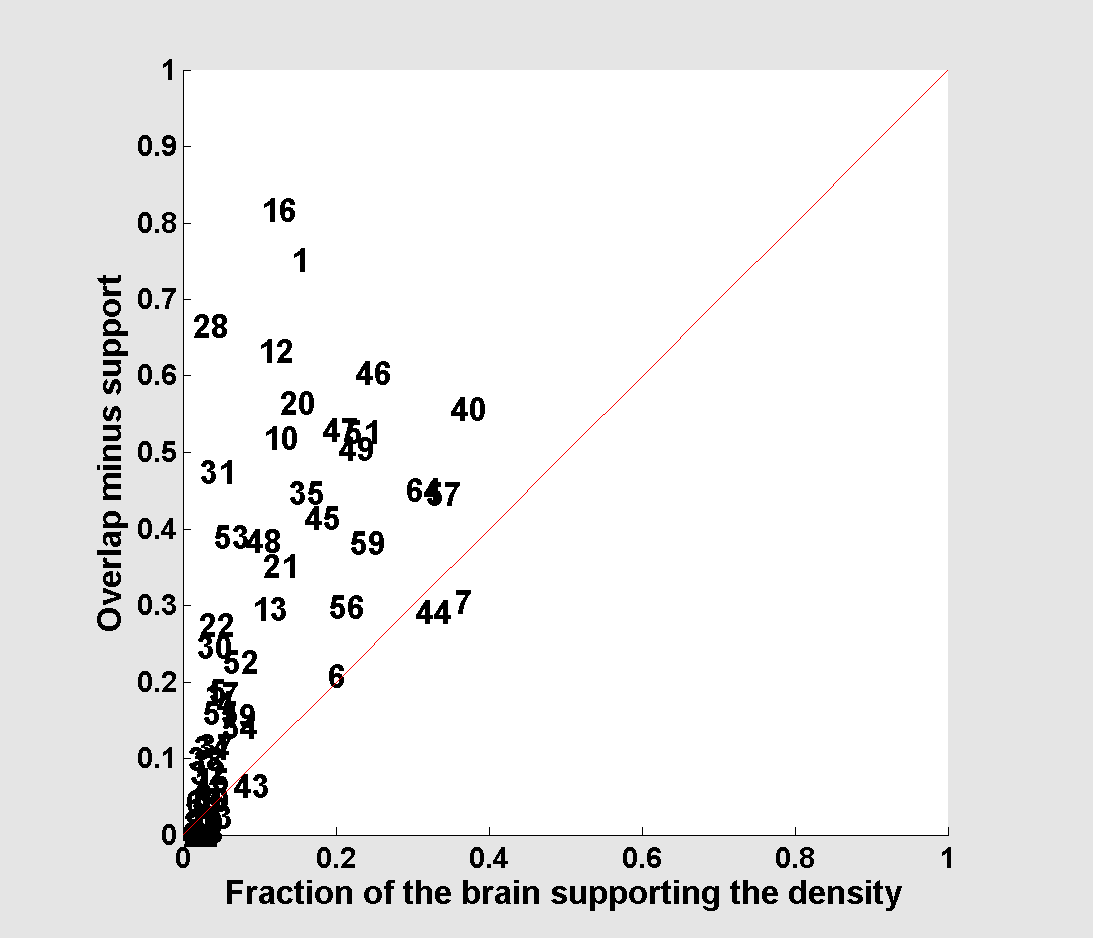}
\caption{Scatter-plot of differences between overlaps and supports $\bar{\mathcal{I}}(t) - \xi(t)$ against fraction of voxels supporting the predicted density, $\xi(t)$.          Medium spiny neurons ($t=16$) appear on top of the scatter-plot.}
\label{overlapsMinusFracsVersusFracs}
\end{figure}

 Taking a look at the cell types $t$ at the first ranks (for example
 the pyramidal neurons $t=40$ and medium spiny neurons $t=16$)
 one notices that their support tends to be larger (even though they are always confined to 
 less than a third of the voxels), and also more localized in 
 neuroanatomical areas defined in the ARA, whereas the ones at the bottom of the 
 ranking tend to have sparser profiles, which can be restricted to just a few voxels. 
  The value of the overlap between a random profile and a fixed profile
 is biased upwards by the size of the support of the profile. If one considers
 the ensemble of random subsets of $[1..V]$ (or equivalently random subsets
 of the voxels in the ARA), of fixed size $W$, denoted by $\nu$ as follows:\\
\begin{equation}
 \nu_v = \pm 1,\;\; v\in [ 1.. V],\;\;\; \sum_{v=1}^V \nu_v= W,
\end{equation}
 and if one considers the ensemble 
 of brain-wide profiles $\chi$  supported on these voxels (with some uniform value $\phi$), that is 
\begin{equation}
\chi(v) = \phi \mathbf{1}( \nu_v = 1 ),
\end{equation}
 then the average overlap of these
 profiles with $\rho_t$ averages at the fraction of the brain supporting $\rho_t$, 
 assuming independence between the random support $\nu$ and the support of $\rho_t$:
\begin{equation}
\mathcal{U}_t^{rand}:=\left\langle \frac{1}{\sum_{v=1}^V\chi(v)}\sum_{w=1}^V\chi(w)\mathbf{1}(\rho_t(w) > 0 )\right\rangle = 
\frac{1}{W}\sum_{w=1}^V\frac{W}{V}\frac{|{\mathrm{Supp}}(t)|}{V} = \xi(t),
\label{randomOverlap}
\end{equation} 
 where $\xi(t)$ is the fraction of voxels in the brain annotation supporting the estimated profile 
  $\rho_t$ (see Eq. \ref{suppDef}) 
 which is one reason for which cell types largest supports tend to be on the left of 
 the heat map of Fig. \ref{figureCDF}. The crude estimate $\mathcal{U}_t^{rand}$ 
 also provides a reference for the 
 expectation by chance to which the values or $\bar{\mathcal{I}}(t)$ can be compared 
 these two quantities, and again rank the cell-type-specific transcriptomes
 by decreasing values of this differences. The types with high positions above the diagonal (such as $t=1,12,16,40,26$)  
 tend to be conserved between scatter-plpots \ref{overlapsVersusFracs} and \ref{overlapsMinusFracsVersusFracs},
 even though the exact shape of the cloud of points change.\\

The indices that are higher above the diagonal in the scatter-plot \ref{overlapsVersusFracs},
 (resp. high in the vertical direction in the scatter-plot \ref{overlapsMinusFracsVersusFracs}),
 are the ones whose densities are the most stable under sub-sampling (resp. the ones
 whose stability propertes deviate more than what would be expected by chance, 
 based on random independent profiles). Not surprisingly, the medium spiny neurons labeled 
$t=16$ are singled out by these scatter-plot.\\

 So far the analysis of results has been very general and abstract. 
The support analysis is just one possible quantitative approach to the 
ranking of cell-type-specific transcriptomes by the amplitude of errors,
 and we have to look at the results for each of the cell types. However, we can notice that the transcriptomes
 that are ranked highly by our procedure tend to be ones that one
would have singled out by eye for anatomical reasons before the sub-sampling procedure
was carried out, because these transcriptomes tend to:\\
 $1.$ have a density profile  $\rho_t$ that highlight a known neuroanatomical region,\\
 $2.$ come from cells that were 
indeed microdessected from these regions.\\

Moreover, we can compute the localization scores of the average sub-sample
on the same bar diagram as the ones of the predicted profile, in order to see
if the fraction of signal in the average sub-sample that is not supported by the prediction
changes radically the ranking of regions in the ARA by density of cell-type $t$ or not.
A bar diagram of these localization scores is presented in the 
figures for every cell type in part (b) of the figure containing the plot 
of the cumulative distribution functions of the overlaps.









\subsection{Error estimates on the anatomical analysis of profiles}

\begin{figure}
\includegraphics[width=1\textwidth,keepaspectratio]{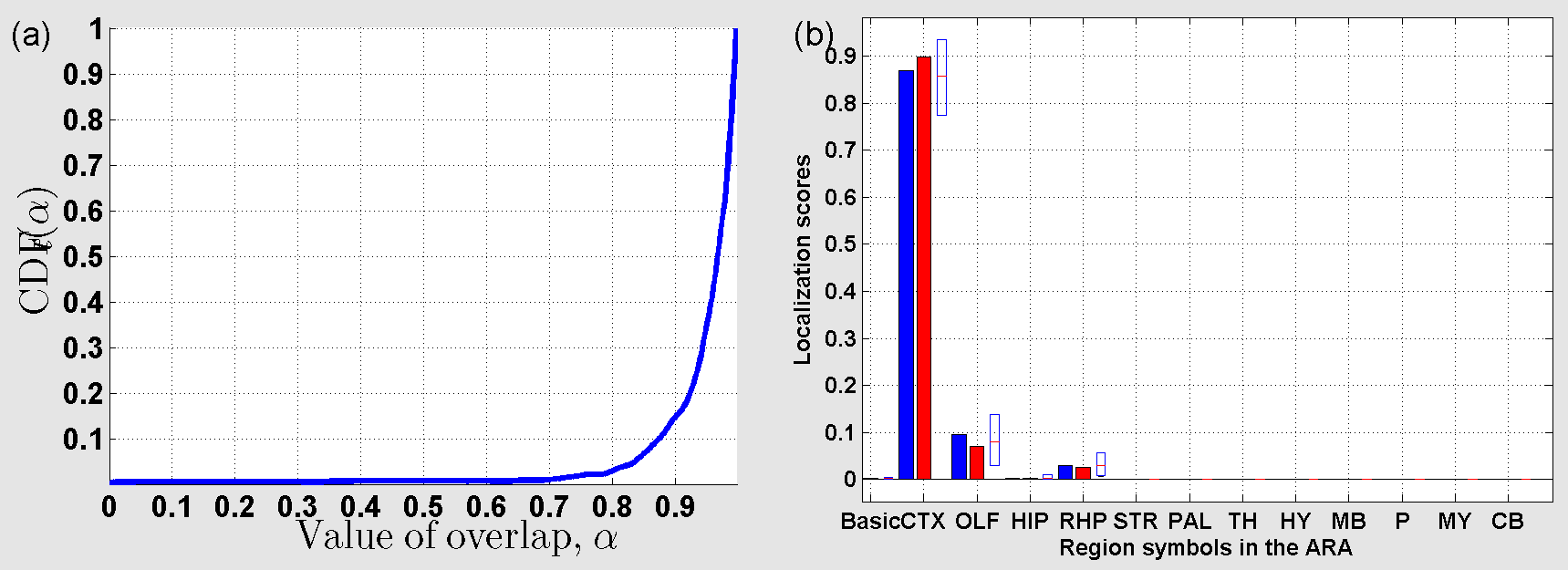}
\caption{{\bf{Cumulative distribution function of overlaps and distribution of localization scores for cortical pyramidal neurons, $t=40$.}} (a) The function ${\mathrm{\sc{CDF}}}_t$ for $t=40$. (b) Localization scores in the 'Big12' annotation. The top region of the density is the cerebral cortex both in the
 original model and in the average sub-sampled density. Moreover, the localization score in the cortex is larger than 78\% in 75\% of the samples, which allows
 to distinguish the cortex from any other region with non-zero signal (such as the olfactory areas and the hippocampus).} 
\label{bestAnatomyPlot40}
\end{figure}
\begin{figure}
\includegraphics[width=1\textwidth,keepaspectratio]{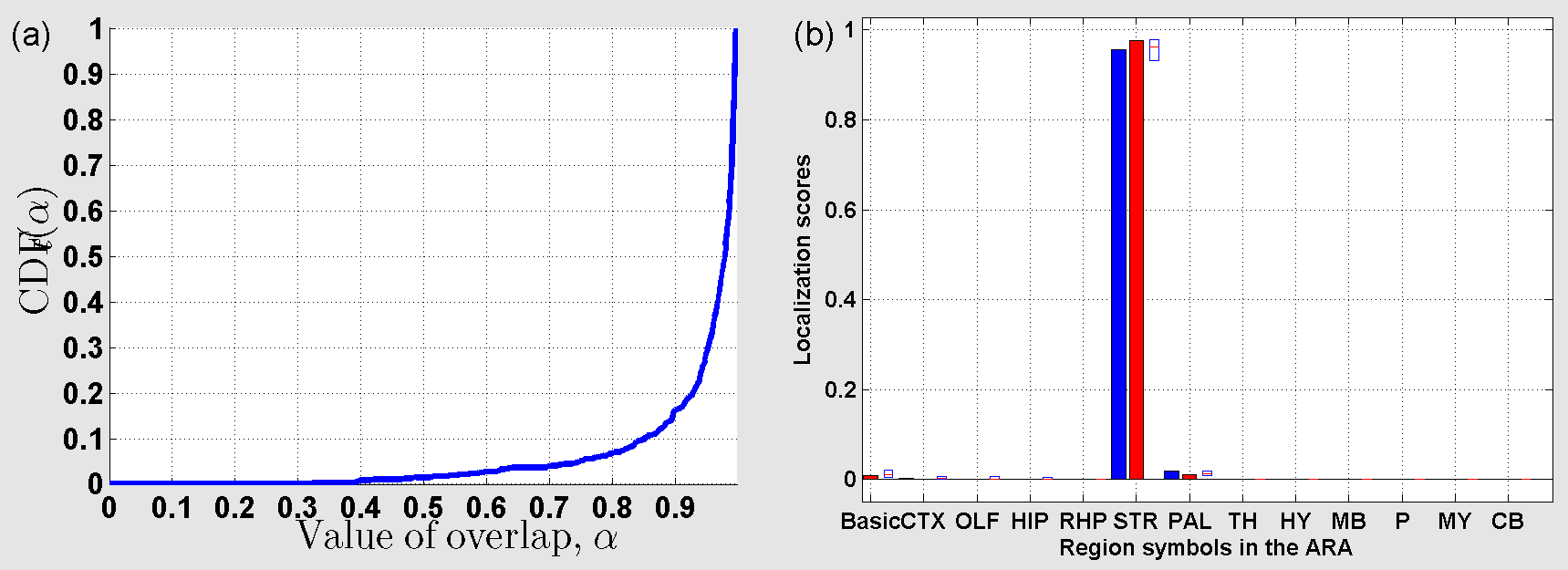}
\caption{{\bf{Cumulative distribution function of overlaps and distribution of localization scores for medium spiny neurons, $t=16$.}} (a) The function ${\mathrm{\sc{CDF}}}_t$ for $t=16$. (b) Localization scores in the 'Big12' annotation. The top region of the density is the striatum both in the
 original model and in the average sub-sampled density. Moreover, the localization score in the striatum is larger than 90\% in 75\% of the samples. All other regions
 score less than 5\% on average.}
\label{bestAnatomyPlot16}
\end{figure}

The brain-wide density profiles $\rho$ (or rather any quantity defined in 
 the left hemisphere, where the so-called {\ttfamily{Big12}} annotation\footnote{This
annotation is the coarsest annotation in the ARA, and consists of 12 regions together with the
 so-called 'Basic cell groups and regions'. The regions are designated by the following 
 symbols in the plots:  Basic cell groups and regions = Basic,
Cerebral cortex = CTX, Olfactory areas = OLF, Hippocampal region =
HIP, Retrohippocampal region = RHP, Striatum = STR, Pallidum = PAL,
Thalamus = TH, Hypothalamus = HY, Midbrain = MB, Pons = P, Medulla =
MY, Cerebellum = CB} is defined),
 can be confronted to classical neuroanatomy using a bar diagram of the 
localization scores of $\rho$ in the regions of the 'Big12' annotation.
 A number of these bar diagrams where presented in \cite{preprintFirstAnalysis}. Let us recall
 the definition of the localization scores \cite{markerGenes} of $\rho$ in region
 $\omega$:
\begin{equation}
\lambda_\omega(\rho) = \frac{\sum_{v\in\omega}\rho_t(v)^2}{\sum_{v\in{\mathrm{Brain\;Annotation}}}\rho_t(v)^2}
\label{localizationDefinition}
\end{equation}
The (squared) $L^2$-norm is only one possible choice of norm in the fraction \ref{localizationDefinition}.
 It happens to have a generalization to sums of functions chosen from a family,
 that maps the maximization of localization scores to a generalized eigenvalue 
 problem (using the $L^1$-norm instead does not lead to major changes
 in the designation of the top region by density). The important thing in the present discussion is that 
  we can repeat the computation of localization scores for each cell type labeled
 $t$. We can complement the bar diagram of localization scores $(\lambda_\omega(\rho_t)_{\omega\in\mathrm{Big12}}$
 by the bar diagram of localization scores of the average sub-sampled profile, 
   $(\lambda_\omega(\bar{\rho}_t)_{\omega\in\mathrm{Big12}}$. Moreover, we can compute the
 whole family of localization scores of all the 
 sub-sampled profiles $(\lambda_\omega(\rho^{(s)}_t))_{\omega\in\mathrm{Big12}, s \in [1..S]}$.
 The more peaked the distribution of these scores is around $(\lambda_\omega(\bar{\rho}_t)_{\omega\in\mathrm{Big12}}$
 for each region $\omega$, and the closer the values $(\lambda_\omega(\bar{\rho}_t)_{\omega\in\mathrm{Big12}}$
 are to  $(\lambda_\omega(\rho_t)_{\omega\in\mathrm{Big12}}$, the more stable the anatomical properties
 of the density profile $\rho_t$ are under sub-sampling. For visualization purposes, the bar diagrams 
 of the localization scores of $\rho_t$ and $\bar{\rho_t}$ are supplemented on the 
 same graph by a box plot of the localization scores of samples, showing the median of the distribution,
 and extending vertically between the first and third quartiles 
of the distribution $(\lambda_\omega(\rho^{(s)}_t))_{s\in [1..S]}$ for each region $\omega$.
 A stable prediction results in two very similar bar diagrams, with very small boxes appended to their right.
 Not surprisingly, the cell types labeled $t=40$ and $t=16$, which were ranked 
 first and second by the overlap analysis, give rise to such plots (see Figs. \ref{bestAnatomyPlot40} and 
 \ref{bestAnatomyPlot16}). Analogous figures for all values of $t$ 
 can be found at the end of the present note (section 6).

\subsection{Discussion of results by brain regions}
\subsubsection{Cerebral cortex}
The cerebral cortex is the largest region in the coarsest 
version of the ARA. It is also the one from which the largest number of cell types
 were dissected (see the subsection on 'Basic cell groups
 and regions' for a separate discussion of some
 cell-type-specific samples dissected from the cerebral cortex, and the subsection on olfactory areas for a separate discussion of indices $t=48$
 and $t=53$, that were dissected from the amygdala).
This subsection discusses a group of 28 cortical cell types
 which contains both (1) samples that have striking cortical profiles $\rho_t$
 in the model and good stability properties under sub-sampling 
and (2) samples that have very sparse and 'amorphous' profiles $\rho_t$ 
 in the model and bad stability properties.\\

 A more detailed composition of these groups is as follows:\\
1. A set of cortical pyramidal neurons, some which with striking
  cortical patterns ($t=40$, Pyramidal Neurons, Callosally projecting, P14, are ranked first by average overlap with the model 
 ($r_{1}^{signal}=40$), and 85\% of the sub-sampled profiles have at least 90 \% overlap with the model, followed by  $t=46$ and $t=47$ in the
 top ten cell types by average overlap), and also some GABAergic interneurons whose density
 profiles is mostly subcortical, and distributed among 
 many brain regions.\\
2. A set of cell types that have very sparse profiles, and tend to have 
 less sparse profiles in the sub-samples. For some cell types such as the mixed neurons (index $t=9$),
the average sub-sample is much more credible as it returns a cortical pattern rather than 
 an isolated voxel (the ranking is of course very low, $r^{signal}_{61}=9$). 
  The low ranking of some cell-type-specific transcriptomes that come
 from non-adult animals could correspond to tha fact that these cell types
 mature early and therefore fit poorly to the entire atlas. This developmental effect 
 may differ from gene to gene, and when 
 subsets of genes are taken into account, they can happen to offer 
 a better fot to the cortex, hence a cortical contribution to $\bar{\rho}_t$. 
 However, this lack of conservation of anatomical properties under sub-sampling 
 is not a sure way to detect non-adult transcriptomes, as the 
 pyramidal neurons at the top of the overlap distribution correspond to P14.\\

\begin{table}
\centering
\input{tableDiscussionTypesCortexPre.tex}
\caption{{\bf{Table of cell types extracted from the cortex, ranked by average overlap.}} The cell types 
whose profile is localized mostly in the 'Basic cell groups and regions' of the ARA are included in Table 
\ref{tableDiscussionTypesBasic}. The index called "rank" (first column), at a fixed value of $t$ (third column), denotes the integer
 $k$ such that $r_k^{signal}=t$.}
\label{tableDiscussionTypesCortex}
\end{table}

\subsubsection{Olfactory areas}
There is no cell-type-specific transcriptome labeled to have been dissected
 from olfactory areas in Tables \ref{metadataAnatomyTable1} and \ref{metadataAnatomyTable2},
 but two (pyramidal neurons labeled $t=48$, and glutamatergic neurons, $t=53$) 
 were extracted from the amygdala, and the amygdala 
 is split between the cortex and the olfactory areas in the finer versions 
 of the ARA. Visual inspection of Figures \ref{subSampledFour48} and \ref{subSampledFour53}
 shows rather striking contrast between the amygdala and the rest of the brain (especially in 
 pyramidal neurons).\\

 However the two cell types are only ranked 19 th and 20th out 0f 64 ($r^{signal}_{19} =48$ and 
 $r^{signal}_{20}=53$), but the procedure is more sensitive to the support 
of $\rho_t$ than to the contrast inside this support (see Figs. \ref{subSampleSplit48} and \ref{subSampleSplit53} in section 7 for thresholded heat maps 
 of $\rho_{48}$ and $\rho_{53}$ at 99\% of conditional probability of detecting cell types $t=48$ and $t=53$; they both show 
 distinct amygdalar sets of voxels). With $t=48$,
 we are clearly in a case where a small set of voxels inside
 the support of $\rho_t$ concentrates a large fraction of the signal.
 In the average sub-samples profiles $\bar{\rho}_{48}$ and $\bar{\rho}_{53}$, the support is more extended
 across the brain (for instance there are more hippocampal voxels with non-zero signal in $\bar{\rho}_{53}$ than in 
 $\rho_{53}$, and they make up for about 3 \% of the signal in $\bar{\rho}_{53}$), but the same set of voxels shows high contrast.
 The olfactory areas are the top region by localization score both in $\bar{\rho}_{53}$ and $\rho_{53}$ at about 60\% and 54\%
 respectively, followed by the cerebral cortex at 30 \% and 33 \% respectively; inspection of Figs. \ref{cdfPlot53} 
 \ref{subSampledFour53} illustrates the fact that in both the original and sub-semplaed profiles,
 these sscores come largely from voxels in the amygdala.\\

\subsubsection{Hippocampus}
The transcriptome of hippocampal pyramidal neurons labeled $t=49$ is one of the cell
types for which the density $\rho_t$ provides the most spectacular guess 
 of the anatomical origin (with more than 85\% of the signal 
 in the hippocampal or retrohippocampal regions of the ARA). These localization 
 scores are even a few percentage points higher for $\bar{\rho}_{49}$, 
 and 75 \% of the sub-samples have more than 80\% overlap with 
 $\rho_{49}$. Moreover, the ranking by average overlap is  9 out of 64 ($r^{signal}_{9}=49$).\\

 Perhaps more surprisingly, the other cell-type-specific transcriptome
extracted from the hippocampus (GABAergic neurons, SST+ $t=57$), which on the contrary 
has low localization score in the hippocampus, ranks similarly 
 high ($r^{signal}_{8} = 57$), and the rather complex sub-cortical pattern 
of localization scores is rather well-conserved (see Figs. \ref{cdfPlot57} and \ref{subSampledFour57}).
 Looking at the results of $T=64$ confirms that there is stronger solidarity between
 densities of GABAergic interneurons than between any of them
 and the particular region from which they were dissected.

\subsubsection{Striatum}

The sample of medium spiny neurons labeled $t=16$ ranks second
 according to $r^{signal}$, and indeed the density profiles
 $\rho_{16}$ and $\bar{\rho}_{16}$ look very similar, with 97 and 98\%
of the signal in the striatum respectively (see Figs. \ref{cdfPlot16} and \ref{subSampledFour16}).
Moreover, 83\% of the samples have an overlap of more than 90\% 
 with $\rho{16}$.\\

 Interestingly, the other sample of medium spiny neurons in the 
 fitting panel ($t=15$), ranks only 39 out of 64. Looking at Figs. \ref{cdfPlot15} and \ref{subSampledFour15})
 shows that $\bar{\rho}_{15}$ is much more localized in the striatum
 that the prediction  $\rho_{15}$ (about 67\% of the signal is in the striatum, compared to 
about 13 \% for $\rho_{15}$, which makes for a correct prediction of the 
anatomical origin of the sample (as was the case for $\rho_{16}$ and $\bar{\rho}_{16}$.
 We showed in \cite{preprintFirstAnalysis} that the densities $\rho_{15}$ and $\rho_{16}$ are anti-correlated 
 at each voxel, in the sense that $C_{15}$ and $C_{16}$, being very close to each other in gene space,
compete for the same signal. Refitting the whole model (using the full set of genes),
 but just one of the two medium spiny neurons present in the fitting panel, lead 
to densities of medium spiny neurons that are close to $\rho_{16}+\rho_{15}$.
 Here we kept all samples but refitted the models repeatedly to various 
sub-samples of the atlas, and therefore gave the two samples more chances
 to compete for the signal, hence a fuller profile for $\bar{\rho}_{15}$ than 
for $\rho_{15}$. The sub-sampling procedure is therefore a useful complement to 
 the full model, as it reveals a source of uncertainty in the predictions (high similarities
 between transcriptomes lead to negative correlations between results): 
 having $\rho_{15}$ and $\bar{\rho}_{15}$ yields an error bar on $\rho_{15}$
 at each voxel.\\
 
 As for the cholinergic neurons $t=13$ extracted from the striatum,
they are ranked $r^{signal}_13 = 25$, and the average profile is more localized
 in the medulla, and less in the pallidum, than the prediction $\rho_{13}$. Just 
as the one of $\rho_{13}$, the spatial profile $\bar{\rho}_{13}$ is quite complex, 
 spread over several brain regions, of which the striatum is far from being the 
most important.

\subsubsection{Pallidum}

Only one cell type (cholinergic projection neurons, $t=11$), was dissected from the pallidum.
Its predicted profile $\rho_{11}$ is  almost identically zero, and 
 the c.d.f. of overlaps with sub-sampled profiles therefore jumps quickly to 1,
 but the results in individual sub-samples are highly sparse and unstable,
 resulting in the (less sparse but still hardly conclusive) profile $\bar{\rho}_{11}$,
 see Figs.  \ref{cdfPlot11} and \ref{subSampledFour11}).

\subsubsection{Thalamus}

 Only one cell type (cholinergic projection neurons, PV+, $t=59$), was dissected from the thalamus
 (which region visibly begs for other cell types in the panel, some of which that would 
 probably be close in gene space to the Purkinje cells ($t=52$, see the first section
 and the discussion of the sub-sampling results in the cerebellum). The thalamus
 is not the top region by density in $\rho_{59}$, as it is ranked below midbrain.
 The larger contribution of the dorsal midbrain is also observed in the 
 average sub-sampled profile $\bar{\rho}_{59}$. About 60\% of the sub-samples
 have an overlap of at least 70\% with $\rho_{59}$.

\subsubsection{Midbrain}

 The A9 and A10 dopaminergic neurons (indices $t=4$ and $t=5$), have
a better visual contrast between ventral midbrain and its neighborhood
 in the average sub-samples $\bar{\rho}_4$ and $\bar{\rho}_5$
 than in $\rho_4$ and $\rho_5$ (this pair of transcriptome profiles
  has a high similarity in gene space, just as the pair of medium spiny neurons,
 hence the better anatomical properties of the average sub-sample, 
 which gives the profiles $C_4$ and $C_5$ more opportunities
 to compete for signal).\\

 The midbrain cholinergic neurons $t=10$ are ranked 13 out of 64 (60\%
of the sub-samples have an overlap of at least 70\% with the original
 profile $\rho_{10}$. The signal is more contrasted and more concentrated
 in the medulla than in the midbrain in $\bar{\rho}_{10}$, but in both cases
 pons, midbrain and medulla concentrate more than 90 \% of the signal.
Interestingly $\bar{\rho}_{10}$ could be visually mistaken for 
 $\bar{\rho}_{12}$. However, the two profiles differ by the component
 in midbrain, which is proper to $\bar{\rho}_{10}$. As both transcriptomes
 $t=10$ and $t=12$ correspond to cholinergic neurons, with $t=10$ extracted
 from the midbrain and $\rho_{12}$ from the medulla, this similarity 
 is consistent with prior biological knowledge.

\subsubsection{Medulla}
Only one cell type in our data set can be assigned to this brain region,
 as it was dissected from the brain stem (Motor Neurons, Cholinergic Interneurons, $t=12$ ).
 The corresponding profile $\rho_{12}$ returns indeed the medulla as the best guess
 for the anatomical origin of the sample (more than 65 \% of the signal is in the 
 medulla, and 30 \% in the neighboring pons). These scores 
 are conserved within 2\%  in the average profile $\bar{\rho}_{12}$ (see Fig. \ref{cdfPlot12}b,
 and the coronal sections through the top-region by density in Fig. \ref{subSampledFour12}a,b,c, which all
 cut through the medulla),
 and this cell type is ranked 7th out of 64 samples ($r^{signal}_{7} = 12$) 
 for the average overlap between sub-samples and the model, and 
 75\% of the sub-samples (resp. 48\%) have at least 
 an overlap of at least 80\% (resp. 90\%) between the
 model and the average sub-sample (see Fig. \ref{cdfPlot12}a).\\

 Moreover, the contrast between groups of voxels
 inside the medulla is stronger in the average sub-sample than in the original
 model, but singles out the same set of voxels, with three  connected 
components in the left hemisphere.\\

\subsubsection{Cerebellum}

The cerebellar cell types are ranked as follows by $r^{signal}$:\\
- rank 3, Purkinje cells, $t=1$,\\
- rank 11, granule cells, $t=20$,\\
- rank 14, astroglia, $t=28$, pattern close to the white-matter pattern (see the next subsection),\\
- rank 23, mature oligodendrocytes, $t=21$,\\
- rank 27, Purkinje cells, $t=52$, in which case $\bar{\rho}_{52}$ has more signal in the 
cerebellum than in the thalamus, hence a correct prediction of the cerebellum as the origin of 
 the sample,\\
- rank 31, stellate basket cells, $t=19$, in which more of the signal is localized in the cerebellum in the 
 average sub-sample than in the original density,\\
- rank 34, Golgi cells, $t=17$, where the medulla is the top region both in the 
 average sub-sample than in the original density,\\
- rank 43, unipolar brush cells (some Bergman glia), $t=18$,\\
- rank 49, Bergman glia, $t=27$, where the cerebellum is the top region
 by density in the average sub-sample $\bar{\rho}_{27}$, and not in $\rho_{27}$\\
- rank 60, Purkinje cells, $t=25$, for which the very low rank is a consequence of the 
 almost zero profile $\rho_{25}$, whereas the average profile $\bar{\rho}_{25}$ has the
 cerebellum as its top region by density,\\
- rank 63, mixed oligodendrocytes, $t=23$, where the original profile $\rho_{63}$ is zero
 and the average sub-sample rather amorphous.\\

\begin{table}
\centering
\input{tableDiscussionTypesCerebellumPre.tex}
\caption{{\bf{Table of cell types extracted from the cerebellum, ranked by average overlap with the 
 predicted profiles.}} The cell types 
whose profile is localized mostly in the 'Basic cell groups and regions' of the ARA are included in Table 
\ref{tableDiscussionTypesBasic}. The index called "rank" (first column), at a fixed value of $t$ (third column), denotes the integer
 $k$ such that $r_k^{signal}=t$.}
\label{tableDiscussionTypesCerebellum}
\end{table}

\subsubsection{"Basic cell groups and regions"}
\begin{figure}
\includegraphics[width=1\textwidth,keepaspectratio]{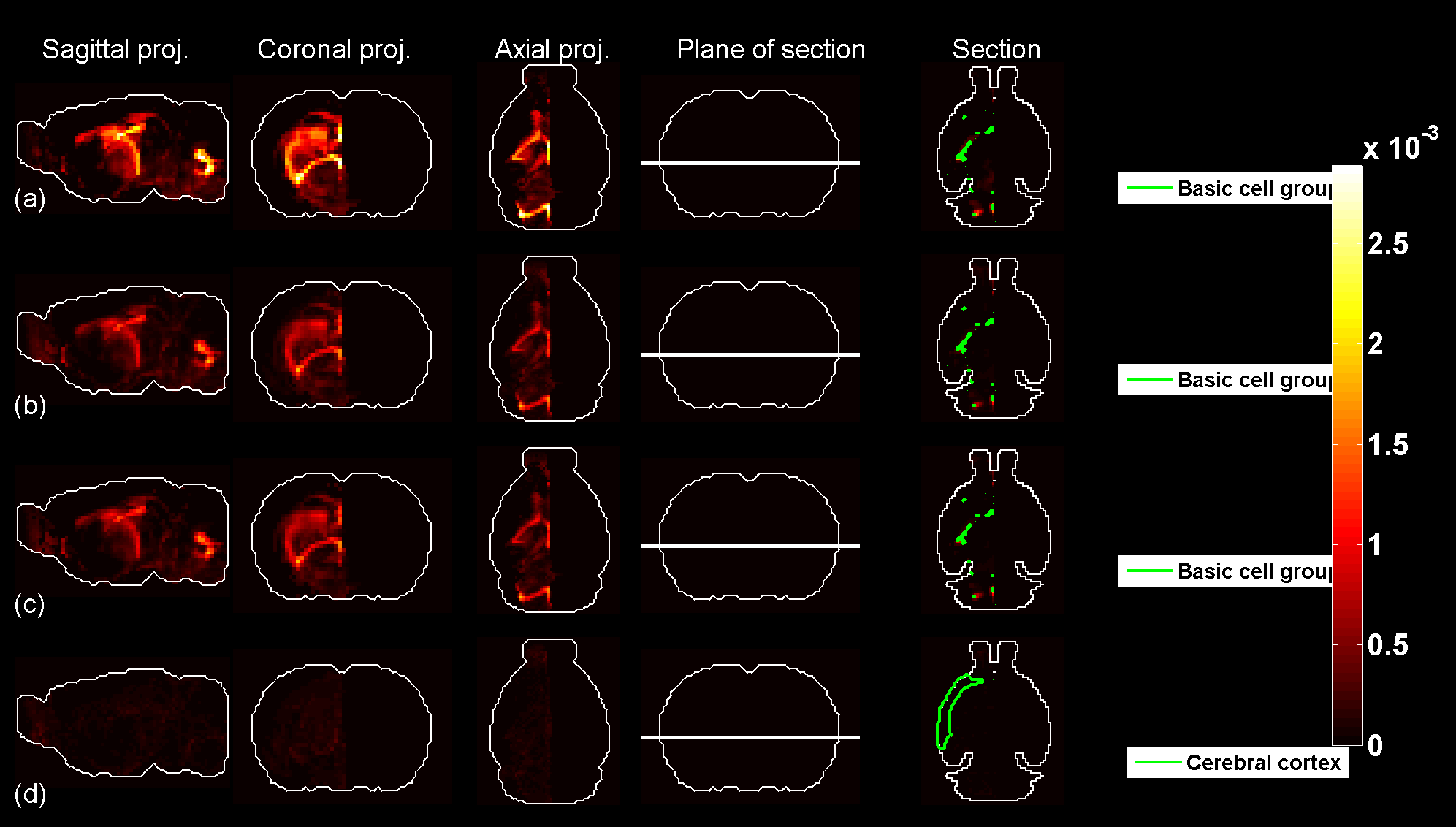}
\caption{{\bf{The sum of density profiles (labeled $t\in\mathcal{T}_{basic}$ of the cell types listed in Table \ref{tableDiscussionTypesBasic}).}}
(a) The sum of profiles in the original model, $\rho_{basic} = \sum_{t\in\mathcal{T}_{basic}} \rho_t$. (b) The sum of average sub-sampled profiles,
$\bar{\rho}_{basic} = \sum_{t\in\mathcal{T}_{basic}} \bar{\rho}_t$. (c) The part of $\bar{\rho}_{basic}$ in the voxels supporting $\rho_{basic}$, or
 $\bar{\rho}^{supported}_{basic}(v) = \sum_{t\in\mathcal{T}_{basic}} \bar{\rho}_t(v){\mathbf{1}}( \rho_{basic}( v ) > 0)$. (d) The difference 
$\bar{\rho}_{basic}-\bar{\rho}^{supported}_{basic}$.}
\label{figureSumBasic}
\end{figure}

 The set of voxels labeled "Basic cell groups of regions" in the ARA contains
 subcortical white matter and the {\emph{arbor vitae}}, and 
several cell types labeled as astrocytes and 
dissected from other regions of the brain, in particular the cerebral cortex
 (see Tables \ref{metadataAnatomyTable1} and \ref{metadataAnatomyTable2}), can be expected
 to be typical of white matter, and indeed a pattern coinciding
 with basic cell groups and regions (and containing the {\emph{arbor vitae}}),
 was noticed to appear for a number of cell types (see Table \ref{tableDiscussionTypesBasic}).
This pattern survives to a certain extent in the average sub-sampled
 profiles, which can be illustrated graphically by plotting
 the sum of the profiles for which "Basic cell groups and regions" is the top region 
 by density for in the original model, namely the cell types labeled $t\in \mathcal{T}_{basic}$
 defined by:
\begin{equation}
 t\in \mathcal{T}_{basic}\; \mathrm{iff} \;{\mathrm{argmax}}_{\omega\in{\mathrm{ARA}}}\left(\lambda_\omega(\rho_t)\right) = 
{\mathrm{'Basic\;cell\;groups'}}
\label{basicTypeDef}
\end{equation}
where $\lambda_\omega(\rho_t)$ is the localization score
 of $\rho_t$ in the brain region $\omega$:
\begin{equation}
\lambda_\omega(\rho_t) = \frac{\sum_{v\in\omega}\rho_t(v)^2}{\sum_{v\in{\mathrm{Brain\;Annotation}}}\rho_t(v)^2}.
\end{equation}
It can be observed on Fig. \ref{figureSumBasic} that the white-matter pattern
 of  'Basic cell groups' is still recognizable in the sum 
of average sub-sampled density profiles plotted in part (b) of the figure, which is almost entirely 
 supported by the support of the original model. Moreover, the contrast between voxels
 in 'Basic cell groups' is less pronounced in the average sub-sampled profile than in the original model
 (the intensity in the {\emph{arbor vitae}} is closer to the intensity in the more frontal component).\\

\begin{table}
\centering
\input{tableDiscussionTypesBasicPre.tex}
\caption{{\bf{Table of cell types for which the top region in the predicted profile $\rho$
is 'Basic cell groups and regions' in the ARA (they are defined by Eq. \ref{basicTypeDef}).}} They are extracted either from the cortex or the
 cerebellum. The index called "rank" (first column), at a fixed value of $t$ (third column), denotes the integer
 $k$ such that $r_k^{signal}=t$.}
\label{tableDiscussionTypesBasic}
\end{table}

\input{appendSplitting.tex}

\section{Numerical experiments with noise added}
 \subsection{Simulation scenario}
Having obtained estimates ${\bar{\mathcal{I}}}(t)$ for the agreement between 
 the estimated profiles and the average result from repeated sub-sampling,
we noticed that these estimates vary across cell types, and tend to be 
lower for transcriptome profiles that have very low density in the 
original model. We mentioned that taking 10 percent of the genes in the 
coronal atlas in each sub-sample probably results in a more severe loss 
 of CNS-specific genes than the one that is incurred by only taking into account 
 the $G=2,131$ in the coronal atlas, because the coronal atlas was 
 designed by prioritizing CNS-specific genes. However, we
 have not given a quantitative estimate of the severity of the sub-sampling
in terms of loss of data.\\
 In this section we estimate the impact of sub-sampling in terms 
of the corresponding intensity of a Gaussian noise added to cell-type-specific transcriptomes.
 To simulate the influence of noise,
 we mix the entries of the matrix $C$
 of cell-type-specific transcriptomes using a 
 random mixing matrix denoted by $M_\sigma$, and impose a positivity constraint to avoid 
 non-realistic negative entries (which are rare if the amplitude of the
 noise is small enough, but would penalize the corresponding cell types
 in the optimization):
\begin{equation}
  M_\sigma(t,s) = \delta_{st} + \sigma \Xi(t,s), 
\label{matrixEnsemble}
\end{equation}
where 
\begin{equation}
\Xi\sim \mathcal{N}(0,1)
\end{equation}
 and $\sigma$ models the 
   amplitude of the noise.   The original fitting panel corresponds to the case where $M$ is the 
 identity matrix. Mixing transcriptome profiles together under the influence of noise 
 amounts to replacing the matrix  $C$ by a matrix $C'_{M_\sigma}$ whose rows
 consist of linear combinations of the original transcriptomes:\\
\begin{equation}
C'_{M_\sigma}(t,g) =  {\mathrm{max}}\left(\sum_{s=1}^T M_\sigma(t,s) C(s,g),0\right).
\label{mixingEquation}
\end{equation}

\begin{figure}
\includegraphics[width=1\textwidth,keepaspectratio]{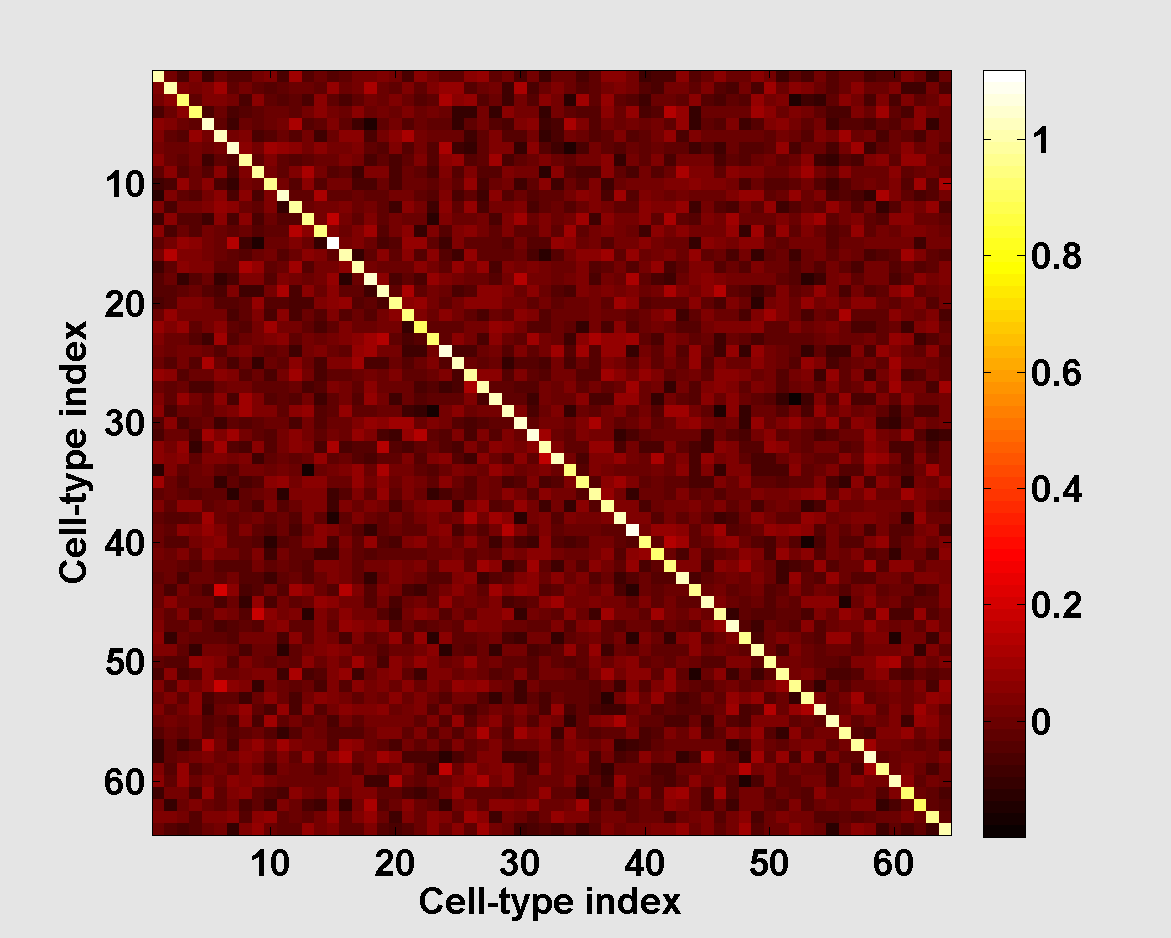}
\caption{{\bf{Heat map of a mixture matrix between cell types}}. It is the sum of the identity 
 matrix of size $T$, and of a Gaussian noise component, as described by Eq. \ref{matrixEnsemble},
 at $\sigma = 0.05$.  
A perfectly diagonal mixture matrix would correspond to the absence of noise (no mixing between cell types). 
 The color map shows that the mixture matrix is still close to the identity.}
\label{mixtureMatrix}
\end{figure}

We refitted the models at growing values of the noise parameter 
 $\sigma$, and found, not surprisingly, that the overlap with the 
 results of the original model is a decreasing
 function of $\sigma$. There is a regime of noise around $\sigma = 0.05$,
which gives rise to overlap values in the same range (with some transcriptomes
 close to zero while others are still close to $90\%$) as the 
 average sub-sample exposed in the previous section.
More precisely, we repeatedly drew random "mixing matrices" from the 
ensemble described by Eq. \ref{matrixEnsemble}, for $\sigma = 0.05$:
\begin{equation}
m_1,\dots,m_U \in \mathbf{R}(T,T),\; i.i.d, m_{i}\sim M_{\sigma}, \; \sigma = 0.05. 
\end{equation} 
 A heat map of one of these mixing matrices is shown in Fig. \ref{mixtureMatrix}.

Again we can summarize the simulations by a pseudocode:\\
{\ttfamily{for $i$ in [ 1..U ]}}\\
 $\;\;\;\;$  {\ttfamily{1. draw a $T$-by-$T$ random matrix $\Xi_i$ with independent, normally distributed entries;}}\\ 
 $\;\;\;\;$ {\ttfamily{2. compute the mixing matrix $m_i = {\mathbf{I}}_T + \sigma \Xi_i$ and the matrix $C'_{m_i}$}}\\
 $\;\;\;\;$  {\ttfamily{3. compute  $(\rho^{(i,noised)}_t(v))_{1\leq t \leq T}$ at each voxel:}}\\
$\;\;\;\;$     $(\rho^{(i,noised)}_t(v))_{1\leq t \leq T} = {\mathrm{argmin}}_{\phi\in\mathbf{R}_+^T}\left(||E(v,.)-\sum_t  C'_{m_i}(t, .)\phi(t)||^2\right).$\\
{\ttfamily{end}}\\
\subsection{Results}

We fitted the model for each of the corresponding matrices $C'_{m_i}$,
using the entire data set of $G$ genes:
\begin{equation}
 \left(\rho^{(i,noised)}_t(v)\right)_{1\leq t \leq T} = {\mathrm{argmin}}_{\nu\in\mathbf{R}_+^T}\left(   ||\sum_{g=1}^G( E(v,g) -\sum_{t=1}^TC'_{m_i}(t,g)\nu(t)||^2\right),
\end{equation}
 hence a family of $U$ random densities of $T$ cell types, for which one can compute
 the overlap with the results $\rho_t$ of the original model, as we did from the random  
 densities computed by sub-sampling the atlas (by substituting $\rho^{i,noised}_t$ to $\rho^{(s)}_t$ in 
\ref{overlapDef}.\\

\begin{figure}
\includegraphics[width=1\textwidth,keepaspectratio]{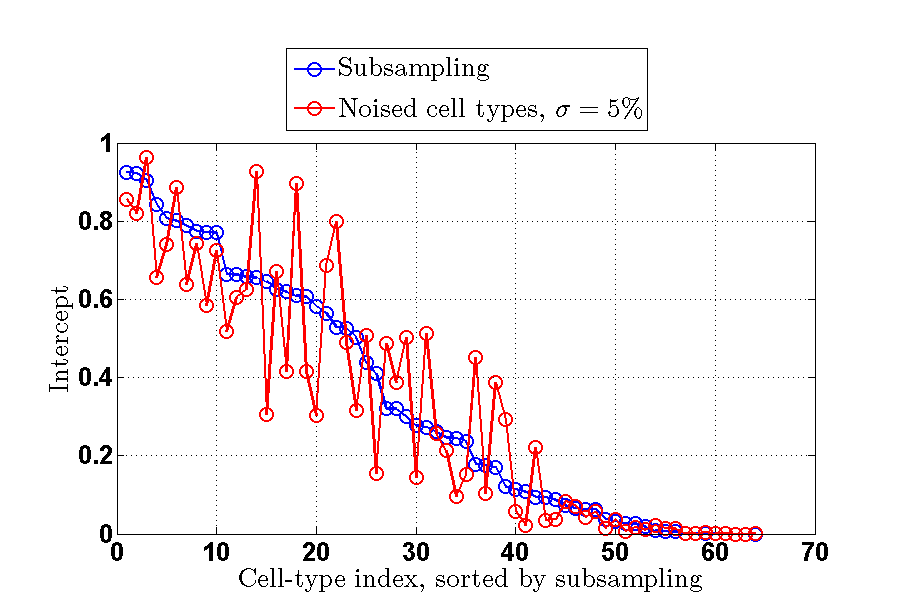}
\caption{{\bf{Overlaps with average sub-sampled densities (with 200 genes per sub-sample), and with average densities
 estimated from random mixings, with $\sigma=5\%$. }}The cell types at the low end of the distribution 
 are the same for both quantities.}
\label{noiseVersussub-sampling}
\end{figure}

 We computed the average cell-type-specific profiles noises as follows:
\begin{equation}
\bar{\rho}^{noised}_t(v) = \frac{1}{U}\sum_{i=1}^U \left(\rho^{i,noised}_t(v)\right)_{1\leq t \leq T}.
\label{meanNoised}
\end{equation}
The overlap between $\bar{\rho}^{noised}_t$ and $\rho_t$ can be computed for 
all indices $t$:
\begin{equation}
 \bar{{\mathcal{I}}}^{noised}(i,t):= \frac{1}{\sum_v  \bar{\rho}^{(i,noised)}_t(v) }\sum_v {\mathbf{1}}(\rho_t(v)>0) \bar{\rho}^{(i,noised)}_t(v) .
\label{overlapDefNoised}
\end{equation}
These quantities take values in the interval [0,1], and a new ranking 
of cell-type-specific transcriptome is induced by sorting them in decreasing 
 order. Is this ranking compatible with the one
induced by the sub-sampling procedure? To look into this 
 graphically, we plotted the values of the average noised overlap  $\bar{{\mathcal{I}}}^{noised} = \sum_{i=1}^U \bar{{\mathcal{I}}}^{noised}(i,t) /U$, ordered 
 according to the ranking of Eq. \ref{rankingSubsampling}, on the same graph as the 
 sorted overlaps with sub-sampled profiles (Fig. \ref{noiseVersussub-sampling}).
Even though the ranking is not exactly the same as the one induced from sub-sampling,
 this numerical experiment illustrates the fact that a mixture
 of cell types deviating from the diagonal by Gaussian random matrices at $\sigma = 5\%$ noise
 is empirically close to the result of sub-sampling to 10\% of the atlas.\\





\clearpage
\section{Tables: cell-type-specific transcriptomes}
\subsection{Description, labeling and anatomical origin of cell-type-specific transcriptomes}
For each of the cell-type-specific samples analyzed in this note, the
following two tables give a brief description of the cell type, the
region from which the samples were extracted according to the
coarsest version of the Allen Reference Atlas, and the
finest region to which it can be assigned according to the data
provided in the studies
\cite{OkatyCells,RossnerCells,CahoyCells,DoyleCells,ChungCells,ArlottaCells,HeimanCells,foreBrainTaxonomy}.
 The indices in the first columns of the tables are the ones refered to as $t$.
\begin{table}
\centering
\begin{tabular}{|m{0.05\textwidth}|m{0.39\textwidth}|m{0.18\textwidth}|m{0.38\textwidth}|}
\hline
\textbf{{\footnotesize{Index}}}&\textbf{Description}&\textbf{{\footnotesize{Region in the ARA (\bigTwelve)}}}&\textbf{Finest label in the ARA}\\\hline
1&Purkinje Cells&Cerebellum&       Cerebellar cortex \\\hline
2&Pyramidal Neurons&Cerebral cortex& Primary motor area; Layer 5 \\\hline
3&Pyramidal Neurons&Cerebral cortex& {\small{Primary somatosensory area; Layer 5}} \\\hline
4&A9 Dopaminergic Neurons&Midbrain& Substantia nigra\_ compact part \\\hline
5&A10 Dopaminergic Neurons&Midbrain& Ventral tegmental area \\\hline
6&Pyramidal Neurons&Cerebral cortex& Cerebral cortex; Layer 5 \\\hline
7&Pyramidal Neurons&Cerebral cortex&Cerebral cortex; Layer 5 \\\hline
8&Pyramidal Neurons&Cerebral cortex&Cerebral cortex; Layer 6\\\hline
9&Mixed Neurons&Cerebral cortex& Cerebral cortex \\\hline
10&{\footnotesize{Motor Neurons, Midbrain Cholinergic Neurons}}&Midbrain& Peduncolopontine nucleus\\\hline
11&Cholinergic Projection Neurons&Pallidum& Pallidum\_ ventral region\\\hline
12&{\footnotesize{Motor Neurons, Cholinergic Interneurons}}&Medulla& Spinal cord\\\hline
13&Cholinergic Neurons&Striatum&Striatum \\\hline
14&Interneurons&Cerebral cortex& Cerebral cortex\\\hline
15&Drd1+ Medium Spiny Neurons&Striatum& Striatum\\\hline
16&Drd2+ Medium Spiny Neurons&Striatum& Striatum\\\hline
17&Golgi Cells&Cerebellum& Cerebellar cortex\\\hline
18&Unipolar Brush cells (some Bergman Glia)&Cerebellum&Cerebellar cortex \\\hline
19&Stellate Basket Cells&Cerebellum& Cerebellar cortex\\\hline
20&Granule Cells&Cerebellum& Cerebellar cortex\\\hline
21&Mature Oligodendrocytes&Cerebellum& Cerebellar cortex\\\hline
22&Mature Oligodendrocytes&Cerebral cortex& Cerebral cortex\\\hline
23&Mixed Oligodendrocytes&Cerebellum& Cerebellar cortex\\\hline
24&Mixed Oligodendrocytes&Cerebral cortex& Cerebral cortex\\\hline
25&Purkinje Cells&Cerebellum& Cerebellar cortex\\\hline
26&Neurons&Cerebral cortex& Cerebral cortex\\\hline
27&Bergman Glia&Cerebellum& Cerebellar cortex\\\hline
28&Astroglia&Cerebellum& Cerebellar cortex\\\hline
29&Astroglia&Cerebral cortex&  Cerebral cortex\\\hline
30&Astrocytes&Cerebral cortex&  Cerebral cortex\\\hline
31&Astrocytes&Cerebral cortex& Cerebral cortex \\\hline
32&Astrocytes&Cerebral cortex& Cerebral cortex \\\hline
33&Mixed Neurons&Cerebral cortex& Cerebral cortex \\\hline
34&Mixed Neurons&Cerebral cortex& Cerebral cortex \\\hline
35&Mature Oligodendrocytes&Cerebral cortex& Cerebral cortex \\\hline
36&Oligodendrocytes&Cerebral cortex&  Cerebral cortex\\\hline
37&Oligodendrocyte Precursors&Cerebral cortex& Cerebral cortex \\\hline
\end{tabular}
\caption{Anatomical origin of the cell-type-specific samples (I).}
\label{metadataAnatomyTable1}
\end{table}

\begin{table}
\centering
\begin{tabular}{|m{0.05\textwidth}|m{0.39\textwidth}|m{0.18\textwidth}|m{0.38\textwidth}|}
\hline
\textbf{{\footnotesize{Index}}}&\textbf{Description}&\textbf{{\footnotesize{Region in the ARA (\bigTwelve)}}}&\textbf{Finest label in the ARA}\\\hline
38&\footnotesize{Pyramidal Neurons, Callosally projecting, P3}&Cerebral cortex& Cerebral cortex \\\hline
39&\footnotesize{Pyramidal Neurons, Callosally projecting, P6}&Cerebral cortex& Cerebral cortex  \\\hline
40&\footnotesize{Pyramidal Neurons, Callosally projecting, P14}&Cerebral cortex& Cerebral cortex\\\hline
41&\footnotesize{Pyramidal Neurons, Corticospinal, P3}&Cerebral cortex& Cerebral cortex \\\hline
42&\footnotesize{Pyramidal Neurons, Corticospinal, P6}&Cerebral cortex& Cerebral cortex \\\hline
43&\footnotesize{Pyramidal Neurons, Corticospinal, P14}&Cerebral cortex& Cerebral cortex \\\hline
44&\footnotesize{Pyramidal Neurons, Corticotectal, P14}&Cerebral cortex& Cerebral cortex \\\hline
45& Pyramidal Neurons&Cerebral cortex& Cerebral cortex, Layer 5 \\\hline
46& Pyramidal Neurons&Cerebral cortex& Cerebral cortex, Layer 5 \\\hline
47& Pyramidal Neurons&Cerebral cortex& {\small{Primary somatosensory area; Layer 5}} \\\hline
48&Pyramidal Neurons&Cerebral cortex& {\small{Prelimbic area and Infralimbic area; Layer 5 (Amygdala)}} \\\hline
49&Pyramidal Neurons&Hippocampal region& Ammon's Horn; Layer 6B\\\hline
50&Pyramidal Neurons&Cerebral cortex& Primary motor area\\\hline
51&{\footnotesize{Tyrosine Hydroxylase Expressing}}&Pons& Pontine central gray\\\hline
52&Purkinje Cells&Cerebellum& Cerebellar cortex \\\hline
53&\footnotesize{Glutamatergic Neuron (not well defined)}&Cerebral cortex& Cerebral cortex; Layer 6B (Amygdala)\\\hline
54&GABAergic Interneurons, VIP+&Cerebral cortex& Prelimbic area and Infralimbic area\\\hline
55&GABAergic Interneurons, VIP+&Cerebral cortex& Primary somatosensory area\\\hline
56&GABAergic Interneurons, SST+&Cerebral cortex&Prelimbic area and Infralimbic area \\\hline
57&GABAergic Interneurons, SST+&Hippocampal region& Ammon's Horn\\\hline
58&GABAergic Interneurons, PV+&Cerebral cortex& Prelimbic area and Infralimbic area\\\hline
59&GABAergic Interneurons, PV+&Thalamus& {\small{Dorsal part of the lateral geniculate complex}}\\\hline
60&GABAergic Interneurons, PV+, P7&Cerebral cortex& Primary somatosensory area \\\hline
61&GABAergic Interneurons, PV+, P10&Cerebral cortex& Primary somatosensory area\\\hline
62&\footnotesize{GABAergic Interneurons, PV+, P13-P15}&Cerebral cortex&Primary somatosensory area \\\hline
63&GABAergic Interneurons, PV+, P25&Cerebral cortex& Primary somatosensory area\\\hline
64&GABAergic Interneurons, PV+&Cerebral cortex& Primary motor area\\\hline
\end{tabular}
\caption{Anatomical origin of the cell-type-specific samples (II).}
\label{metadataAnatomyTable2}
\end{table}

\clearpage
\subsection{Cell-type-specific trascriptomes, ordered by overlap between 
 estimated density and average sub-sampled density}

\begin{table}
\begin{tabular}{|m{0.08\textwidth}|m{0.32\textwidth}|m{0.08\textwidth}|m{0.08\textwidth}|m{0.08\textwidth}|m{0.15\textwidth}|}
\hline
\textbf{rank by signal}&\textbf{Cell type}&\textbf{{\footnotesize{Index}} $t$}&\textbf{{\tiny{Overlap}} $\bar{\mathcal{I}}(t)$ (\%)}&\textbf{$p_{t, 0.75}$ (\%)}&\textbf{$\mathcal{I}_{thresh}(t,0.75)$ (\%)}\\\hline
1&\tiny{Pyramidal Neurons, Callosally projecting, P14}&40&94.8&98&98.8\\\hline
2&\tiny{Drd2+ Medium Spiny Neurons}&16&94.5&94.9&99.4\\\hline
3&\tiny{Purkinje Cells}&1&93.7&92.8&99.4\\\hline
4&\tiny{Pyramidal Neurons}&46&84.1&81.6&96.7\\\hline
5&\tiny{Tyrosine Hydroxylase Expressing}&51&87.5&84&97.3\\\hline
6&\tiny{GABAergic Interneurons, PV+}&64&82.9&76.3&95.5\\\hline
7&\tiny{Motor Neurons, Cholinergic Interneurons}&12&86.2&81.1&97.5\\\hline
8&\tiny{GABAergic Interneurons, SST+}&57&80.3&70.4&93\\\hline
9&\tiny{Pyramidal Neurons}&49&85.9&78.2&97.7\\\hline
10&\tiny{Pyramidal Neurons}&47&82&68.2&95.1\\\hline
11&\tiny{Granule Cells}&20&80.1&69.1&96.7\\\hline
12&\tiny{Pyramidal Neurons}&7&69.5&38.3&79.5\\\hline
13&\tiny{Motor Neurons, Midbrain Cholinergic Neurons}&10&70&48.4&83.8\\\hline
14&\tiny{Astroglia}&28&70.5&56.5&90.2\\\hline
15&\tiny{Pyramidal Neurons, Corticotectal, P14}&44&71.4&48.3&85.5\\\hline
16&\tiny{GABAergic Interneurons, PV+}&59&70.7&48.4&84.2\\\hline
17&\tiny{Pyramidal Neurons}&45&67.4&40.2&84.8\\\hline
18&\tiny{Mature Oligodendrocytes}&35&71.8&50.7&87.5\\\hline
19&\tiny{Pyramidal Neurons}&48&69.4&42.9&88.7\\\hline
20&\tiny{Glutamatergic Neuron (not well defined)}&53&62.5&38&83.8\\\hline
21&\tiny{GABAergic Interneurons, SST+}&56&60.5&28.2&76.8\\\hline
22&\tiny{Astrocytes}&31&61.6&36.2&83\\\hline
23&\tiny{Mature Oligodendrocytes}&21&59.2&27.6&77.3\\\hline
24&\tiny{Pyramidal Neurons}&6&66&42.1&85.4\\\hline
25&\tiny{Cholinergic Neurons}&13&49.6&16.3&67.2\\\hline
26&\tiny{Mature Oligodendrocytes}&22&53.4&25.9&76.8\\\hline
27&\tiny{Purkinje Cells}&52&39.7&6.5&52.3\\\hline
28&\tiny{Astrocytes}&30&40.8&7.9&54.1\\\hline
29&\tiny{A10 Dopaminergic Neurons}&5&40.5&9.6&59.8\\\hline
30&\tiny{A9 Dopaminergic Neurons}&4&39.5&15.4&60.2\\\hline
31&\tiny{Stellate Basket Cells}&19&33.2&3.2&43.4\\\hline
32&\tiny{GABAergic Interneurons, VIP+}&55&34.1&13.6&55.9\\\hline
\end{tabular}
\caption{Ranking of cell types by overlap and agreement for top region (I).}
\label{tableCompare1}
\end{table}

\begin{table}
\begin{tabular}{|m{0.08\textwidth}|m{0.32\textwidth}|m{0.08\textwidth}|m{0.12\textwidth}|m{0.08\textwidth}|m{0.15\textwidth}|}
\hline
\textbf{rank by signal}&\textbf{Cell type}&\textbf{{\footnotesize{Index}} $t$}&\textbf{{\tiny{Overlap}} $\bar{\mathcal{I}}(t)$ (\%)}&\textbf{$p_{t, 0.75}$ (\%)}&\textbf{$\mathcal{I}_{thresh}(t,0.75)$ (\%)}\\\hline
33&\tiny{GABAergic Interneurons, VIP+}&54&28.7&2.7&40.2\\\hline
34&\tiny{Golgi Cells}&17&27.7&5.1&39.5\\\hline
35&\tiny{Mixed Neurons}&34&35&10.4&52.1\\\hline
36&\tiny{Pyramidal Neurons, Corticospinal, P14}&43&28.1&4.3&37.7\\\hline
37&\tiny{Oligodendrocyte Precursors}&37&24.7&1.7&34\\\hline
38&\tiny{Pyramidal Neurons, Callosally projecting, P3}&38&22.4&1.9&35.9\\\hline
39&\tiny{Drd1+ Medium Spiny Neurons}&15&19.3&1.4&27.9\\\hline
40&\tiny{Astrocytes}&32&18.4&0.5&25.6\\\hline
41&\tiny{Pyramidal Neurons}&8&13.6&3.2&17\\\hline
42&\tiny{Pyramidal Neurons}&50&12.8&1.3&16\\\hline
43&\tiny{Unipolar Brush cells (some Bergman Glia)}&18&11.6&0.4&16.8\\\hline
44&\tiny{GABAergic Interneurons, PV+, P7}&60&12.8&1.5&20.3\\\hline
45&\tiny{Mixed Neurons}&33&11.8&2.4&14.3\\\hline
46&\tiny{Oligodendrocytes}&36&8.4&-0.2&12.3\\\hline
47&\tiny{GABAergic Interneurons, PV+, P25}&63&12.7&0.4&15\\\hline
48&\tiny{Pyramidal Neurons, Callosally projecting, P6}&39&13&1.5&17.4\\\hline
49&\tiny{Bergman Glia}&27&6.6&0.5&6.8\\\hline
50&\tiny{Astroglia}&29&6.4&0.7&6.4\\\hline
51&\tiny{Pyramidal Neurons, Corticospinal, P6}&42&7.7&1.9&7.4\\\hline
52&\tiny{GABAergic Interneurons, PV+}&58&3.1&0&2.9\\\hline
53&\tiny{GABAergic Interneurons, PV+, P10}&61&3.4&0.2&1.2\\\hline
54&\tiny{Pyramidal Neurons}&3&3&0&0.4\\\hline
55&\tiny{Pyramidal Neurons}&2&1.4&0&0.2\\\hline
56&\tiny{Pyramidal Neurons, Corticospinal, P3}&41&0.8&-0.2&0.2\\\hline
57&\tiny{Neurons}&26&0.4&-0.2&0.2\\\hline
58&\tiny{GABAergic Interneurons, PV+, P13-P15}&62&0.1&-0.2&0.2\\\hline
59&\tiny{Interneurons}&14&0&-0.2&0.2\\\hline
60&\tiny{Purkinje Cells}&25&0.2&0&0.2\\\hline
61&\tiny{Mixed Neurons}&9&0.3&-0.2&0.2\\\hline
62&\tiny{Cholinergic Projection Neurons}&11&0.1&-0.2&0.2\\\hline
63&\tiny{Mixed Oligodendrocytes}&23&0&22.4&72.3\\\hline
64&\tiny{Mixed Oligodendrocytes}&24&0&22.4&72.3\\\hline
\end{tabular}
\caption{Ranking of cell types by overlap and agreement for top region (II).}
\label{tableCompare2}
\end{table}

%% file: tableDiscussionTypesCortexPre.tex
\begin{tabular}{|l|l|l|l|}
\hline
\textbf{Rank}&\textbf{Cell type}&\textbf{Index $t$}&\textbf{Overlap $\bar{\mathcal{I}}(t)$ (\%)}\\\hline
1&\tiny{Pyramidal Neurons, Callosally projecting, P14}&40&94.1\\\hline
4&\tiny{Pyramidal Neurons}&46&84.5\\\hline
6&\tiny{GABAergic Interneurons, PV+}&64&80.7\\\hline
10&\tiny{Pyramidal Neurons}&47&76.2\\\hline
12&\tiny{Pyramidal Neurons}&7&67.3\\\hline
15&\tiny{Pyramidal Neurons, Corticotectal, P14}&44&66\\\hline
17&\tiny{Pyramidal Neurons}&45&64.8\\\hline
21&\tiny{GABAergic Interneurons, SST+}&56&55.6\\\hline
24&\tiny{Pyramidal Neurons}&6&53.6\\\hline
28&\tiny{Astrocytes}&30&32.7\\\hline
32&\tiny{GABAergic Interneurons, VIP+}&55&25.5\\\hline
33&\tiny{GABAergic Interneurons, VIP+}&54&25.3\\\hline
36&\tiny{Pyramidal Neurons, Corticospinal, P14}&43&18.8\\\hline
37&\tiny{Oligodendrocyte Precursors}&37&17.4\\\hline
38&\tiny{Pyramidal Neurons, Callosally projecting, P3}&38&16.7\\\hline
41&\tiny{Pyramidal Neurons}&8&10.7\\\hline
42&\tiny{Pyramidal Neurons}&50&9.4\\\hline
44&\tiny{GABAergic Interneurons, PV+, P7}&60&8.5\\\hline
45&\tiny{Mixed Neurons}&33&8\\\hline
47&\tiny{GABAergic Interneurons, PV+, P25}&63&7.3\\\hline
48&\tiny{Pyramidal Neurons, Callosally projecting, P6}&39&7.1\\\hline
51&\tiny{Pyramidal Neurons, Corticospinal, P6}&42&3.4\\\hline
52&\tiny{GABAergic Interneurons, PV+}&58&1.8\\\hline
53&\tiny{GABAergic Interneurons, PV+, P10}&61&1.5\\\hline
54&\tiny{Pyramidal Neurons}&3&1\\\hline
55&\tiny{Pyramidal Neurons}&2&0.6\\\hline
56&\tiny{Pyramidal Neurons, Corticospinal, P3}&41&0.3\\\hline
57&\tiny{Neurons}&26&0.2\\\hline
58&\tiny{GABAergic Interneurons, PV+, P13-P15}&62&0.1\\\hline
59&\tiny{Interneurons}&14&0.1\\\hline
61&\tiny{Mixed Neurons}&9&0\\\hline
64&\tiny{Mixed Oligodendrocytes}&24&0\\\hline
\end{tabular}

%% file: tableDiscussionTypesCerebellumPre.tex
\begin{tabular}{|l|l|l|l|}
\hline
\textbf{Rank}&\textbf{Cell type}&\textbf{Index $t$}&\textbf{Overlap $\bar{\mathcal{I}}(t)$ (\%)}\\\hline
3&\tiny{Purkinje Cells}&1&91\\\hline
11&\tiny{Granule Cells}&20&73.1\\\hline
23&\tiny{Mature Oligodendrocytes}&21&53.6\\\hline
27&\tiny{Purkinje Cells}&52&33.1\\\hline
31&\tiny{Stellate Basket Cells}&19&25.9\\\hline
34&\tiny{Golgi Cells}&17&24.6\\\hline
43&\tiny{Unipolar Brush cells (some Bergman Glia)}&18&9.1\\\hline
49&\tiny{Bergman Glia}&27&3.8\\\hline
60&\tiny{Purkinje Cells}&25&0\\\hline
63&\tiny{Mixed Oligodendrocytes}&23&0\\\hline
\end{tabular}

%% file: tableDiscussionTypesBasicPre.tex
\begin{tabular}{|l|l|l|l|}
\hline
\textbf{Rank}&\textbf{Cell type}&\textbf{Index $t$}&\textbf{Overlap $\bar{\mathcal{I}}(t)$ (\%)}\\\hline
14&\tiny{Astroglia}&28&67.2\\\hline
18&\tiny{Mature Oligodendrocytes}&35&63.5\\\hline
22&\tiny{Astrocytes}&31&54.8\\\hline
26&\tiny{Mature Oligodendrocytes}&22&44\\\hline
35&\tiny{Mixed Neurons}&34&23.6\\\hline
40&\tiny{Astrocytes}&32&13.3\\\hline
46&\tiny{Oligodendrocytes}&36&7.2\\\hline
50&\tiny{Astroglia}&29&3.6\\\hline
\end{tabular}

%% file: appendSplitting.tex
\subsection{Repeated sub-sampling of gene space (II): random splitting of the data set in two equal parts}
So far we have obtained a ranking of cell types based on random sub-sampling
that provided error estimates from the loss of 90 \% of data, which proportion we chose 
 to replicate the scaling of our data set with respect to the full genome of the mouse.
 An alternative, more symmetric sub-sampling procedure is the following: 
repeated splitting of the data set into two sets of genes of equal sizes,
followed by reftitting of the model in each of the resulting sub-samples,
 alllows to study the probability of detecting a cell type at each voxel (conditional on the detection of it 
 from the symmetric sub-sample).\\

 The following pseudo-code helps establish notations:\\
{\ttfamily{
for j in [1..J]\\
draw a random permutation $\sigma$ of [1..G] (without replacing);\\
construct the following matrices extracted from the Allen Atlas and cell-type-specific transcriptomes:\\
fit the model to the two sets of genes $\{g_{\sigma_1},\dots,g_{\sigma_{[G/2]}}\}$ and
$\{g_{\sigma_{[G/2]+1}},\dots,g_{\sigma_G}\}$:\\
\begin{equation}
E^{(j,1)} = \left(E(v, g_{\sigma_i}\right)_{1\leq v \leq V,1\leq i \leq [G/2]},\;E^{(j,2)} = \left(E(v, g_{\sigma_i}\right)_{1\leq v \leq V,[G/2]+1\leq i \leq G},
\end{equation}
\begin{equation}
C^{(j,1)} = \left(C(t, g_{\sigma_i}\right)_{1\leq t \leq T,1\leq i \leq [G/2]},\;
C^{(j,2)} = \left(C(t, g_{\sigma_i}\right)_{1\leq t \leq T,[G/2]+1\leq i \leq G},
\end{equation}
fit the model to these two data sets
\begin{equation}
(\rho^{(j,1)}_t(v))_{1\leq t \leq T} = {\mathrm{argmin}}_{\phi\in\mathbf{R}_+^T}\left(E^{(j,1)}(v,.)-\sum_t \phi_t C^{(j,1)}(t, .)\right).
\end{equation}
\begin{equation}
(\rho^{(j,2)}_t(v))_{1\leq t \leq T} = {\mathrm{argmin}}_{\phi\in\mathbf{R}_+^T}\left(E^{(j,2)}(v,.)-\sum_t \phi_t C^{(j,2)}(t, .)\right).
\end{equation}
end\\
}}

This approach has one less parameter than the sub-sampling procedure described above,
and the fact that the sub-samples all have the same size (and come in pairs), invites us to 
 estimate stability properties of the results in terms of conditional probabilities. For each voxel labeled $v$ and each cell type
 labeled $t$, there are two random variables, 
 namely $\rho^{(j,1)}_t( v )$ and $\rho^{(j,2)}_t( v )$, corresponding to the random permutation of genes
 labeled $j$. The cell-type label $t$ is detected at voxel $v$ (an element of the set $S$ in the terminology of \cite{Meinshausen2013}, see Eq. \ref{SDef})
 in sub-sample $(j,1)$ (resp. $(j,2)$) if $\rho^{(j,1)}_t( v ) > 0$ (resp. $\rho^{(j,2)}_t( v ) > 0$.
  We can compute the probability of cell type $t$ being a detected at voxel $v$ in a
 sub-sample $(j,1)$
conditional on the probability of it being detected in the complementary sub-sample $(j,2)$:
\begin{equation}
\mathcal{P}_{12}(t,v) = \frac{|j \in [1..J], \rho^{(j,1)}_t( v )\rho^{(j,2)}_t( v )> 0 |}{|j \in [1..J], \rho^{(j,2)}_t( v )> 0 |}
\label{condiProba12}
\end{equation}
\begin{equation}
\mathcal{P}_{21}(t,v) = \frac{|j \in [1..J], \rho^{(j,1)}_t( v )\rho^{(j,2)}_t( v )> 0 |}{|j \in [1..J], \rho^{(j,1)}_t( v )> 0 |}
\label{condiProba21}
\end{equation}
 If $J$ equalled the total number of permutations of $[1..G]$, the two quantities $\mathcal{P}_{12}$ 
 and $\mathcal{P}_{12}$ defined above 
would be equal for symmetry reasons, but for small values of $J$ they will be different (but hopefully 
close), and we will evaluate both of them.  For a fixed value of $t$, they can be plotted 
 as brain-wide profiles (or hemisphere-wide profiles, as will be the case again for the sake of maximizing the
 number of samples in limited time, again assuming sufficient degree of left-right symmetry), in order 
 to give a visual impression of the voxels at which the cell-type-specific transcriptome labeled $t$ is
a predicting variable. Again, the visual impression of these profiles should be close to the 
 one of $\rho_t$. Moreover, the probabilistic nature of the entries of $\mathcal{P}_{12}$ provides
 confidence intervals at each voxel for wich $\rho_t$ is strictly positive. Thresholding $\mathcal{P}_{12}$ 
at a fixed threshold between $0$ and $1$ yields a mask that can be applied to $\rho_t$ in
 order to keep only the signal that corresponds to predicting variables 
 with probability larger than the threshold.\\

 Going through the plots one by one, one notices than indeed the visual impression is
conserved, and that there seems to be a positive correlation between the values of the conditional probabilities
 and the values of the density profiles in the original model (the thresholded profiles defined 
 above tend to coincide with areas of strongest signal for the cell types that
 were distinguished for their striking density patterns). To test this idea, let us compute the correlation 
 coefficients between conditional predictor densities and density profiles for each cell types:\\
\begin{equation}
\mathrm{Corr}_{12}(t) = {\mathrm{corr}}\left( \rho_t, \mathcal{P}_{12}(t,.)  \right),\;\;\;\mathrm{Corr}_{21}(t) = {\mathrm{corr}}\left( \rho_t, \mathcal{P}_{21}(t,.)  \right)
\label{corrProbaInt}
\end{equation}
 The values are quite large indeed for most cell types (see Fig.  \ref{corrProbaDensity}).
 This suggests a thresholding procedure of voxels for each cell 
 type based on the conditional probabilities and a chosen threshold $\eta$ on conditional probabilities:
\begin{equation}
\rho^{thresh,\eta}_t(v) = \rho_t(v) \mathbf{1}\left( {\mathrm{min}}({P}_{12}(t,v),{P}_{21}(t,v) )\geq \eta \right)
\end{equation}
 For graphical presentation in the figures of section 7, where the threshold $\eta = 99\%$ and $\eta=75\%$ are used.



%% file: appendSubsampleFigures.tex
\clearpage
\section{Figures: predicted and sub-sampled profiles: statistics of overlaps and localization scores in the ARA,
 and visualization}

\clearpage
\begin{figure}
\includegraphics[width=1\textwidth,keepaspectratio]{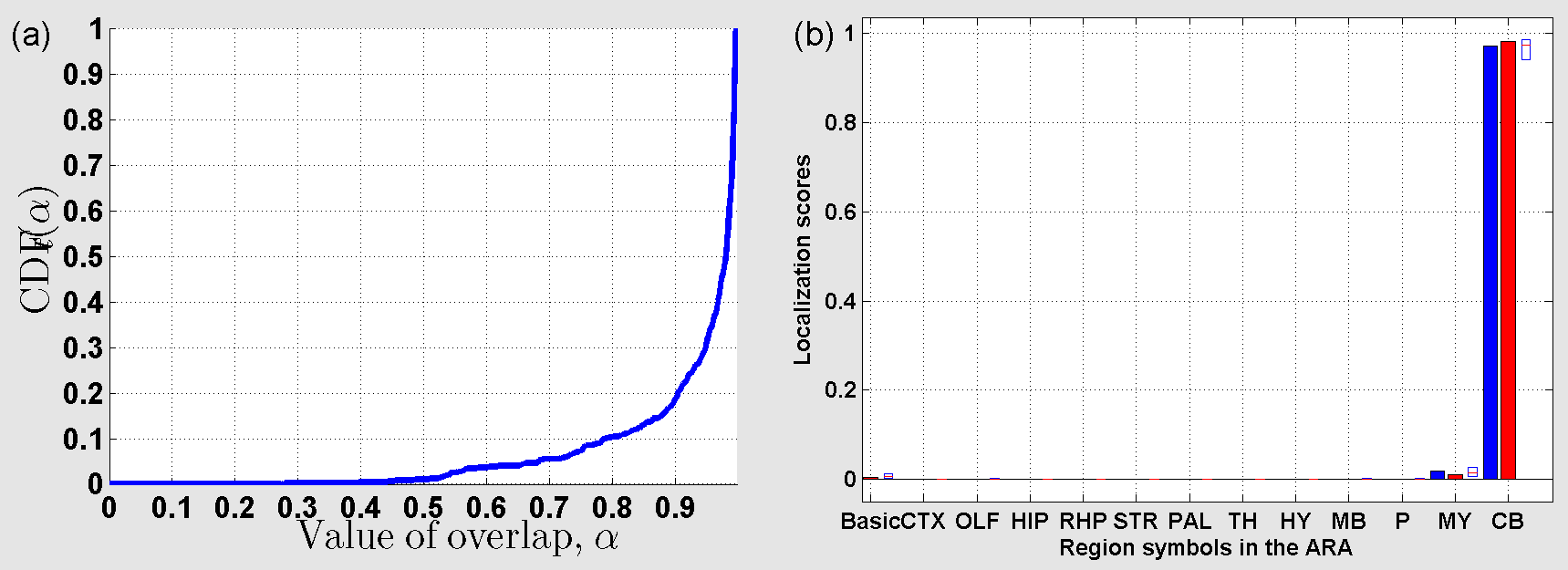}
\caption{(a) Cumulative distribution function (${\mathrm{\sc{CDF}}}_t$) of the overlap between $\rho_t$ and
 sub-sampled profiles for $t=1$. (b) Localization scores in the coarsest version of the ARA for $\rho_t$ (in blue), and 
 $\bar{\rho}_t$ (in red).}
\label{cdfPlot1}
\end{figure}
\begin{figure}
\includegraphics[width=1\textwidth,keepaspectratio]{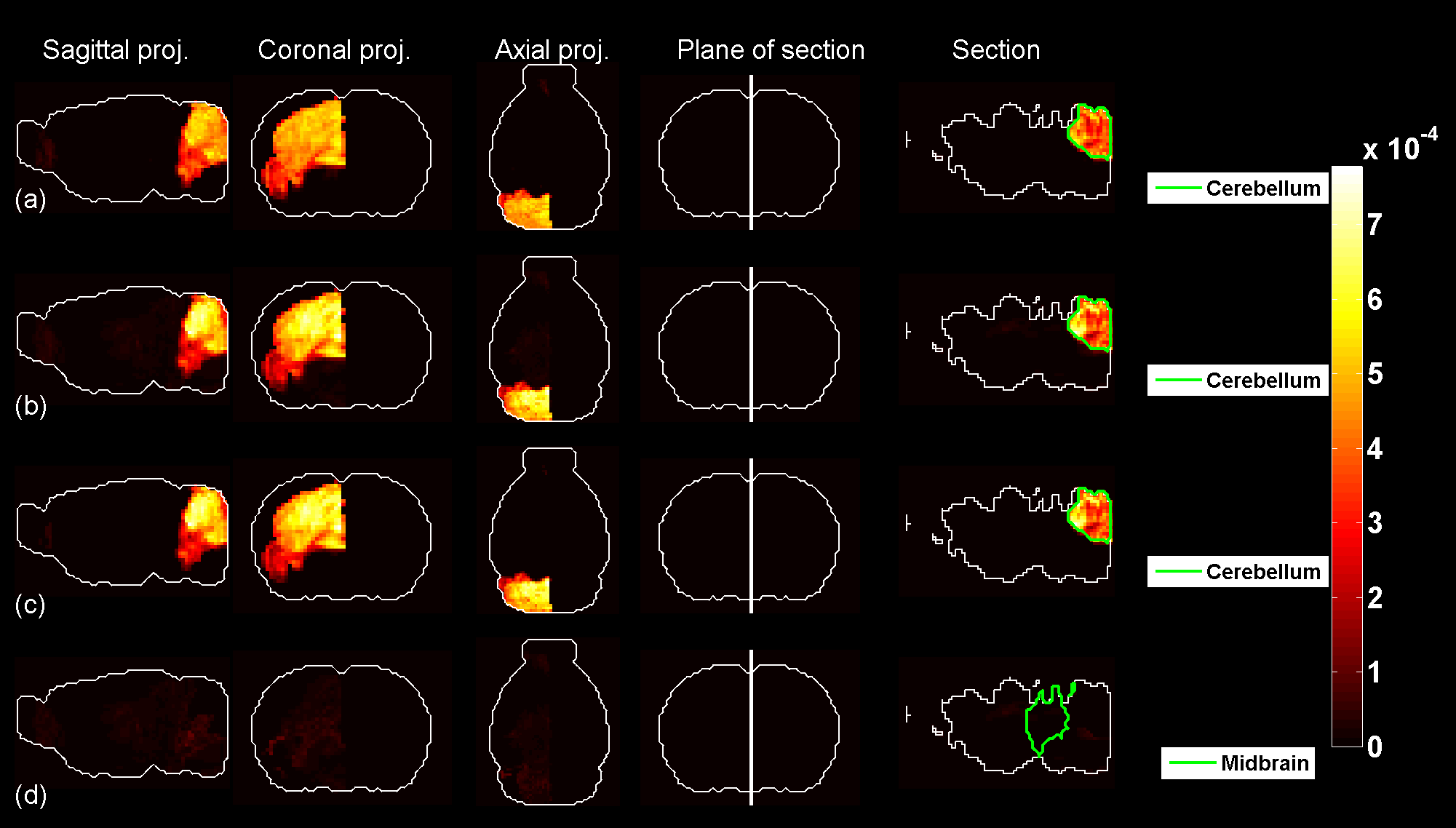}
\caption{Predicted profile and average sub-sampled profile for $t=1$.}
\label{subSampledFour1}
\end{figure}
\clearpage
\begin{figure}
\includegraphics[width=1\textwidth,keepaspectratio]{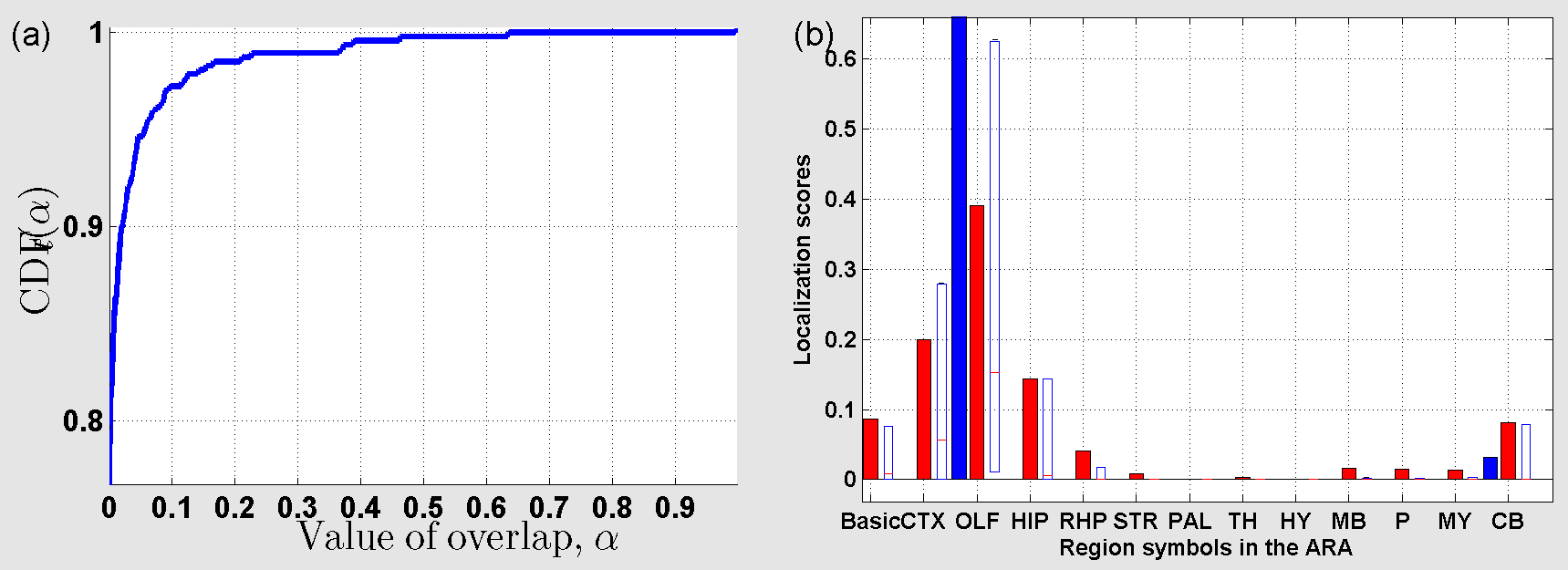}
\caption{(a) Cumulative distribution function (${\mathrm{\sc{CDF}}}_t$) of the overlap between $\rho_t$ and
 sub-sampled profiles for $t=2$. (b) Localization scores in the coarsest version of the ARA for $\rho_t$ (in blue), and 
 $\bar{\rho}_t$ (in red).}
\label{cdfPlot2}
\end{figure}
\begin{figure}
\includegraphics[width=1\textwidth,keepaspectratio]{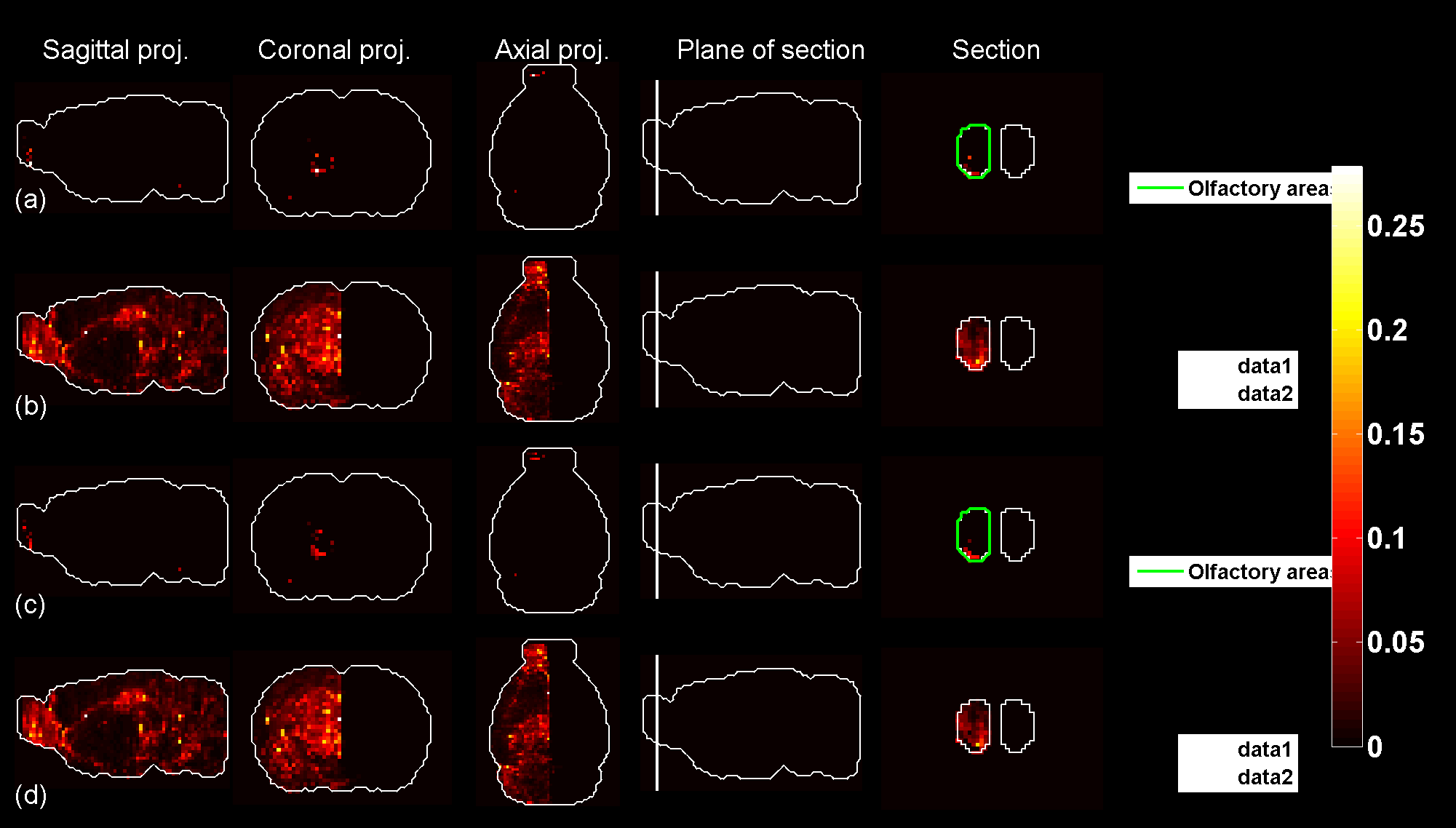}
\caption{Predicted profile and average sub-sampled profile for $t=2$.}
\label{subSampledFour2}
\end{figure}
\clearpage
\begin{figure}
\includegraphics[width=1\textwidth,keepaspectratio]{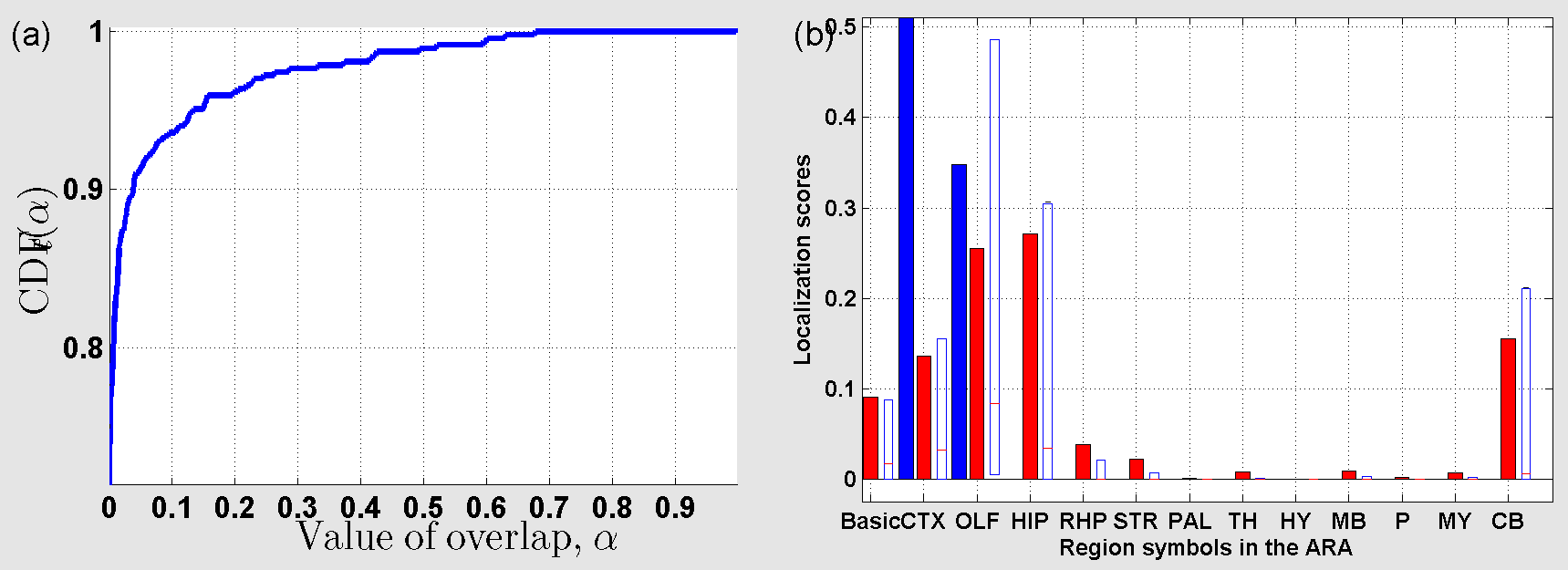}
\caption{(a) Cumulative distribution function (${\mathrm{\sc{CDF}}}_t$) of the overlap between $\rho_t$ and
 sub-sampled profiles for $t=3$. (b) Localization scores in the coarsest version of the ARA for $\rho_t$ (in blue), and 
 $\bar{\rho}_t$ (in red).}
\label{cdfPlot3}
\end{figure}
\begin{figure}
\includegraphics[width=1\textwidth,keepaspectratio]{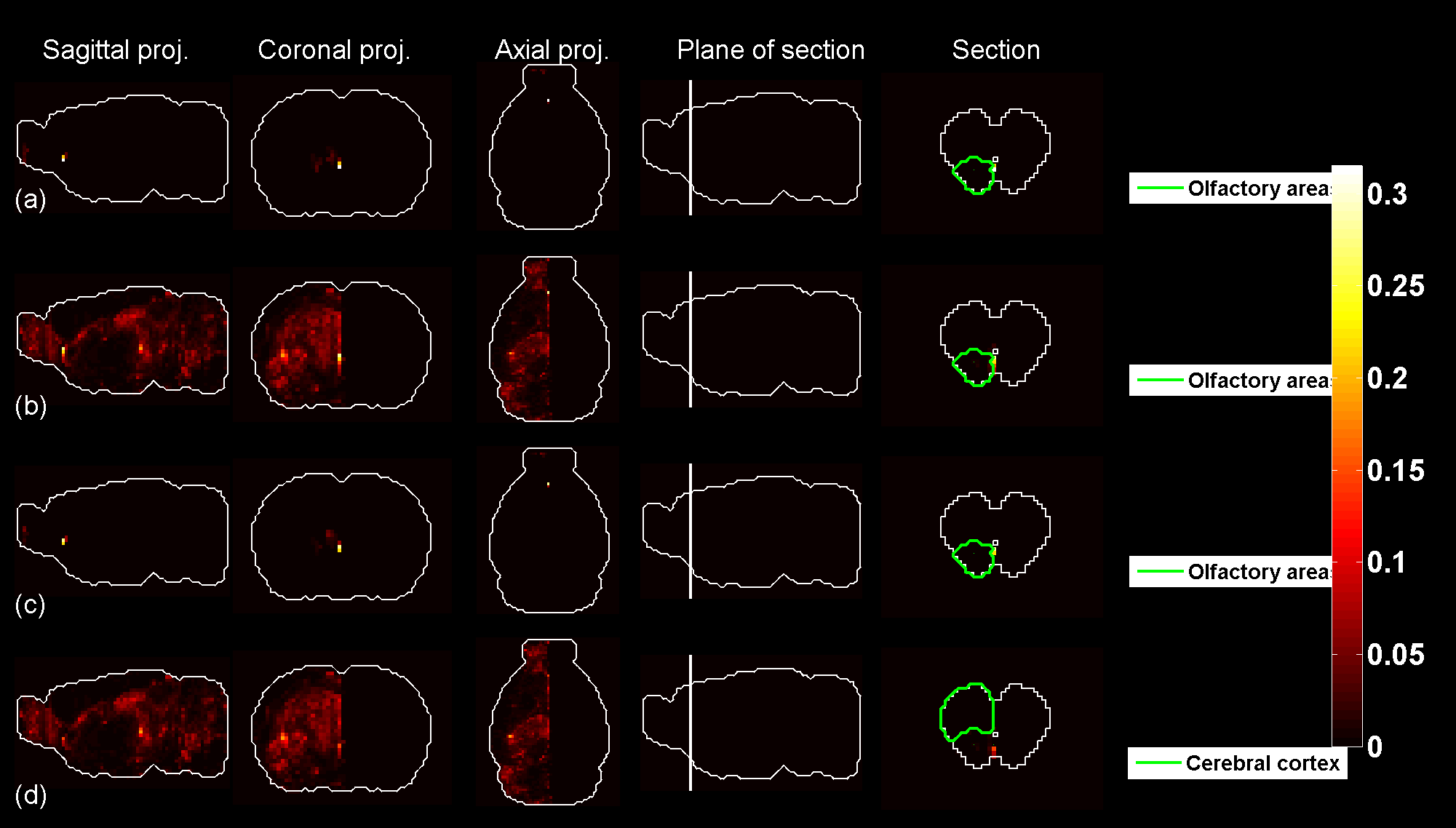}
\caption{Predicted profile and average sub-sampled profile for $t=3$.}
\label{subSampledFour3}
\end{figure}
\clearpage
\begin{figure}
\includegraphics[width=1\textwidth,keepaspectratio]{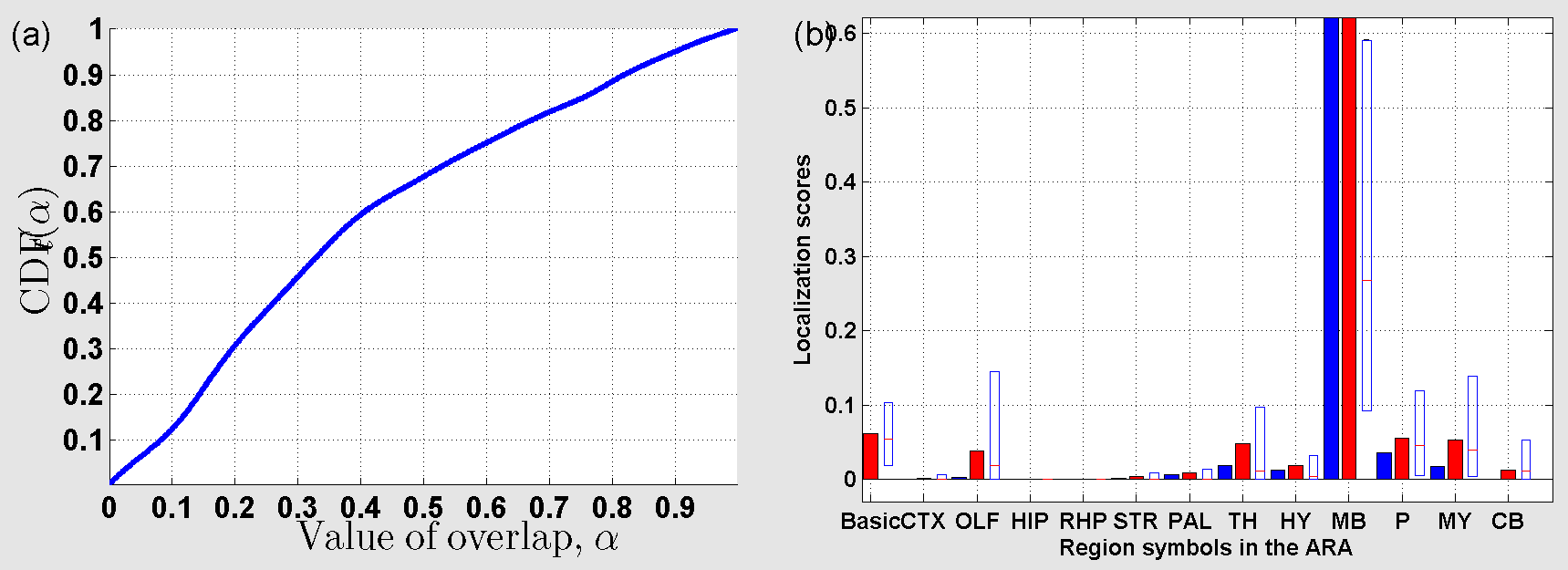}
\caption{(a) Cumulative distribution function (${\mathrm{\sc{CDF}}}_t$) of the overlap between $\rho_t$ and
 sub-sampled profiles for $t=4$. (b) Localization scores in the coarsest version of the ARA for $\rho_t$ (in blue), and 
 $\bar{\rho}_t$ (in red).}
\label{cdfPlot4}
\end{figure}
\begin{figure}
\includegraphics[width=1\textwidth,keepaspectratio]{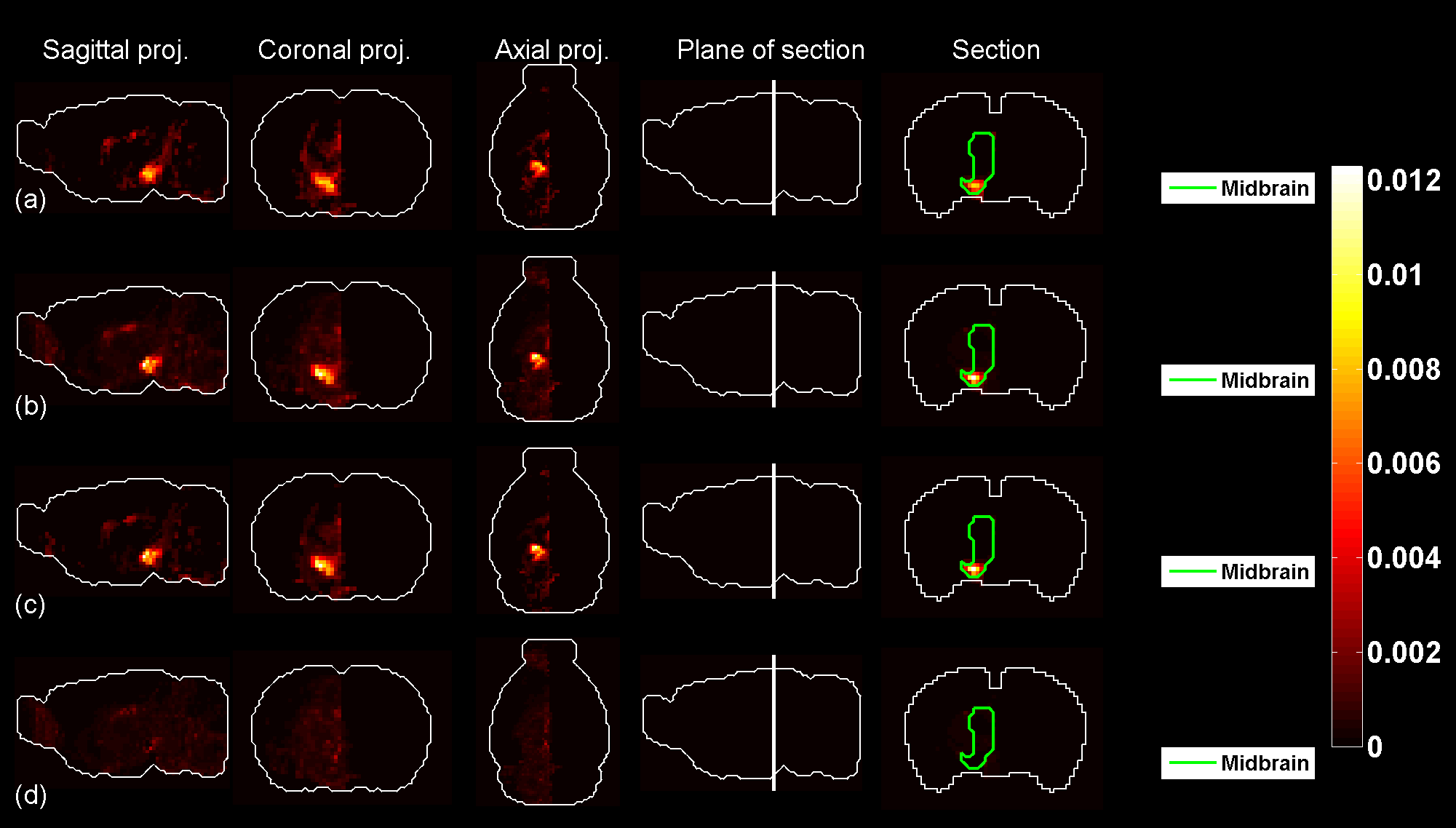}
\caption{Predicted profile and average sub-sampled profile for $t=4$.}
\label{subSampledFour4}
\end{figure}
\clearpage
\begin{figure}
\includegraphics[width=1\textwidth,keepaspectratio]{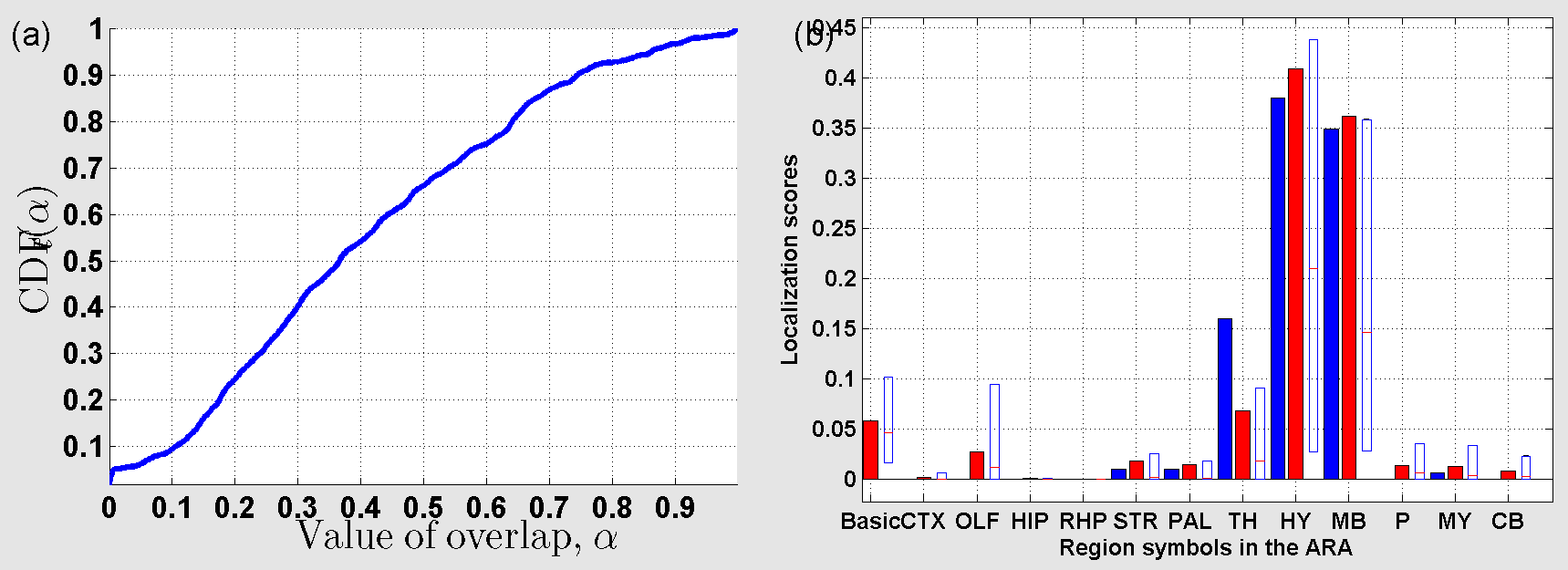}
\caption{(a) Cumulative distribution function (${\mathrm{\sc{CDF}}}_t$) of the overlap between $\rho_t$ and
 sub-sampled profiles for $t=5$. (b) Localization scores in the coarsest version of the ARA for $\rho_t$ (in blue), and 
 $\bar{\rho}_t$ (in red).}
\label{cdfPlot5}
\end{figure}
\begin{figure}
\includegraphics[width=1\textwidth,keepaspectratio]{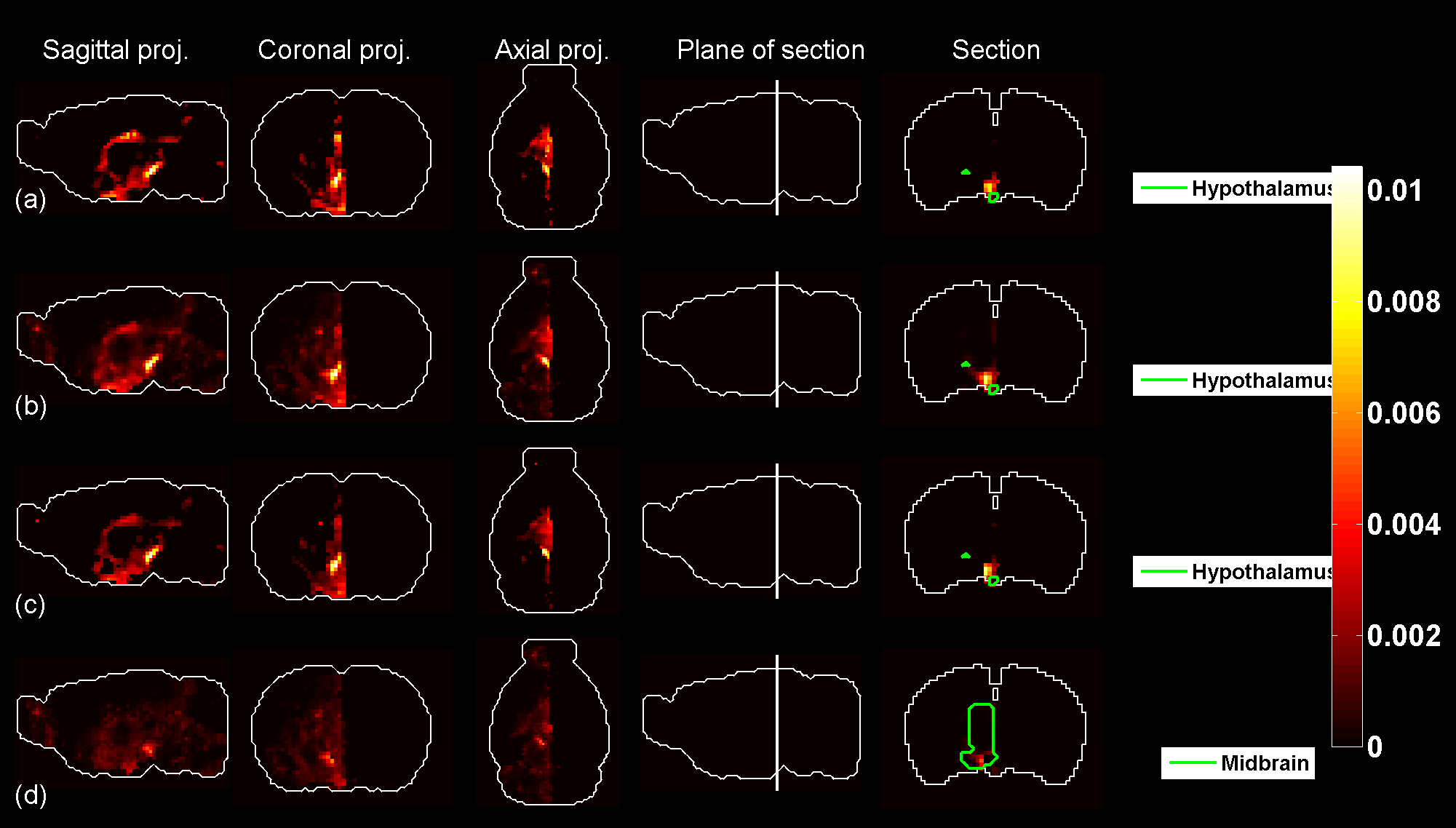}
\caption{Predicted profile and average sub-sampled profile for $t=5$.}
\label{subSampledFour5}
\end{figure}
\clearpage
\begin{figure}
\includegraphics[width=1\textwidth,keepaspectratio]{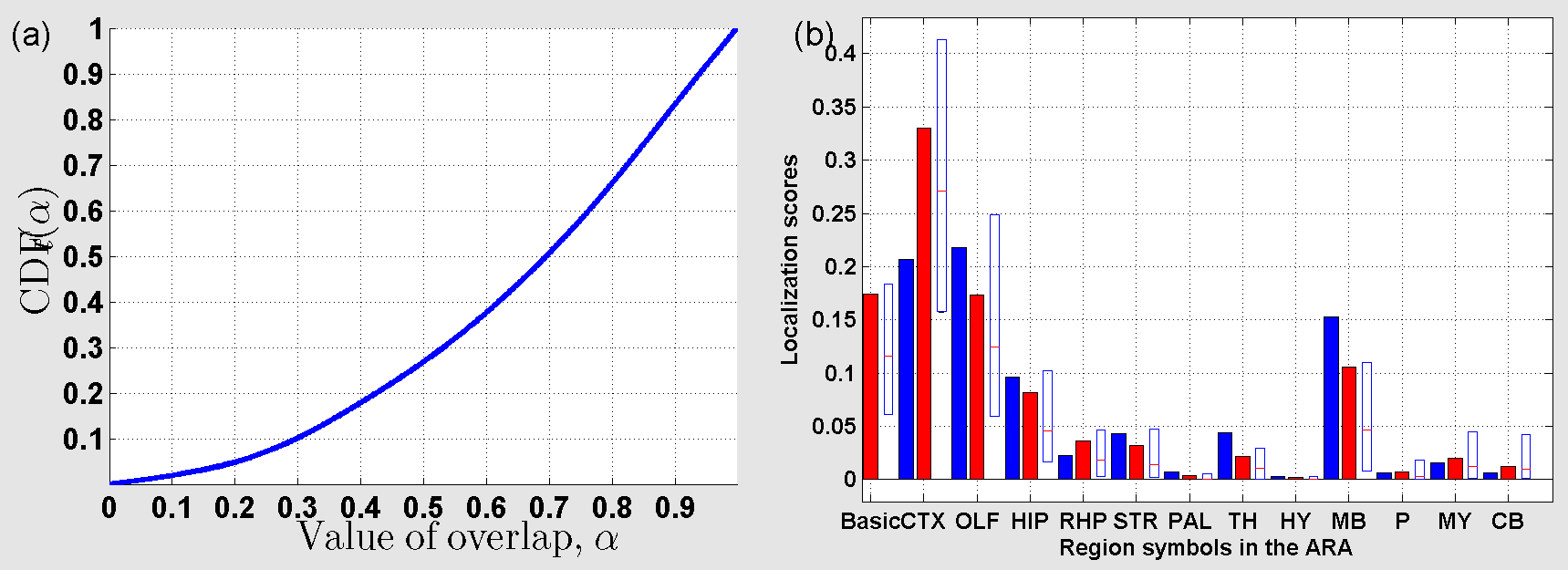}
\caption{(a) Cumulative distribution function (${\mathrm{\sc{CDF}}}_t$) of the overlap between $\rho_t$ and
 sub-sampled profiles for $t=6$. (b) Localization scores in the coarsest version of the ARA for $\rho_t$ (in blue), and 
 $\bar{\rho}_t$ (in red).}
\label{cdfPlot6}
\end{figure}
\begin{figure}
\includegraphics[width=1\textwidth,keepaspectratio]{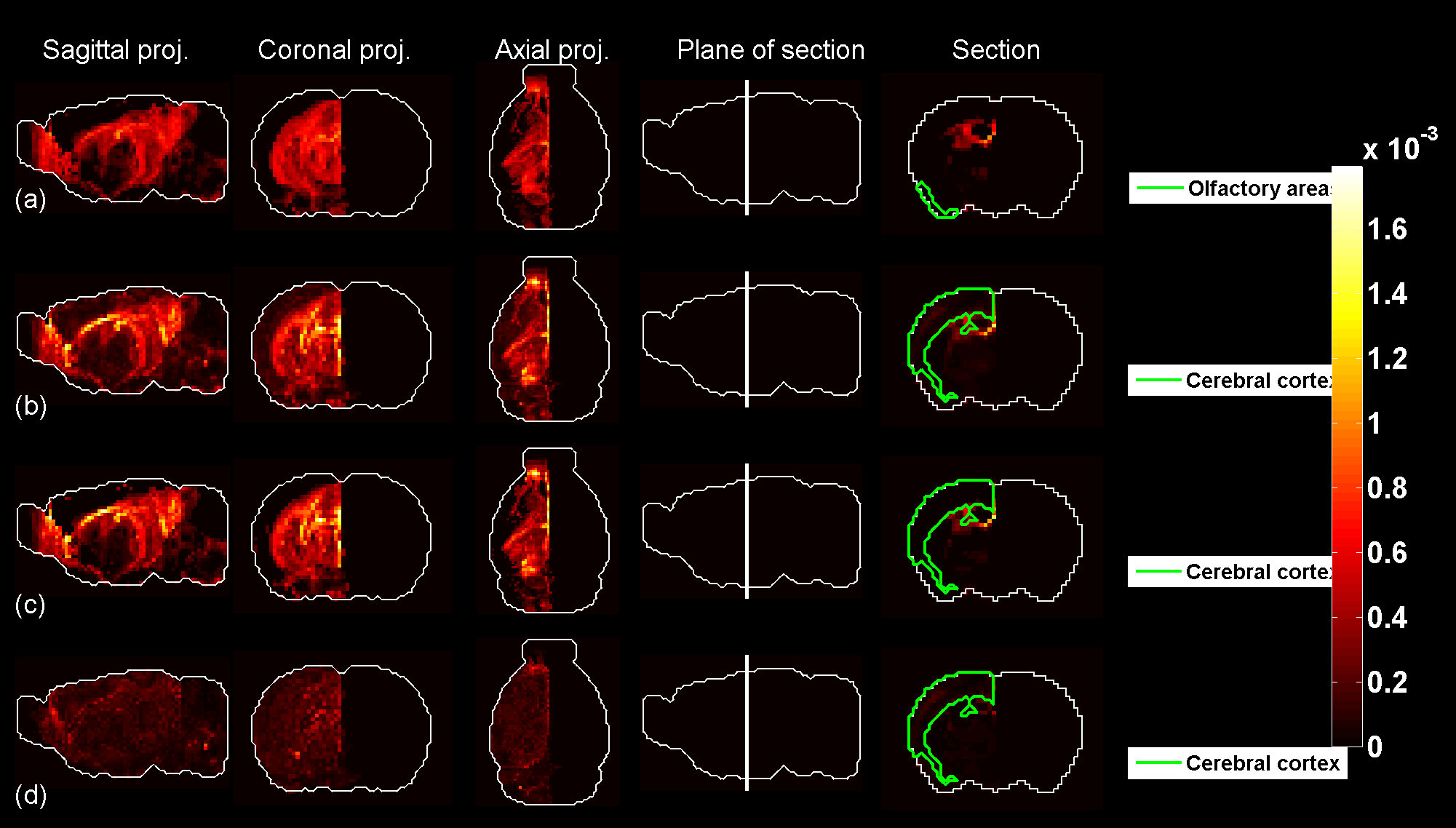}
\caption{Predicted profile and average sub-sampled profile for $t=6$.}
\label{subSampledFour6}
\end{figure}
\clearpage
\begin{figure}
\includegraphics[width=1\textwidth,keepaspectratio]{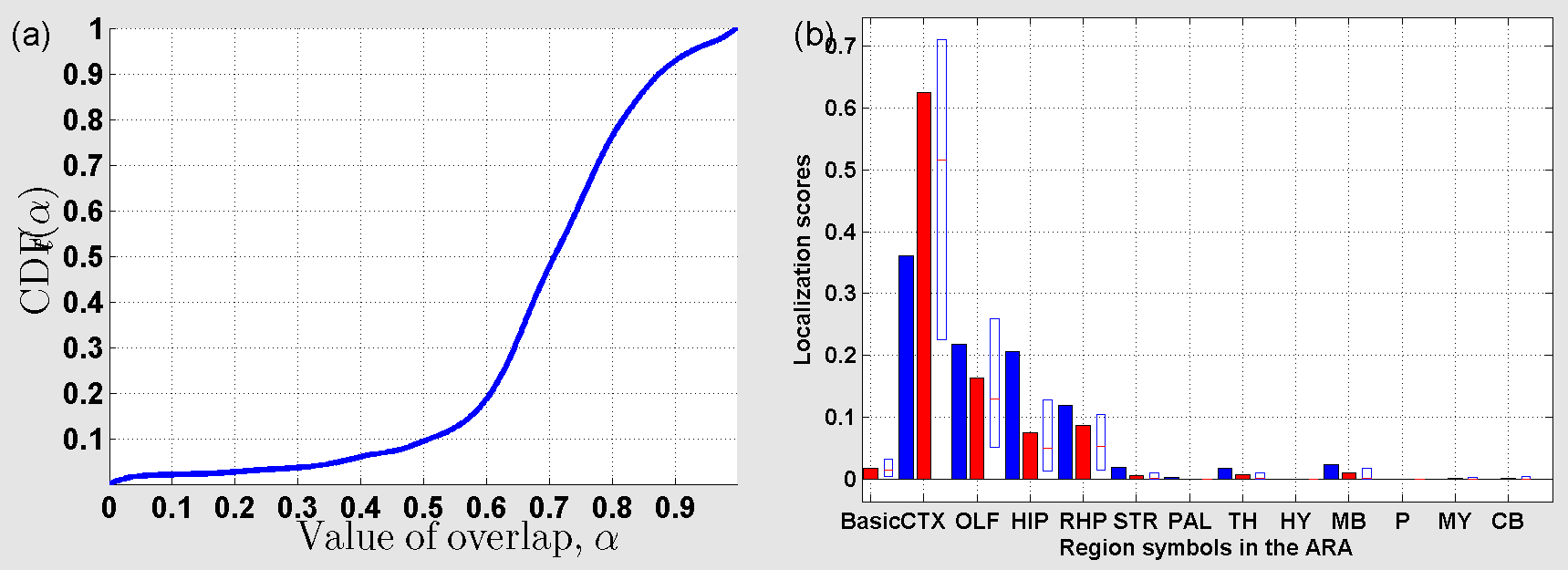}
\caption{(a) Cumulative distribution function (${\mathrm{\sc{CDF}}}_t$) of the overlap between $\rho_t$ and
 sub-sampled profiles for $t=7$. (b) Localization scores in the coarsest version of the ARA for $\rho_t$ (in blue), and 
 $\bar{\rho}_t$ (in red).}
\label{cdfPlot7}
\end{figure}
\begin{figure}
\includegraphics[width=1\textwidth,keepaspectratio]{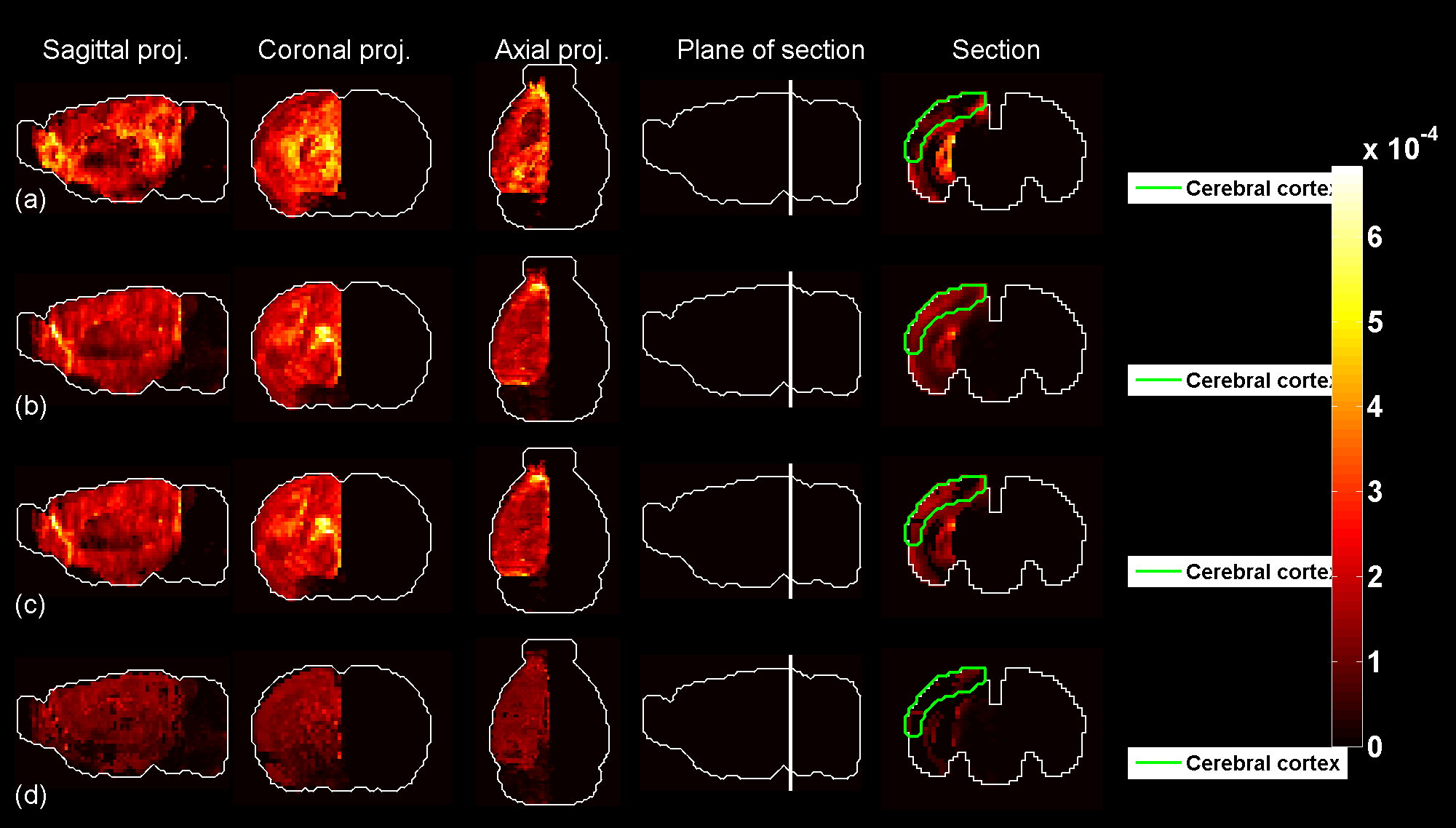}
\caption{Predicted profile and average sub-sampled profile for $t=7$.}
\label{subSampledFour7}
\end{figure}
\clearpage
\begin{figure}
\includegraphics[width=1\textwidth,keepaspectratio]{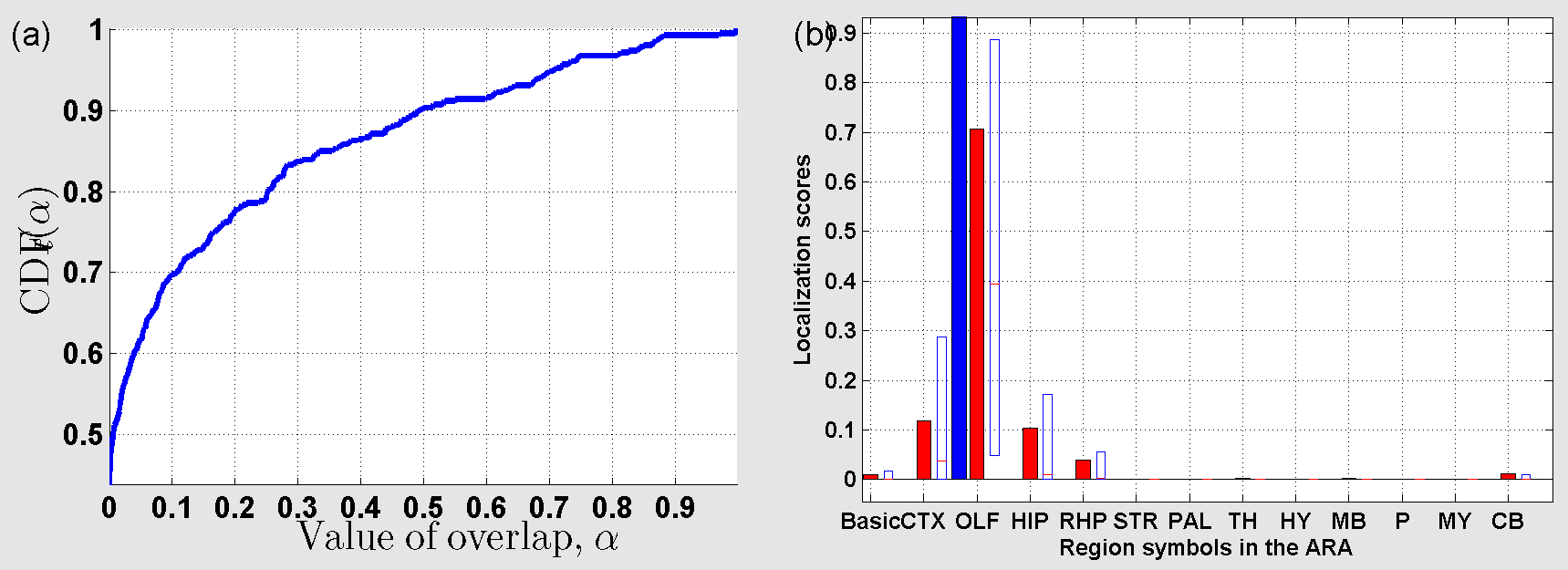}
\caption{(a) Cumulative distribution function (${\mathrm{\sc{CDF}}}_t$) of the overlap between $\rho_t$ and
 sub-sampled profiles for $t=8$. (b) Localization scores in the coarsest version of the ARA for $\rho_t$ (in blue), and 
 $\bar{\rho}_t$ (in red).}
\label{cdfPlot8}
\end{figure}
\begin{figure}
\includegraphics[width=1\textwidth,keepaspectratio]{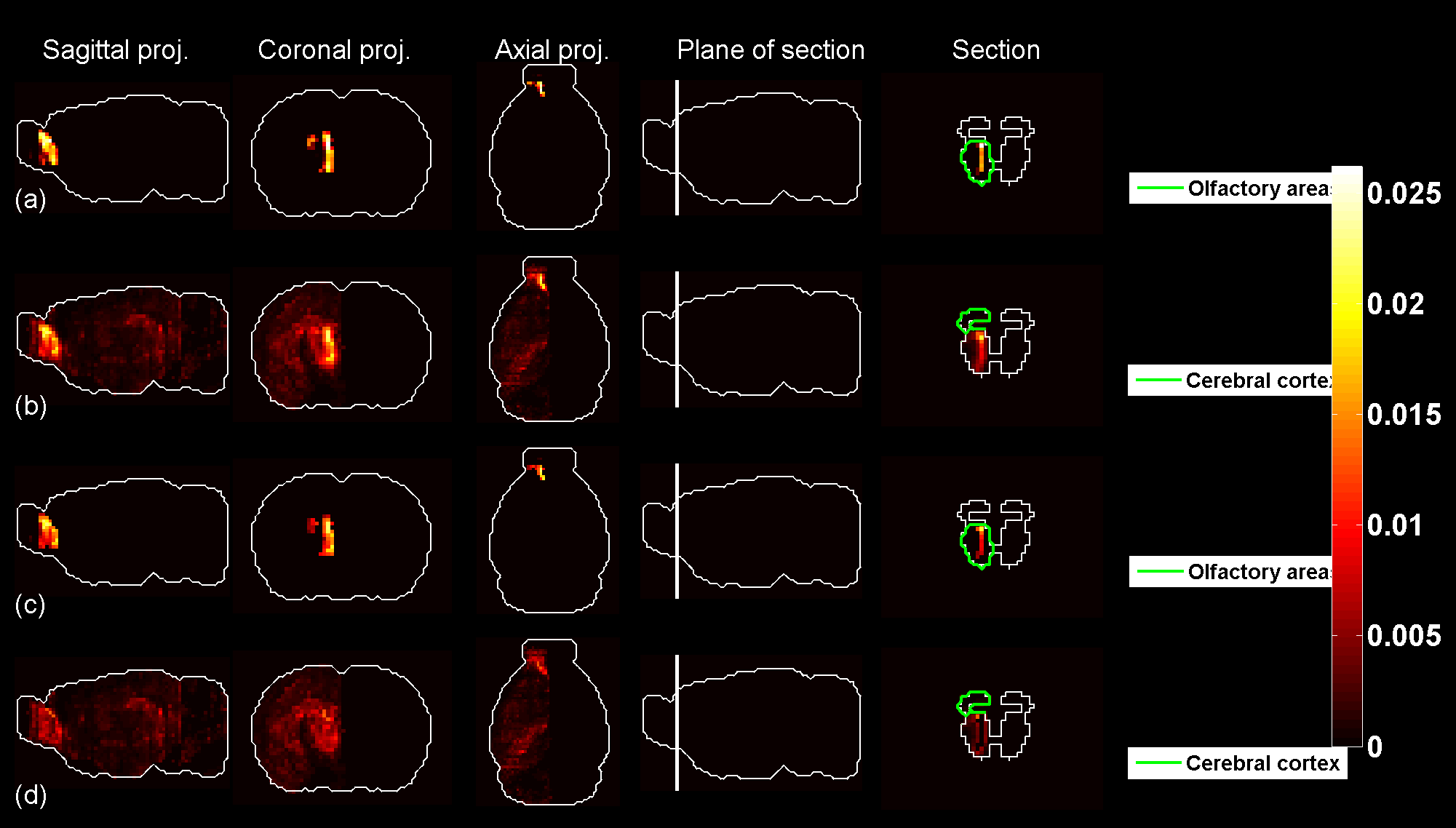}
\caption{Predicted profile and average sub-sampled profile for $t=8$.}
\label{subSampledFour8}
\end{figure}
\clearpage
\begin{figure}
\includegraphics[width=1\textwidth,keepaspectratio]{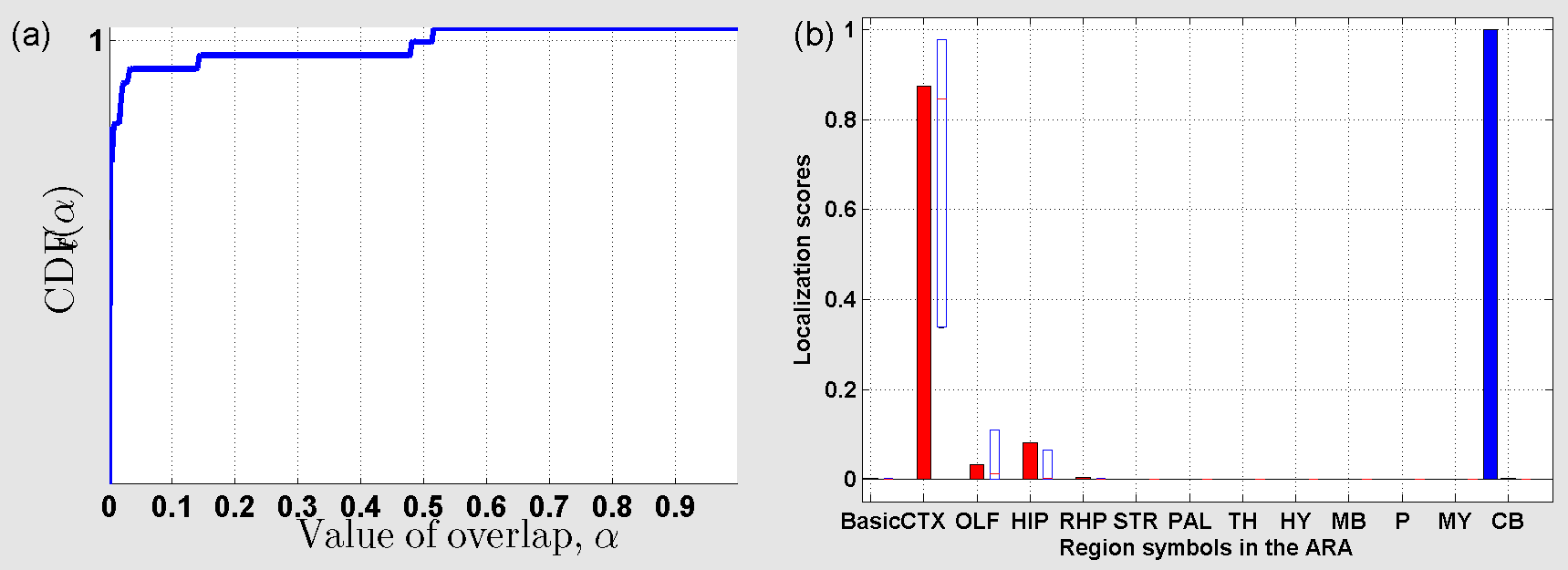}
\caption{(a) Cumulative distribution function (${\mathrm{\sc{CDF}}}_t$) of the overlap between $\rho_t$ and
 sub-sampled profiles for $t=9$. (b) Localization scores in the coarsest version of the ARA for $\rho_t$ (in blue), and 
 $\bar{\rho}_t$ (in red).}
\label{cdfPlot9}
\end{figure}
\begin{figure}
\includegraphics[width=1\textwidth,keepaspectratio]{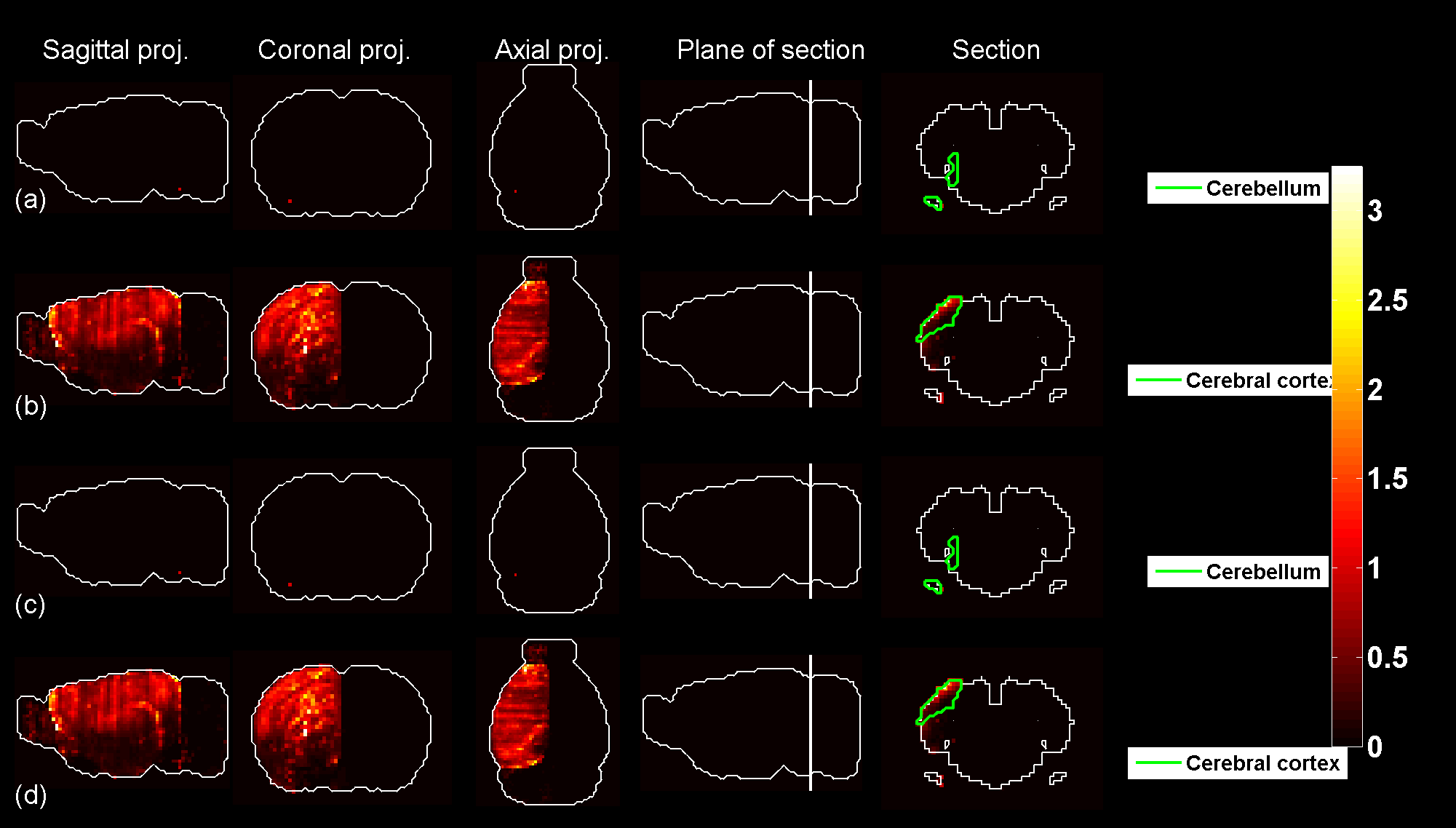}
\caption{Predicted profile and average sub-sampled profile for $t=9$.}
\label{subSampledFour9}
\end{figure}
\clearpage
\begin{figure}
\includegraphics[width=1\textwidth,keepaspectratio]{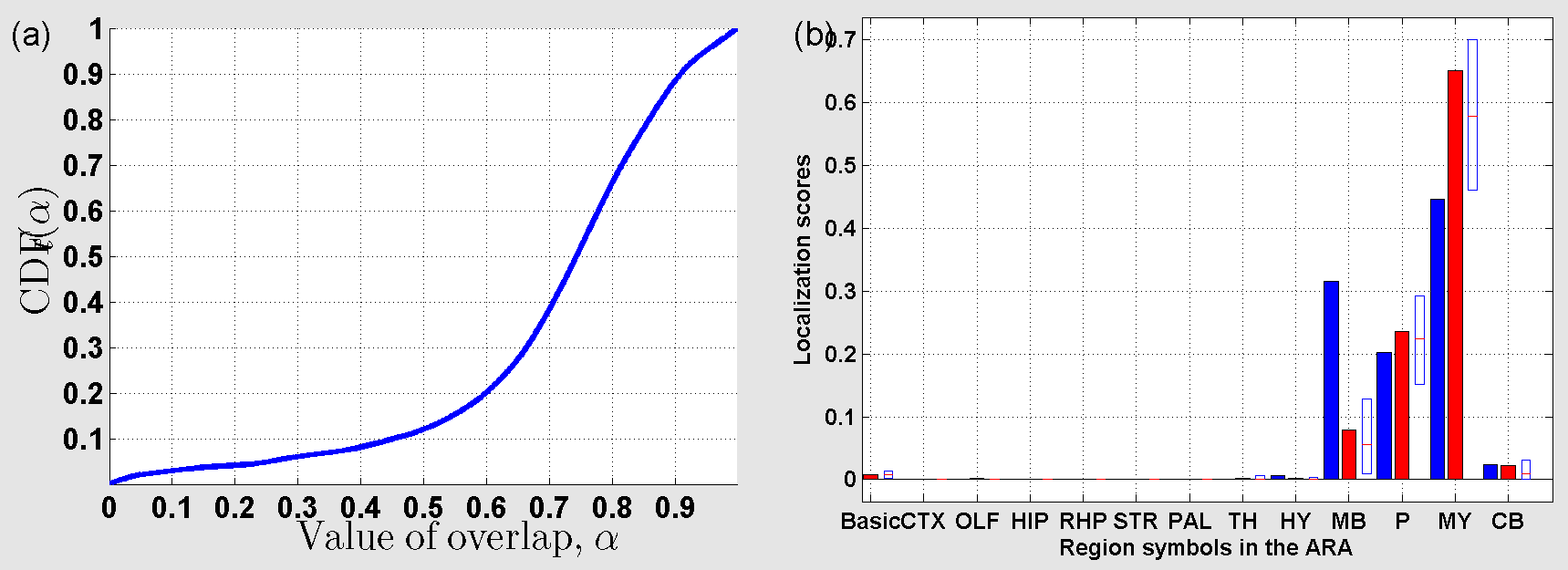}
\caption{(a) Cumulative distribution function (${\mathrm{\sc{CDF}}}_t$) of the overlap between $\rho_t$ and
 sub-sampled profiles for $t=10$. (b) Localization scores in the coarsest version of the ARA for $\rho_t$ (in blue), and 
 $\bar{\rho}_t$ (in red).}
\label{cdfPlot10}
\end{figure}
\begin{figure}
\includegraphics[width=1\textwidth,keepaspectratio]{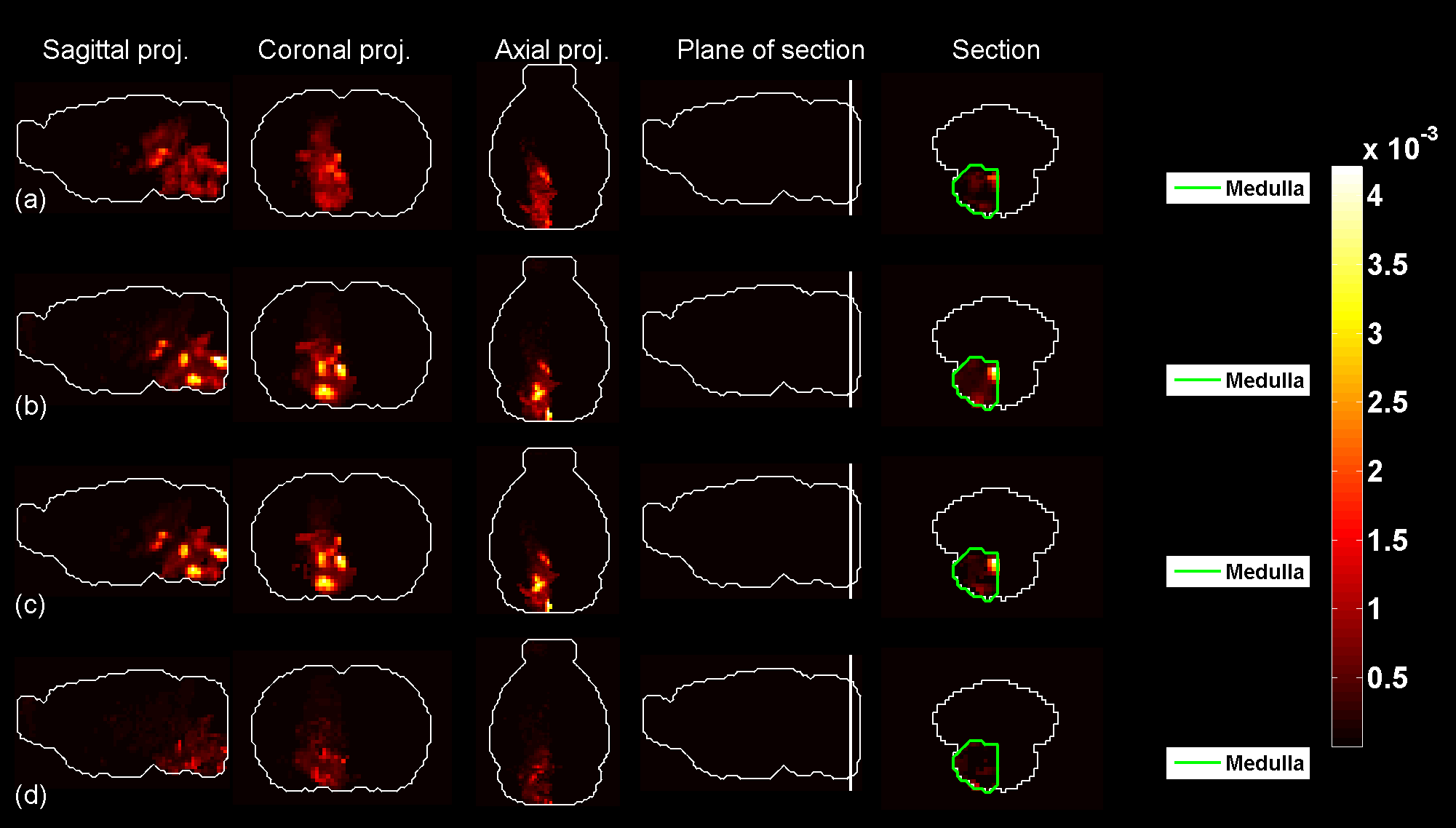}
\caption{Predicted profile and average sub-sampled profile for $t=10$.}
\label{subSampledFour10}
\end{figure}

%% file: appendSubsampleFiguresExtra.tex
\clearpage
\begin{figure}
\includegraphics[width=1\textwidth,keepaspectratio]{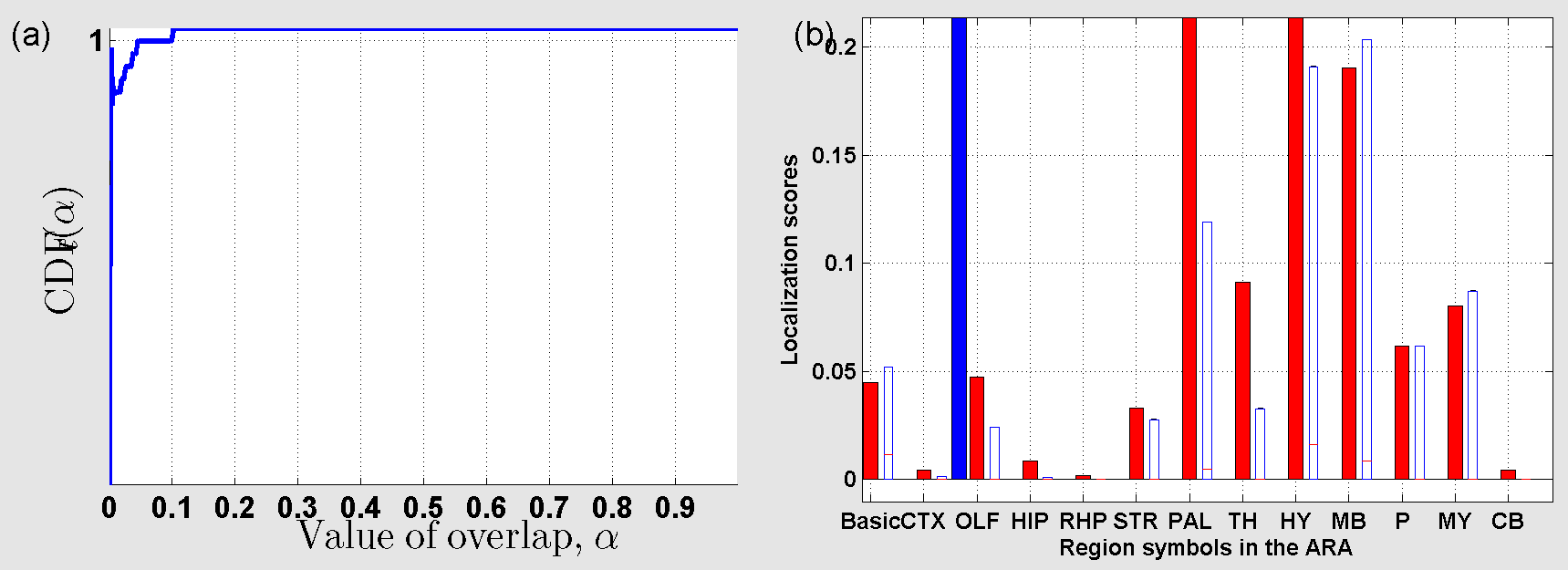}
\caption{(a) Cumulative distribution function (${\mathrm{\sc{CDF}}}_t$) of the overlap between $\rho_t$ and
 sub-sampled profiles for $t=11$. (b) Localization scores in the coarsest version of the ARA for $\rho_t$ (in blue), and 
 $\bar{\rho}_t$ (in red).}
\label{cdfPlot11}
\end{figure}
\begin{figure}
\includegraphics[width=1\textwidth,keepaspectratio]{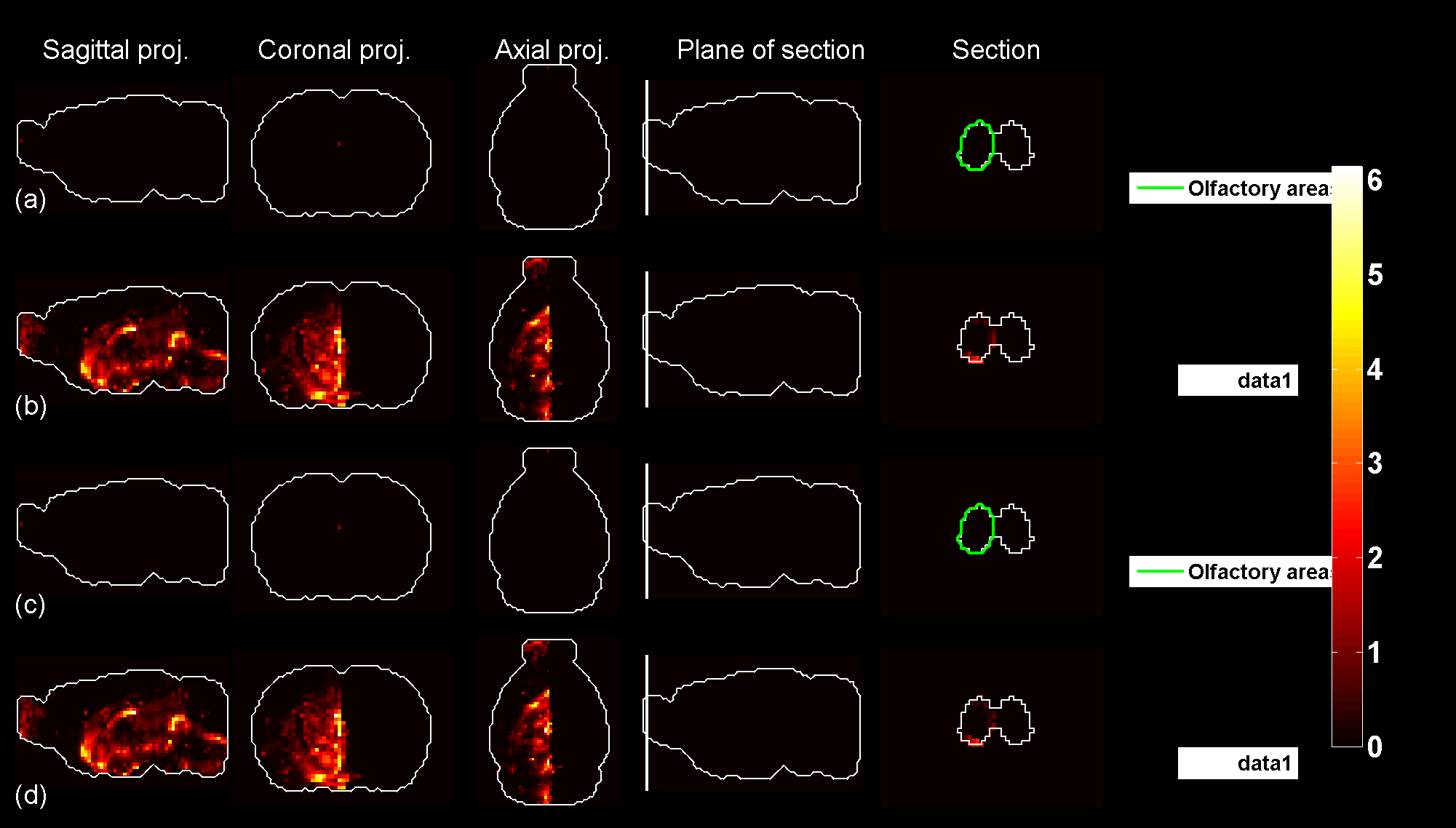}
\caption{Predicted profile and average sub-sampled profile for $t=11$.}
\label{subSampledFour11}
\end{figure}
\clearpage
\begin{figure}
\includegraphics[width=1\textwidth,keepaspectratio]{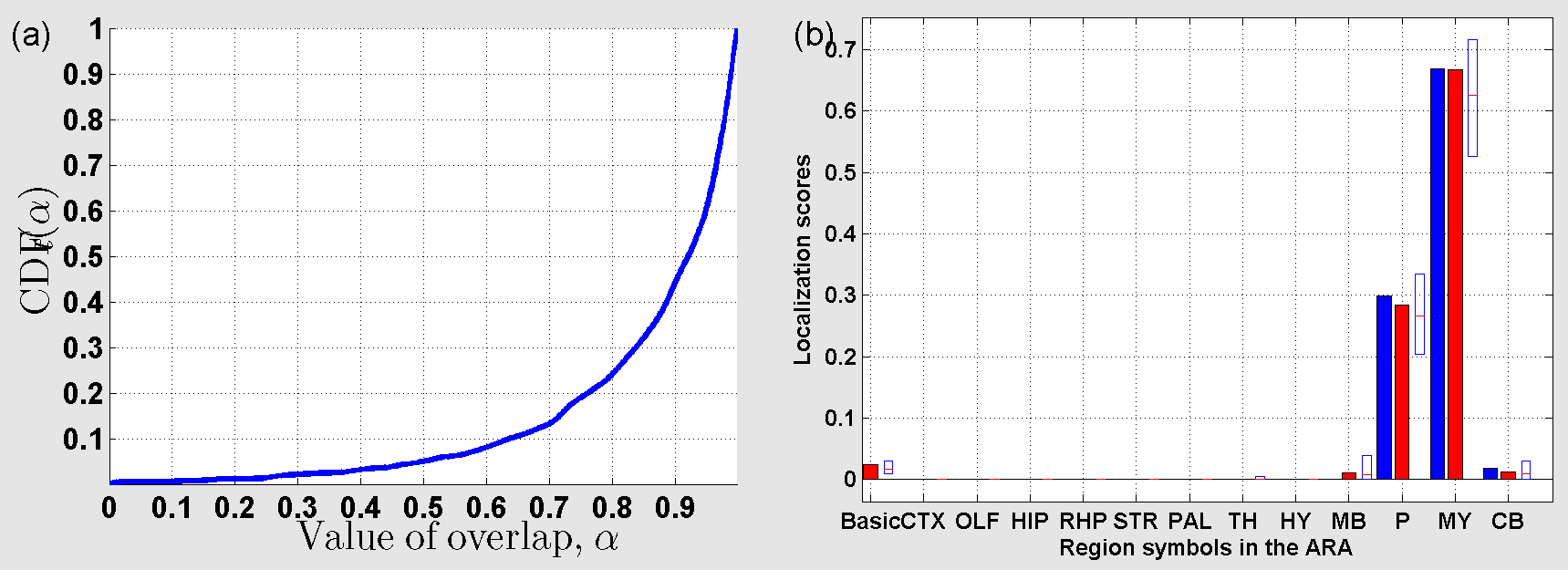}
\caption{(a) Cumulative distribution function (${\mathrm{\sc{CDF}}}_t$) of the overlap between $\rho_t$ and
 sub-sampled profiles for $t=12$. (b) Localization scores in the coarsest version of the ARA for $\rho_t$ (in blue), and 
 $\bar{\rho}_t$ (in red).}
\label{cdfPlot12}
\end{figure}
\begin{figure}
\includegraphics[width=1\textwidth,keepaspectratio]{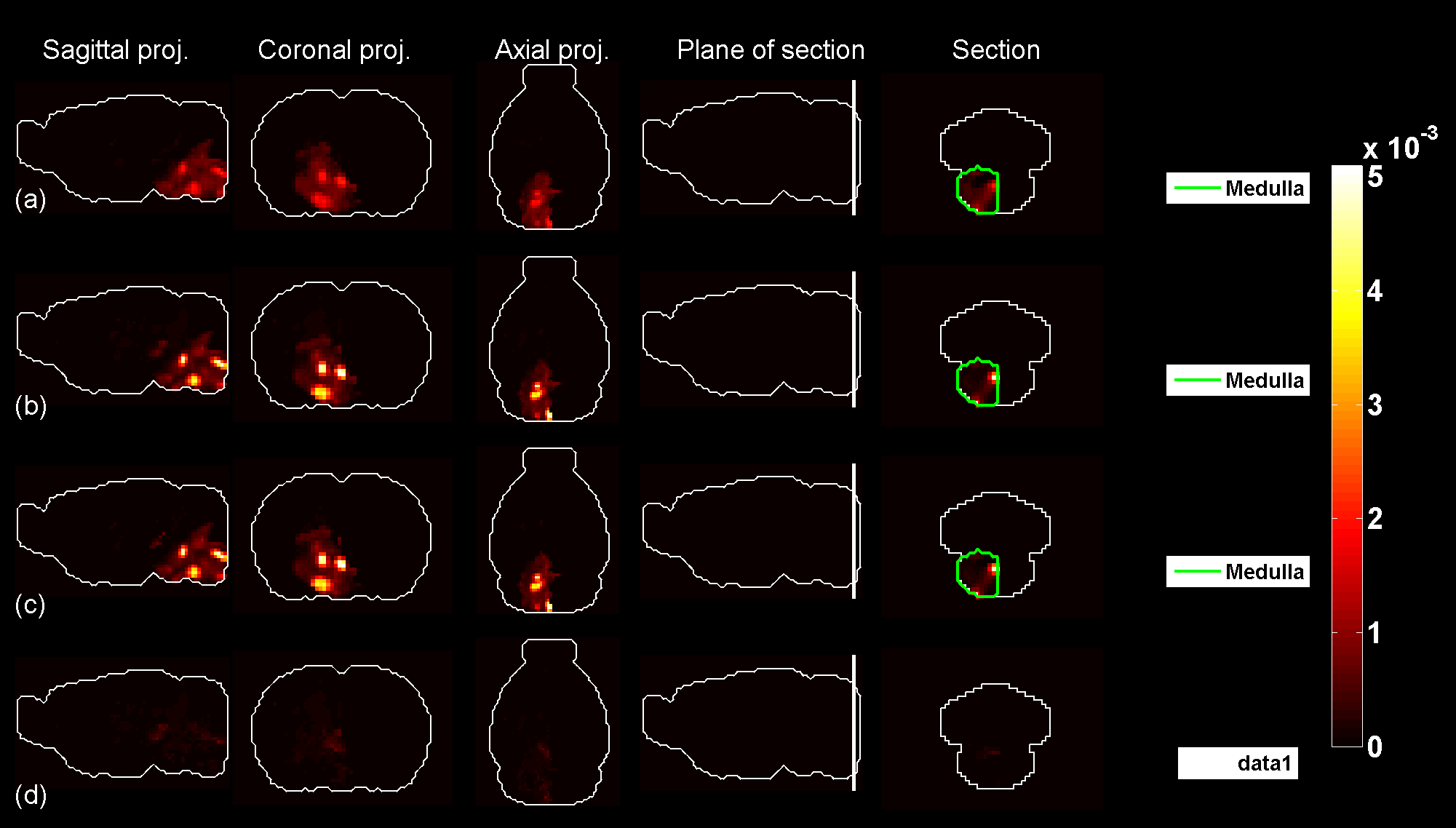}
\caption{Predicted profile and average sub-sampled profile for $t=12$.}
\label{subSampledFour12}
\end{figure}
\clearpage
\begin{figure}
\includegraphics[width=1\textwidth,keepaspectratio]{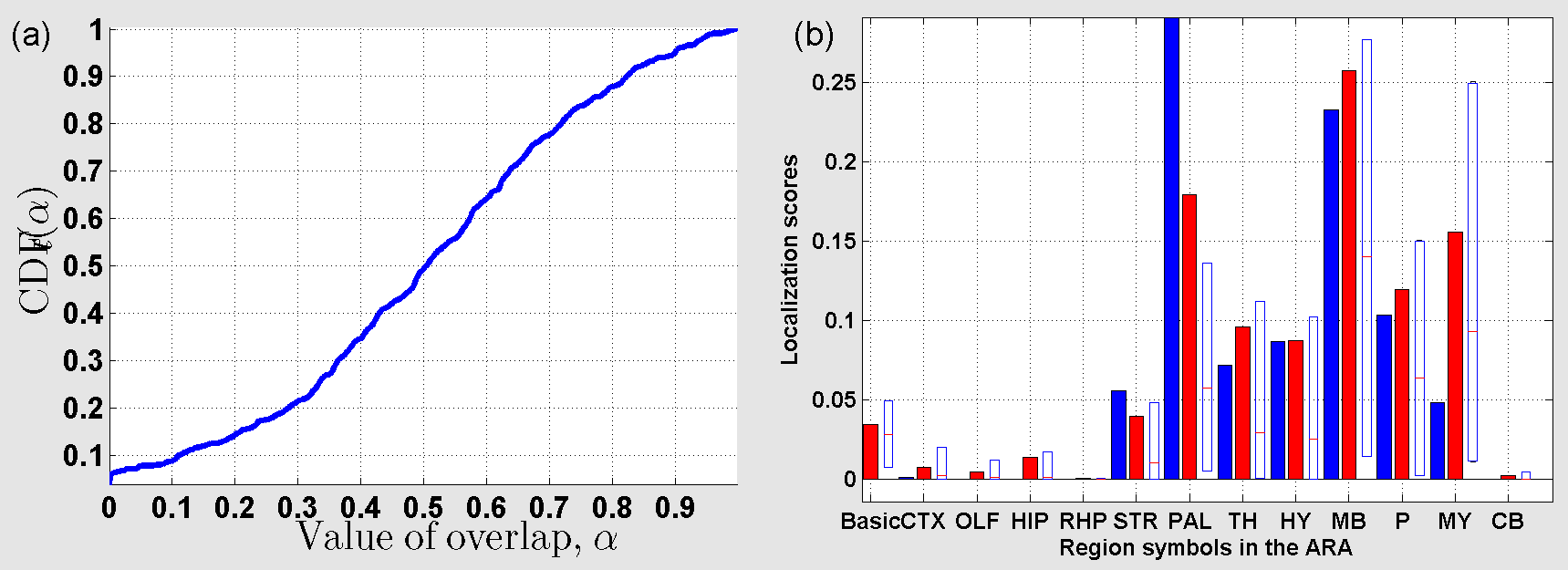}
\caption{(a) Cumulative distribution function (${\mathrm{\sc{CDF}}}_t$) of the overlap between $\rho_t$ and
 sub-sampled profiles for $t=13$. (b) Localization scores in the coarsest version of the ARA for $\rho_t$ (in blue), and 
 $\bar{\rho}_t$ (in red).}
\label{cdfPlot13}
\end{figure}
\begin{figure}
\includegraphics[width=1\textwidth,keepaspectratio]{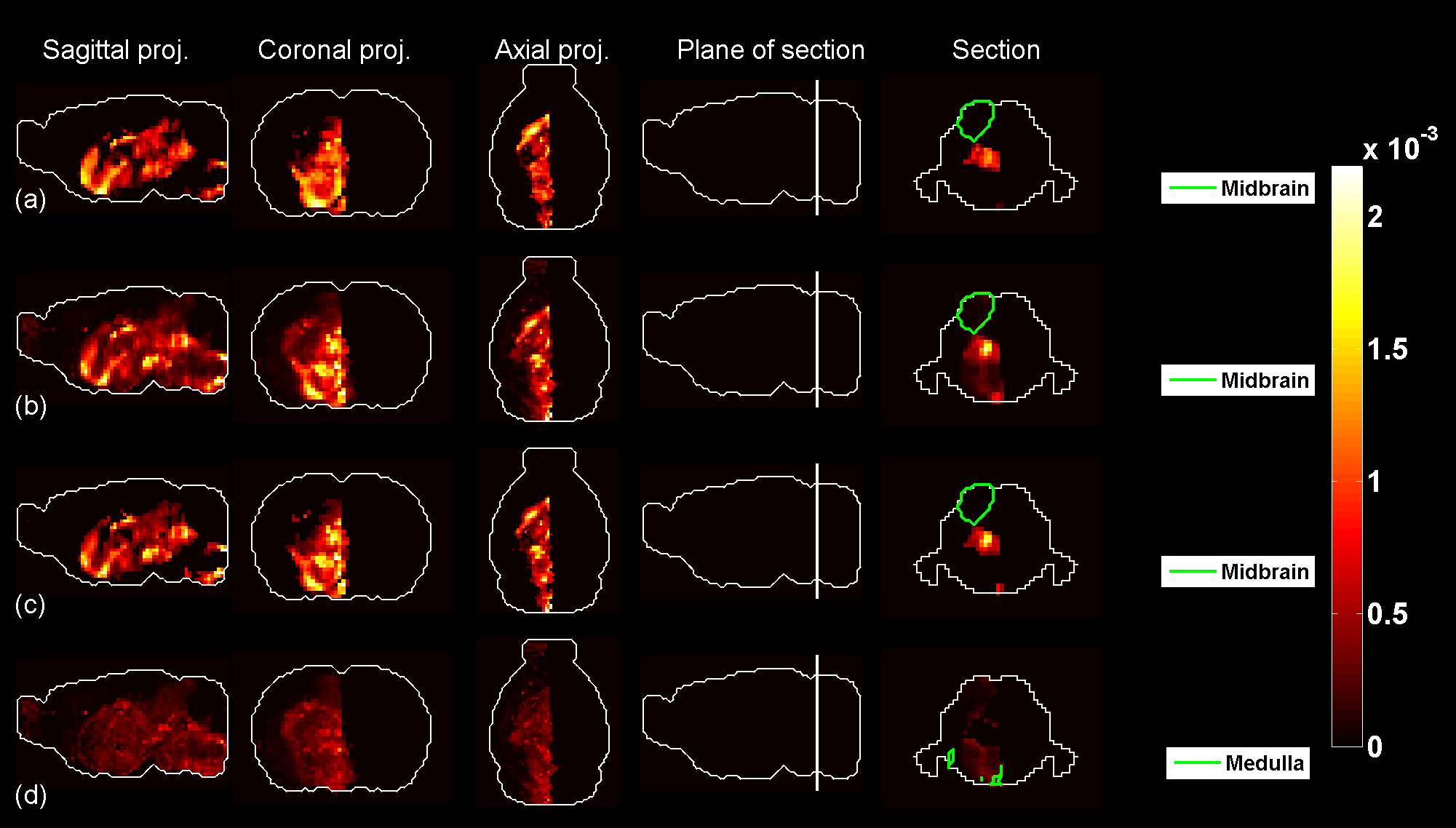}
\caption{Predicted profile and average sub-sampled profile for $t=13$.}
\label{subSampledFour13}
\end{figure}
\clearpage
\begin{figure}
\includegraphics[width=1\textwidth,keepaspectratio]{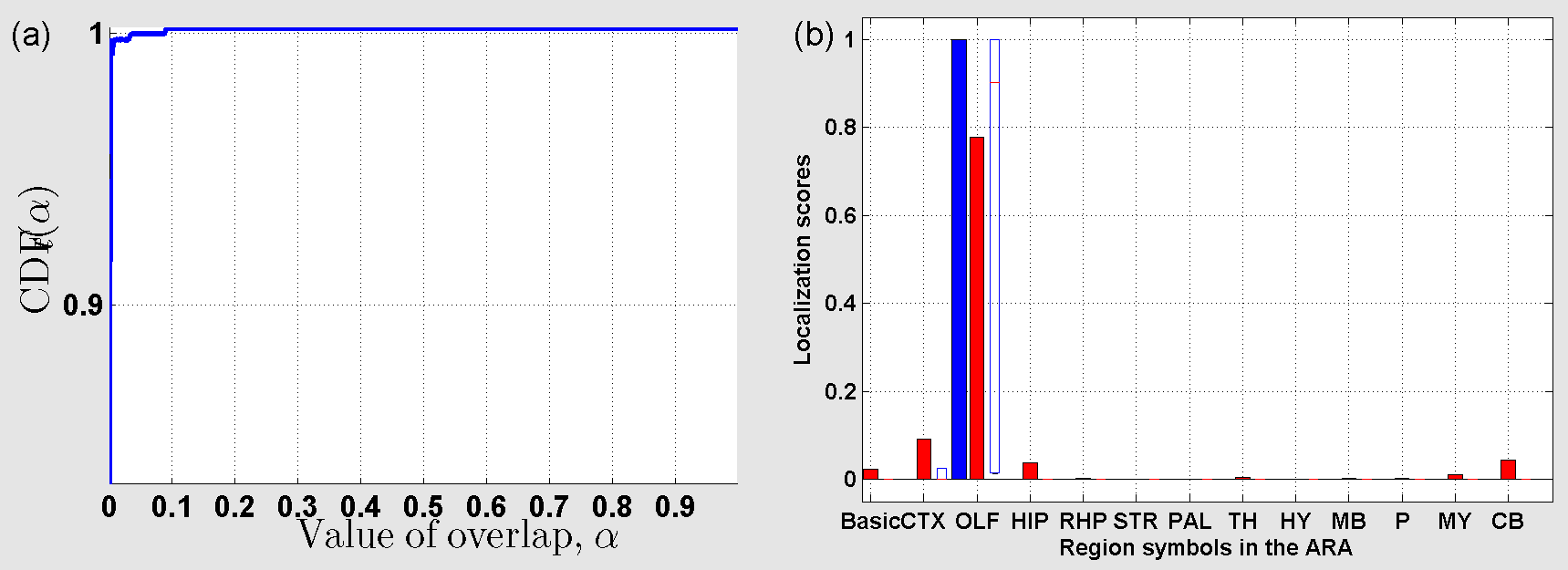}
\caption{(a) Cumulative distribution function (${\mathrm{\sc{CDF}}}_t$) of the overlap between $\rho_t$ and
 sub-sampled profiles for $t=14$. (b) Localization scores in the coarsest version of the ARA for $\rho_t$ (in blue), and 
 $\bar{\rho}_t$ (in red).}
\label{cdfPlot14}
\end{figure}
\begin{figure}
\includegraphics[width=1\textwidth,keepaspectratio]{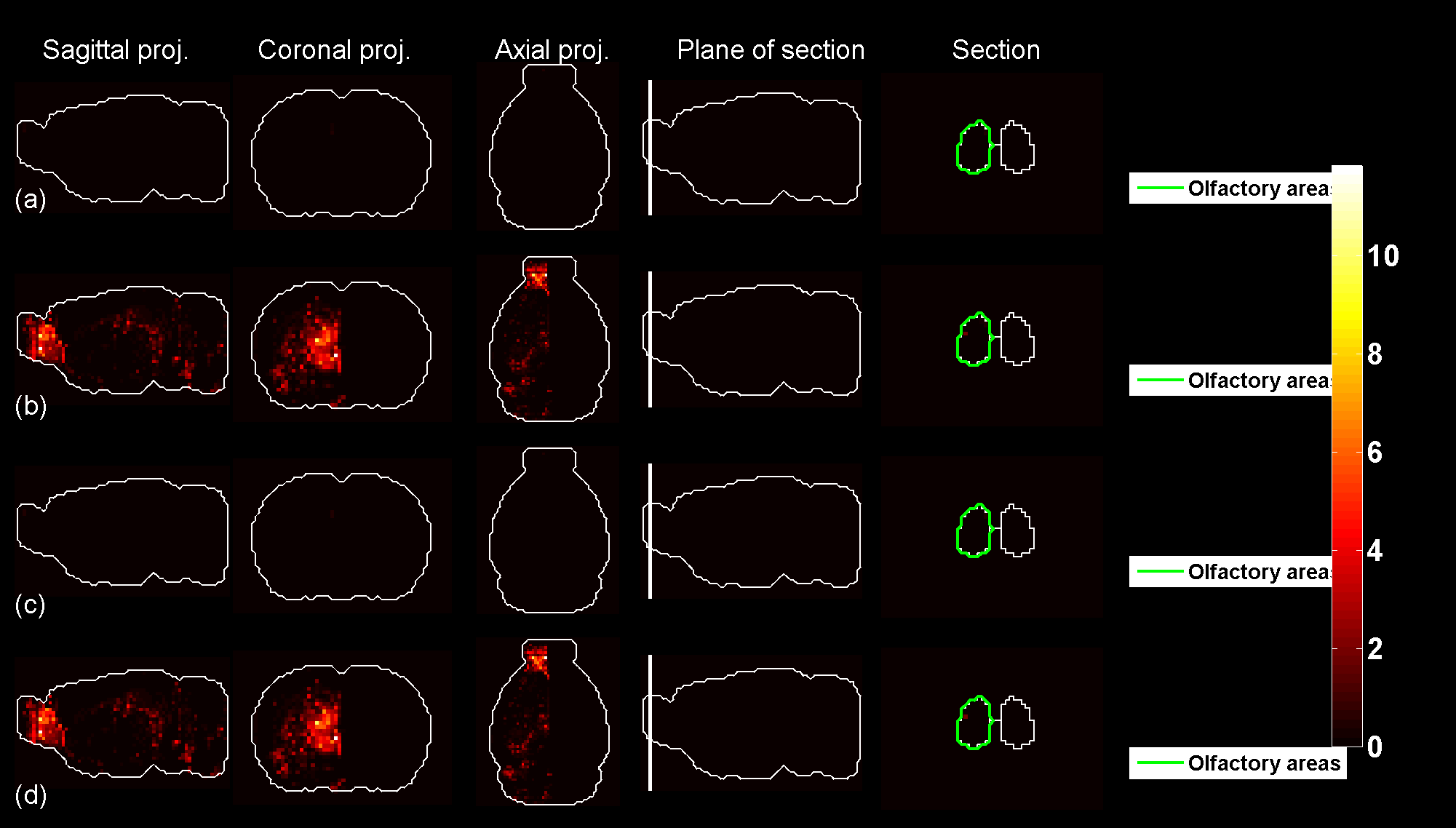}
\caption{Predicted profile and average sub-sampled profile for $t=14$.}
\label{subSampledFour14}
\end{figure}
\clearpage
\begin{figure}
\includegraphics[width=1\textwidth,keepaspectratio]{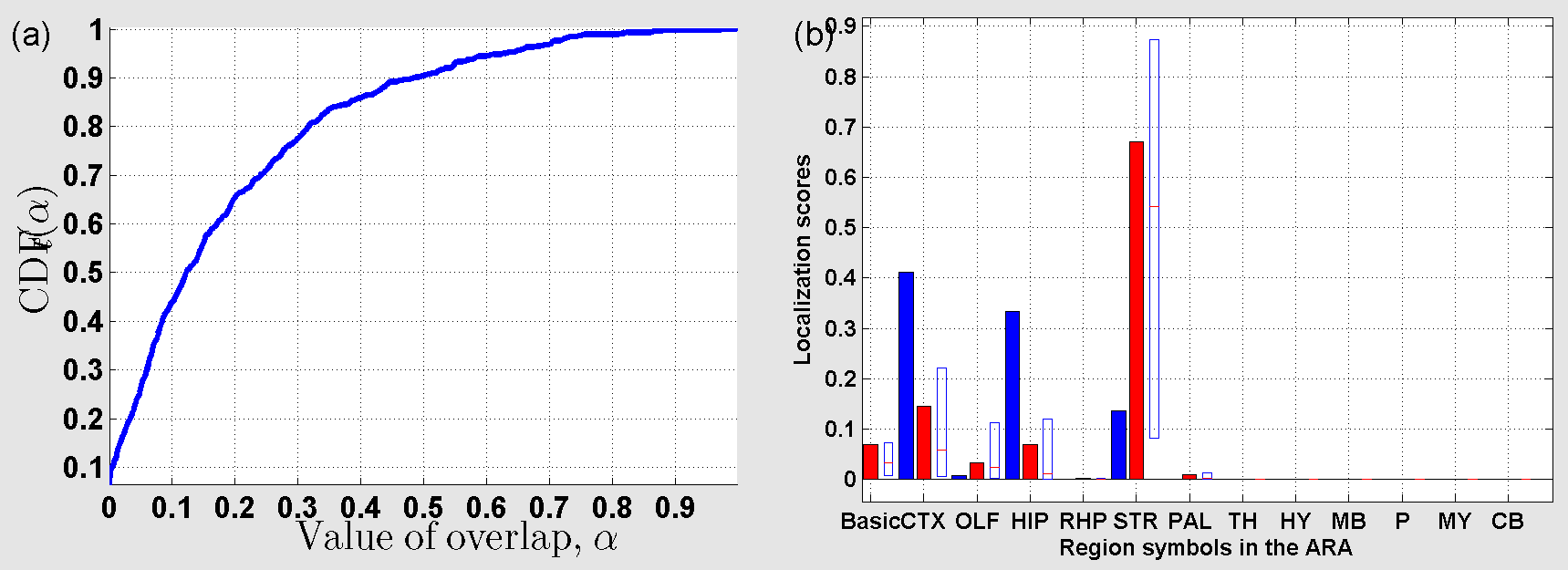}
\caption{(a) Cumulative distribution function (${\mathrm{\sc{CDF}}}_t$) of the overlap between $\rho_t$ and
 sub-sampled profiles for $t=15$. (b) Localization scores in the coarsest version of the ARA for $\rho_t$ (in blue), and 
 $\bar{\rho}_t$ (in red).}
\label{cdfPlot15}
\end{figure}
\begin{figure}
\includegraphics[width=1\textwidth,keepaspectratio]{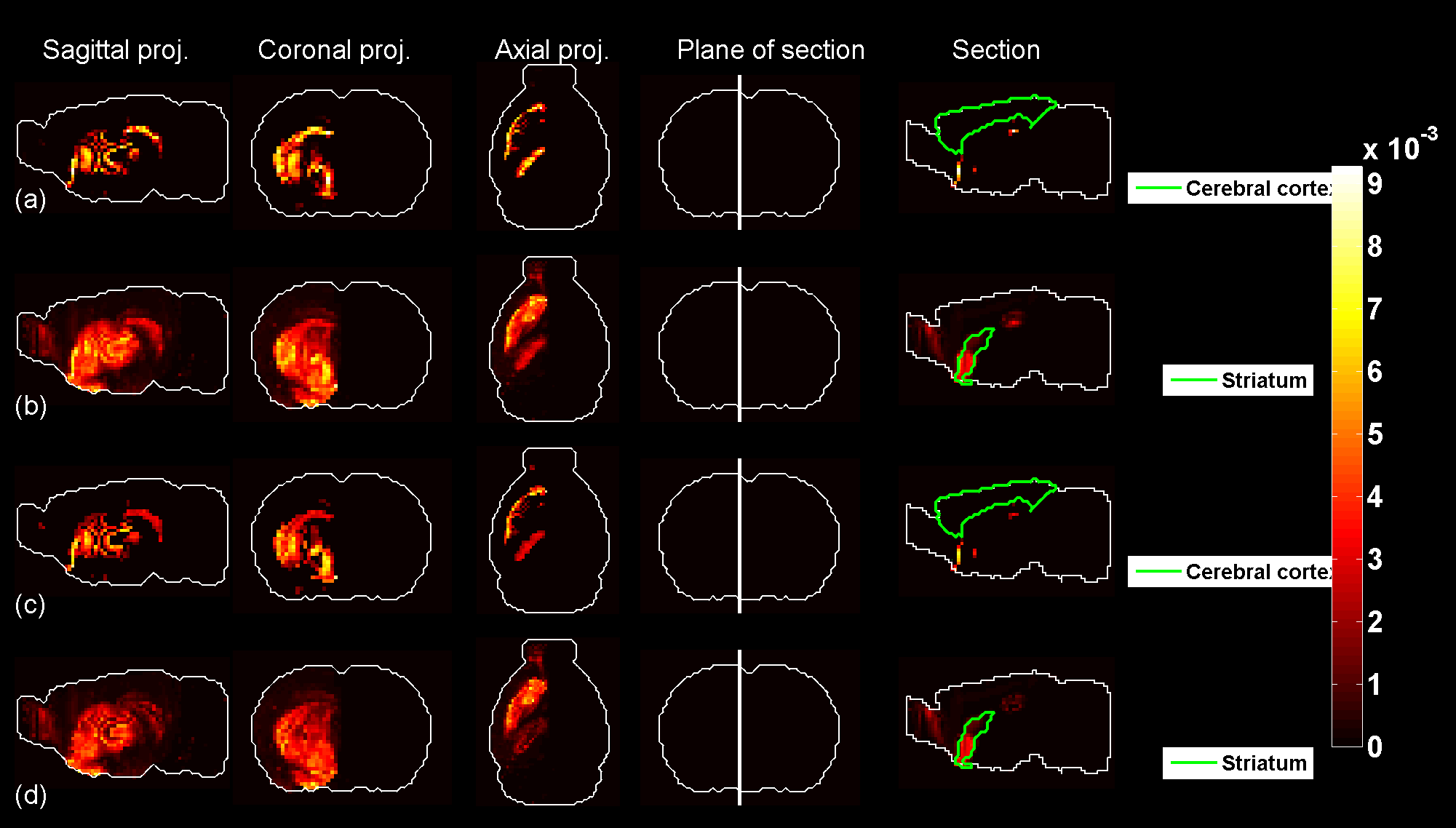}
\caption{Predicted profile and average sub-sampled profile for $t=15$.}
\label{subSampledFour15}
\end{figure}
\clearpage
\begin{figure}
\includegraphics[width=1\textwidth,keepaspectratio]{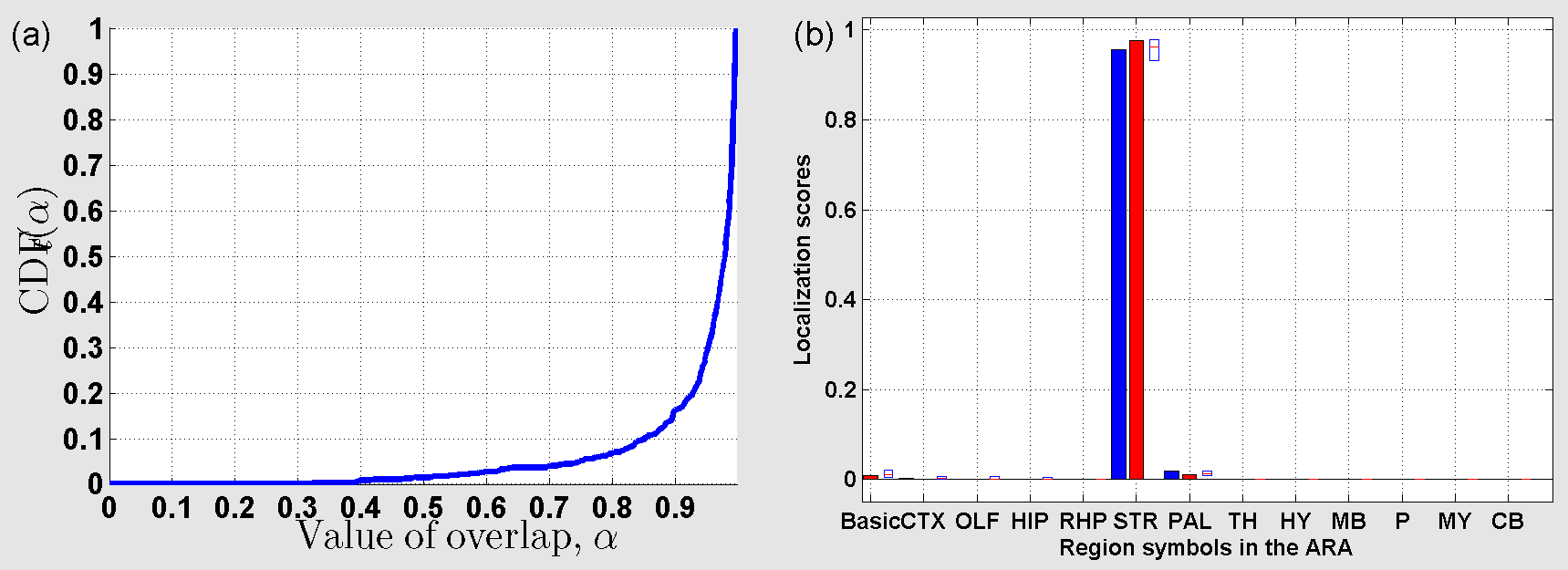}
\caption{(a) Cumulative distribution function (${\mathrm{\sc{CDF}}}_t$) of the overlap between $\rho_t$ and
 sub-sampled profiles for $t=16$. (b) Localization scores in the coarsest version of the ARA for $\rho_t$ (in blue), and 
 $\bar{\rho}_t$ (in red).}
\label{cdfPlot16}
\end{figure}
\begin{figure}
\includegraphics[width=1\textwidth,keepaspectratio]{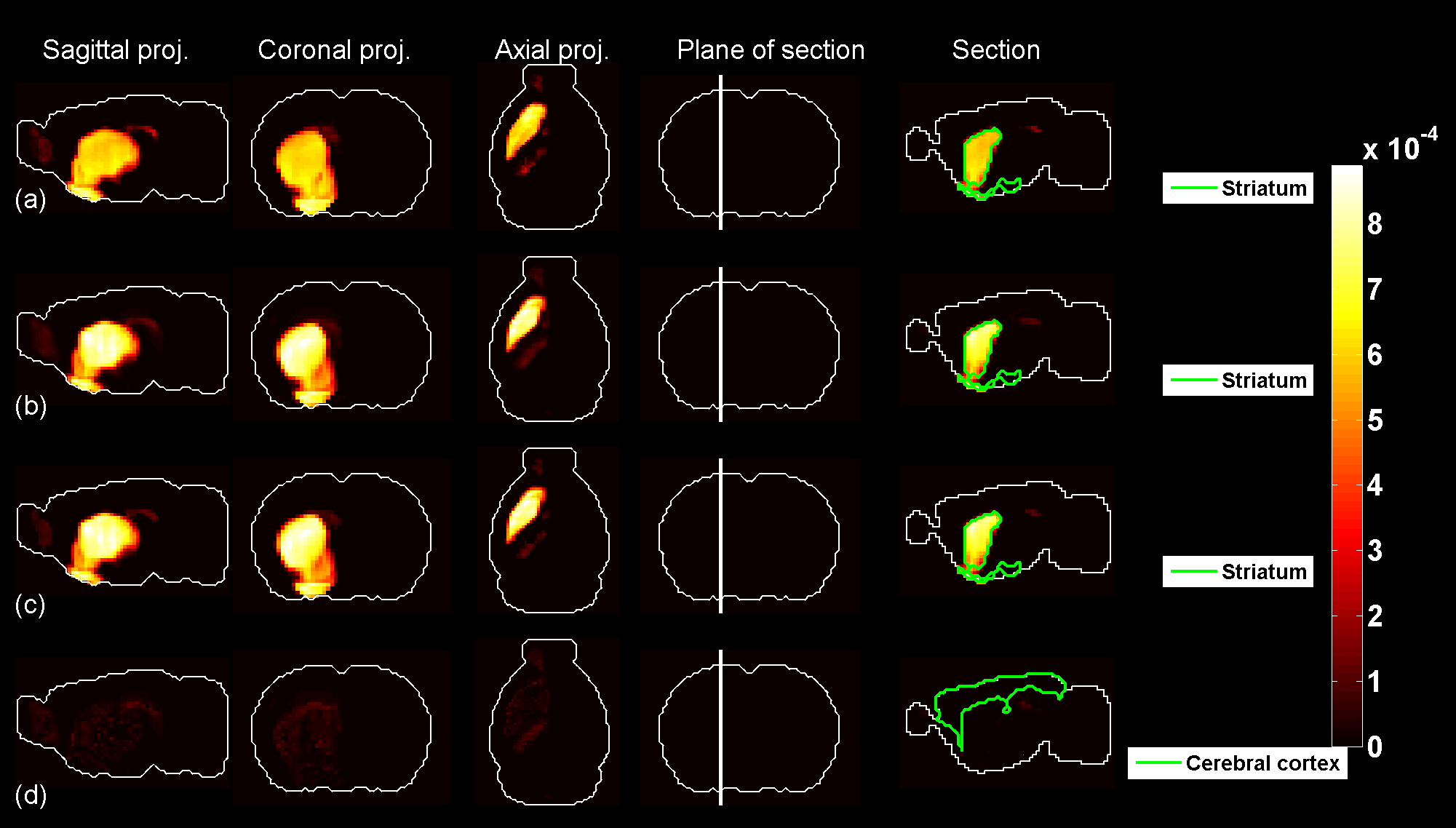}
\caption{Predicted profile and average sub-sampled profile for $t=16$.}
\label{subSampledFour16}
\end{figure}
\clearpage
\begin{figure}
\includegraphics[width=1\textwidth,keepaspectratio]{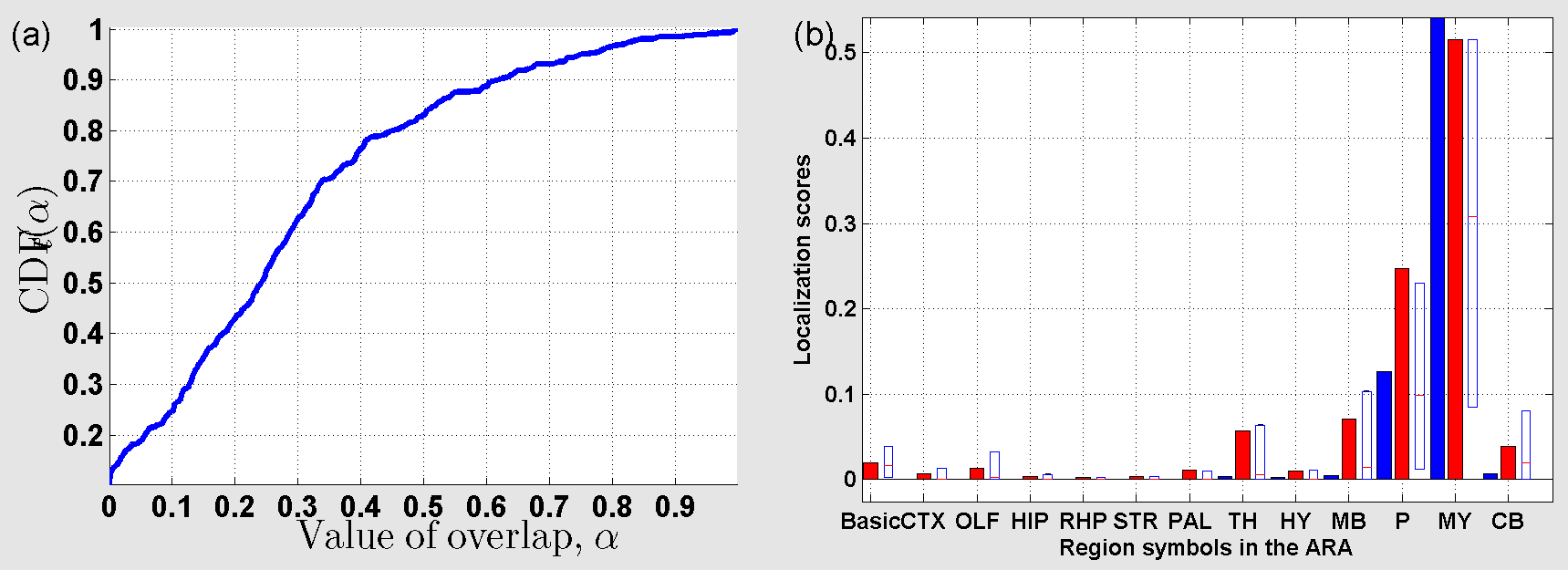}
\caption{(a) Cumulative distribution function (${\mathrm{\sc{CDF}}}_t$) of the overlap between $\rho_t$ and
 sub-sampled profiles for $t=17$. (b) Localization scores in the coarsest version of the ARA for $\rho_t$ (in blue), and 
 $\bar{\rho}_t$ (in red).}
\label{cdfPlot17}
\end{figure}
\begin{figure}
\includegraphics[width=1\textwidth,keepaspectratio]{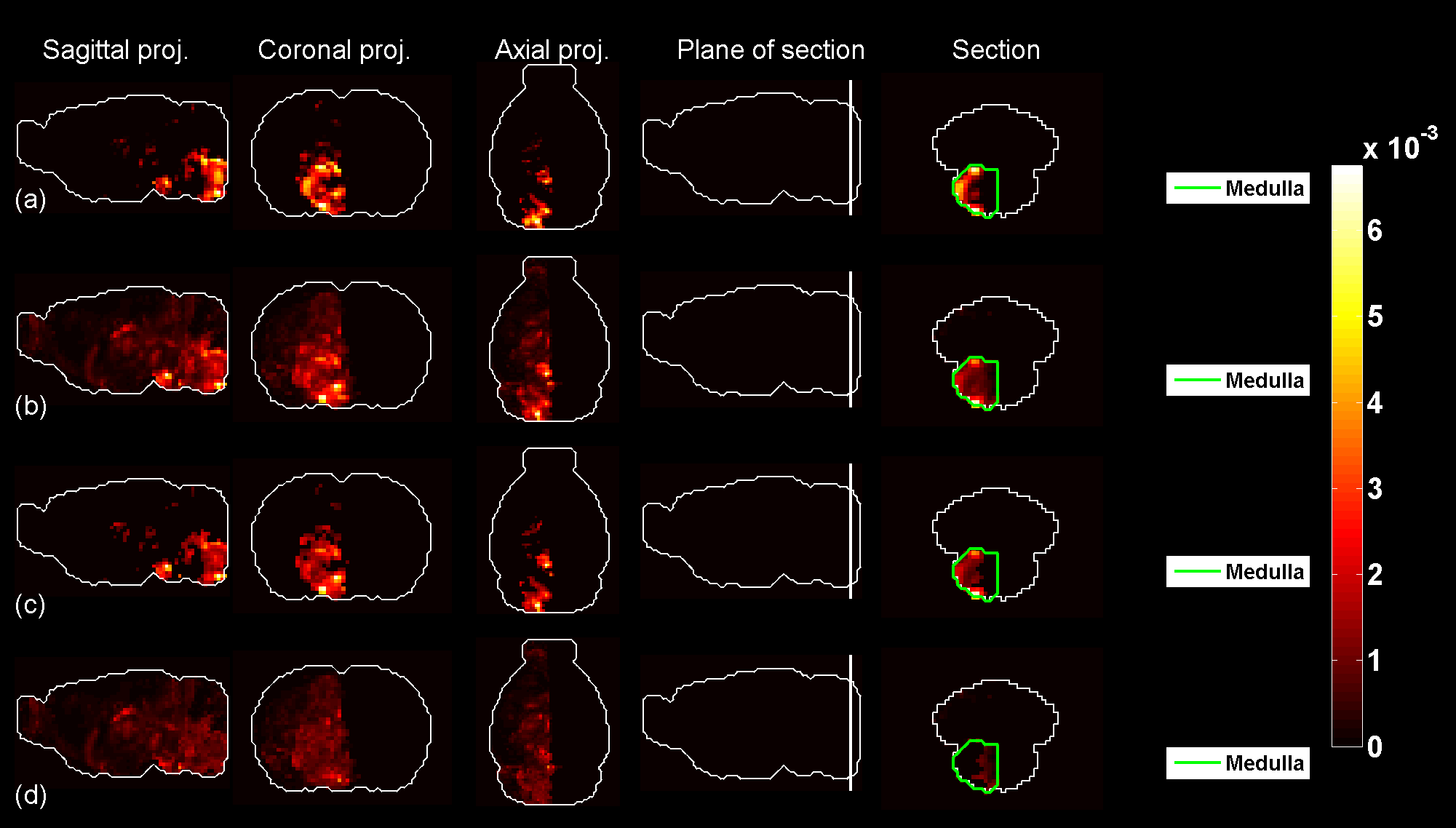}
\caption{Predicted profile and average sub-sampled profile for $t=17$.}
\label{subSampledFour17}
\end{figure}
\clearpage
\begin{figure}
\includegraphics[width=1\textwidth,keepaspectratio]{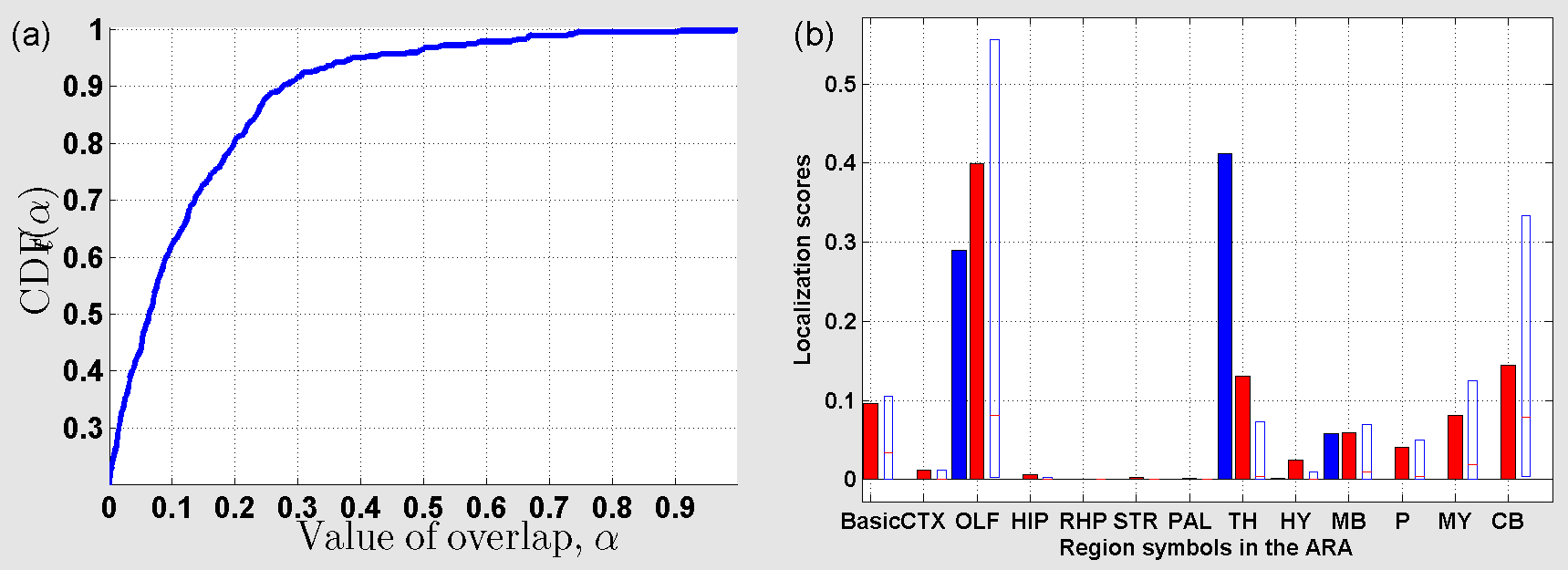}
\caption{(a) Cumulative distribution function (${\mathrm{\sc{CDF}}}_t$) of the overlap between $\rho_t$ and
 sub-sampled profiles for $t=18$. (b) Localization scores in the coarsest version of the ARA for $\rho_t$ (in blue), and 
 $\bar{\rho}_t$ (in red).}
\label{cdfPlot18}
\end{figure}
\begin{figure}
\includegraphics[width=1\textwidth,keepaspectratio]{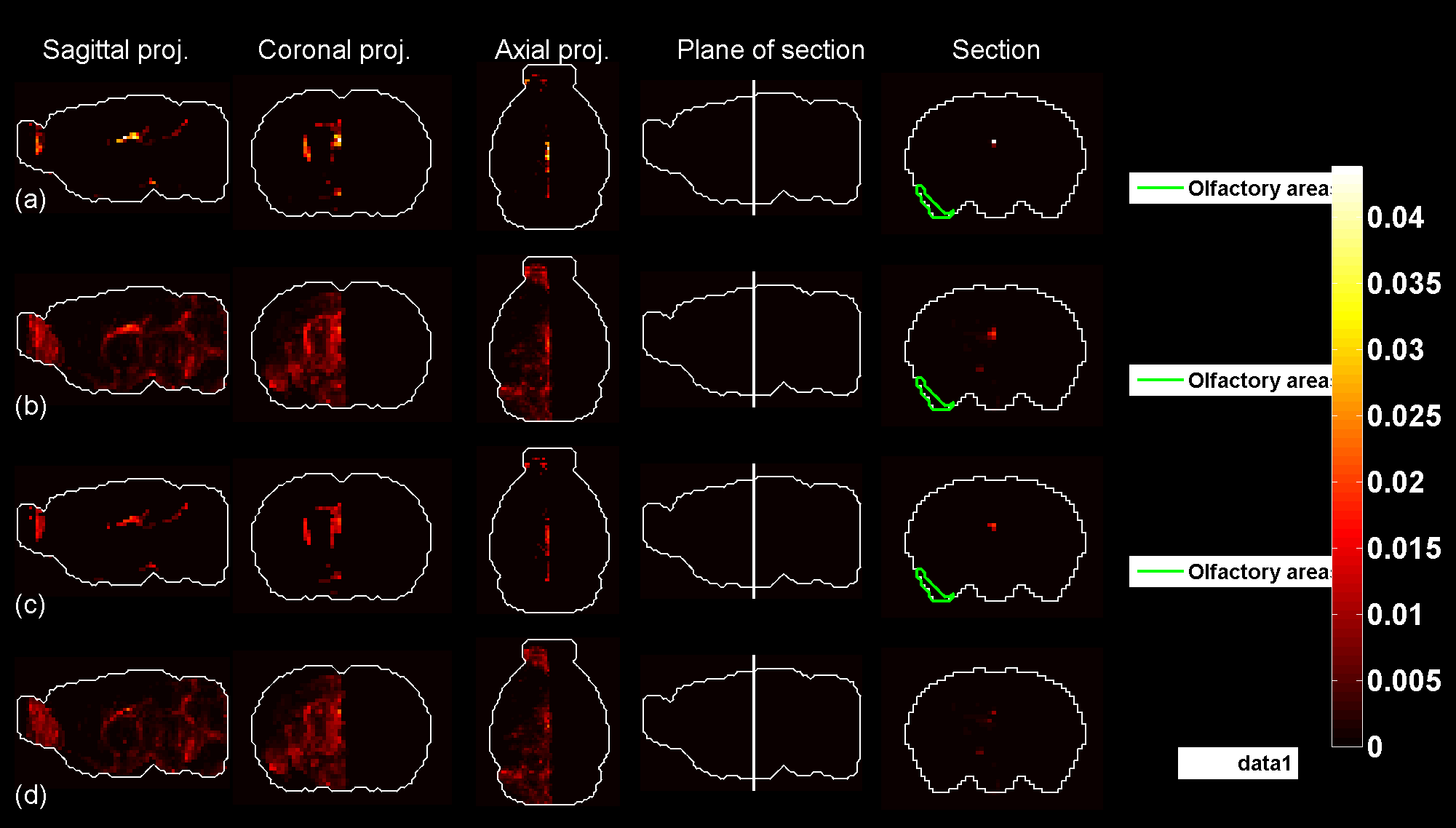}
\caption{Predicted profile and average sub-sampled profile for $t=18$.}
\label{subSampledFour18}
\end{figure}
\clearpage
\begin{figure}
\includegraphics[width=1\textwidth,keepaspectratio]{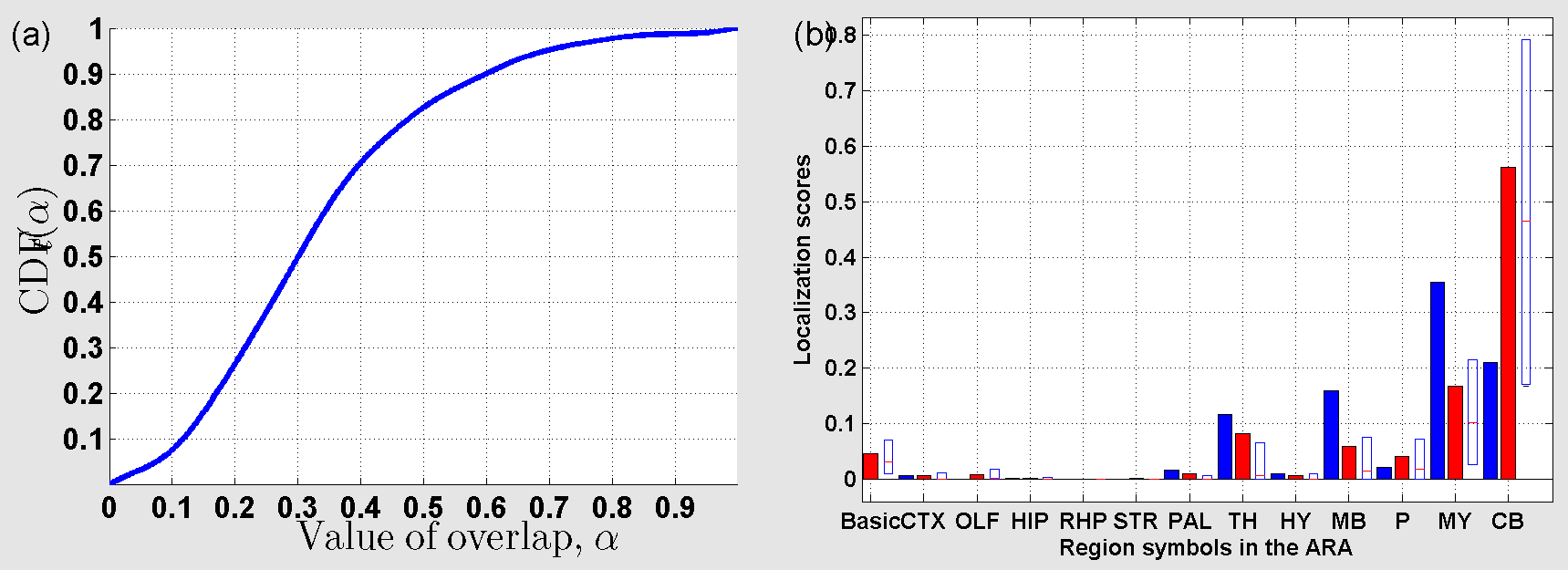}
\caption{(a) Cumulative distribution function (${\mathrm{\sc{CDF}}}_t$) of the overlap between $\rho_t$ and
 sub-sampled profiles for $t=19$. (b) Localization scores in the coarsest version of the ARA for $\rho_t$ (in blue), and 
 $\bar{\rho}_t$ (in red).}
\label{cdfPlot19}
\end{figure}
\begin{figure}
\includegraphics[width=1\textwidth,keepaspectratio]{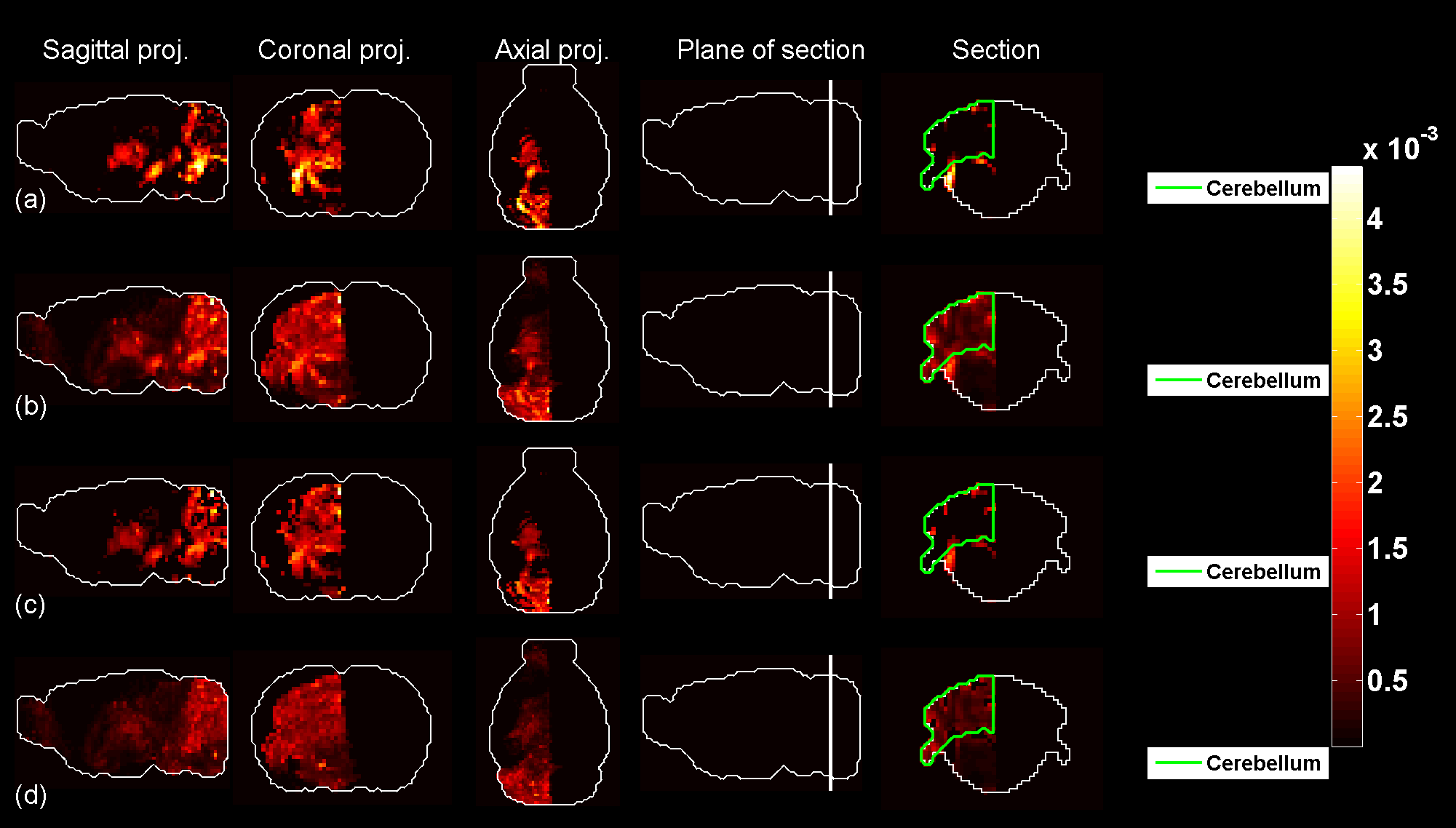}
\caption{Predicted profile and average sub-sampled profile for $t=19$.}
\label{subSampledFour19}
\end{figure}
\clearpage
\begin{figure}
\includegraphics[width=1\textwidth,keepaspectratio]{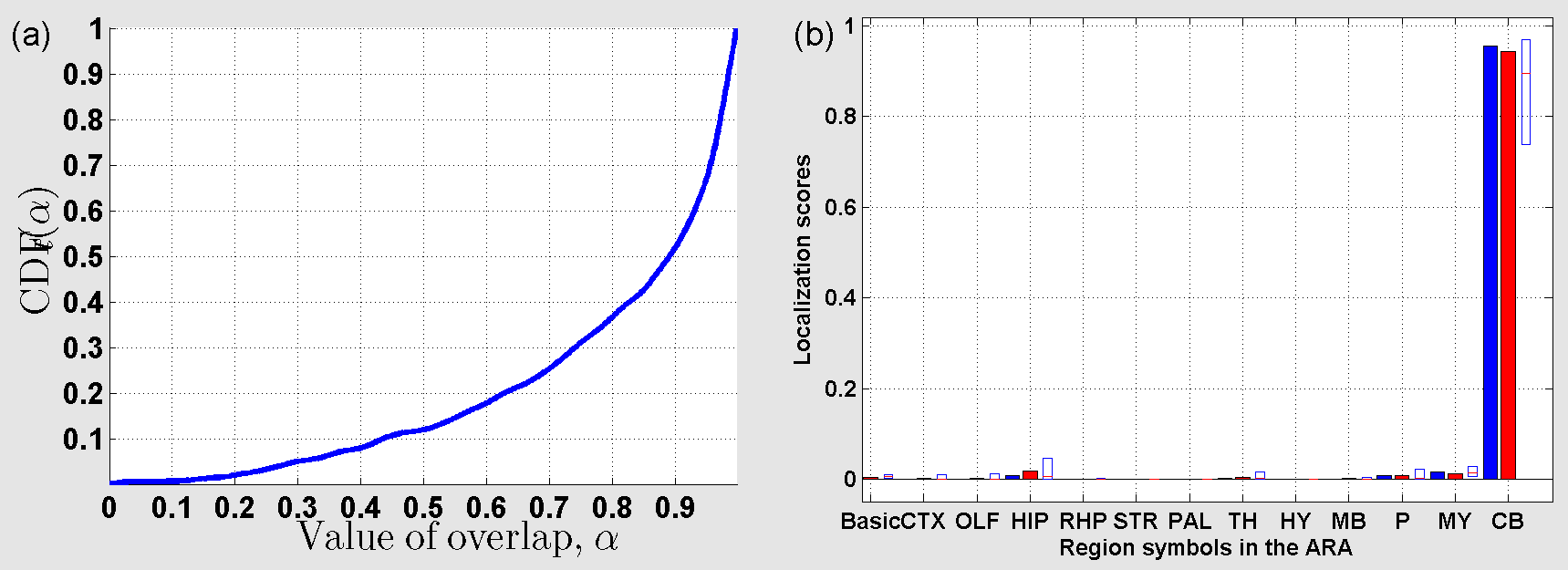}
\caption{(a) Cumulative distribution function (${\mathrm{\sc{CDF}}}_t$) of the overlap between $\rho_t$ and
 sub-sampled profiles for $t=20$. (b) Localization scores in the coarsest version of the ARA for $\rho_t$ (in blue), and 
 $\bar{\rho}_t$ (in red).}
\label{cdfPlot20}
\end{figure}
\begin{figure}
\includegraphics[width=1\textwidth,keepaspectratio]{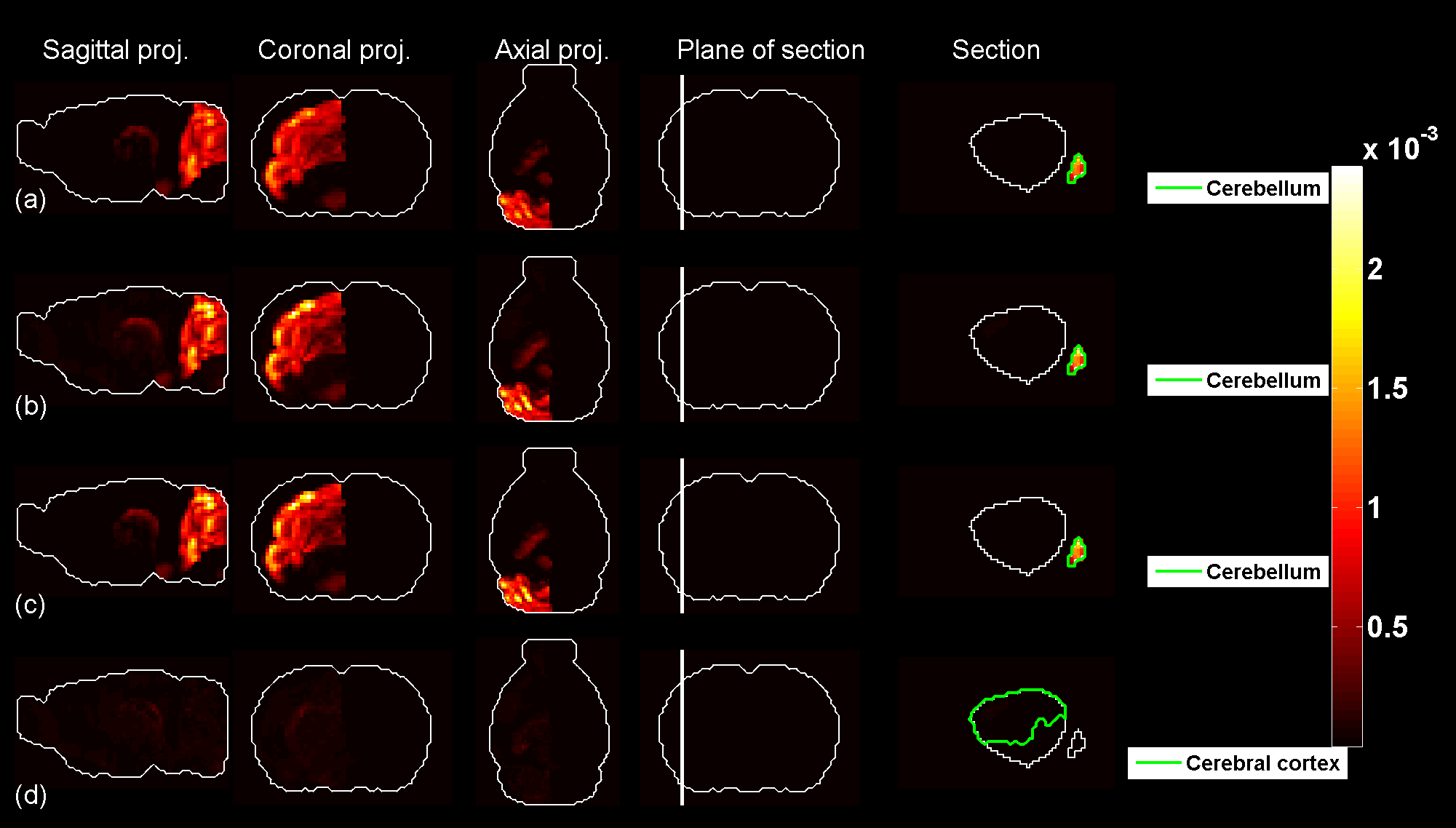}
\caption{Predicted profile and average sub-sampled profile for $t=20$.}
\label{subSampledFour20}
\end{figure}
\clearpage
\begin{figure}
\includegraphics[width=1\textwidth,keepaspectratio]{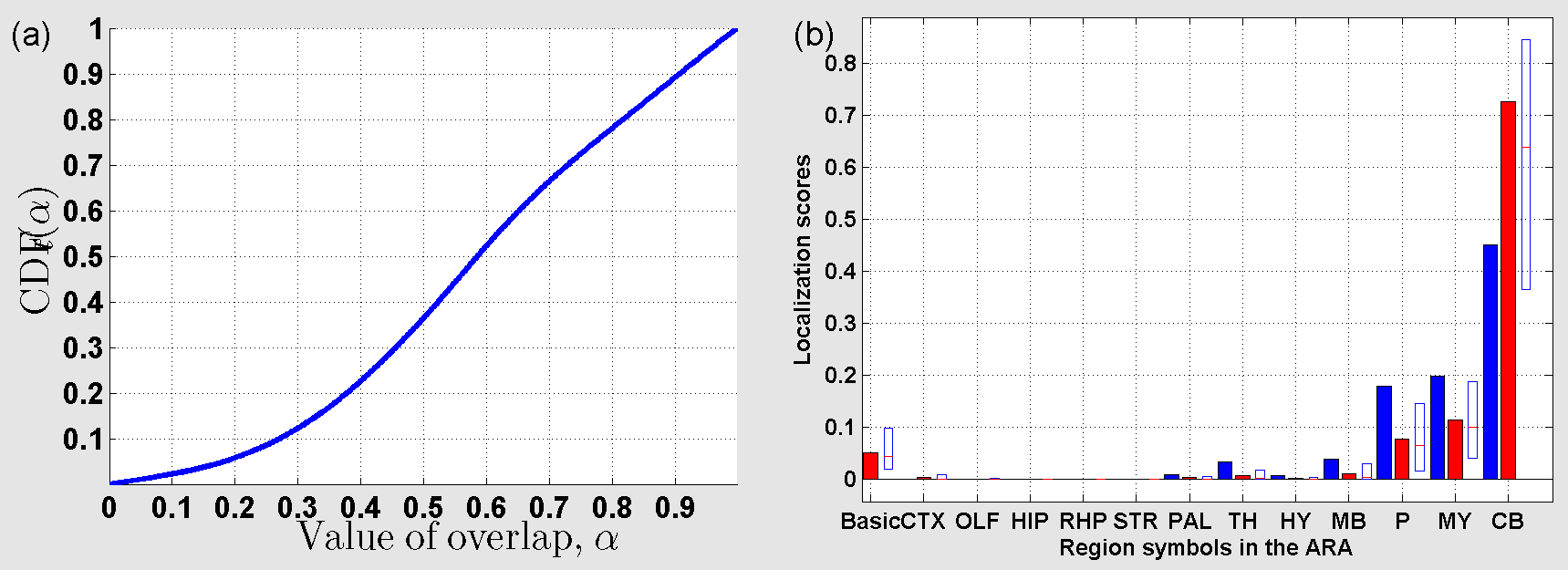}
\caption{(a) Cumulative distribution function (${\mathrm{\sc{CDF}}}_t$) of the overlap between $\rho_t$ and
 sub-sampled profiles for $t=21$. (b) Localization scores in the coarsest version of the ARA for $\rho_t$ (in blue), and 
 $\bar{\rho}_t$ (in red).}
\label{cdfPlot21}
\end{figure}
\begin{figure}
\includegraphics[width=1\textwidth,keepaspectratio]{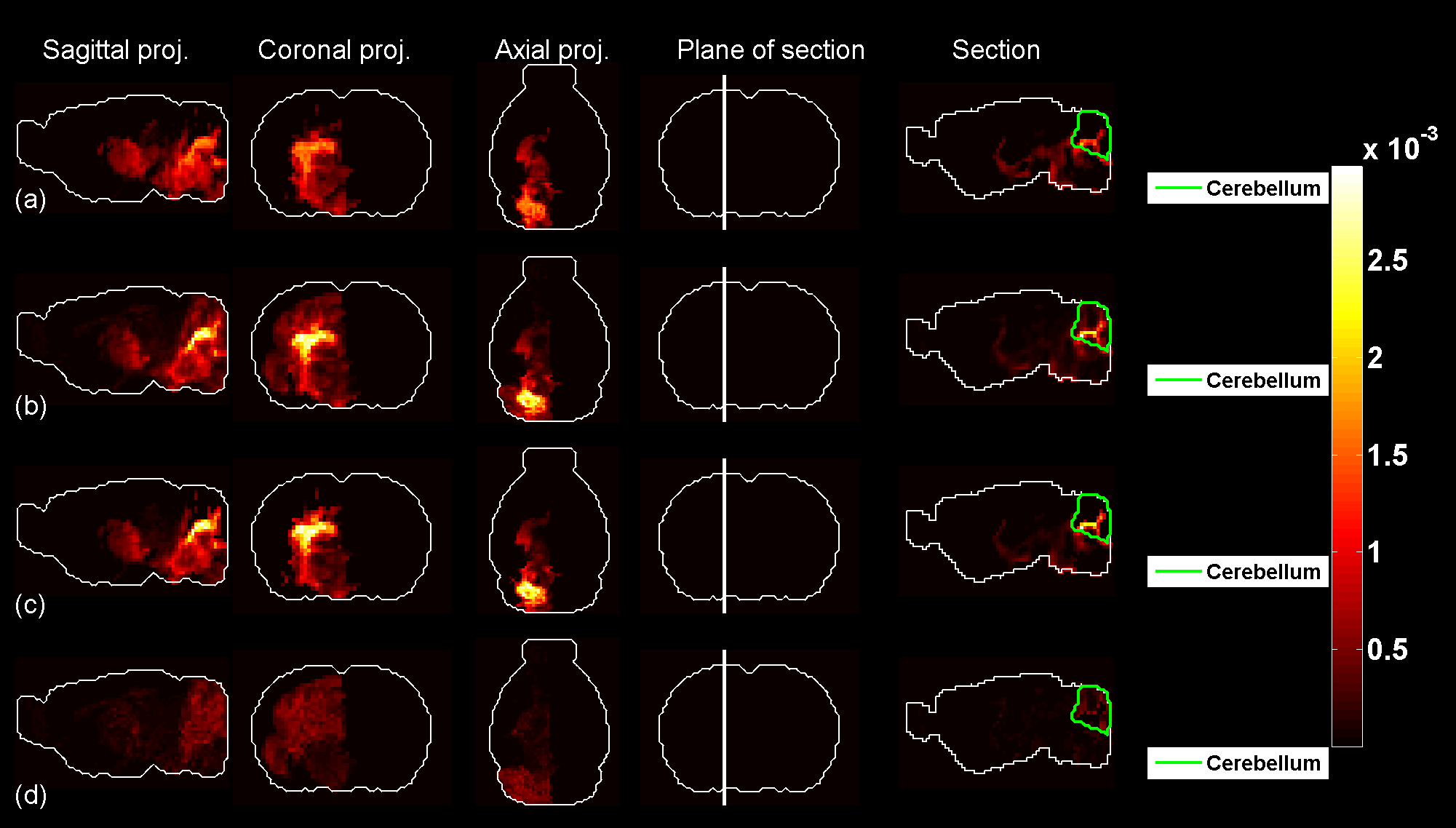}
\caption{Predicted profile and average sub-sampled profile for $t=21$.}
\label{subSampledFour21}
\end{figure}
\clearpage
\begin{figure}
\includegraphics[width=1\textwidth,keepaspectratio]{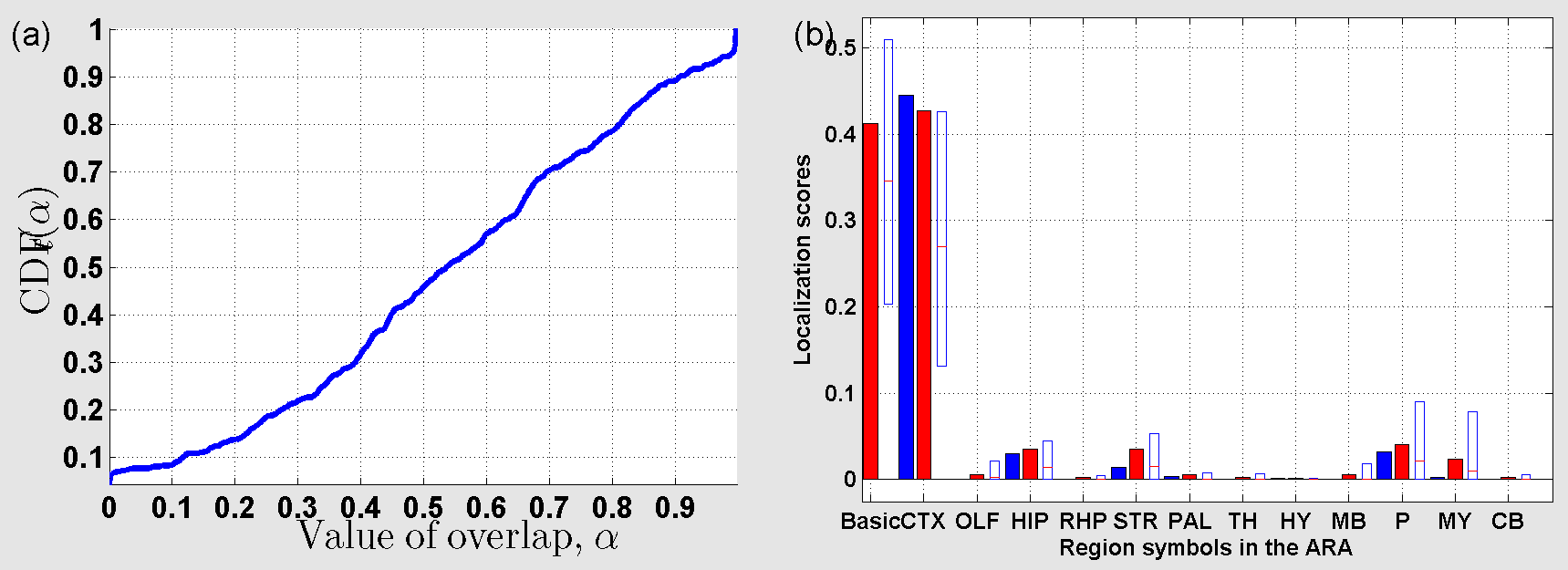}
\caption{(a) Cumulative distribution function (${\mathrm{\sc{CDF}}}_t$) of the overlap between $\rho_t$ and
 sub-sampled profiles for $t=22$. (b) Localization scores in the coarsest version of the ARA for $\rho_t$ (in blue), and 
 $\bar{\rho}_t$ (in red).}
\label{cdfPlot22}
\end{figure}
\begin{figure}
\includegraphics[width=1\textwidth,keepaspectratio]{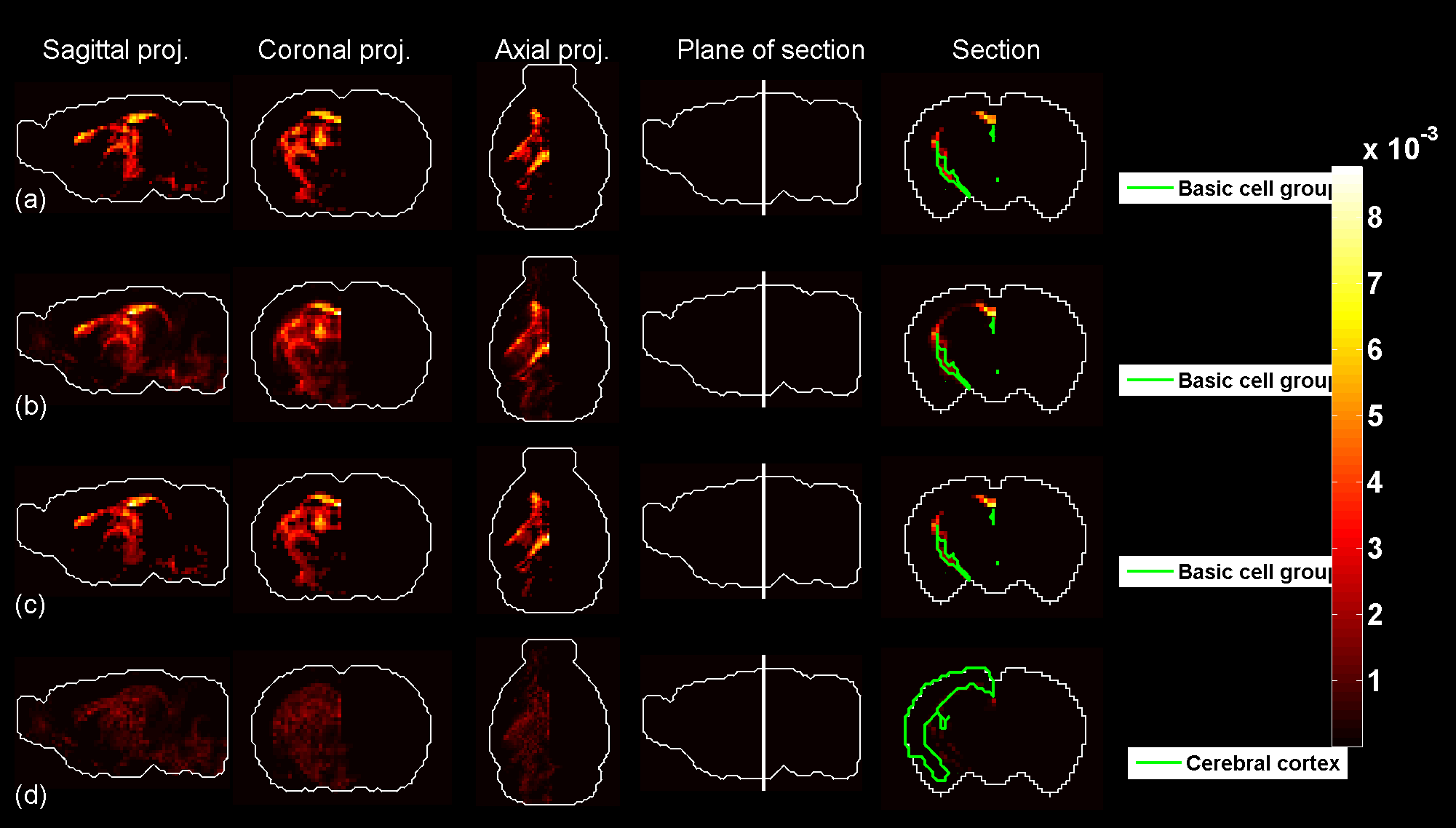}
\caption{Predicted profile and average sub-sampled profile for $t=22$.}
\label{subSampledFour22}
\end{figure}
\clearpage
\begin{figure}
\includegraphics[width=1\textwidth,keepaspectratio]{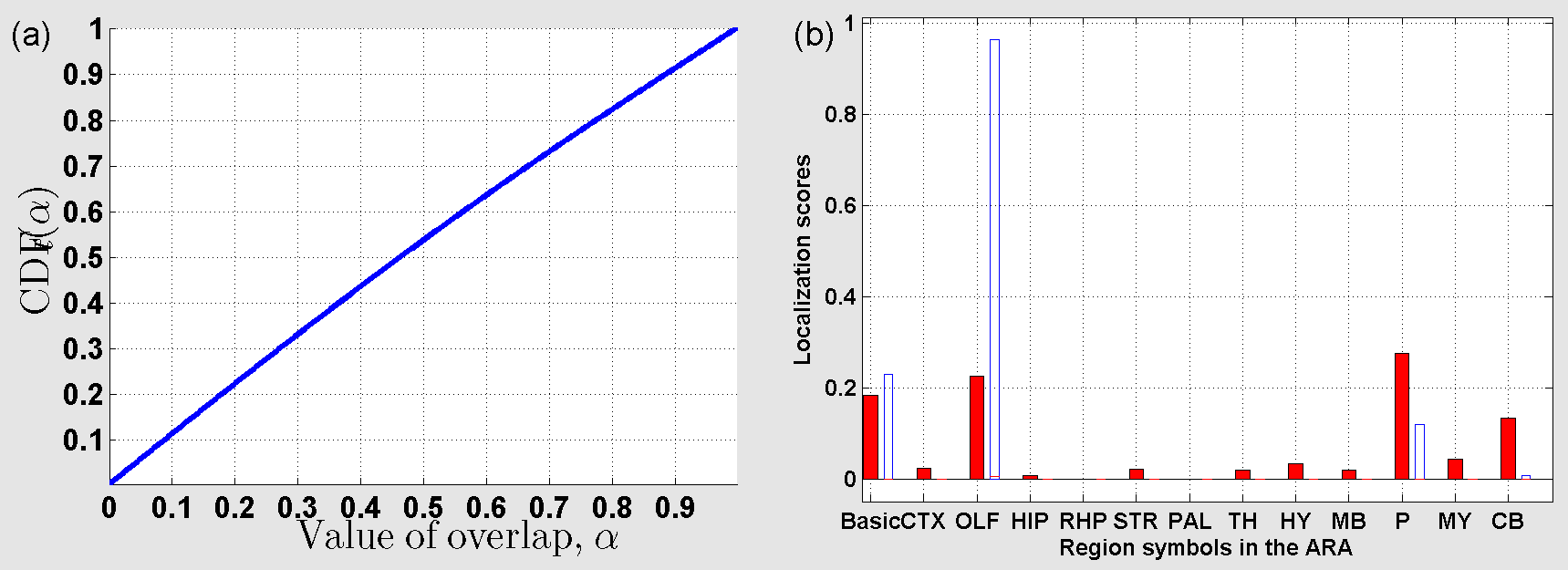}
\caption{(a) Cumulative distribution function (${\mathrm{\sc{CDF}}}_t$) of the overlap between $\rho_t$ and
 sub-sampled profiles for $t=23$. (b) Localization scores in the coarsest version of the ARA for $\rho_t$ (in blue), and 
 $\bar{\rho}_t$ (in red).}
\label{cdfPlot23}
\end{figure}
\begin{figure}
\includegraphics[width=1\textwidth,keepaspectratio]{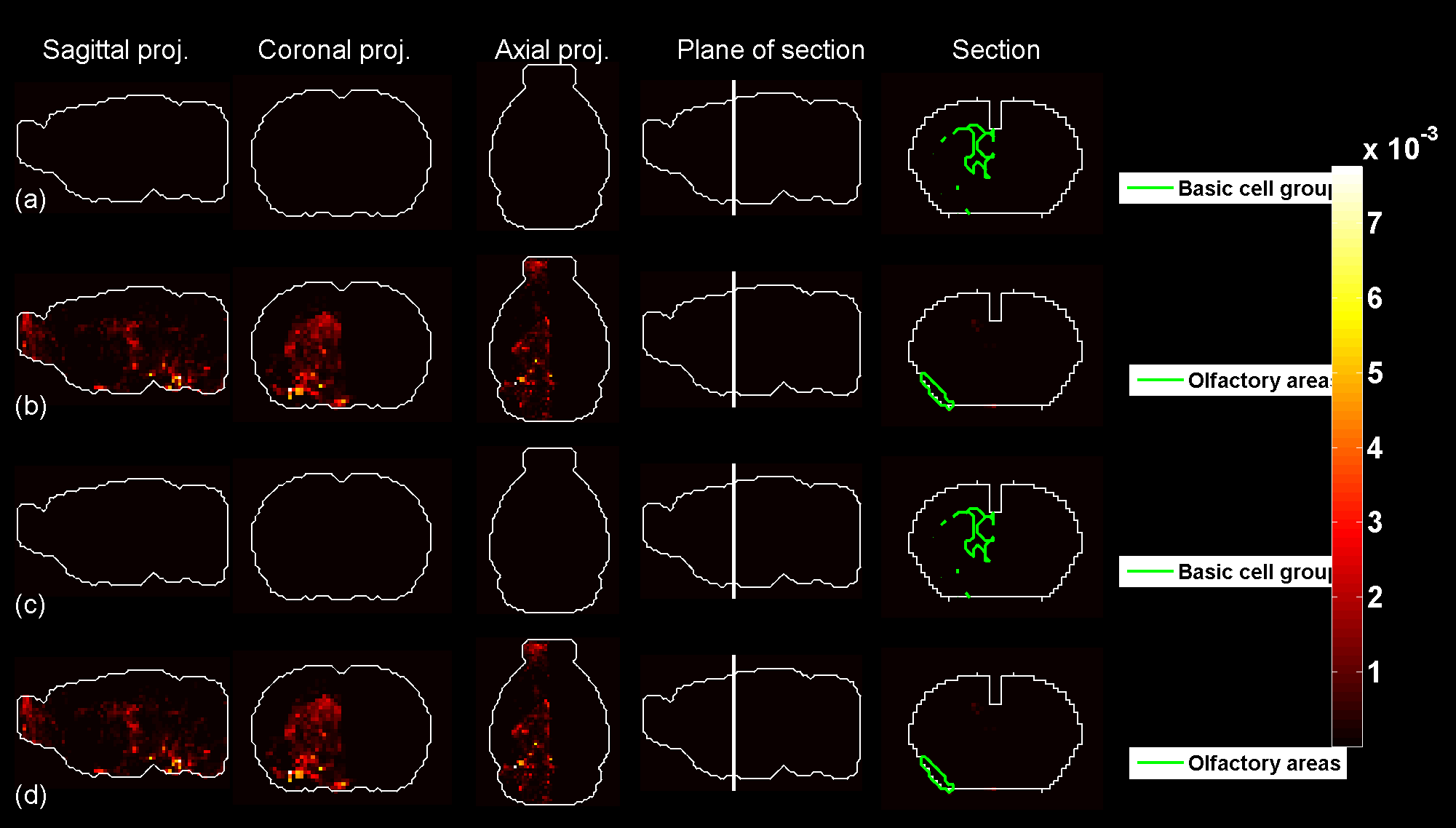}
\caption{Predicted profile and average sub-sampled profile for $t=23$.}
\label{subSampledFour23}
\end{figure}
\clearpage
\begin{figure}
\includegraphics[width=1\textwidth,keepaspectratio]{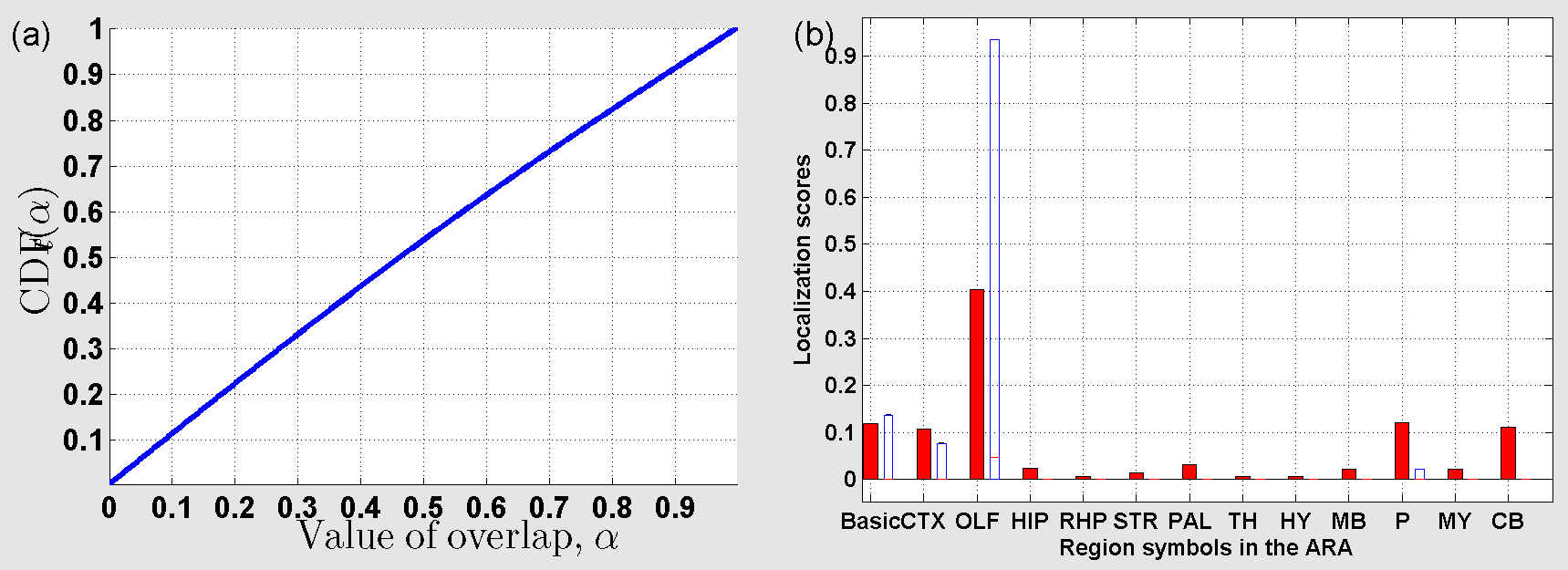}
\caption{(a) Cumulative distribution function (${\mathrm{\sc{CDF}}}_t$) of the overlap between $\rho_t$ and
 sub-sampled profiles for $t=24$. (b) Localization scores in the coarsest version of the ARA for $\rho_t$ (in blue), and 
 $\bar{\rho}_t$ (in red).}
\label{cdfPlot24}
\end{figure}
\begin{figure}
\includegraphics[width=1\textwidth,keepaspectratio]{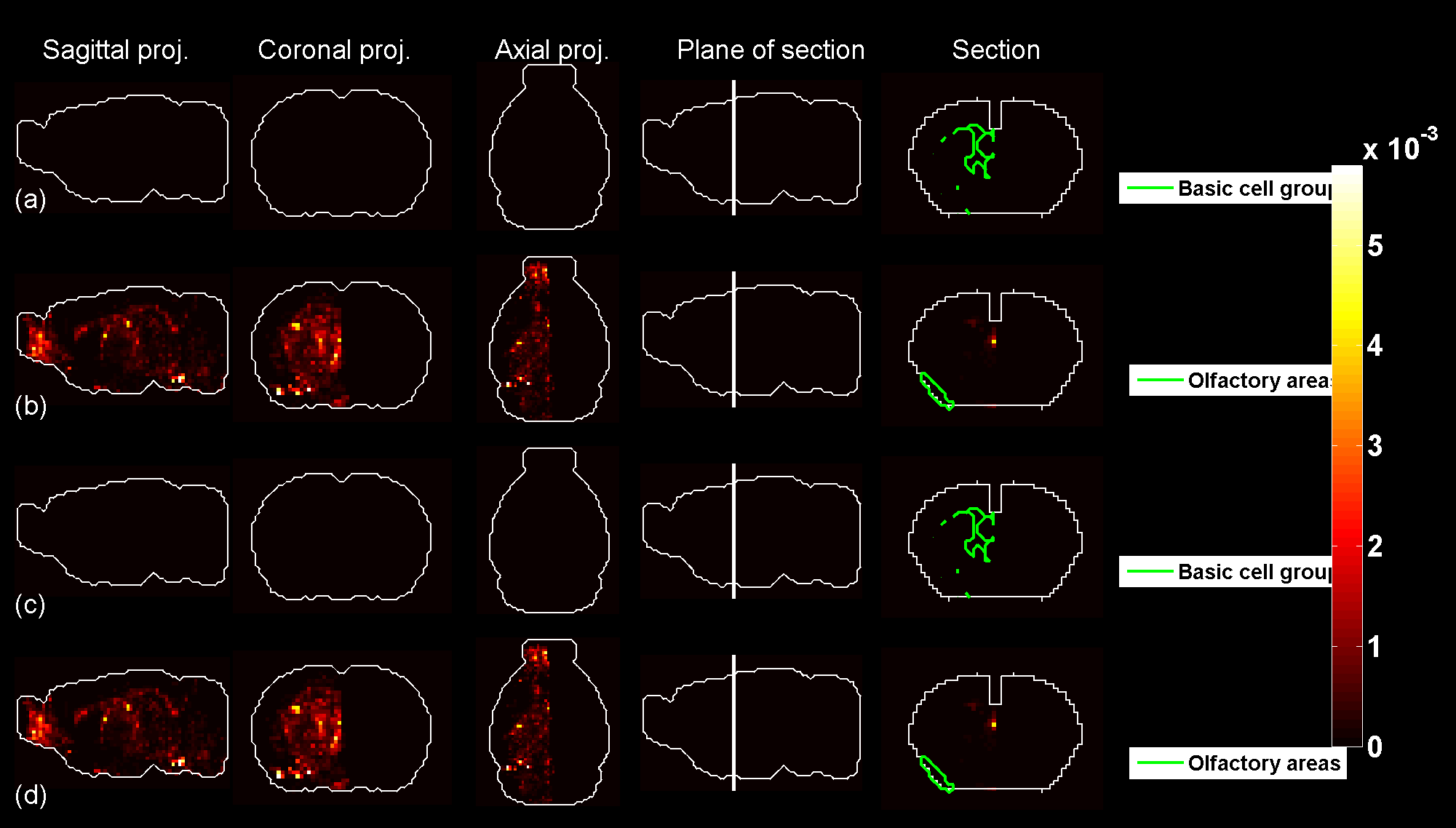}
\caption{Predicted profile and average sub-sampled profile for $t=24$.}
\label{subSampledFour24}
\end{figure}
\clearpage
\begin{figure}
\includegraphics[width=1\textwidth,keepaspectratio]{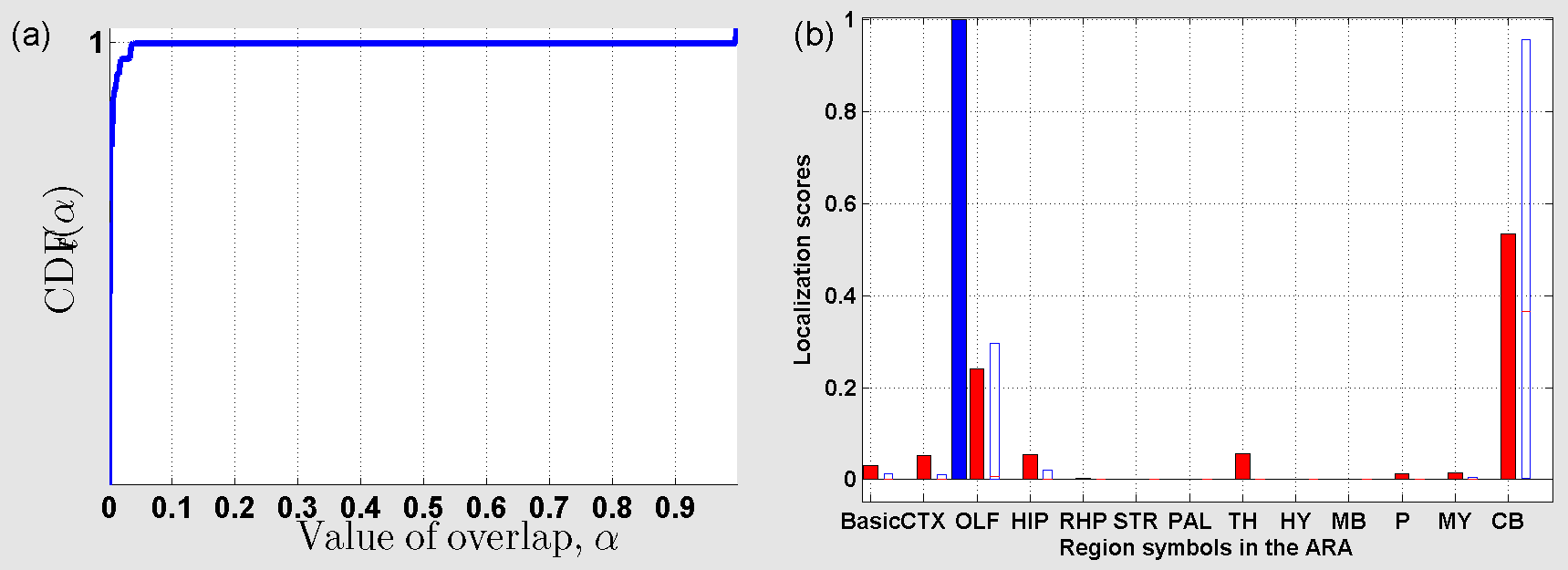}
\caption{(a) Cumulative distribution function (${\mathrm{\sc{CDF}}}_t$) of the overlap between $\rho_t$ and
 sub-sampled profiles for $t=25$. (b) Localization scores in the coarsest version of the ARA for $\rho_t$ (in blue), and 
 $\bar{\rho}_t$ (in red).}
\label{cdfPlot25}
\end{figure}
\begin{figure}
\includegraphics[width=1\textwidth,keepaspectratio]{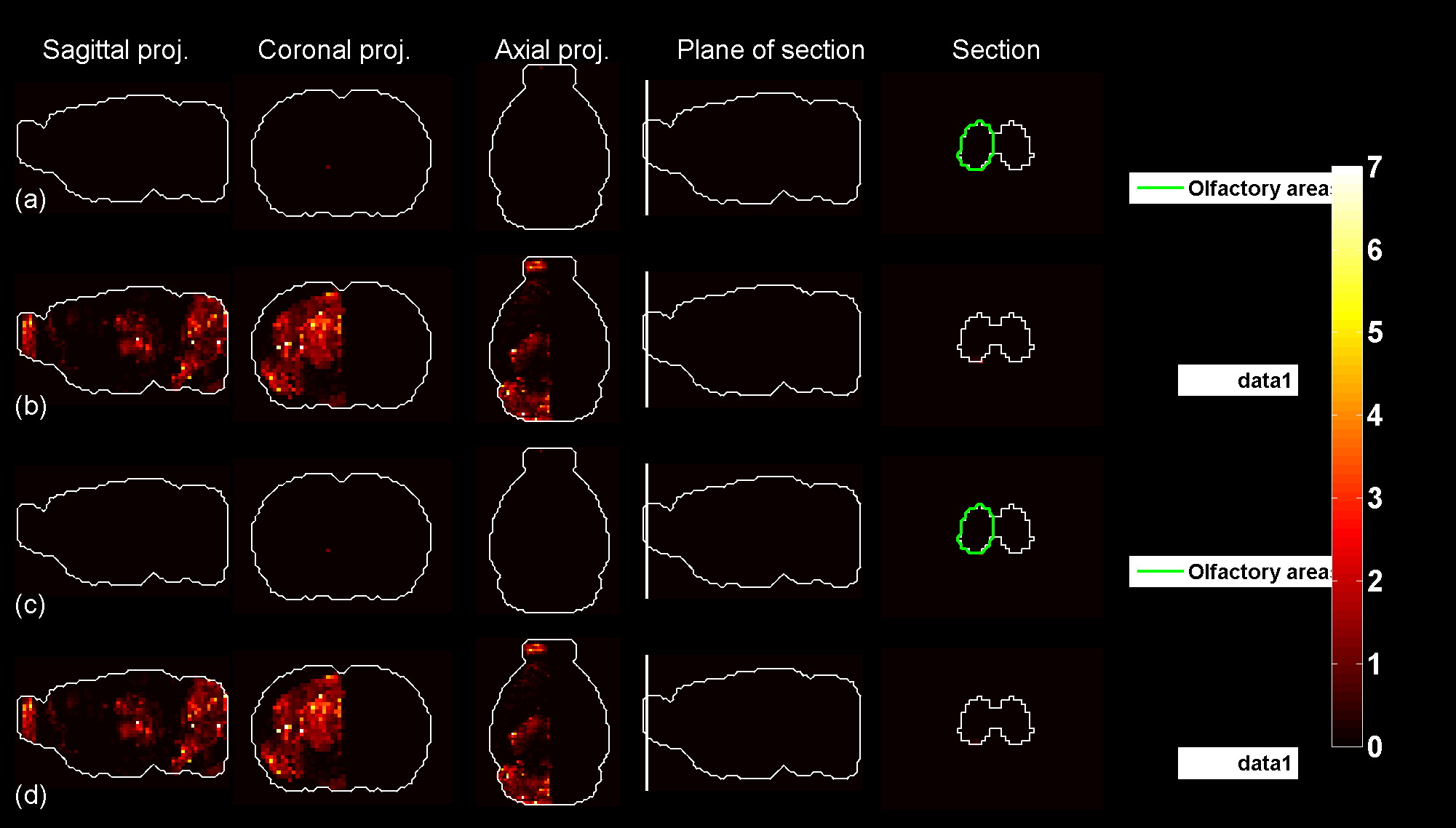}
\caption{Predicted profile and average sub-sampled profile for $t=25$.}
\label{subSampledFour25}
\end{figure}
\clearpage
\begin{figure}
\includegraphics[width=1\textwidth,keepaspectratio]{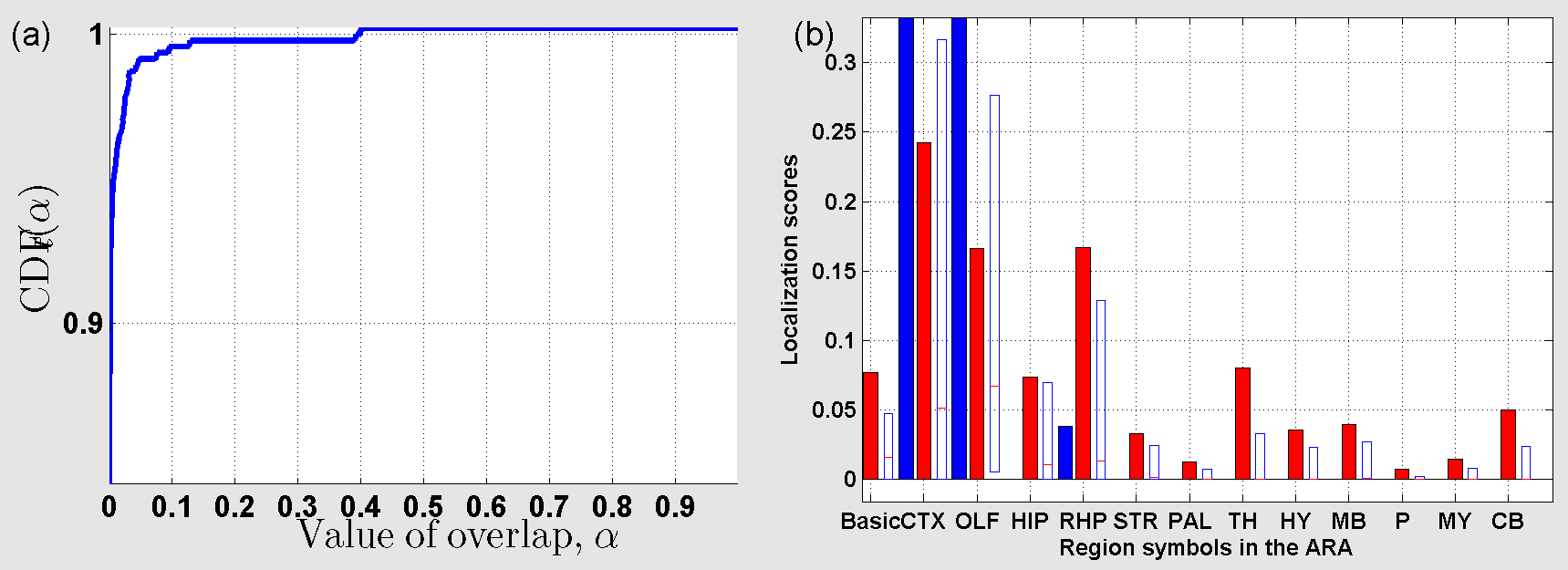}
\caption{(a) Cumulative distribution function (${\mathrm{\sc{CDF}}}_t$) of the overlap between $\rho_t$ and
 sub-sampled profiles for $t=26$. (b) Localization scores in the coarsest version of the ARA for $\rho_t$ (in blue), and 
 $\bar{\rho}_t$ (in red).}
\label{cdfPlot26}
\end{figure}
\begin{figure}
\includegraphics[width=1\textwidth,keepaspectratio]{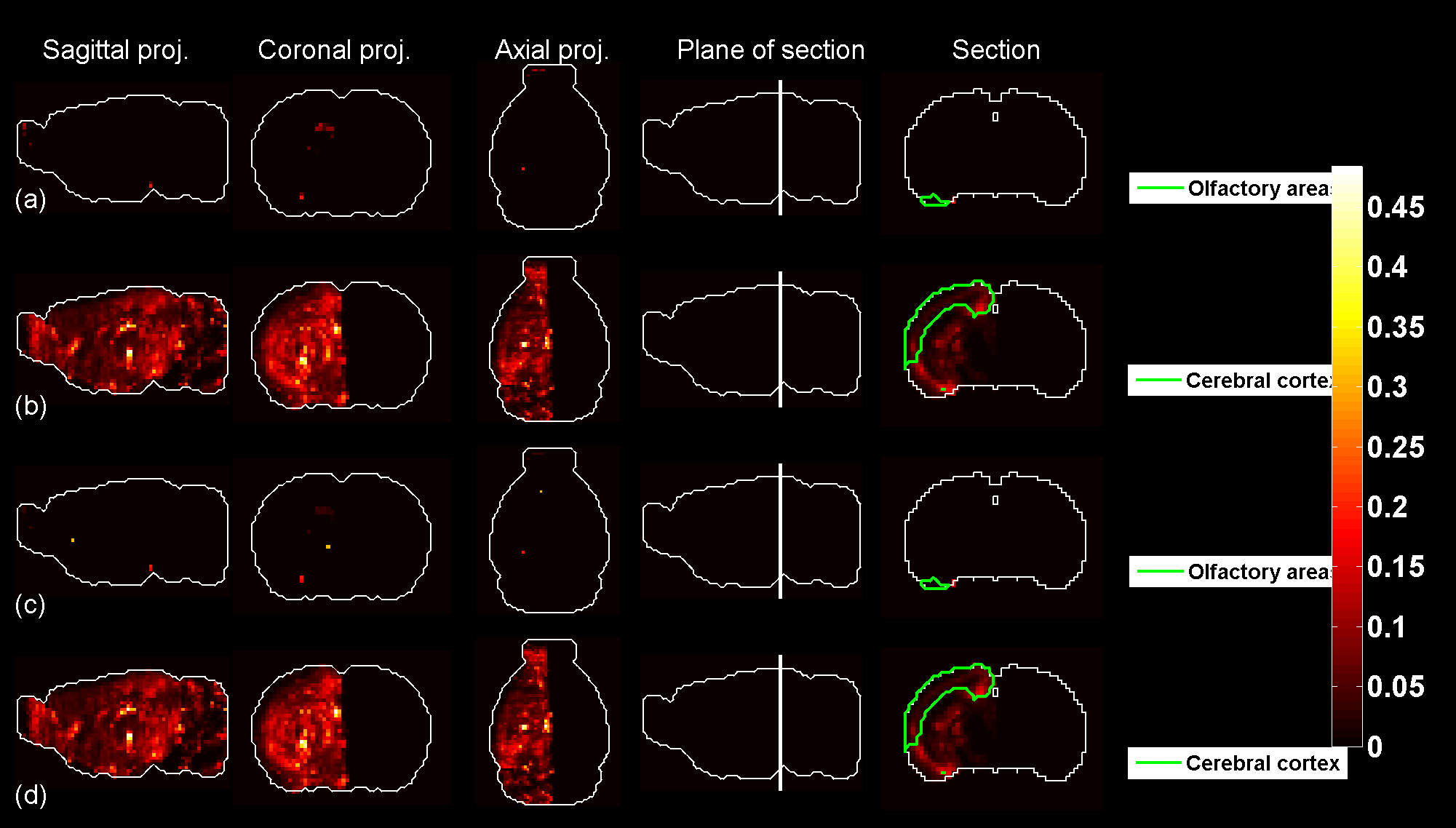}
\caption{Predicted profile and average sub-sampled profile for $t=26$.}
\label{subSampledFour26}
\end{figure}
\clearpage
\begin{figure}
\includegraphics[width=1\textwidth,keepaspectratio]{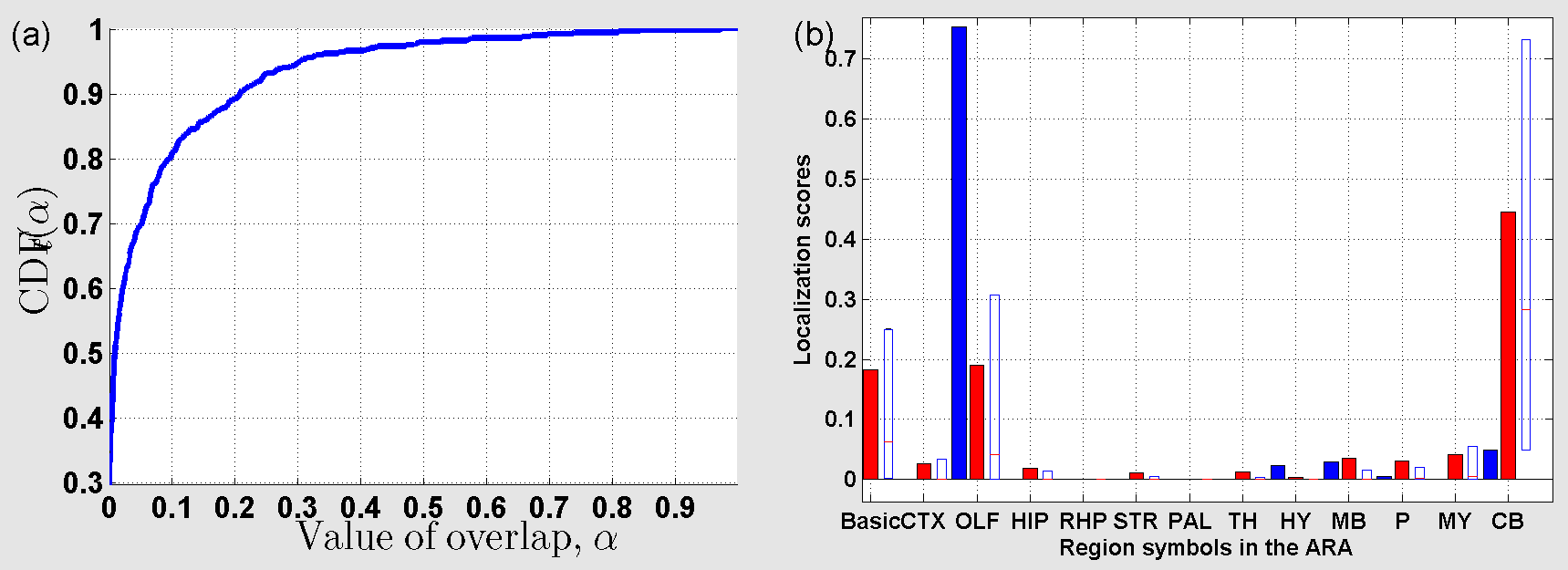}
\caption{(a) Cumulative distribution function (${\mathrm{\sc{CDF}}}_t$) of the overlap between $\rho_t$ and
 sub-sampled profiles for $t=27$. (b) Localization scores in the coarsest version of the ARA for $\rho_t$ (in blue), and 
 $\bar{\rho}_t$ (in red).}
\label{cdfPlot27}
\end{figure}
\begin{figure}
\includegraphics[width=1\textwidth,keepaspectratio]{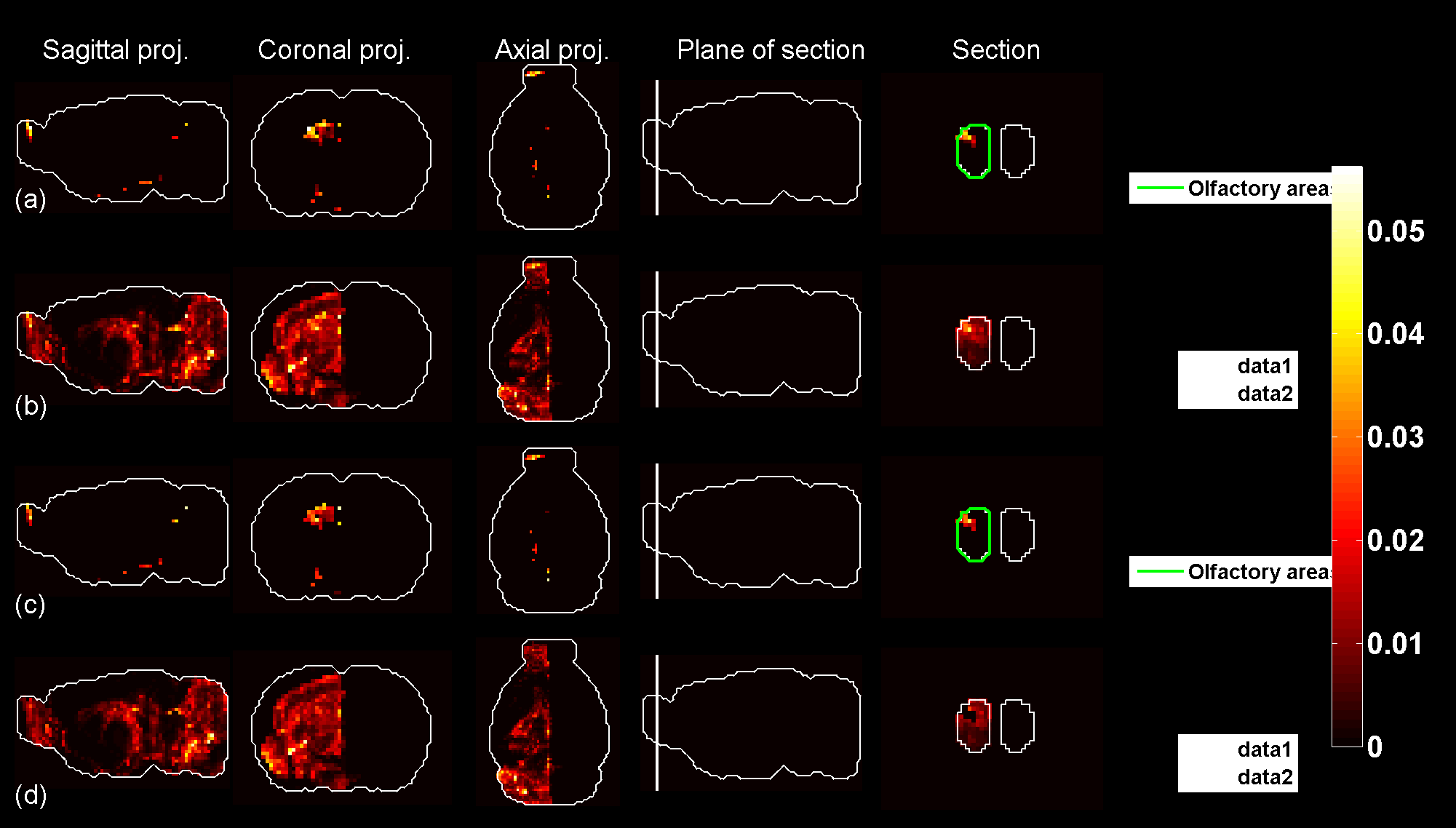}
\caption{Predicted profile and average sub-sampled profile for $t=27$.}
\label{subSampledFour27}
\end{figure}
\clearpage
\begin{figure}
\includegraphics[width=1\textwidth,keepaspectratio]{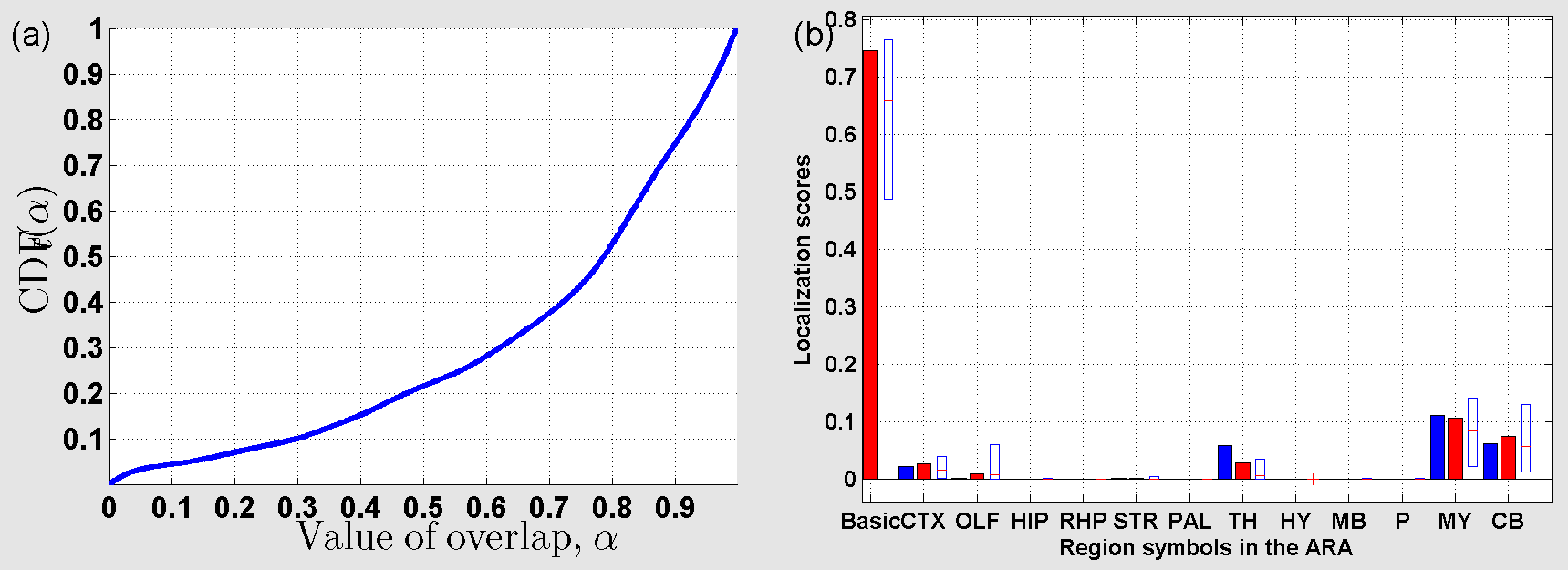}
\caption{(a) Cumulative distribution function (${\mathrm{\sc{CDF}}}_t$) of the overlap between $\rho_t$ and
 sub-sampled profiles for $t=28$. (b) Localization scores in the coarsest version of the ARA for $\rho_t$ (in blue), and 
 $\bar{\rho}_t$ (in red).}
\label{cdfPlot28}
\end{figure}
\begin{figure}
\includegraphics[width=1\textwidth,keepaspectratio]{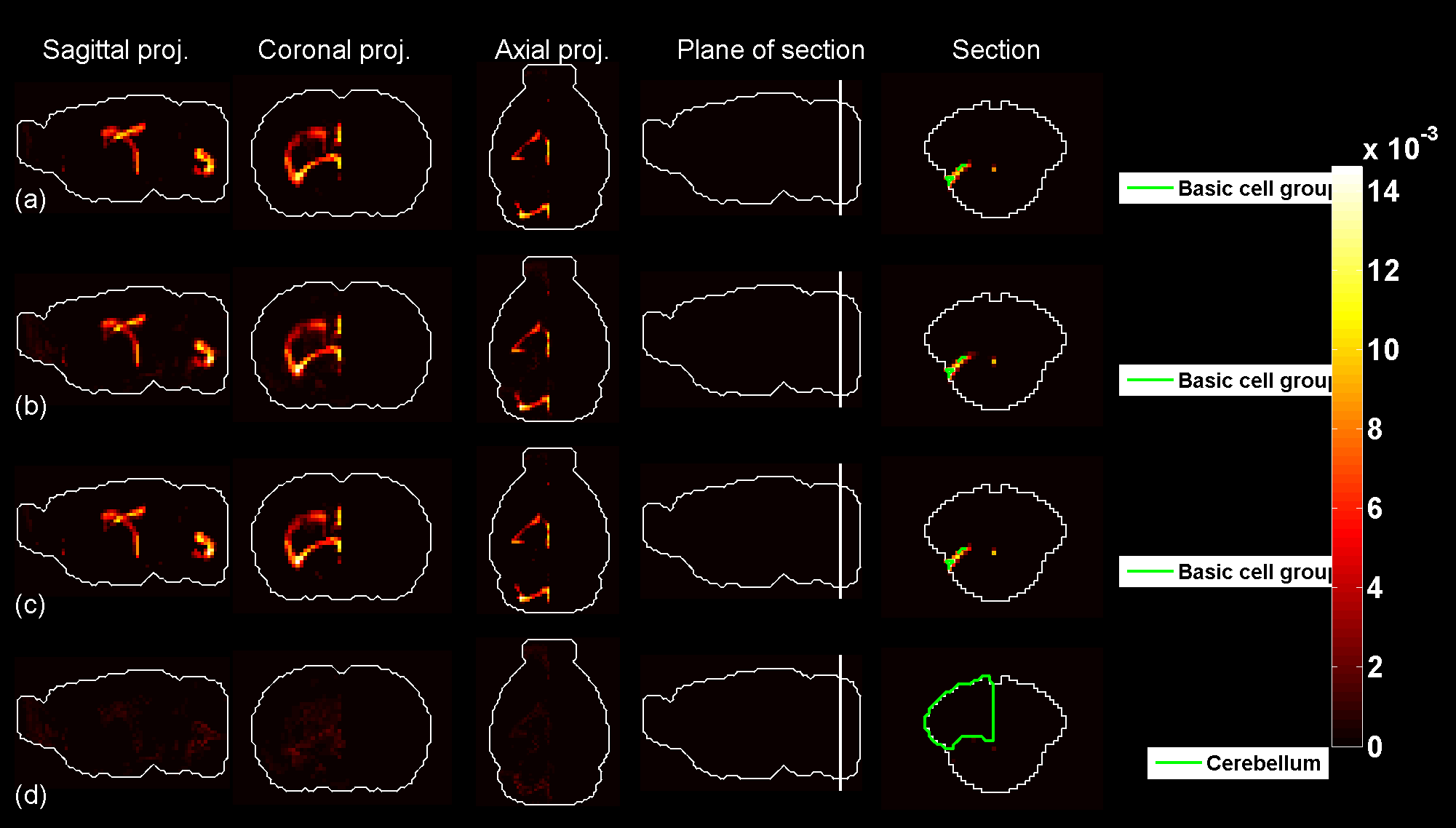}
\caption{Predicted profile and average sub-sampled profile for $t=28$.}
\label{subSampledFour28}
\end{figure}
\clearpage
\begin{figure}
\includegraphics[width=1\textwidth,keepaspectratio]{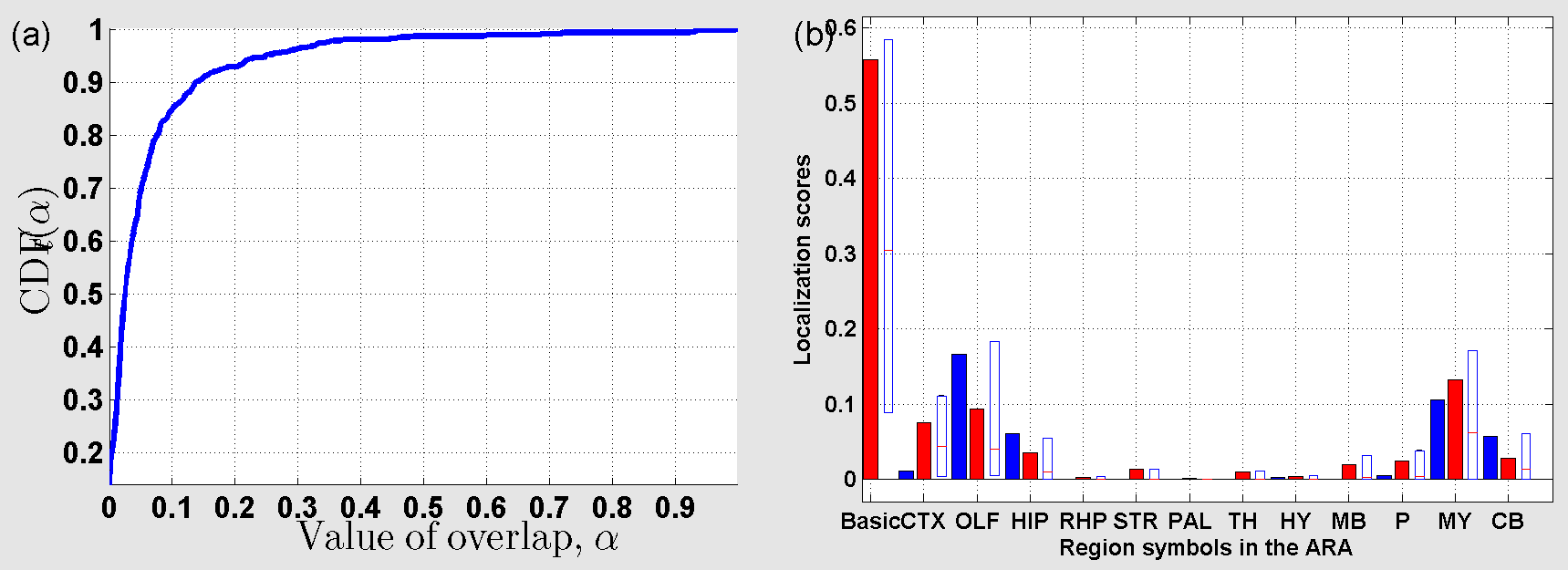}
\caption{(a) Cumulative distribution function (${\mathrm{\sc{CDF}}}_t$) of the overlap between $\rho_t$ and
 sub-sampled profiles for $t=29$. (b) Localization scores in the coarsest version of the ARA for $\rho_t$ (in blue), and 
 $\bar{\rho}_t$ (in red).}
\label{cdfPlot29}
\end{figure}
\begin{figure}
\includegraphics[width=1\textwidth,keepaspectratio]{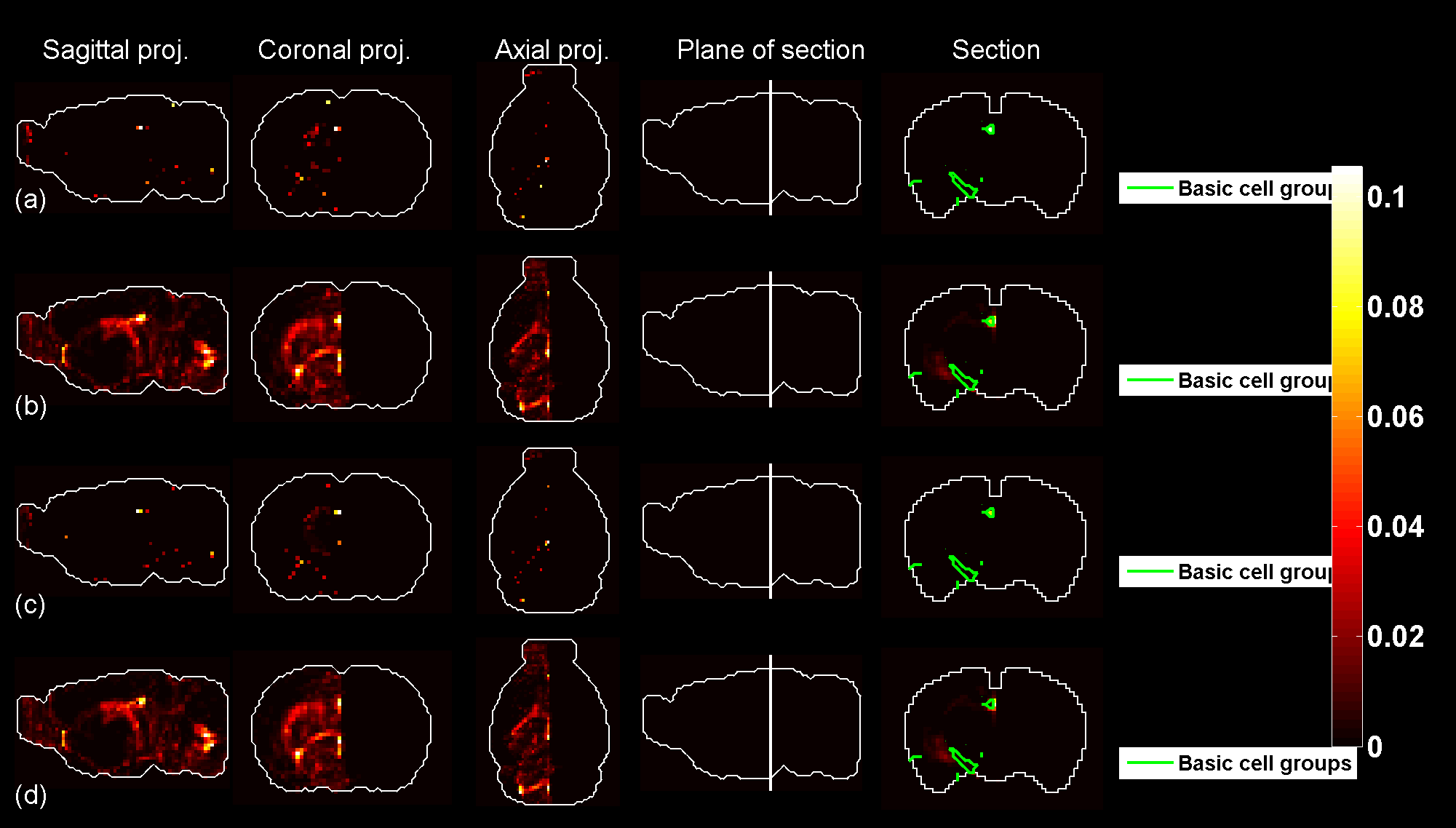}
\caption{Predicted profile and average sub-sampled profile for $t=29$.}
\label{subSampledFour29}
\end{figure}
\clearpage
\begin{figure}
\includegraphics[width=1\textwidth,keepaspectratio]{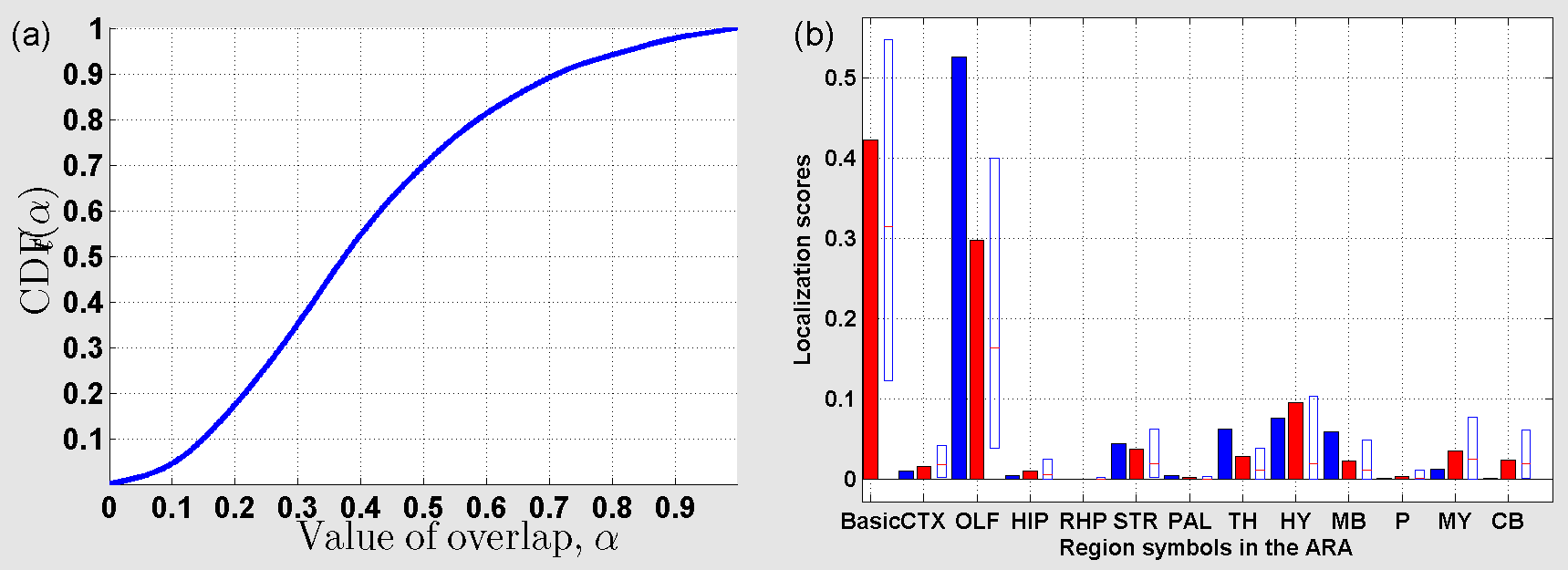}
\caption{(a) Cumulative distribution function (${\mathrm{\sc{CDF}}}_t$) of the overlap between $\rho_t$ and
 sub-sampled profiles for $t=30$. (b) Localization scores in the coarsest version of the ARA for $\rho_t$ (in blue), and 
 $\bar{\rho}_t$ (in red).}
\label{cdfPlot30}
\end{figure}
\begin{figure}
\includegraphics[width=1\textwidth,keepaspectratio]{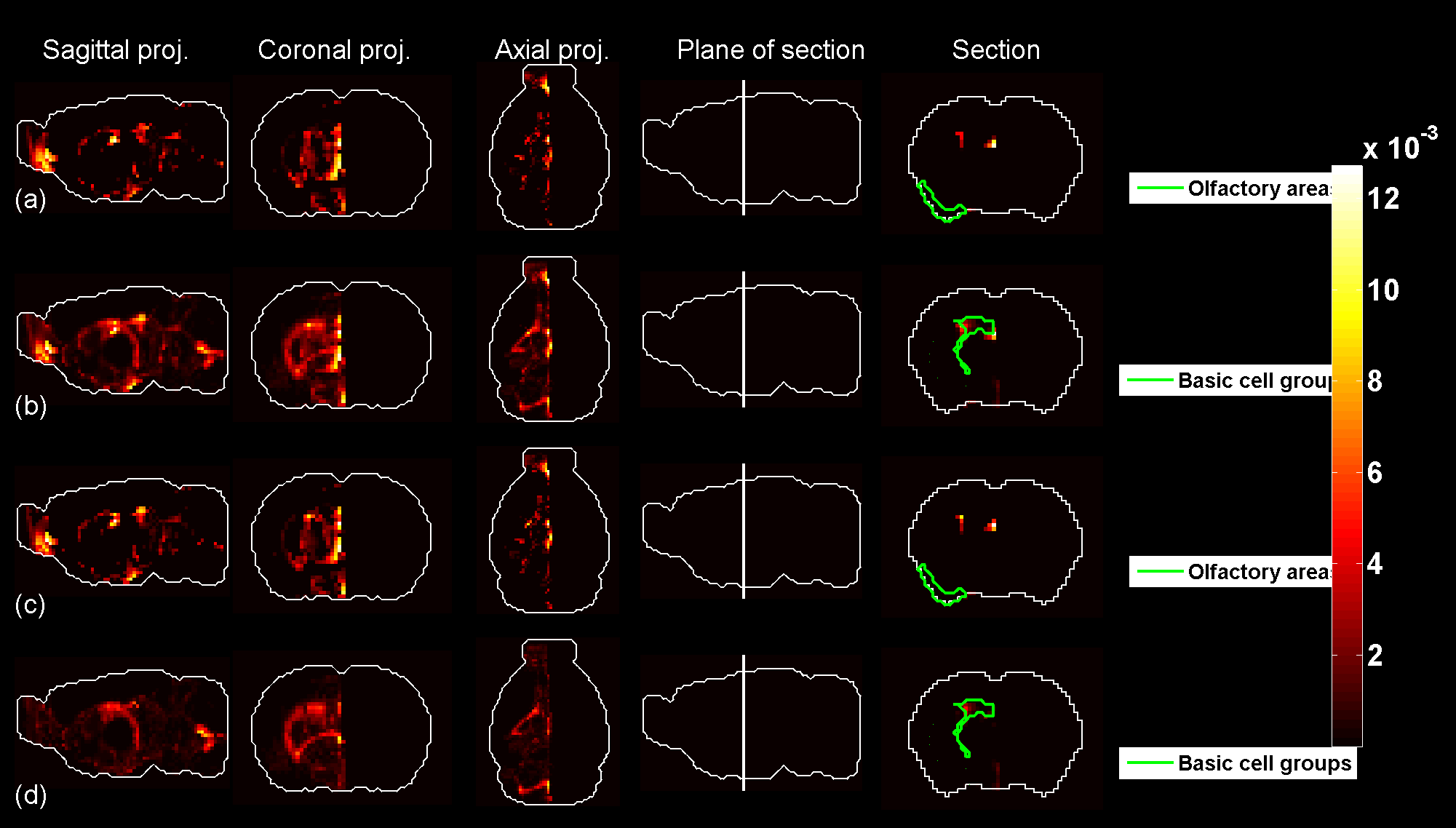}
\caption{Predicted profile and average sub-sampled profile for $t=30$.}
\label{subSampledFour30}
\end{figure}
\clearpage
\begin{figure}
\includegraphics[width=1\textwidth,keepaspectratio]{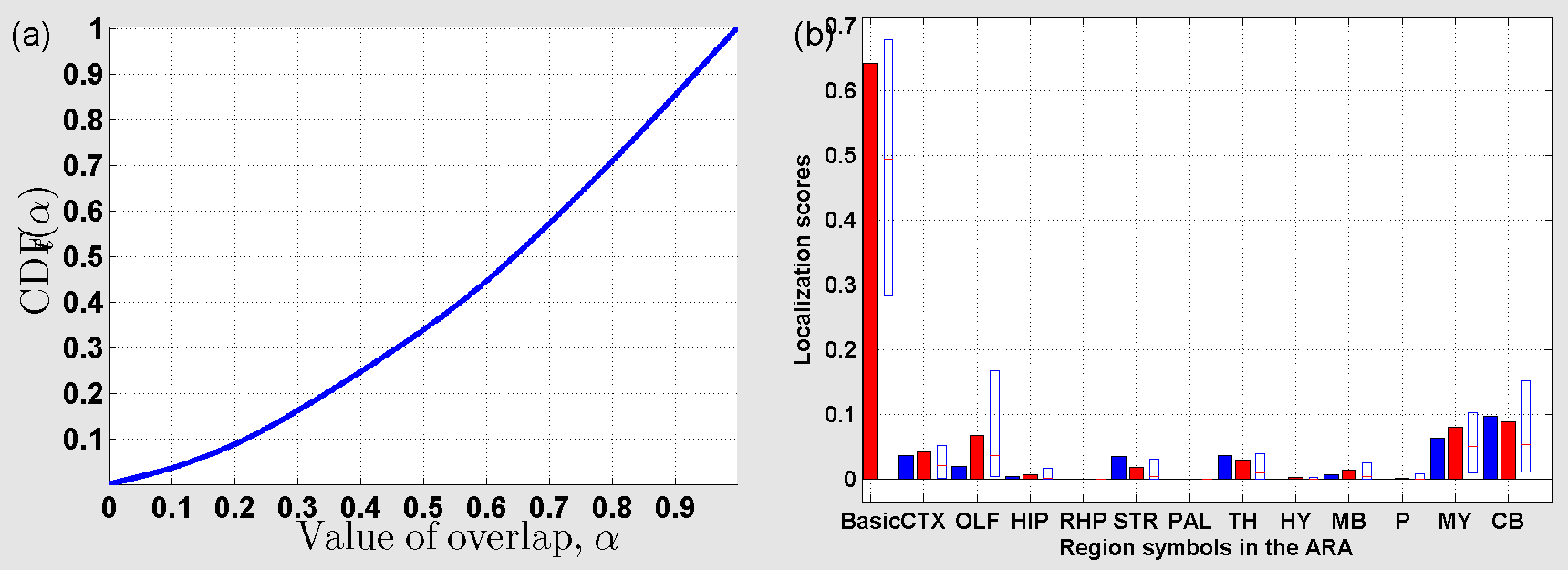}
\caption{(a) Cumulative distribution function (${\mathrm{\sc{CDF}}}_t$) of the overlap between $\rho_t$ and
 sub-sampled profiles for $t=31$. (b) Localization scores in the coarsest version of the ARA for $\rho_t$ (in blue), and 
 $\bar{\rho}_t$ (in red).}
\label{cdfPlot31}
\end{figure}
\begin{figure}
\includegraphics[width=1\textwidth,keepaspectratio]{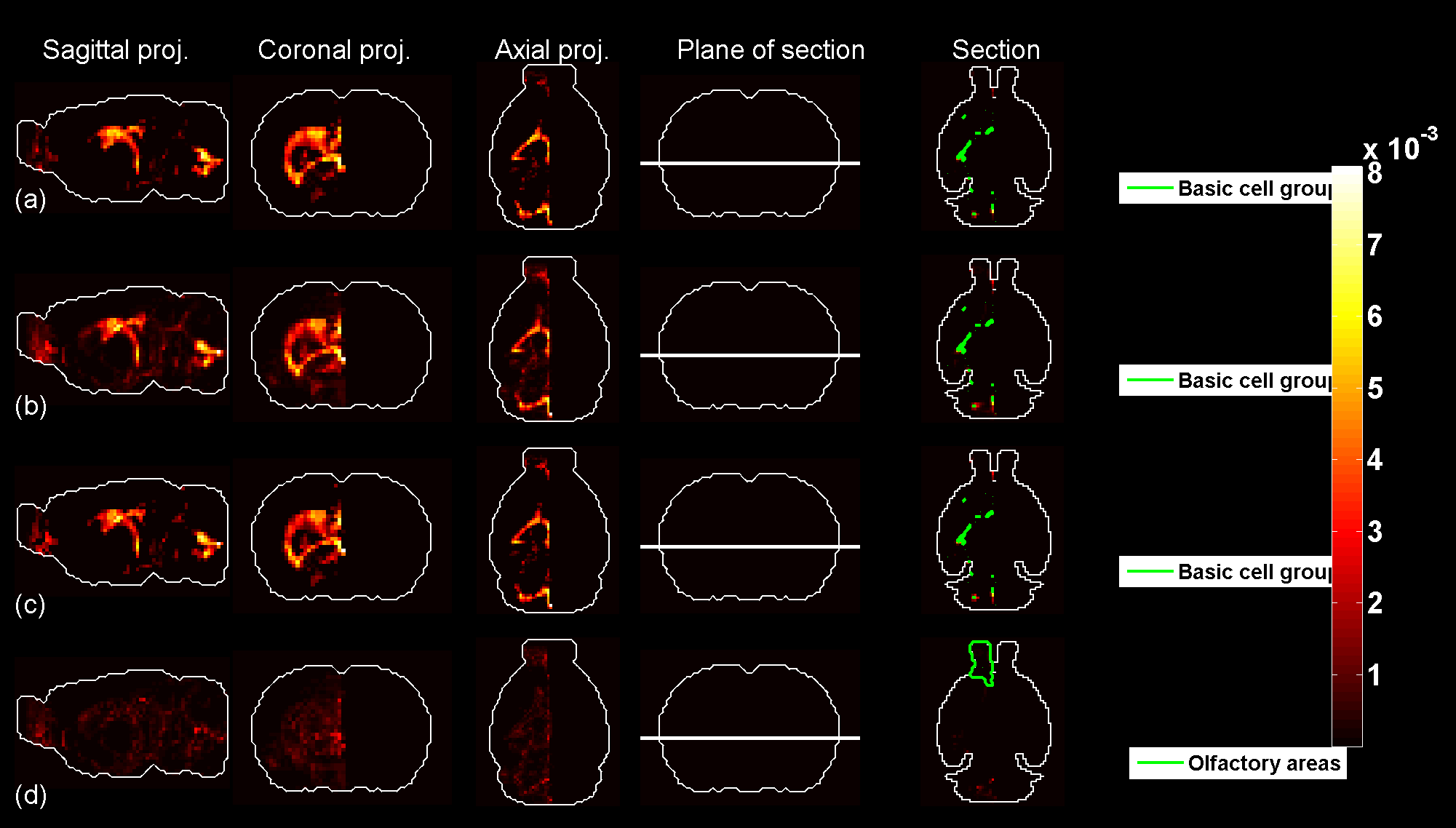}
\caption{Predicted profile and average sub-sampled profile for $t=31$.}
\label{subSampledFour31}
\end{figure}
\clearpage
\begin{figure}
\includegraphics[width=1\textwidth,keepaspectratio]{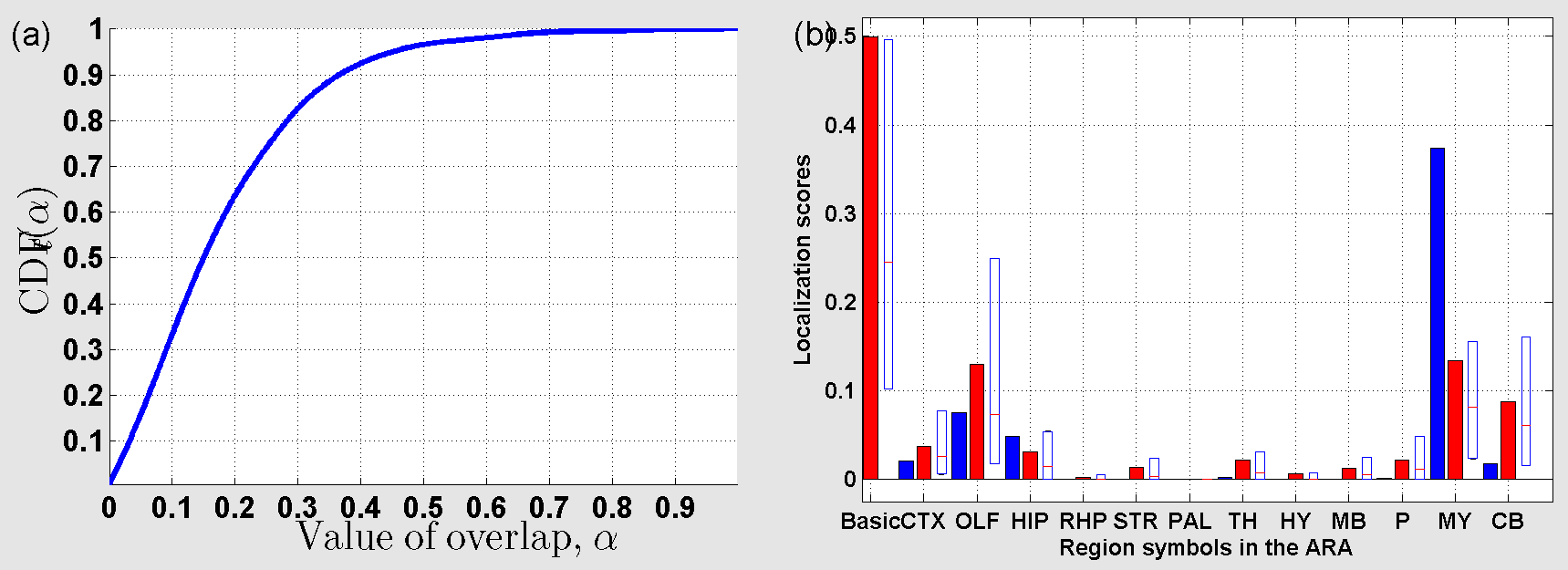}
\caption{(a) Cumulative distribution function (${\mathrm{\sc{CDF}}}_t$) of the overlap between $\rho_t$ and
 sub-sampled profiles for $t=32$. (b) Localization scores in the coarsest version of the ARA for $\rho_t$ (in blue), and 
 $\bar{\rho}_t$ (in red).}
\label{cdfPlot32}
\end{figure}
\begin{figure}
\includegraphics[width=1\textwidth,keepaspectratio]{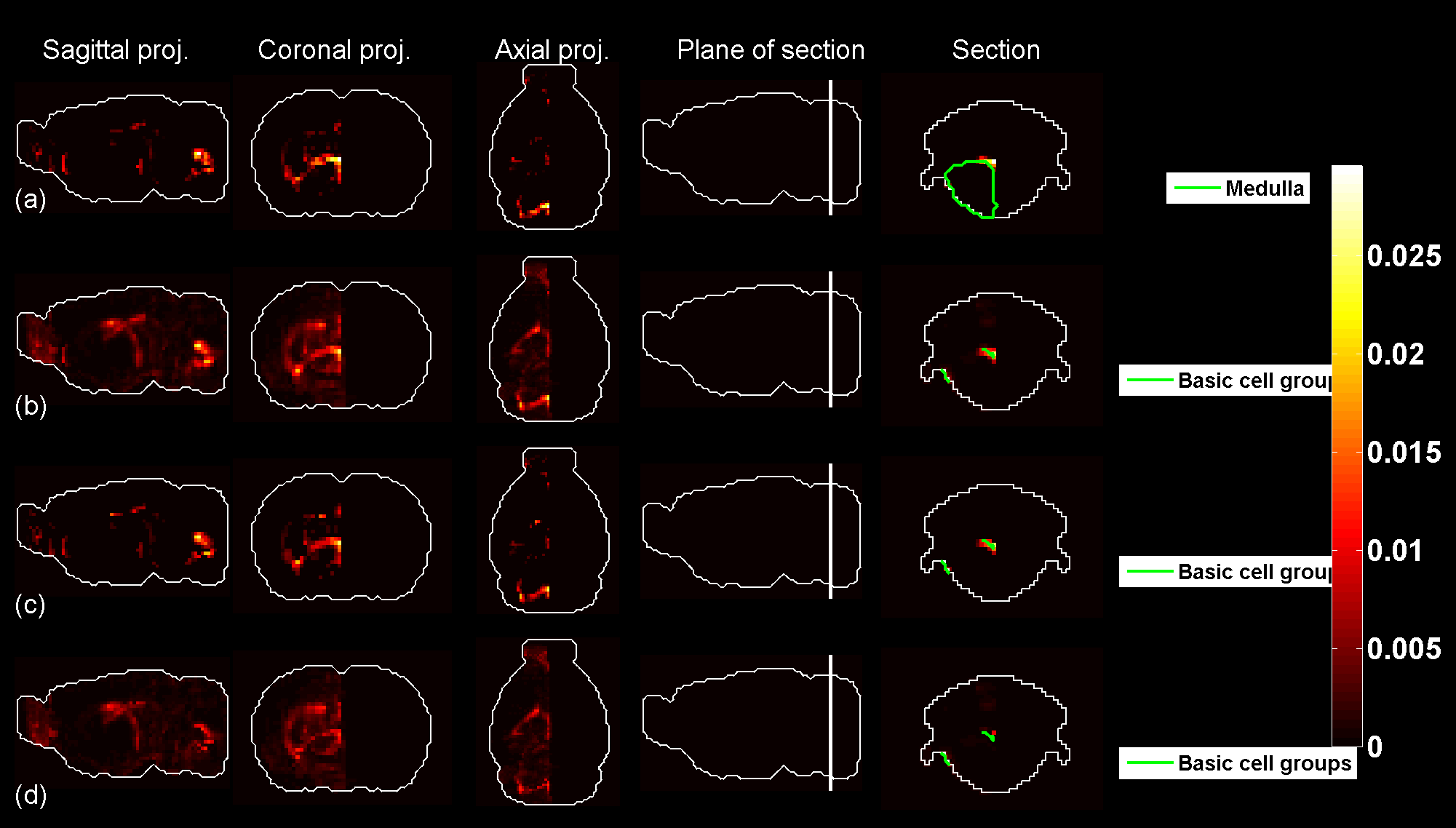}
\caption{Predicted profile and average sub-sampled profile for $t=32$.}
\label{subSampledFour32}
\end{figure}
\clearpage
\begin{figure}
\includegraphics[width=1\textwidth,keepaspectratio]{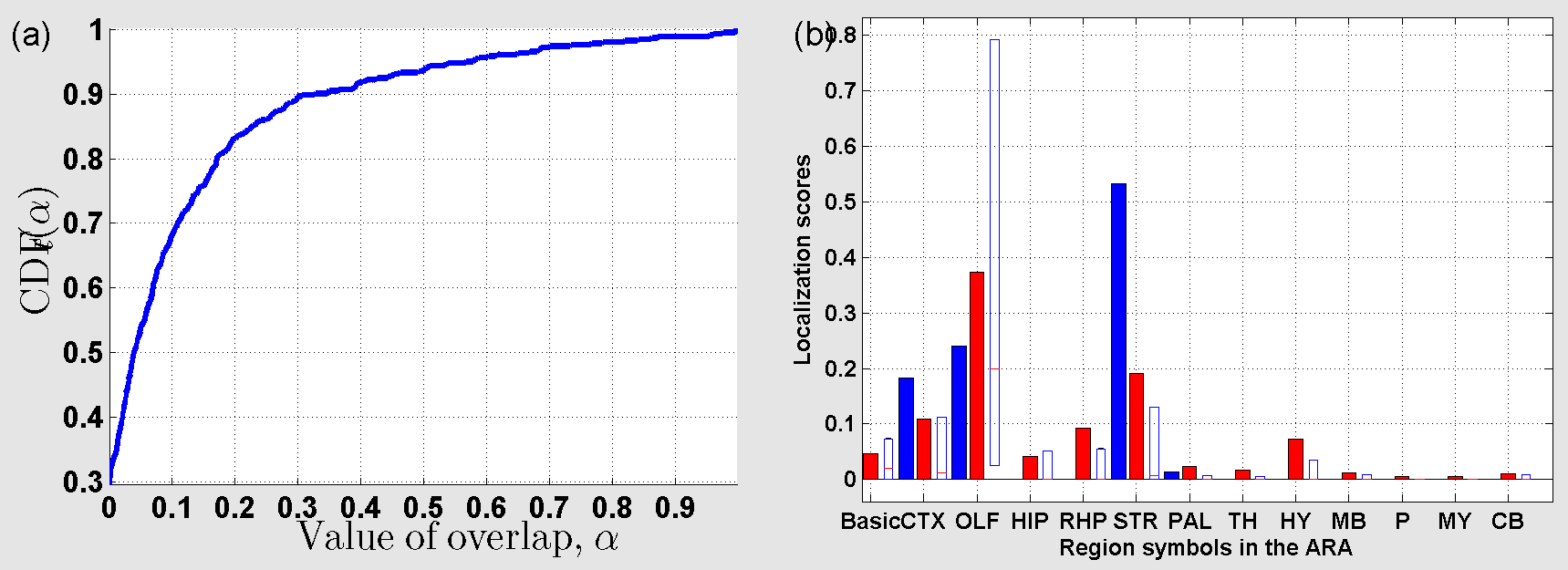}
\caption{(a) Cumulative distribution function (${\mathrm{\sc{CDF}}}_t$) of the overlap between $\rho_t$ and
 sub-sampled profiles for $t=33$. (b) Localization scores in the coarsest version of the ARA for $\rho_t$ (in blue), and 
 $\bar{\rho}_t$ (in red).}
\label{cdfPlot33}
\end{figure}
\begin{figure}
\includegraphics[width=1\textwidth,keepaspectratio]{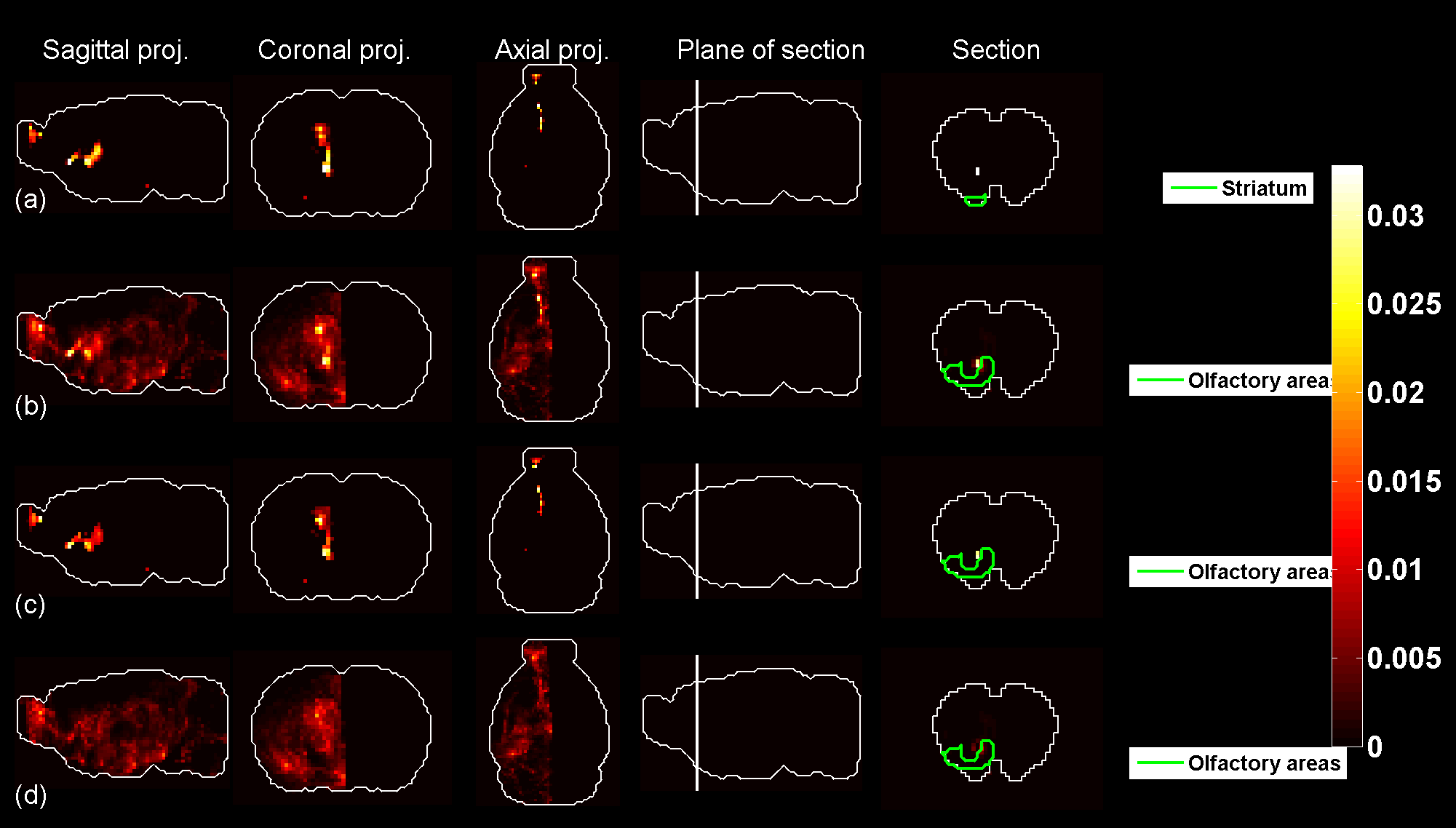}
\caption{Predicted profile and average sub-sampled profile for $t=33$.}
\label{subSampledFour33}
\end{figure}
\clearpage
\begin{figure}
\includegraphics[width=1\textwidth,keepaspectratio]{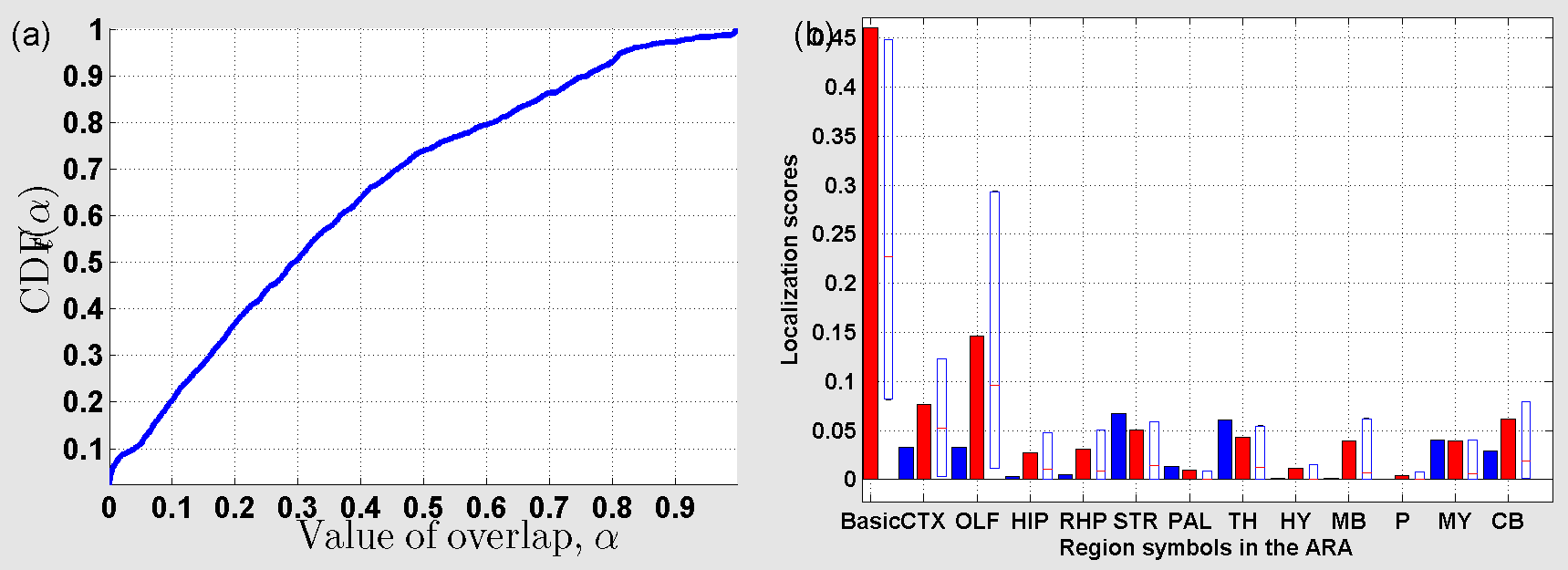}
\caption{(a) Cumulative distribution function (${\mathrm{\sc{CDF}}}_t$) of the overlap between $\rho_t$ and
 sub-sampled profiles for $t=34$. (b) Localization scores in the coarsest version of the ARA for $\rho_t$ (in blue), and 
 $\bar{\rho}_t$ (in red).}
\label{cdfPlot34}
\end{figure}
\begin{figure}
\includegraphics[width=1\textwidth,keepaspectratio]{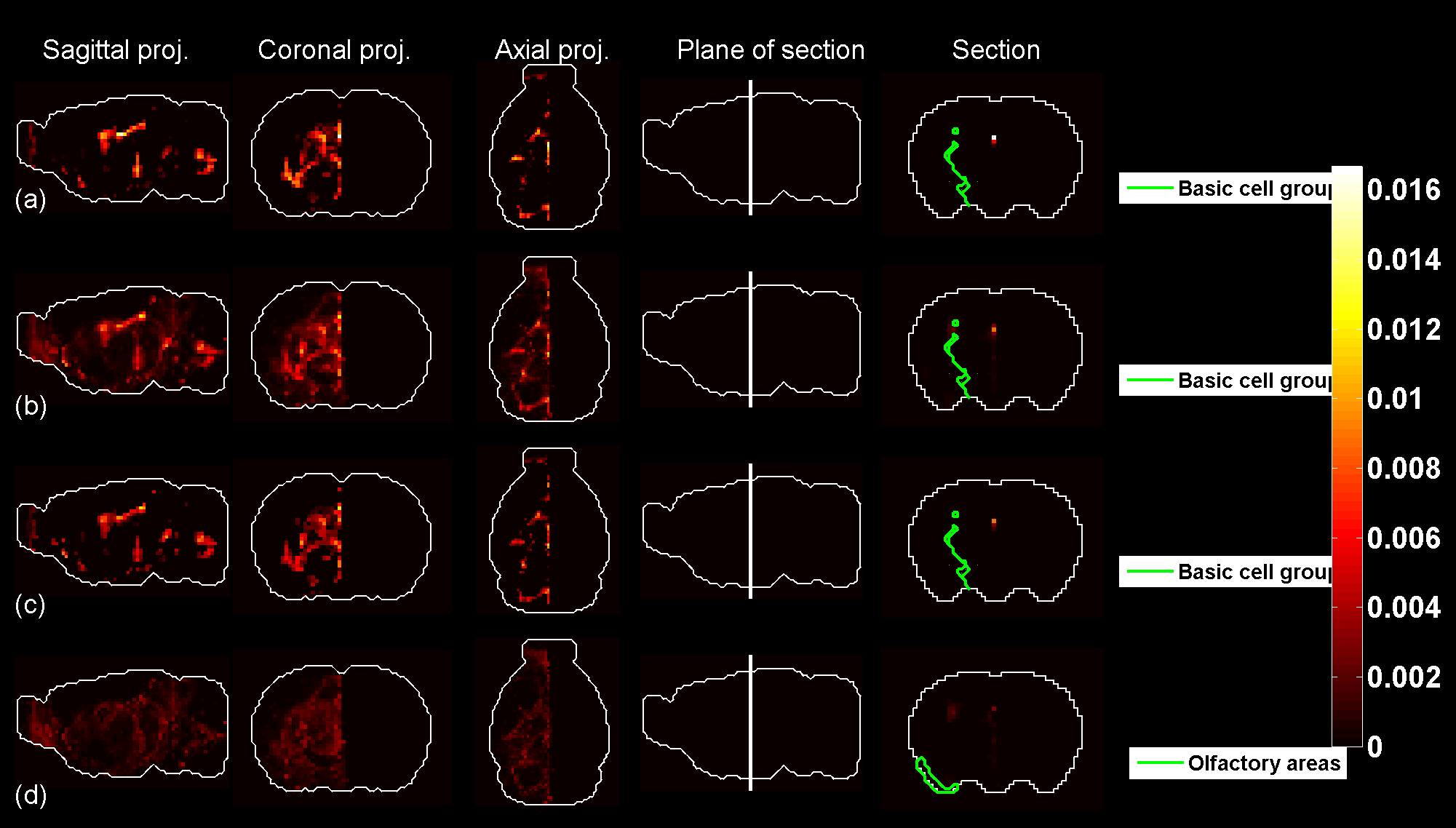}
\caption{Predicted profile and average sub-sampled profile for $t=34$.}
\label{subSampledFour34}
\end{figure}
\clearpage
\begin{figure}
\includegraphics[width=1\textwidth,keepaspectratio]{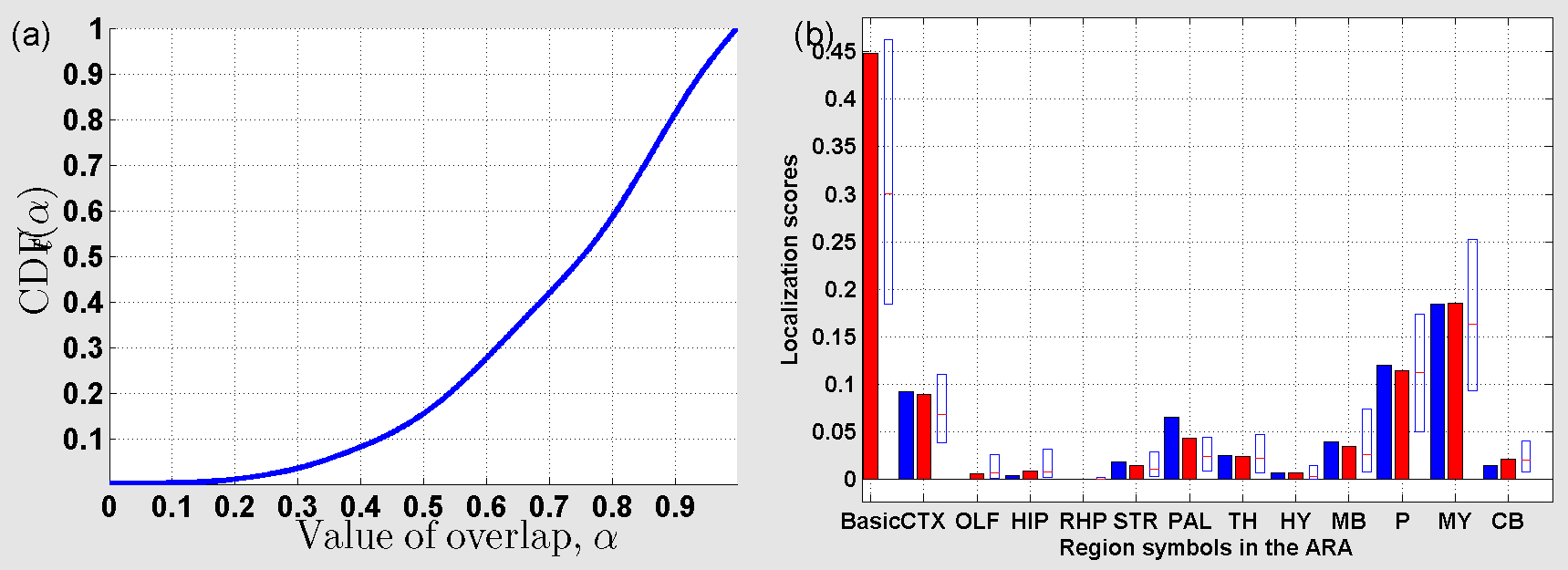}
\caption{(a) Cumulative distribution function (${\mathrm{\sc{CDF}}}_t$) of the overlap between $\rho_t$ and
 sub-sampled profiles for $t=35$. (b) Localization scores in the coarsest version of the ARA for $\rho_t$ (in blue), and 
 $\bar{\rho}_t$ (in red).}
\label{cdfPlot35}
\end{figure}
\begin{figure}
\includegraphics[width=1\textwidth,keepaspectratio]{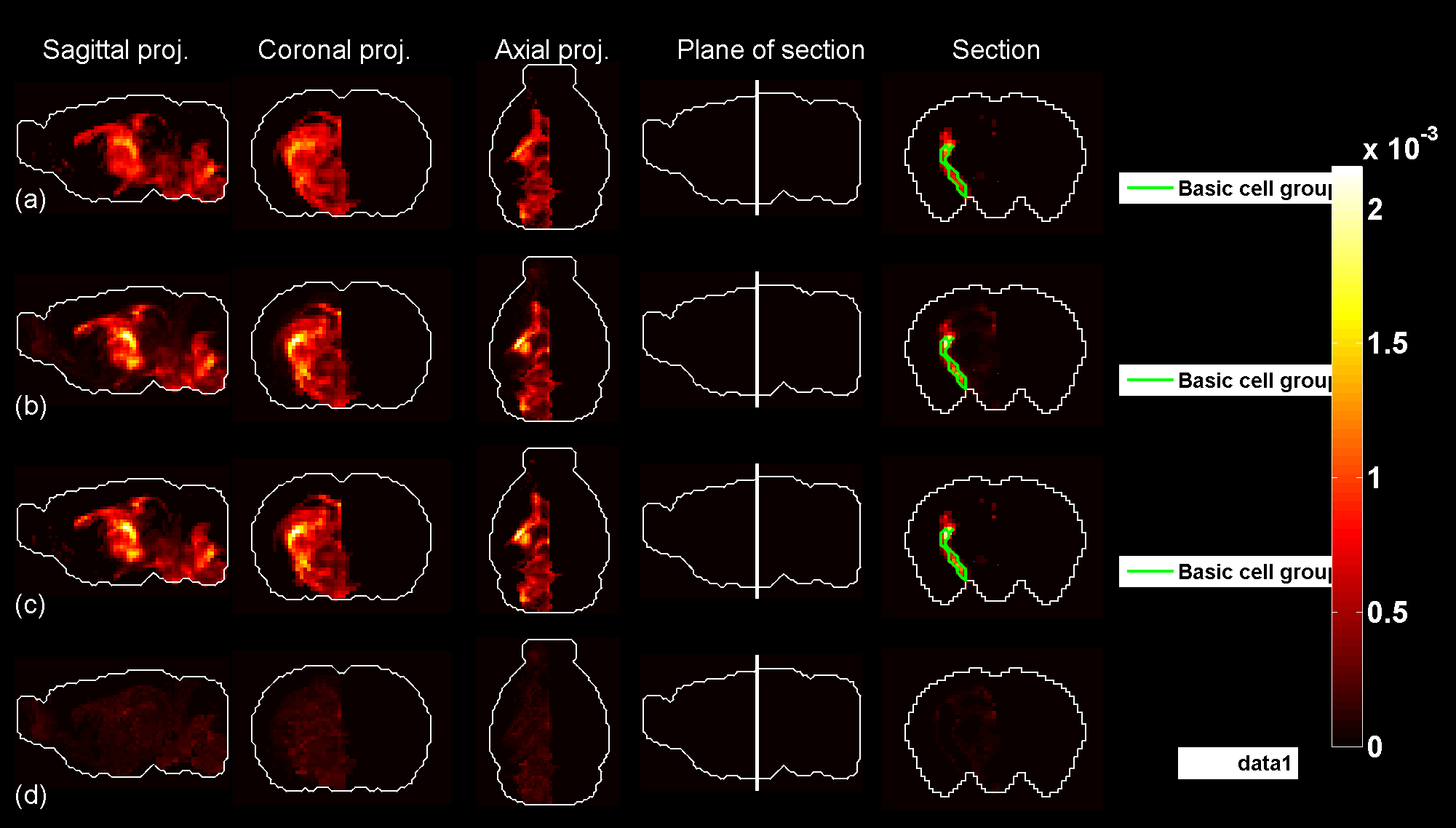}
\caption{Predicted profile and average sub-sampled profile for $t=35$.}
\label{subSampledFour35}
\end{figure}
\clearpage
\begin{figure}
\includegraphics[width=1\textwidth,keepaspectratio]{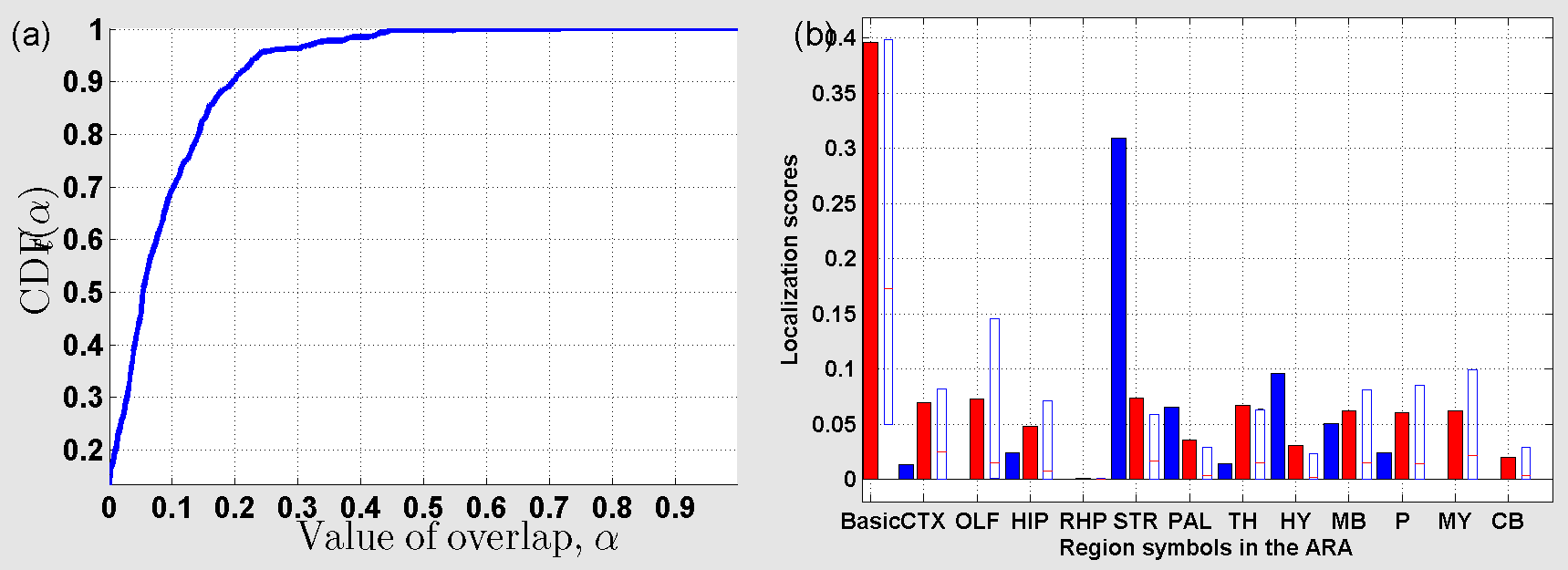}
\caption{(a) Cumulative distribution function (${\mathrm{\sc{CDF}}}_t$) of the overlap between $\rho_t$ and
 sub-sampled profiles for $t=36$. (b) Localization scores in the coarsest version of the ARA for $\rho_t$ (in blue), and 
 $\bar{\rho}_t$ (in red).}
\label{cdfPlot36}
\end{figure}
\begin{figure}
\includegraphics[width=1\textwidth,keepaspectratio]{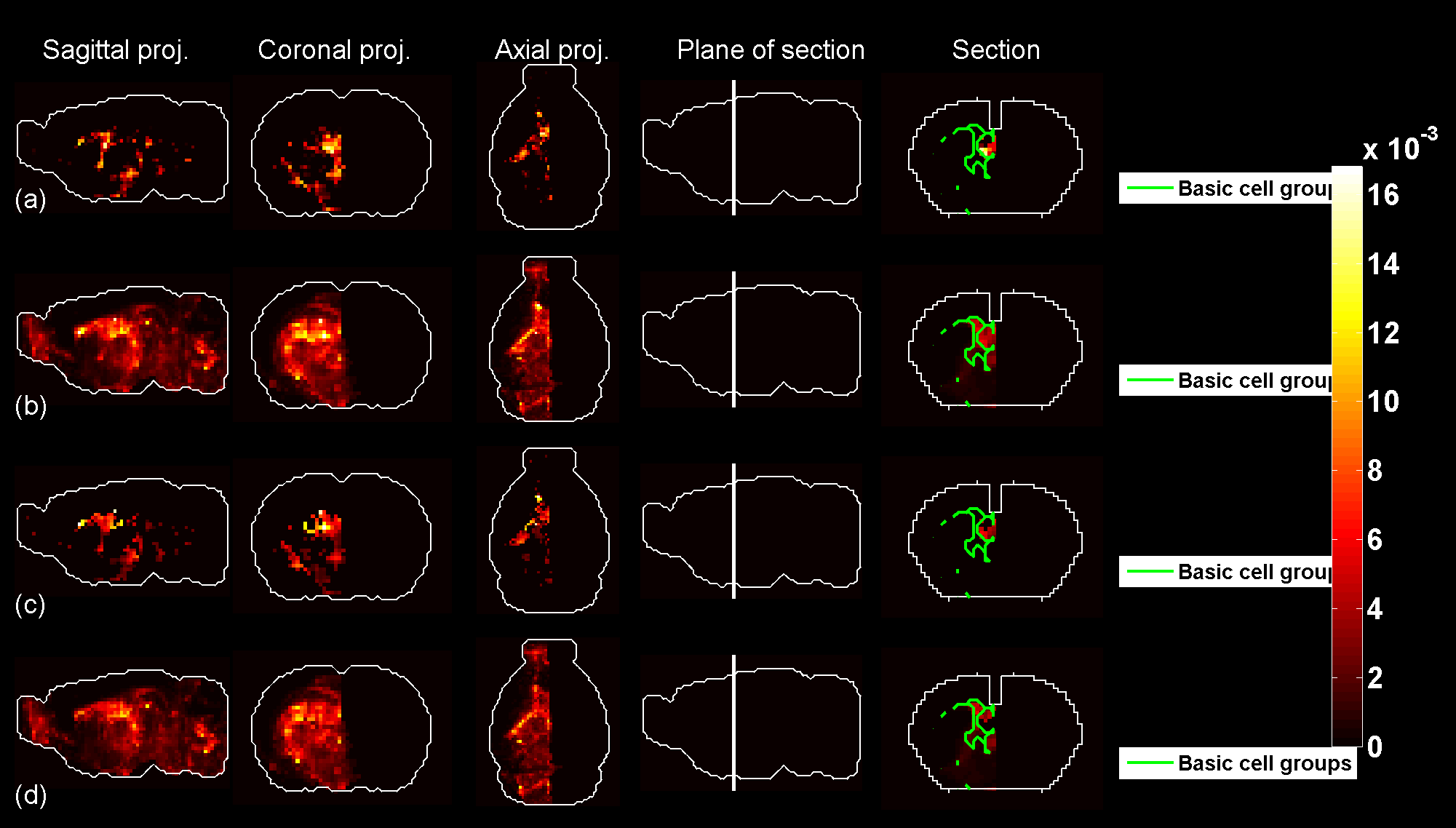}
\caption{Predicted profile and average sub-sampled profile for $t=36$.}
\label{subSampledFour36}
\end{figure}
\clearpage
\begin{figure}
\includegraphics[width=1\textwidth,keepaspectratio]{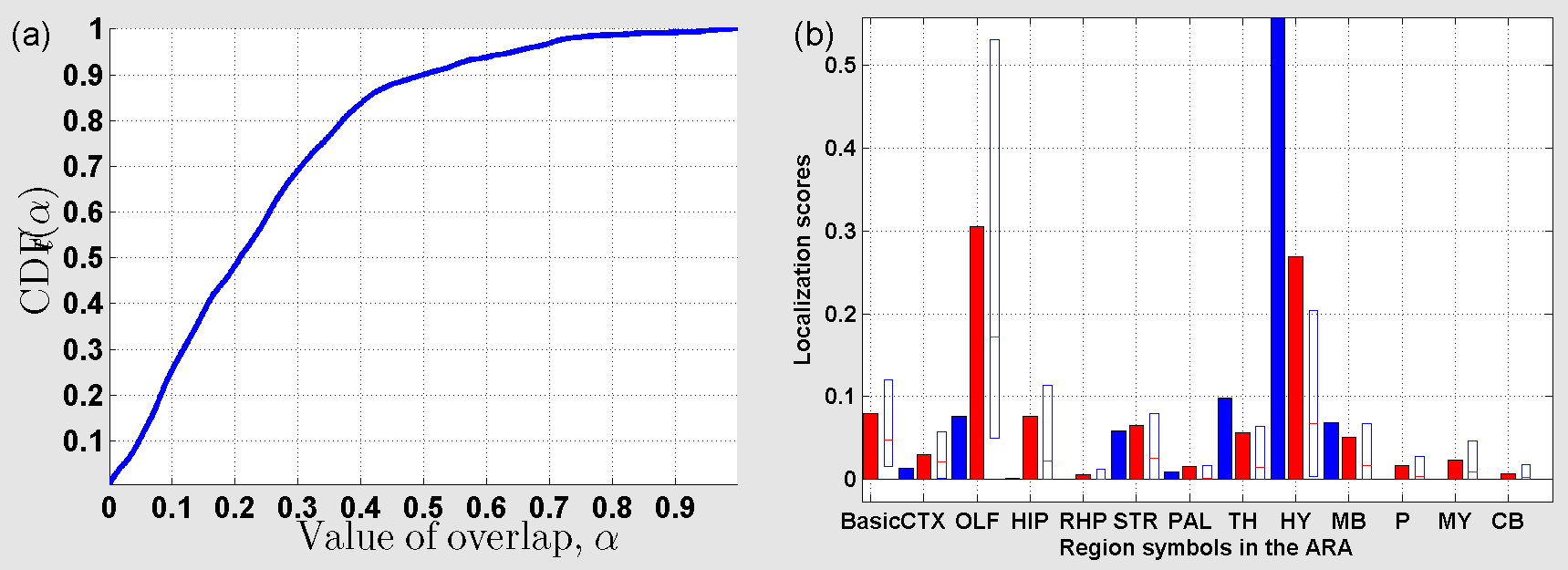}
\caption{(a) Cumulative distribution function (${\mathrm{\sc{CDF}}}_t$) of the overlap between $\rho_t$ and
 sub-sampled profiles for $t=37$. (b) Localization scores in the coarsest version of the ARA for $\rho_t$ (in blue), and 
 $\bar{\rho}_t$ (in red).}
\label{cdfPlot37}
\end{figure}
\begin{figure}
\includegraphics[width=1\textwidth,keepaspectratio]{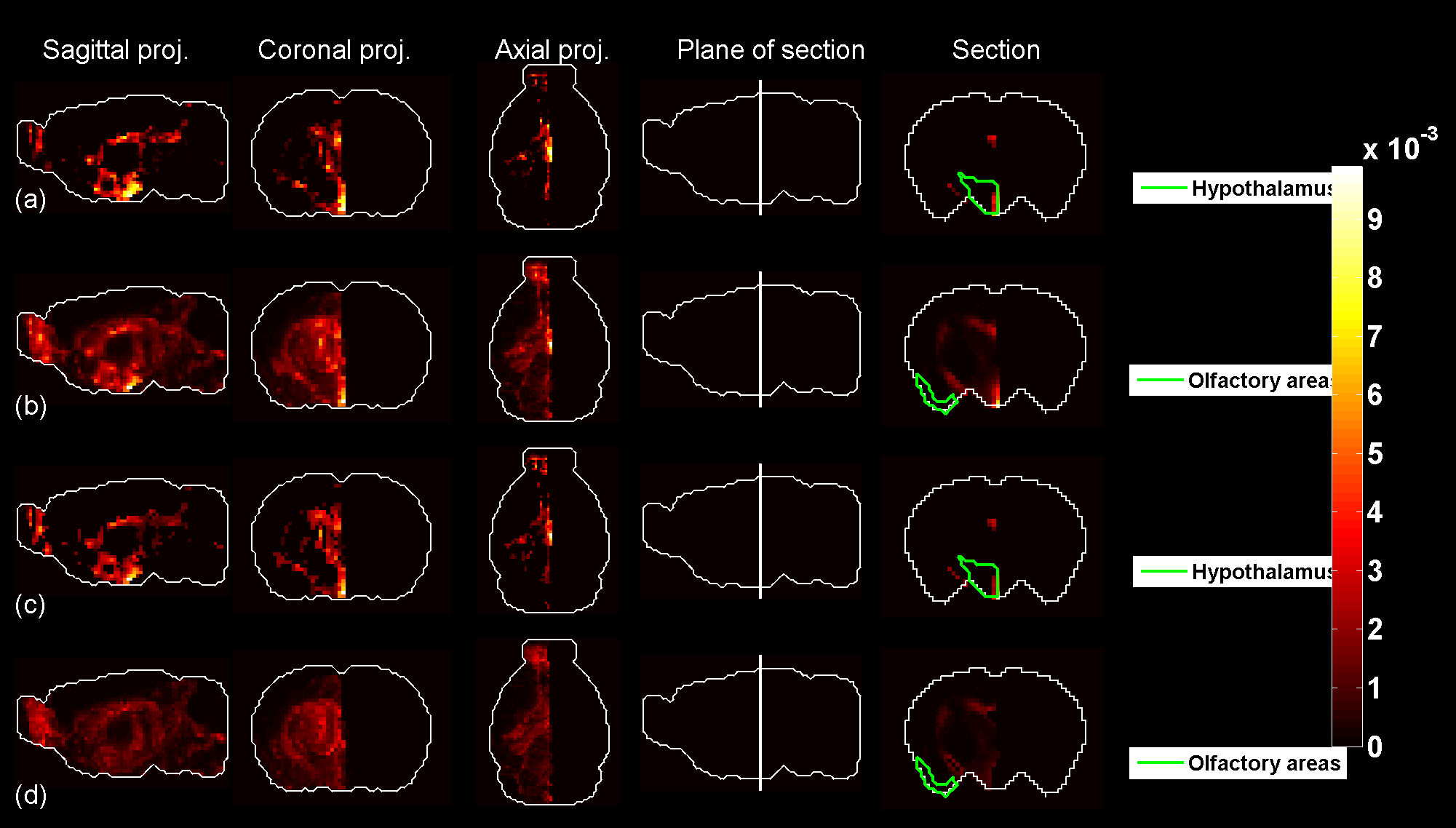}
\caption{Predicted profile and average sub-sampled profile for $t=37$.}
\label{subSampledFour37}
\end{figure}
\clearpage
\begin{figure}
\includegraphics[width=1\textwidth,keepaspectratio]{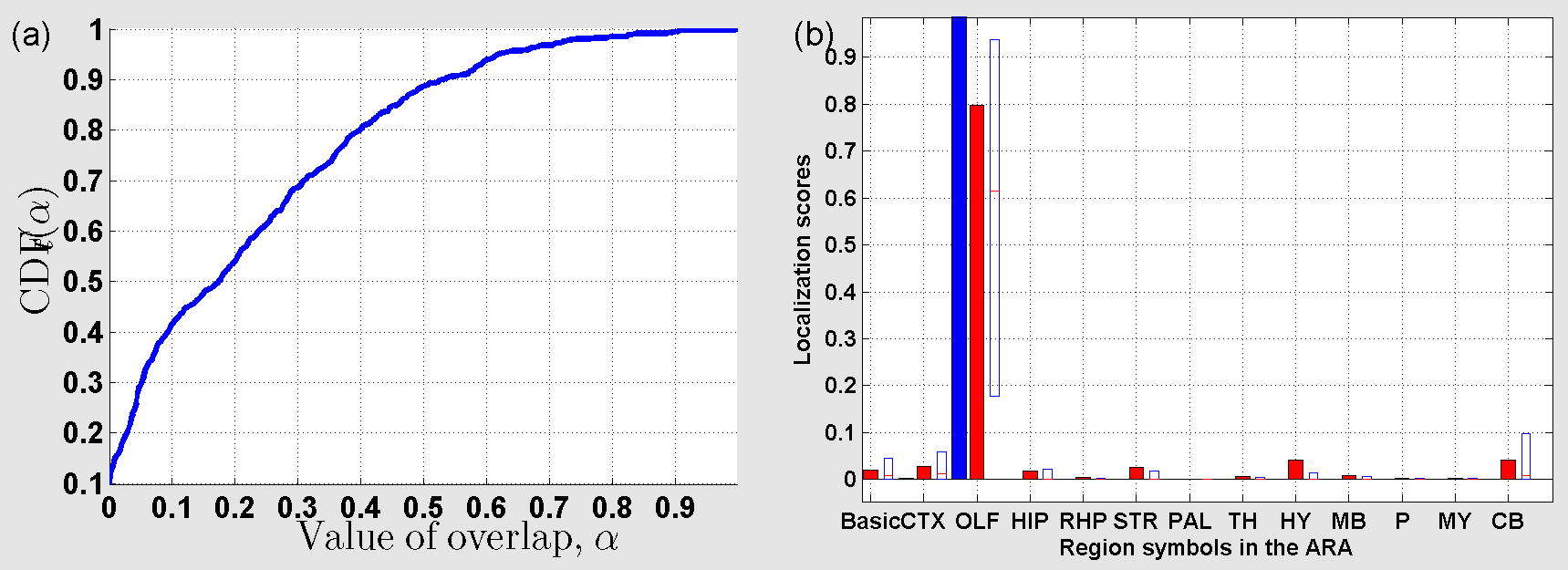}
\caption{(a) Cumulative distribution function (${\mathrm{\sc{CDF}}}_t$) of the overlap between $\rho_t$ and
 sub-sampled profiles for $t=38$. (b) Localization scores in the coarsest version of the ARA for $\rho_t$ (in blue), and 
 $\bar{\rho}_t$ (in red).}
\label{cdfPlot38}
\end{figure}
\begin{figure}
\includegraphics[width=1\textwidth,keepaspectratio]{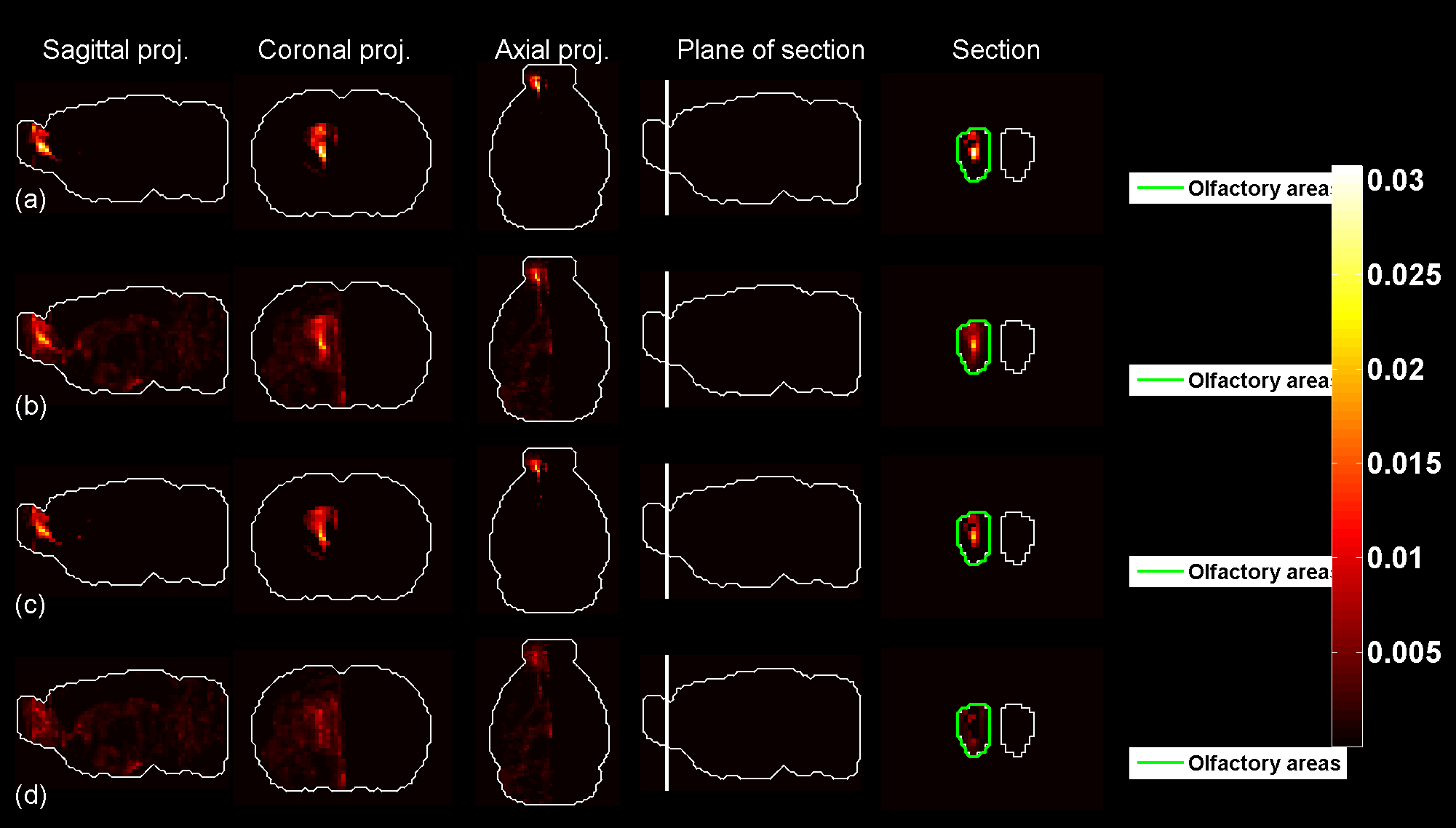}
\caption{Predicted profile and average sub-sampled profile for $t=38$.}
\label{subSampledFour38}
\end{figure}
\clearpage
\begin{figure}
\includegraphics[width=1\textwidth,keepaspectratio]{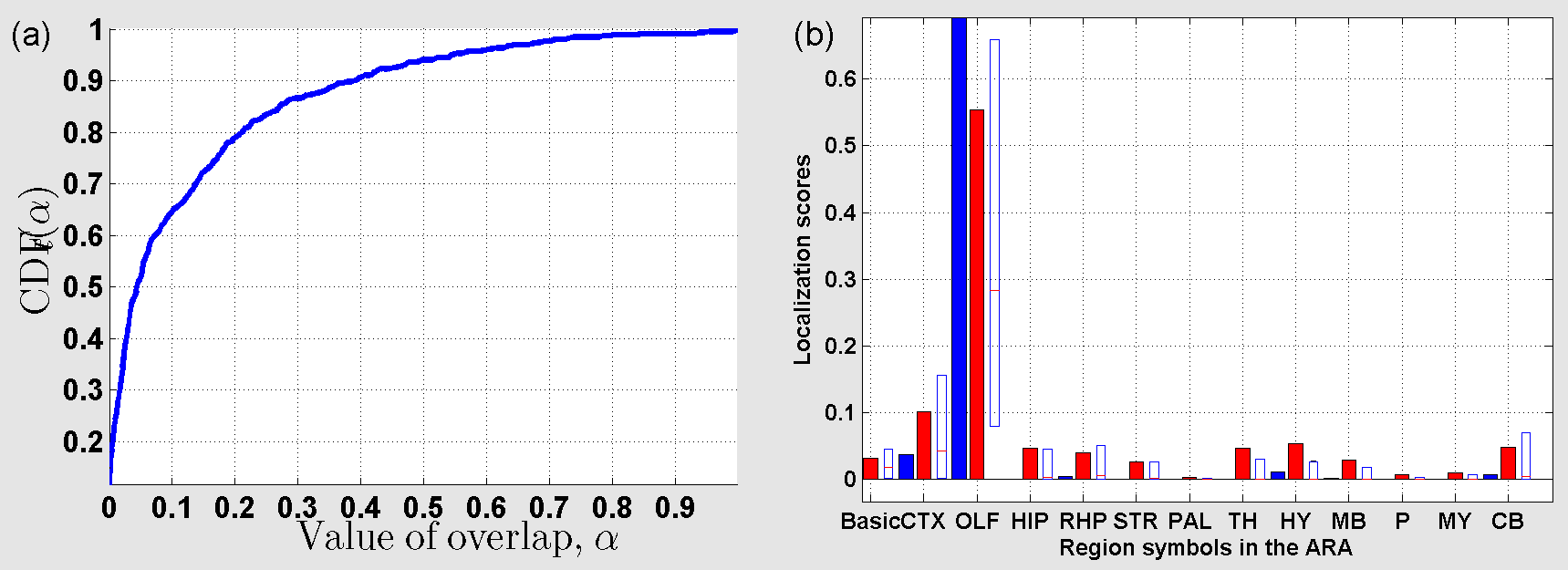}
\caption{(a) Cumulative distribution function (${\mathrm{\sc{CDF}}}_t$) of the overlap between $\rho_t$ and
 sub-sampled profiles for $t=39$. (b) Localization scores in the coarsest version of the ARA for $\rho_t$ (in blue), and 
 $\bar{\rho}_t$ (in red).}
\label{cdfPlot39}
\end{figure}
\begin{figure}
\includegraphics[width=1\textwidth,keepaspectratio]{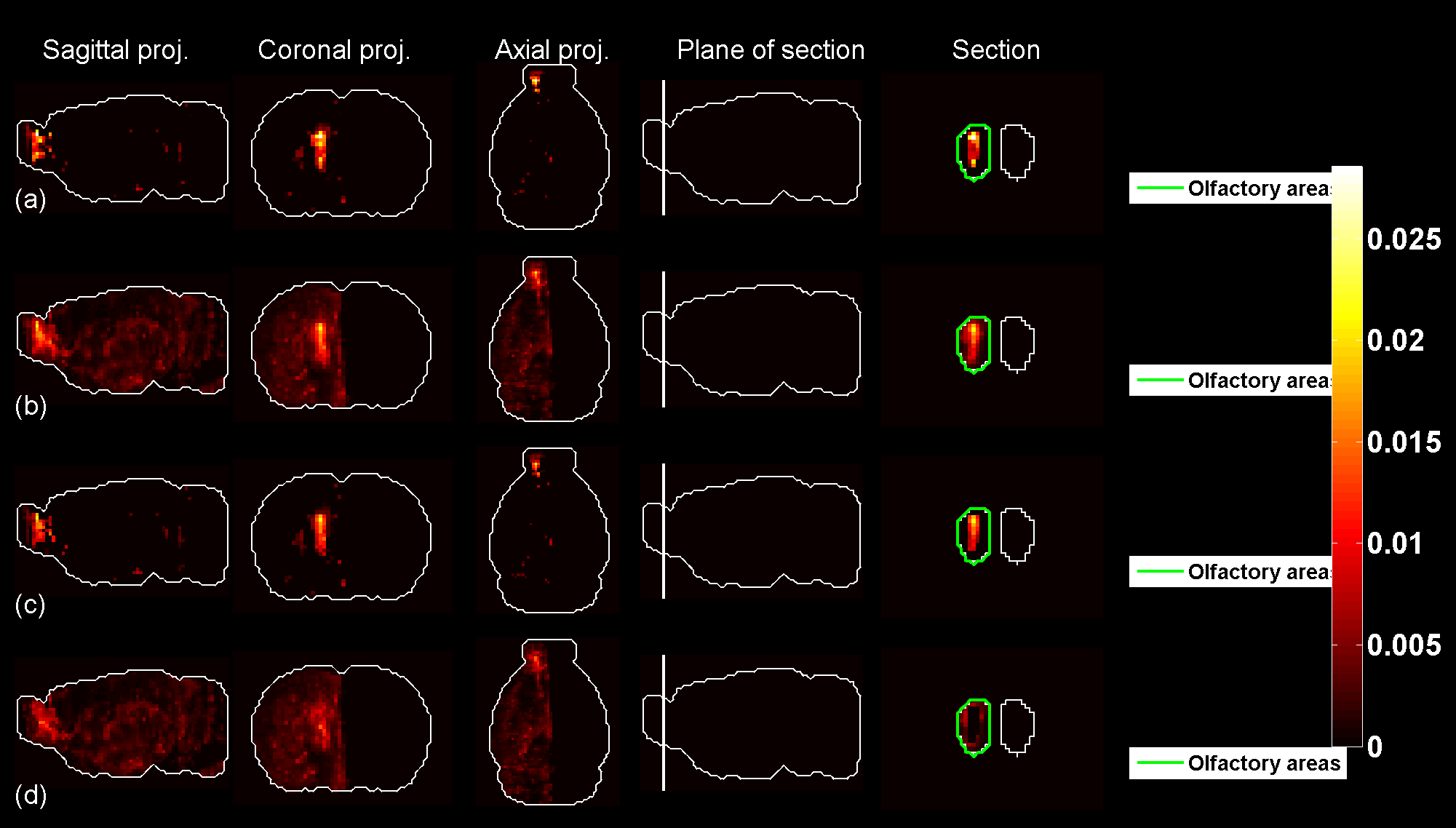}
\caption{Predicted profile and average sub-sampled profile for $t=39$.}
\label{subSampledFour39}
\end{figure}
\clearpage
\begin{figure}
\includegraphics[width=1\textwidth,keepaspectratio]{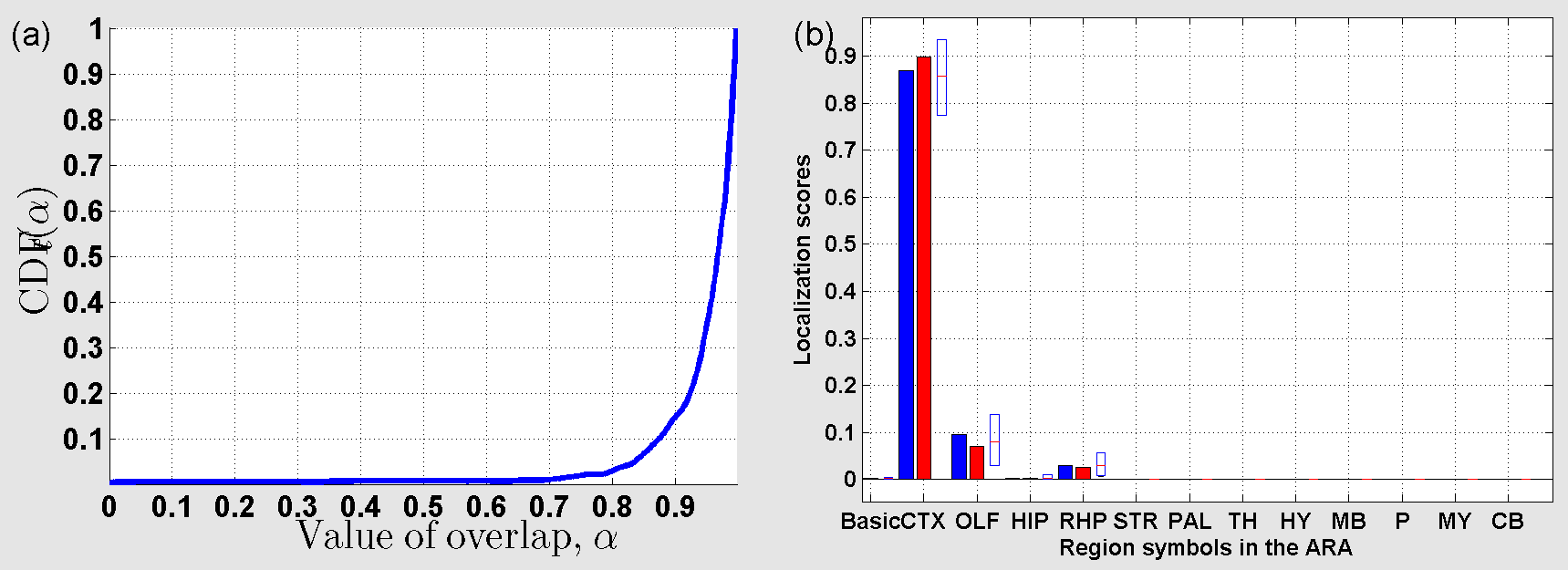}
\caption{(a) Cumulative distribution function (${\mathrm{\sc{CDF}}}_t$) of the overlap between $\rho_t$ and
 sub-sampled profiles for $t=40$. (b) Localization scores in the coarsest version of the ARA for $\rho_t$ (in blue), and 
 $\bar{\rho}_t$ (in red).}
\label{cdfPlot40}
\end{figure}
\begin{figure}
\includegraphics[width=1\textwidth,keepaspectratio]{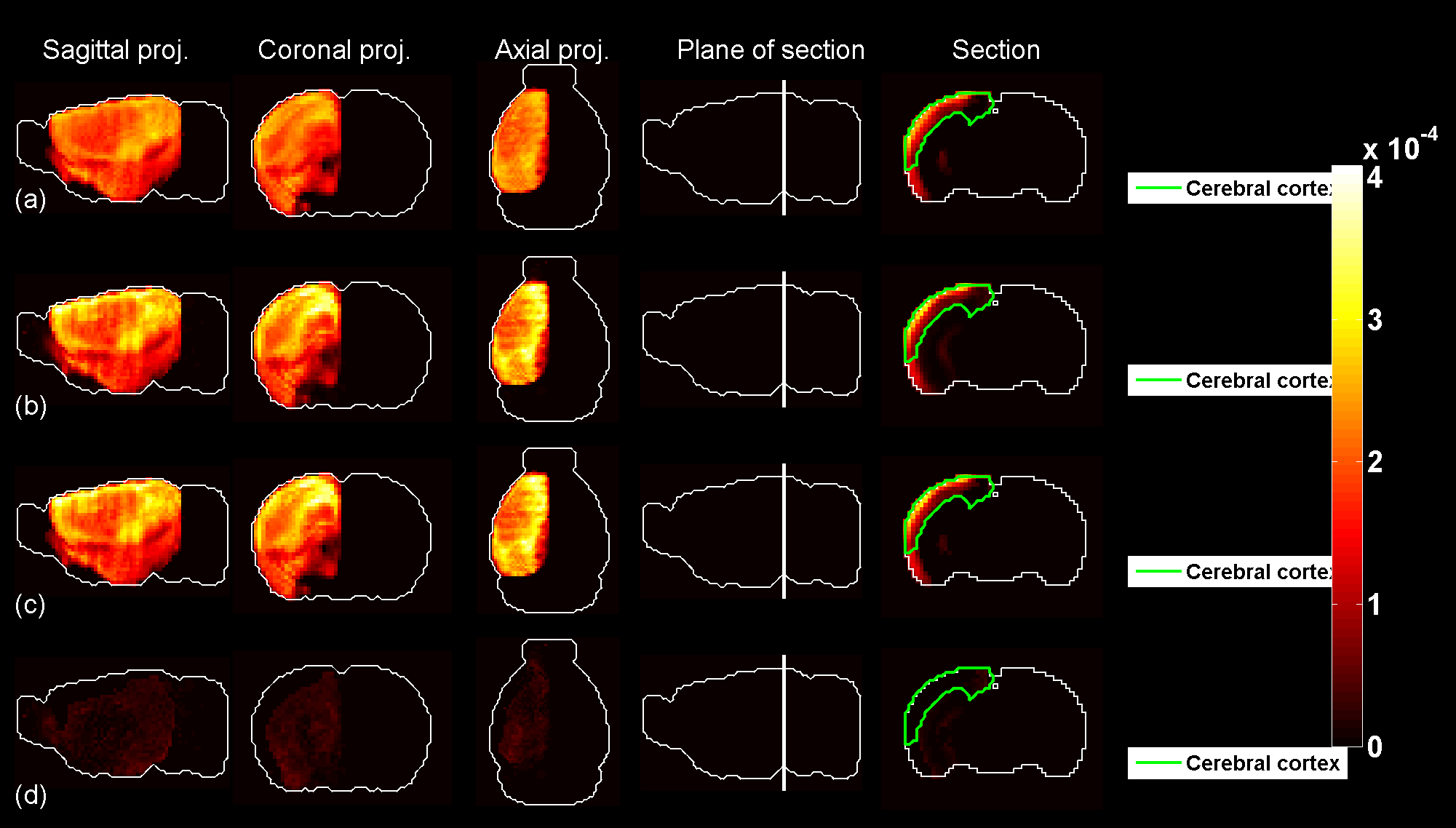}
\caption{Predicted profile and average sub-sampled profile for $t=40$.}
\label{subSampledFour40}
\end{figure}
\clearpage
\begin{figure}
\includegraphics[width=1\textwidth,keepaspectratio]{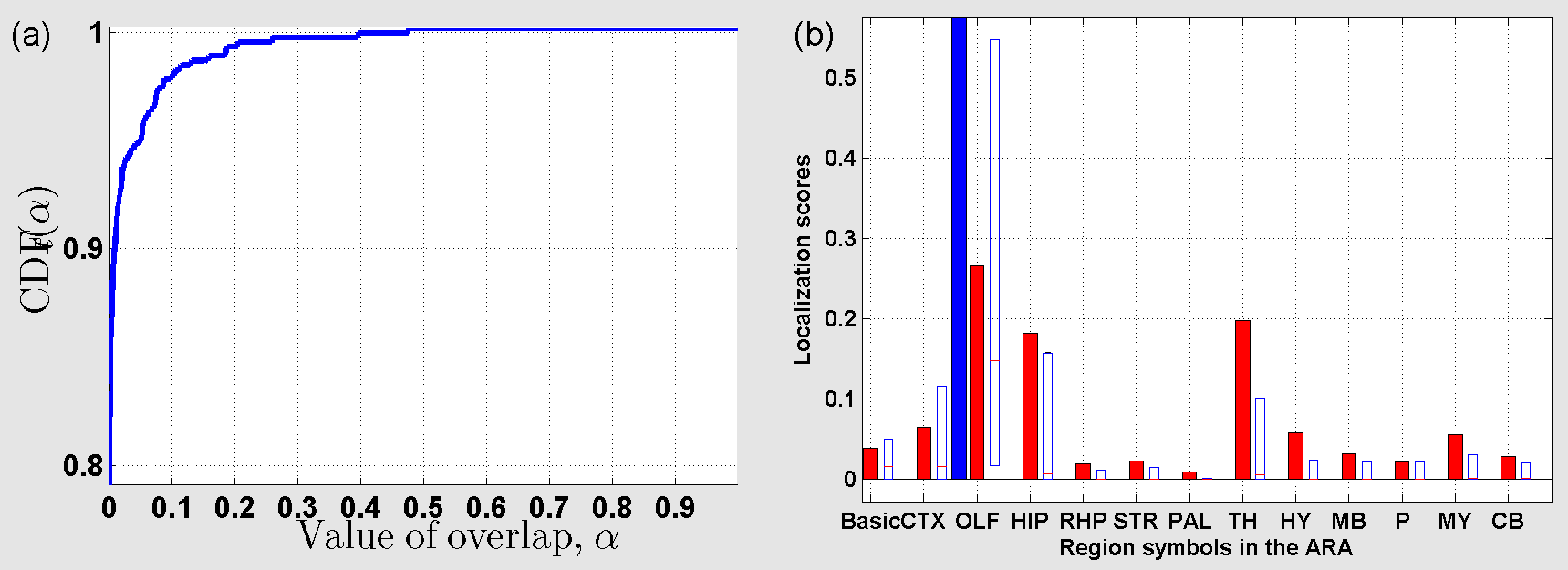}
\caption{(a) Cumulative distribution function (${\mathrm{\sc{CDF}}}_t$) of the overlap between $\rho_t$ and
 sub-sampled profiles for $t=41$. (b) Localization scores in the coarsest version of the ARA for $\rho_t$ (in blue), and 
 $\bar{\rho}_t$ (in red).}
\label{cdfPlot41}
\end{figure}
\begin{figure}
\includegraphics[width=1\textwidth,keepaspectratio]{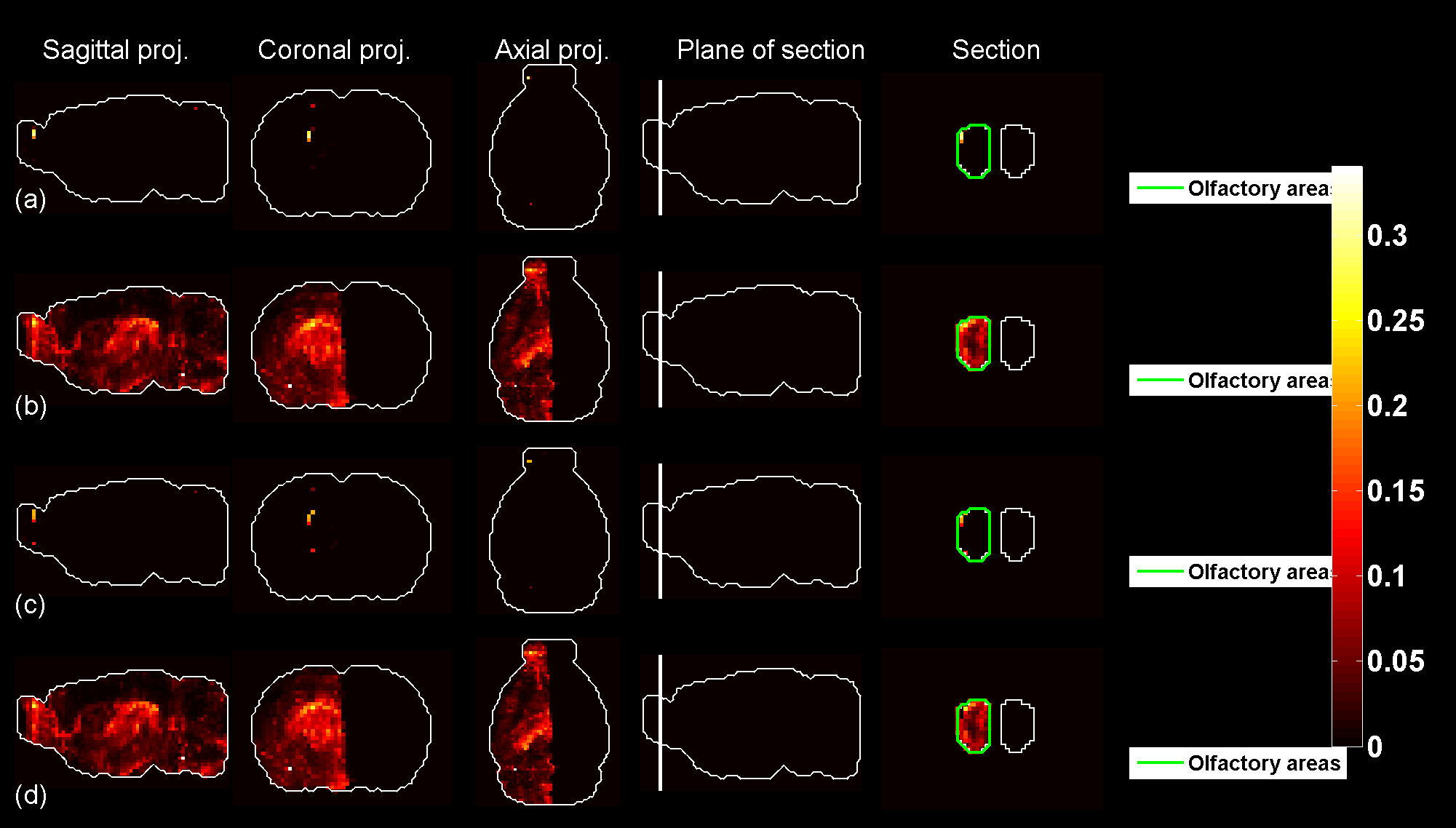}
\caption{Predicted profile and average sub-sampled profile for $t=41$.}
\label{subSampledFour41}
\end{figure}
\clearpage
\begin{figure}
\includegraphics[width=1\textwidth,keepaspectratio]{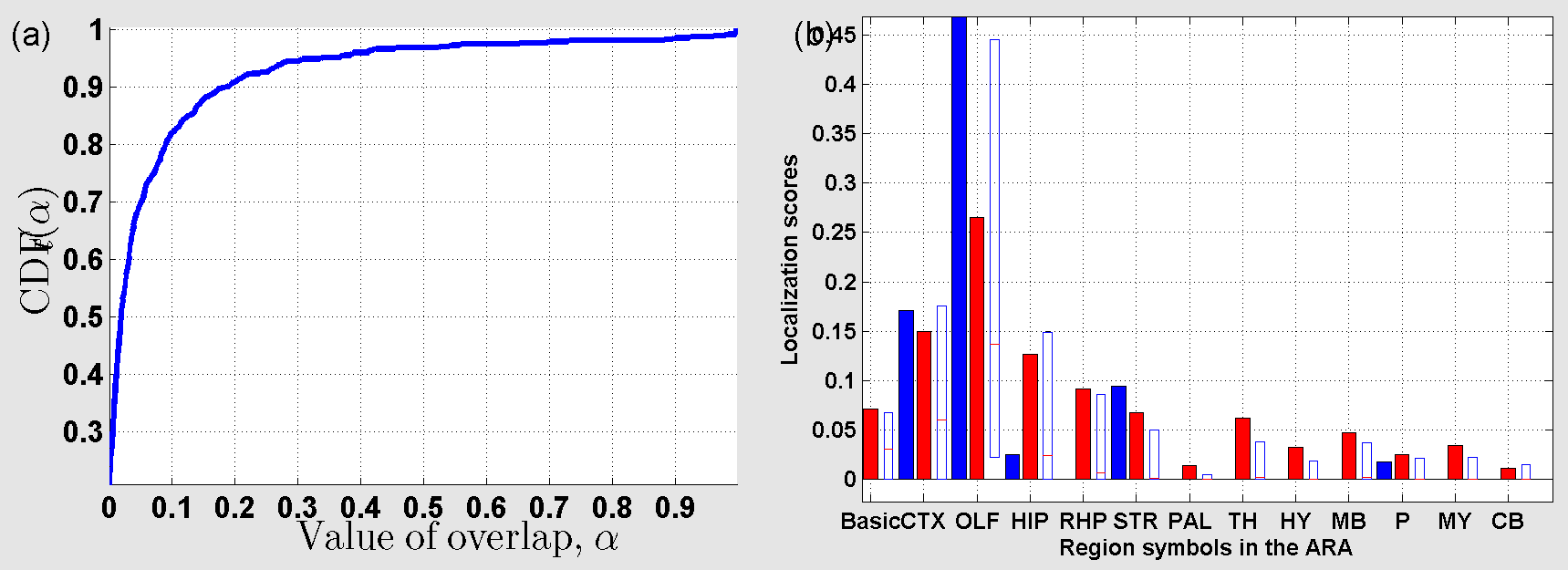}
\caption{(a) Cumulative distribution function (${\mathrm{\sc{CDF}}}_t$) of the overlap between $\rho_t$ and
 sub-sampled profiles for $t=42$. (b) Localization scores in the coarsest version of the ARA for $\rho_t$ (in blue), and 
 $\bar{\rho}_t$ (in red).}
\label{cdfPlot42}
\end{figure}
\begin{figure}
\includegraphics[width=1\textwidth,keepaspectratio]{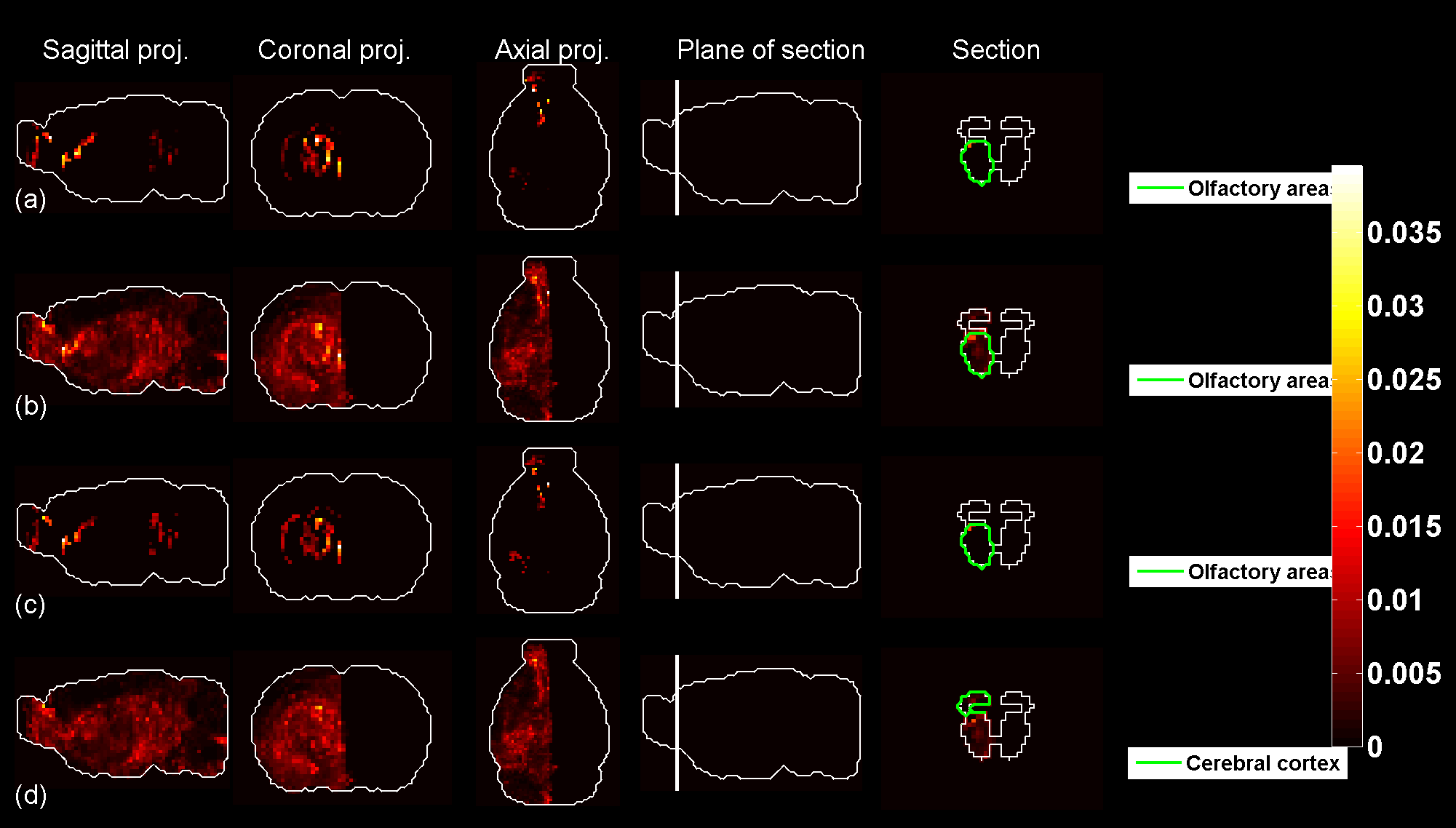}
\caption{Predicted profile and average sub-sampled profile for $t=42$.}
\label{subSampledFour42}
\end{figure}
\clearpage
\begin{figure}
\includegraphics[width=1\textwidth,keepaspectratio]{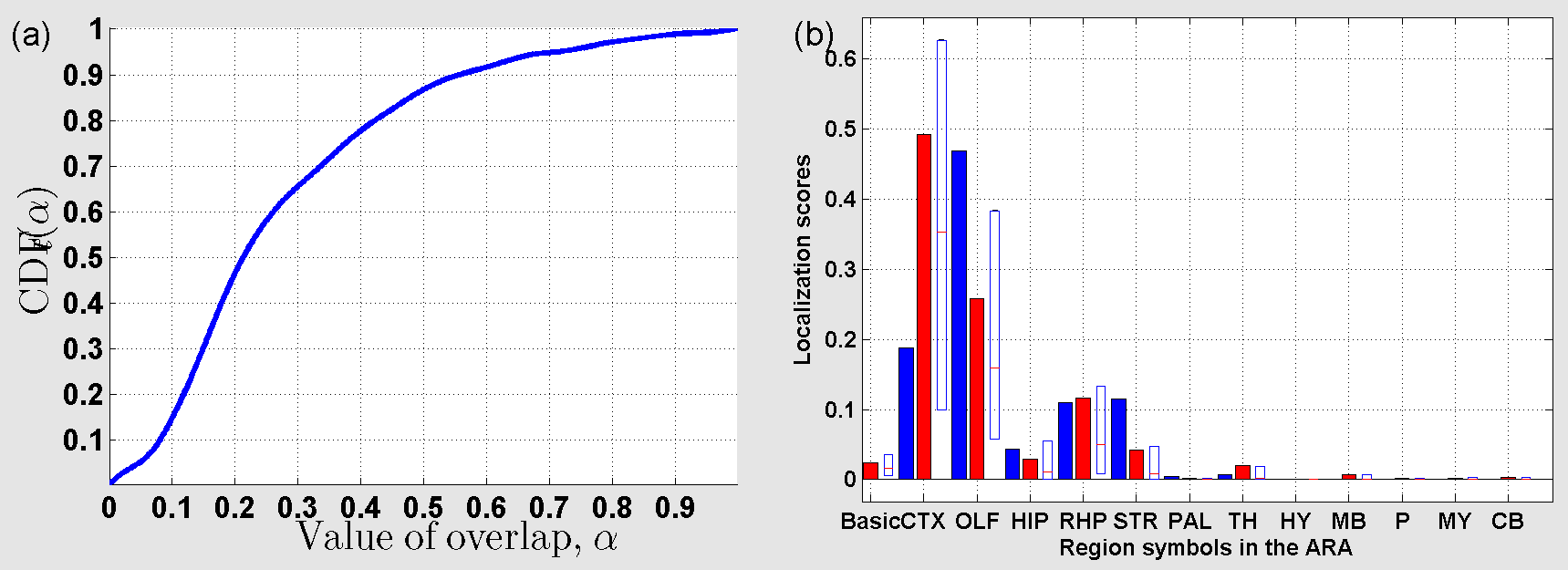}
\caption{(a) Cumulative distribution function (${\mathrm{\sc{CDF}}}_t$) of the overlap between $\rho_t$ and
 sub-sampled profiles for $t=43$. (b) Localization scores in the coarsest version of the ARA for $\rho_t$ (in blue), and 
 $\bar{\rho}_t$ (in red).}
\label{cdfPlot43}
\end{figure}
\begin{figure}
\includegraphics[width=1\textwidth,keepaspectratio]{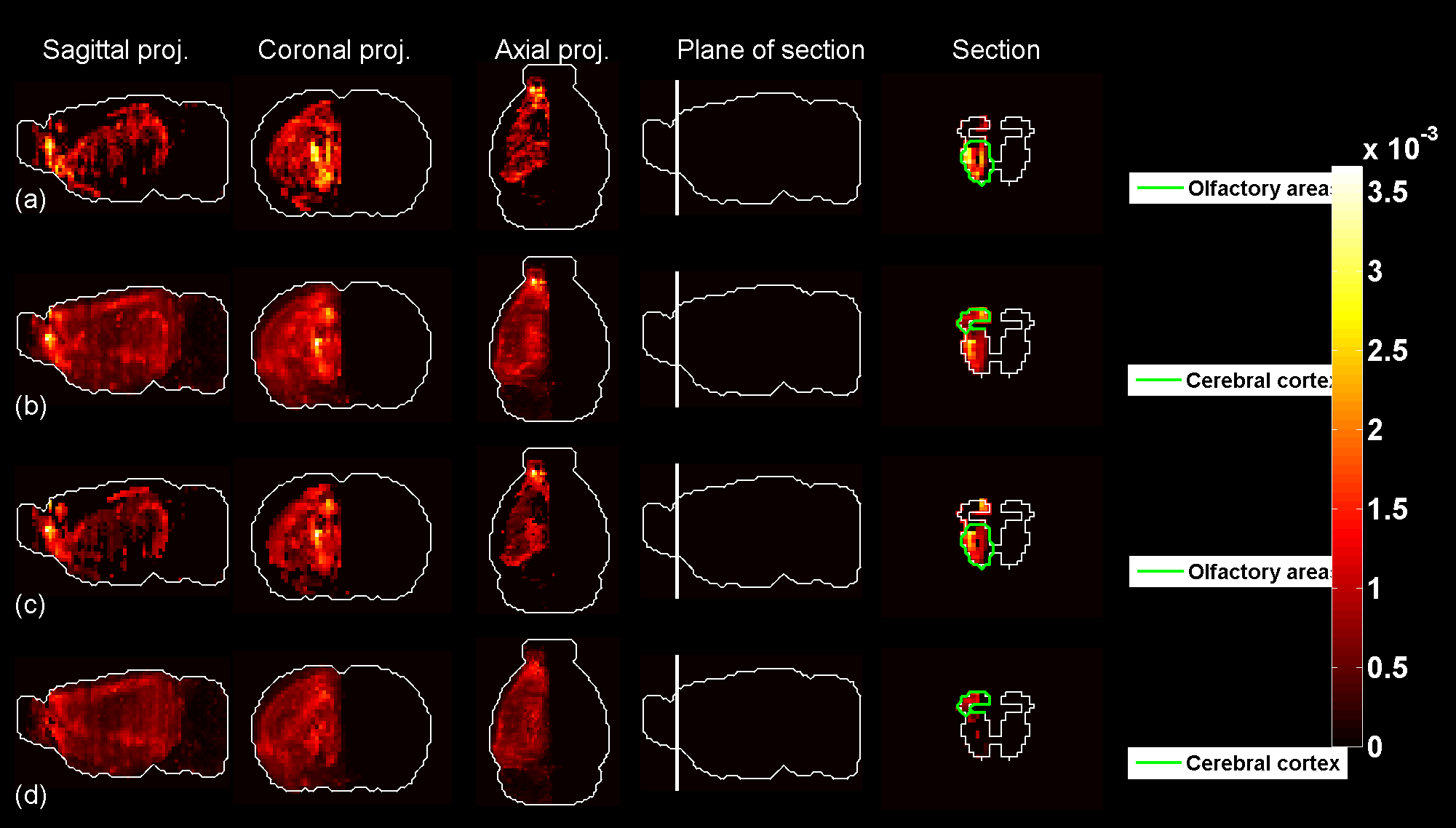}
\caption{Predicted profile and average sub-sampled profile for $t=43$.}
\label{subSampledFour43}
\end{figure}
\clearpage
\begin{figure}
\includegraphics[width=1\textwidth,keepaspectratio]{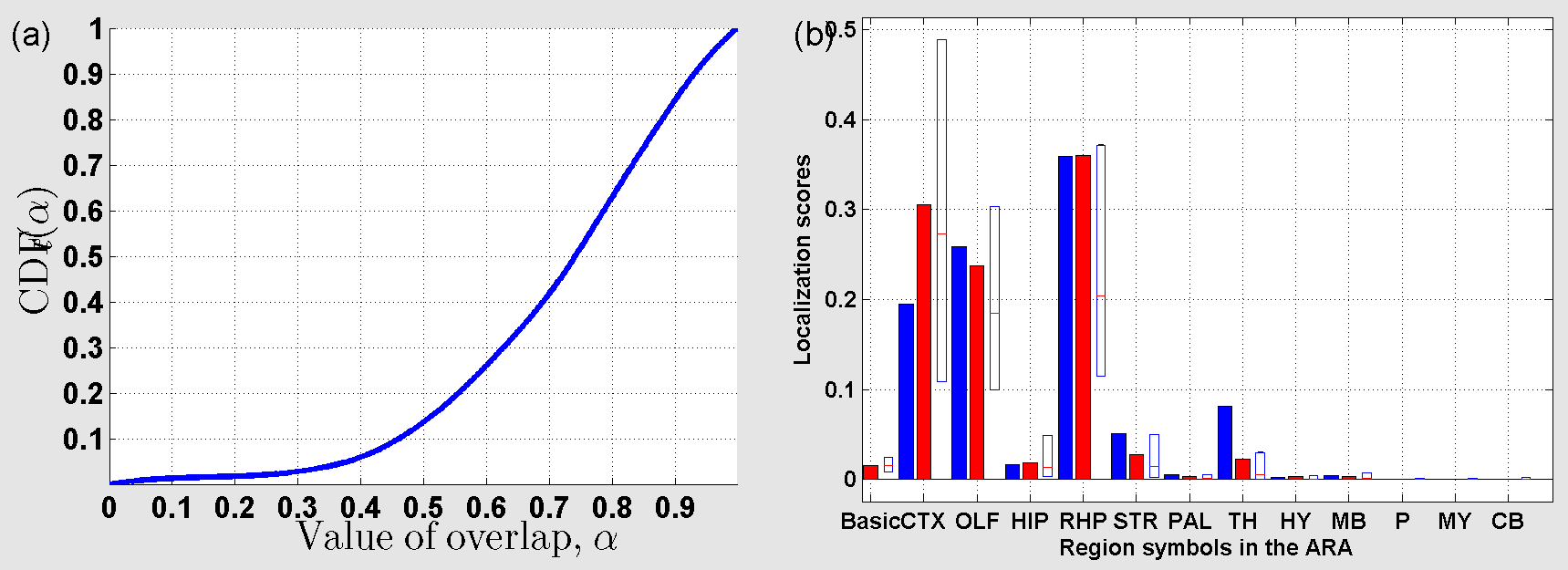}
\caption{(a) Cumulative distribution function (${\mathrm{\sc{CDF}}}_t$) of the overlap between $\rho_t$ and
 sub-sampled profiles for $t=44$. (b) Localization scores in the coarsest version of the ARA for $\rho_t$ (in blue), and 
 $\bar{\rho}_t$ (in red).}
\label{cdfPlot44}
\end{figure}
\begin{figure}
\includegraphics[width=1\textwidth,keepaspectratio]{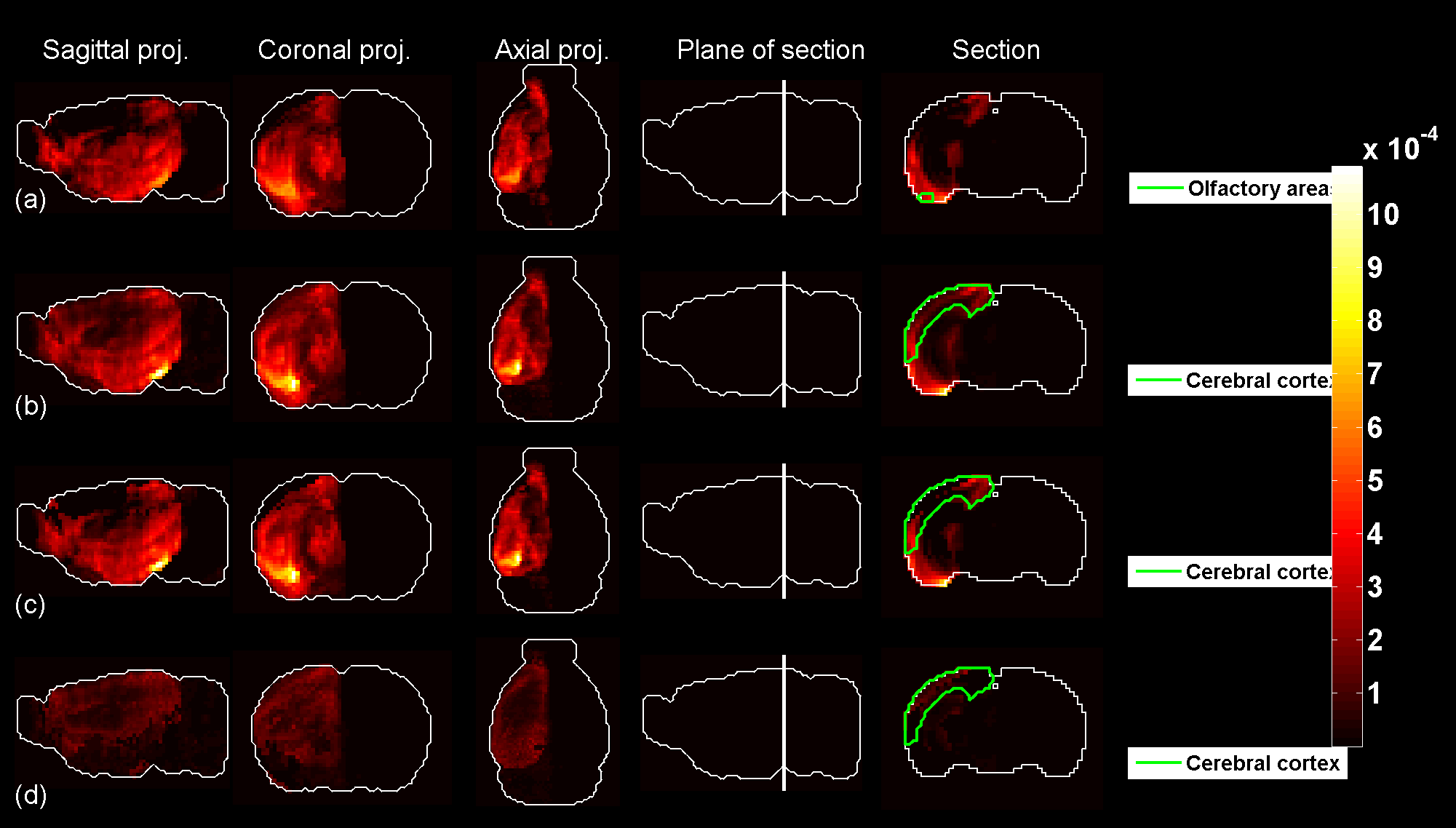}
\caption{Predicted profile and average sub-sampled profile for $t=44$.}
\label{subSampledFour44}
\end{figure}
\clearpage
\begin{figure}
\includegraphics[width=1\textwidth,keepaspectratio]{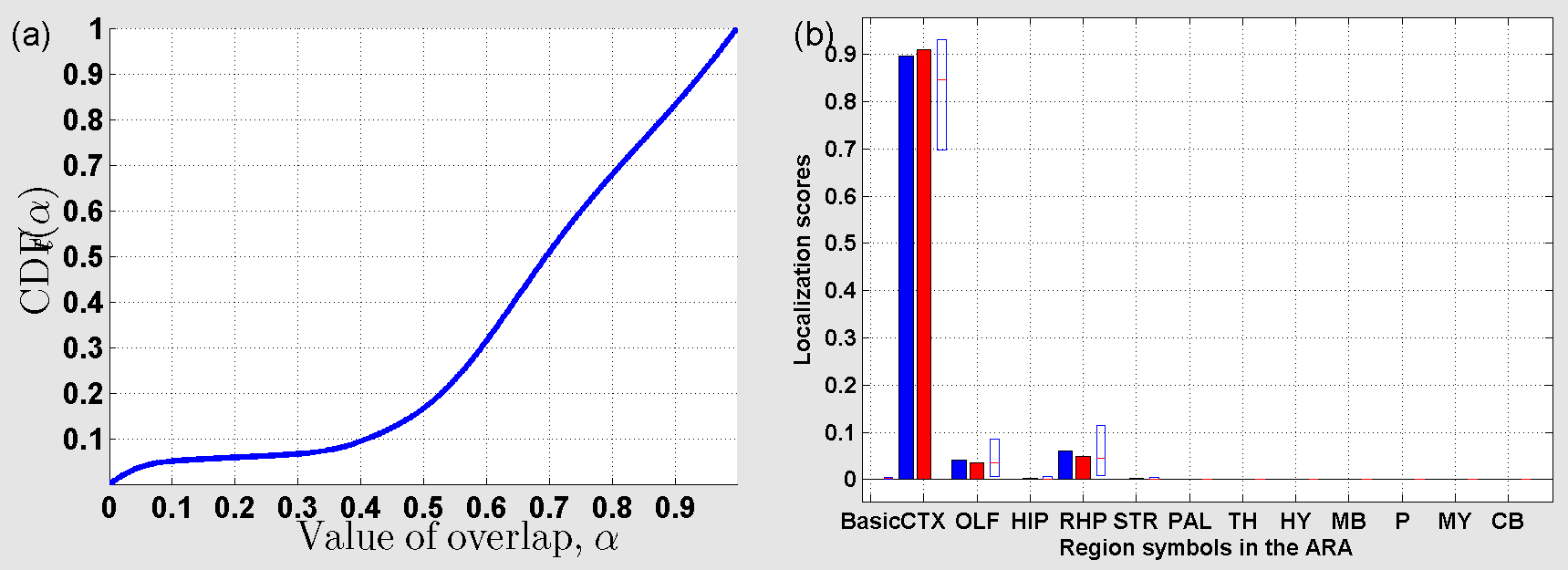}
\caption{(a) Cumulative distribution function (${\mathrm{\sc{CDF}}}_t$) of the overlap between $\rho_t$ and
 sub-sampled profiles for $t=45$. (b) Localization scores in the coarsest version of the ARA for $\rho_t$ (in blue), and 
 $\bar{\rho}_t$ (in red).}
\label{cdfPlot45}
\end{figure}
\begin{figure}
\includegraphics[width=1\textwidth,keepaspectratio]{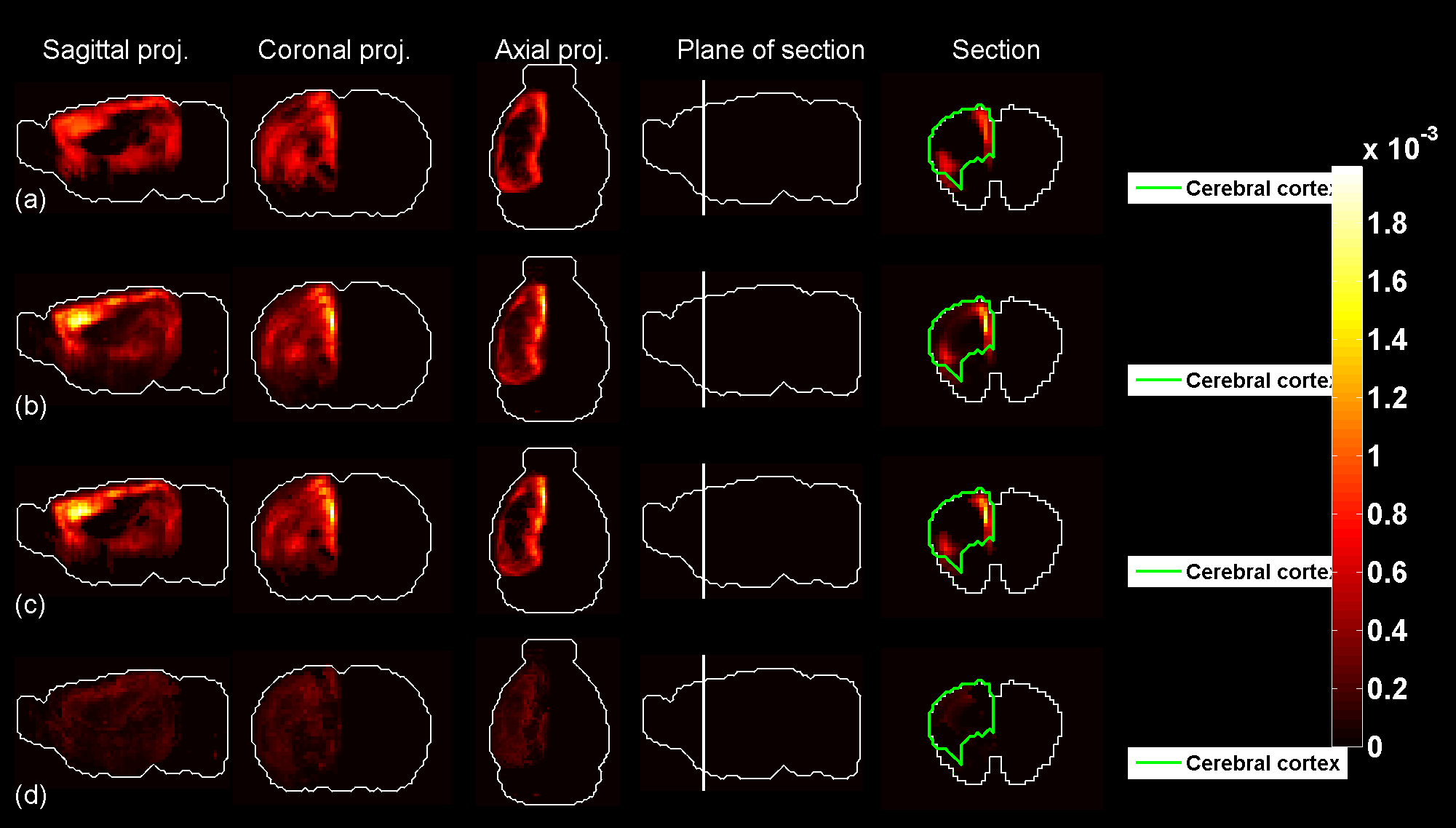}
\caption{Predicted profile and average sub-sampled profile for $t=45$.}
\label{subSampledFour45}
\end{figure}
\clearpage
\begin{figure}
\includegraphics[width=1\textwidth,keepaspectratio]{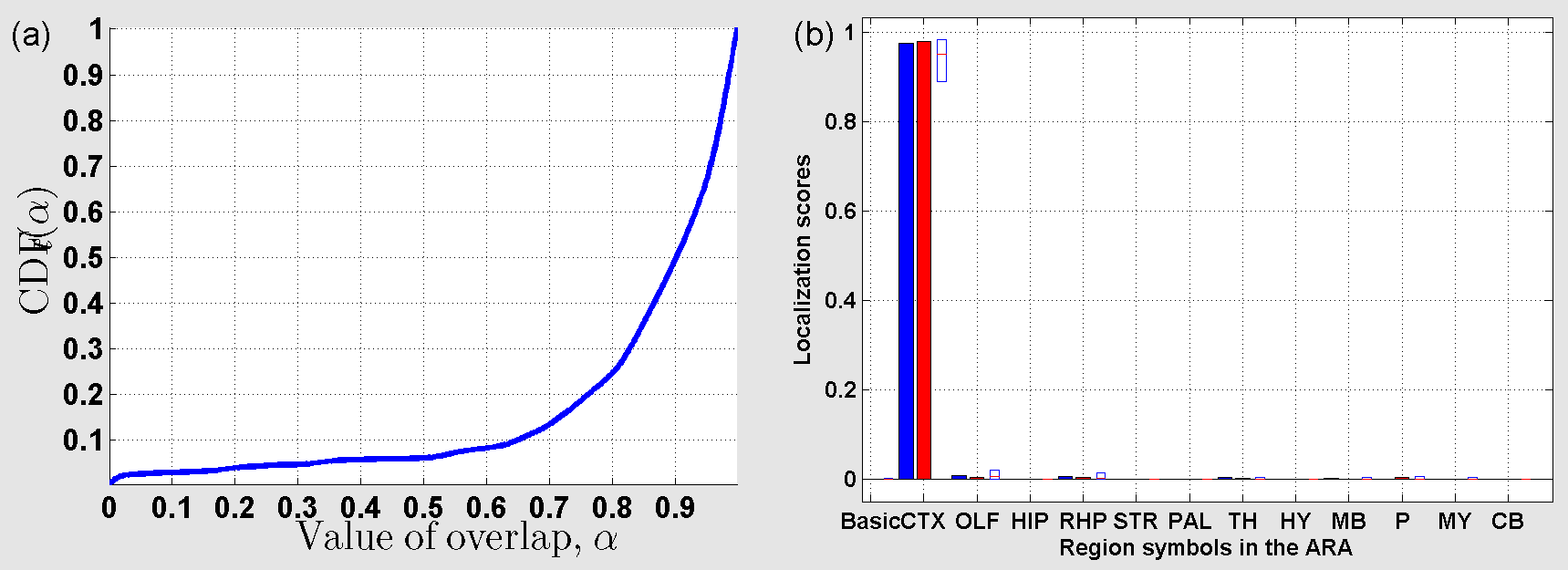}
\caption{(a) Cumulative distribution function (${\mathrm{\sc{CDF}}}_t$) of the overlap between $\rho_t$ and
 sub-sampled profiles for $t=46$. (b) Localization scores in the coarsest version of the ARA for $\rho_t$ (in blue), and 
 $\bar{\rho}_t$ (in red).}
\label{cdfPlot46}
\end{figure}
\begin{figure}
\includegraphics[width=1\textwidth,keepaspectratio]{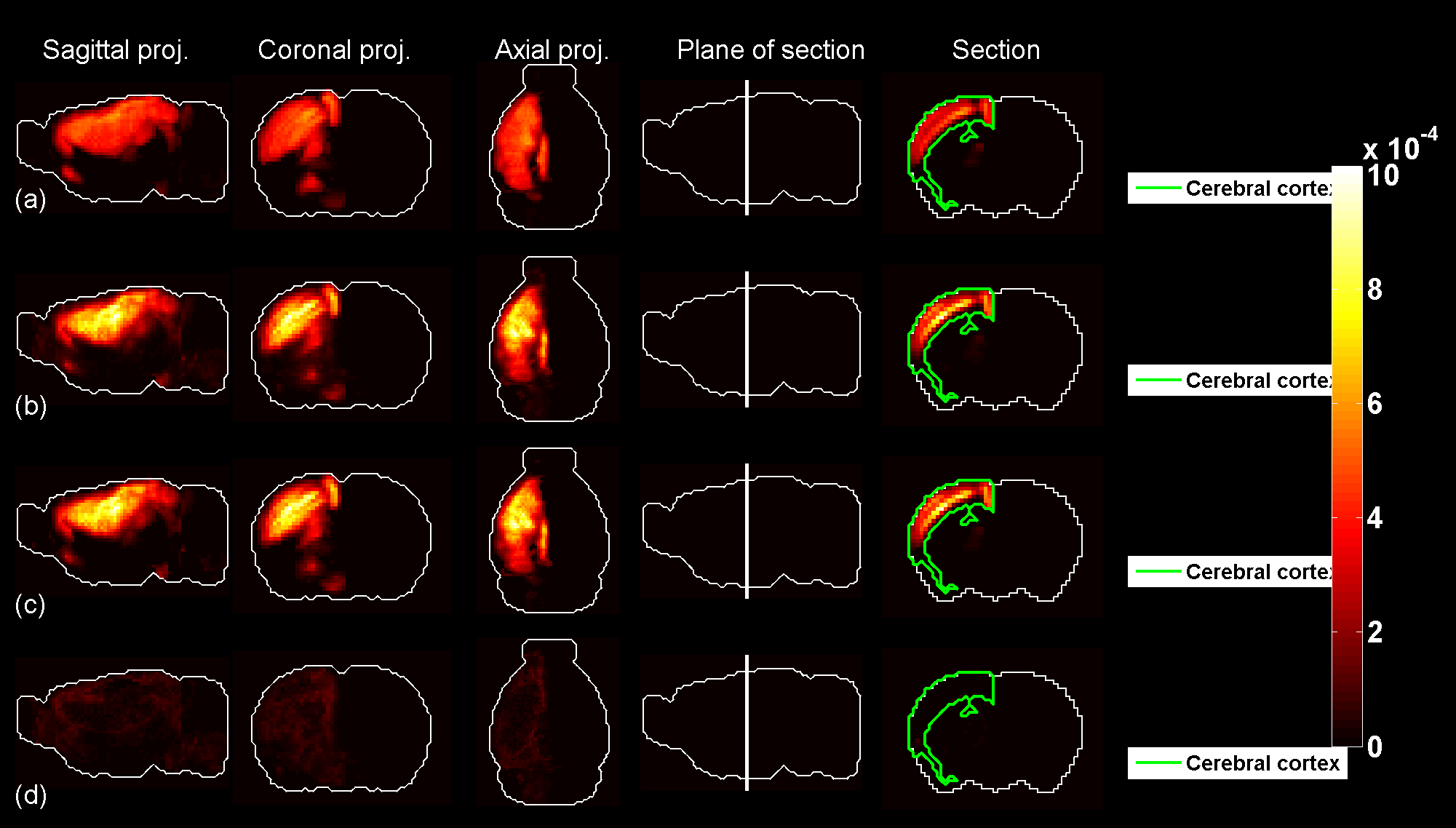}
\caption{Predicted profile and average sub-sampled profile for $t=46$.}
\label{subSampledFour46}
\end{figure}
\clearpage
\begin{figure}
\includegraphics[width=1\textwidth,keepaspectratio]{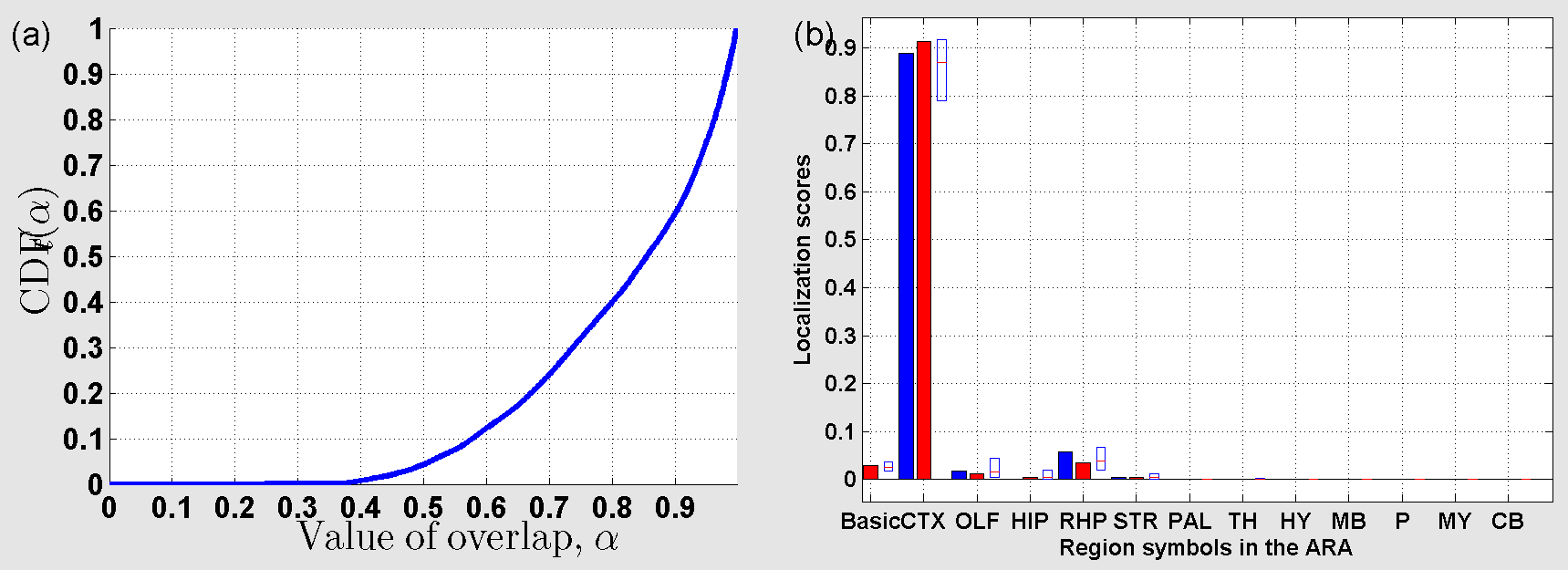}
\caption{(a) Cumulative distribution function (${\mathrm{\sc{CDF}}}_t$) of the overlap between $\rho_t$ and
 sub-sampled profiles for $t=47$. (b) Localization scores in the coarsest version of the ARA for $\rho_t$ (in blue), and 
 $\bar{\rho}_t$ (in red).}
\label{cdfPlot47}
\end{figure}
\begin{figure}
\includegraphics[width=1\textwidth,keepaspectratio]{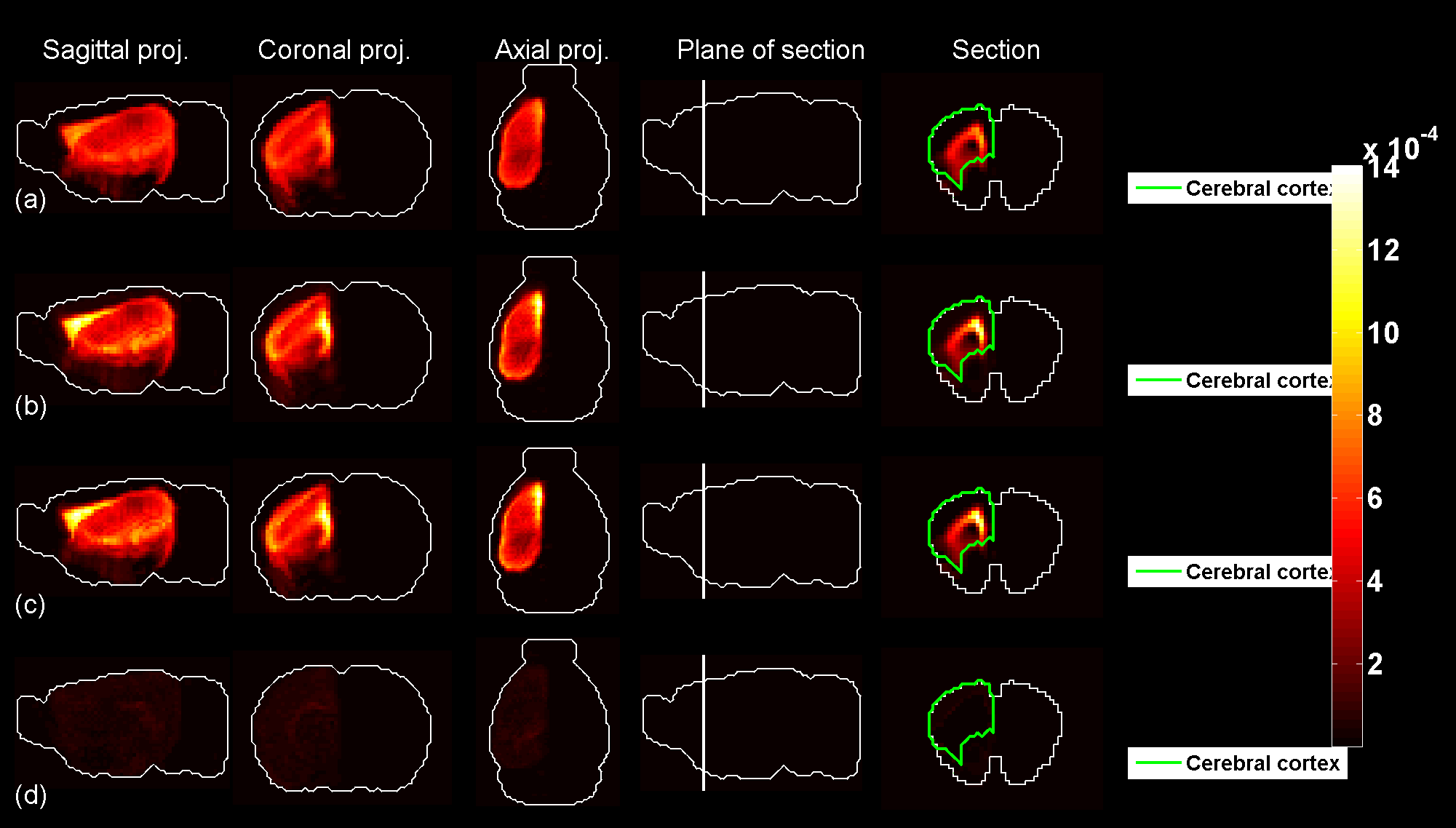}
\caption{Predicted profile and average sub-sampled profile for $t=47$.}
\label{subSampledFour47}
\end{figure}
\clearpage
\begin{figure}
\includegraphics[width=1\textwidth,keepaspectratio]{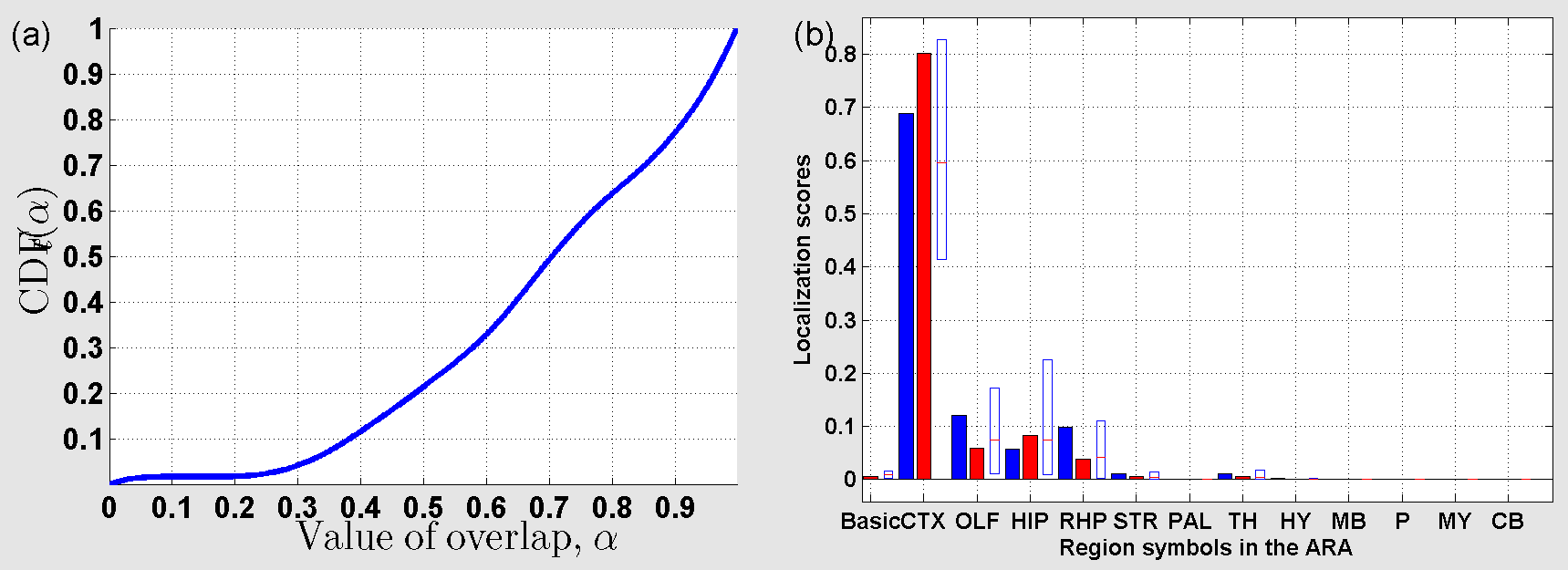}
\caption{(a) Cumulative distribution function (${\mathrm{\sc{CDF}}}_t$) of the overlap between $\rho_t$ and
 sub-sampled profiles for $t=48$. (b) Localization scores in the coarsest version of the ARA for $\rho_t$ (in blue), and 
 $\bar{\rho}_t$ (in red).}
\label{cdfPlot48}
\end{figure}
\begin{figure}
\includegraphics[width=1\textwidth,keepaspectratio]{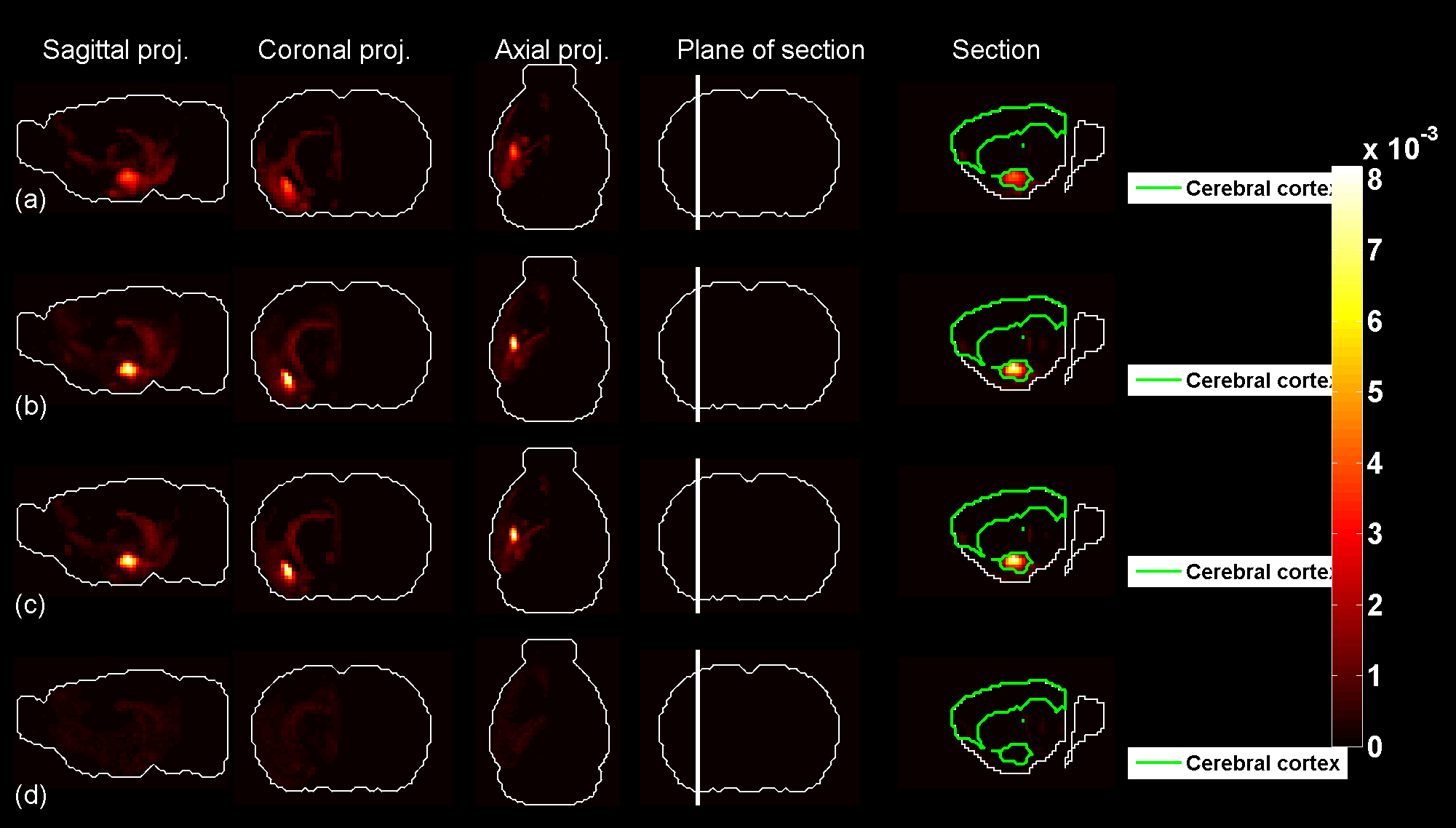}
\caption{Predicted profile and average sub-sampled profile for $t=48$.}
\label{subSampledFour48}
\end{figure}
\clearpage
\begin{figure}
\includegraphics[width=1\textwidth,keepaspectratio]{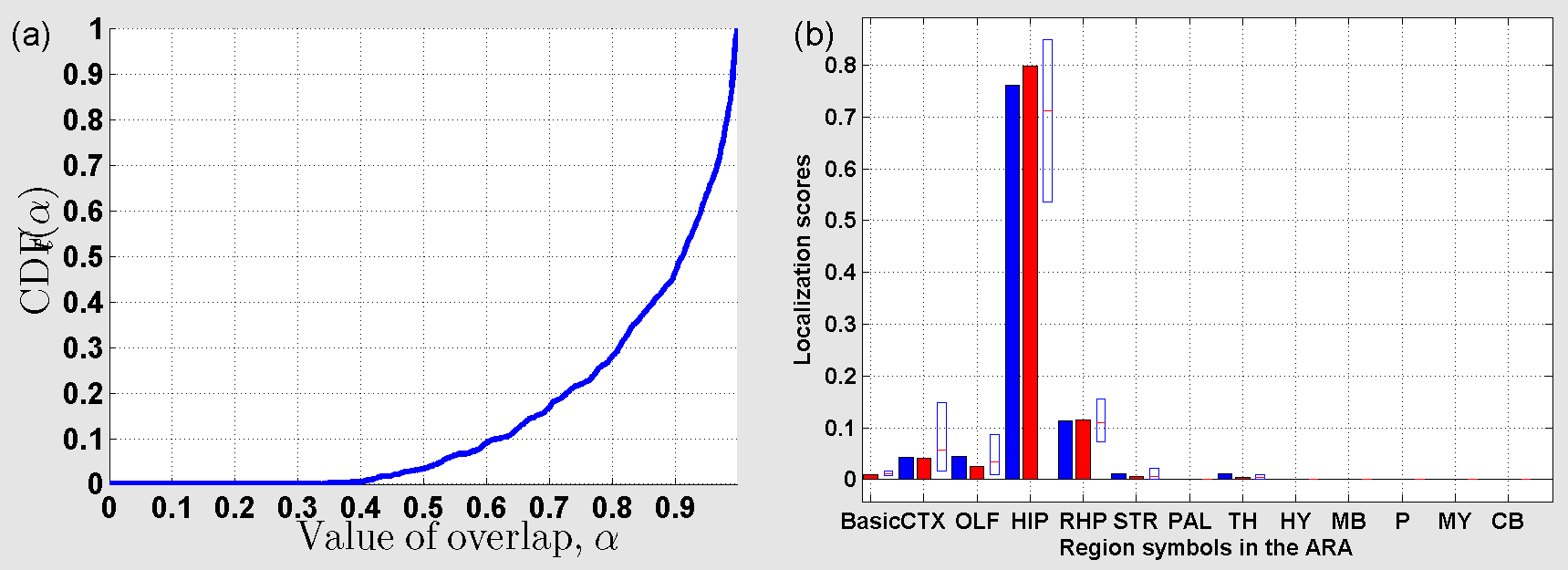}
\caption{(a) Cumulative distribution function (${\mathrm{\sc{CDF}}}_t$) of the overlap between $\rho_t$ and
 sub-sampled profiles for $t=49$. (b) Localization scores in the coarsest version of the ARA for $\rho_t$ (in blue), and 
 $\bar{\rho}_t$ (in red).}
\label{cdfPlot49}
\end{figure}
\begin{figure}
\includegraphics[width=1\textwidth,keepaspectratio]{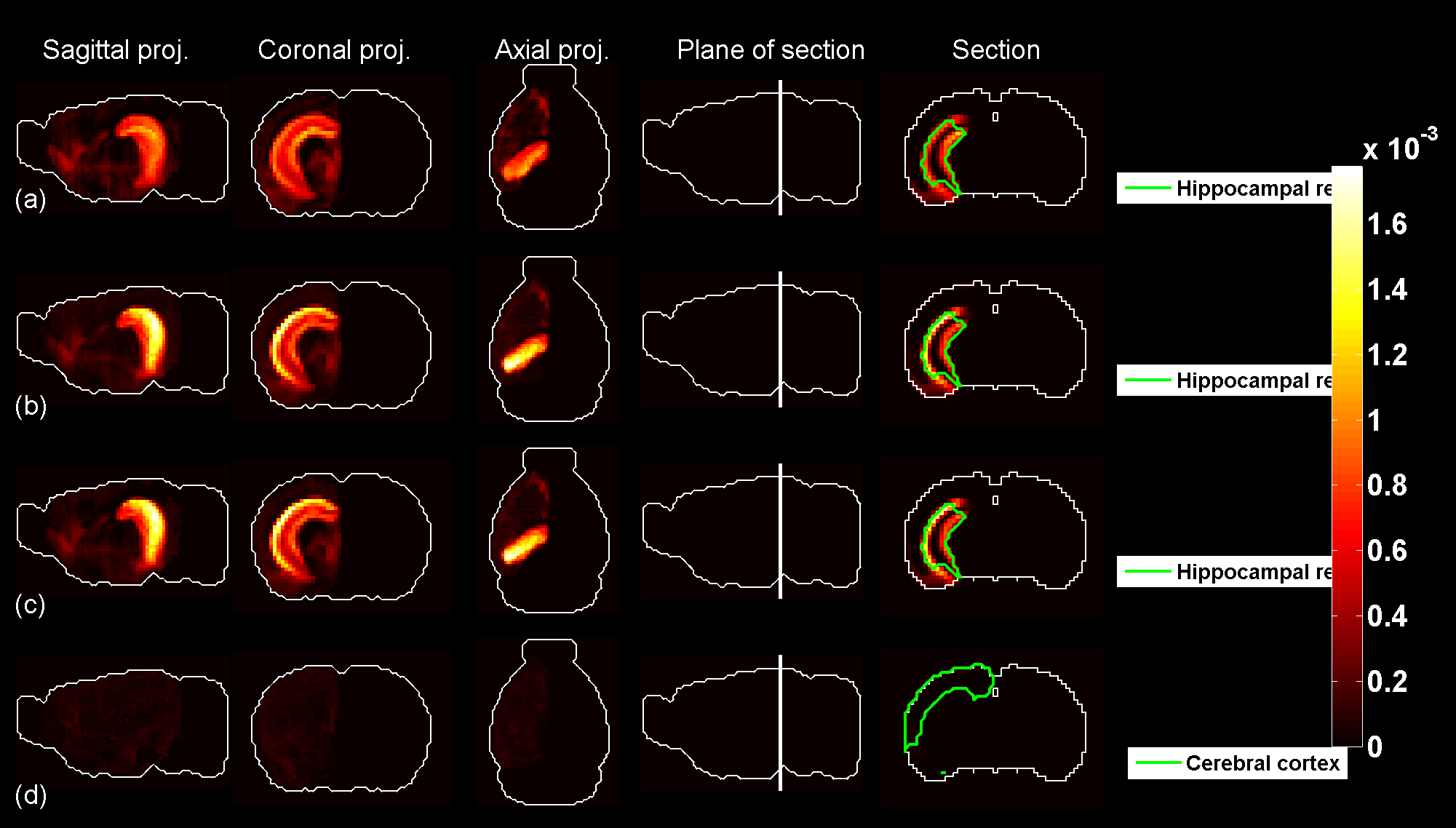}
\caption{Predicted profile and average sub-sampled profile for $t=49$.}
\label{subSampledFour49}
\end{figure}
\clearpage
\begin{figure}
\includegraphics[width=1\textwidth,keepaspectratio]{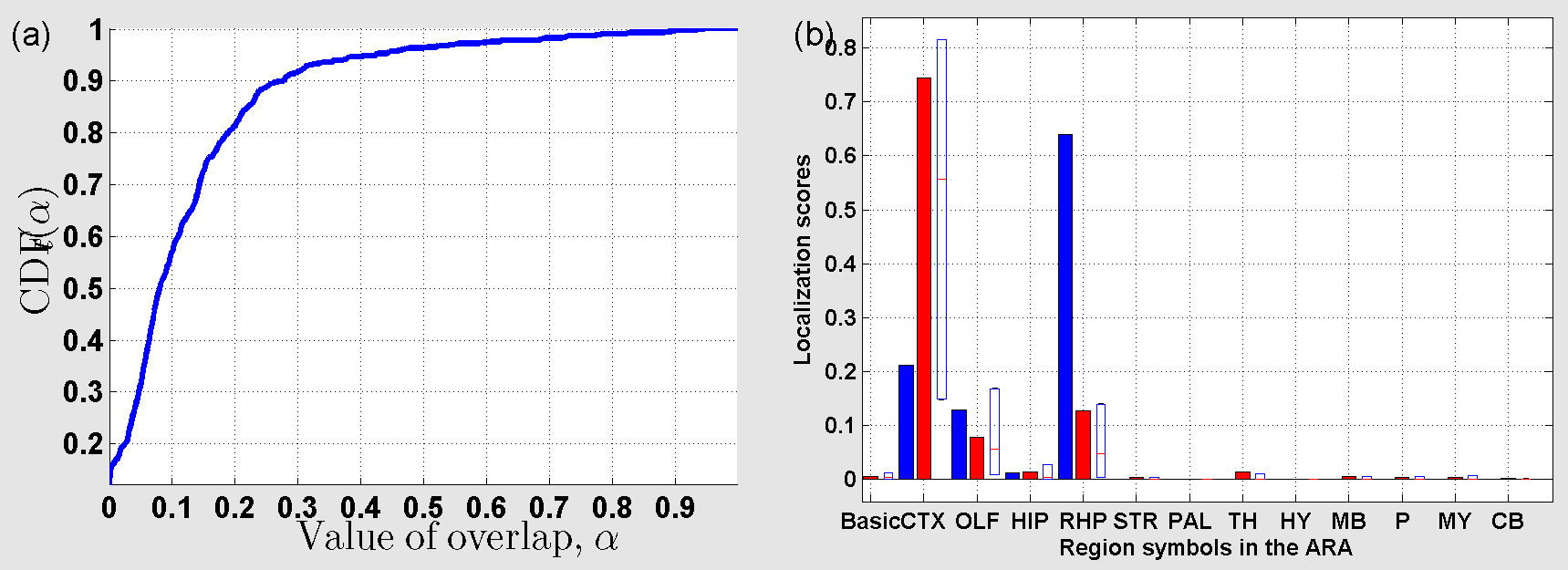}
\caption{(a) Cumulative distribution function (${\mathrm{\sc{CDF}}}_t$) of the overlap between $\rho_t$ and
 sub-sampled profiles for $t=50$. (b) Localization scores in the coarsest version of the ARA for $\rho_t$ (in blue), and 
 $\bar{\rho}_t$ (in red).}
\label{cdfPlot50}
\end{figure}
\begin{figure}
\includegraphics[width=1\textwidth,keepaspectratio]{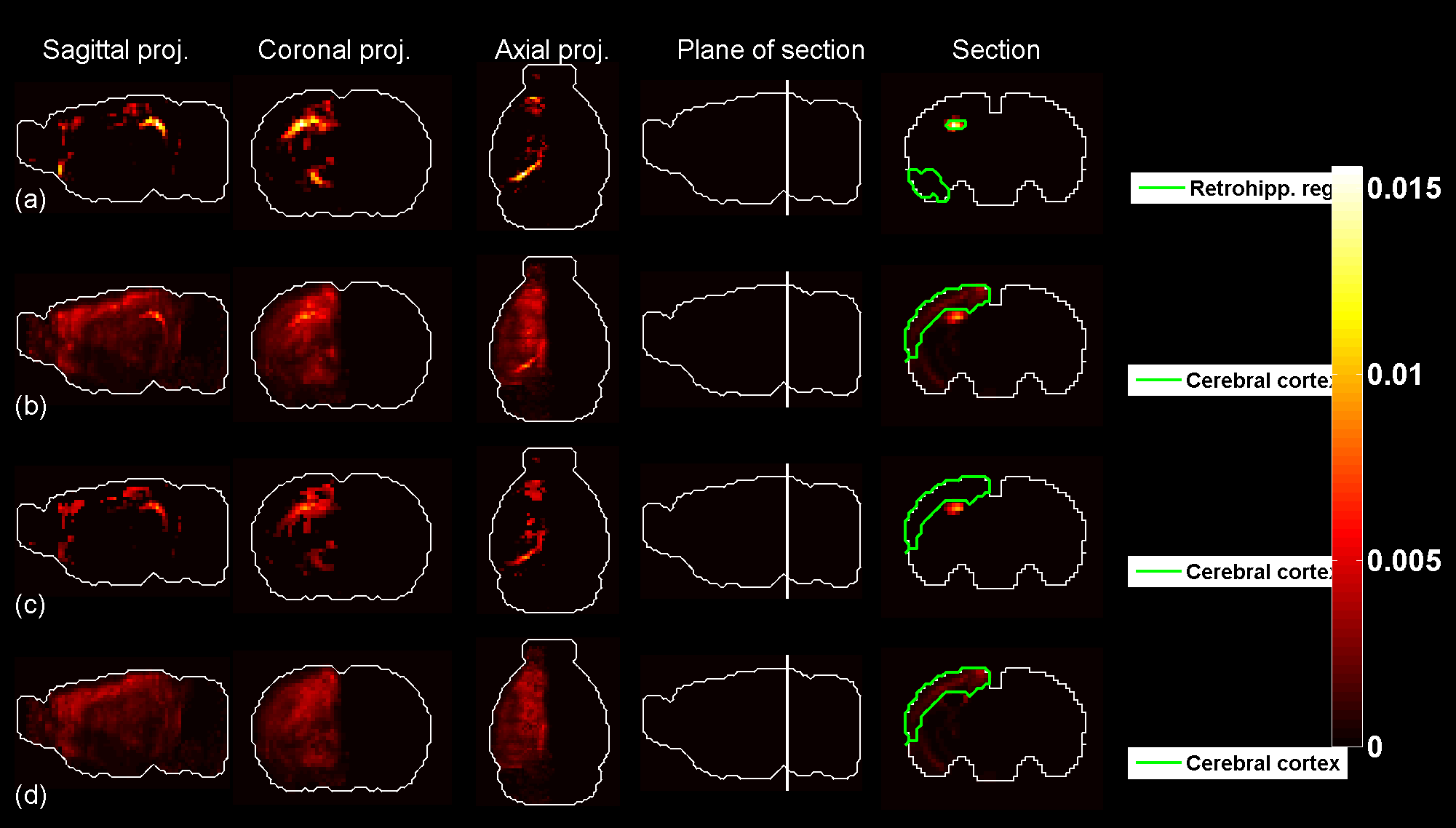}
\caption{Predicted profile and average sub-sampled profile for $t=50$.}
\label{subSampledFour50}
\end{figure}
\clearpage
\begin{figure}
\includegraphics[width=1\textwidth,keepaspectratio]{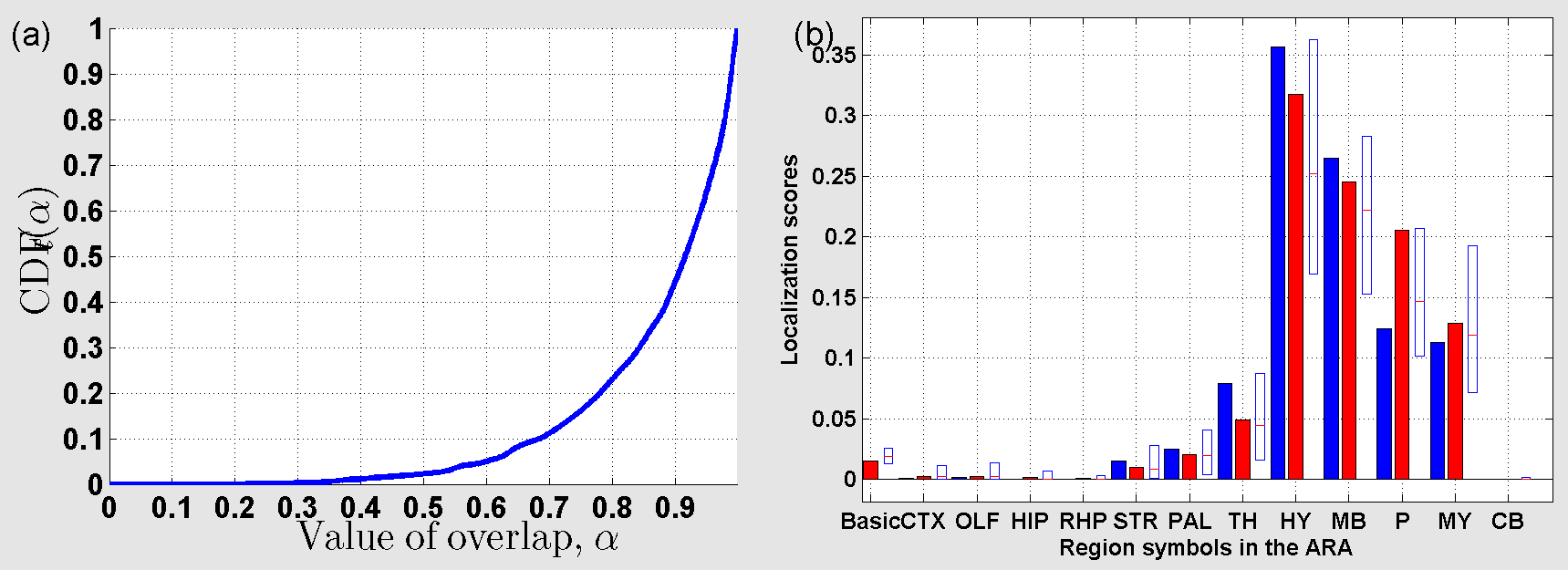}
\caption{(a) Cumulative distribution function (${\mathrm{\sc{CDF}}}_t$) of the overlap between $\rho_t$ and
 sub-sampled profiles for $t=51$. (b) Localization scores in the coarsest version of the ARA for $\rho_t$ (in blue), and 
 $\bar{\rho}_t$ (in red).}
\label{cdfPlot51}
\end{figure}
\begin{figure}
\includegraphics[width=1\textwidth,keepaspectratio]{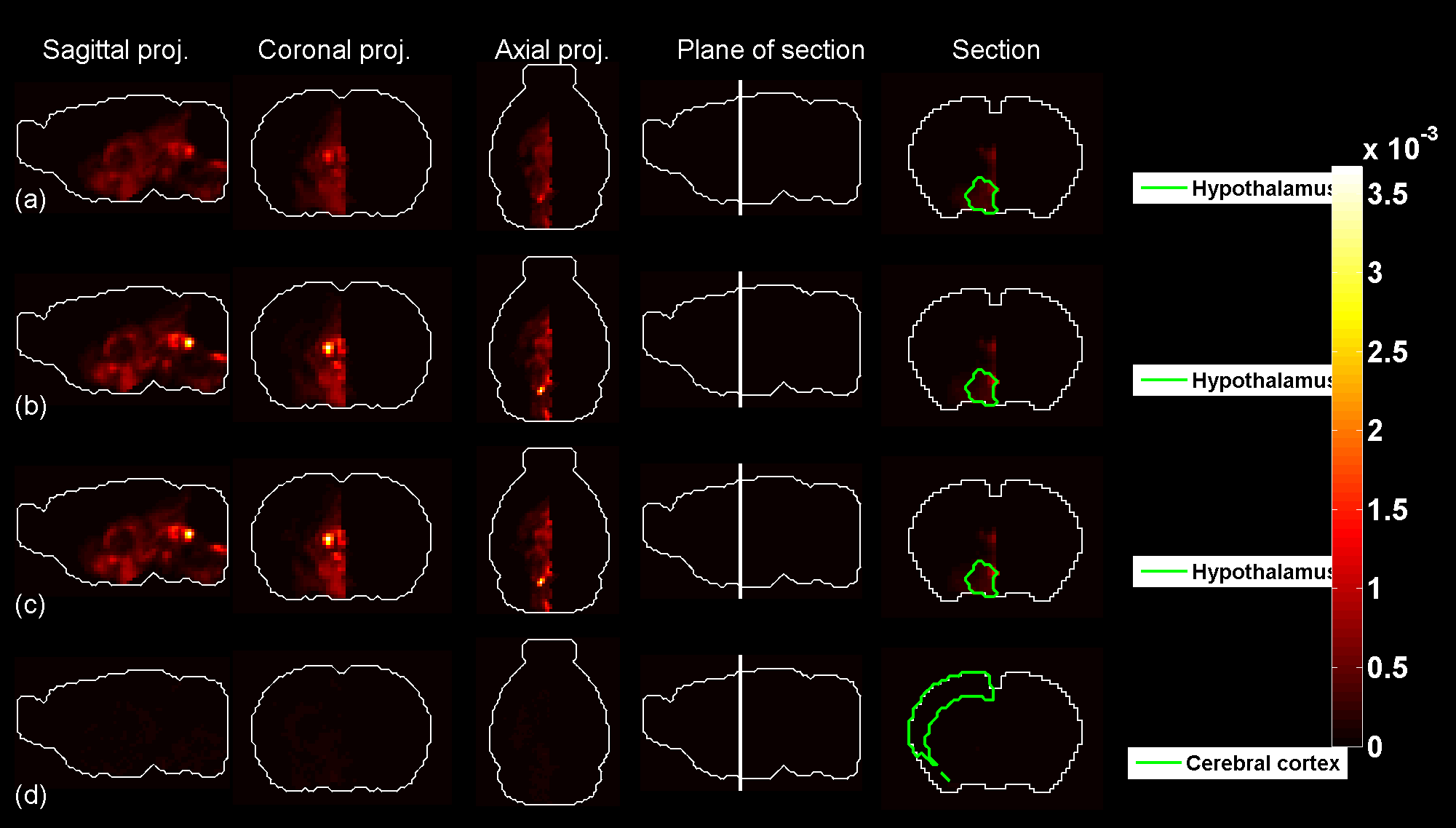}
\caption{Predicted profile and average sub-sampled profile for $t=51$.}
\label{subSampledFour51}
\end{figure}
\clearpage
\begin{figure}
\includegraphics[width=1\textwidth,keepaspectratio]{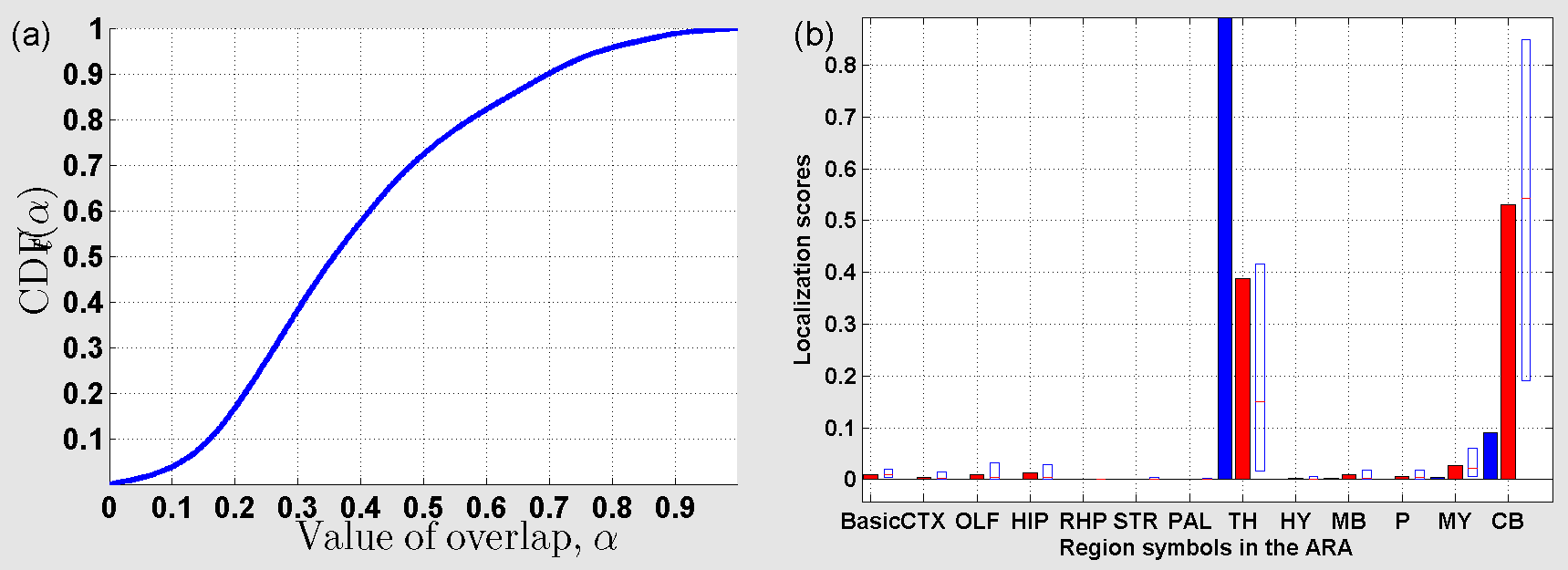}
\caption{(a) Cumulative distribution function (${\mathrm{\sc{CDF}}}_t$) of the overlap between $\rho_t$ and
 sub-sampled profiles for $t=52$. (b) Localization scores in the coarsest version of the ARA for $\rho_t$ (in blue), and 
 $\bar{\rho}_t$ (in red).}
\label{cdfPlot52}
\end{figure}
\begin{figure}
\includegraphics[width=1\textwidth,keepaspectratio]{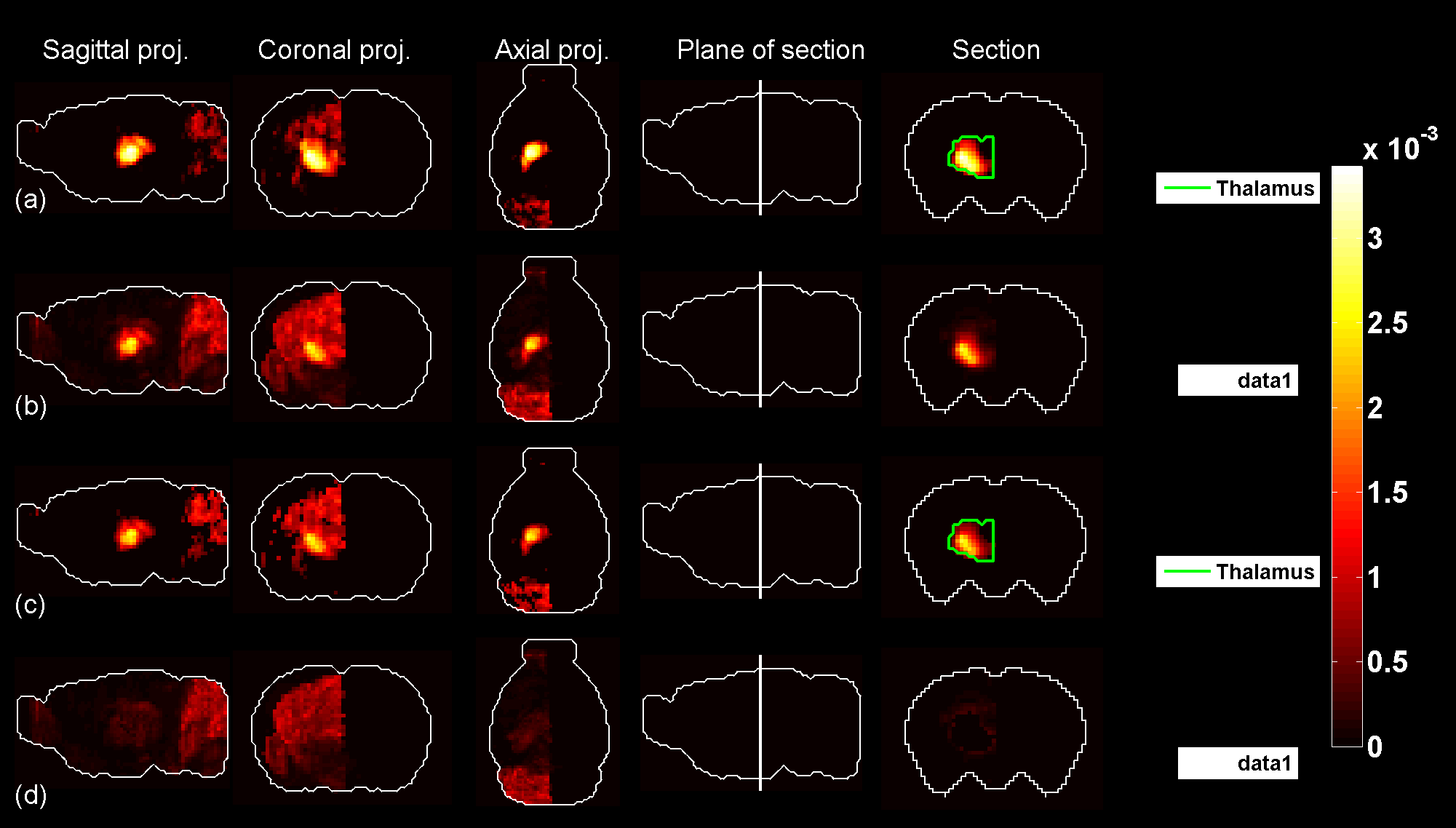}
\caption{Predicted profile and average sub-sampled profile for $t=52$.}
\label{subSampledFour52}
\end{figure}
\clearpage
\begin{figure}
\includegraphics[width=1\textwidth,keepaspectratio]{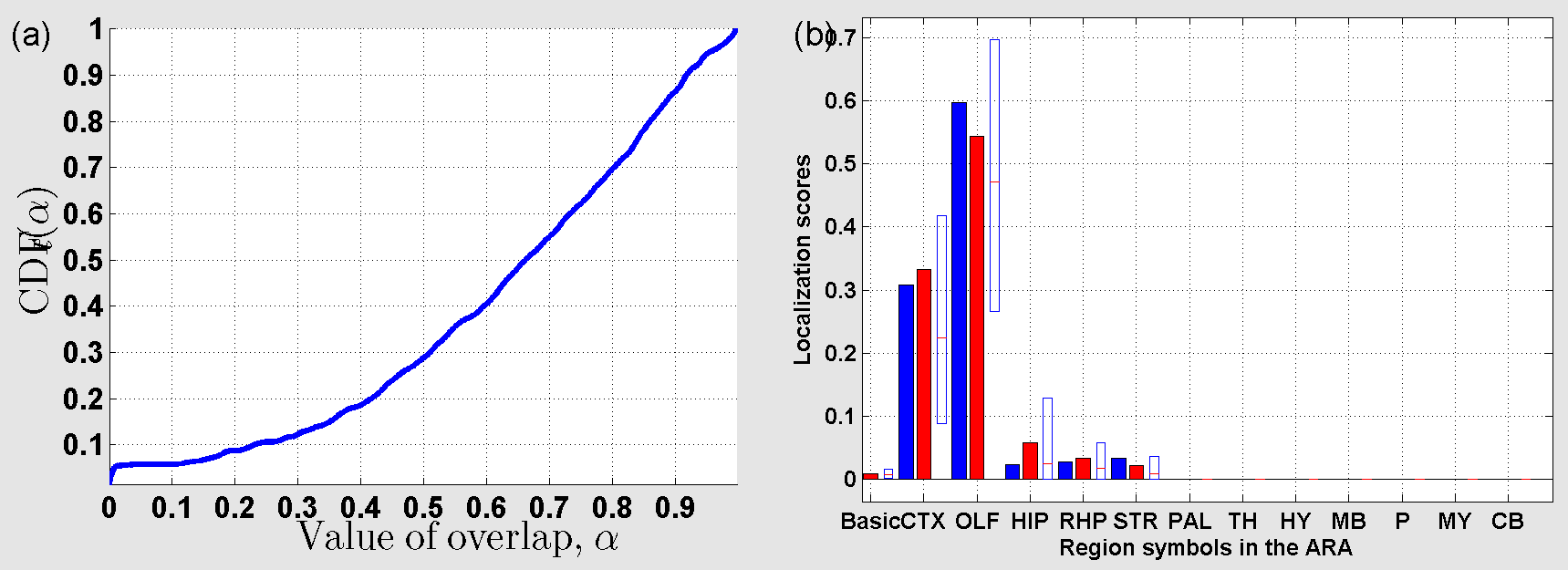}
\caption{(a) Cumulative distribution function (${\mathrm{\sc{CDF}}}_t$) of the overlap between $\rho_t$ and
 sub-sampled profiles for $t=53$. (b) Localization scores in the coarsest version of the ARA for $\rho_t$ (in blue), and 
 $\bar{\rho}_t$ (in red).}
\label{cdfPlot53}
\end{figure}
\begin{figure}
\includegraphics[width=1\textwidth,keepaspectratio]{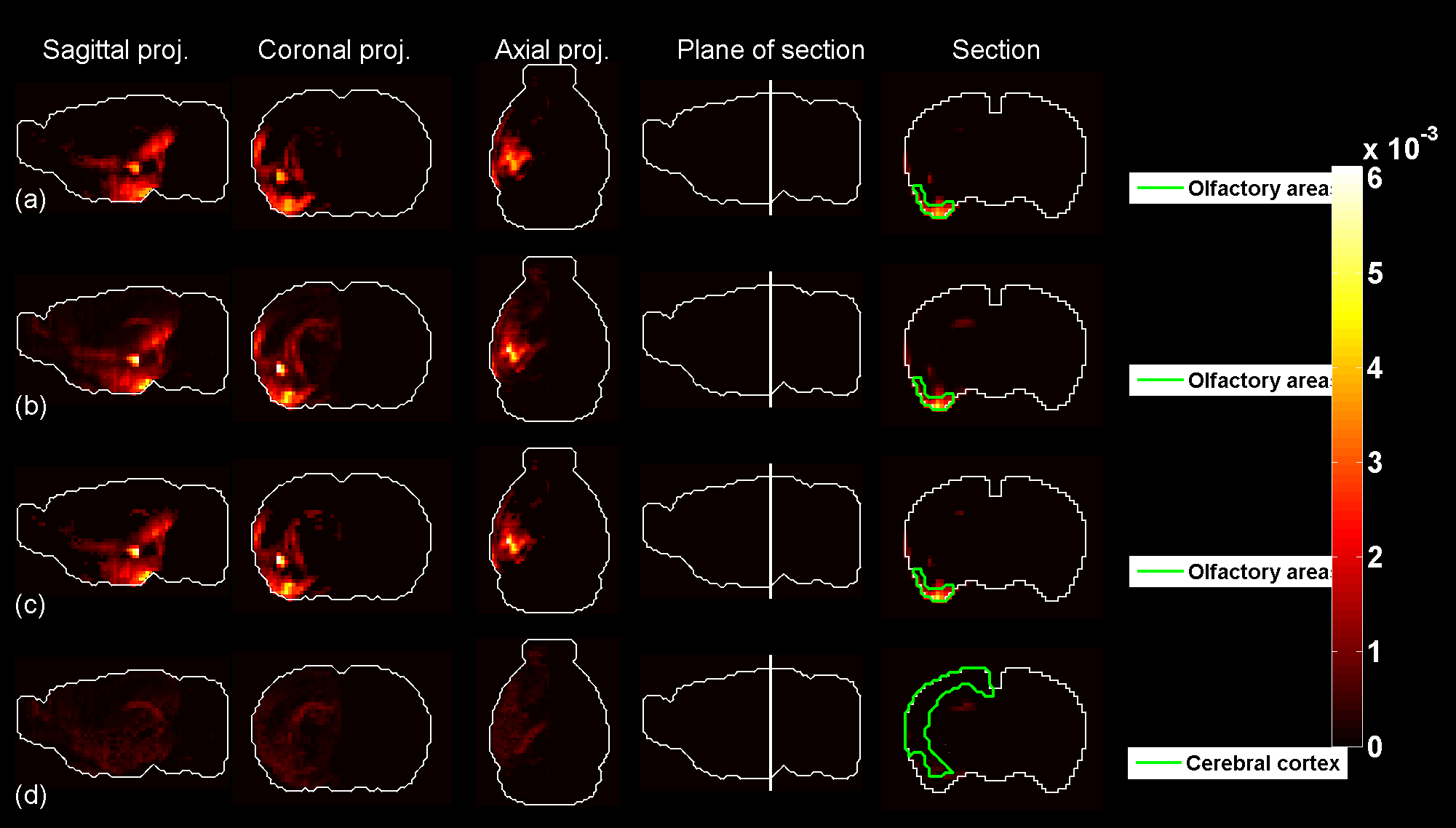}
\caption{Predicted profile and average sub-sampled profile for $t=53$.}
\label{subSampledFour53}
\end{figure}
\clearpage
\begin{figure}
\includegraphics[width=1\textwidth,keepaspectratio]{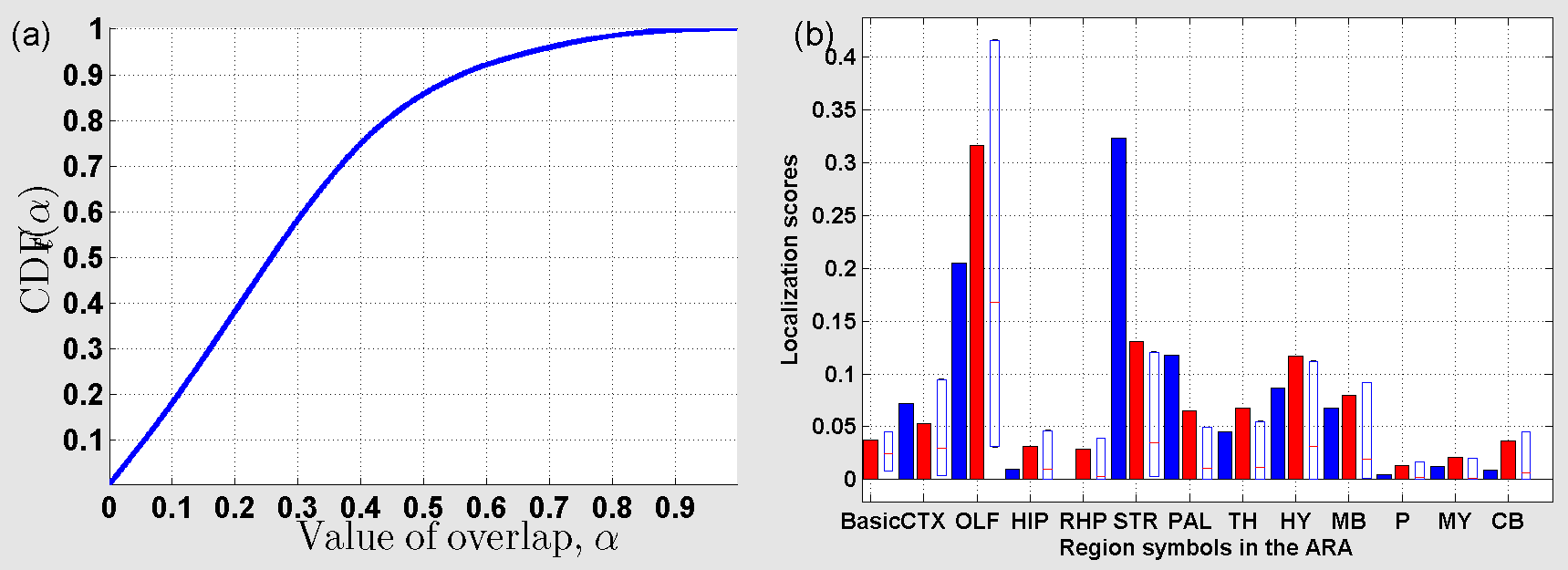}
\caption{(a) Cumulative distribution function (${\mathrm{\sc{CDF}}}_t$) of the overlap between $\rho_t$ and
 sub-sampled profiles for $t=54$. (b) Localization scores in the coarsest version of the ARA for $\rho_t$ (in blue), and 
 $\bar{\rho}_t$ (in red).}
\label{cdfPlot54}
\end{figure}
\begin{figure}
\includegraphics[width=1\textwidth,keepaspectratio]{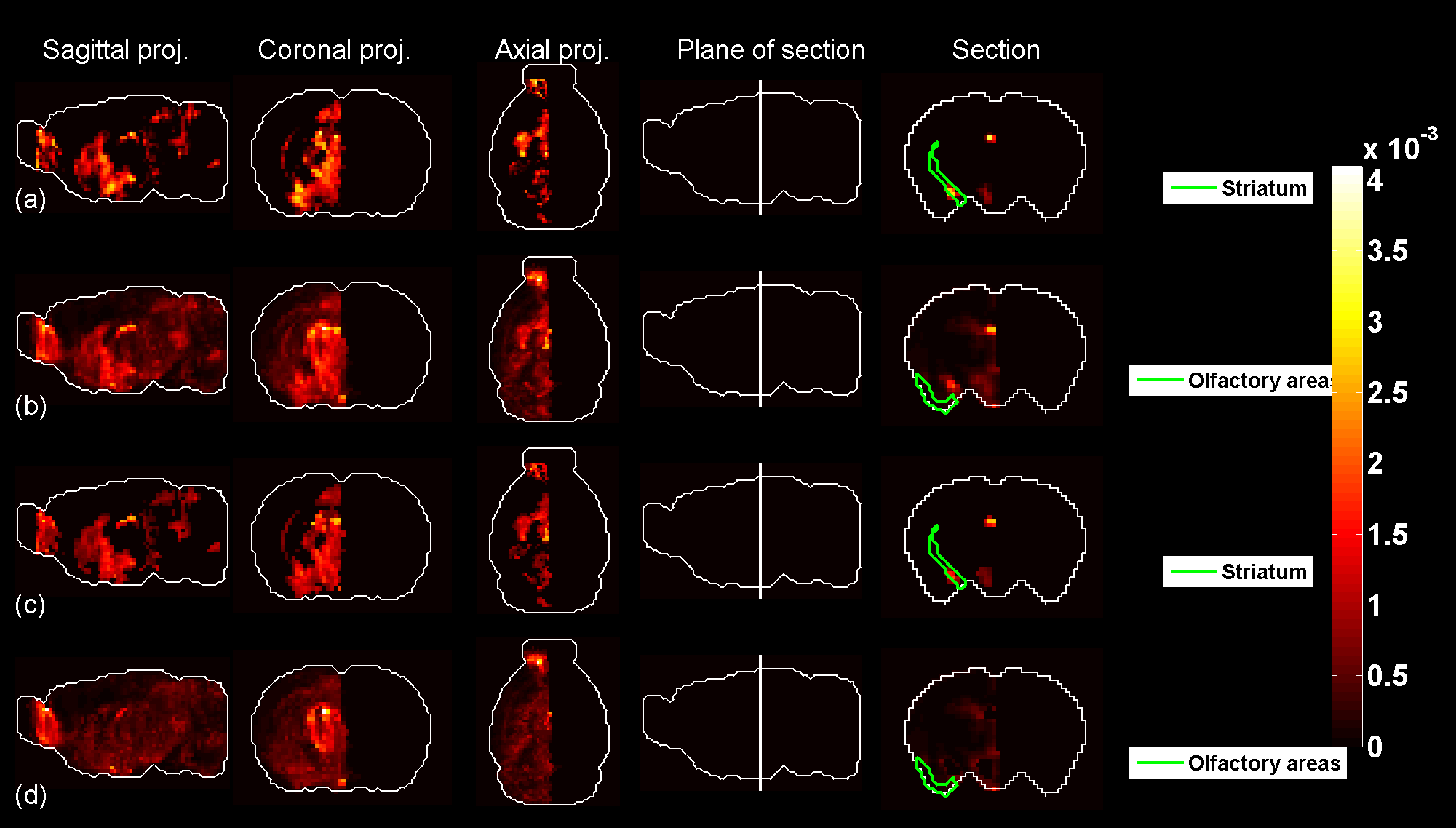}
\caption{Predicted profile and average sub-sampled profile for $t=54$.}
\label{subSampledFour54}
\end{figure}
\clearpage
\begin{figure}
\includegraphics[width=1\textwidth,keepaspectratio]{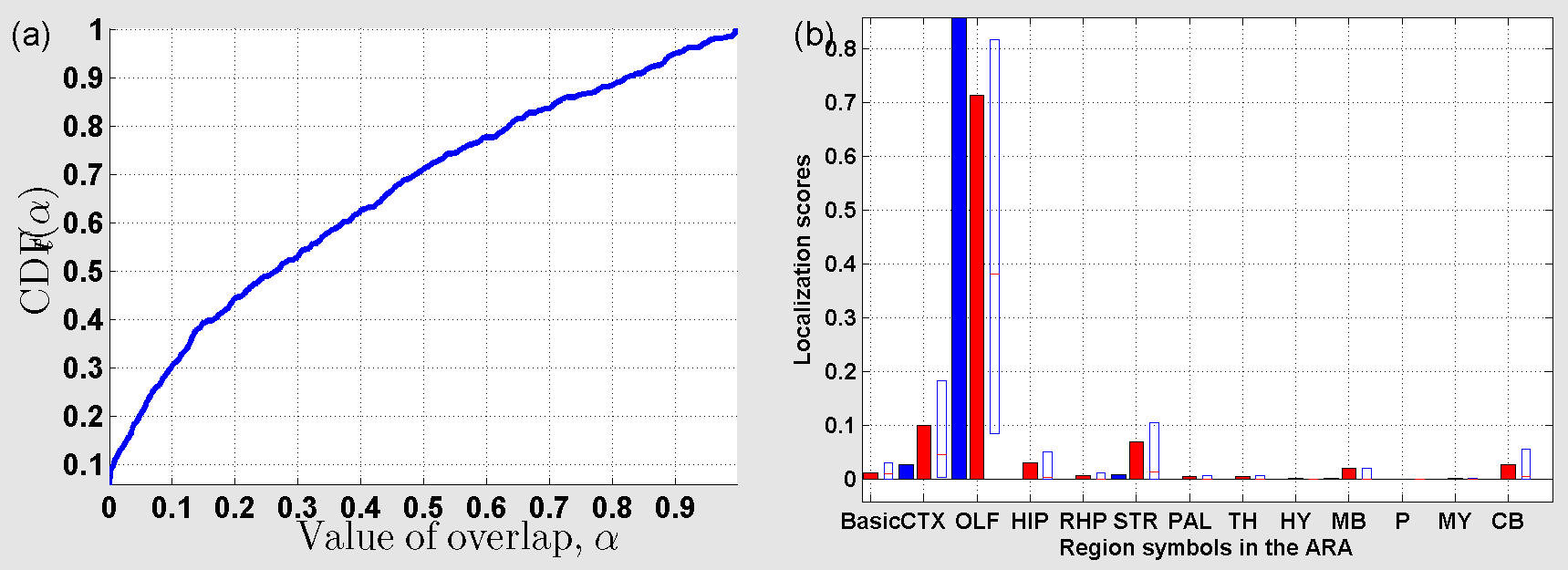}
\caption{(a) Cumulative distribution function (${\mathrm{\sc{CDF}}}_t$) of the overlap between $\rho_t$ and
 sub-sampled profiles for $t=55$. (b) Localization scores in the coarsest version of the ARA for $\rho_t$ (in blue), and 
 $\bar{\rho}_t$ (in red).}
\label{cdfPlot55}
\end{figure}
\begin{figure}
\includegraphics[width=1\textwidth,keepaspectratio]{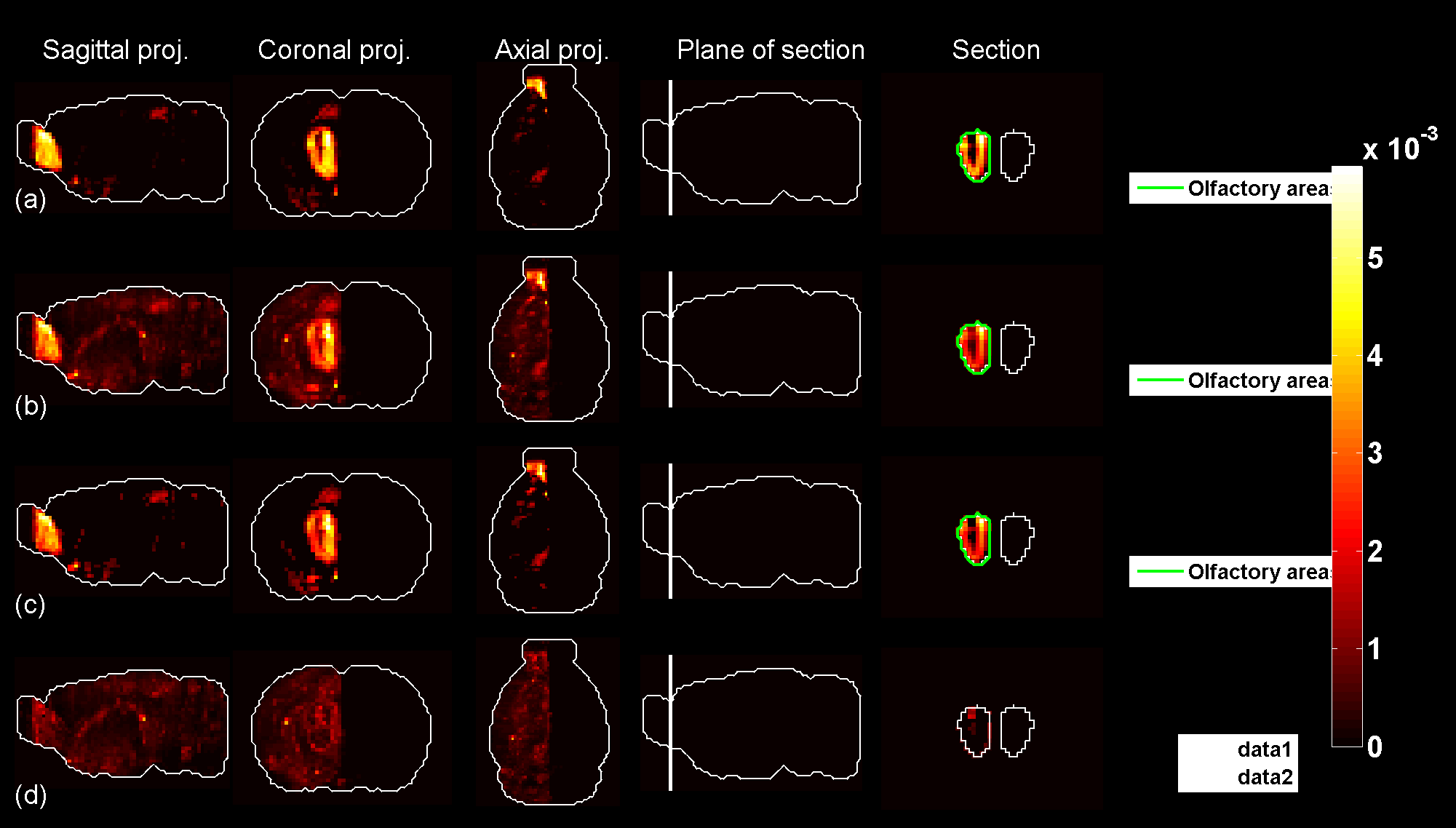}
\caption{Predicted profile and average sub-sampled profile for $t=55$.}
\label{subSampledFour55}
\end{figure}
\clearpage
\begin{figure}
\includegraphics[width=1\textwidth,keepaspectratio]{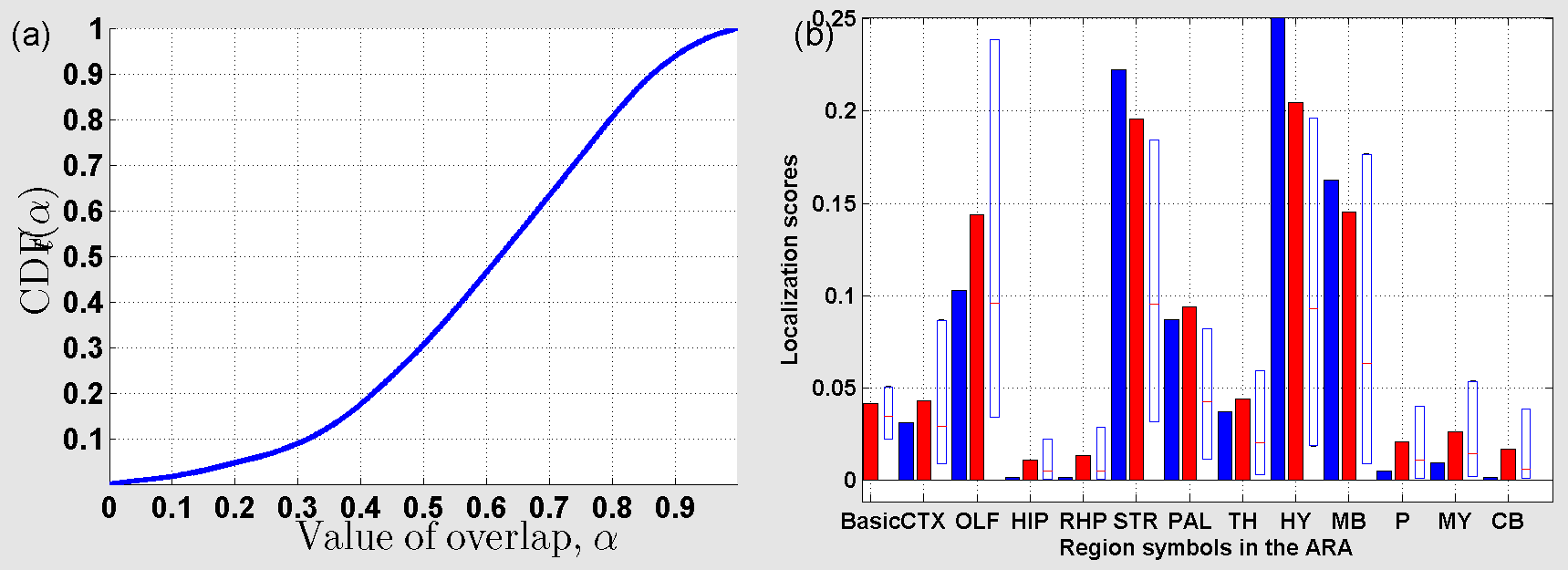}
\caption{(a) Cumulative distribution function (${\mathrm{\sc{CDF}}}_t$) of the overlap between $\rho_t$ and
 sub-sampled profiles for $t=56$. (b) Localization scores in the coarsest version of the ARA for $\rho_t$ (in blue), and 
 $\bar{\rho}_t$ (in red).}
\label{cdfPlot56}
\end{figure}
\begin{figure}
\includegraphics[width=1\textwidth,keepaspectratio]{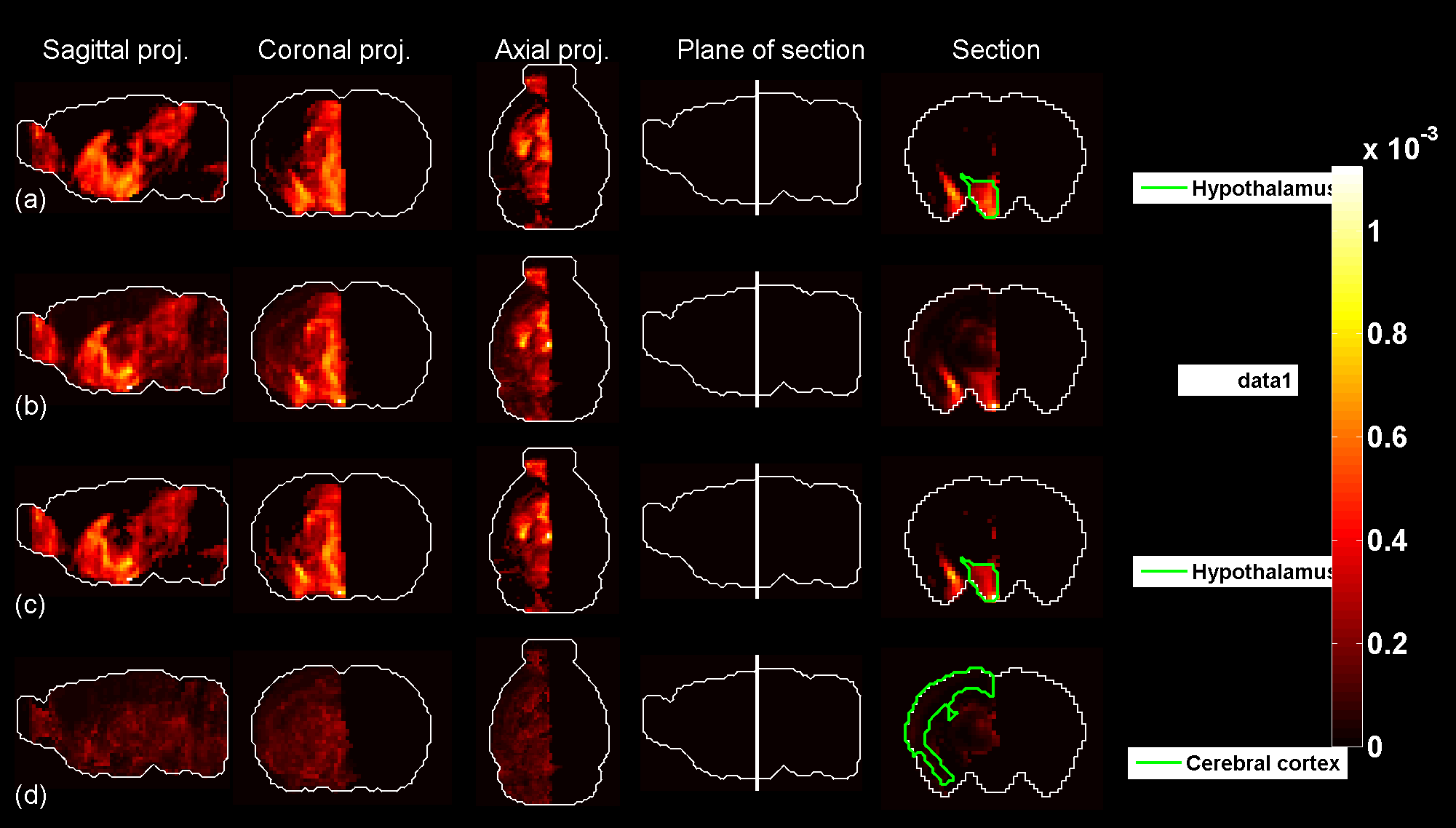}
\caption{Predicted profile and average sub-sampled profile for $t=56$.}
\label{subSampledFour56}
\end{figure}
\clearpage
\begin{figure}
\includegraphics[width=1\textwidth,keepaspectratio]{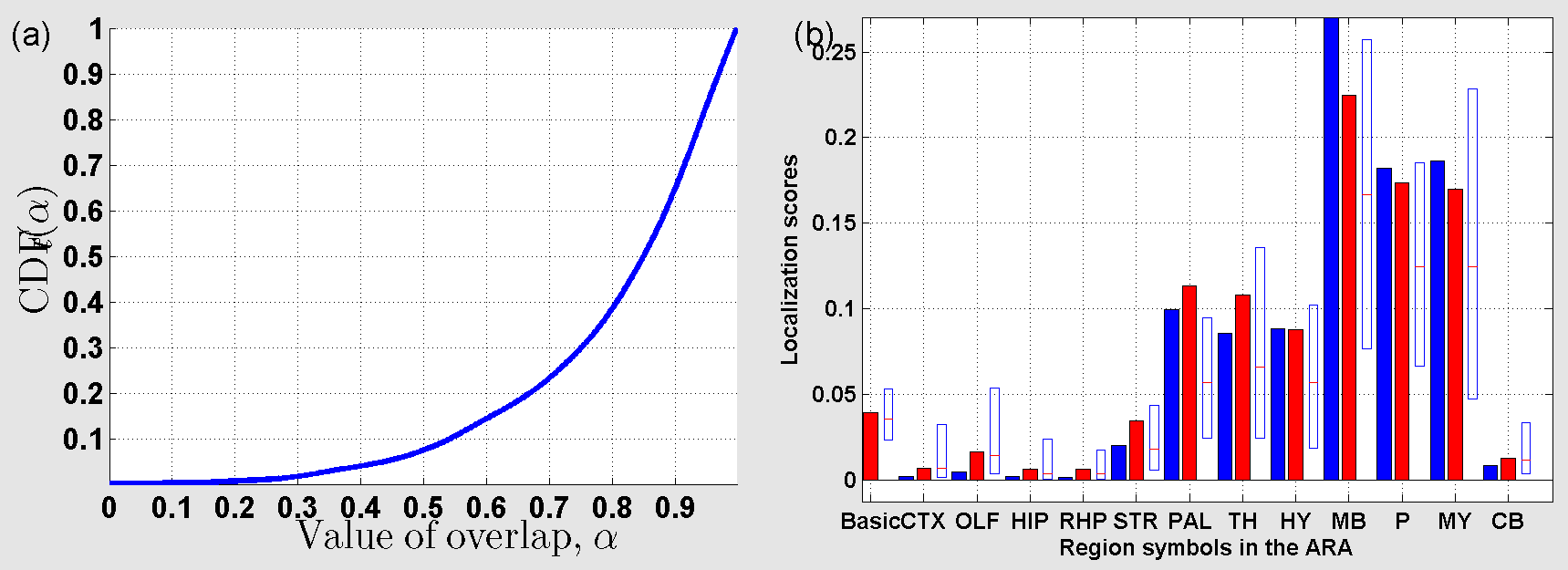}
\caption{(a) Cumulative distribution function (${\mathrm{\sc{CDF}}}_t$) of the overlap between $\rho_t$ and
 sub-sampled profiles for $t=57$. (b) Localization scores in the coarsest version of the ARA for $\rho_t$ (in blue), and 
 $\bar{\rho}_t$ (in red).}
\label{cdfPlot57}
\end{figure}
\begin{figure}
\includegraphics[width=1\textwidth,keepaspectratio]{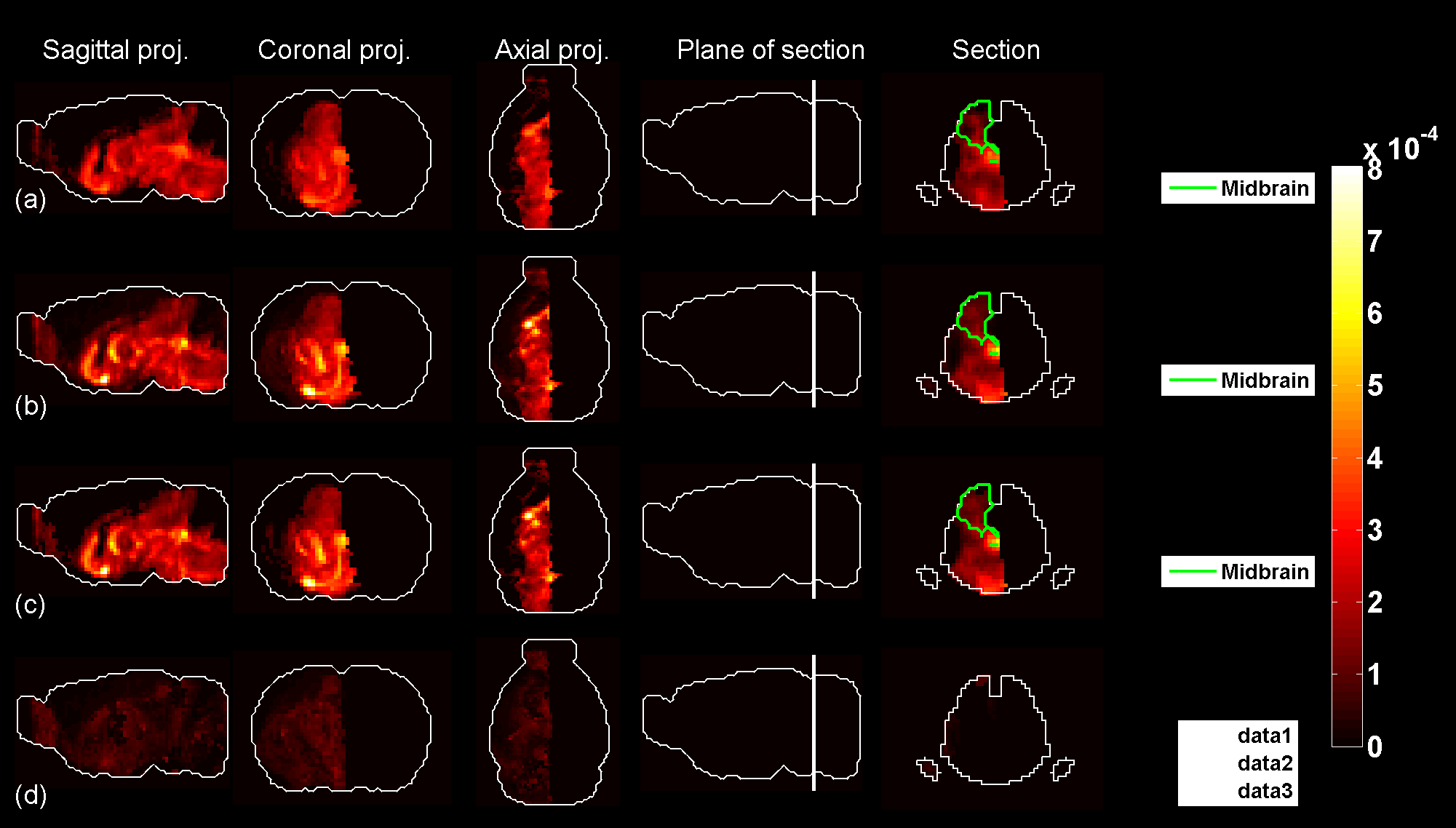}
\caption{Predicted profile and average sub-sampled profile for $t=57$.}
\label{subSampledFour57}
\end{figure}
\clearpage
\begin{figure}
\includegraphics[width=1\textwidth,keepaspectratio]{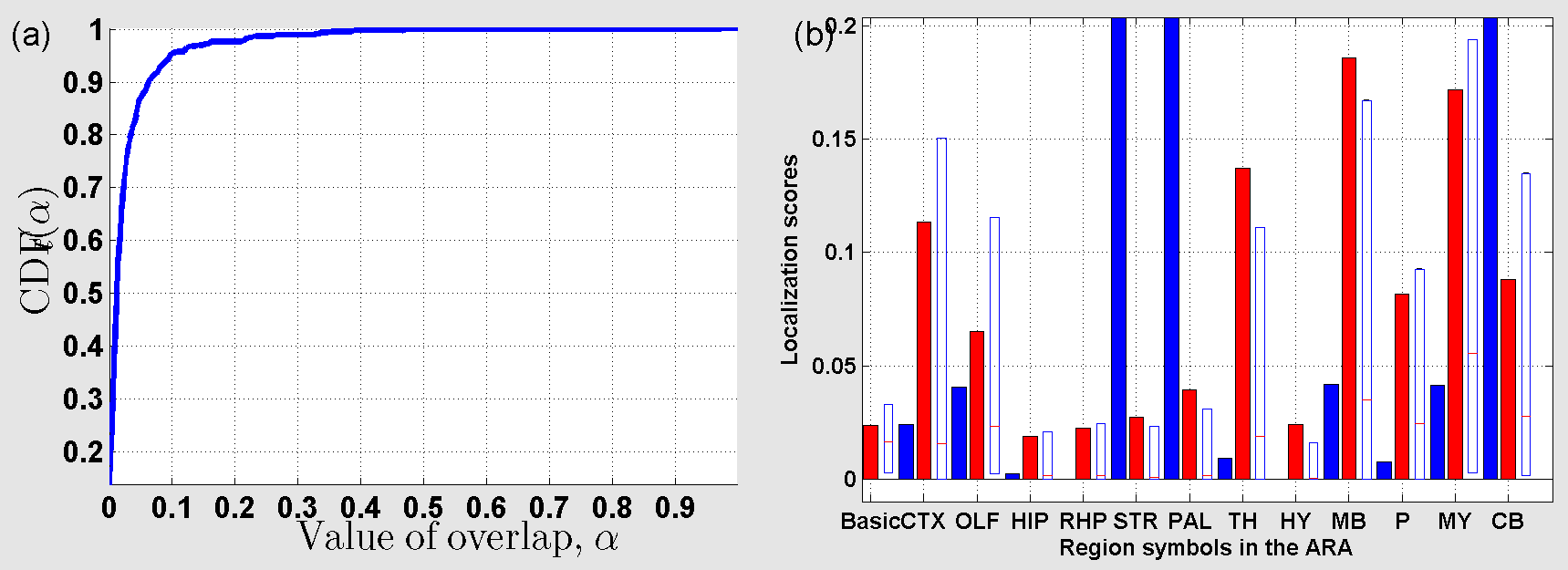}
\caption{(a) Cumulative distribution function (${\mathrm{\sc{CDF}}}_t$) of the overlap between $\rho_t$ and
 sub-sampled profiles for $t=58$. (b) Localization scores in the coarsest version of the ARA for $\rho_t$ (in blue), and 
 $\bar{\rho}_t$ (in red).}
\label{cdfPlot58}
\end{figure}
\begin{figure}
\includegraphics[width=1\textwidth,keepaspectratio]{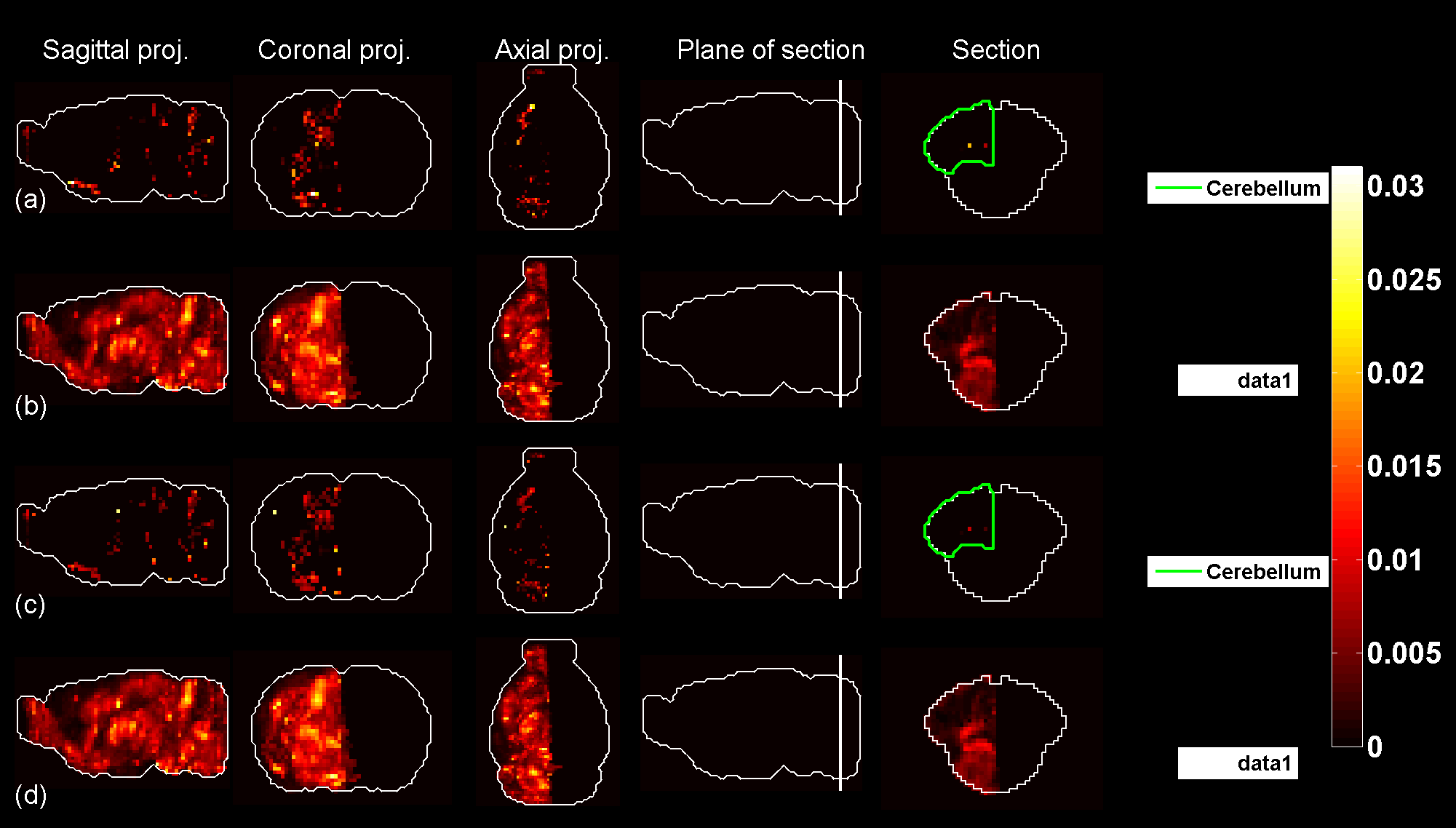}
\caption{Predicted profile and average sub-sampled profile for $t=58$.}
\label{subSampledFour58}
\end{figure}
\clearpage
\begin{figure}
\includegraphics[width=1\textwidth,keepaspectratio]{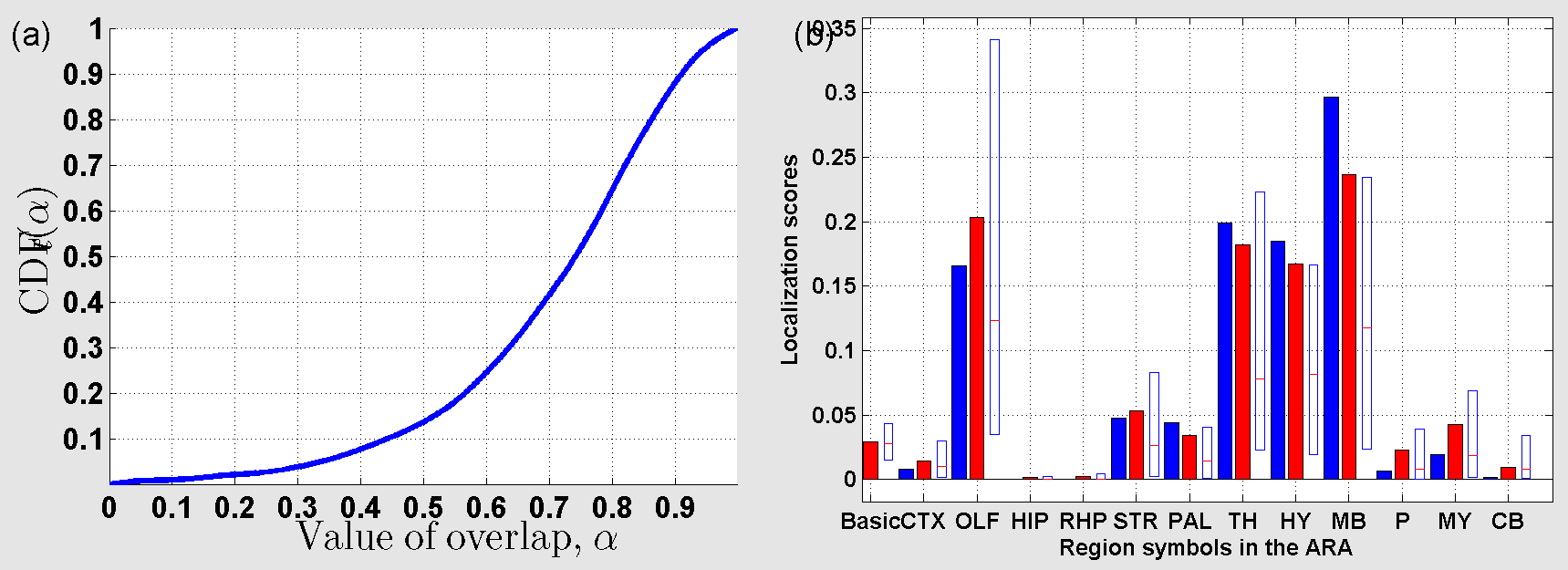}
\caption{(a) Cumulative distribution function (${\mathrm{\sc{CDF}}}_t$) of the overlap between $\rho_t$ and
 sub-sampled profiles for $t=59$. (b) Localization scores in the coarsest version of the ARA for $\rho_t$ (in blue), and 
 $\bar{\rho}_t$ (in red).}
\label{cdfPlot59}
\end{figure}
\begin{figure}
\includegraphics[width=1\textwidth,keepaspectratio]{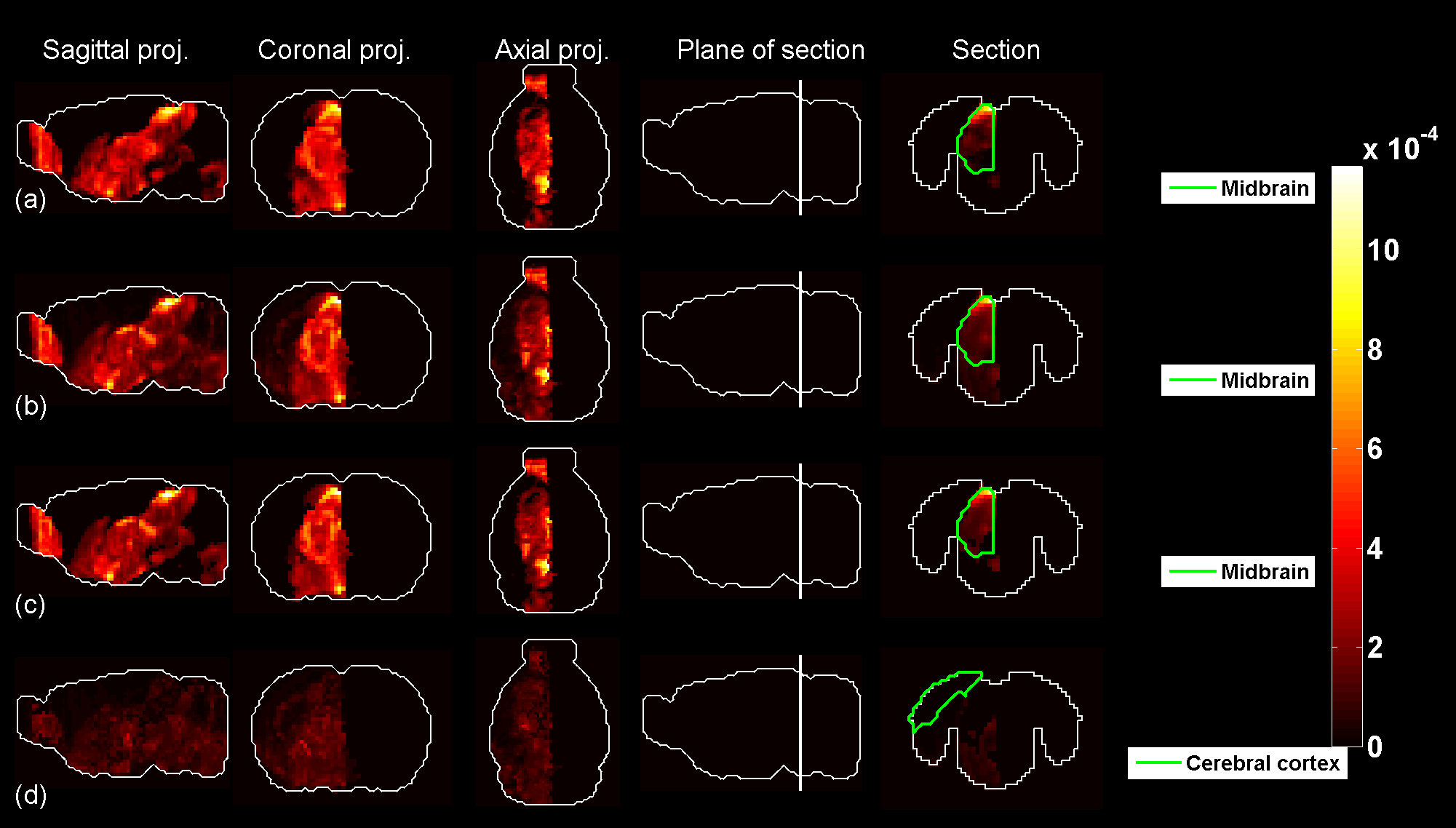}
\caption{Predicted profile and average sub-sampled profile for $t=59$.}
\label{subSampledFour59}
\end{figure}
\clearpage
\begin{figure}
\includegraphics[width=1\textwidth,keepaspectratio]{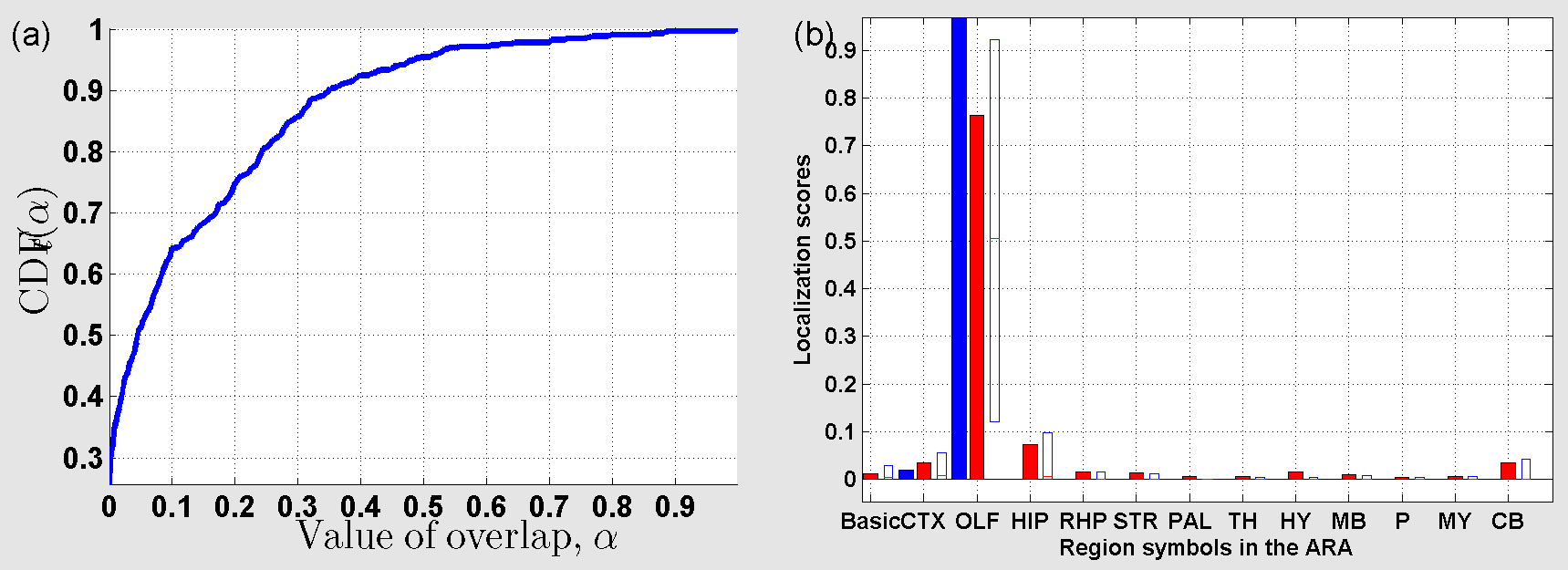}
\caption{(a) Cumulative distribution function (${\mathrm{\sc{CDF}}}_t$) of the overlap between $\rho_t$ and
 sub-sampled profiles for $t=60$. (b) Localization scores in the coarsest version of the ARA for $\rho_t$ (in blue), and 
 $\bar{\rho}_t$ (in red).}
\label{cdfPlot60}
\end{figure}
\begin{figure}
\includegraphics[width=1\textwidth,keepaspectratio]{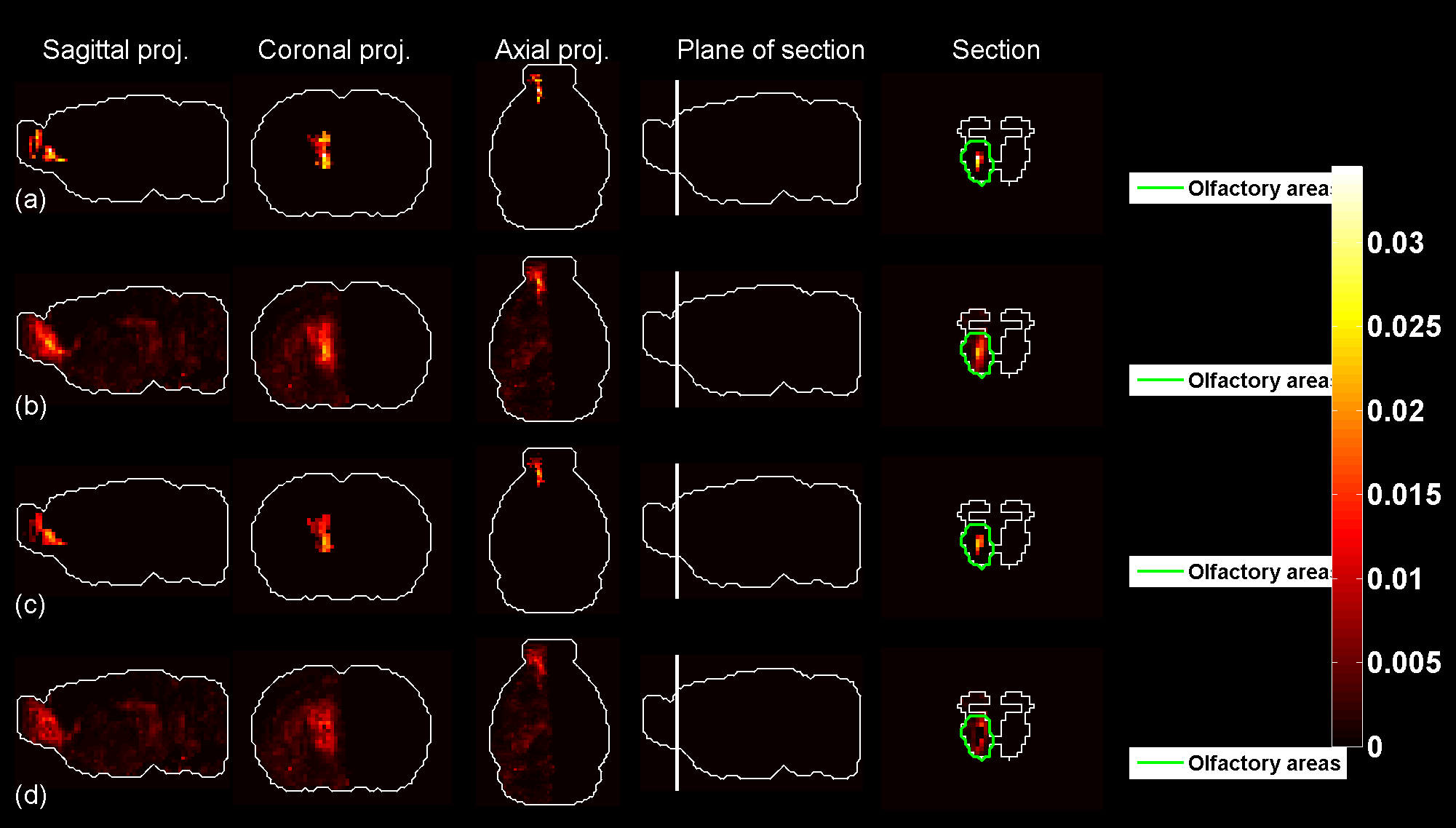}
\caption{Predicted profile and average sub-sampled profile for $t=60$.}
\label{subSampledFour60}
\end{figure}
\clearpage
\begin{figure}
\includegraphics[width=1\textwidth,keepaspectratio]{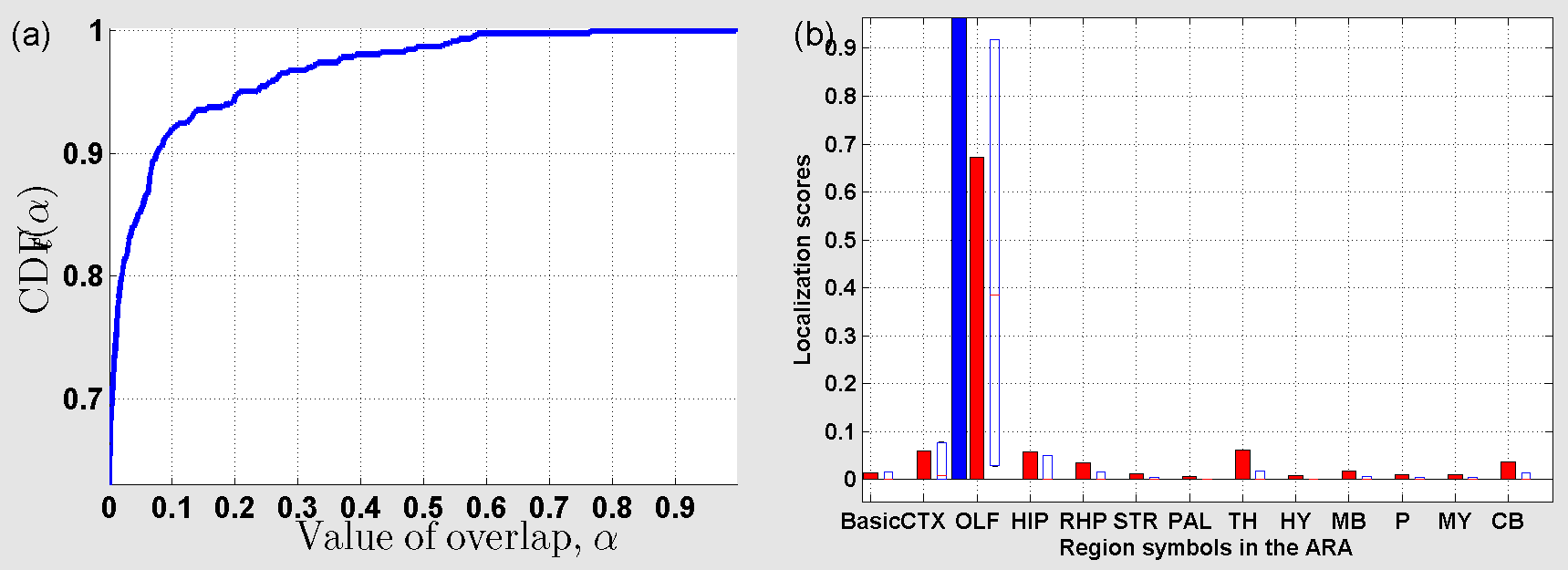}
\caption{(a) Cumulative distribution function (${\mathrm{\sc{CDF}}}_t$) of the overlap between $\rho_t$ and
 sub-sampled profiles for $t=61$. (b) Localization scores in the coarsest version of the ARA for $\rho_t$ (in blue), and 
 $\bar{\rho}_t$ (in red).}
\label{cdfPlot61}
\end{figure}
\begin{figure}
\includegraphics[width=1\textwidth,keepaspectratio]{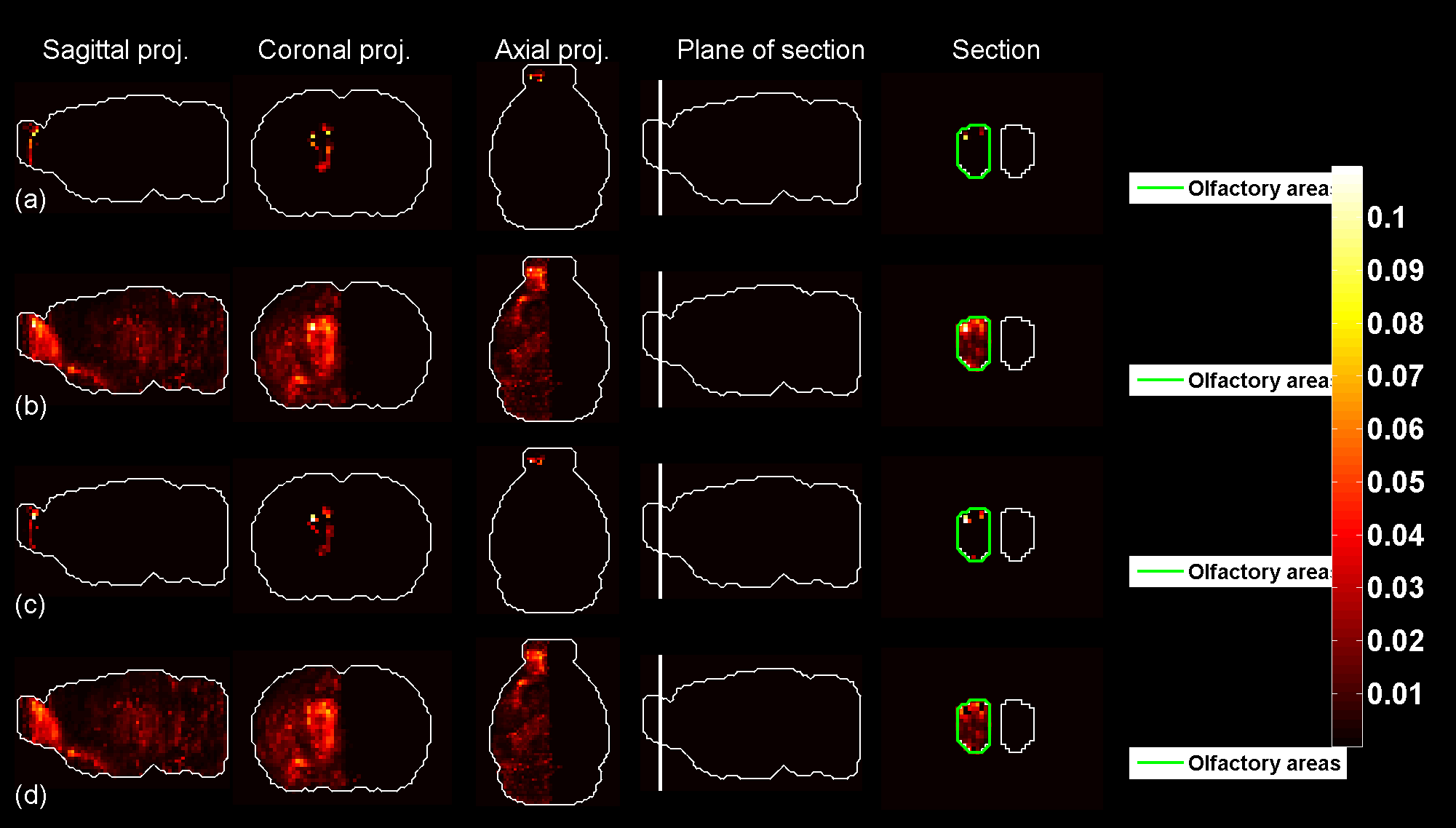}
\caption{Predicted profile and average sub-sampled profile for $t=61$.}
\label{subSampledFour61}
\end{figure}
\clearpage
\begin{figure}
\includegraphics[width=1\textwidth,keepaspectratio]{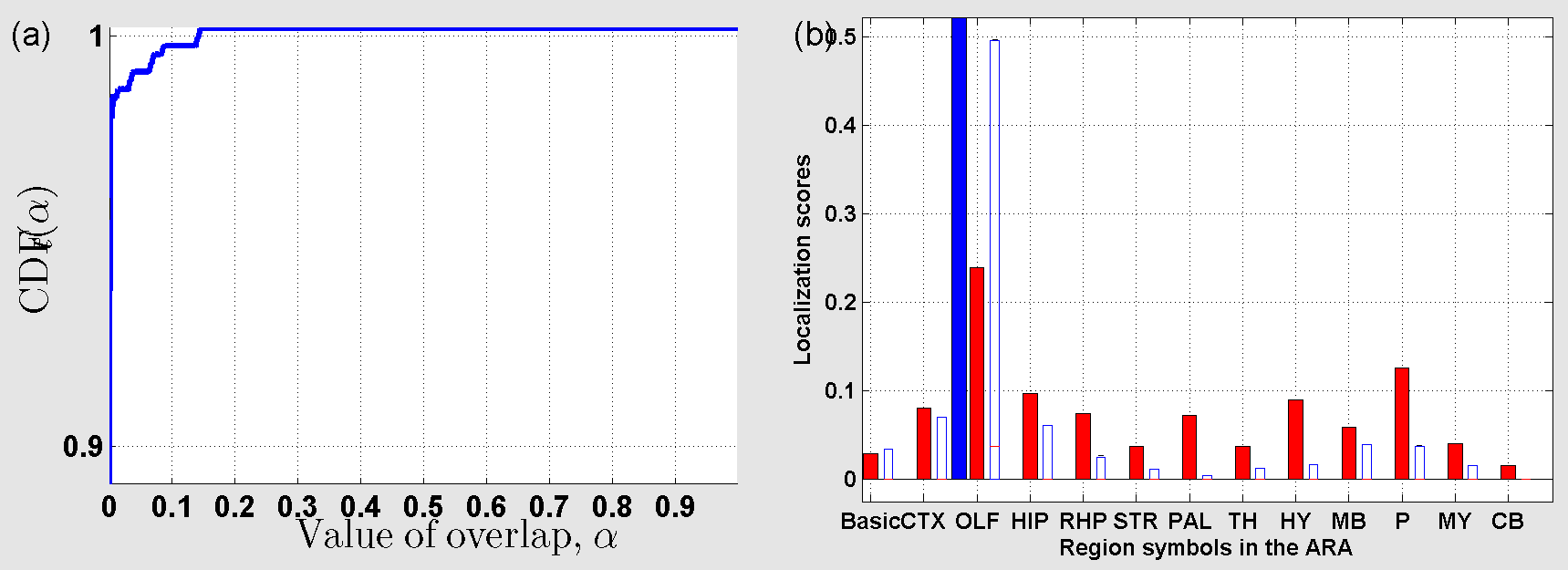}
\caption{(a) Cumulative distribution function (${\mathrm{\sc{CDF}}}_t$) of the overlap between $\rho_t$ and
 sub-sampled profiles for $t=62$. (b) Localization scores in the coarsest version of the ARA for $\rho_t$ (in blue), and 
 $\bar{\rho}_t$ (in red).}
\label{cdfPlot62}
\end{figure}
\begin{figure}
\includegraphics[width=1\textwidth,keepaspectratio]{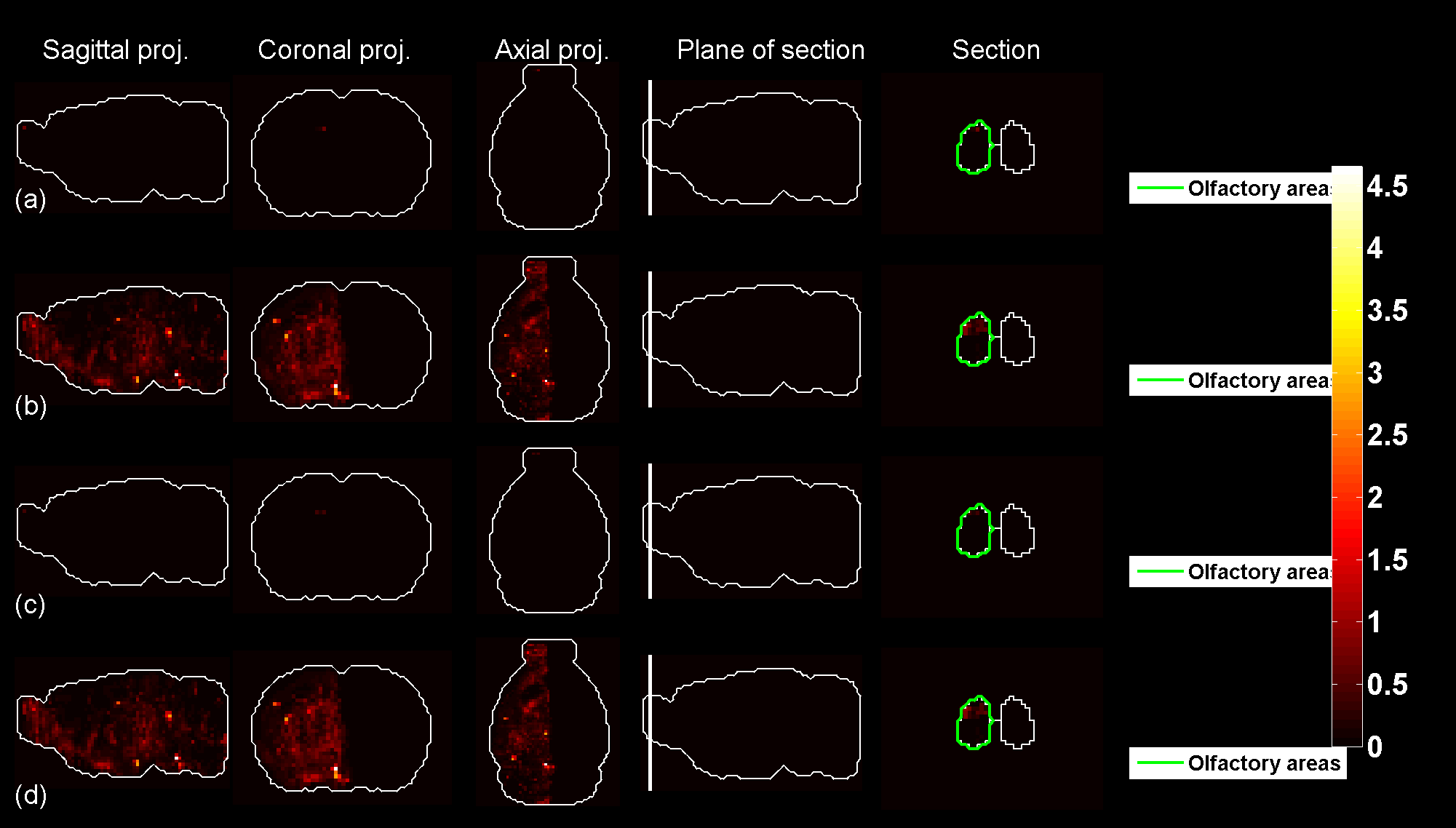}
\caption{Predicted profile and average sub-sampled profile for $t=62$.}
\label{subSampledFour62}
\end{figure}
\clearpage
\begin{figure}
\includegraphics[width=1\textwidth,keepaspectratio]{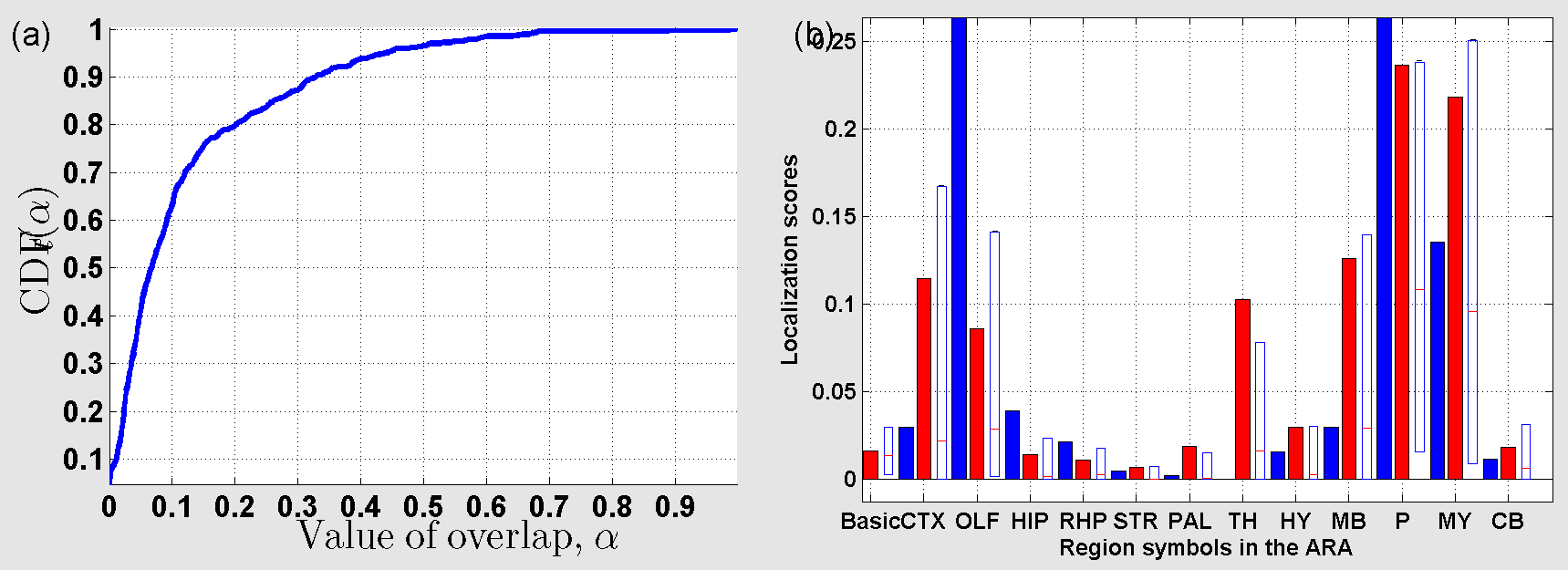}
\caption{(a) Cumulative distribution function (${\mathrm{\sc{CDF}}}_t$) of the overlap between $\rho_t$ and
 sub-sampled profiles for $t=63$. (b) Localization scores in the coarsest version of the ARA for $\rho_t$ (in blue), and 
 $\bar{\rho}_t$ (in red).}
\label{cdfPlot63}
\end{figure}
\begin{figure}
\includegraphics[width=1\textwidth,keepaspectratio]{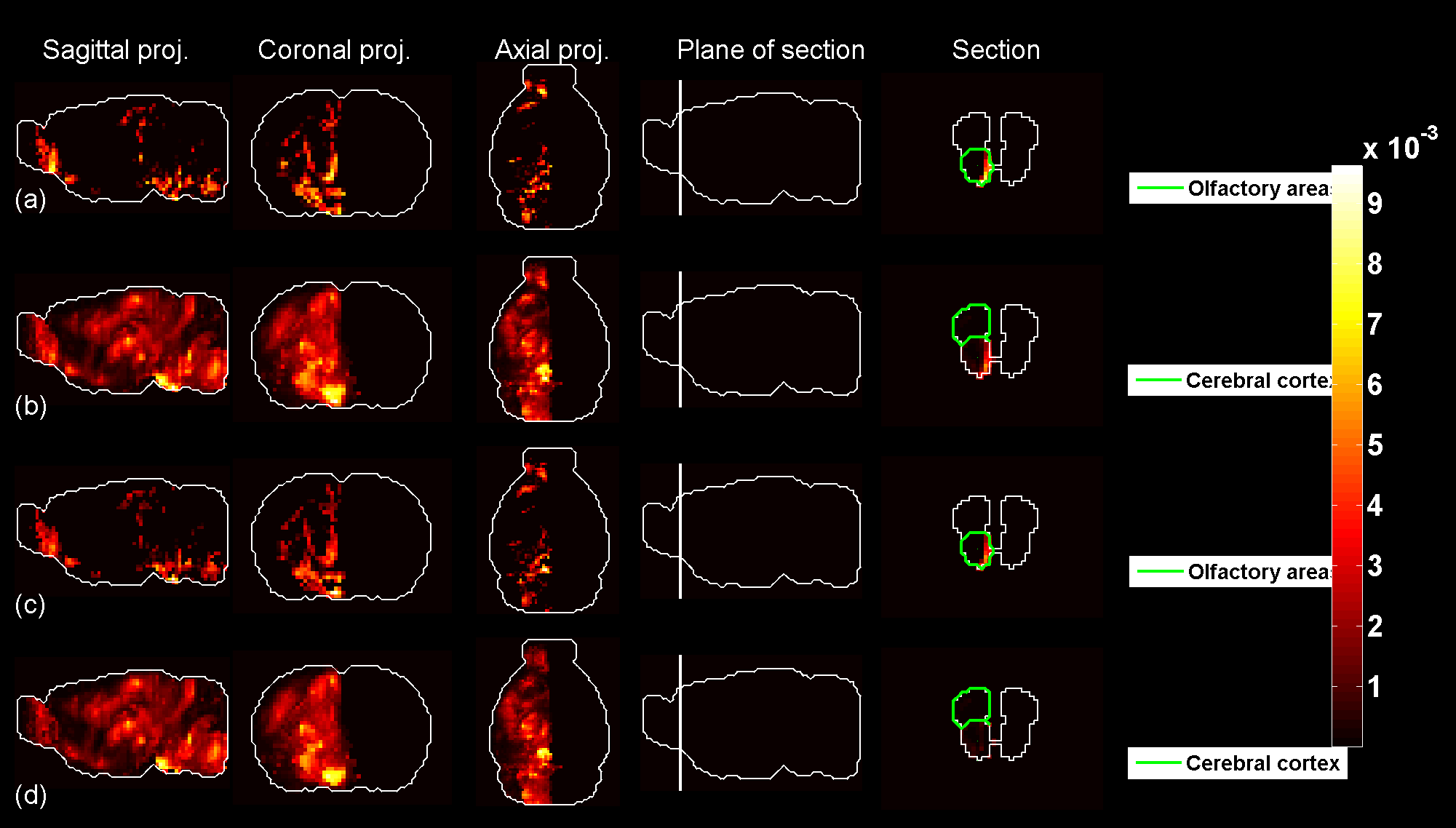}
\caption{Predicted profile and average sub-sampled profile for $t=63$.}
\label{subSampledFour63}
\end{figure}
\clearpage
\begin{figure}
\includegraphics[width=1\textwidth,keepaspectratio]{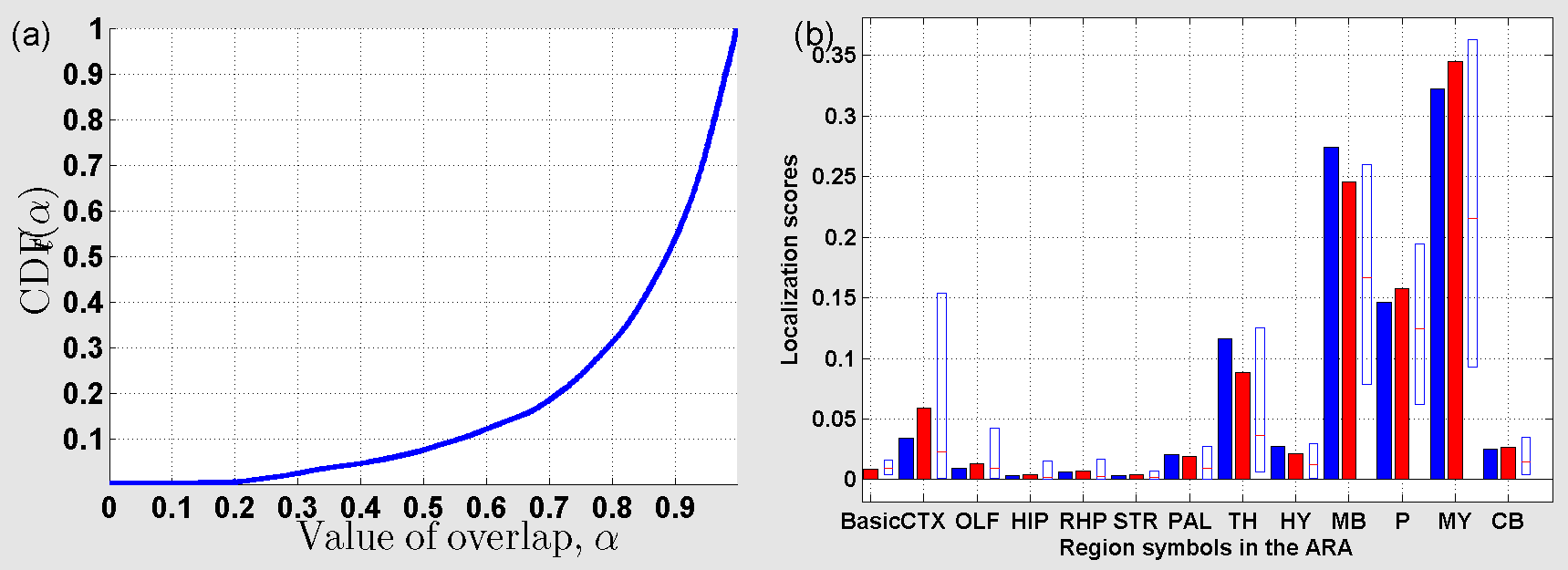}
\caption{(a) Cumulative distribution function (${\mathrm{\sc{CDF}}}_t$) of the overlap between $\rho_t$ and
 sub-sampled profiles for $t=64$. (b) Localization scores in the coarsest version of the ARA for $\rho_t$ (in blue), and 
 $\bar{\rho}_t$ (in red).}
\label{cdfPlot64}
\end{figure}
\begin{figure}
\includegraphics[width=1\textwidth,keepaspectratio]{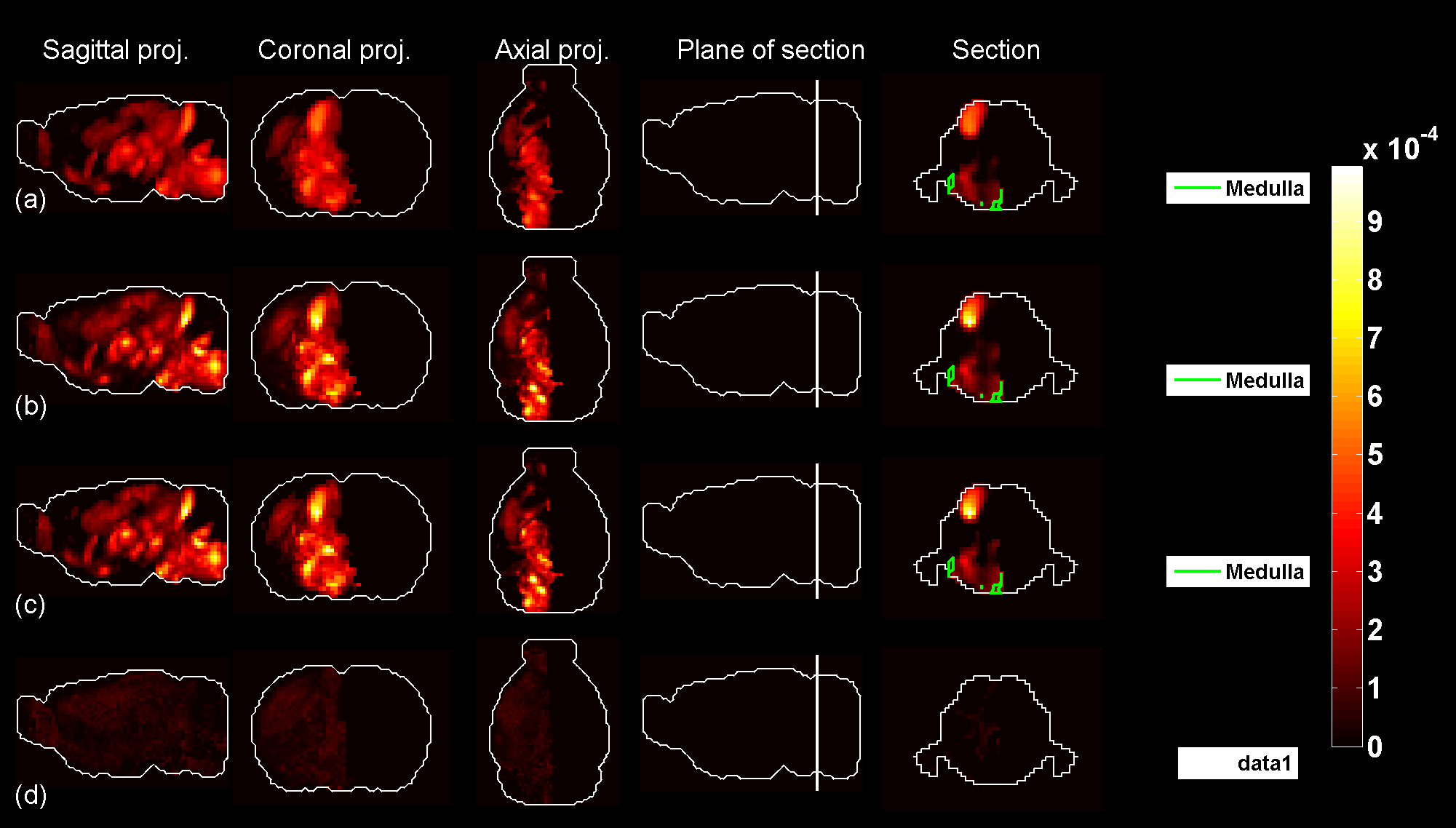}
\caption{Predicted profile and average sub-sampled profile for $t=64$.}
\label{subSampledFour64}
\end{figure}

%% file: appendSplittingFigures.tex
\clearpage
\begin{figure}
\includegraphics[width=1\textwidth,keepaspectratio]{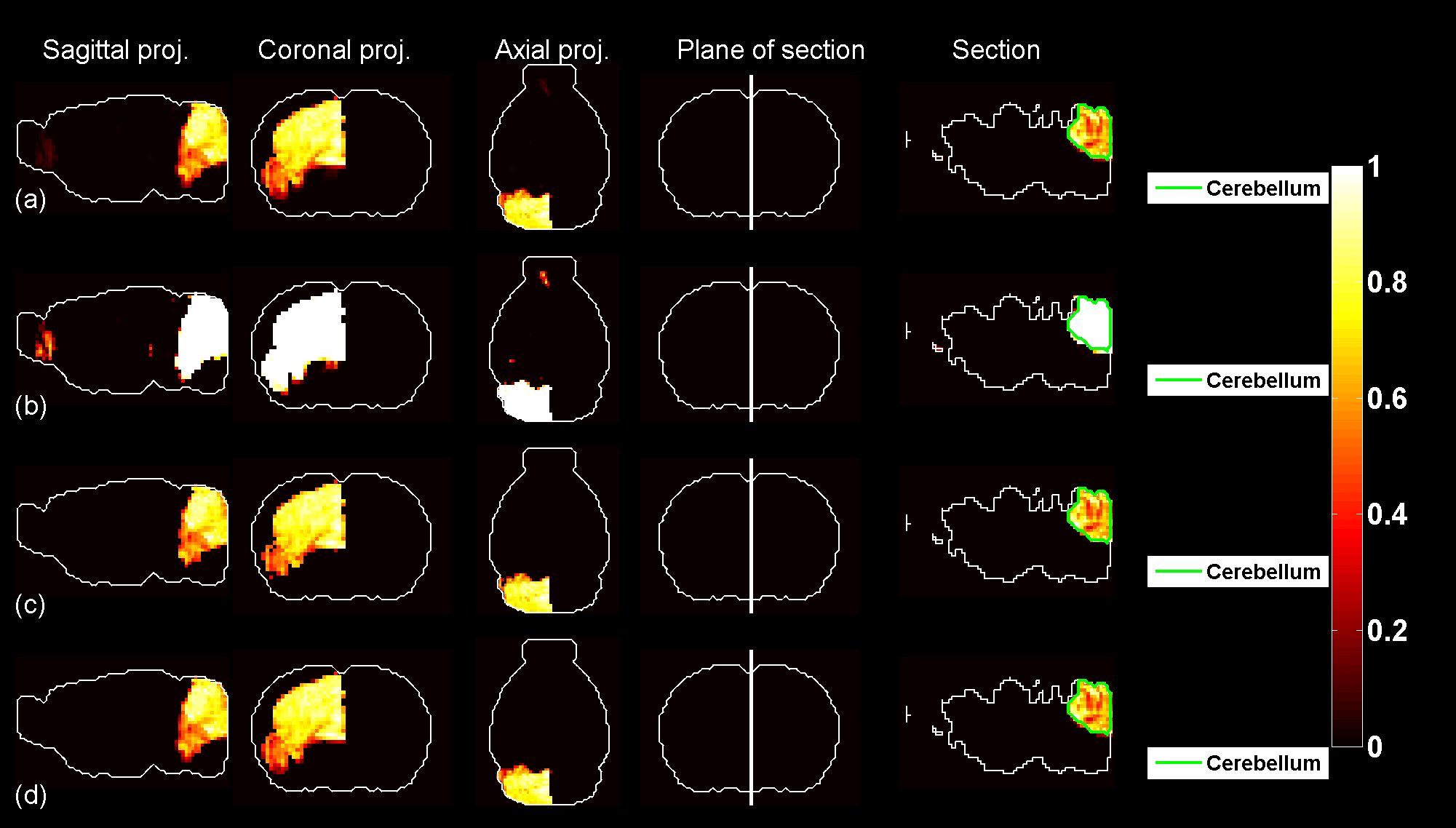}
\caption{Predicted profile, probability profile and thresholded profiles for $t=1$.}
\label{subSampleSplit1}
\end{figure}
\clearpage
\begin{figure}
\includegraphics[width=1\textwidth,keepaspectratio]{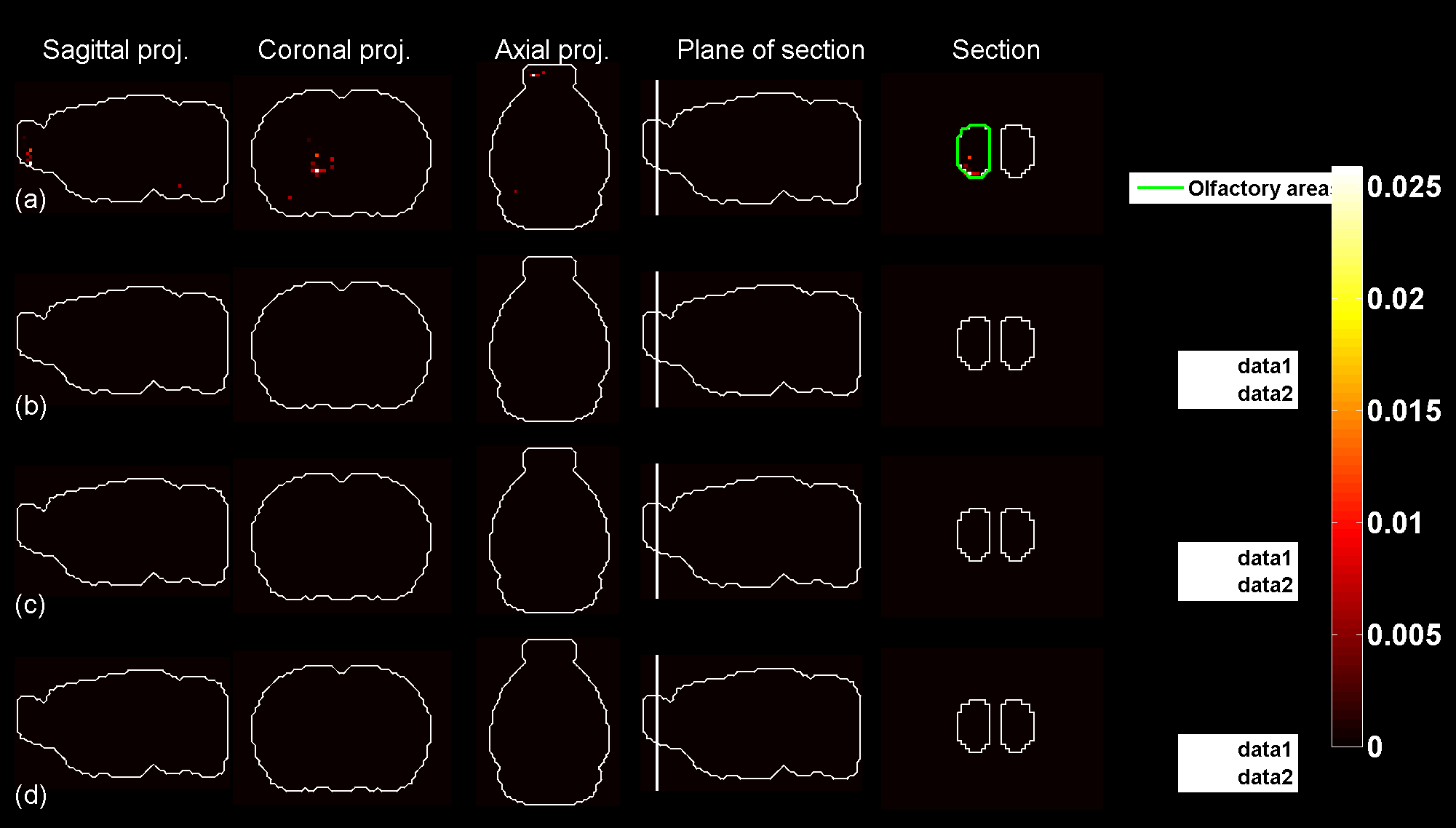}
\caption{Predicted profile, probability profile and thresholded profiles for $t=2$.}
\label{subSampleSplit2}
\end{figure}
\clearpage
\begin{figure}
\includegraphics[width=1\textwidth,keepaspectratio]{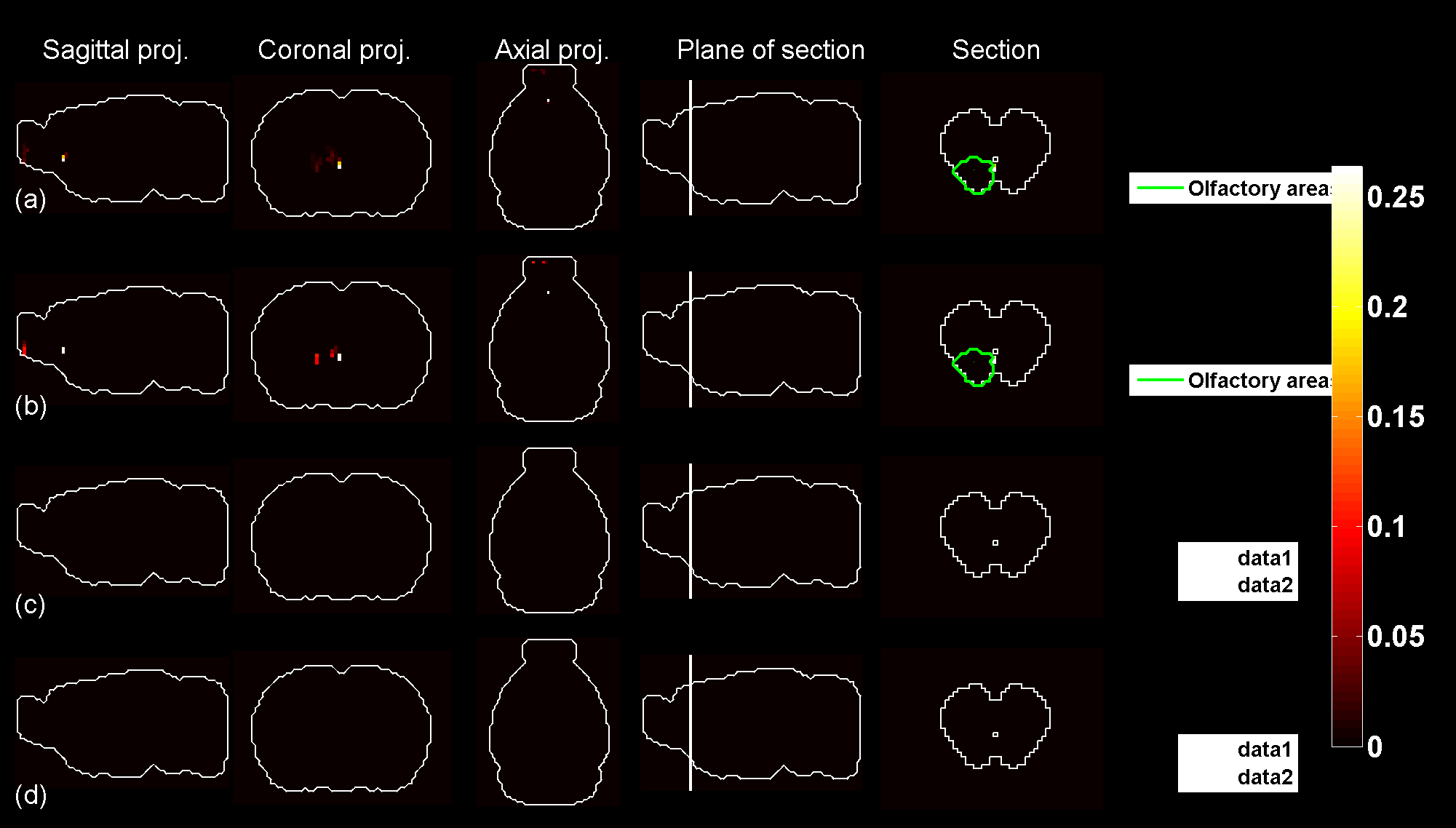}
\caption{Predicted profile, probability profile and thresholded profiles for $t=3$.}
\label{subSampleSplit3}
\end{figure}
\clearpage
\begin{figure}
\includegraphics[width=1\textwidth,keepaspectratio]{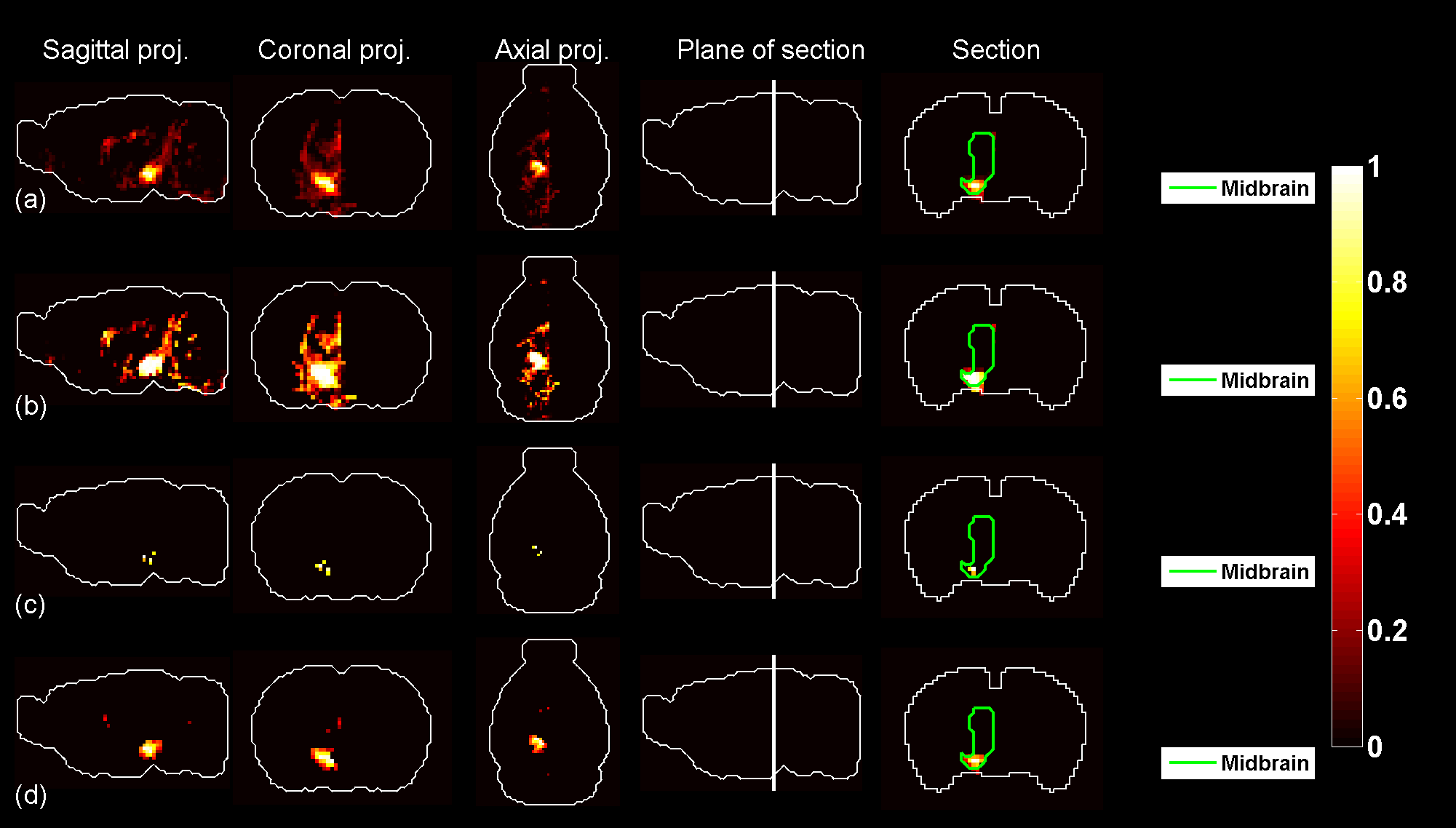}
\caption{Predicted profile, probability profile and thresholded profiles for $t=4$.}
\label{subSampleSplit4}
\end{figure}
\clearpage
\begin{figure}
\includegraphics[width=1\textwidth,keepaspectratio]{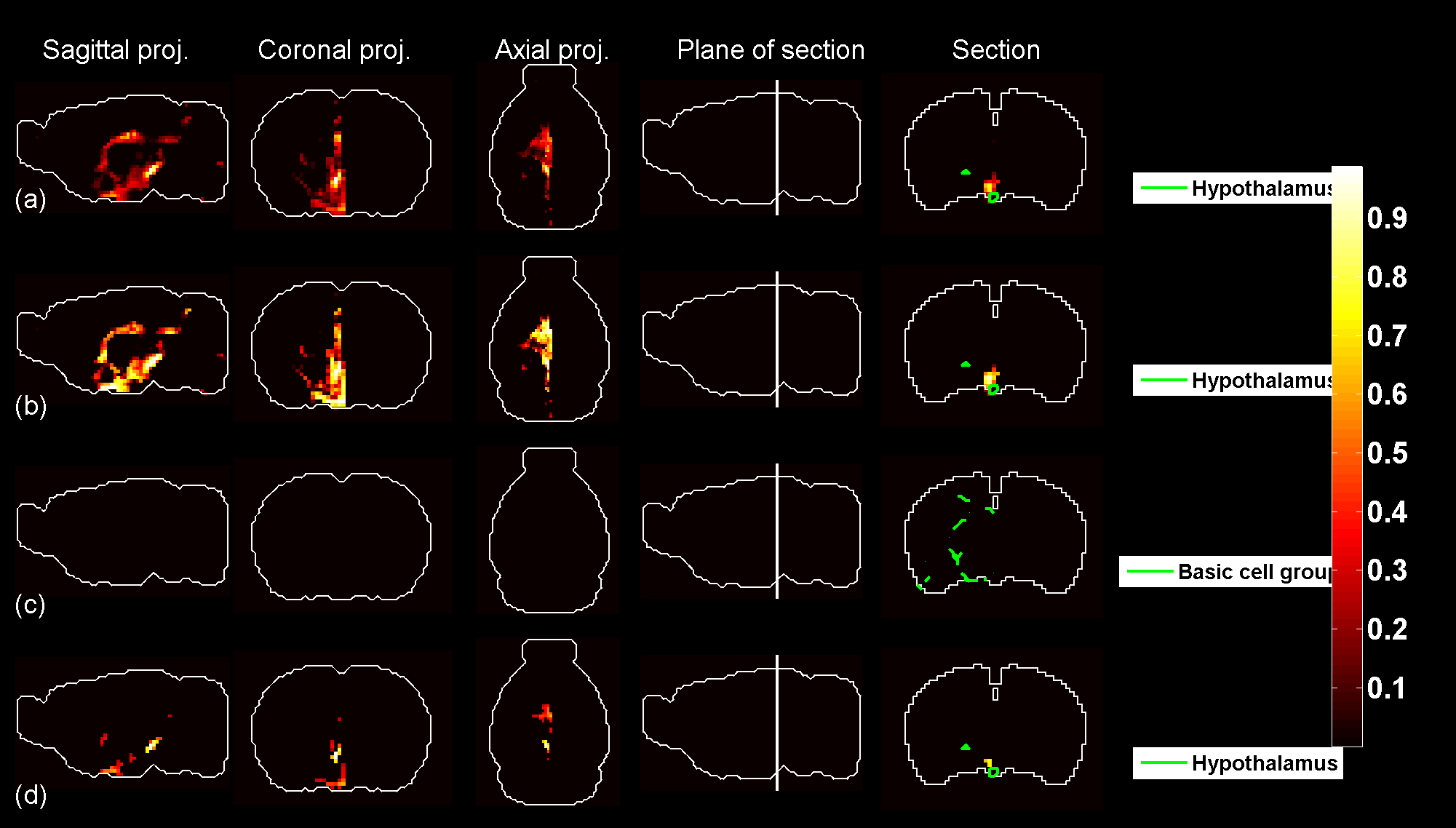}
\caption{Predicted profile, probability profile and thresholded profiles for $t=5$.}
\label{subSampleSplit5}
\end{figure}
\clearpage
\begin{figure}
\includegraphics[width=1\textwidth,keepaspectratio]{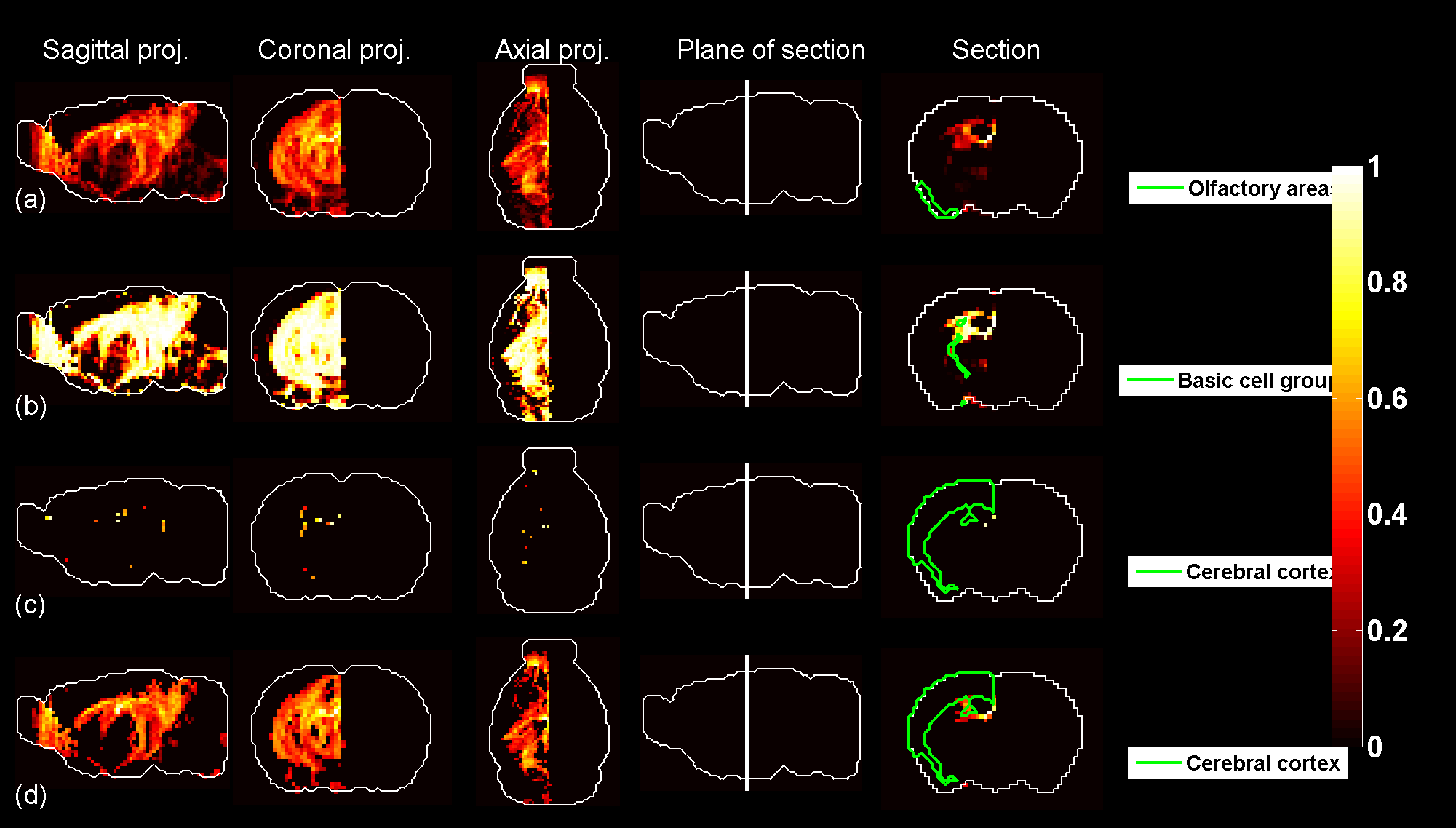}
\caption{Predicted profile, probability profile and thresholded profiles for $t=6$.}
\label{subSampleSplit6}
\end{figure}
\clearpage
\begin{figure}
\includegraphics[width=1\textwidth,keepaspectratio]{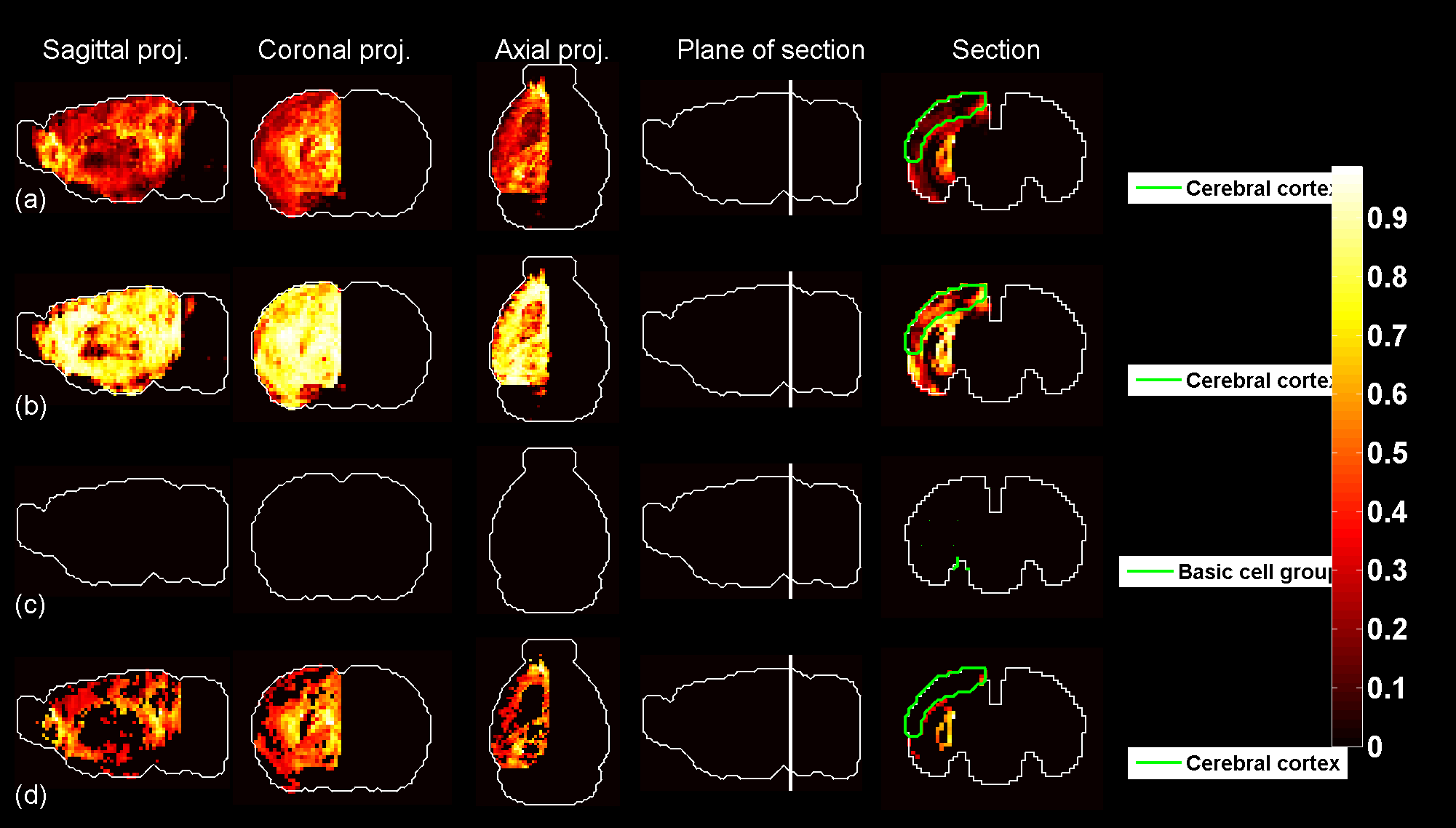}
\caption{Predicted profile, probability profile and thresholded profiles for $t=7$.}
\label{subSampleSplit7}
\end{figure}
\clearpage
\begin{figure}
\includegraphics[width=1\textwidth,keepaspectratio]{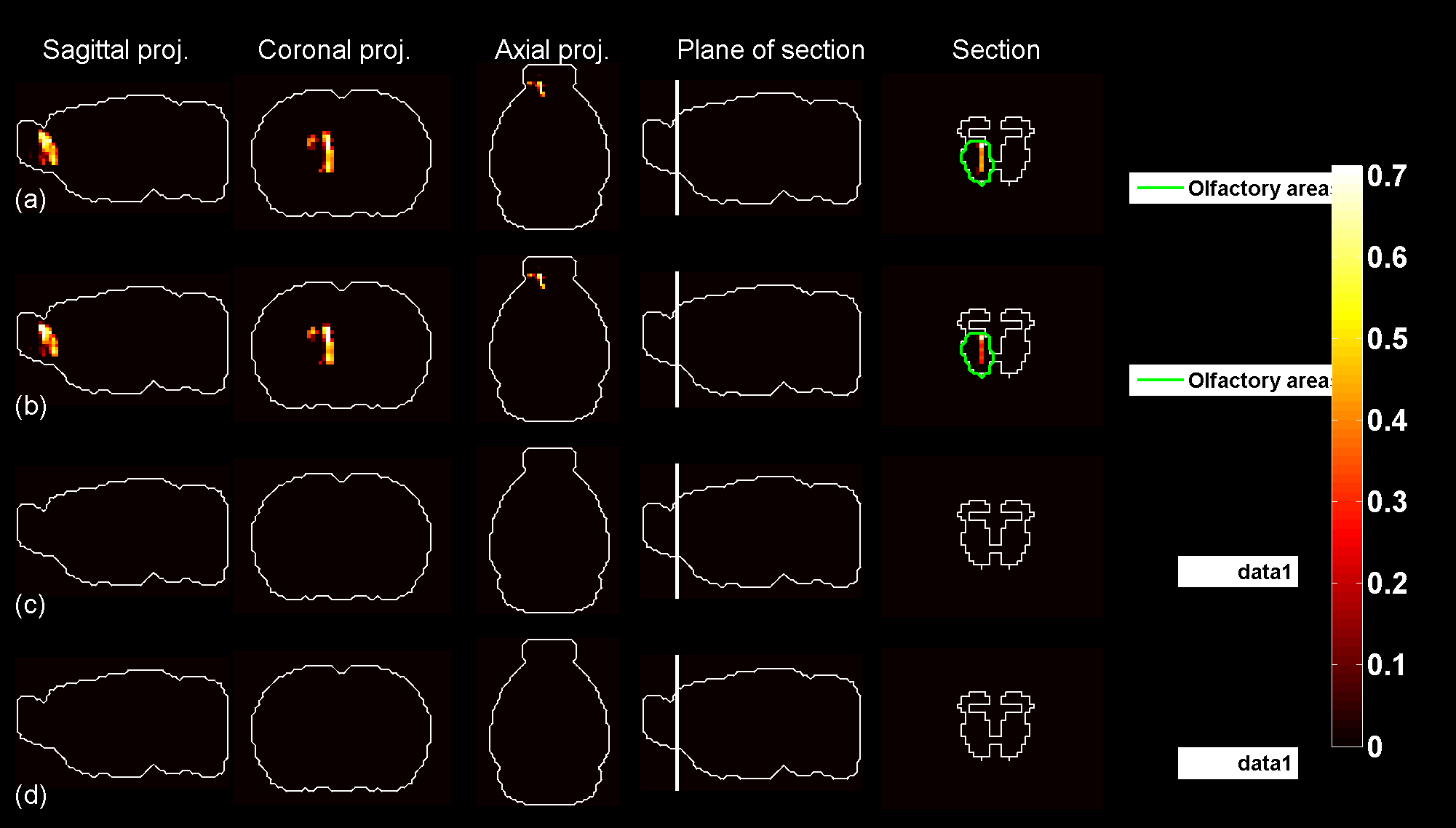}
\caption{Predicted profile, probability profile and thresholded profiles for $t=8$.}
\label{subSampleSplit8}
\end{figure}
\clearpage
\begin{figure}
\includegraphics[width=1\textwidth,keepaspectratio]{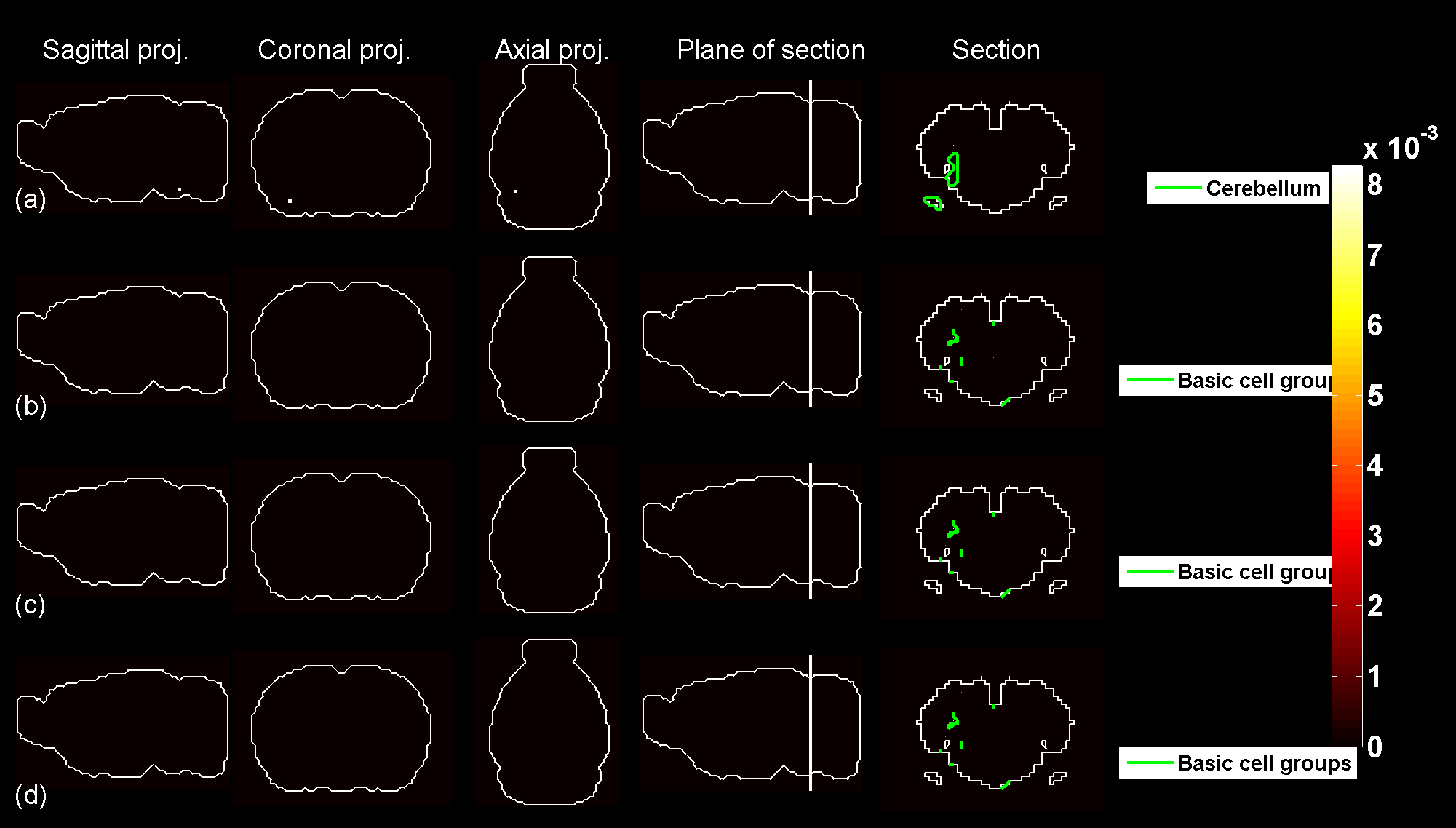}
\caption{Predicted profile, probability profile and thresholded profiles for $t=9$.}
\label{subSampleSplit9}
\end{figure}
\clearpage
\begin{figure}
\includegraphics[width=1\textwidth,keepaspectratio]{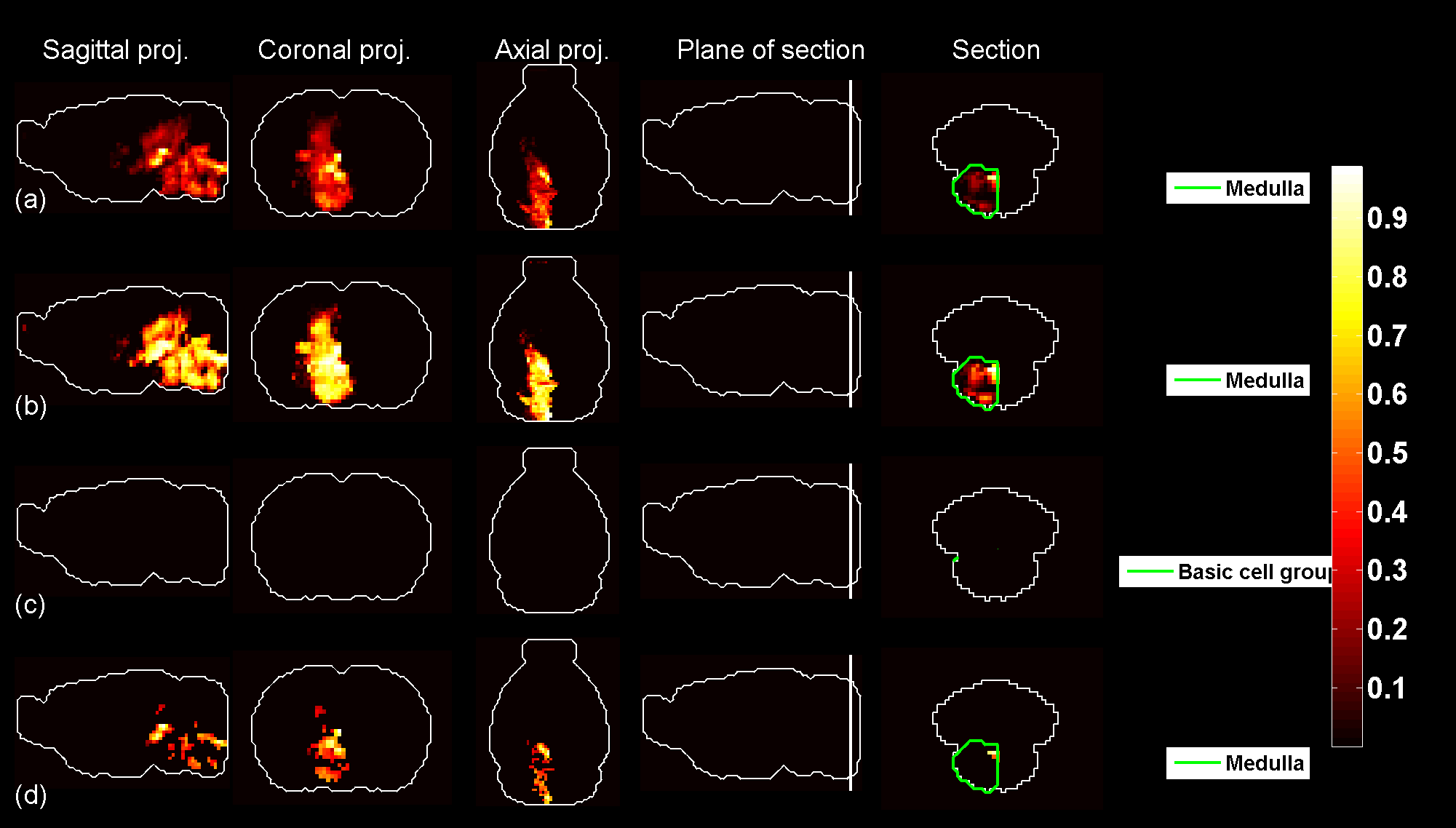}
\caption{Predicted profile, probability profile and thresholded profiles for $t=10$.}
\label{subSampleSplit10}
\end{figure}
\clearpage
\begin{figure}
\includegraphics[width=1\textwidth,keepaspectratio]{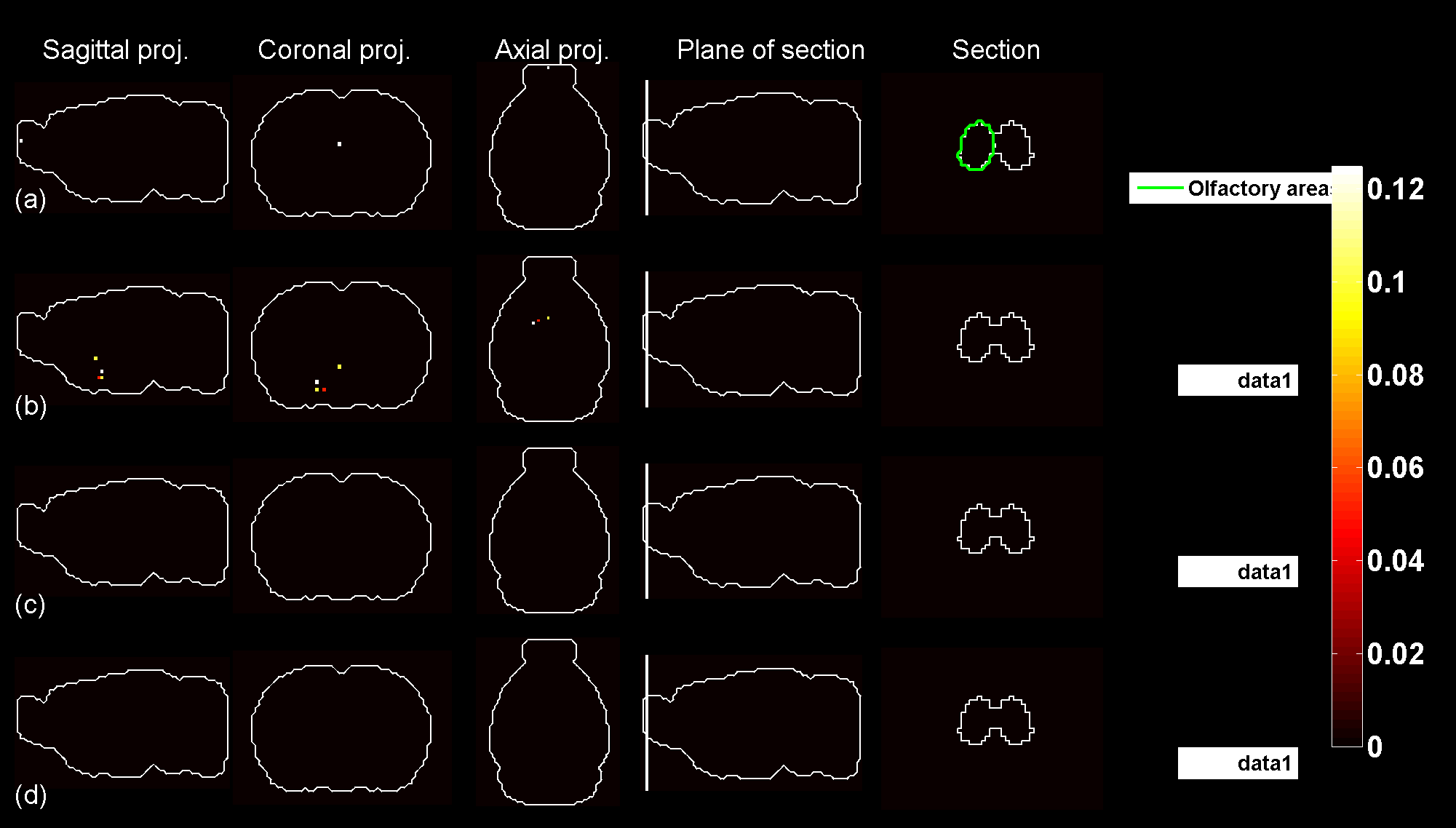}
\caption{Predicted profile, probability profile and thresholded profiles for $t=11$.}
\label{subSampleSplit11}
\end{figure}
\clearpage
\begin{figure}
\includegraphics[width=1\textwidth,keepaspectratio]{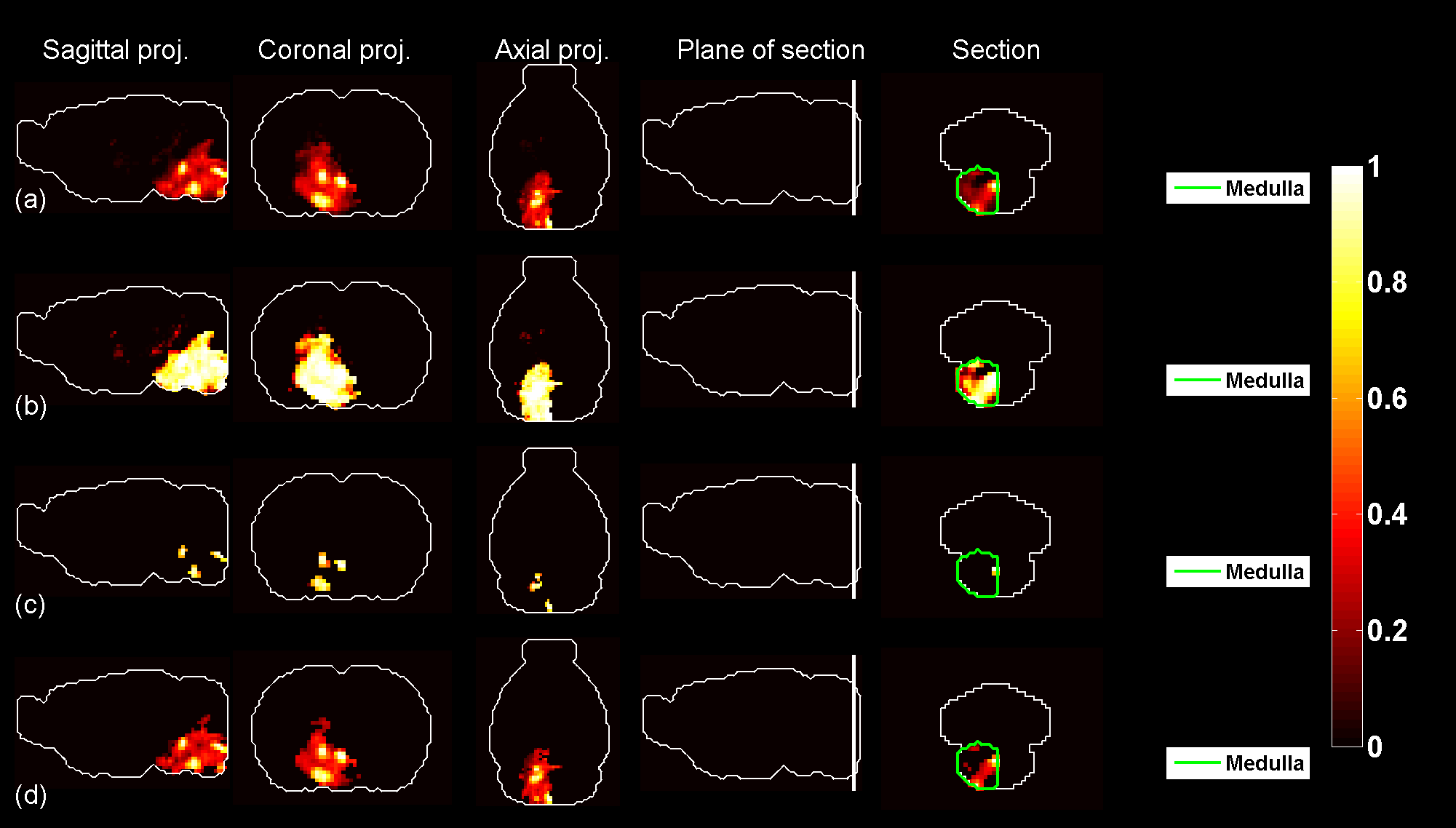}
\caption{Predicted profile, probability profile and thresholded profiles for $t=12$.}
\label{subSampleSplit12}
\end{figure}
\clearpage
\begin{figure}
\includegraphics[width=1\textwidth,keepaspectratio]{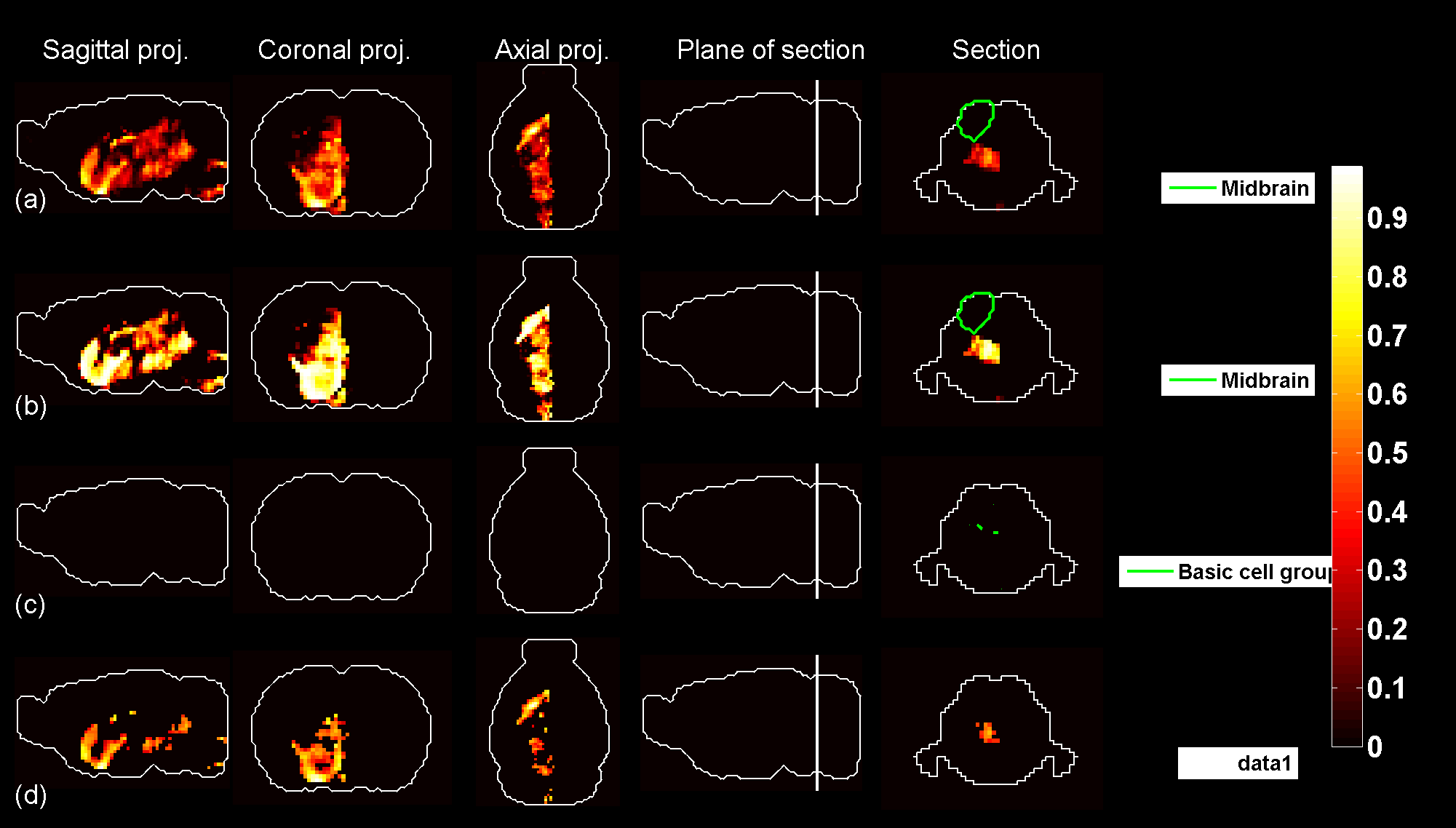}
\caption{Predicted profile, probability profile and thresholded profiles for $t=13$.}
\label{subSampleSplit13}
\end{figure}
\clearpage
\begin{figure}
\includegraphics[width=1\textwidth,keepaspectratio]{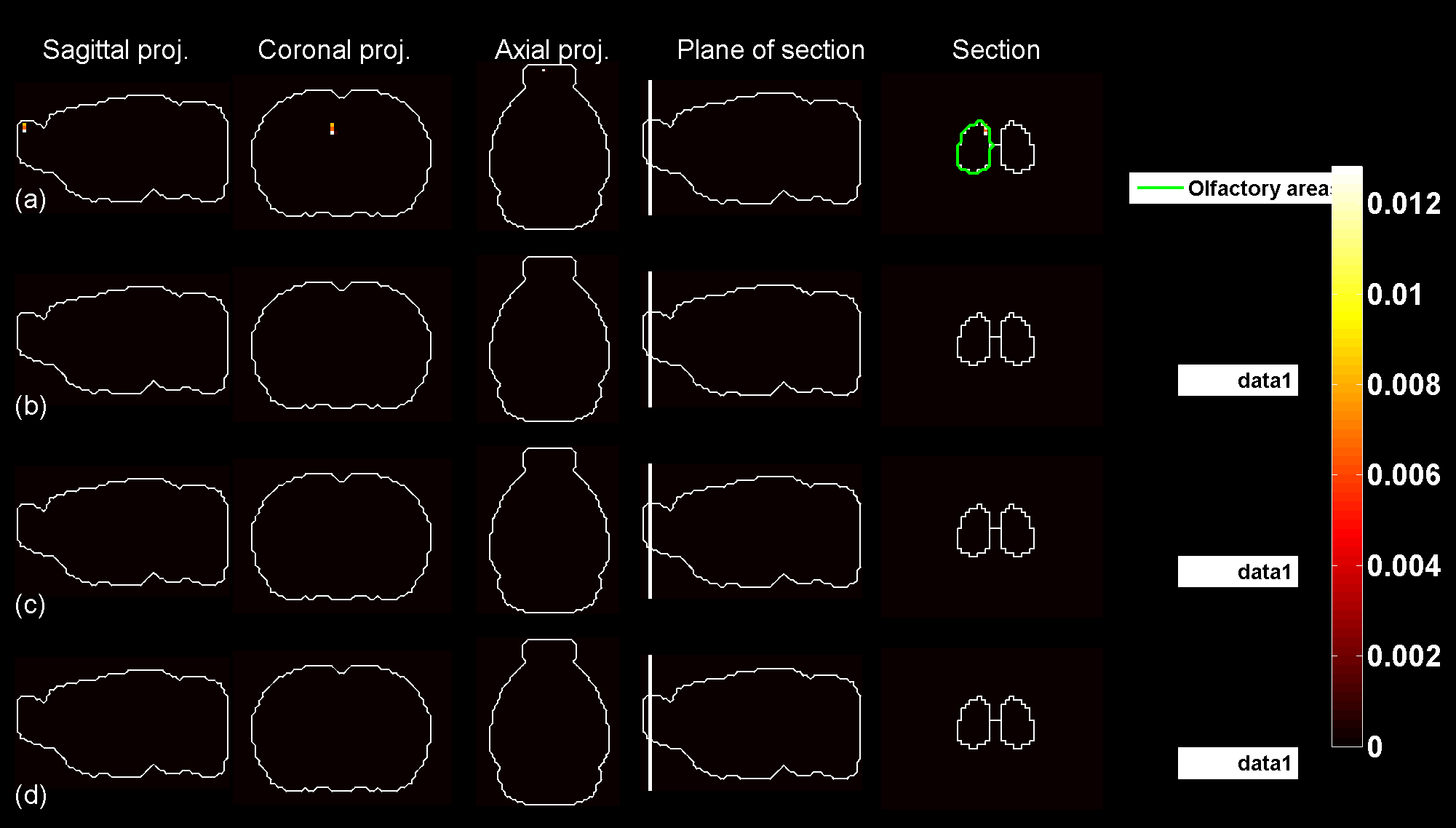}
\caption{Predicted profile, probability profile and thresholded profiles for $t=14$.}
\label{subSampleSplit14}
\end{figure}
\clearpage
\begin{figure}
\includegraphics[width=1\textwidth,keepaspectratio]{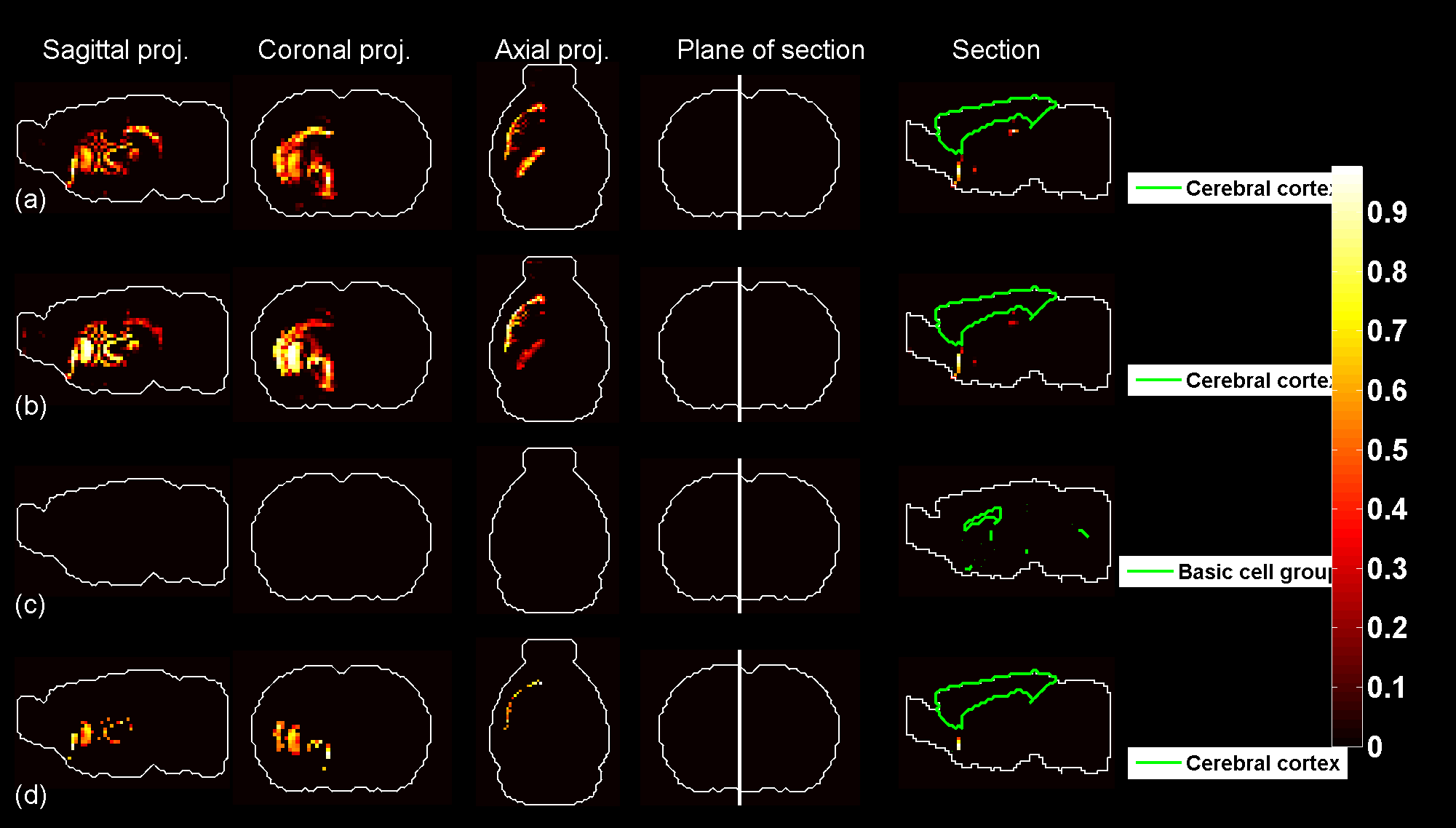}
\caption{Predicted profile, probability profile and thresholded profiles for $t=15$.}
\label{subSampleSplit15}
\end{figure}
\clearpage
\begin{figure}
\includegraphics[width=1\textwidth,keepaspectratio]{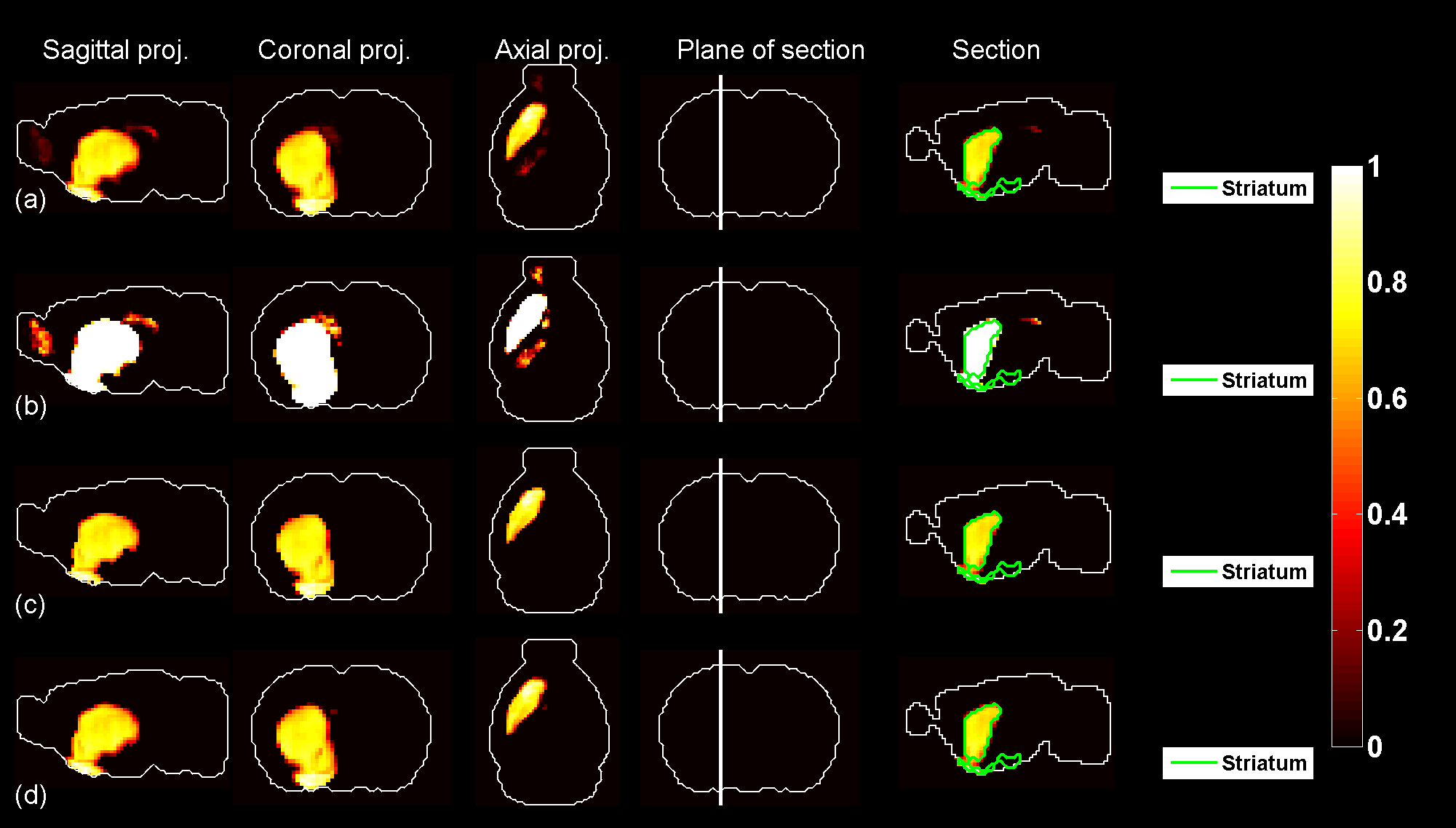}
\caption{Predicted profile, probability profile and thresholded profiles for $t=16$.}
\label{subSampleSplit16}
\end{figure}
\clearpage
\begin{figure}
\includegraphics[width=1\textwidth,keepaspectratio]{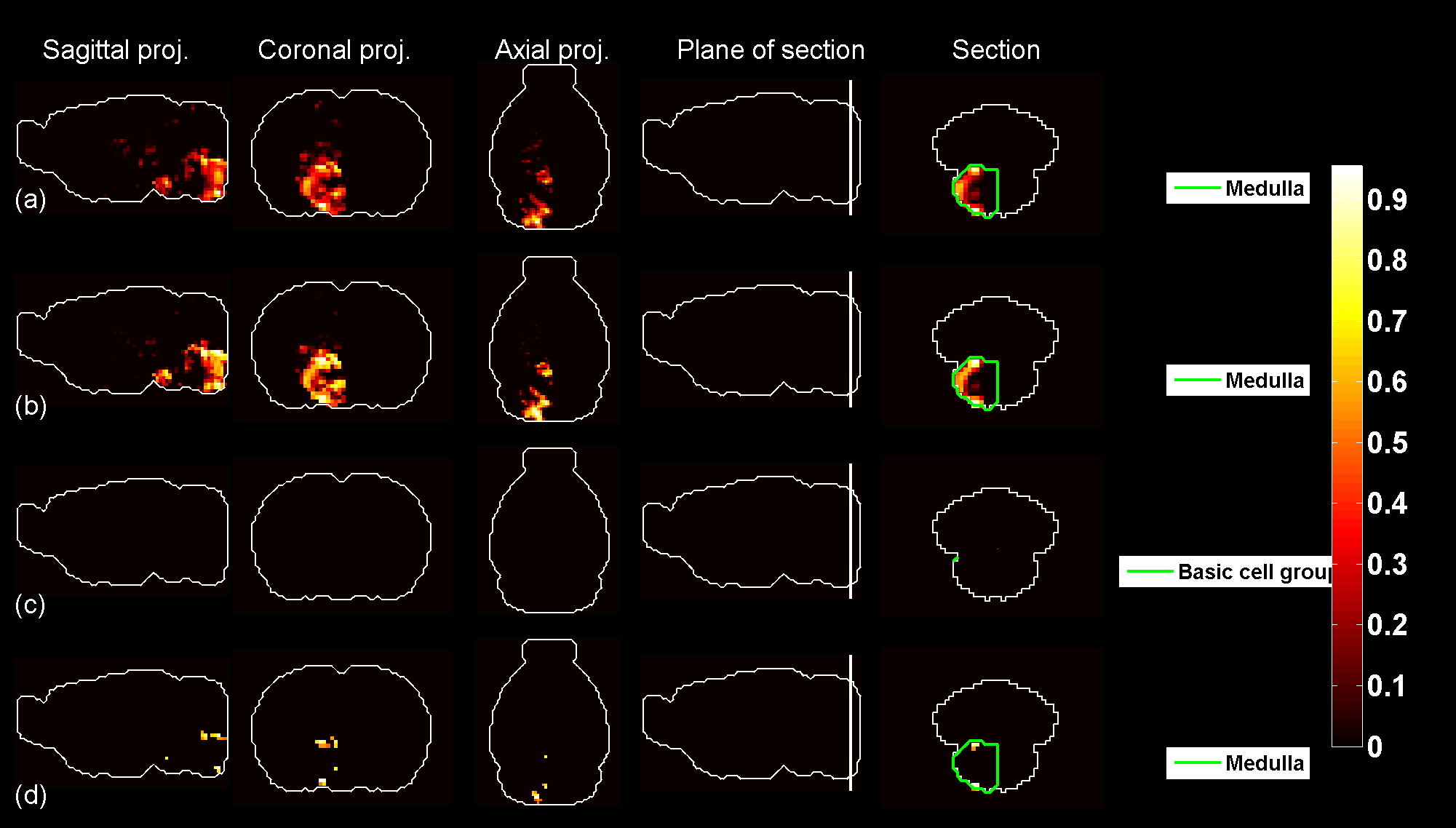}
\caption{Predicted profile, probability profile and thresholded profiles for $t=17$.}
\label{subSampleSplit17}
\end{figure}
\clearpage
\begin{figure}
\includegraphics[width=1\textwidth,keepaspectratio]{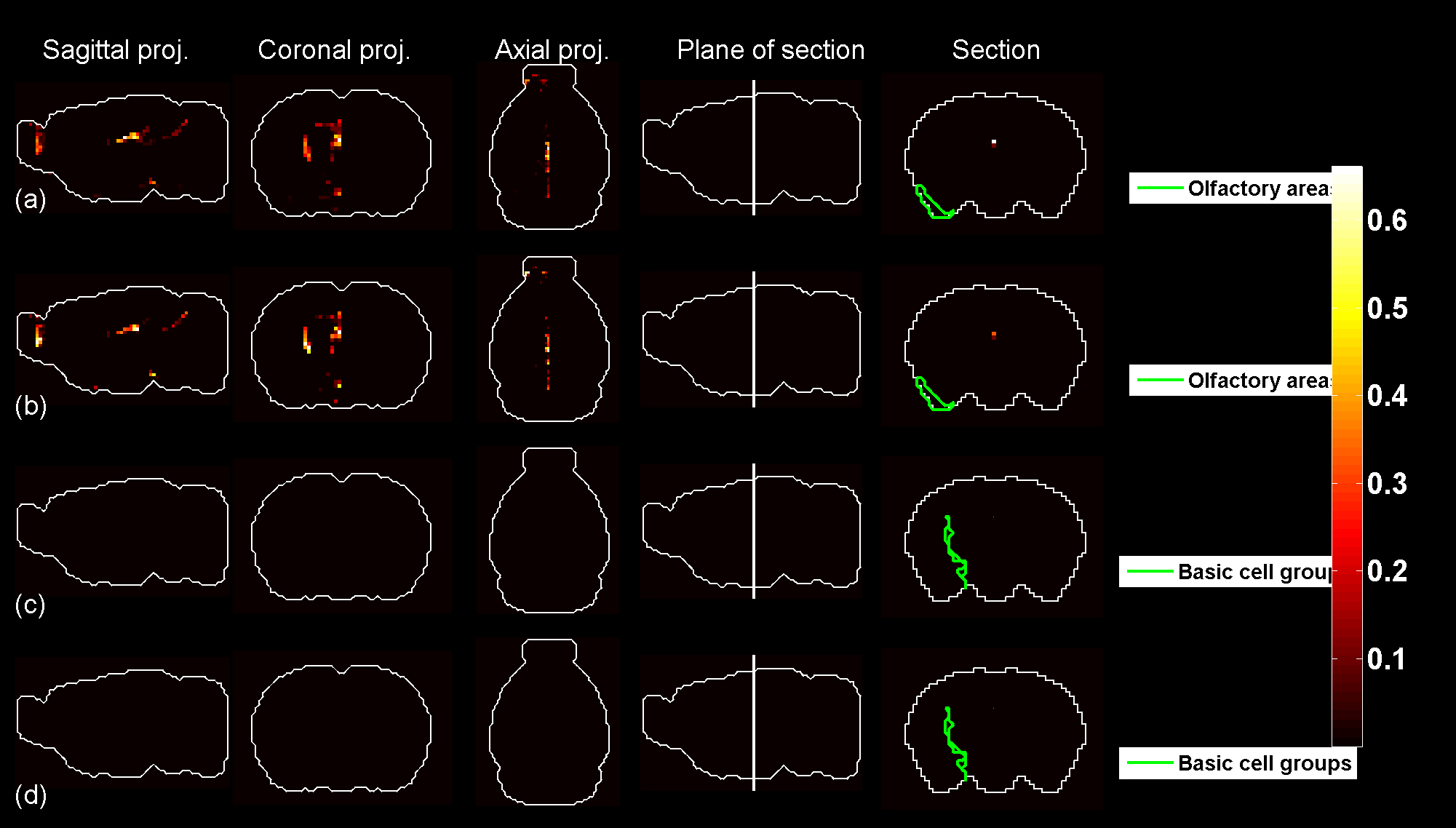}
\caption{Predicted profile, probability profile and thresholded profiles for $t=18$.}
\label{subSampleSplit18}
\end{figure}
\clearpage
\begin{figure}
\includegraphics[width=1\textwidth,keepaspectratio]{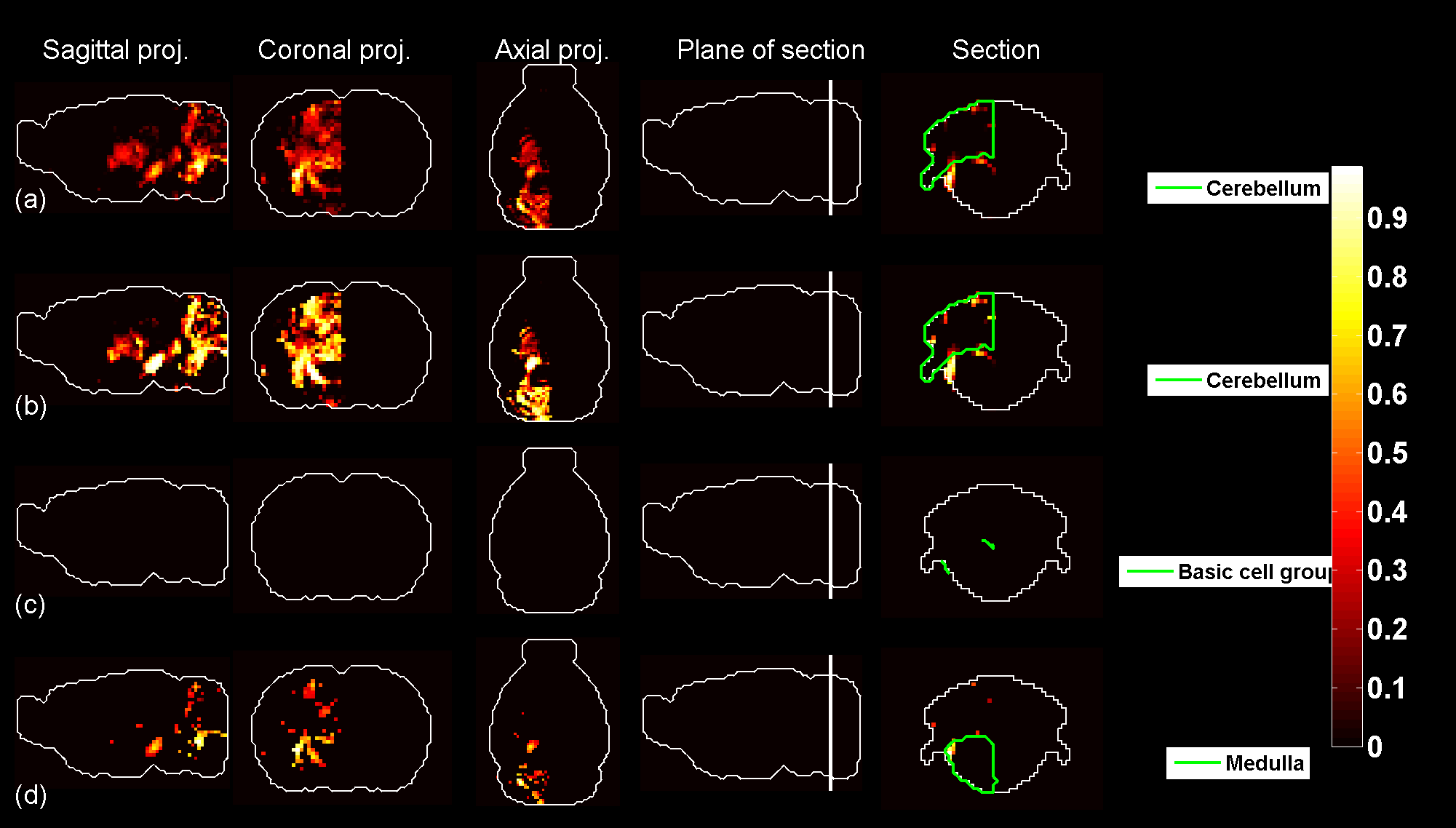}
\caption{Predicted profile, probability profile and thresholded profiles for $t=19$.}
\label{subSampleSplit19}
\end{figure}
\clearpage
\begin{figure}
\includegraphics[width=1\textwidth,keepaspectratio]{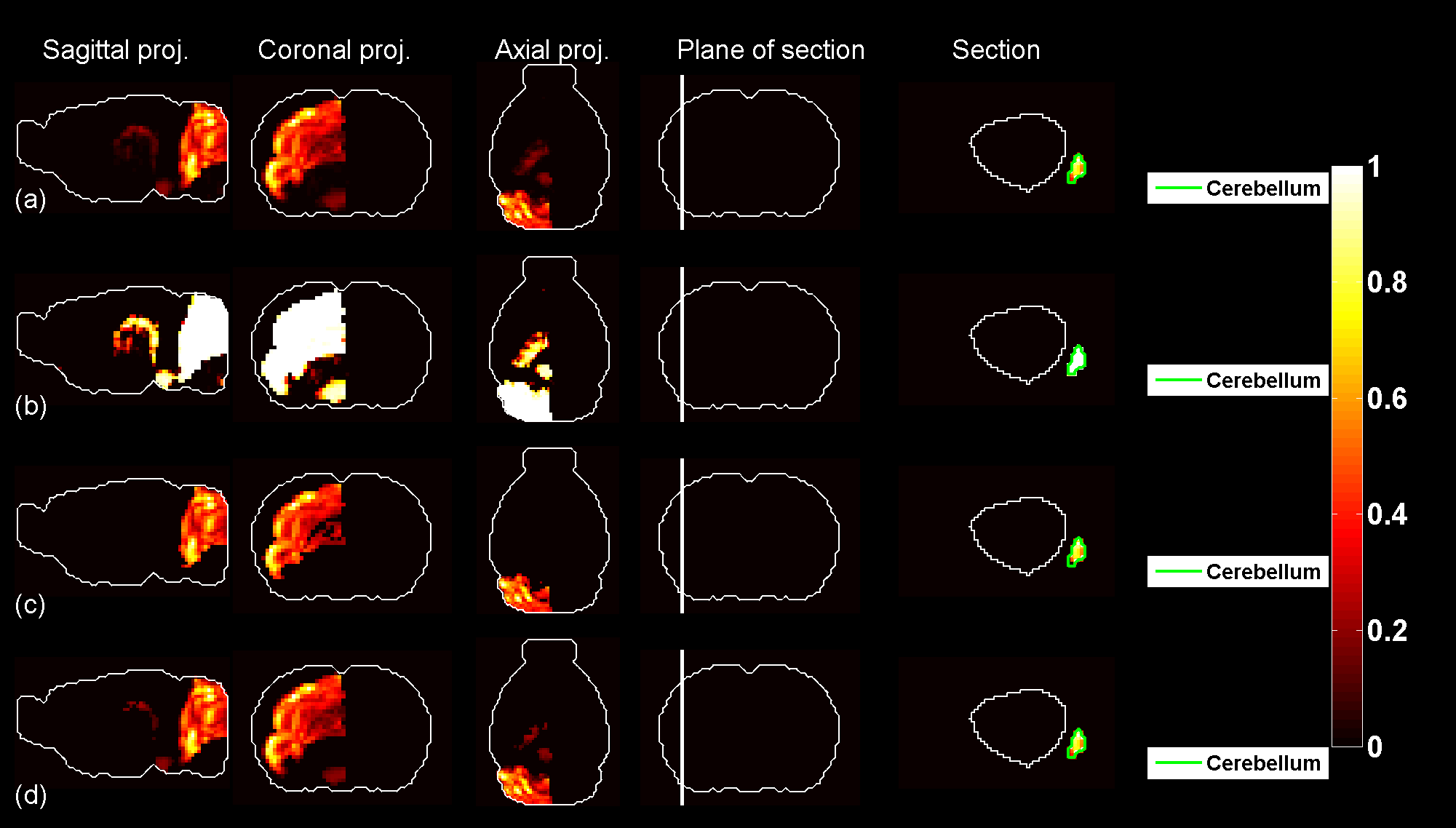}
\caption{Predicted profile, probability profile and thresholded profiles for $t=20$.}
\label{subSampleSplit20}
\end{figure}
\clearpage
\begin{figure}
\includegraphics[width=1\textwidth,keepaspectratio]{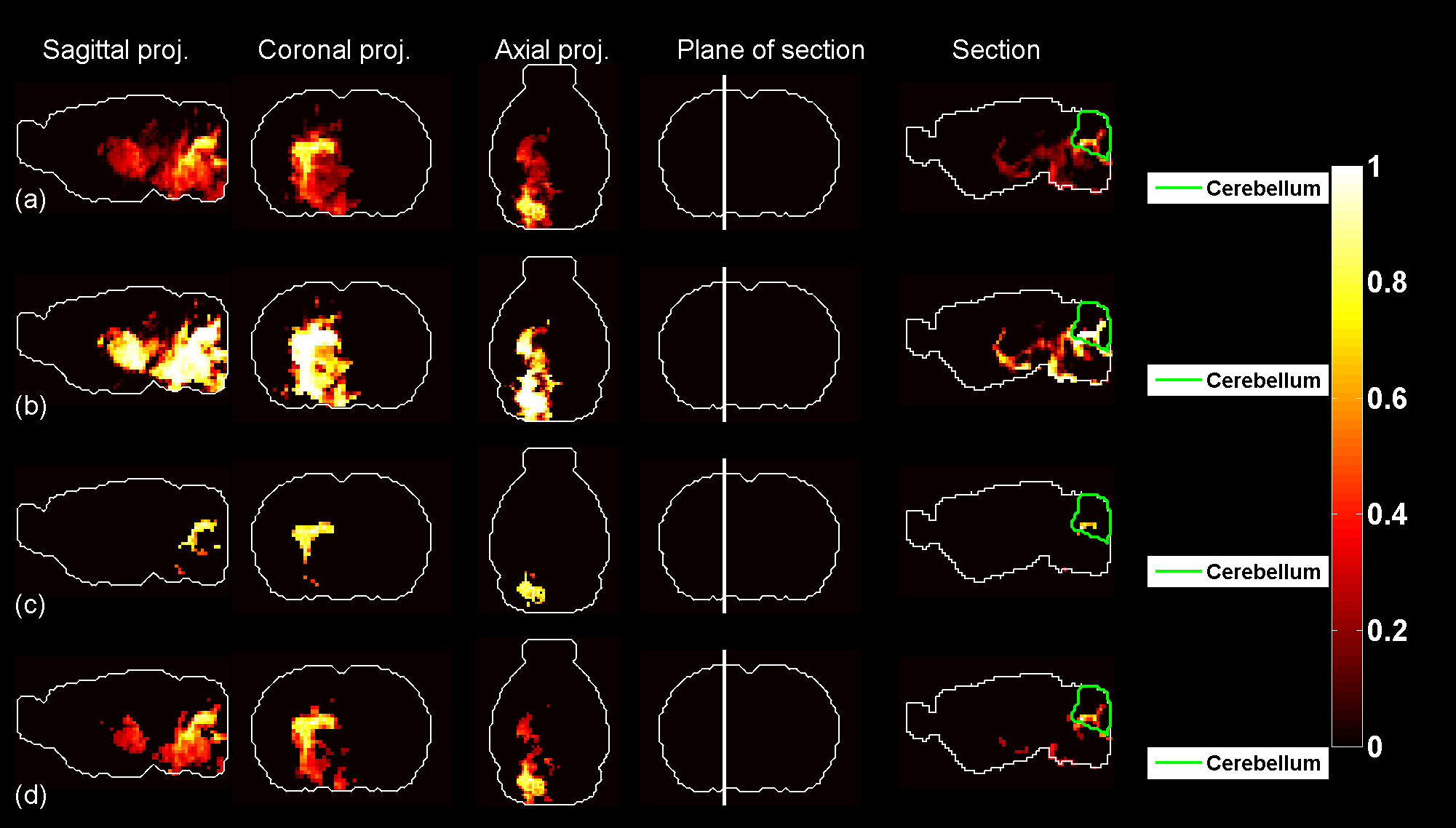}
\caption{Predicted profile, probability profile and thresholded profiles for $t=21$.}
\label{subSampleSplit21}
\end{figure}
\clearpage
\begin{figure}
\includegraphics[width=1\textwidth,keepaspectratio]{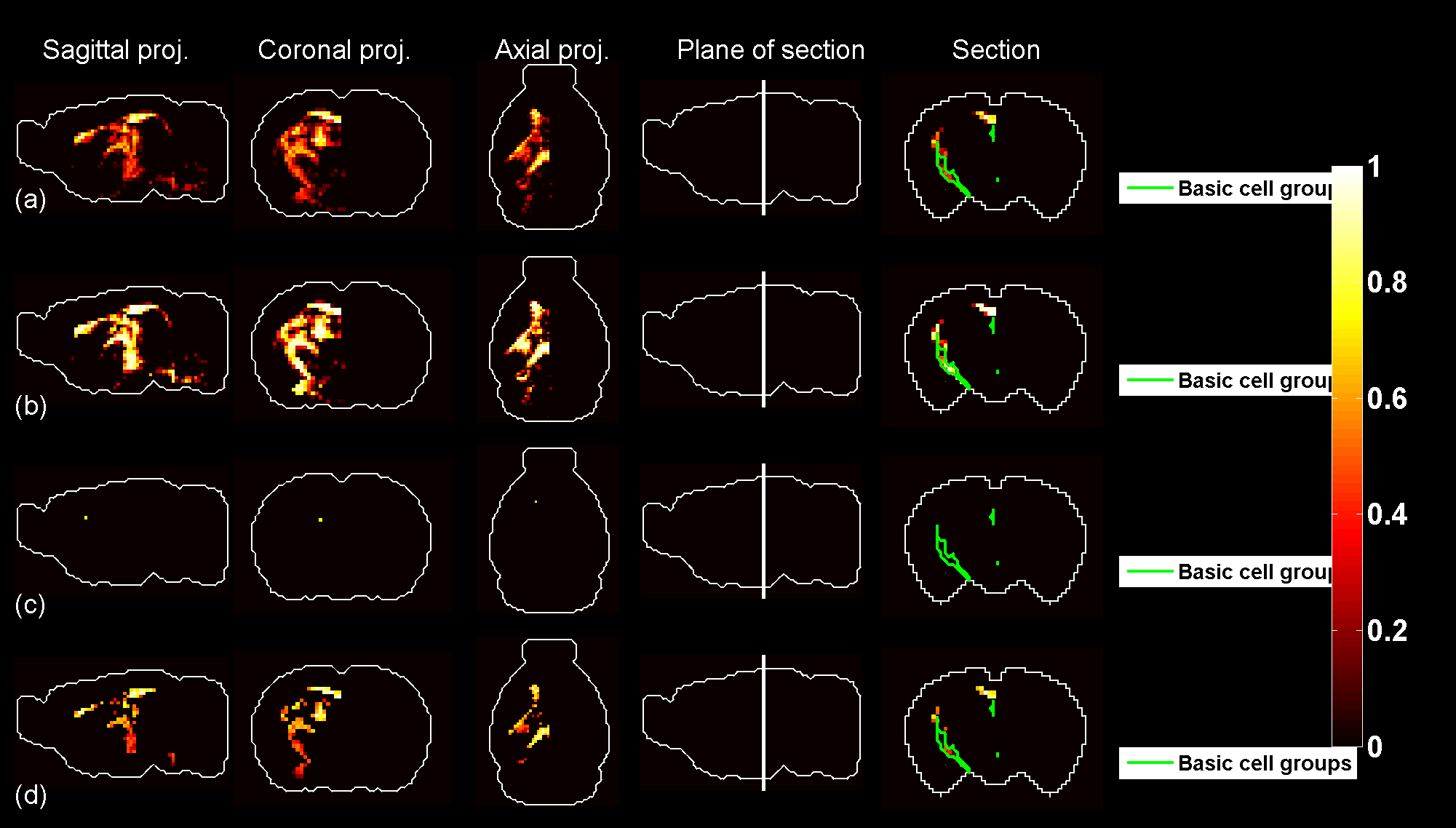}
\caption{Predicted profile, probability profile and thresholded profiles for $t=22$.}
\label{subSampleSplit22}
\end{figure}
\clearpage
\begin{figure}
\includegraphics[width=1\textwidth,keepaspectratio]{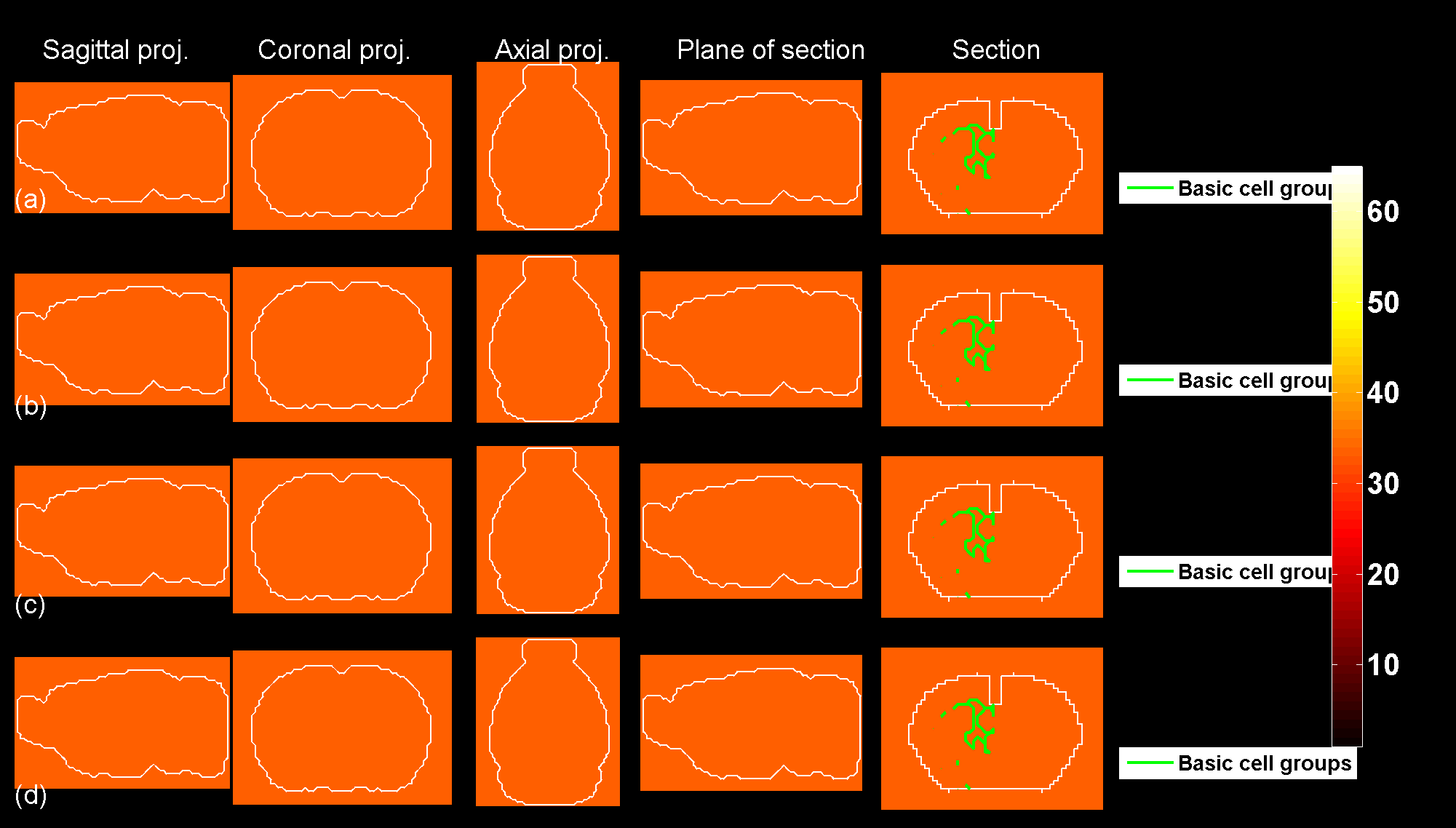}
\caption{Predicted profile, probability profile and thresholded profiles for $t=23$.}
\label{subSampleSplit23}
\end{figure}
\clearpage
\begin{figure}
\includegraphics[width=1\textwidth,keepaspectratio]{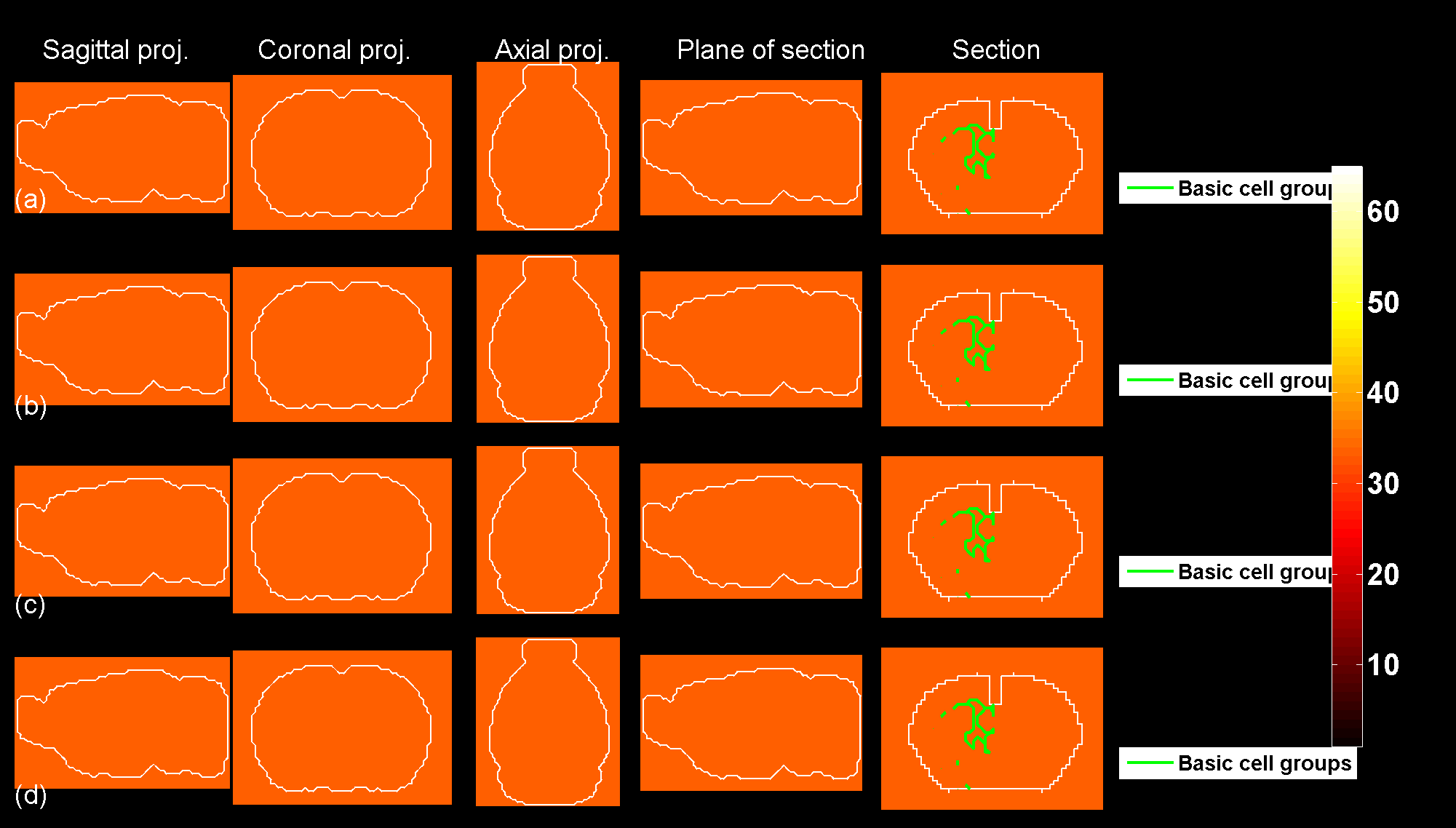}
\caption{Predicted profile, probability profile and thresholded profiles for $t=24$.}
\label{subSampleSplit24}
\end{figure}
\clearpage
\begin{figure}
\includegraphics[width=1\textwidth,keepaspectratio]{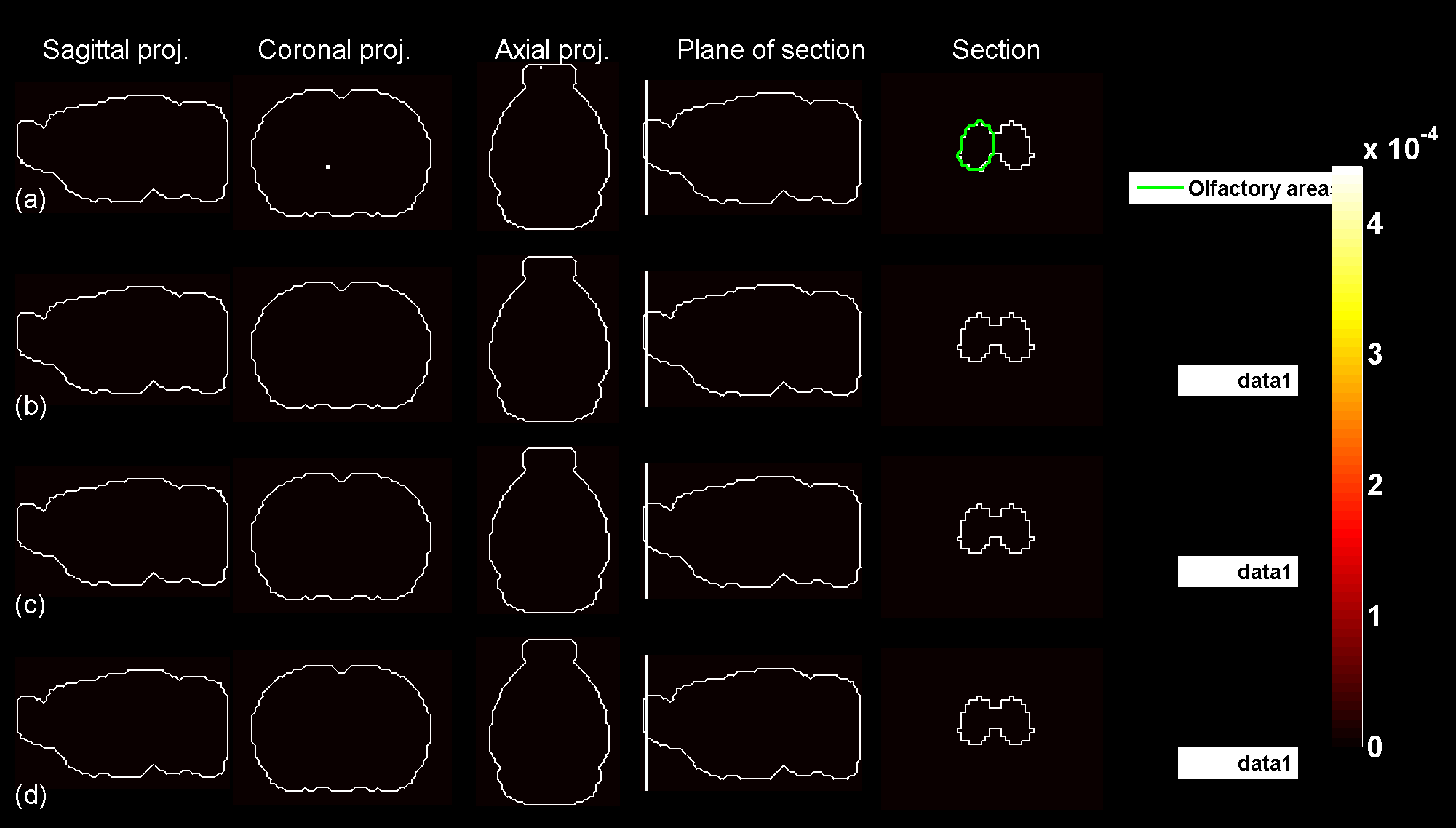}
\caption{Predicted profile, probability profile and thresholded profiles for $t=25$.}
\label{subSampleSplit25}
\end{figure}
\clearpage
\begin{figure}
\includegraphics[width=1\textwidth,keepaspectratio]{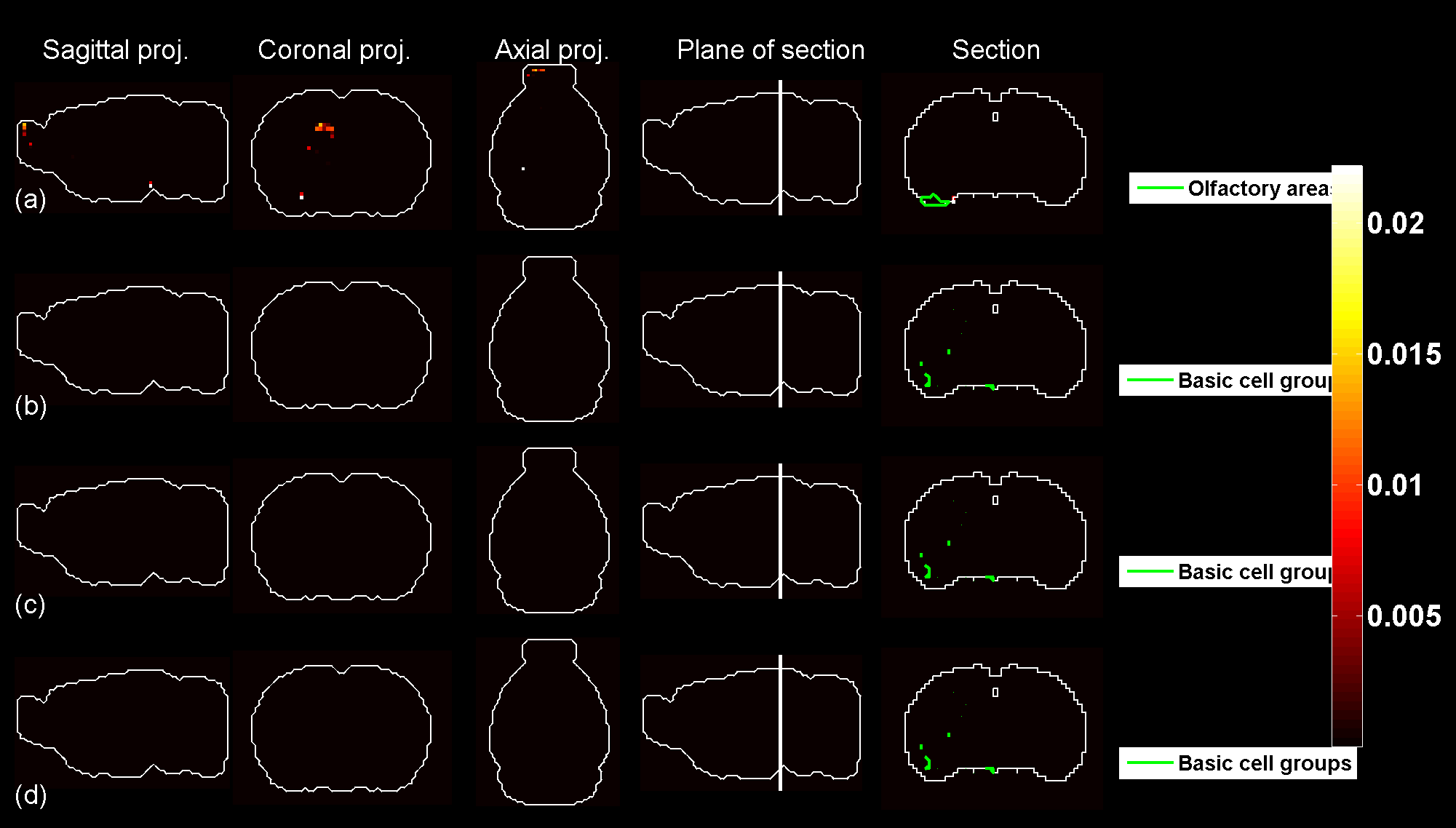}
\caption{Predicted profile, probability profile and thresholded profiles for $t=26$.}
\label{subSampleSplit26}
\end{figure}
\clearpage
\begin{figure}
\includegraphics[width=1\textwidth,keepaspectratio]{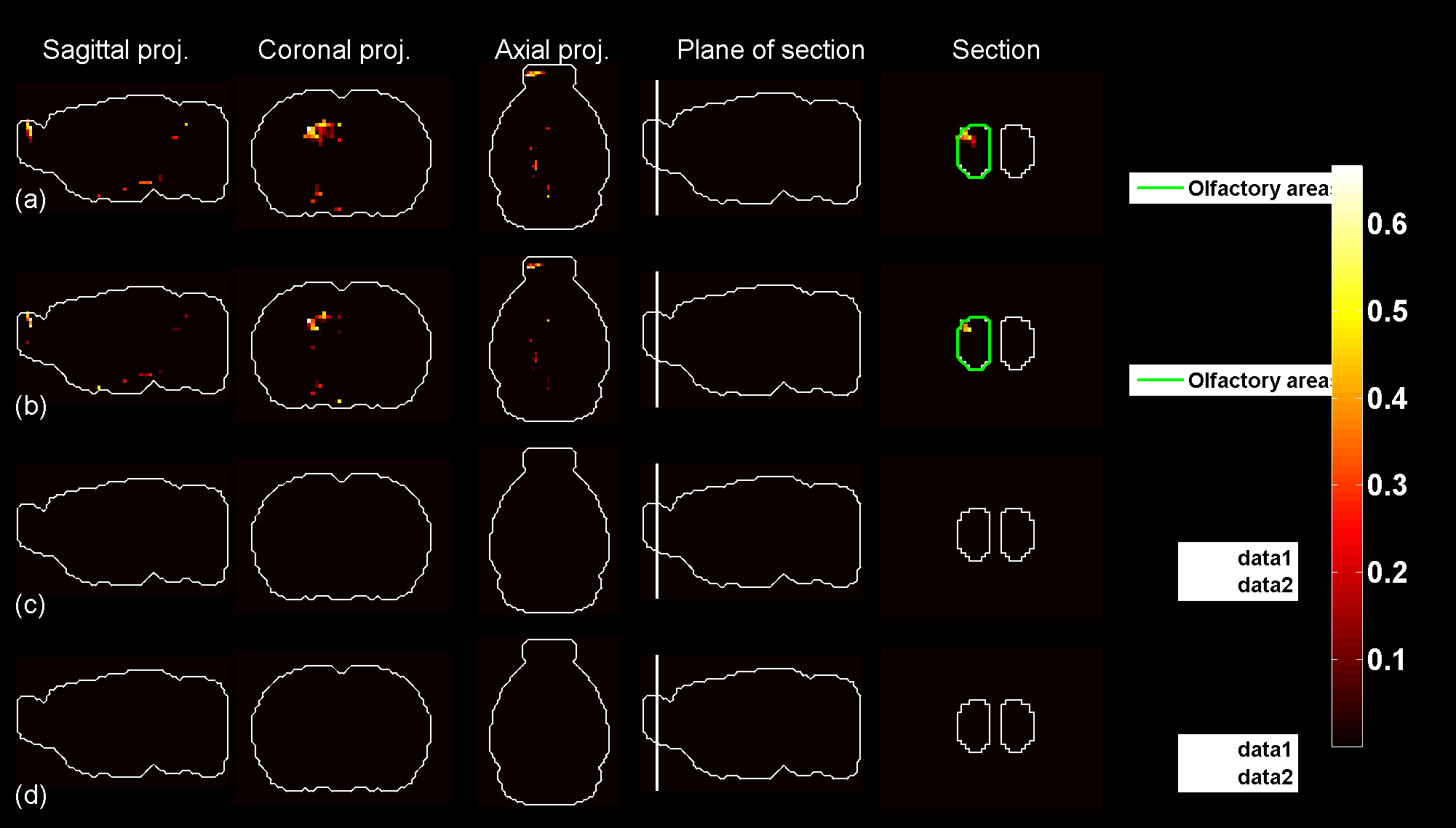}
\caption{Predicted profile, probability profile and thresholded profiles for $t=27$.}
\label{subSampleSplit27}
\end{figure}
\clearpage
\begin{figure}
\includegraphics[width=1\textwidth,keepaspectratio]{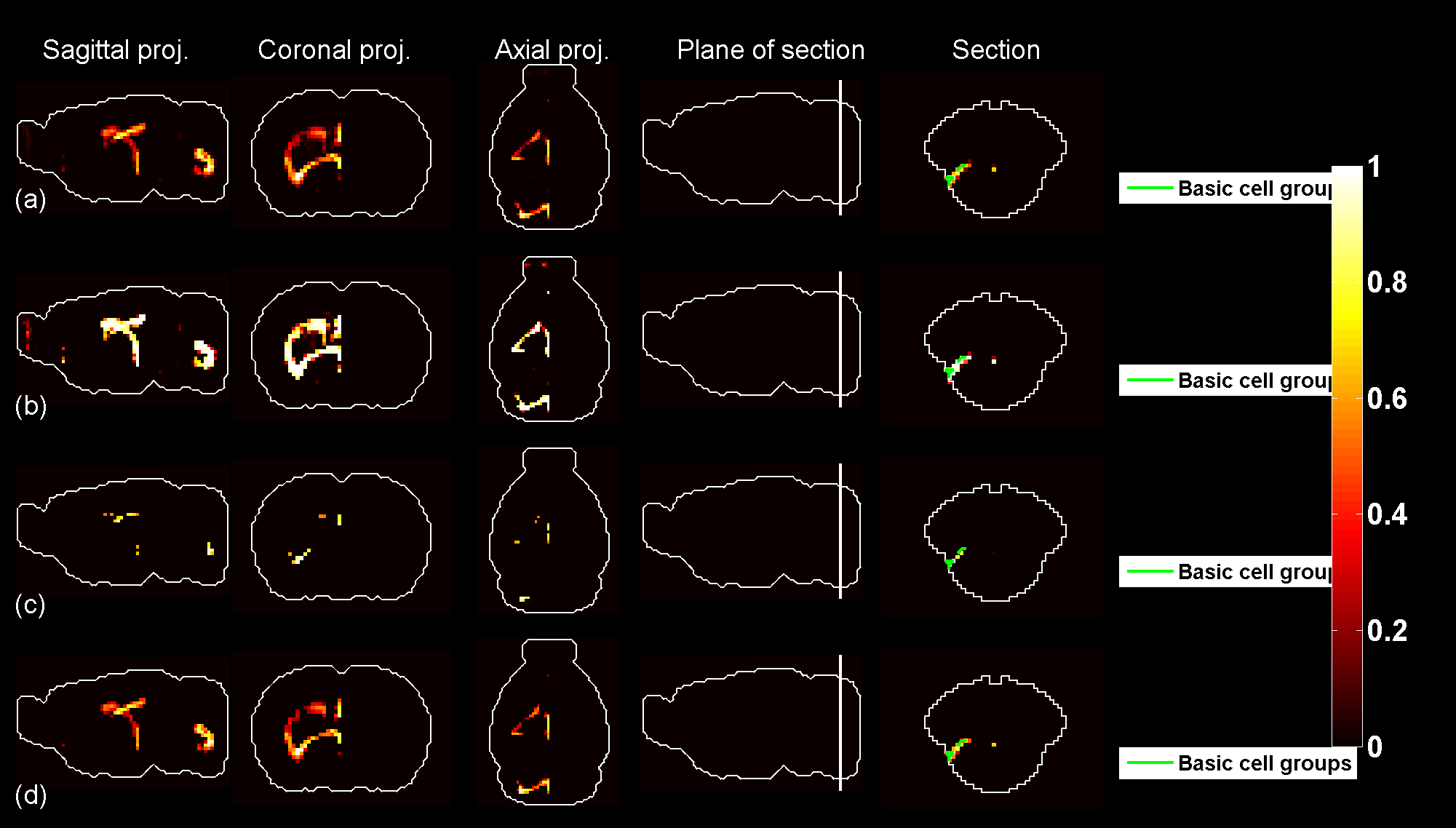}
\caption{Predicted profile, probability profile and thresholded profiles for $t=28$.}
\label{subSampleSplit28}
\end{figure}
\clearpage
\begin{figure}
\includegraphics[width=1\textwidth,keepaspectratio]{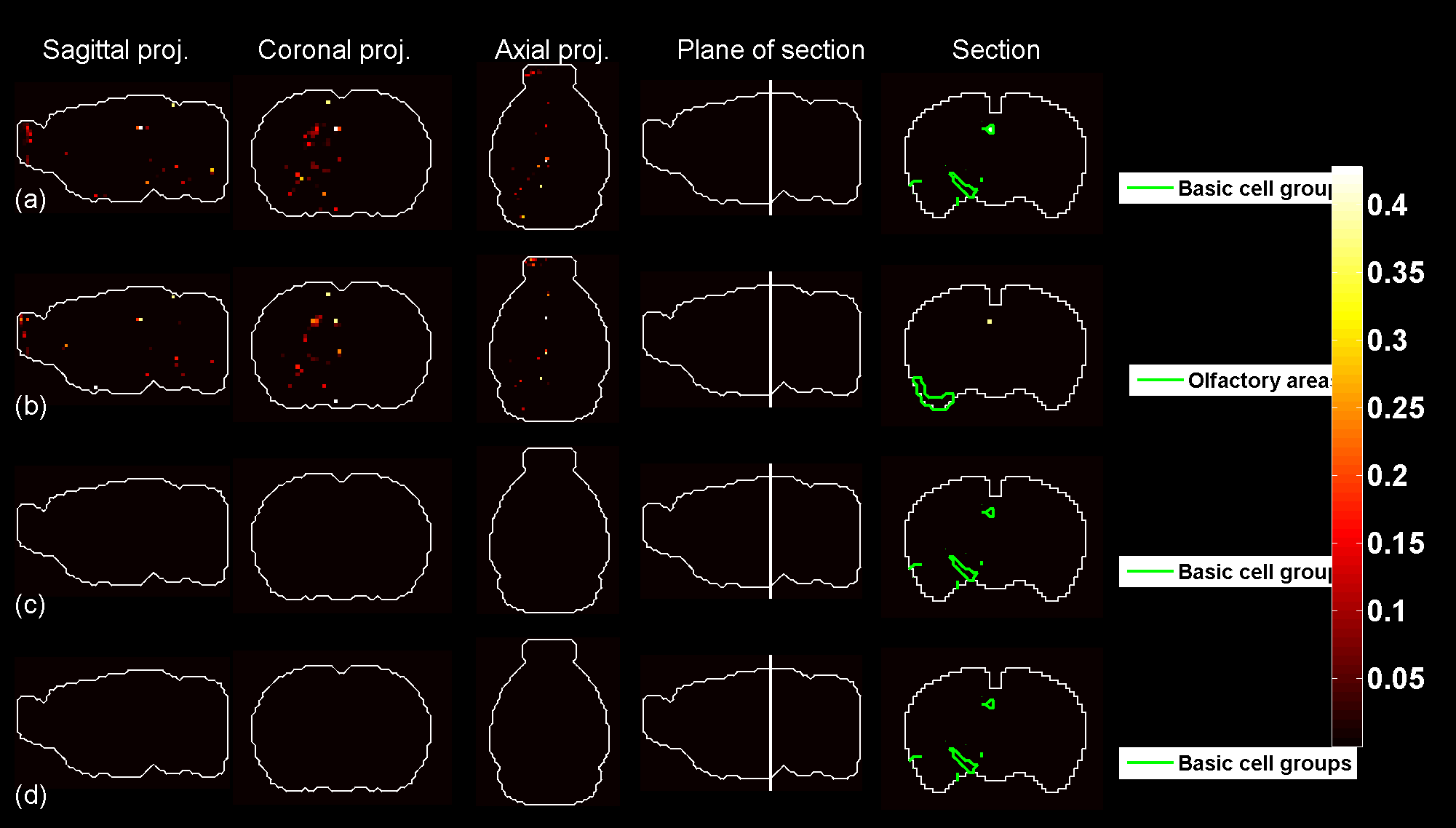}
\caption{Predicted profile, probability profile and thresholded profiles for $t=29$.}
\label{subSampleSplit29}
\end{figure}
\clearpage
\begin{figure}
\includegraphics[width=1\textwidth,keepaspectratio]{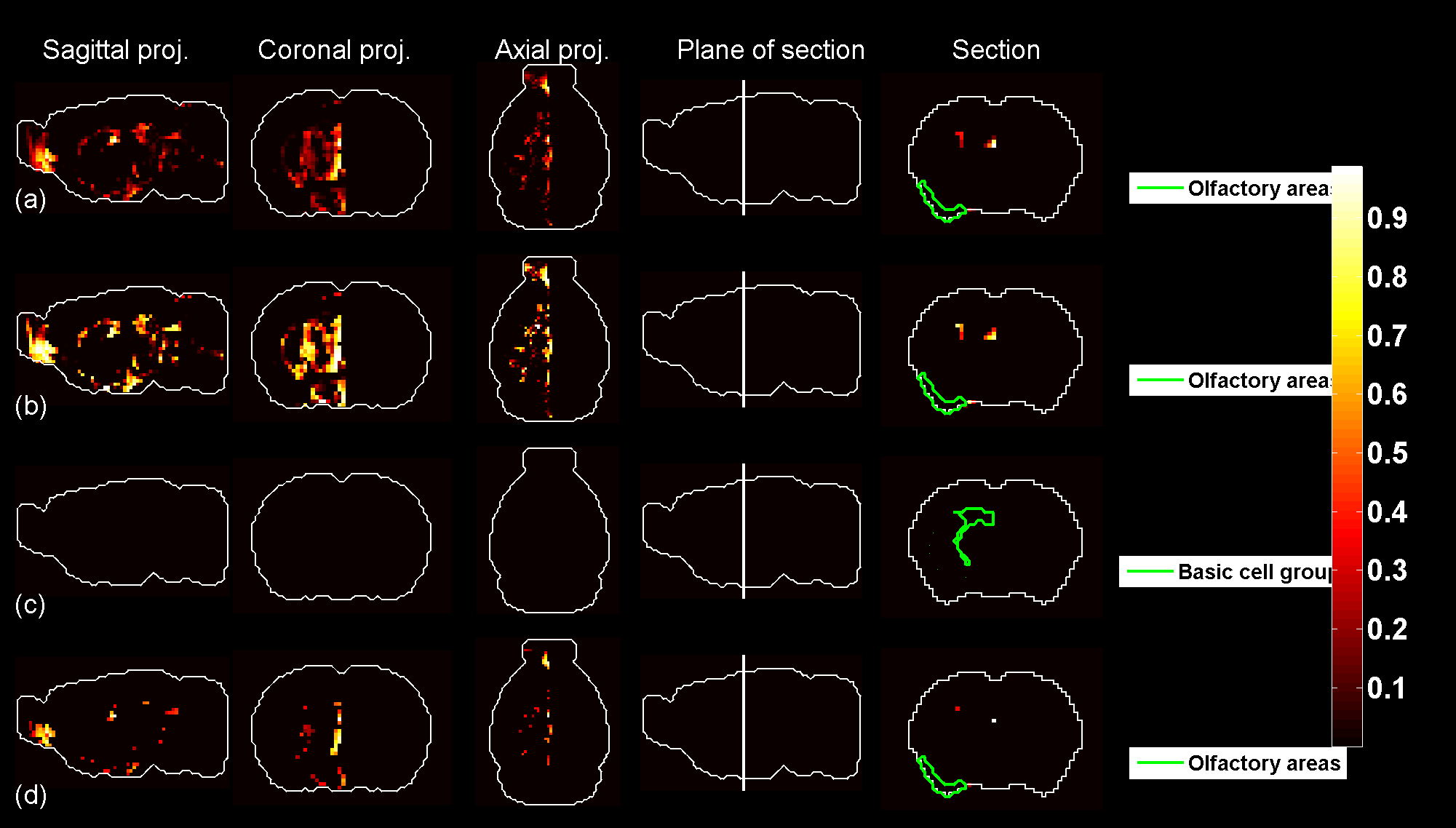}
\caption{Predicted profile, probability profile and thresholded profiles for $t=30$.}
\label{subSampleSplit30}
\end{figure}
\clearpage
\begin{figure}
\includegraphics[width=1\textwidth,keepaspectratio]{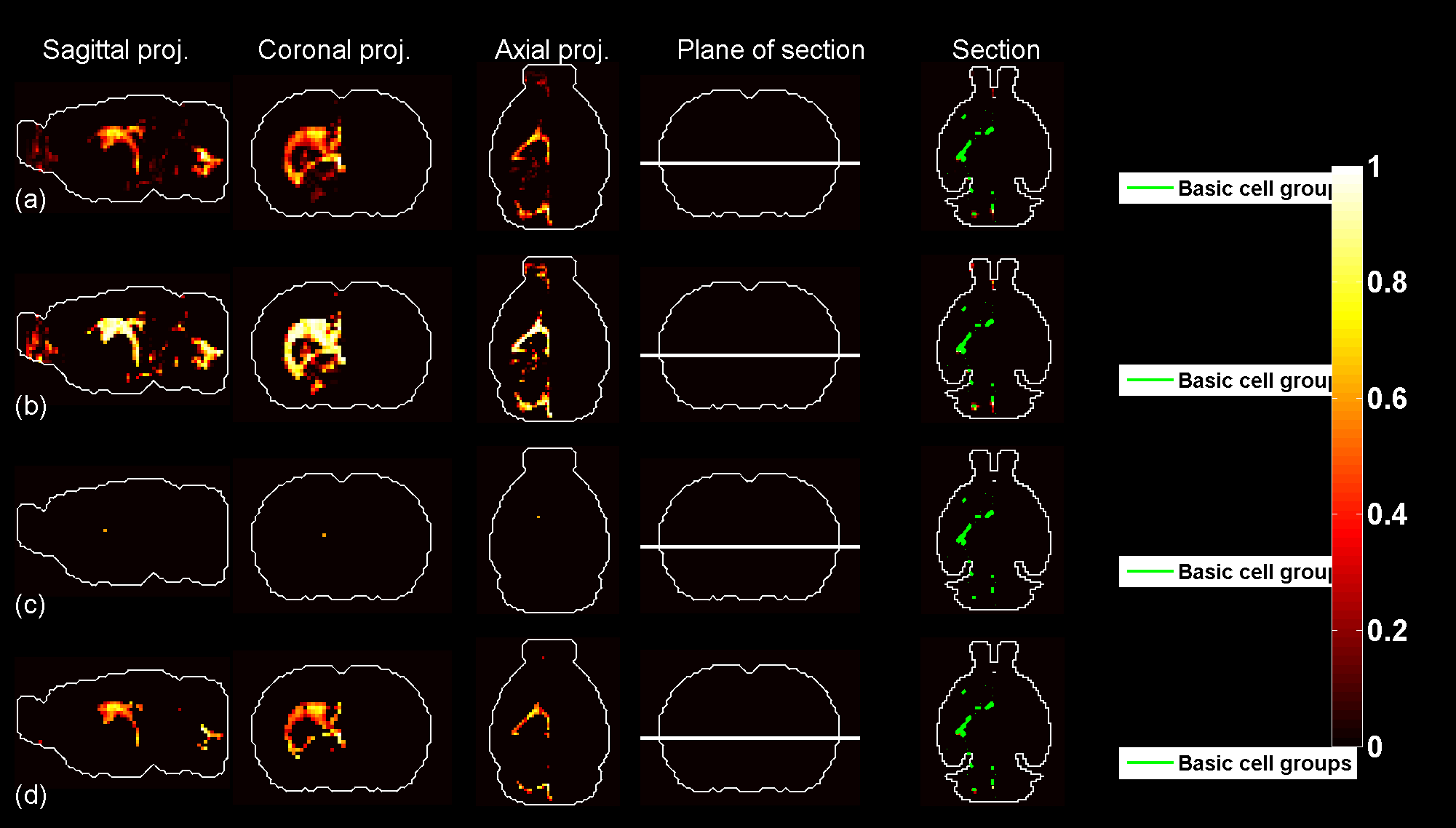}
\caption{Predicted profile, probability profile and thresholded profiles for $t=31$.}
\label{subSampleSplit31}
\end{figure}
\clearpage
\begin{figure}
\includegraphics[width=1\textwidth,keepaspectratio]{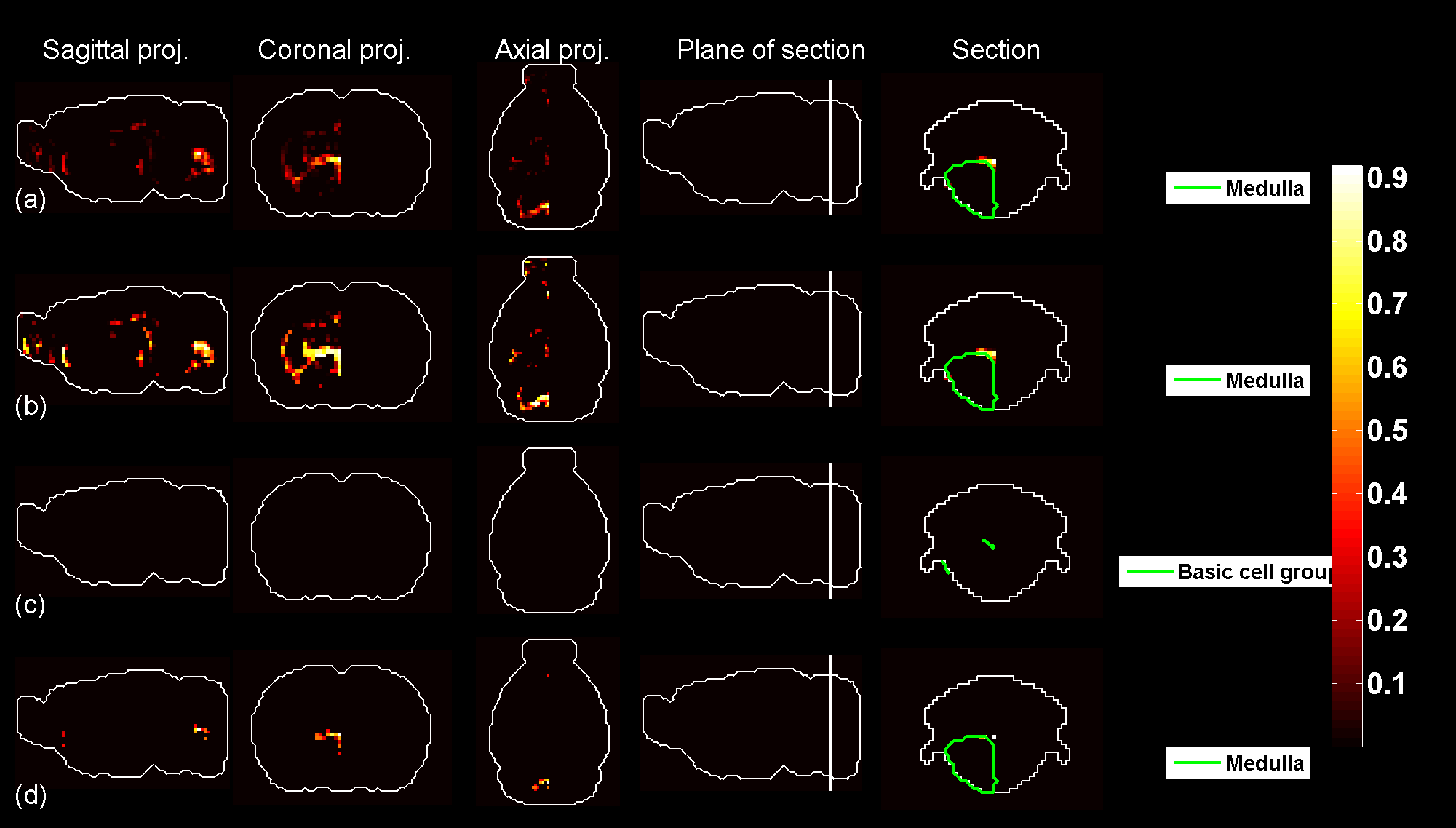}
\caption{Predicted profile, probability profile and thresholded profiles for $t=32$.}
\label{subSampleSplit32}
\end{figure}
\clearpage
\begin{figure}
\includegraphics[width=1\textwidth,keepaspectratio]{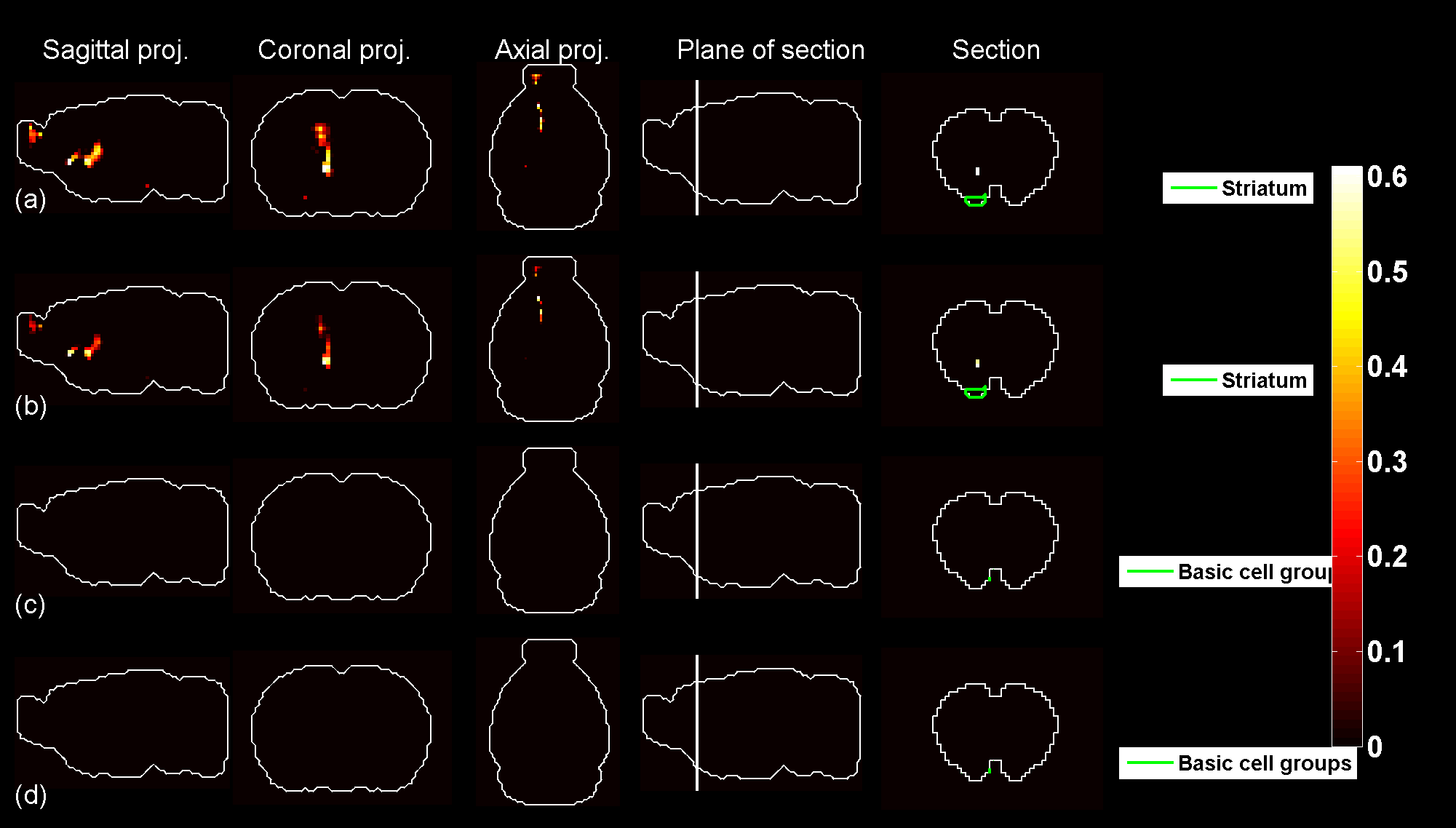}
\caption{Predicted profile, probability profile and thresholded profiles for $t=33$.}
\label{subSampleSplit33}
\end{figure}
\clearpage
\begin{figure}
\includegraphics[width=1\textwidth,keepaspectratio]{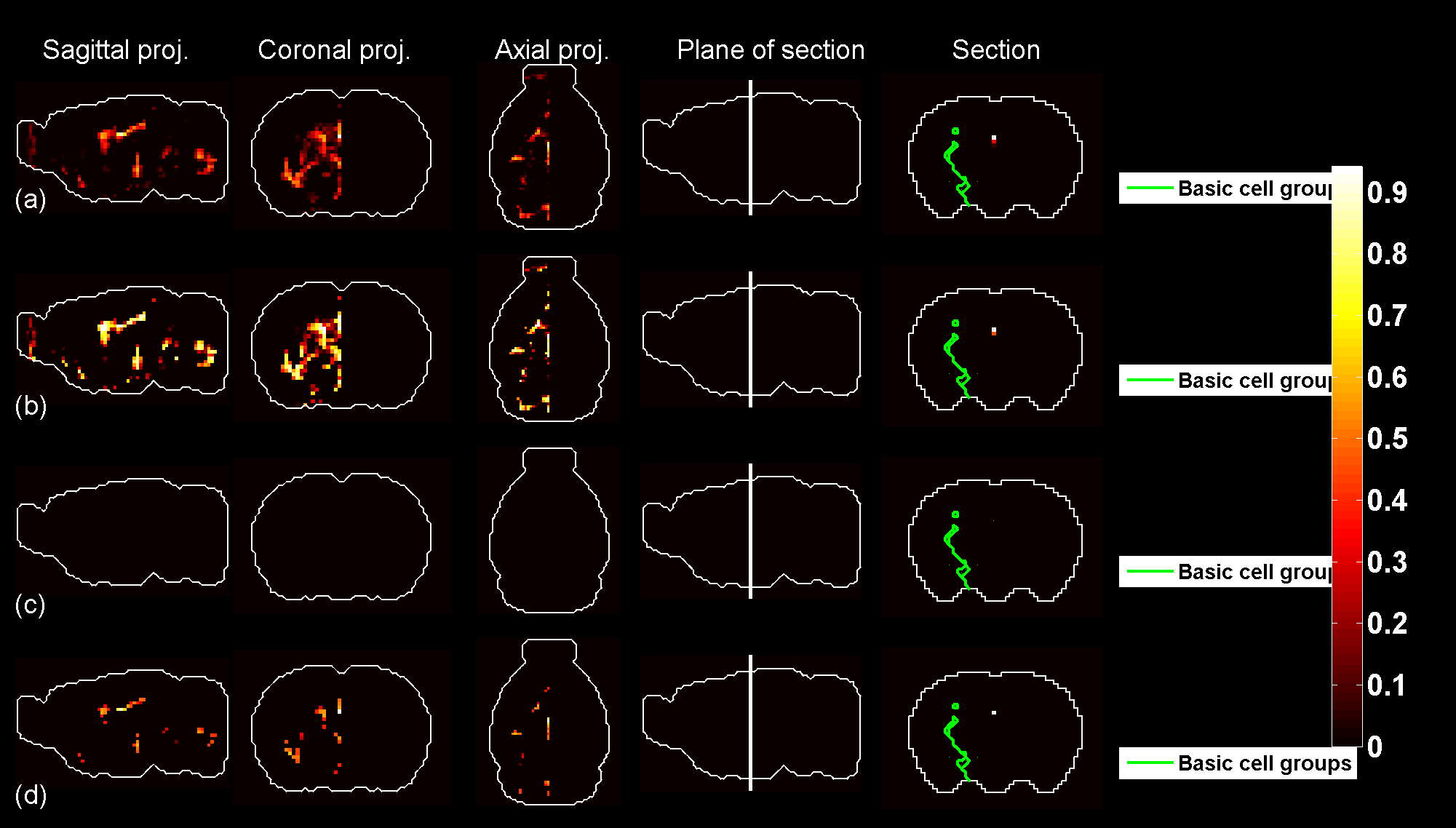}
\caption{Predicted profile, probability profile and thresholded profiles for $t=34$.}
\label{subSampleSplit34}
\end{figure}
\clearpage
\begin{figure}
\includegraphics[width=1\textwidth,keepaspectratio]{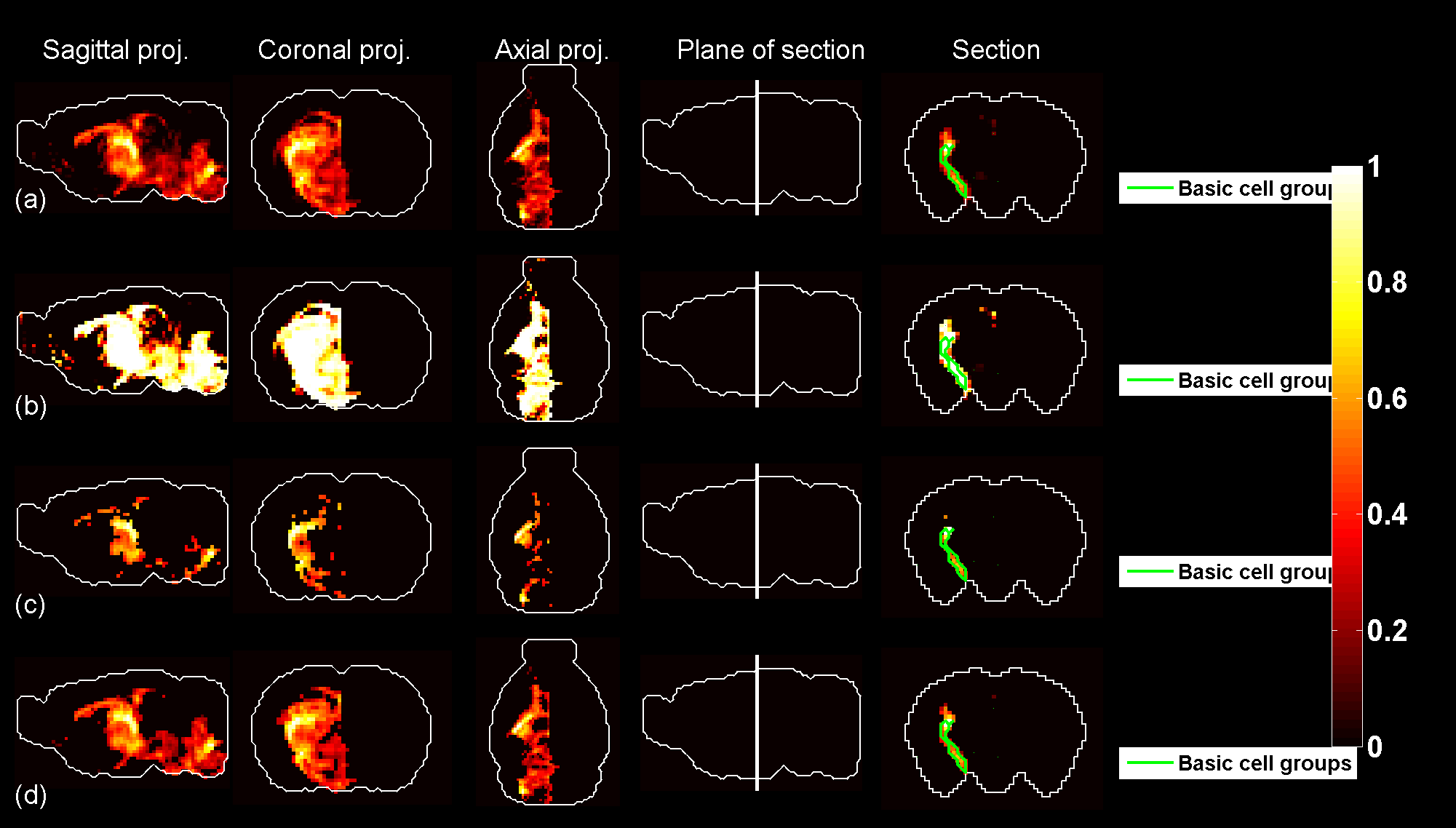}
\caption{Predicted profile, probability profile and thresholded profiles for $t=35$.}
\label{subSampleSplit35}
\end{figure}
\clearpage
\begin{figure}
\includegraphics[width=1\textwidth,keepaspectratio]{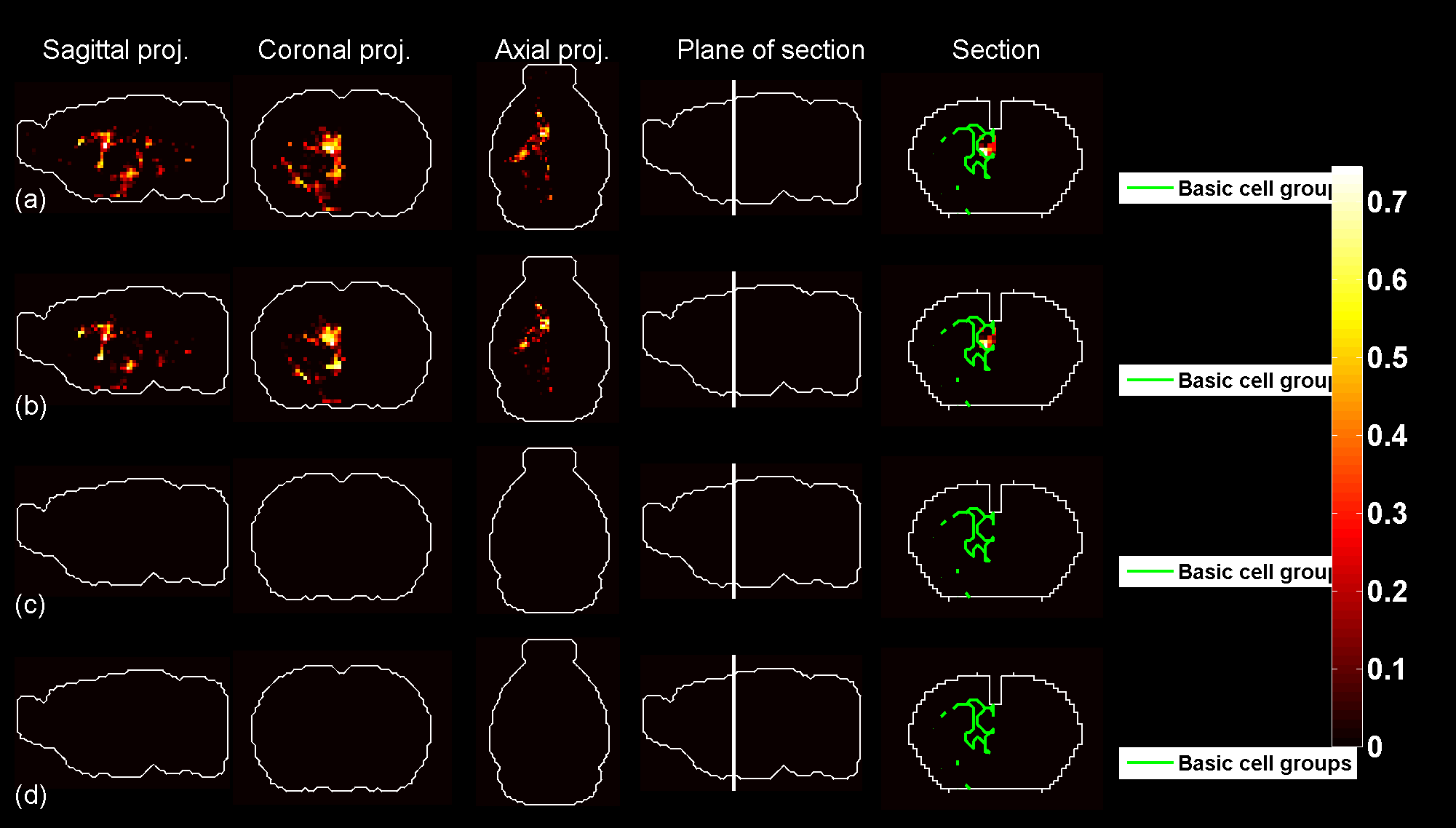}
\caption{Predicted profile, probability profile and thresholded profiles for $t=36$.}
\label{subSampleSplit36}
\end{figure}
\clearpage
\begin{figure}
\includegraphics[width=1\textwidth,keepaspectratio]{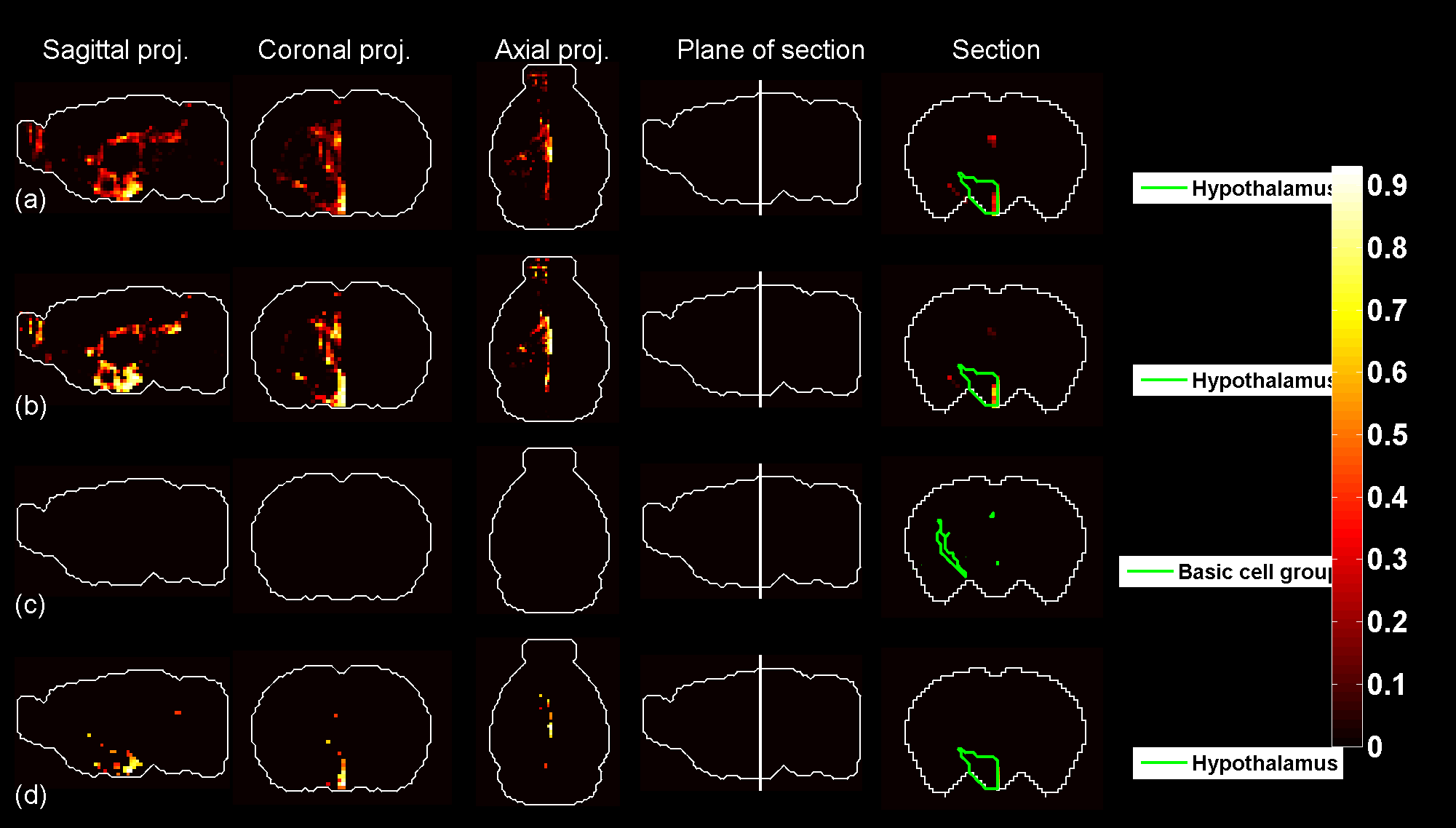}
\caption{Predicted profile, probability profile and thresholded profiles for $t=37$.}
\label{subSampleSplit37}
\end{figure}
\clearpage
\begin{figure}
\includegraphics[width=1\textwidth,keepaspectratio]{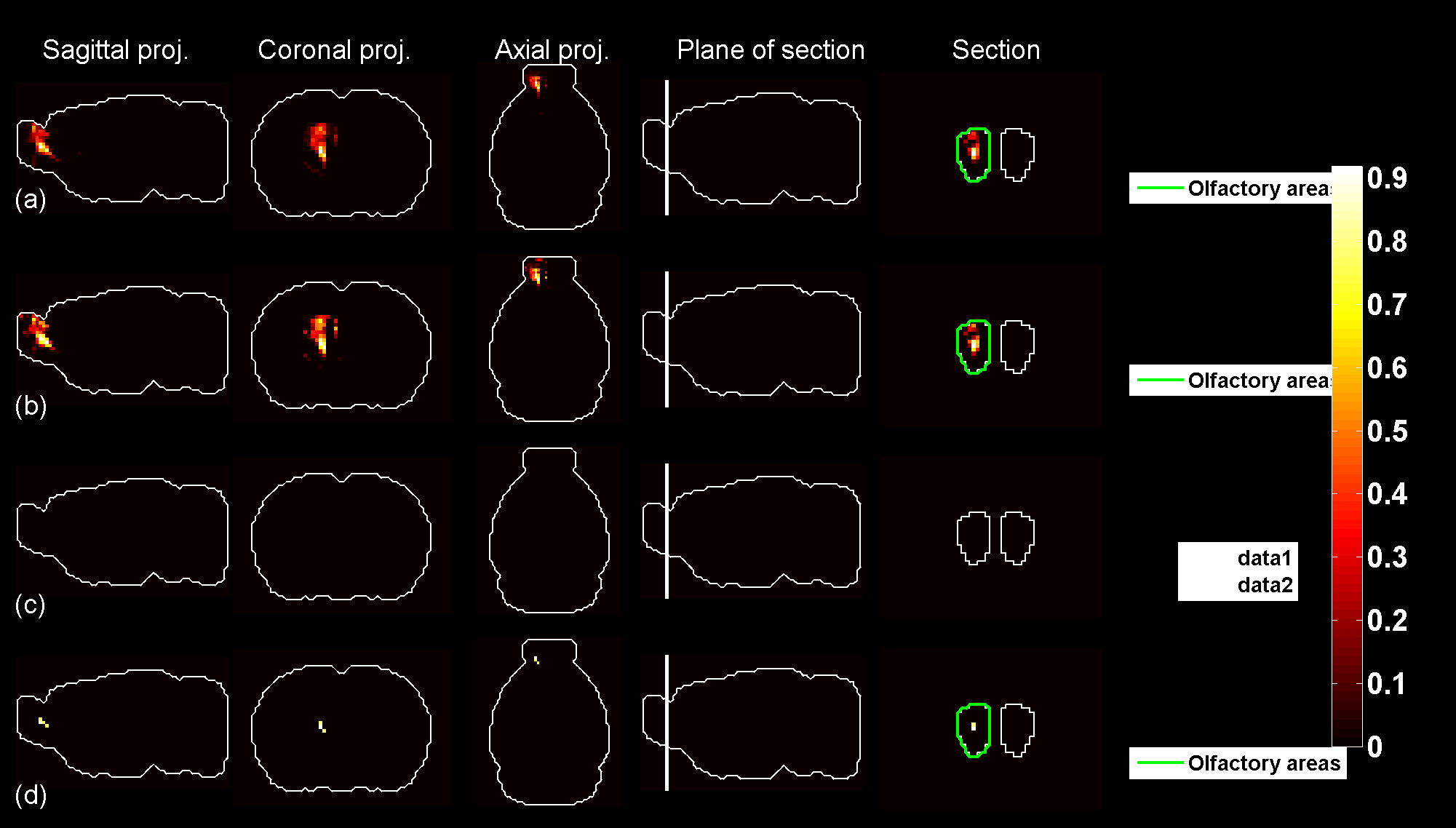}
\caption{Predicted profile, probability profile and thresholded profiles for $t=38$.}
\label{subSampleSplit38}
\end{figure}
\clearpage
\begin{figure}
\includegraphics[width=1\textwidth,keepaspectratio]{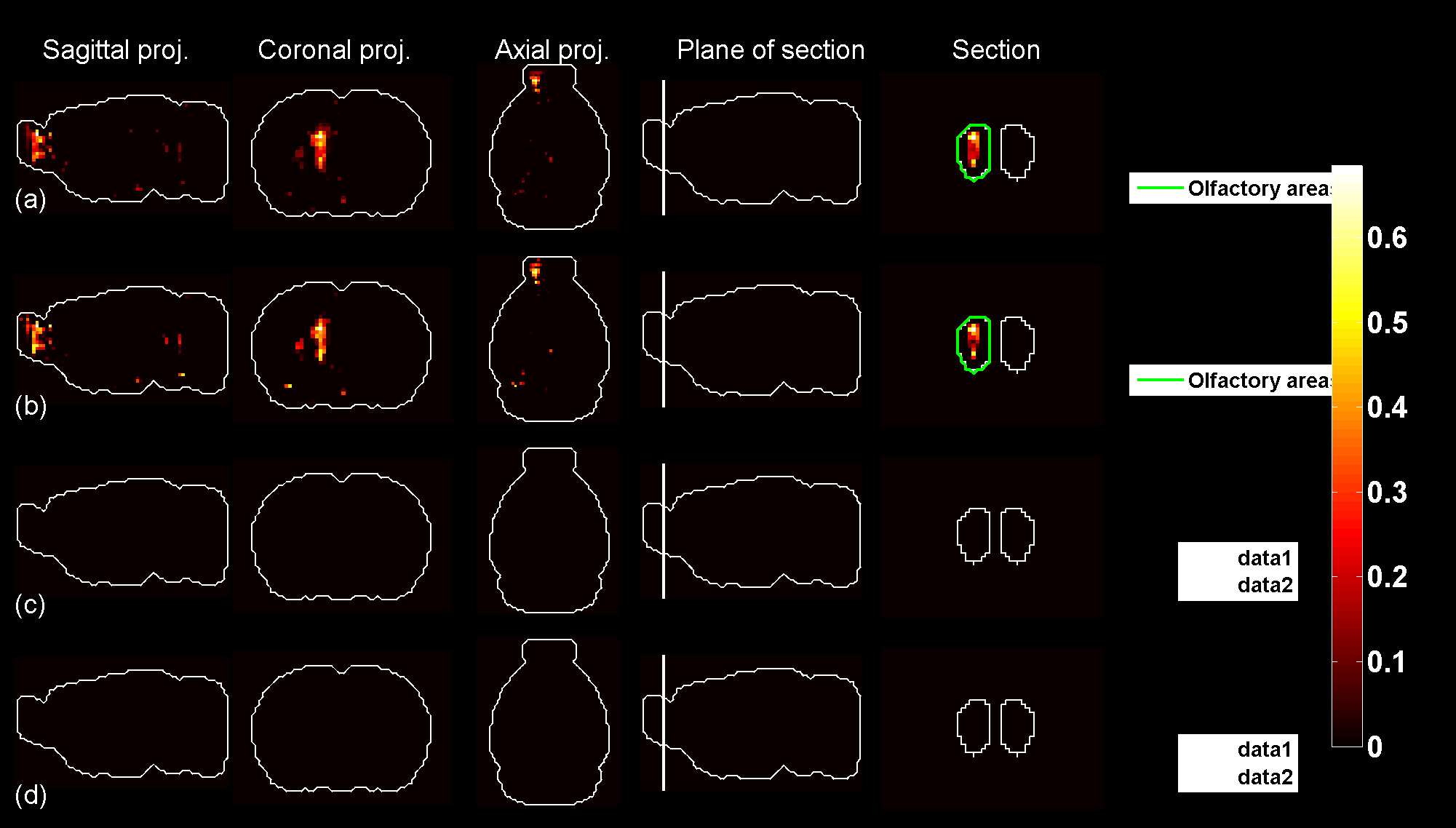}
\caption{Predicted profile, probability profile and thresholded profiles for $t=39$.}
\label{subSampleSplit39}
\end{figure}
\clearpage
\begin{figure}
\includegraphics[width=1\textwidth,keepaspectratio]{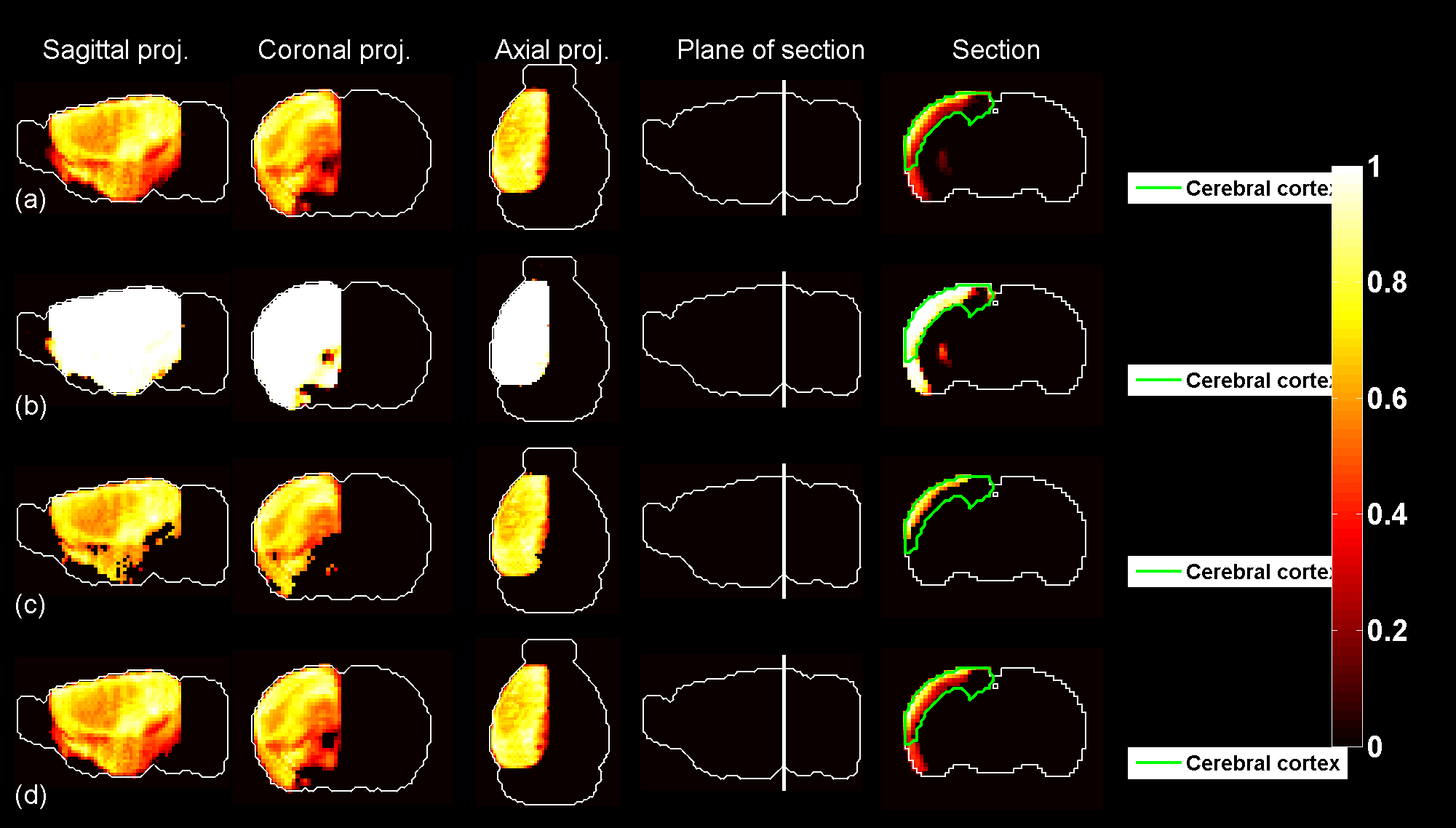}
\caption{Predicted profile, probability profile and thresholded profiles for $t=40$.}
\label{subSampleSplit40}
\end{figure}
\clearpage
\begin{figure}
\includegraphics[width=1\textwidth,keepaspectratio]{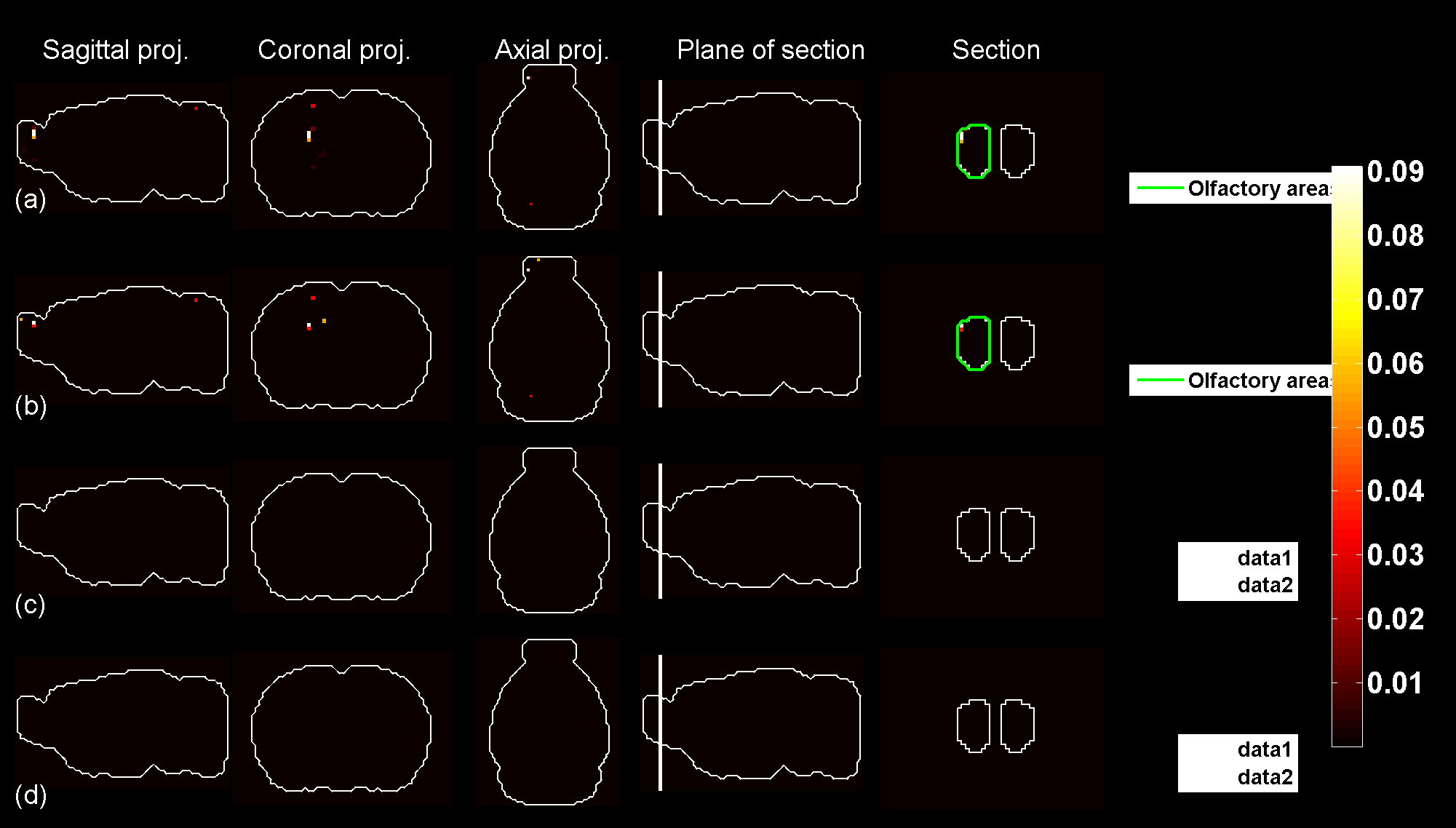}
\caption{Predicted profile, probability profile and thresholded profiles for $t=41$.}
\label{subSampleSplit41}
\end{figure}
\clearpage
\begin{figure}
\includegraphics[width=1\textwidth,keepaspectratio]{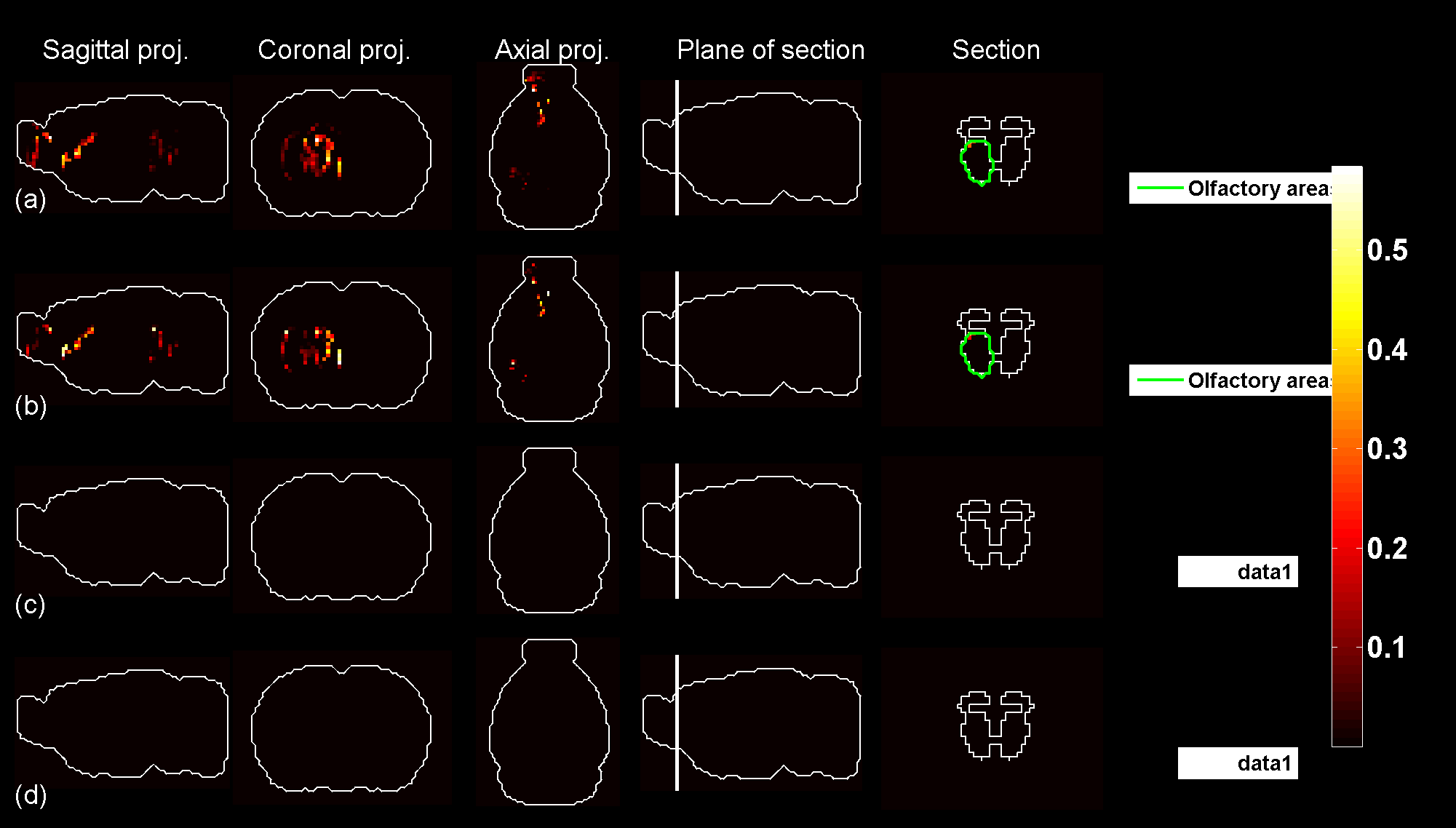}
\caption{Predicted profile, probability profile and thresholded profiles for $t=42$.}
\label{subSampleSplit42}
\end{figure}
\clearpage
\begin{figure}
\includegraphics[width=1\textwidth,keepaspectratio]{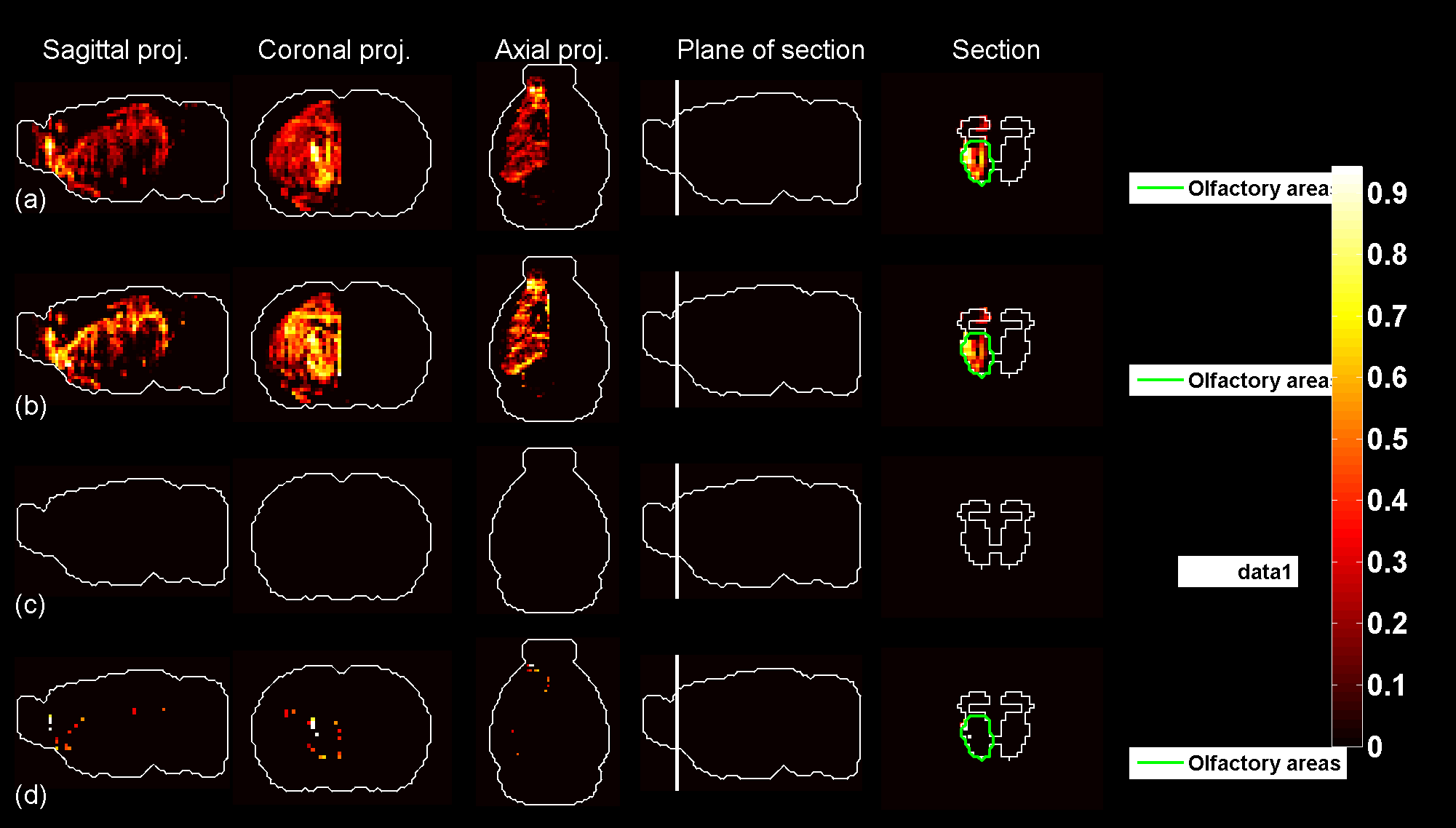}
\caption{Predicted profile, probability profile and thresholded profiles for $t=43$.}
\label{subSampleSplit43}
\end{figure}
\clearpage
\begin{figure}
\includegraphics[width=1\textwidth,keepaspectratio]{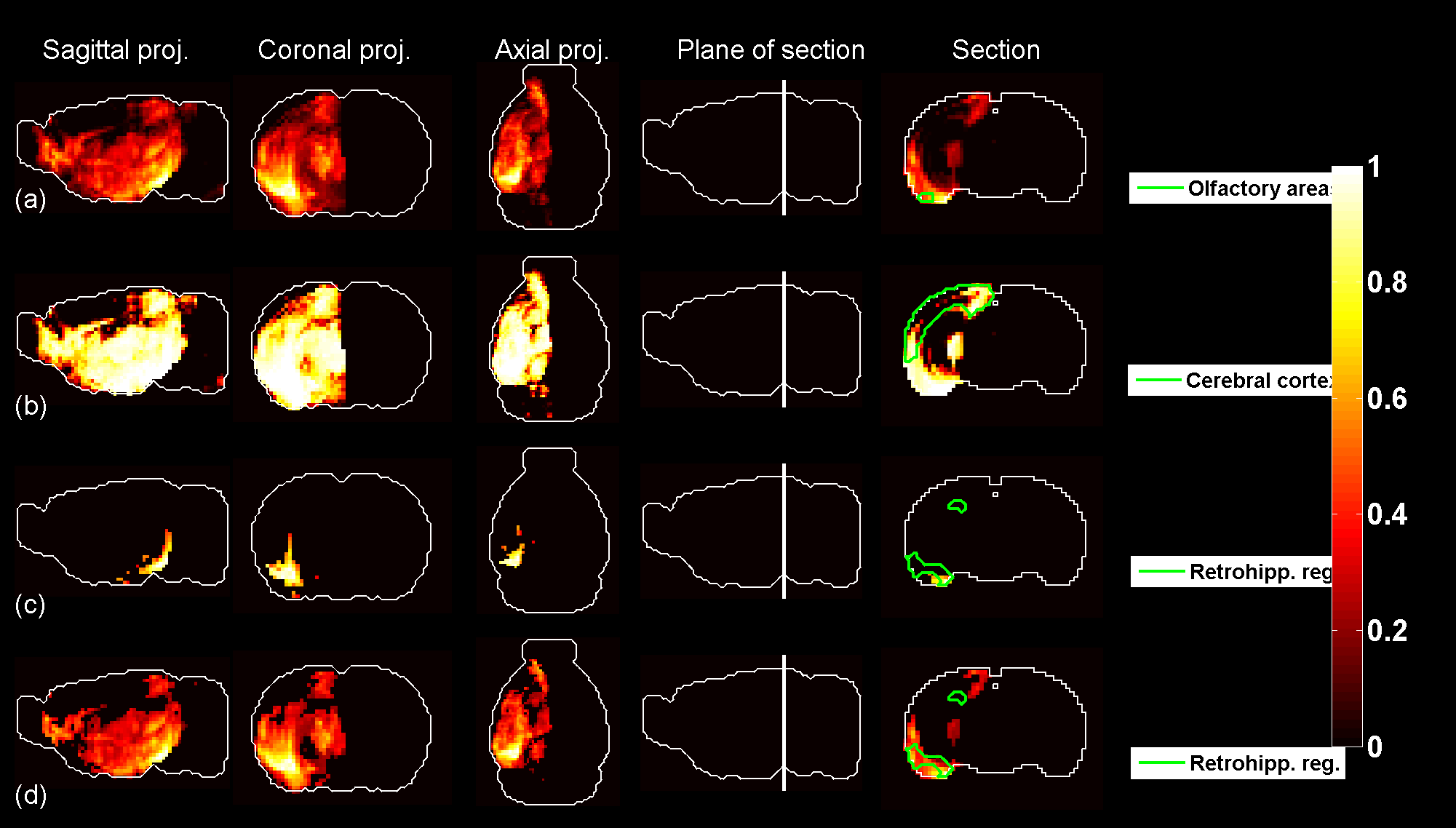}
\caption{Predicted profile, probability profile and thresholded profiles for $t=44$.}
\label{subSampleSplit44}
\end{figure}
\clearpage
\begin{figure}
\includegraphics[width=1\textwidth,keepaspectratio]{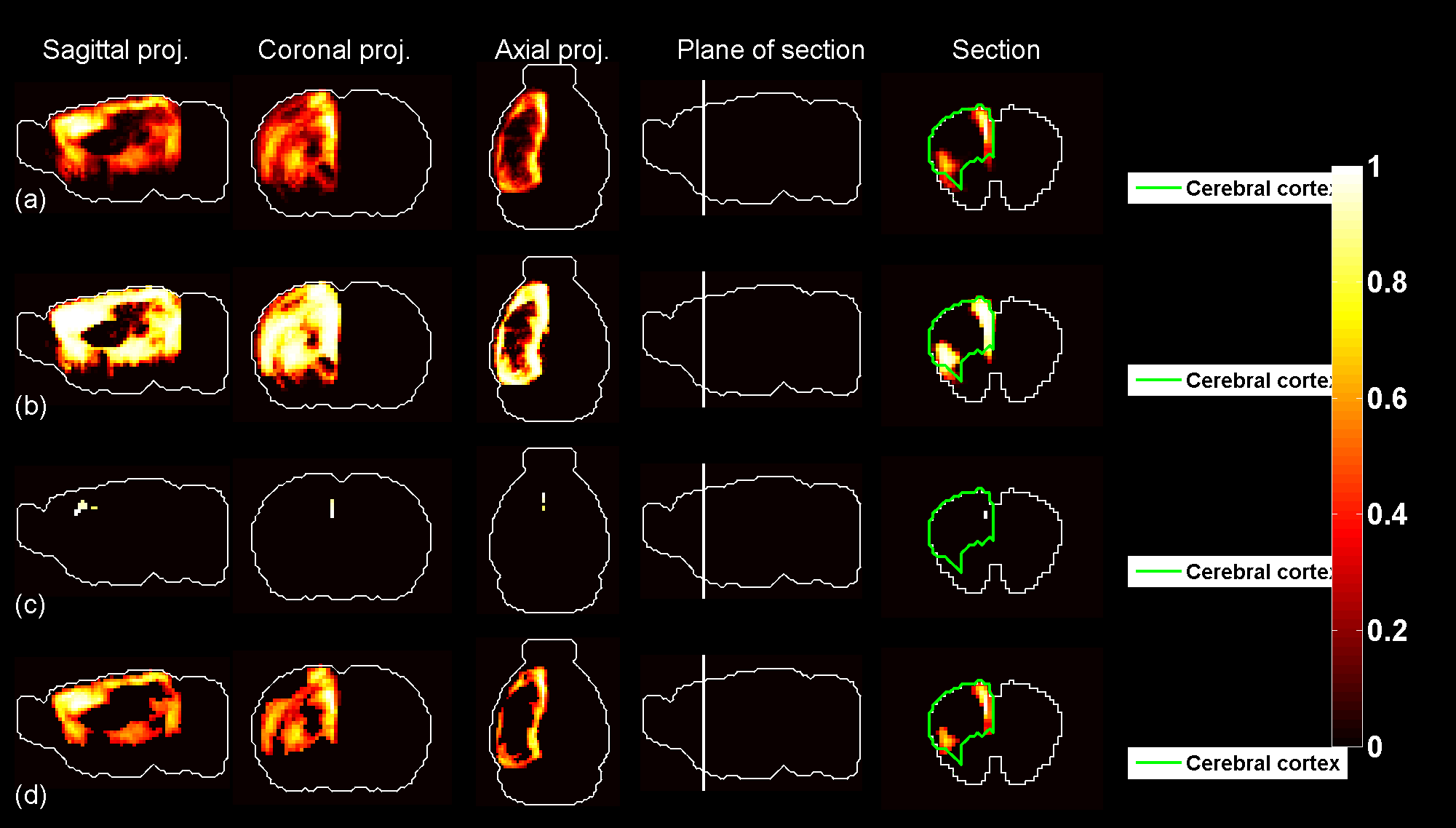}
\caption{Predicted profile, probability profile and thresholded profiles for $t=45$.}
\label{subSampleSplit45}
\end{figure}
\clearpage
\begin{figure}
\includegraphics[width=1\textwidth,keepaspectratio]{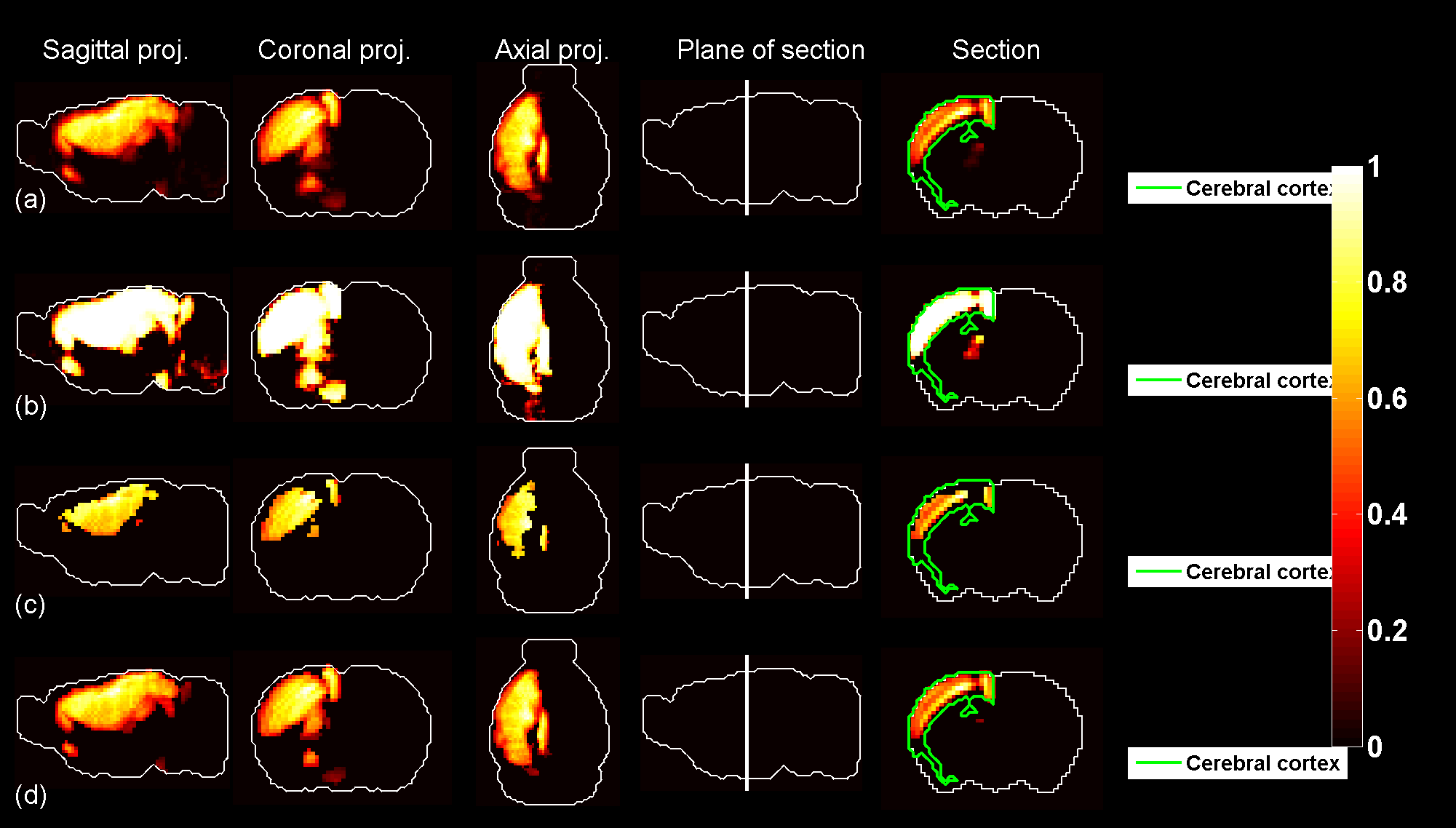}
\caption{Predicted profile, probability profile and thresholded profiles for $t=46$.}
\label{subSampleSplit46}
\end{figure}
\clearpage
\begin{figure}
\includegraphics[width=1\textwidth,keepaspectratio]{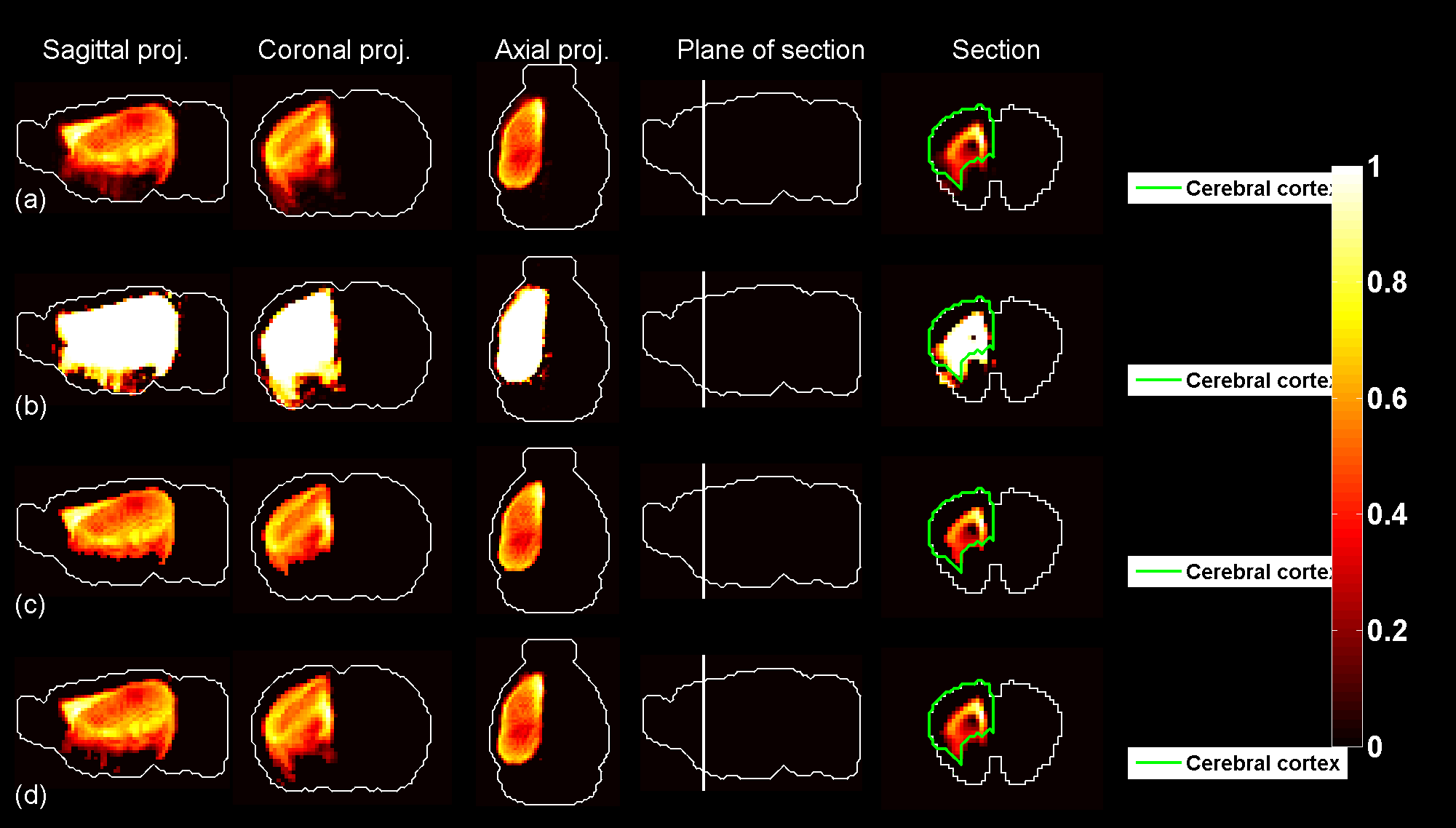}
\caption{Predicted profile, probability profile and thresholded profiles for $t=47$.}
\label{subSampleSplit47}
\end{figure}
\clearpage
\begin{figure}
\includegraphics[width=1\textwidth,keepaspectratio]{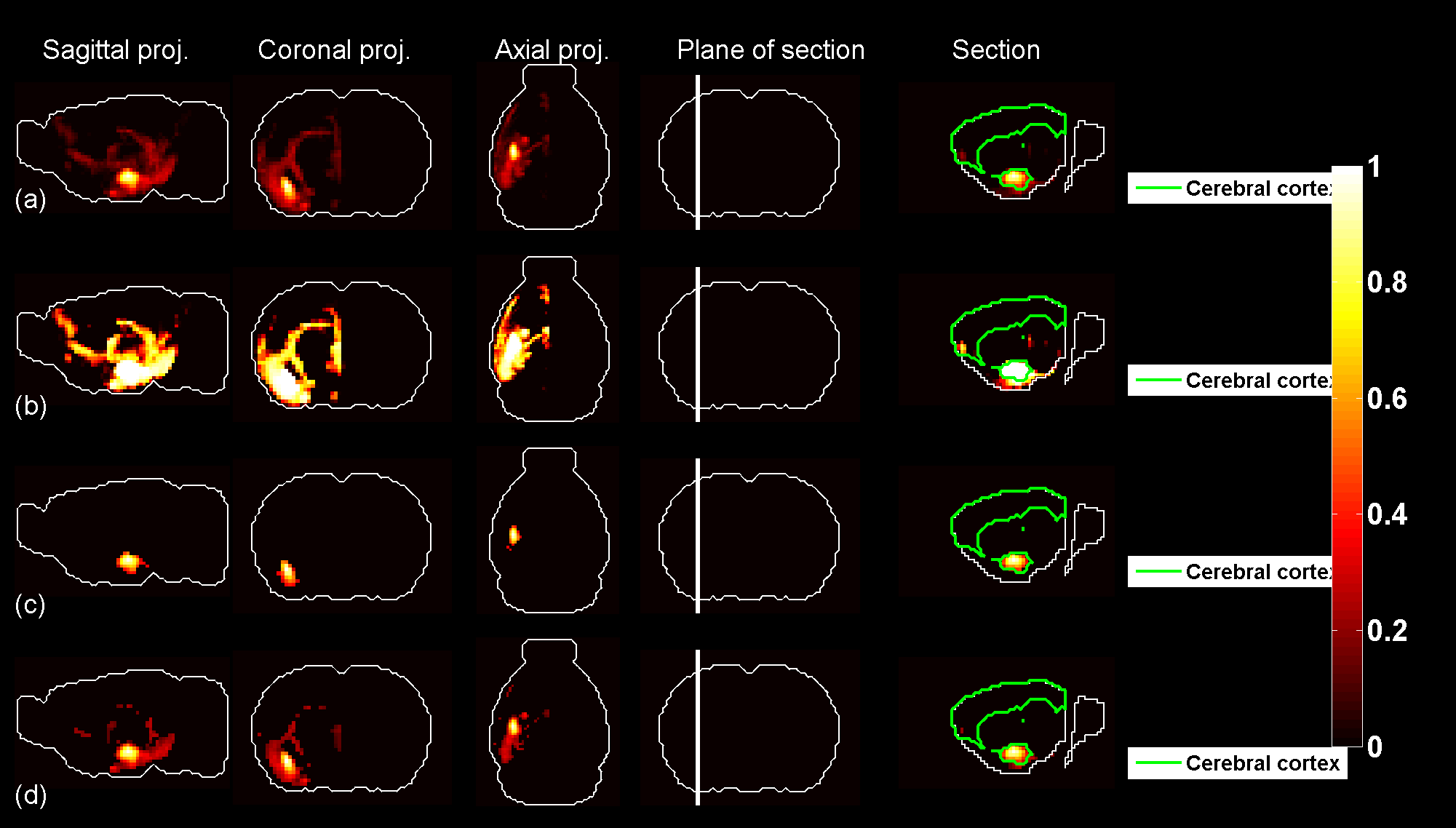}
\caption{Predicted profile, probability profile and thresholded profiles for $t=48$.}
\label{subSampleSplit48}
\end{figure}
\clearpage
\begin{figure}
\includegraphics[width=1\textwidth,keepaspectratio]{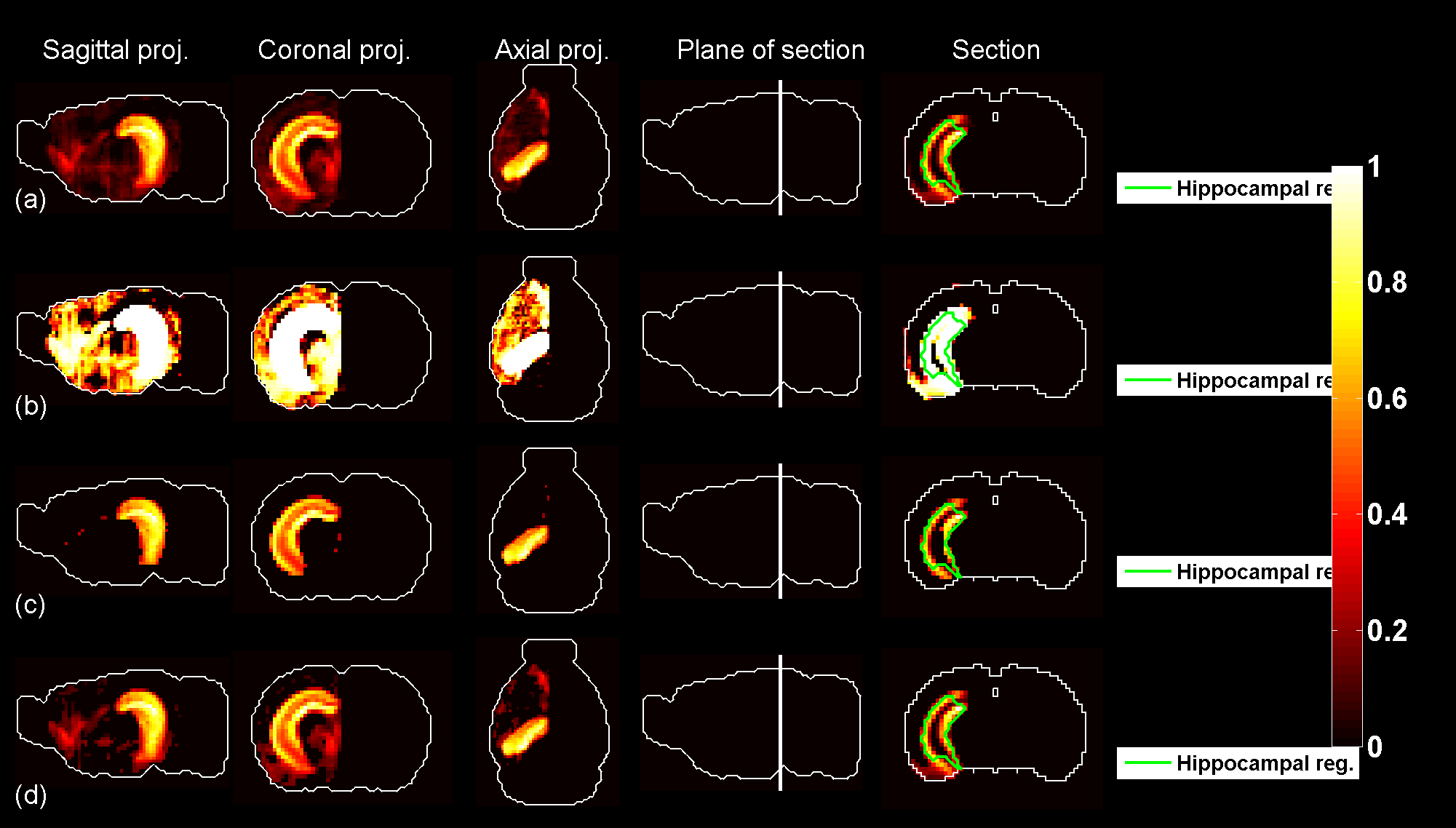}
\caption{Predicted profile, probability profile and thresholded profiles for $t=49$.}
\label{subSampleSplit49}
\end{figure}
\clearpage
\begin{figure}
\includegraphics[width=1\textwidth,keepaspectratio]{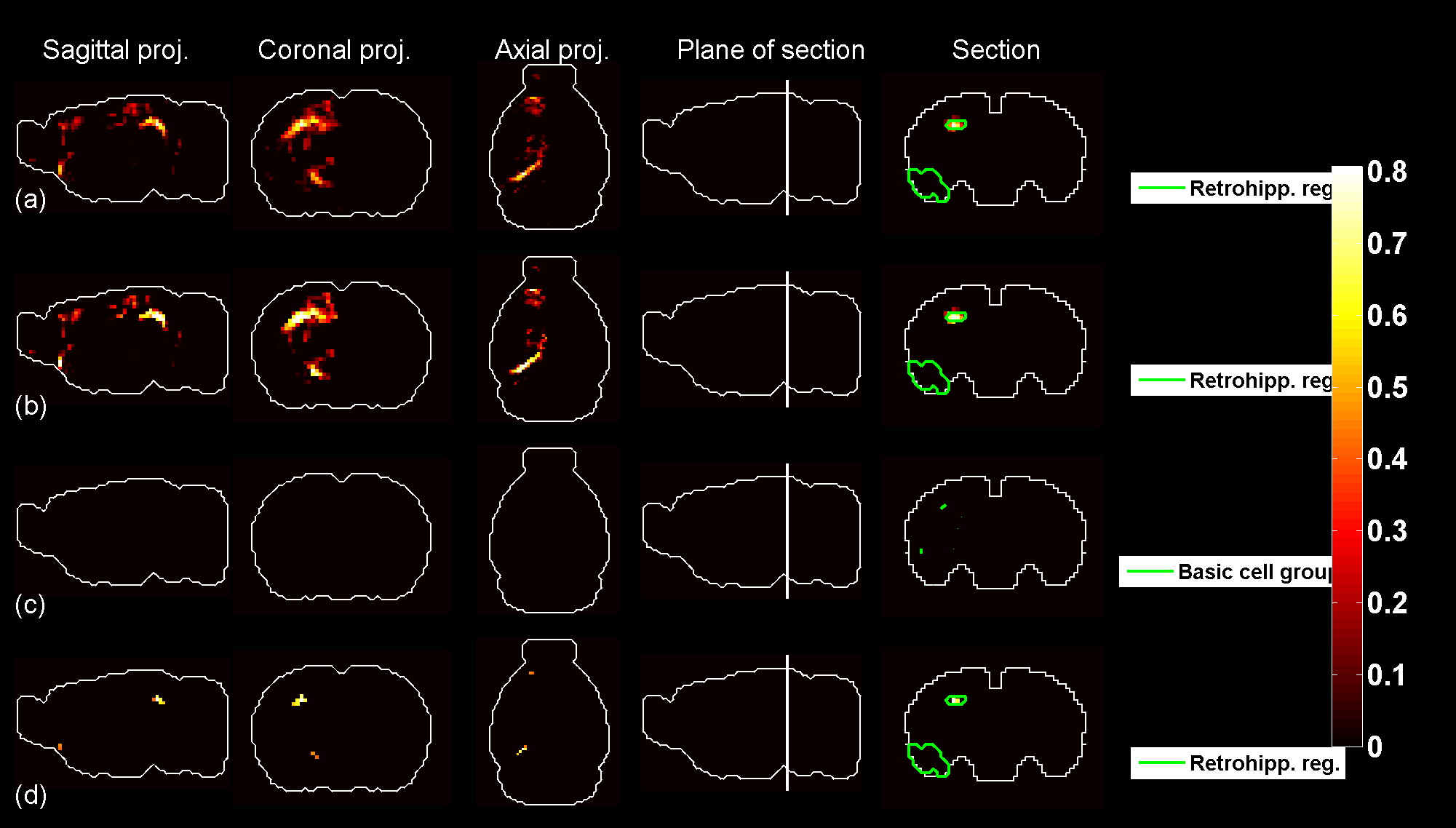}
\caption{Predicted profile, probability profile and thresholded profiles for $t=50$.}
\label{subSampleSplit50}
\end{figure}
\clearpage
\begin{figure}
\includegraphics[width=1\textwidth,keepaspectratio]{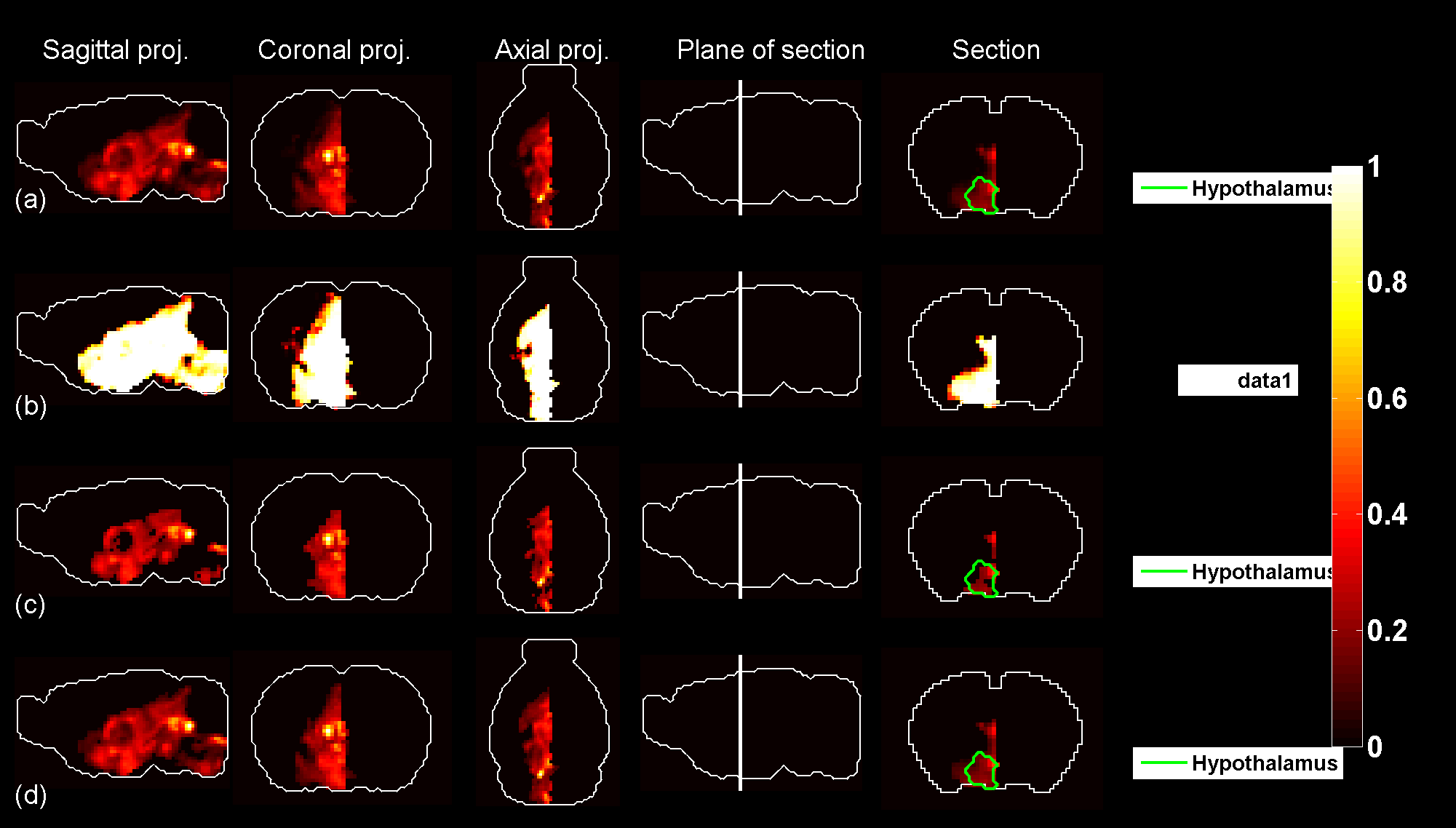}
\caption{Predicted profile, probability profile and thresholded profiles for $t=51$.}
\label{subSampleSplit51}
\end{figure}
\clearpage
\begin{figure}
\includegraphics[width=1\textwidth,keepaspectratio]{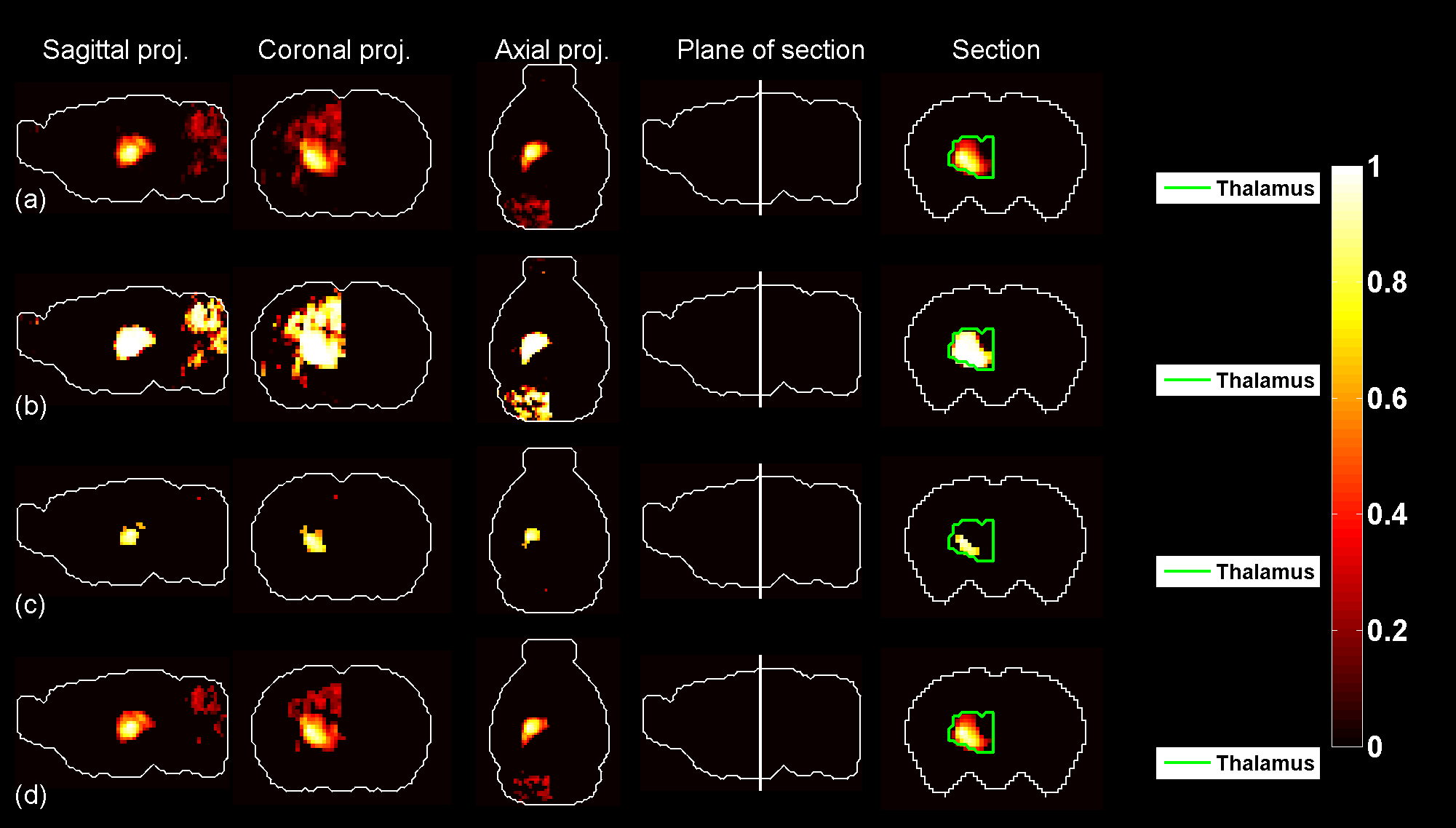}
\caption{Predicted profile, probability profile and thresholded profiles for $t=52$.}
\label{subSampleSplit52}
\end{figure}
\clearpage
\begin{figure}
\includegraphics[width=1\textwidth,keepaspectratio]{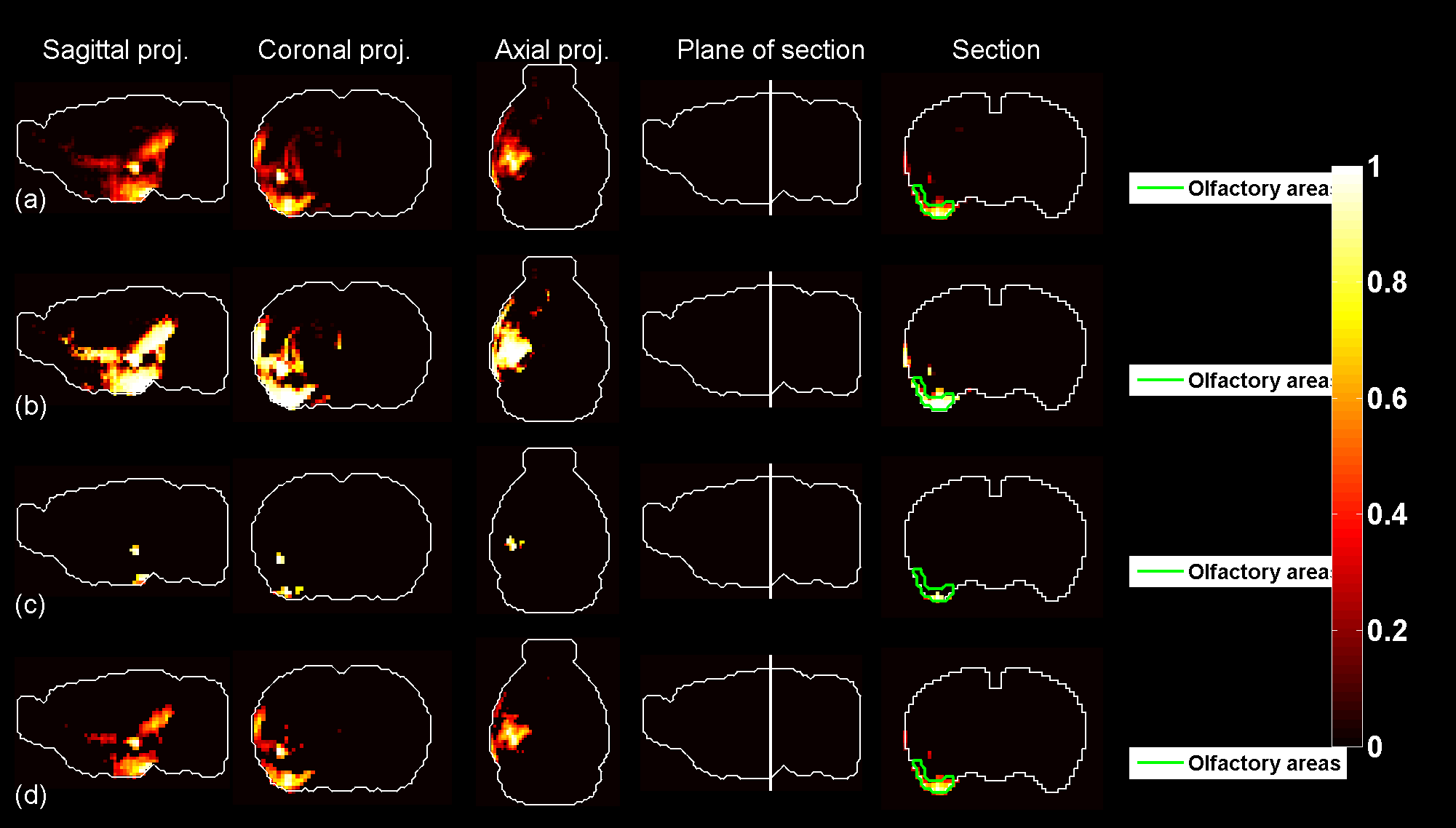}
\caption{Predicted profile, probability profile and thresholded profiles for $t=53$.}
\label{subSampleSplit53}
\end{figure}
\clearpage
\begin{figure}
\includegraphics[width=1\textwidth,keepaspectratio]{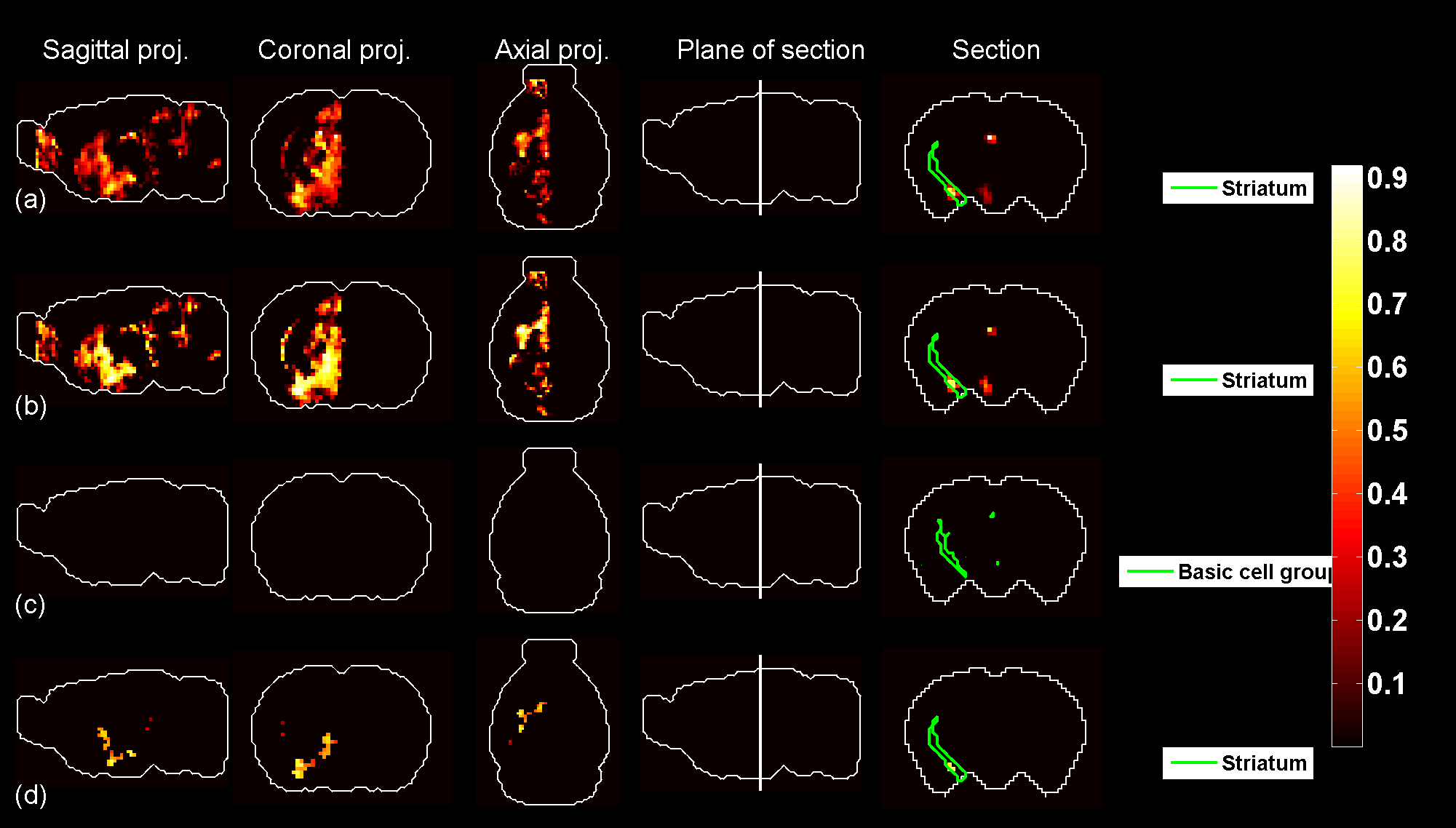}
\caption{Predicted profile, probability profile and thresholded profiles for $t=54$.}
\label{subSampleSplit54}
\end{figure}
\clearpage
\begin{figure}
\includegraphics[width=1\textwidth,keepaspectratio]{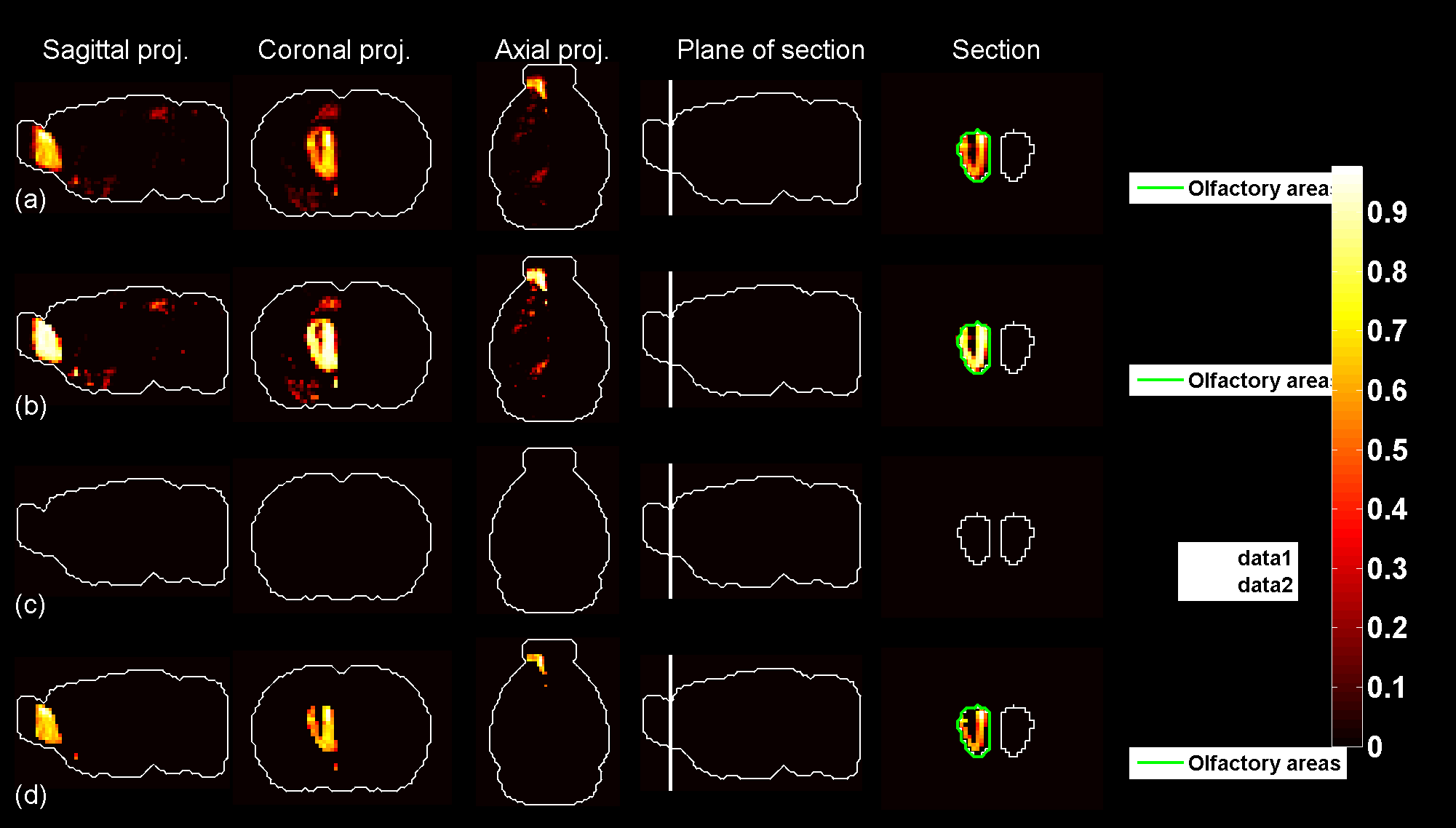}
\caption{Predicted profile, probability profile and thresholded profiles for $t=55$.}
\label{subSampleSplit55}
\end{figure}
\clearpage
\begin{figure}
\includegraphics[width=1\textwidth,keepaspectratio]{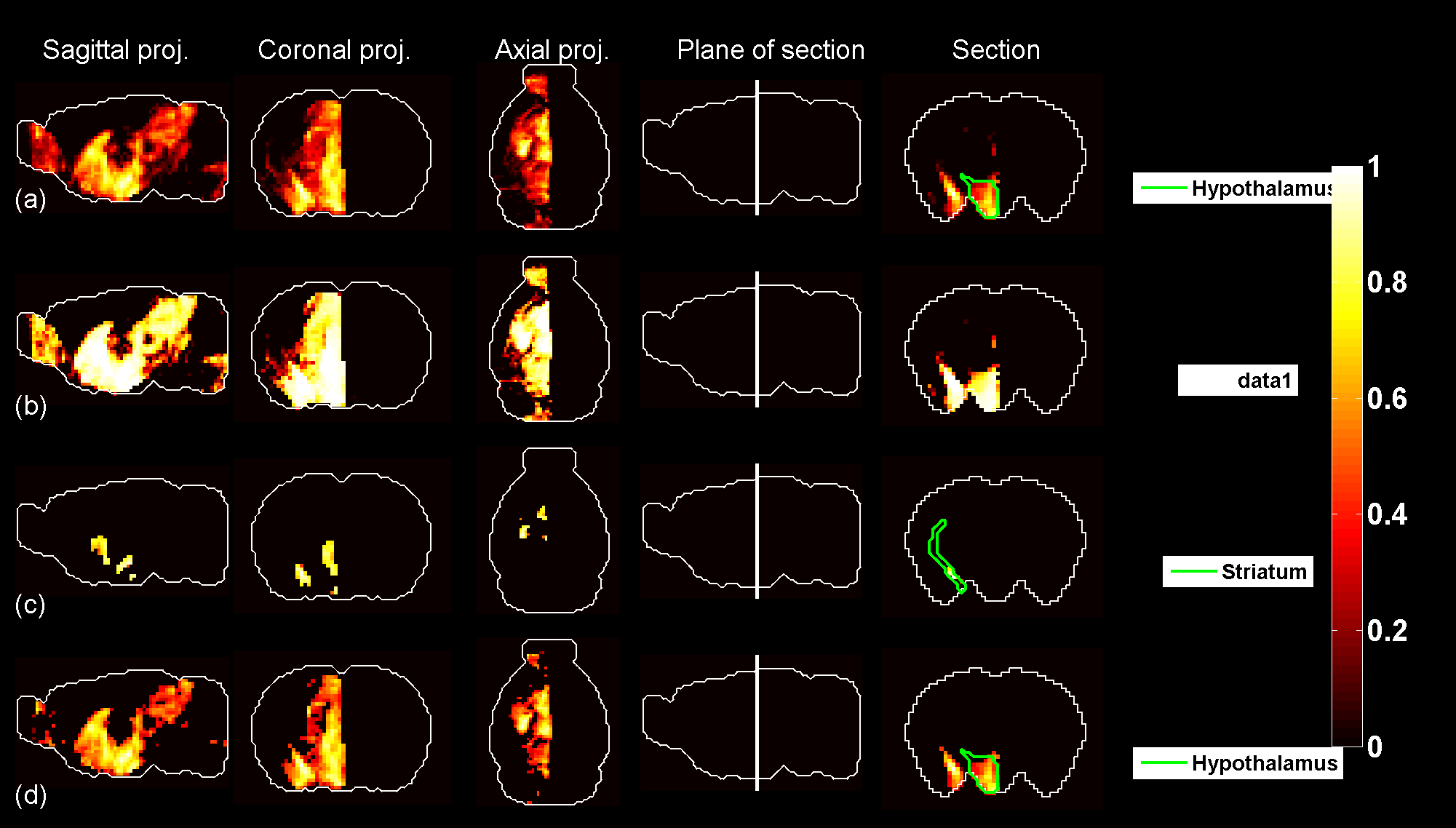}
\caption{Predicted profile, probability profile and thresholded profiles for $t=56$.}
\label{subSampleSplit56}
\end{figure}
\clearpage
\begin{figure}
\includegraphics[width=1\textwidth,keepaspectratio]{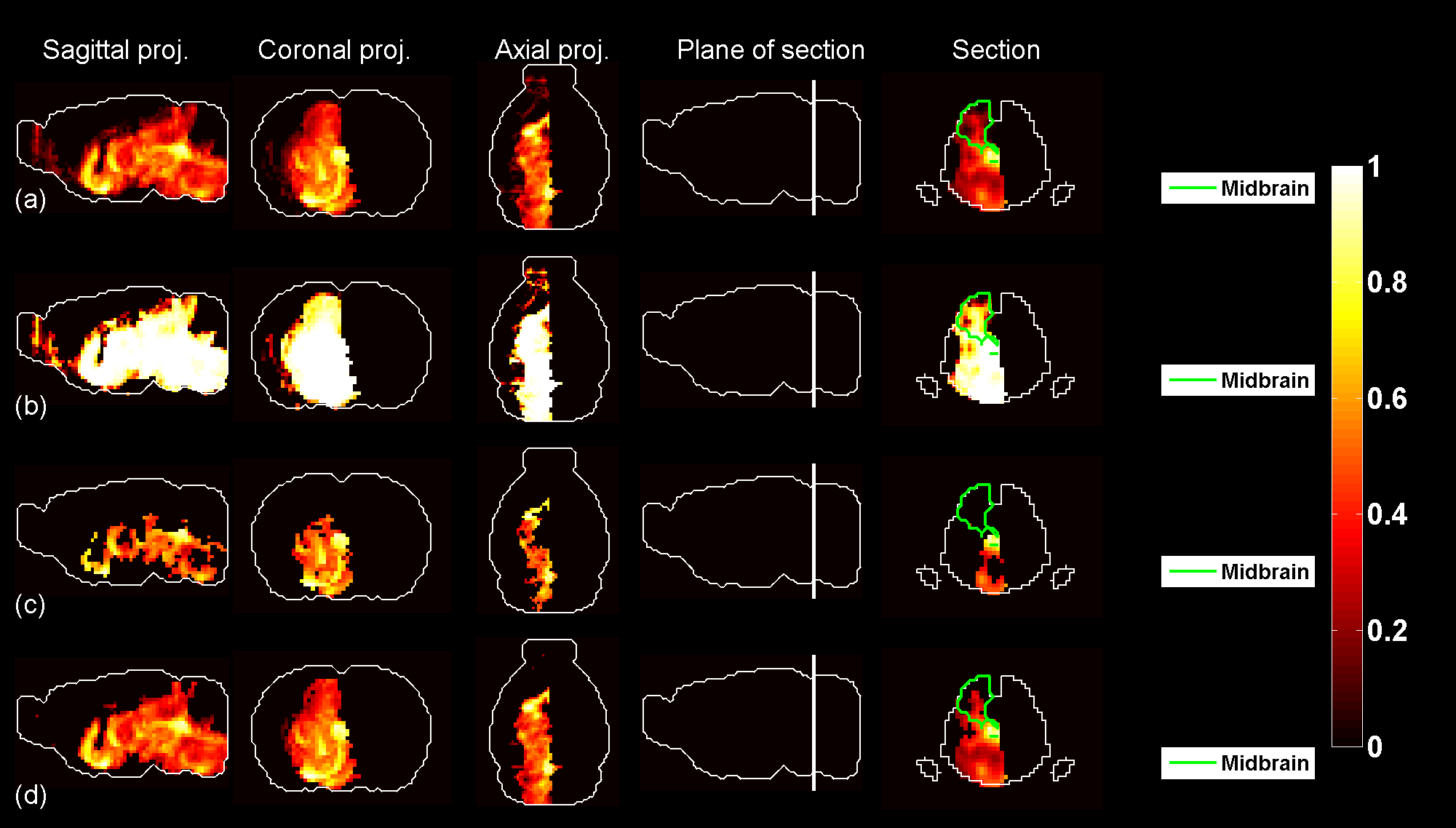}
\caption{Predicted profile, probability profile and thresholded profiles for $t=57$.}
\label{subSampleSplit57}
\end{figure}
\clearpage
\begin{figure}
\includegraphics[width=1\textwidth,keepaspectratio]{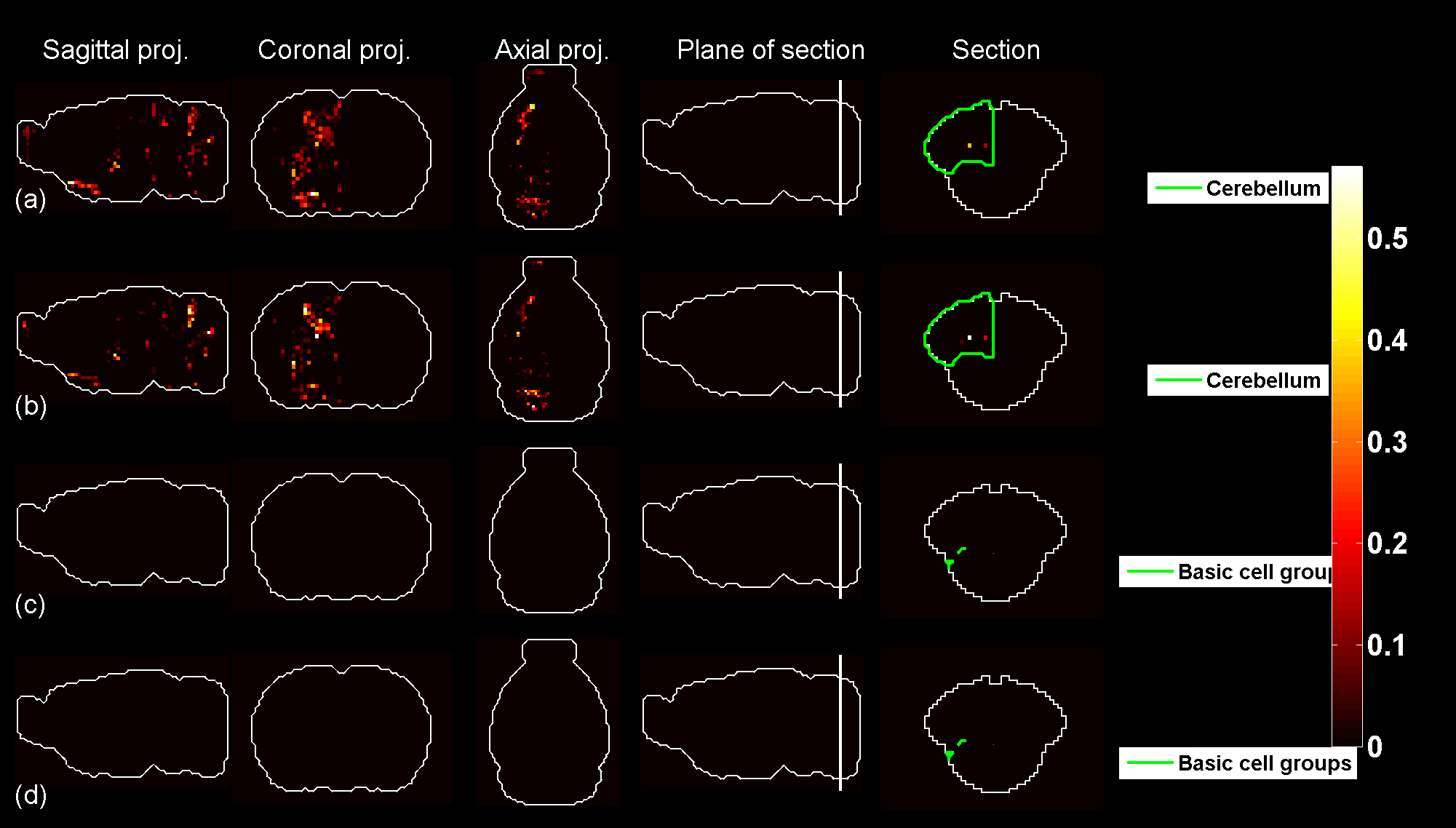}
\caption{Predicted profile, probability profile and thresholded profiles for $t=58$.}
\label{subSampleSplit58}
\end{figure}
\clearpage
\begin{figure}
\includegraphics[width=1\textwidth,keepaspectratio]{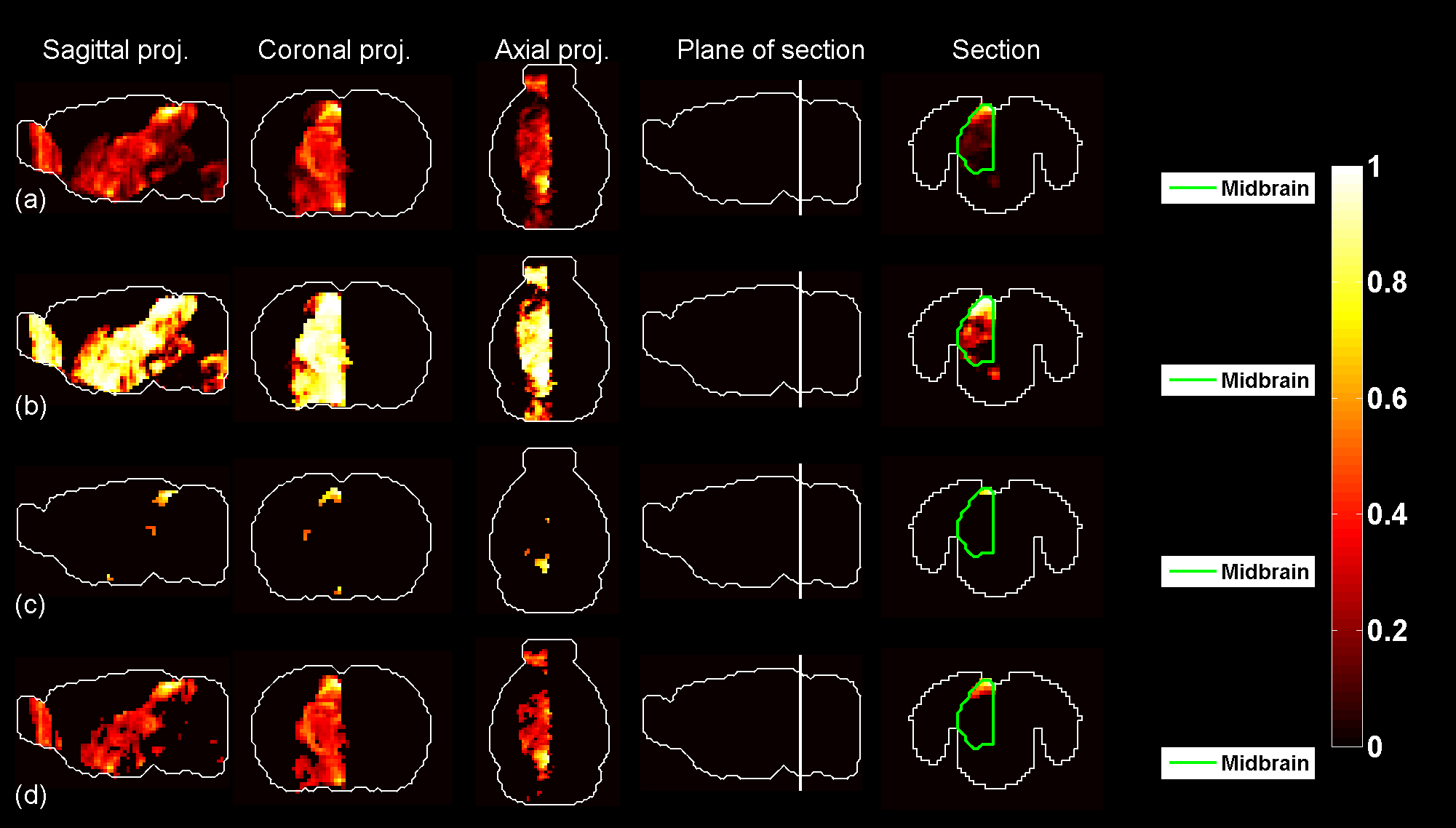}
\caption{Predicted profile, probability profile and thresholded profiles for $t=59$.}
\label{subSampleSplit59}
\end{figure}
\clearpage
\begin{figure}
\includegraphics[width=1\textwidth,keepaspectratio]{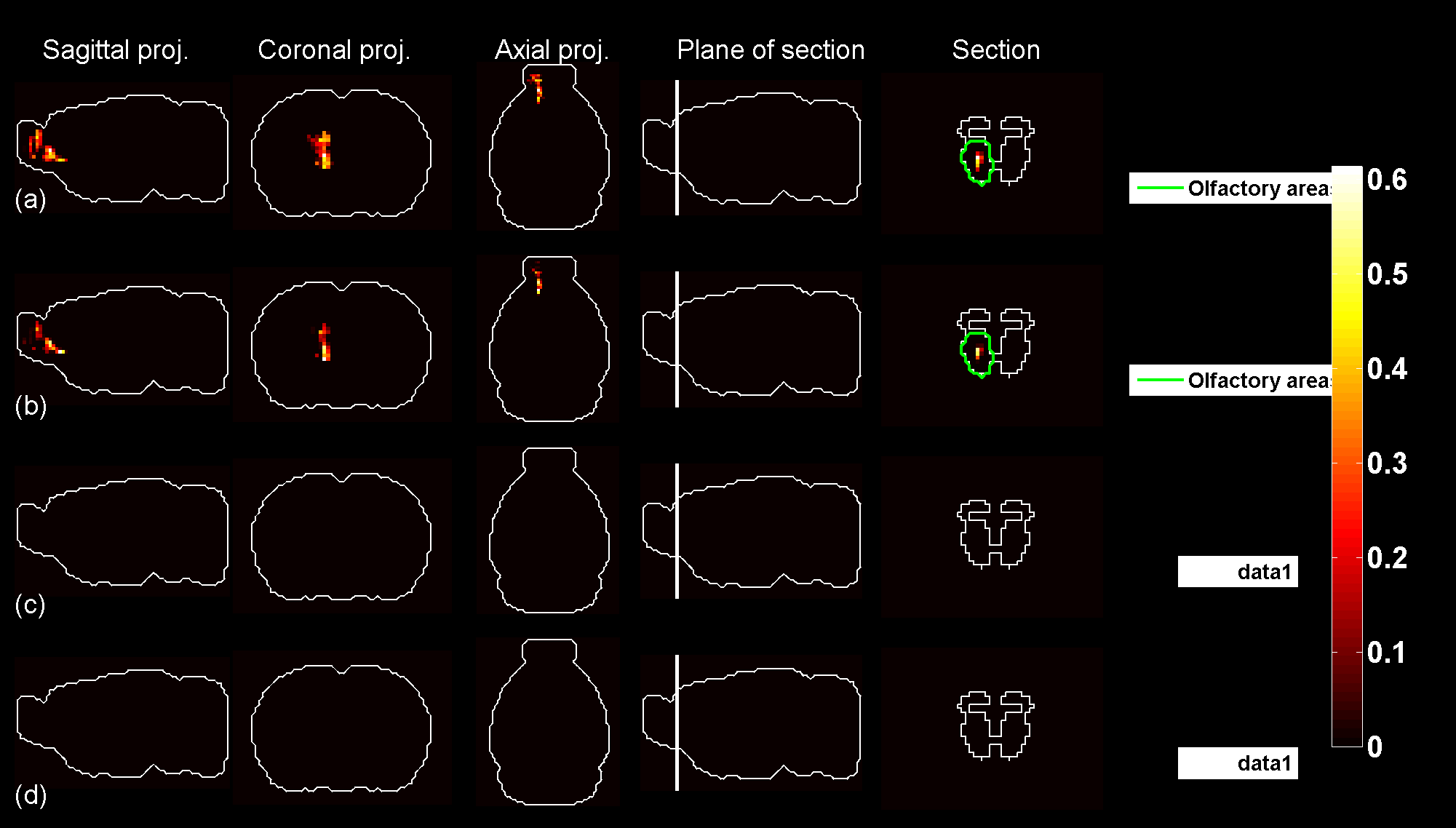}
\caption{Predicted profile, probability profile and thresholded profiles for $t=60$.}
\label{subSampleSplit60}
\end{figure}
\clearpage
\begin{figure}
\includegraphics[width=1\textwidth,keepaspectratio]{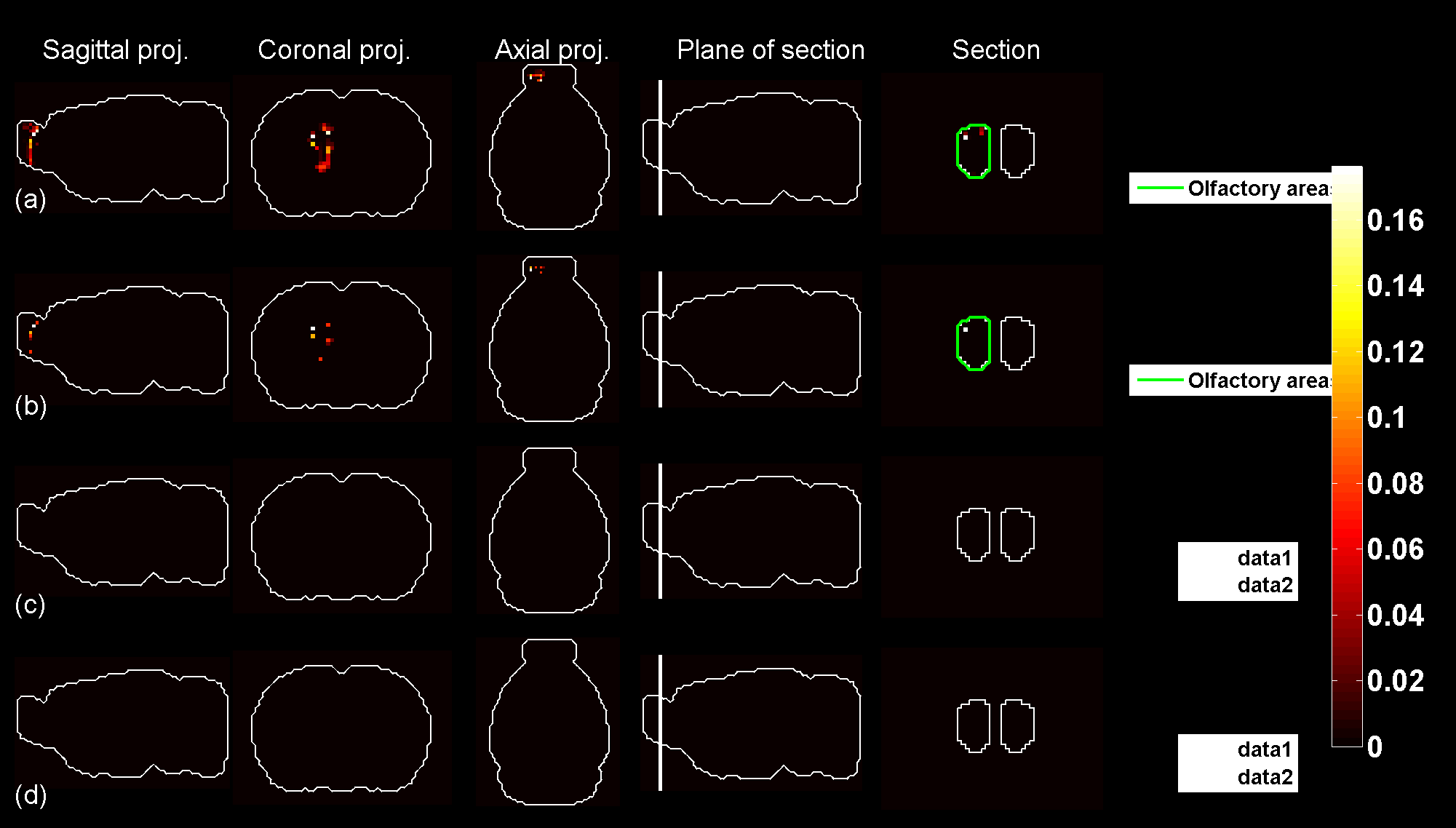}
\caption{Predicted profile, probability profile and thresholded profiles for $t=61$.}
\label{subSampleSplit61}
\end{figure}
\clearpage
\begin{figure}
\includegraphics[width=1\textwidth,keepaspectratio]{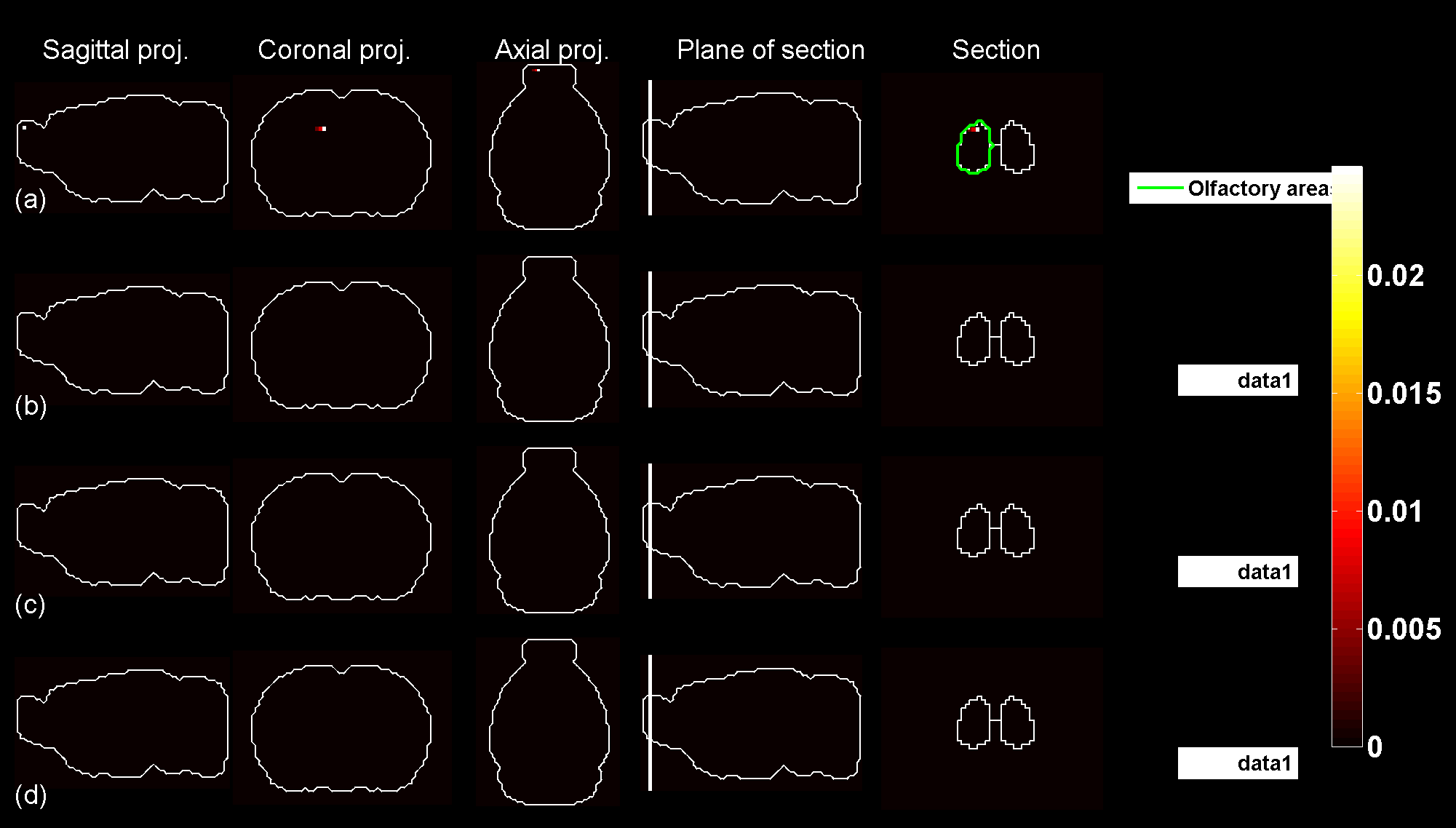}
\caption{Predicted profile, probability profile and thresholded profiles for $t=62$.}
\label{subSampleSpli652}
\end{figure}
\clearpage
\begin{figure}
\includegraphics[width=1\textwidth,keepaspectratio]{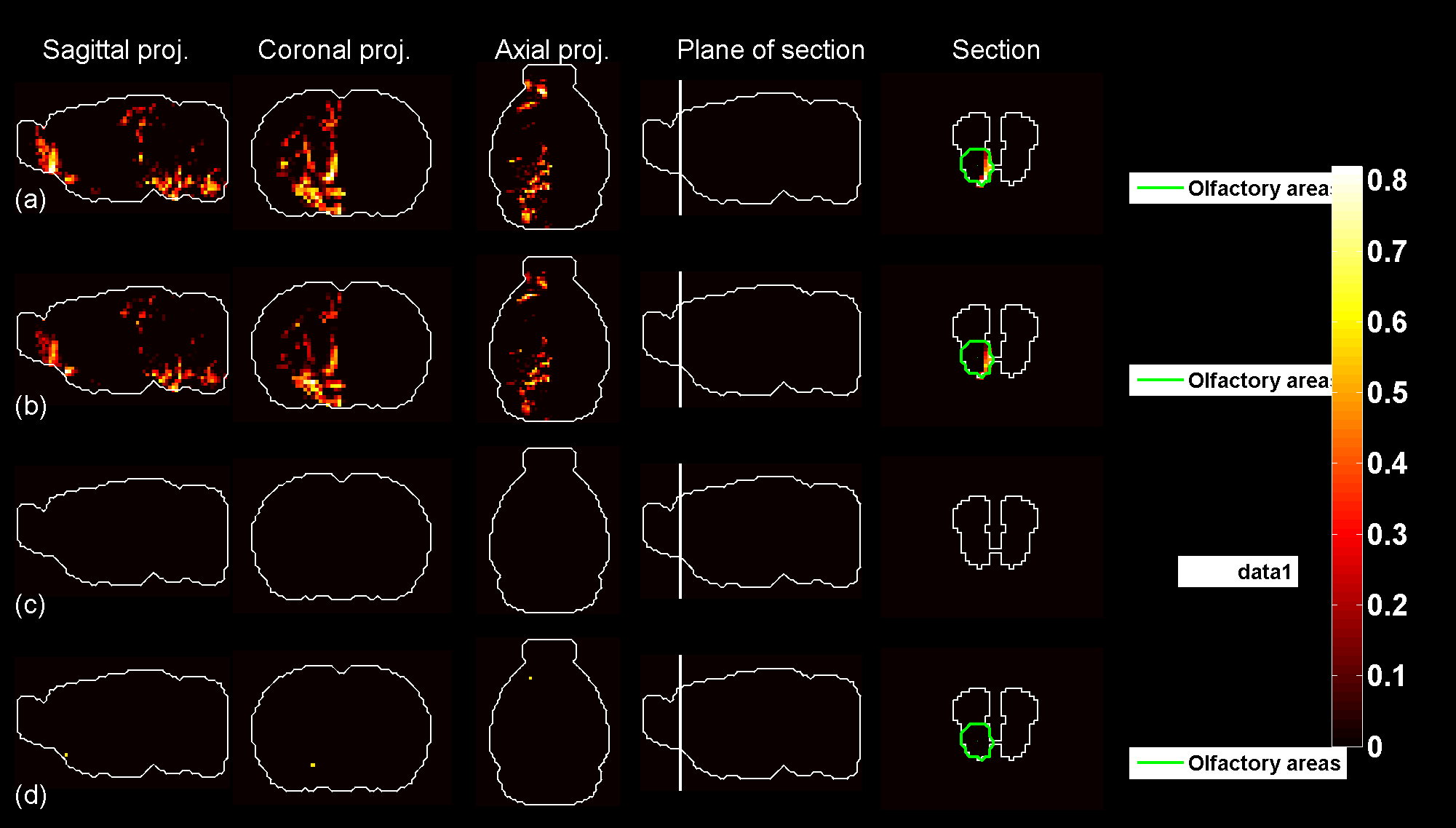}
\caption{Predicted profile, probability profile and thresholded profiles for $t=63$.}
\label{subSampleSplit63}
\end{figure}
\clearpage
\begin{figure}
\includegraphics[width=1\textwidth,keepaspectratio]{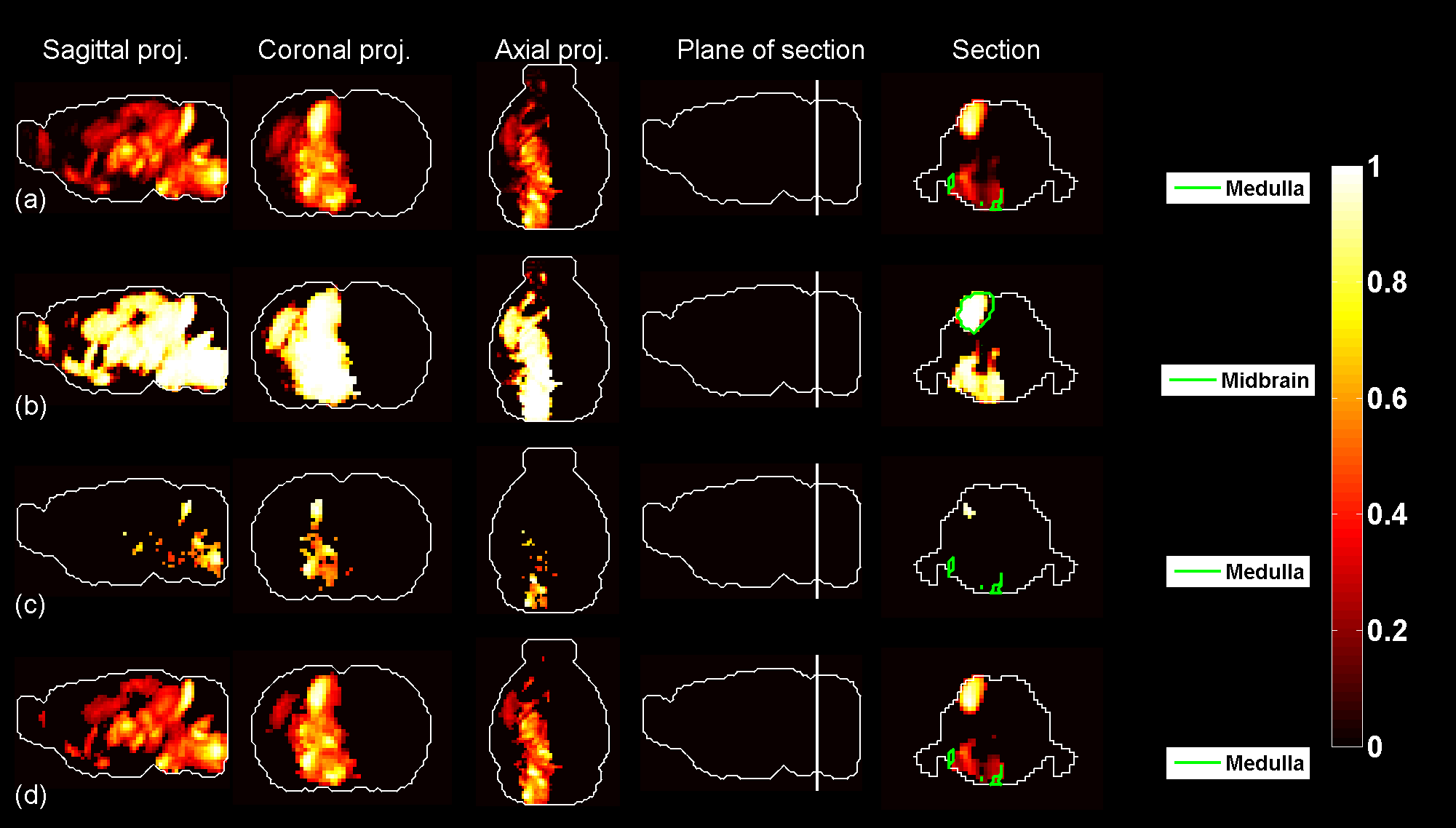}
\caption{Predicted profile, probability profile and thresholded profiles for $t=64$.}
\label{subSampleSplit64}
\end{figure}

%% file: preprintSecond.bbl
\begin{thebibliography}{1}

\bibitem{preprintFirstAnalysis} P. Grange, M. Hawrylycz, P.P. Mitra, {\emph{Cell-type-specific microarray data and the Allen atlas:
quantitative analysis of brain-wide patterns of correlation and density}}, {\ttfamily{[arXiv:1303.0013]}}.

\bibitem{GeneToBrain} M. Bota, H.-W. Dong, L.W. Swanson, {\emph{From gene networks to brain networks}}, Nature neuroscience (2003) {\bf{6}} (8), 795--9.
         
\bibitem{AllenGenome}
 E.S. Lein, M. Hawrylycz, N. Ao, M. Ayres, A. Bensinger, A. Bernard, A.F. Boe,
M.S. Boguski, K.S. Brockway, E.J. Byrnes, L. Chen, L. Chen, T.M. Chen, M.C. Chin, 
J. Chong, B.E. Crook, A. Czaplinska, C.N. Dang, S. Datta, N.R. Dee, {\emph{et al.}},
 {\emph{Genome-wide atlas of gene expression in the adult mouse brain.}}
Nature {\bf{445}}, 168--176 (2007).


\bibitem{images}  L. Ng, M. Hawrylycz, D. Haynor, {\emph{Automated high-throughput registration for localizing 3D mouse brain gene expression using ITK}}, Insight-Journal (2005).

\bibitem{BrainAtlasInsights} S.M. Sunkin and J.G. Hohmann,
  {\emph{Insights from spatially mapped gene expression in the mouse
      brain}}, Human Molecular Genetics, 2007, Vol. 16, Review Issue 2.



\bibitem{neufoAllen} L. Ng, S.D. Pathak, C. Kuan, C. Lau, H. Dong, A. Sodt, C. Dang,
 B. Avants, P. Yushkevich, J.C. Gee, D. Haynor, E. Lein, A. Jones and M. Hawrylycz,
{\emph{Neuroinformatics for genome-wide 3D gene expression mapping in the mouse brain}},
IEEE/ACM Trans. Comput. Biol. Bioinform. (2007), Jul-Sep {\bf{4(3)}} 382--93.


\bibitem{digitalAtlasing} M. Hawrylycz, R.A. Baldock, A. Burger, T. Hashikawa, G.A. Johnson, M. Martone, 
L. Ng, C. Lau, S.D. Larsen, J. Nissanov, L. Puelles, S. Ruffins, F. Verbeek, I. Zaslavsky1, J. Boline, {\emph{Digital Atlasing and Standardization in the Mouse Brain}}, PLoS  Computational Biology {\bf{7}} (2) (2011).


\bibitem{AllenFiveYears} A.R. Jones, C.C. Overly and S.M. Sunkin,
  {\emph{The Allen Brain Atlas: 5 years and beyond}}, Nature Reviews
  (Neuroscience), Volume {\bf{10}} (November 2009), {\bf{1}}.


\bibitem{corrStructureAllen} M. Hawrylycz, L. Ng, D. Page, J. Morris, C. Lau, S. Faber, V. Faber, S. Sunkin,
 V. Menon, E.S. Lein, A. Jones, {\emph{Multi-scale correlation structure of gene expression in the brain}}, Neural Networks {\bf{24}} (2011) 933--942.


\bibitem{AllenAtlas} H.-W. Dong, \emph{The Allen reference atlas: a
  digital brain atlas of the C57BL/6J male mouse}, Wiley, 2007.



\bibitem{AllenAtlasMol}
L. Ng, A. Bernard, C. Lau, C.C. Overly, H.-W. Dong, C. Kuan, S. Pathak, S.M. Sunkin, C. Dang, J.W. Bohland, H. Bokil, P.P. Mitra, L. Puelles, J. Hohmann, D.J. Anderson, E.S. Lein, A.R. Jones, M. Hawrylycz,
{\emph{An anatomic gene expression atlas of the adult mouse brain}},
Nature Neuroscience {\bf{12}}, 356--362 (2009).



\bibitem{NeuroBlast} L. Ng {\emph{et al.}}, {\emph{NeuroBlast: a 3D spatial homology search tool for gene
 expression}}, BMC Neuroscience 2007, {\bf{8}}(Suppl 2):P11.


\bibitem{toolboxManual} P. Grange, J.W. Bohland, M. Hawrylycz and
  P.P. Mitra, {\emph{Brain Gene Expression Analysis: a MATLAB toolbox
      for the analysis of brain-wide gene-expression data}},
  {\ttfamily{[arXiv:1211.6177 [q-bio.QM]]}}.


\bibitem{BGEA} The {\ttfamily{Brain Gene Expression Atlas}}  MATLAB toolbox
 is downloadable from {\ttfamily{http://brainarchitecture.org/allen-atlas-brain-toolbox}}. 



\bibitem{qbCoExpression} P. Grange, M. Hawrylycz, P.P. Mitra, {\emph{Computational neuroanatomy and co-
expression of genes in the adult mouse brain, analysis tools for the Allen Brain Atlas}},
Quantitative Biology (2013), in press, {\ttfamily{[arXiv:1301.1730 [q-bio.QM]]}}.


\bibitem{markerGenes} P. Grange, P.P. Mitra, {\emph{Computational neuroanatomy and gene expression: optimal sets of marker genes for brain regions}}, IEEE, in CISS 2012, 46th annual conference on Information Science and Systems (Princeton).




\bibitem{OkatyCells} B.W. Okaty, M.N. Miller, K. Sugino, C.M. Hempel, S.B. Nelson,
{\emph{Transcriptional and electrophysiological maturation of neocortical fast-spiking GABAergic interneurons}}, J. Neurosci. (2009) {\bf{29(21)}} 7040-52.


\bibitem{RossnerCells} M.J. Rossner, J. Hirrlinger, S.P. Wichert, C. Boehm, D. Newrzella, H. Hiemisch, G. Eisenhardt, C. Stuenkel, O. von Ahsen, K.A. Nave, {\emph{Global transcriptome analysis of genetically identified neurons in the adult cortex}},
J. Neurosci. 2006 {\bf{26(39)}} 9956-66.



\bibitem{CahoyCells} J.D. Cahoy, B. Emery, A. Kaushal, L.C. Foo,
  J.L. Zamanian, K.S. Christopherson, Y. Xing, 
  J.L. Lubischer, P.A. Krieg,
  S.A. Krupenko, W.J. Thompson, B.A. Barres, {\emph{A transcriptome database
      for astrocytes, neurons, and oligodendrocytes: a new resource
      for understanding brain development and function}},
  J. Neurosci. 2008 {\bf{28(1)}} 264-78.




\bibitem{DoyleCells} J.P. Doyle, J.D. Dougherty, M. Heiman, E.F. Schmidt, T.R. Stevens, G. Ma, S. Bupp,
 P. Shrestha, R.D. Shah, M.L. Doughty, S. Gong, P. Greengard, N. Heintz, 
{\emph{Application of a translational profiling approach for the comparative
analysis of CNS cell types}},
Cell (2008) {\bf{135(4)}} 749-62. 


\bibitem{ChungCells} C.Y. Chung, H. Seo, K.C. Sonntag, A. Brooks, L. Lin, O. Isacson
 {\emph{Cell type-specific gene expression of midbrain dopaminergic neurons
reveals molecules involved in their vulnerability and protection}}. Hum. Mol. Genet. (2005)
{\bf{14}}: 1709--1725.


\bibitem{ArlottaCells} P. Arlotta, B.J. Molyneaux, J. Chen, J. Inoue,
 R. Kominami {\emph{et al.}} (2005) {\emph{Neuronal subtype-specific genes
 that control corticospinal motor neuron development in vivo}},
 Neuron {\bf{45}}: 207--221.


\bibitem{HeimanCells} M. Heiman, A. Schaefer, S. Gong,  Peterson JD, Day M, Ramsey KE, Suárez-Fariñas M, Schwarz C, Stephan DA, Surmeier DJ, P. Greengard, N. Heintz, (2008)
{\emph{A translational profiling approach for the molecular characterization of
of CNS cell types}}, Cell {\bf{135}}: 738--748.


\bibitem{foreBrainTaxonomy} K. Sugino, C.M. Hempel, M.N. Miller, A.M. Hattox, P. Shapiro, C. Wu, Z.J. Huang,
 S.B. Nelson, {\emph{Molecular taxonomy of major neuronal classes in the adult mouse forebrain}},
 Nature Neuroscience {\bf{9}}, 99-107 (2005). 


 
\bibitem{OkatyComparison} B.W. Okaty, K. Sugino, S.B. Nelson,
{\emph{A Quantitative Comparison of Cell-Type-Specific Microarray
 Gene Expression Profiling Methods in the Mouse Brain}}, PLoS One (2011) {\bf{6(1)}}.



\bibitem{ISHVsMicroarray} C.K. Lee, S.M. Sunkin, C.C. Kuan, C.L. Thompson, S. Pathak,
  L. Ng, C. Lau, S. Fischer, M. Mortrud, C. Slaughterbeck, A. Jones, E. Lein, M. Hawrylycz,
 {\emph{Quantitative methods for genome-scale analysis of in situ hybridization and correlation with microarray data}}, Genome Biol. (2008); {\bf{9(1)}}: R23.

\bibitem{cvxLink} Research, Inc. CVX: {\emph{Matlab software for disciplined convex programming}}, version 2.0 beta. {\ttfamily{http://cvxr.com/cvx, September 2012}}.

\bibitem{cvxLecture}  M. Grant and S. Boyd, {\emph{Graph implementations for nonsmooth convex programs, Recent Advances in Learning and Control (a tribute to M. Vidyasagar)}}, V. Blondel, S. Boyd, and H. Kimura, editors, pages 95-110, Lecture Notes in Control and Information Sciences, Springer, 2008. {\ttfamily{http://stanford.edu/$\sim$boyd/graph$\_$dcp.html}}.


\bibitem{concordance} J.W. Bohland, H. Bokil, C.B. Allen, P.P. Mitra, {\emph{The Brain Atlas Concordance Problem: Quantitative Comparison of Anatomical Parcellations}}, PLoS ONE (2009).


\bibitem{methodsPaper}
J.W. Bohland, H. Bokil, C.-K. Lee, L. Ng, C. Lau, C. Kuan, M. Hawrylycz, P.P. Mitra,
{\emph{Clustering of spatial gene expression patterns in the mouse brain and comparison with classical neuroanatomy}},
Methods, Volume {\bf{50}}, Issue 2, February 2010, Pages 105-112.


\bibitem{MenasheCoExpr} I. Menashe, P. Grange, E.C. Larsen,
  S. Banerjee-Basu and P.P. Mitra, {\emph{Co-expression profiling of
      autism genes in the mouse brain}}, PLoS Comput. Biol. 9(7): e1003128.


















\bibitem{Meinshausen2013} 
N. Meinshausen (2013). Sign-constrained least squares estimation for high-dimensional regression. Electronic Journal of Statistics, 7, 1607--1631.

\end{thebibliography}
